\def\versionno{ fuRs8 -- version 17.9 -- by ir -- 15.12.03   }

\catcode`\@=11
\newif\if@fewtab\@fewtabtrue
{\count255=\time\divide\count255 by 60
\xdef\hourmin{\number\count255}
\multiply\count255 by-60\advance\count255 by\time
\xdef\hourmin{\hourmin:\ifnum\count255<10 0\fi\the\count255}}
\def\ps@draft{\let\@mkboth\@gobbletwo
    \def\@oddfoot{\hbox to 7 cm{\tiny \versionno
       \hfil}\hskip -7cm\hfil\rm\thepage \hfil {\tiny\draftdate}}
    \def\@oddhead{}
    \def\@evenhead{}\let\@evenfoot\@oddfoot}
\def\draftdate{\number\month/\number\day/\number\year\ \ \ \hourmin }

\def\citen#1{\if@filesw \immediate\write \@auxout {\string\citation{#1}}\fi%
\@tempcntb\m@ne \let\@h@ld\relax \def\@citea{}%
\@for \@citeb:=#1\do {\@ifundefined {b@\@citeb}%
    {\@h@ld\@citea\@tempcntb\m@ne{\bf ?}%
    \@warning {Citation `\@citeb ' on page \thepage \space undefined}}%
    {\@tempcnta\@tempcntb \advance\@tempcnta\@ne
    \setbox\z@\hbox\bgroup\ifcat0\csname b@\@citeb \endcsname \relax
    \egroup \@tempcntb\number\csname b@\@citeb \endcsname \relax
    \else \egroup \@tempcntb\m@ne \fi \ifnum\@tempcnta=\@tempcntb
    \ifx\@h@ld\relax \edef \@h@ld{\@citea\csname b@\@citeb\endcsname}%
    \else \edef\@h@ld{\hbox{--}\penalty\@highpenalty
    \csname b@\@citeb\endcsname}\fi
    \else \@h@ld\@citea\csname b@\@citeb \endcsname \let\@h@ld\relax \fi}%
\def\@citea{,\penalty\@highpenalty\hskip.13em plus.13em minus.13em}}\@h@ld}
\def\@citex[#1]#2{\@cite{\citen{#2}}{#1}}%
\def\@cite#1#2{\leavevmode\unskip\ifnum\lastpenalty=\z@\penalty\@highpenalty\fi%
  \ [{\multiply\@highpenalty 3 #1%
  \if@tempswa,\penalty\@highpenalty\ #2\fi}]}   %
\makeatother 
\catcode`\@=12

\newcounter{defthm}

\newcommand\Ai[1]  {\langle #1{,}A\rangle}

\def\alg           {algebra}

\newcommand\apppicture[2] {\put(#2,0) {\begin{picture}(0,0)(0,0)
                          \scalebox{.29}{\includegraphics{a3-#1.eps}} \end{picture}}}
\def\Atop          {\mbox{$A_{\rm top}$}}
\def\ATop          {{A_{\rm top}}}

\def\bc            {boundary condition}

\def\be            {\begin{equation}}
\def\bea           {\begin{equation}\begin{array}l}
\def\bearl         {\begin{array}{l}}
\def\bearll        {\begin{array}{ll}}

\def\boxelement    {reversing move}
\def\bs            {boundary state}

\def\calc          {\mbox{$\mathcal C$}}
\def\calca         {\mbox{${\mathcal C}_{\!A}$}}

\def\calh          {\mbox{$\mathcal H$}}
\def\calH          {{\mathcal H}}

\def\calm          {\mbox{$\mathcal M$}}
\def\cats          {categories}

\def\cft           {conformal field theory}

\def\cfts          {conformal field theories}
\def\Cfts          {Conformal field theories}
\def\chii          {\raisebox{.15em}{$\chi$}}
\def\chiI          {\raisebox{.15em}{$\scriptstyle\chi$}}
\def\cir           {\,{\circ}\,}

\def\complex       {\mathbbm C}
\def\con           {conformal }
\def\Con           {Conformal }
\def\corfu         {correlation function}
\def\Corfu         {Correlation function}

\def\dimA          {{\rm dim}(A)}
\def\dimc          {{\rm dim}_{\scriptscriptstyle\complex}}
\def\dimM          {{\rm dim}(\dot M)}

\def\dim           {{\rm dim}}
\def\dsty          {\displaystyle }

\def\eE            {{\rm e}}
\def\ee            {\end{equation}}
\def\eear          {\end{array}}

\def\End           {{\rm End}}
\newcommand\epicture[2] {\end{picture}\\{}\\[#1.#2em]\end{array}}
\def\eps           {\varepsilon}
\def\Eps           {\epsilon}

\def\eq            {\,{=}\,}

\newcommand\erf[1] {(\ref{#1})}

\def\FF            {{\sf F}}
\def\findim        {fini\-te-di\-men\-si\-o\-nal}
\newcommand\foodnode[1] {\,\footnote{~#1}}
\newcommand\Frac[2]{\mbox{\large$\frac{#1}{#2}$}}
\newcommand\Fs[6]  {{\sf F}_{\,{#5}\,{#6}}^{\,({#1}\,{#2}\,{#3})\,{#4}}}
\def\fsi           {Fro\-be\-ni\-us\hy Schur indicator}

\newcommand\goodnode[1] { }

\newcommand\glue[1]{\stackrel{#1}{\leftarrowtail}}
\def\GG            {{\sf G}}
\newcommand\Gs[6]  {{\sf G}_{\,{#5}\,{#6}}^{\,({#1}\,{#2}\,{#3})\,{#4}}}
\def\Gz            {Generalised }
\def\Hom           {{\rm Hom}}
\def\HomA          {{\rm Hom}_{\!A}}

\newcommand\hsp[1] {\mbox{\hspace{#1 em}}}
\def\hy            {$\mbox{-\hspace{-.66 mm}-}$}
\def\I             {\mbox{$\II$}}

\def\id            {\mbox{\sl id}}
\def\idsmall       {\mbox{\scriptsize\sl id}}

\def\ii            {{\rm i}}
\def\II            {\mathcal I}

\def\iN            {\,{\in}\,}
\def\In            {\prec}
\def\IN            {\,{\In}\,}
\newcommand\includeourbeautifulpicture[1] {{\begin{picture}(0,0)(0,0)
                   \scalebox{.38}{\includegraphics{#1.eps}} \end{picture}}}
\def\inda          {{\rm Ind}_{A\!}}

\newcommand{\Ising}{{\mathcal{C}_{\sf Is}}}
\def\J             {\mbox{$\JJ$}}
\def\JA            {Jandl algebra}
\def\JJ            {\mathcal J}
\def\K             {{\rm K}}
\def\kappai        {\kappa}
\def\kappaj        {{\kappa'}}

\newcommand\labl[1]{\label{#1}\ee}

\def\lhs           {left hand side}

\def\llb           {\mbox{\large[}}

\def\Loc           {{\rm Hom}_{\rm loc}}
\def\LocUp         {{\rm Hom}^{\rm loc}}
\newcommand\lr[2]  {\langle #1{,}#2\rangle}
\newcommand\lra[2] {\langle #1{,}#2\rangle_{\!A}^{}}
\def\lrb           {\mbox{\large]}}

\newcommand\m[7]   {m_{#1#2,#3#4}^{\;#5#6;\,#7}}
\def\M             {{\dot M}}

\def\mocla         {\mbox{$\mathcal S$}}

\def\mtc           {modular tensor category}

\def\mtop          {{m_{\rm top}}}
\def\MX            {{\rm M}_X}

\newcommand\N[3]   {{N_{#1#2}}^{\!\!#3}}
\def\nxt           {\raisebox{.08em}{\rule{.44em}{.44em}}\hsp{.4}}
\def\Nxt           {\raisebox{.08em}{\rule{.44em}{.44em}}}

\def\objc          {{\mathcal O}bj(\calc)}

\def\one           {{\bf1}}

\def\onemat        {{\mbox{\small1}\!\!{1}}}

\def\otA           {\,{\otimes}_{\!A}^{}\,}
\def\OtA           {{\otimes}_{\!A}^{}}
\def\otB           {\,{\otimes}_{B}^{}\,}
\def\OtB           {{\otimes}_{B}^{}}
\def\oti           {\,{\otimes}\,}
\def\Oti           {{\otimes}}
\def\parfu         {partition function}
\newcommand\ppmatrix[4]{\mbox{{\Large(}$\!\!\begin{array}{cc}{}\\[-1.8em]
                   \scs #1\!&\!\!\scs #2\\[-.39em]\scs #3\!&\!\!\scs #4
                   \\[-.19em]\eear\!\!${\Large)}}}
\newcommand\ppvector[2]{\mbox{{\Large(}$\!\!\begin{array}{cc}{}\\[-1.8em]
                   \!\scs #1\!\\[-.39em]\!\scs #2\!\\[-.19em]\eear\!\!${\Large)}}}
\def\q             {quantum }
\def\Q             {Quantum }
\def\qed           {\hfill\checkmark\medskip}
\def\QFT           {Quantum Field Theory}        

\def\r             {{\rho}}
\newcommand\R[5]   {{\sf R}^{(#1\,#2)#3}_{#4\,#5}}

\def\reals         {\mbox{$\mathbb R$}}
\def\rep           {representation}

\def\rhs           {right hand side}
\def\rmm           {{\rm m\ddot o}}
\def\rmM           {{\rm M\ddot o}}
\def\rmT           {{\rm T}}
\newcommand\RP     {\mathbb{RP}^3}
\def\RR            {{\sf R}} 
\newcommand\Rs[4]  {{\sf R}^{#1\,(#2\,#3)#4}}        

\newcommand\Rss[3] {{\sf R}^{(#1\,#2)#3}}
\def\scs           {\scriptstyle} 
\newcommand\sect[1]{\section{#1}\setcounter{equation}0\setcounter{defthm}0} 
\def\sigmaA        {\sigma_{\!A}}
\def\sigmaB        {\sigma_{\!B}}

\def\ssFA          {symmetric special Frobenius algebra}

\def\staralgebra   {algebra with reversion}
\def\staralgebras  {algebras with reversion}
 
\def\stt           {string theory}
\def\stts          {string theories}
\def\su            {\mathfrak{su}}
\def\sym           {symmetry} 
\def\syms          {sym\-me\-tries}

\def\tcs           {tensor categories}
\newcommand\tfrac[2]{\textstyle{\frac{#1}{#2}}}
\def\thetaA        {\theta_{\!A}}
\def\tft           {topological field theory}

\def\tfts          {topological field theories}

\def\top           {_{\rm top}}

\def\tr            {{\rm tr}\,}
\newcommand\Tr[1]  {{\rm tr}_{#1}\,}
\def\twodim        {two-di\-men\-si\-o\-nal}

\def\Vect          {{\mathcal V}\mbox{\sl ect}}

\newcommand\void[1]{}
\def\wrt           {with respect to }
\def\wzwm          {WZW model}
\def\X             {D}

\def\zet           {\mbox{$\mathbb Z$}}
\newcommand\orr    {{\rm or}}

\documentclass[12pt]{article}
 
\usepackage{latexsym,amssymb,amsfonts,epsf,color,colordvi}
\usepackage{bbm}

\usepackage[dvips]{graphics} %
\newtheorem{defthm}{\whattheorem}[section]
\newcommand\dt[1]  {\noindent\def\whattheorem{#1}\pagebreak[0]\begin{defthm}{}%
                   \samepage{$\!\!${\rm:}\nopagebreak\\[-1.91em]{}}\end{defthm}}
\newcommand\dtl[2] {\noindent\def\whattheorem{#1}\pagebreak[0]\begin{defthm}{}%
                   \samepage{$\!\!${\rm:}\label{#2}\nopagebreak\\[-1.91em]{}}%
                   \end{defthm}}

\begin{document}
\begin{flushright}  {~} \\[-12mm]
{\sf hep-th/0306164}\\[1mm]{\sf HU-EP-03/23}\\[1mm]
{\sf Hamburger$\;$Beitr\"age$\;$zur$\;$Mathematik$\;$Nr.$\;$177}\\[1mm]
{\sf June 2003} \end{flushright}
\begin{center} \vskip 13mm
{\Large\bf TFT CONSTRUCTION OF RCFT CORRELATORS}\\[4mm]
{\Large\bf II: UNORIENTED WORLD SHEETS}\\[16mm]
{\large 
J\"urgen Fuchs$\;^1$ \ \ \ Ingo Runkel$\;^2$ \ \ \ Christoph Schweigert$\;^{3,4}$}
\\[8mm]
$^1\;$ Institutionen f\"or fysik, \ Karlstads Universitet\\
Universitetsgatan 5, \ S\,--\,651\,88\, Karlstad\\[5mm]
$^2\;$ Institut f\"ur Physik, Humboldt Universit\"at Berlin\\
Newtonstra\ss{}e 15, \ D\,--\,12\,489\, Berlin\\[5mm]
$^3\;$ Institut f\"ur Theoretische Physik, RWTH Aachen\\
Sommerfeldstra\ss{}e 28, \ D\,--\,52\,074\, Aachen\\[5mm]
$^4\;$ Fachbereich Mathematik, \ Universit\"at Hamburg\\
Schwerpunkt Algebra und Zahlentheorie\\
Bundesstra\ss e 55, \ D\,--\,20\,146\, Hamburg
\end{center}
\vskip 12mm

\begin{quote}{\bf Abstract}\\[1mm]
A full rational CFT, consistent on all orientable world sheets, can be
constructed from the underlying chiral CFT, i.e.\ a vertex algebra, its
representation category $\calc$, and the system of chiral blocks,
once we select a symmetric special Frobenius algebra $A$ in the category
$\calc$ [I].
Here we show that the construction of [I] can be extended to unoriented
world sheets by specifying one additional datum: a reversion $\sigma$ on
$A$ -- an isomorphism from the opposed algebra of $A$ to $A$ that squares to
the twist. A given full CFT on oriented surfaces can admit inequivalent
reversions, which give rise to different amplitudes on unoriented
surfaces, in particular to different Klein bottle amplitudes.
\\
We study the classification of reversions, work out the construction of
the annulus, M\"obius strip and Klein bottle partition functions, and
discuss properties of defect lines on non-orientable world sheets.
As an illustration, the Ising model is treated in detail.
\end{quote}
\vfill
\newpage


\begin{flushright}
{\sl
          manche meinen
\\[-1pt]  lechts und rinks
\\[-1pt]  kann man nicht
\\[-1pt]  velwechsern.
\\[-1pt]  werch ein illtum!
}\\[.4em]
{\small Ernst Jandl \cite{jndl}} 
\end{flushright}
\tableofcontents

\newpage

\sect{Introduction and summary} 

Understanding conformal field theory on unoriented surfaces 
is not necessarily regarded as a fundamental goal in the analysis
of CFT. But there is at least one crucial application where, together with
world sheets having a boundary, non-orientable world sheets appear 
naturally: type I string theory. Accordingly, it is not surprising that the 
investigation of CFT on unoriented surfaces started out, and for some 
time was largely confined to, studies in string theory; see e.g.\
\cite{bumo,clny2,rodr,poca,bcdcd,bisa,bisa2,hora2,roVa,bisa3,fips,gipo}.
However, apart from dealing with inherently string theoretic aspects, such
as the Klein bottle and M\"obius strip projections \cite{bisa},
these studies eventually also led to a better understanding of
CFT on non-orientable surfaces in its own right. For instance, it was
realised \cite{bisa2} that it is necessary to have a left-right symmetric
bulk spectrum, and concepts like the $P$-matrix which relates the open
and closed string channels of the M\"obius strip amplitude \cite{bisa4}
and the crosscap constraint \cite{fips} were introduced. (The crosscap
state itself -- that is, the conformal block for a one-point correlation
function on the crosscap $\mathbb{RP}^2$ -- was also described in
\cite{ishi,onis}.) One crucial insight \cite{masa3,sagn,bisa3}
was that starting from a given CFT defined on orientable world
sheets, in particular its torus and annulus partition functions,
there can be several inequivalent ways to extend it to non-orientable
surfaces, giving rise to distinct Klein bottle and M\"obius amplitudes.
Fixing this freedom was termed the `choice of Klein bottle projection'.

The knowledge gathered in this research finally made it possible to
proceed to a more intrinsic study of rational CFT on unoriented surfaces. 
This development was pushed forward in particular by the Rome 
\cite{bips,prss,prss2,prss3,sasT2} and Amsterdam 
\cite{huss,husc,huss2,sche11,huis,sosc} groups. In the latter works the 
role of simple currents for the construction of different Klein bottle 
projections was emphasised.
These fields are particularly important for the study of non-orientable
surfaces; for example, all known Klein bottle amplitudes are related
to the standard one by a simple current. In the orientable case simple
currents appear in connection with symmetry breaking boundary conditions
\cite{fuSc10,fuSc11}; combining the results for orientable and
non-orientable world sheets finally led to
universal formulas for the boundary and crosscap coefficients
for all (symmetric) simple current modifications of the
charge conjugation invariant of any rational CFT \cite{fhssw}.

Meanwhile the interest in these matters has increased a lot --
for reviews as well as further references see \cite{stan4,Ansa}.
To name just a few results, let us mention that various properties of
the $Y$-tensor (which appears in the M\"obius strip and Klein bottle
amplitudes) were uncovered \cite{prss,sasT2,huss,bohs,gann14,bant11,brho},
that the standard Klein bottle coefficients were seen to coincide with
Frobenius-Schur indicators \cite{huss,bant11}, and that
polynomial equations relating the annulus, M\"obius and Klein bottle
coefficients were derived \cite{scst}. Many more results have been obtained
in the study of specific (classes of) models, ranging from the critical
Ising model \cite{bips,yAma} and free boson orbifolds \cite{scso} to
WZW \cite{prss,prss2,brun3,huSs,bacw,hahe2} and coset \cite{hiki,brho} 
models, permutation orbifolds \cite{kadA}, and
$N\eq2$ superconformal theories \cite{brho2,huSc}.

\medskip

The amount of information obtained so far is certainly impressive. But
one should also be aware of the fact that results were typically obtained
by solving only a relatively small number of necessary consistency
conditions, involving only the most basic correlation functions such as
the crosscap states and the M\"obius strip and Klein bottle amplitudes.
In the present paper we address rational CFT on arbitrary unoriented
world sheets within a general approach that allows for a uniform
construction of arbitrary correlation functions and which is guaranteed
to satisfy all locality, modular invariance and factorisation constraints.
Thus the resulting correlators
are automatically physical, unlike e.g.\ the ones obtained by solving
only the modular invariance constraint for the torus or the NIM-rep
property for the annulus.

\subsubsection*{Summary of the construction of \cite{fuRs4}}

The starting point of the construction of correlation functions in a full
rational CFT that was initiated in \cite{fuRs,fuRs4} is an underlying
chiral CFT: a chiral algebra (which contains the
Virasoro algebra), its \rep s, and the associated vector spaces $\calh(S)$ of
conformal blocks for every closed Riemann surface $S$.
A correlator $C$ of the full CFT on a world sheet $X$ is a specific element
of the space $\calh(\hat X)$ of conformal blocks on the complex double
$\hat X$ of $X$. The central idea \cite{fffs2,fffs3} is to use tools from
three-dimensional topological field theory (TFT) to describe this vector
$C\iN \calh(\hat X)$. The TFT in question is constructed from the category
$\calc$ of representations of the chiral algebra, which is a modular tensor
category.

{}From a given chiral CFT one can, in general, construct several inequivalent
full CFTs; well-known examples are the A-D-E series of minimal models and
the free boson compactified at different radii. In the construction
of \cite{fuRs,fuRs4} this is reflected by the choice of a
{\em symmetric special Frobenius algebra\/} $A$ in $\calc$. Accordingly,
for the construction of correlators the TFT methods are complemented by
non-commutative algebra in tensor categories.

To apply the TFT tools we need to fix a three-manifold $\MX$ and a ribbon
graph $R$ in $\MX$. $\MX$ is the {\em connecting manifold\/}
\cite{hora7,fffs2}; the world sheet $X$ is naturally embedded in $\MX$
(in fact, $X$ is a retract of $\MX$) and the boundary $\partial\MX$
is the complex double $\hat X$. To the data $\MX$ and $R$ the TFT associates 
a particular element of $\calh(\partial\MX)$; this is the specific vector
that gives the correlator $C$ on the world sheet $X$.

To construct the ribbon graph $R\,{\subset}\,\MX$ one chooses a (dual)
triangulation of $X$. Along the edges of the triangulation ribbons are placed
that are labelled by the algebra $A$, while the trivalent vertices are 
constructed from the multiplication of $A$.
Thinking of the world sheet $X$ as embedded in $\MX$, the part $R_A$ of
the ribbon graph $R$ built in this way lies entirely in $X$. The ribbons
have an orientation as a two-manifold; when $X$ is oriented, then there
is a canonical choice for the orientation of $R_A$, namely the one induced
by the embedding $R_A \,{\subset}\,\MX$. As shown in \cite{fuRs4}, the 
correlator $C\iN\calh(\hat X)$ obtained by this construction is independent
of the chosen triangulation of $X$. However, for a general algebra $A$, $C$
does depend on the orientation of the world sheet $X$.

\subsubsection*{Summary of results}

To build correlators on unoriented (or even non-orientable) world sheets
one needs an additional structure on the algebra $A$ that ensures 
independence of the (local) orientation of the world sheet.
For the particular case of \twodim\ {\em topological\/} field theory,
for which the relevant tensor category $\calc$ is the category $\Vect$ of 
complex vector spaces (so that $A$ is an algebra in the ordinary sense), it 
was found in \cite{kamo,alNa} that what is required is an isomorphism of $A$
to its opposed algebra that squares to the identity.

Algebras for which an involutive isomorphism to the opposed algebra
exists have been discussed in the literature under the name `algebras
with an involution'. However, in the setting of two-dimensional
{\em conformal\/} field theory, the braiding in the category $\calc$ is, in 
general, not symmetric. As a consequence, in the genuinely braided case one 
deals with a whole family of algebras $A^{(n)}$, labelled by integers, rather
than just an algebra and its opposed algebra; the twist $\thetaA$ turns out to
provide an algebra isomorphism from $A^{(n)}$ to $A^{(n+2)}$.

It turns out that the notion of algebras with involution generalises to the
braided setting as follows. For $A\,{\equiv}\,A^{(0)}$,
the opposed algebra $A_{\rm op}\,{\equiv}\,A^{(-1)}$,
whose multiplication is obtained from the multiplication
in $A$ by composition with the inverse braiding,
must be isomorphic, as an \alg, to $A$,
and furthermore this isomorphism must square to the twist.
That is, we need an algebra isomorphism $\sigma\iN\Hom(A_{\rm op},A)$
such that $\sigma\cir\sigma\eq\thetaA$. Such an isomorphism will be
called a {\em reversion\/} on $A$, and a symmetric special
Frobenius \alg\ with reversion will be called a {\em\JA\/}, see 
definition \ref{df:star}. The basic result of this paper is:
\begin{quote}
   {\sl Given a chiral CFT and a \JA\ -- a symmetric special Frobenius algebra
   with reversion -- one can construct a full CFT defined on all world
   sheets, including unoriented world sheets which may have a boundary.}
\end{quote}

\noindent
As already mentioned, to be able to formulate the CFT on unoriented world 
sheets the bulk spectrum must be left-right symmetric, i.e.\ the matrix
describing the torus partition function must be symmetric, $Z^{\rm t}\eq Z$.
Now $Z\eq Z(A)$ is uniquely determined by $A$, and as shown in \cite{fuRs4} 
one has $Z(A_{\rm op})\eq Z(A)^{\rm t}$. Thus $Z(A)^{\rm t}\eq Z(A)$ is 
indeed a {\em necessary\/} condition for the existence of a reversion 
on $A$. The existence of a reversion, in turn, gives a precise {\em
sufficient\/} condition ensuring that one can formulate the CFT
on unoriented world sheets as well.

{}From the point of view of 
type II string theory, a closed string 
background is characterised by a full CFT on closed oriented world
sheets. Given such a background, one can consider several D-brane 
configurations. This is reflected in the fact that Morita equivalent 
algebras give rise to the same full CFT defined on oriented world sheets. 
Thus in the oriented case a full CFT is associated
to a whole Morita class $\mocla$, not just a single algebra
$A\iN\mocla$. Once we have chosen a particular D-brane configuration
by picking a representative $A\iN\mocla$ we can try to find
an orientifold projection to describe a type I background. This
corresponds to finding reversions on $A$, and hence in this sense
a \JA\ encodes a type I background. Schematically we have
    \\[.52em]
\begin{tabular}{lrcl} $\quad$&
choice of Morita class $\mocla$ of algebras
    & $\longleftrightarrow$ & choice of closed string background \\[1.4mm]&
choice of representative $A$ in $\mocla$
    & $\longleftrightarrow$ & choice of D-brane configuration\\[1.4mm]&
choice of reversion $\sigma$ on $A$
    & $\longleftrightarrow$ & choice of orientifold projection
\end{tabular}
    \\[.55em]
{}From the string perspective it is intuitively clear that the number of 
possible orientifold projections does depend on the choice of D-brane 
configuration. This rephrases another crucial insight of this paper:
\begin{quote}
   {\sl The number of possible reversions does depend on the choice of 
   representative\\
   in a Morita class of symmetric special Frobenius algebras.}
\end{quote}
This result immediately poses the question of classification of
reversions; this issue will be addressed in section \ref{sec:*-class}.
The outcome is that every Morita class $\mocla$ contains a {\em finite\/} 
number of representatives $A_i$ with reversions $\sigma_{i,j}$ 
such that every reversion $\sigma$ on any $A\iN\mocla$
can be related to one of the \JA s $(A_i,\sigma_{i,j})$, in
a way described in proposition \ref{pr:B-star}. In CFT terminology,
this means that a reversion on an arbitrary D-brane configuration in 
a given background can be related to a reversion on either an 
elementary D-brane or else a superposition of two distinct elementary 
D-branes with the same boundary field content.

This result is in accordance with the geometric interpretation of
D-branes and orientifold planes in string theory. Either an elementary 
D-brane is already invariant with respect to reflection about the
orientifold plane, or it is not, in which case one must take the
configuration consisting of the D-brane and its image under the
reflection about the orientifold plane. We want to stress, however, that
our classification of reversions results from an intrinsic
CFT computation which does not rely on the existence of any space-time
interpretation of the CFT. Thus our analysis shows that the geometric
picture retains some validity in the deep quantum regime.

Also recall from \cite{fuRs4} that boundary conditions are described by
left $A$-modules. The existence of a reversion
$\sigma$ on $A$ allows us to associate to any left $A$-module $M$
another left $A$-module $M^\sigma$, and thereby to define a
($\sigma$-dependent) conjugation $C^\sigma$ on left $A$-modules, i.e.\ on
the space of \bc s (see definition \ref{df:conj-mod} and proposition
\ref{pr:CB-prop}). Note that, in contrast, the structure of a \ssFA\ alone
does not allow one to endow the category of left $A$-modules with a conjugation.

\medskip

The prescription for the three-manifold $\MX$ and ribbon graph that describe
a correlation function is given, for the case without field insertions,
in section \ref{sec:crg}. Three examples are worked out in detail:
the annulus (section \ref{sec:ann}), M\"obius strip (section \ref{sec:moebius})
and Klein bottle (section \ref{sec:klein}) partition functions. For the annulus,
both boundaries are chosen to have the same orientation. The corresponding
annulus coefficients, giving the multiplicities of open string states,
are written with two lower indices, ${\rm A}_{kMN}$; we show that they are
symmetric in the boundary conditions $M$ and $N$, and that the
coefficients ${\rm A}_{0MN}$ coincide with the boundary conjugation matrix
(equation \erf{eq:Aknm-sym} and lemma \ref{lem:C-A}).
The ribbon graphs for the three partitions functions are presented in
the figures \erf{eq:Akmn}, \erf{eq:MkR} and \erf{eq:Kk}, respectively.
The coefficients $\rmm_{kR}$ of the M\"obius strip amplitude are shown to be
integers and satisfy $\frac12\,(\rmm_{kR}\,{+}\,{\rm A}_{kRR}) \iN \zet_{\ge 0}$
(theorem \ref{thm:m-prop}). The latter property is necessary for a
consistent interpretation in type I string theory, where
this combination of annulus and M\"obius strip counts the number of open
string states. Similarly, for the Klein bottle coefficients $K_k$
it is shown that ${\rm K}_k \iN \zet$ and
$\frac12\,(Z_{kk}\,{+}\,{\rm K}_k) \iN \zet_{\ge 0}$, where $Z_{ij}$ are
the coefficients of the torus partition function (theorem \ref{thm:K-prop}).

Another important constraint is the consistency of the M\"obius strip and Klein
bottle amplitudes in the crossed channel, where in both cases only closed
strings propagate. Note that all these constraints are automatically
satisfied on general grounds, because the amplitudes constructed with the
present method are consistent with factorisation.
Nevertheless we investigate the crossed channels for M\"obius strip
and Klein bottle explicitly; in particular we establish the relations
  \be
  \rmm_{kR} = \Frac{1}{S_{00}}\sum_{l,\alpha,\beta} P_{kl}\, S^A_{R,l\alpha}\,
  g^{\bar l\,l}_{\alpha\beta} \; \gamma^\sigma_{l\beta}
  \qquad{\rm and}\qquad
  \K_k =  \Frac{1}{S_{00}} \sum_{l,\alpha,\beta} S_{kl} \,
  \gamma^\sigma_{l\alpha} \, g^{\bar l\,l}_{\alpha\beta} \,
  \gamma^\sigma_{l\beta}  \labl{eq:introcross}
(see equations \erf{eq:M-crossed}, \erf{eq:K-crossed} and sections
\ref{sec:S3/Z2}\,--\,\ref{CcMK}). This is done for several reasons. First, 
the details of the general proof of factorisation have not been published
yet, and it is illustrative to discuss these particular cases separately. 
Second, the relations \erf{eq:introcross} are used to actually compute 
$\rmm_{kR}$ and $K_k$. This is done by giving explicit expressions for the 
constituents $S^A$ (formulas \erf{eq:SA-inv-1} and \erf{eq:SA-inv-2}), 
$\gamma^\sigma$ (\erf{eq:Ga-bas}, \erf{eq:Gamma-inv-1} and 
\erf{eq:Gamma-inv-2}) and $g$ (formula \erf{eq:g-inv}). The latter expressions 
involve only the defining data of the modular tensor category (the braiding and
fusing data $\RR$ and $\FF$), the \JA\ (the product $m$ and the reversion 
$\sigma$), the $A$-modules (the representation morphisms $\rho$), and
a basis $\mu$ in the space of local morphisms (definition
\ref{def:loc} and equation \erf{eq:mu-expand}).

\medskip

We also discuss defect lines on unoriented world sheets (section 
\ref{sec:defect}). Here a new feature arises. Suppose a conformal defect is 
wrapping a non-orientable cycle on the world sheet, that is, a cycle whose
neighbourhood has the topology of a M\"obius strip. For a generic
defect this is only possible when inserting a disorder field, because
the two ends of a defect wrapping a non-orientable cycle 
generally cannot be joined in a conformally invariant manner. 
While a conformal defect can be deformed continously without changing
the value of the correlator, the position of the disorder field
insertion must remain fixed, i.e.\ the insertion of a disorder field marks 
a point on the world sheet. This way one can single out an interesting 
subclass of conformal defects, namely those which can wrap a
non-orientable cycle without the need to insert any disorder field.
A defect $Y$ belongs to this subclass iff $\Hom_{A|A}(Y^s,Y)$ has nonzero
dimension, see the end of section \ref{sec:defect} and equation 
\erf{eq:defect-Hom}. (Of course $Y^s$ depends on the choice of reversion.)

Finally we work out two examples in detail. First, the classification of 
reversions is carried out for the case $\calc\eq\Vect$, illustrating the
simplifications that occur when working with algebras in the ordinary 
sense (section \ref{sec:vect}).
Second, as simplest non-trivial CFT model, the critical Ising model is 
treated in great detail (section \ref{sec:ising-ex}). The defining data of 
the modular category $\calc$ are presented, and the \ssFA s are classified 
and shown to form a single Morita class. Then all reversions are 
classified; it is found that all reversions can be related to two 
fundamental ones. The boundary conjugation matrix and the M\"obius strip 
and Klein bottle amplitudes are computed. On the basis of these
a lattice interpretation is offered for the two different reversions.

\medskip

In four appendices we provide further details. 
We display a collection of moves on ribbon graphs which are useful in 
the evaluation of various invariants (appendix \ref{sec:app-moves}), 
provide the slightly lengthy proof of the algorithm for finding
reversions (appendix \ref{approof}), comment on the recursion formula 
\erf{eq:rec-rel} for computing invariants obtained by glueing two solid 
tori (appendix \ref{sec:app-tor}), and give an auxiliary 
calculation for the classification of reversions in the Ising model that
is omitted in the main text (appendix \ref{sec:ising-star}).


\vskip3em \noindent {\bf Acknowledgement.}\\
Part of this work was carried out at the ETH Z\"urich; we are grateful 
to J\"urg Fr\"ohlich for his hospitality and continued interest. J.F.\ is 
supported in part by VR under contract no.\ F\,5102\,--\,20005368/2000, 
and I.R.\ is supported by the DFG project KL1070/2--1.

\newpage

\sect{\JA s} \label{sec:sao} 

To be able to deal with non-orientable world sheets, 
an additional structure on algebras -- a reversion -- 
must be introduced. To see why this structure is indeed needed, let us 
have a look at the prescription on how to construct the ribbon graph for 
an oriented world sheet and point out 
what is needed to extend that prescription to unoriented world sheets.
 
Recall from section I:5.1 that in the oriented case we
choose a dual triangulation of the world sheet and then place $A$-ribbons 
along its edges. These ribbons must be oriented as 
two-manifolds. In a graphical calculus the two possible orientations 
are indicated by drawing the ribbon with two different colours, say with 
a white and a black side, respectively. For the case of oriented world sheets,
in \cite{fuRs4} we have chosen the orientation inherited from
the orientation of the world sheet. For non-orientable world sheets, it is
still possible to choose a dual triangulation, but after doing so
we face two problems. First, we do not know how to orient the ribbons. 
Of course, we can still choose local orientations. But every
non-orientable surface contains a cycle that, after fattening, gives a ribbon 
with the topology of the M\"obius strip. As a consequence, we are inevitably 
faced with the task to join two $A$-ribbons with opposite orientations,
or in other words, we must join the white side of an $A$-ribbon to the black 
side of another $A$-ribbon. 
A priori it is unclear how this should be done; let us symbolise this
unknown operation by a dashed box and refer to it as a {\em \boxelement\/}.
The simplest manipulation one could carry out at a \boxelement\ is the
purely geometric operation of just performing a ``half-twist", i.e.\
rotating the ribbon by an angle of $\pi$, according to
  \bea \begin{picture}(60,48)(0,28)
  \put(0,0)     {\includeourbeautifulpicture{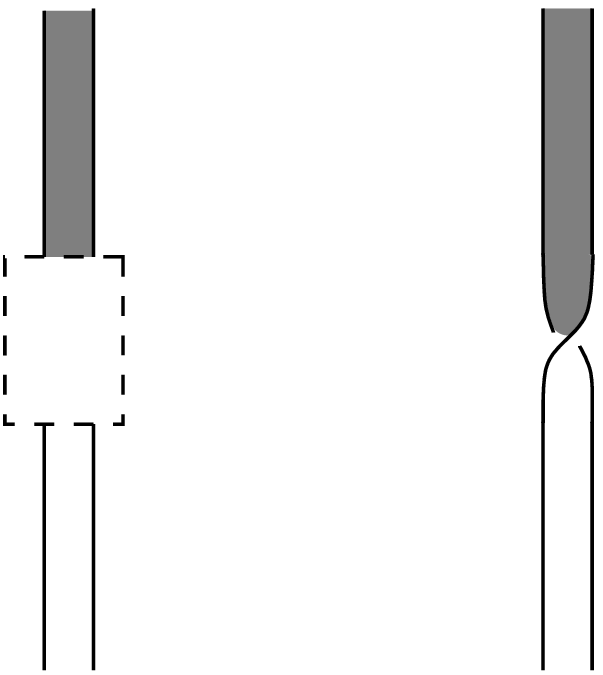}}
  \setlength{\unitlength}{24810sp}
  \setlength{\unitlength}{1pt}
  \put(32,33)  {$ \stackrel?= $}
  \epicture13 \labl{eq:half-twist}
Already at this point, it is apparent that there is too much arbitrariness
in this prescription: it is not clear that turning the ribbon in the
opposite direction will give an equivalent result.

To determine what properties the \boxelement\ must possess in order to avoid
such ambiguities, we must analyze the situation in more detail.
Let us choose, right from the start, at every vertex of the
triangulation a local orientation of the world sheet. This also fixes,
around that vertex, what is to be considered as the white and the black side
of the ribbon. Independence of the choices of local orientation and of the
triangulation can then be seen to impose the conditions
  \bea \begin{picture}(350,65)(0,31)
  \put(0,0)     {\includeourbeautifulpicture{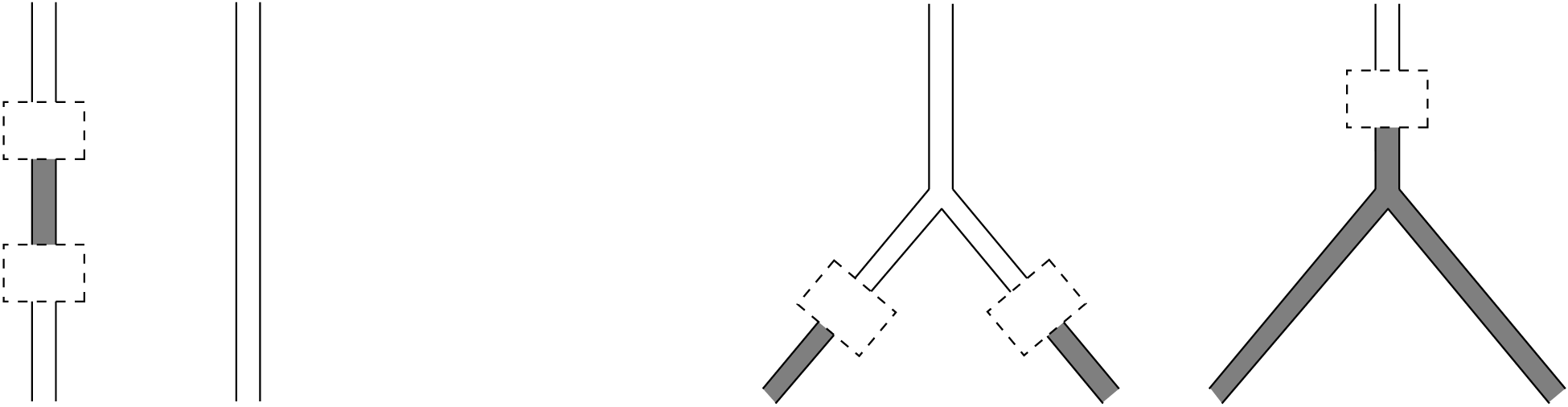}}
  \setlength{\unitlength}{24810sp}
  \setlength{\unitlength}{1pt}
  \put(32,44)  {$=$}
  \put(116,44) {and}
  \put(261,44) {$=$}
  \epicture14 \labl{eq:box-prop}
on the \boxelement. The second of these equalities ensures that two
different choices of the local orientation at the vertex give equivalent
results, while the first equality allows us to cancel two reversing moves.

One can quickly convince oneself that the first ansatz \erf{eq:half-twist}
of simply turning the ribbons by an angle $\pi$ does not obey these
conditions\foodnode{%
  For an interpretation of half-twists in the context of quantum
  groups see remark 7.2 of \cite{kirI9}.}.
Instead a more general ansatz is required. To this end we include in the 
\boxelement\ a suitable `compensating' morphism $\sigma\iN\Hom(A,A)$, i.e.\ 
every half-twist is accompanied by the morphism $\sigma$, according to
  \bea \begin{picture}(60,45)(0,30)
  \put(0,0)     {\includeourbeautifulpicture{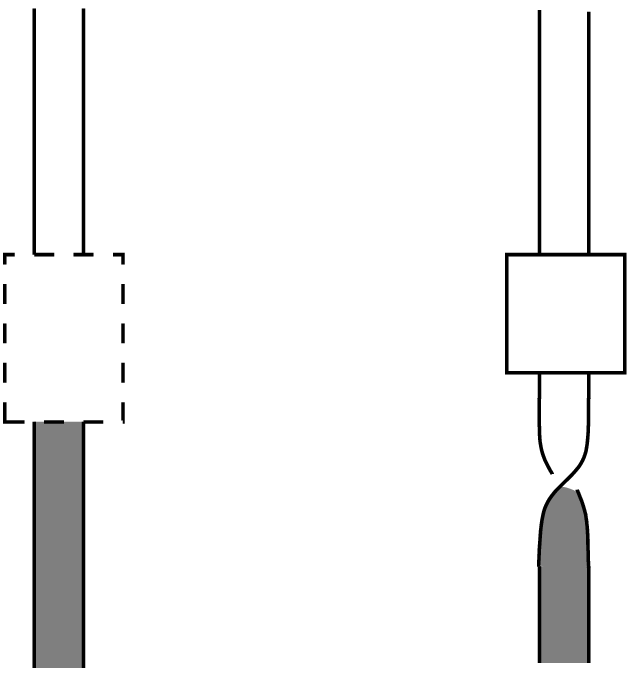}}
  \setlength{\unitlength}{24810sp}
  \put(156,98) {$\sigma$}
  \put(77,95)  {$=$}
  \setlength{\unitlength}{1pt}
  \epicture13 \labl{eq:box-el}
The conditions \erf{eq:box-prop} then translate into statements about the
morphism $\sigma$ -- they imply that $\sigma$ is a {\em reversion\/} on $A$,
see definition \ref{df:star} below (one also needs proposition \ref{pr:*ssFA}).

As a matter of fact, the reasoning above just repeats what has already been
done in \cite{kamo} for \twodim\ lattice topological field theory. A
2d TFT is defined by an algebra over $\complex$, i.e.\ an algebra in 
$\Vect_{\complex}$. The line of arguments just described lead the authors 
of \cite{kamo} to introduce an involutive algebra anti-homomorphism
$\sigma$, i.e.\ a linear map satisfying
$\sigma(ab)\eq\sigma(b)\,\sigma(a)$ and $\sigma\cir\sigma\eq\id$.

\subsection{Reversions}\label{sec:str}

For general modular tensor categories the only modification comes from the 
fact that the twist and the braiding can be non-trivial. This suggests the
\\[-2.5em]

\dtl{Definition}{df:star}
(i)~~A {\em reversion\/} on an algebra $A\eq(A,m,\eta)$ is an
endomorphism $\sigma \iN \Hom(A,A)$ that is an
algebra anti-homomorphism and squares to the twist, i.e.
  \be
  \sigma \circ \eta = \eta \, , \qquad
  \sigma \circ m = m \circ c_{A,A} \circ (\sigma\oti\sigma)
  \, , \qquad
  \sigma \circ \sigma = \thetaA \,.  \labl{eq:sig-def}
If the algebra $A$ is also a coalgebra,
$A\eq(A,m,\eta,\Delta,\eps)$, then we demand that in addition
  \be
  \eps \circ \sigma = \eps \qquad {\rm and} \qquad
  \Delta \circ \sigma = (\sigma\oti\sigma) \circ c_{A,A}
  \circ \Delta  \labl{eq:sig-coprod}
hold. In pictures:
  \bea \begin{picture}(420,44)(9,22)
  \put(0,0)     {\includeourbeautifulpicture{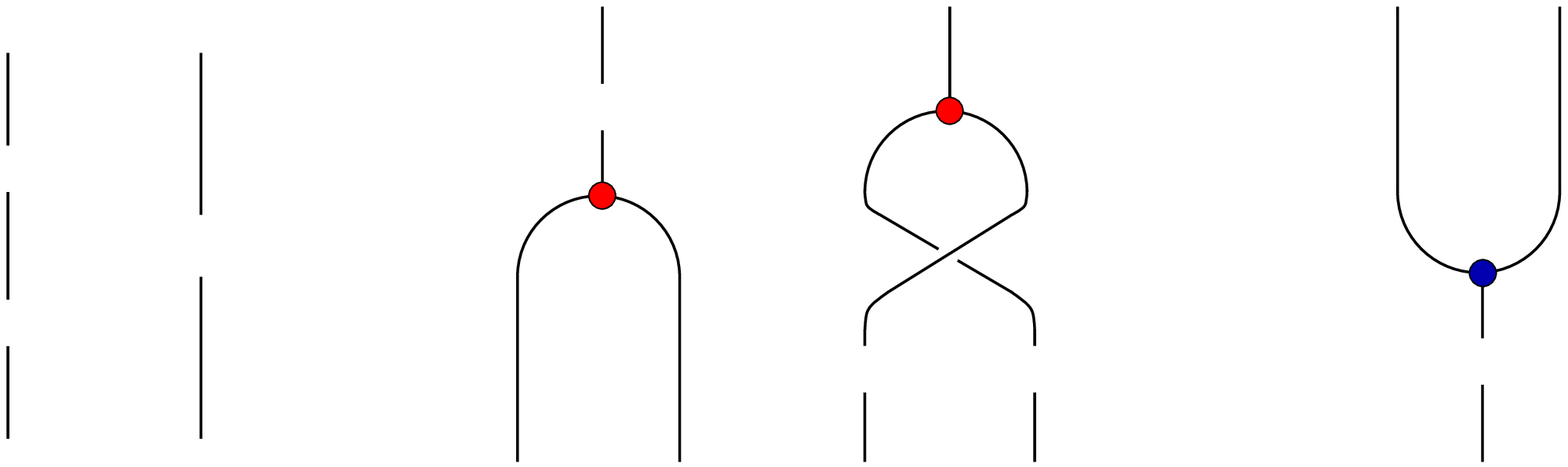}}
  \setlength{\unitlength}{24810sp}
  \put(0,-100){
    \put(36,172) {$=$}
    \put(4.5,205.5) {\scriptsize$\sigma$}
    \put(4.5,148.5) {\scriptsize$\sigma$}
    \put(75,174) {\scriptsize$\theta$}}
  \put(-2,-206){
    \put(279,276) {$=$}
    \put(226,334) {\scriptsize$\sigma$}
    \put(323,238) {\scriptsize$\sigma$}
    \put(386,238) {\scriptsize$\sigma$}}
  \put(320,0){
    \put(278,77) {$=$}
    \put(227.5,35) {\scriptsize$\sigma$}
    \put(324,131) {\scriptsize$\sigma$}
    \put(386,131) {\scriptsize$\sigma$}}
  \put(314,-208){
    \put(540,294) {$=$}
    \put(509.6,296.5) {\scriptsize$\sigma$}}
  \put(510,-2){
    \put(540,73) {$=$}
    \put(513,79.3) {\scriptsize$\sigma$}}
  \setlength{\unitlength}{1pt}
  \put(-.5,-7.5)   {\scriptsize$A$}
  \put(-1,61.5)    {\scriptsize$A$}
  \put(26,-7.5)    {\scriptsize$A$}
  \put(26.5,61.5)  {\scriptsize$A$}
  \put(70.5,-9.5)  {\scriptsize$A$}
  \put(83,67.5)    {\scriptsize$A$}
  \put(93.5,-9.5)  {\scriptsize$A$}
  \put(118.5,-9.5) {\scriptsize$A$}
  \put(132,67.5)   {\scriptsize$A$}
  \put(143.5,-9.5) {\scriptsize$A$}
  \put(194,67.5)   {\scriptsize$A$}
  \put(206.5,-9.5) {\scriptsize$A$}
  \put(216,67.5)   {\scriptsize$A$}
  \put(243,67.5)   {\scriptsize$A$}
  \put(254.5,-9.5) {\scriptsize$A$}
  \put(266,67.5)   {\scriptsize$A$}
  \put(310,63.5)   {\scriptsize$A$}
  \put(337,63.5)   {\scriptsize$A$}
  \put(385,-4.5)   {\scriptsize$A$}
  \put(413,-4.5)   {\scriptsize$A$}
  \epicture11 \ee
(ii)~\,The quadruple $A\eq(A,m,\eta,\sigma)$ consisting of an algebra and
a reversion is called an {\em \staralgebra\/}.
\\[.3em]
(iii)~A symmetric special Frobenius \alg\ with reversion will 
also be called a {\em \JA\/}.

\medskip

Recall from corollary I:3.19 and proposition I:3.20 that from a given 
algebra $A$ one can obtain a whole series of algebras $A^{(n)}$ by 
composing the multiplication with some power $c_{A,A}^n$ of the braiding.
The twist $\thetaA$ provides algebra isomorphisms
from $A^{(n)}$ to $A^{(n+2)}$. With respect to this property, finding a 
reversion $\sigma$ on $A$ amounts to taking a square root of the twist:
\\[-2.3em]

\dtl{Proposition}{pr:An-An+1}
Let $A$ be an algebra and $n$ an integer. A morphism $\sigma \iN\Hom(A,A)$ is
a reversion for $A$ if and only if it is an algebra morphism
from $A^{(n)}$ to $A^{(n+1)}$ and squares to $\thetaA$.
\\
An analogous statement holds if $A$ is in addition a coalgebra.

\medskip\noindent
Proof:\\
Let $A\eq(A,m,\eta)$ be an algebra. By definition,
  \be
  A^{(n)} = (A, m \cir(c_{A,A})^n, \eta) \,.  \labl{eq:An-def}
That $\sigma\iN\Hom(A,A)$ is an algebra
morphism from $A^{(n)}$ to $A^{(n+1)}$ squaring to the twist thus means
  \be
  \sigma \circ m^{(n)} =  m^{(n+1)} \circ (\sigma\oti\sigma) \,, \qquad
  \sigma \circ \eta^{(n)} = \eta^{(n+1)} \qquad {\rm and} \qquad
  \sigma \circ \sigma = \thetaA \,.  \labl{eq:alg-hom-cond}
By functoriality of the braiding we have
$c_{A,A} \cir (\sigma\oti\sigma)\eq
(\sigma\oti\sigma)\cir c_{A,A}$. Hence by
substituting the definition \erf{eq:An-def}
into \erf{eq:alg-hom-cond} one sees that 
\erf{eq:sig-def} and \erf{eq:alg-hom-cond} are indeed equivalent.
\\[5pt]
If in addition $A$ is coalgebra
with coproduct $\Delta$ and counit $\eps$,
then the coproduct and counit of $A^{(n)}$ are given by
$\Delta^{(n)}\eq(c_{A,A})^{-n} \cir \Delta$
and $\eps^{(n)}\eq\eps$. Then by the same reasoning as above one
checks the equivalence of
  \be
  (\sigma\oti\sigma) \cir \Delta^{(n)} = \Delta^{(n+1)} \cir \sigma 
  \qquad {\rm and} \qquad
  \eps^{(n)} = \eps^{(n+1)} \cir \sigma \ee
with the conditions \erf{eq:sig-coprod}.
\qed

If we are given two reversions $\sigma_1$ and $\sigma_2$ 
of an algebra $A$, the above proposition implies that
$\omega\eq\sigma_1^{-1}{\circ}\, \sigma_2$ is an algebra 
automorphism of $A$. Conversely, all possible 
reversions for a given algebra $A$ can be obtained by composing 
a single reversion $\sigma_0$ with suitable algebra automorphisms
$\omega$ of $A$. Note, however, that while the first two
properties in \erf{eq:sig-def} for $\sigma\,{:=}\,\omega \cir \sigma_0$
being a reversion follow directly from the fact that $\omega$ is
an algebra morphism, the requirement $\sigma\cir\sigma\eq\thetaA$ places
an additional condition on $\omega$. Writing 
$\thetaA\eq\sigma_0 \cir\sigma_0$ yields the following result.
\\[-2.3em]

\dtl{Proposition}{prop:alg-aut}
Let $A$ be an algebra with reversion $\sigma_0$ and
$\omega \iN \Hom(A,A)$ be an automorphism of $A$. Then
$\sigma\eq\omega \cir\sigma_0$ is a reversion on $A$ if and only if 
  \be
  \omega \circ \sigma_0 \circ \omega = \sigma_0 \,.  \labl{eq:aut-sig}

Not all reversions obtained from algebra automorphisms
obeying \erf{eq:aut-sig} should be treated as distinct. 
Let $A_1$ and $A_2$ be two algebras and $f\in\Hom(A_1,A_2)$
be an algebra isomorphism. If $A_1$ and $A_2$ have reversions 
$\sigma_1$ and $\sigma_2$, then $f$ is an isomorphism of \staralgebras\ if 
$f \cir \sigma_1\eq \sigma_2 \cir f$. In particular
two reversions $(A,\sigma)$ and $(A,\omega \cir\sigma)$
on the same algebra are isomorphic if there exists an automorphism $f$ 
of $A$ such that $\omega \cir\sigma\eq f\cir \sigma \cir f^{-1}$.

\medskip

If $A$ is a symmetric special Frobenius algebra, the
coproduct and counit are determined through the multiplication.
This reduces the number of conditions
$\sigma$ has to satisfy to be a reversion on $A$. In fact,
due to the proposition below it is enough to check that 
  \be
  \sigma\circ\sigma=\thetaA \labl{eq:*ssFA1}
and
  \be
  \sigma \circ m = m \circ c_{A,A} \circ (\sigma \oti \sigma) \,.
  \labl{eq:*ssFA2}
These are precisely the conditions that correspond to 
the properties displayed graphically in picture \erf{eq:box-prop}.

\dtl{Proposition}{pr:*ssFA}
Let $A$ be an algebra, and $\sigma \iN \Hom(A,A)$ satisfy the relation
\erf{eq:*ssFA2}. Then $\sigma$ also obeys
  \be
  \sigma \circ \eta = \eta \,.  \ee
If in addition $\sigma$ has the property \erf{eq:*ssFA1} and $A$ is 
symmetric special Frobenius, then
$\sigma$ is a reversion on $A$, i.e.\ we also have
  \be
  \eps = \eps \circ \sigma \,, \qquad
  (\sigma\oti\sigma) \circ \Delta = c_{A,A}^{-1} \circ \Delta 
  \circ \sigma \,.  \ee

\medskip\noindent
Proof:\\
The equality $\sigma\cir\eta\eq\eta$ is obtained as
  \be
  \sigma \circ \eta  
  = m \circ (\sigma \otimes \id_A) \circ (\eta\otimes\eta)  
  = \sigma \circ m \circ c_{A,A}^{-1} \circ (\eta \otimes \sigma^{-1})
    \circ \eta 
  = \eta \,,  \ee
where the second step uses \erf{eq:*ssFA2}.
\\[3pt]
To show that $\eps\cir\sigma\eq\eps$ we use that, by lemma I:3.11,
the counit $\eps$ of a symmetric Frobenius algebra  is of the form 
$\eps\eq\beta\,\eps_\natural$ for some $\beta\iN\complex^\times$, with 
the morphism $\eps_\natural$ as defined in (I:3.46). It follows that
  \bea \begin{picture}(260,46)(0,35)
  \put(122,0)     {\includeourbeautifulpicture{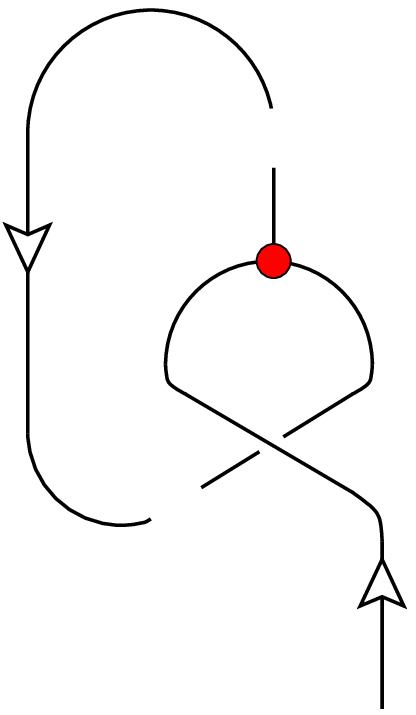}}
  \put(0,40)       {$ \eps \circ \sigma \;=\; \beta \, \eps_\natural
                      \circ \sigma \;=\; \beta$}
  \put(175,40)     {$ =\; \beta \, \eps_\rho \;=\; \eps \,.$}
  \setlength{\unitlength}{24810sp}
  \setlength{\unitlength}{1pt}
  \put(120,73)     {\scriptsize$A$}
  \put(138.9,19.3) {\scriptsize$\sigma^{\!-\!1}$}
  \put(150,61.5)   {\scriptsize$\sigma$}
  \put(159.5,-8.5) {\scriptsize$A$}
  \epicture21 \ee
Here the second step uses \erf{eq:*ssFA2}. In the third step $\sigma^{-1}$
is cancelled against $\sigma$ and the
$A$-loop is moved to the right side of the ingoing $A$-ribbon;
the two resulting twists in the $A$-loop cancel, so that one is left
with $\eps_\rho$ as defined in (I:3.46). The last step
uses lemma I:3.9(ii) which states that $\eps_\rho\eq\eps_\natural$.
\\
Finally we must check that $(\sigma\oti\sigma)\cir\Delta\eq c_{A,A}^{-1}
\cir\Delta\cir\sigma$. To this end we use the morphism
$\Phi_1\iN\Hom(A,A^\vee)$ in (I:3.33). We have
  \bea \begin{picture}(280,68)(0,33)
  \put(70,0)     {\includeourbeautifulpicture{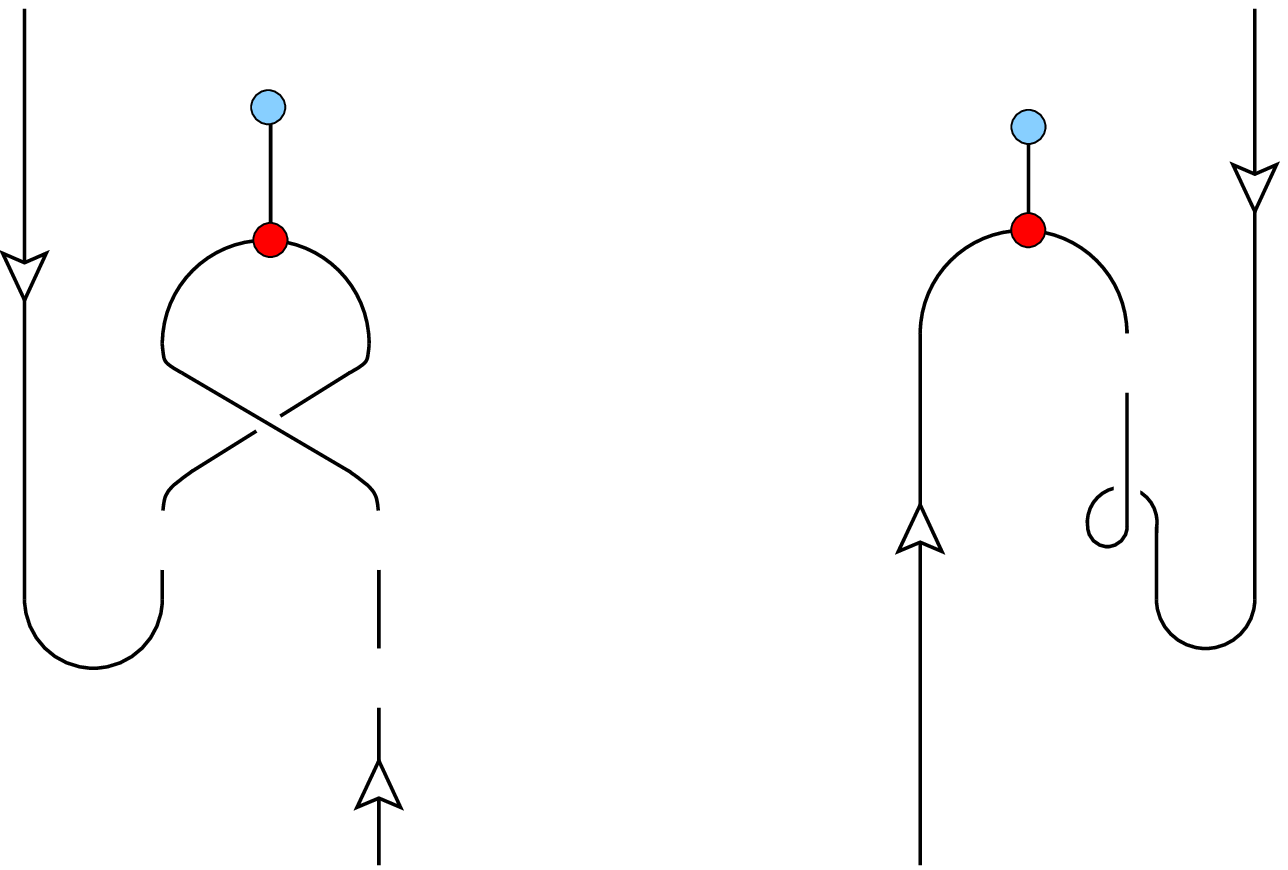}}
  \put(0,45)       {$ \Phi_1 \circ \sigma \; =$}
  \put(138,45)     {$=$}
  \put(230,45)     {$ =\; \sigma^\vee \circ \Phi_1 \,.$}
  \setlength{\unitlength}{24810sp}
  \setlength{\unitlength}{1pt}
  \put(68.5,98.5)  {\scriptsize$A^\vee_{}$}
  \put(83.9,33.7)  {\scriptsize$\sigma^{\!-\!1}$}
  \put(107.5,-9)   {\scriptsize$A$}
  \put(106.9,33.7) {\scriptsize$\sigma^{\!-\!1}$}
  \put(109.3,19.4) {\scriptsize$\sigma$}
  \put(166.5,-9)   {\scriptsize$A$}
  \put(190.8,53.4) {\scriptsize$\sigma^{\!-\!1}$}
  \put(203.5,98.5) {\scriptsize$A^\vee_{}$}
  \epicture19 \labl{eq:sig-phi}
In the first step the definition of $\Phi_1$ is inserted and
$\eps\eq\eps\cir\sigma^{-1}$ together with \erf{eq:*ssFA2} is used. 
The second step is a deformation of the $A$-ribbon.
In the last step the twist is absorbed via property \erf{eq:*ssFA1}, 
i.e.\ $\sigma^{-1} \cir\thetaA\eq\sigma$. We are then left with 
$\sigma^\vee \,{\circ}\, \Phi_2$ (the dual $f^\vee$ of a morphism $f$ 
has been defined in (I:2.8) and (I:2.12)), and finally we can use that 
$\Phi_2\eq\Phi_1$ because $A$ is symmetric (definition I:3.4). 
Now from the expression (I:3.42) for the coproduct $\Delta$
we have
  \begin{eqnarray}  \begin{picture}(420,112)(6,0)
  \put(89,0)     {\includeourbeautifulpicture{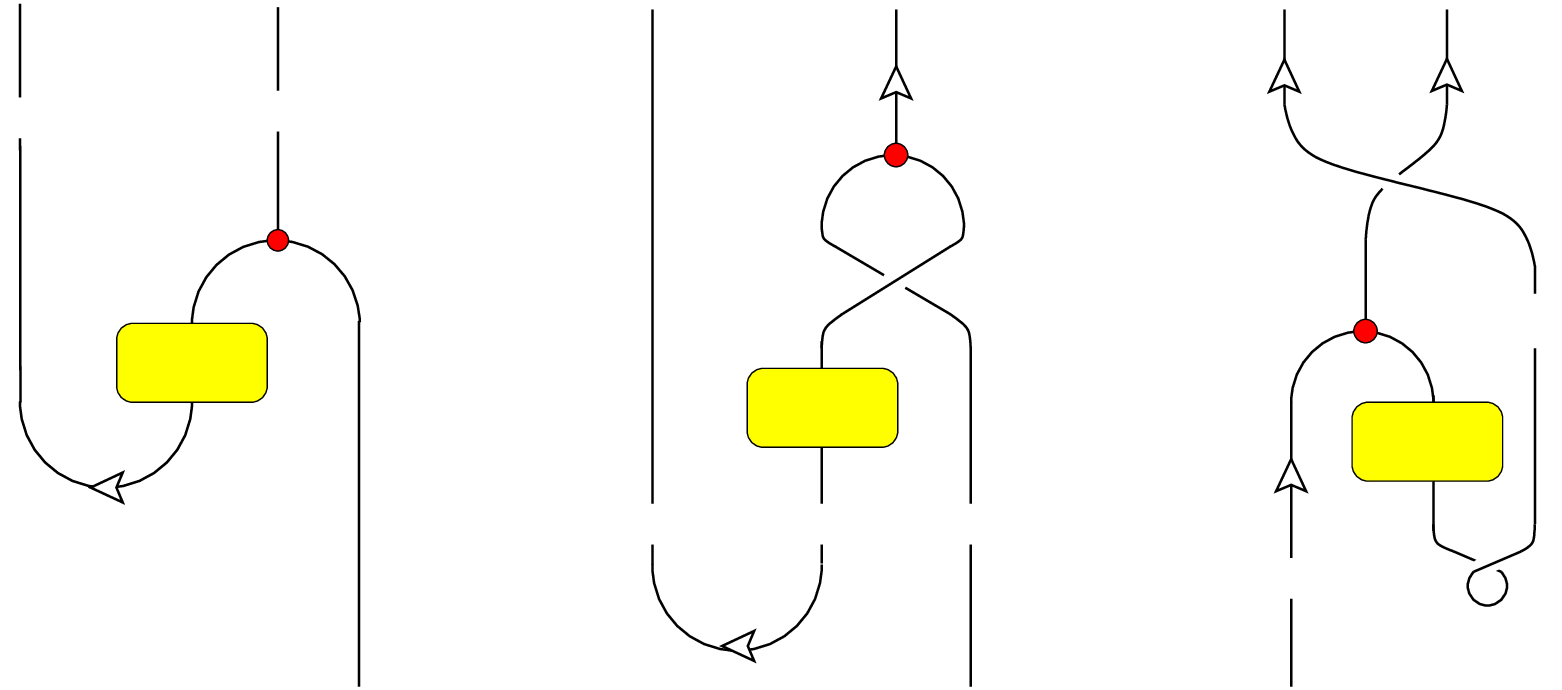}}
  \put(0,50)       {$ (\sigma\oti\sigma) \circ \Delta \; = $}
  \put(90.3,90.5)  {\scriptsize$\sigma$}
  \put(114,49.5)   {\scriptsize$\Phi_1^{\!-\!1}$}
  \put(131.3,90.5) {\scriptsize$\sigma$}
  \put(166,50)     {$=$}
  \put(190.9,24.8) {\scriptsize$\sigma$}
  \put(215,42.5)   {\scriptsize$\Phi_1^{\!-\!1}$}
  \put(217.4,24.4) {\scriptsize$\sigma^{\!\vee}$}
  \put(242.1,24.8) {\scriptsize$\sigma$}
  \put(266,50)     {$=$}
  \put(293.3,16.1) {\scriptsize$\sigma$}
  \put(312,37.5)   {\scriptsize$\Phi_1^{\!-\!1}$}
  \put(332.3,55.8) {\scriptsize$\theta$}
  \put(350,50)     {$ =\; (c_{A,A})^{-1} \circ \Delta \circ \sigma \,,$}
  \setlength{\unitlength}{24810sp}
  \setlength{\unitlength}{1pt}
  \end{picture} \nonumber\\[-1.5em]{} \label{pic7}
  \\[-1.3em]{}\nonumber\end{eqnarray}
where the second step combines the result \erf{eq:sig-phi} with property 
\erf{eq:*ssFA2}. In the third step the two $\sigma$'s are combined 
to a twist $\thetaA$ via \erf{eq:*ssFA1} and the leftmost $A$-ribbon
is moved to the right, resulting in another twist. Then the two twists 
cancel and one recognises another expression in the list (I:3.42) for the 
coproduct (recall $\Phi_1\eq\Phi_2$). 
\qed

\dt{Remark}
(i) According to proposition I:5.3, the matrices $Z_{ij}$ that give the 
torus partition functions of the full CFTs for $A$ and $A_{\rm op}\,
{\equiv}\,A^{(-1)}$ are related by $Z(A_{\rm op})\eq Z(A)^{\rm t}$.
Since the morphism $\sigma$ provides an algebra isomorphism
between $A_{\rm op}$ and $A$, it follows that $Z(A)\eq Z(A)^{\rm t}$ is
a necessary condition for $A$ to admit a reversion.
\\[5pt]
(ii) Recall from theorem I:3.6(i) that an algebra $(A,m,\eta)$ can have, 
up to isomorphism, only a single symmetric special Frobenius structure. 
The qualifiers `symmetric', `special' and `Frobenius' contain thus no
additional information, but rather indicate a restriction on the class 
of algebras one can use.
\\
For reversions the situation is different. While the existence
of a reversion further restricts the class of algebras
(see point (i)), a given symmetric special Frobenius 
algebra can allow for several inequivalent reversions. 
Section \ref{sec:ex} will provide examples for this phenomenon. 
To construct a conformal field theory on unoriented surfaces, one must thus
choose {\em both\/} a (symmetric special Frobenius) algebra {\em and\/} a 
reversion on that algebra; different reversions on the same 
algebra can give inequivalent conformal field theories.
\\[5pt]
(iii)~There can exist several distinct morphisms $\sigma_{(i)}$ that turn a
given \ssFA\ into a \JA. An important point to notice is that the number of 
such morphisms is, in general, {\em not\/} the same for every member of a
given Morita class of algebras.
Again we will meet examples in section \ref{sec:ex}. (We even cannot 
exclude the possibility that some representatives of the Morita class
allow for reversions while others cannot carry any reversion
at all, though we are not aware of an example of this phenomenon.)

\newpage
\subsection{Expressing reversions in a basis}\label{sec:*bas}

When introducing bases for the vector spaces $\Hom(U_i,A)$ and
$\Hom(A,U_i)$, with $U_i$ any simple object, the morphism $\sigma$ is 
encoded in a collection of matrices $\sigma(i)$. The conditions 
\erf{eq:*ssFA1} and \erf{eq:*ssFA2} can then be formulated as relations 
among those matrices. Let us describe this in detail.

As an object of $\calc$ the algebra $A$ is a direct sum
$A \,{\cong}\, \bigoplus_{i\in\II} n_i\, U_i$ with some
multiplicities $\Ai i\eq n_i\iN\zet_{\ge 0}$, where $\{U_i\,|\,i\iN\II\}$ 
is a (finite) set of representatives for the isomorphism classes of
irreducible objects of $\calc$. We choose a basis $\{\alpha\}$
in the morphism space $\Hom(U_i,A)$ and a dual
basis $\{\bar\alpha\}\,{\subset}\,\Hom(A, U_i)$, as in (I:3.4). 
In this basis the multiplication on $A$ is given by numbers
$\m a\alpha b\beta c\gamma\delta$ as in (I:3.7). Similarly,
the reversion $\sigma$ is then
encoded in matrices $\sigma(i)_\alpha^{\;\beta}$ via
  \bea \begin{picture}(170,50)(0,42)
  \put(0,0)     {\includeourbeautifulpicture{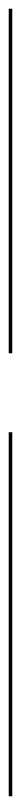}}
  \put(147,0)   {\includeourbeautifulpicture{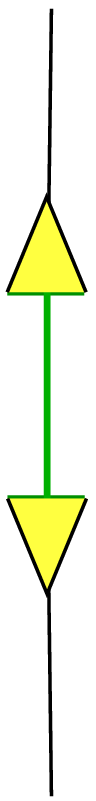}}
  \put(30,41)      {$ =\; \dsty\sum_{i\in\II}\, \sum_{\alpha=1}^{\Ai i}
                   \sum_{\beta=1}^{\Ai i}\, \sigma(i)_\alpha^{\;\beta} $}
  \setlength{\unitlength}{24810sp}
  \setlength{\unitlength}{1pt}
  \put(0.1,-9.5)   {\scriptsize$A$}
  \put(.8,90.5)    {\scriptsize$A$}
  \put(2.4,42.9)   {\scriptsize$\sigma$}
  \put(148.2,-9.5) {\scriptsize$A$}
  \put(148.8,90.5) {\scriptsize$A$}
  \put(154.2,43.3) {\scriptsize$i$}
  \put(156.7,26.5) {\tiny$\overline\alpha$}
  \put(156.7,60.5) {\tiny$\beta$}
  \epicture25 \labl{eq:*bas}
where we use the short-hand $\langle X,Y\rangle\,{\equiv}\,\dimc\,\Hom(X,Y)$.
Inserting \erf{eq:*bas} into \erf{eq:*ssFA1} and \erf{eq:*ssFA2} 
we get the conditions 
  \be  \bearl
  \dsty\sum_{\beta=1}^{\Ai i} \sigma(i)_\alpha^{\;\beta}\,
  \sigma(i)_\beta^{\;\gamma} = \delta_{\alpha,\gamma}\, \theta_i
  \qquad {\rm and}
  \\{}\\[-.7em]
  \dsty\sum_{\rho = 1}^{\Ai k}
  \m i\alpha j\beta k\rho\delta \; \sigma(k)_\rho^{\;\gamma}
  = \sum_{\mu = 1}^{\Ai i} \sum_{\nu = 1}^{\Ai j}
  \sigma(i)_\alpha^{\;\mu}\, \sigma(j)_\beta^{\;\nu}
  \sum_{\eps=1}^{\N jik}\, \m j\nu i\mu k\gamma\eps\, \R ijk\eps\delta
  \eear  \labl{eq:*ssFA-bas}
for the matrices $\sigma(i)$, where \RR\ are the braiding matrices which 
express the braiding after a choice of bases, see (I:2.41). If we restrict 
to the particular subclass of modular tensor categories for which 
$\N ijk \iN\{0,1\}$ and to algebras with multiplicities $\Ai i \iN\{0,1\}$, 
the equations \erf{eq:*ssFA-bas} simplify considerably:
  \be
  \sigma(i)^2 = \theta_i \qquad {\rm and} \qquad
  m_{ij}^k \, \sigma(k) = \sigma(i) \, \sigma(j) \, m_{ji}^k
  \,\Rs{}ijk \,.  \labl{eq:*ssFA-bas-simple}
(Thus $\sigma(i)$ is a square root of the number $\theta_i$; note that its 
value does not depend on any gauge choices.) However, the simplifying 
assumption $\Ai i \iN \{0,1\}$ is very restrictive. Indeed, one finds that 
certain types of reversions can only be realised on algebras having 
$\Ai i \,{>}\, 1$ for some simple objects $U_i$ (in fact this happens 
already for $\su(2)_k$ WZW models with $k$ odd).

\subsection{Conjugate modules}\label{sec:com}

The reversion $\sigma$ can be combined with a braiding to turn left 
modules into right modules. This is achieved via the mapping
  \bea \begin{picture}(175,55)(0,32)
  \put(0,0)     {\includeourbeautifulpicture{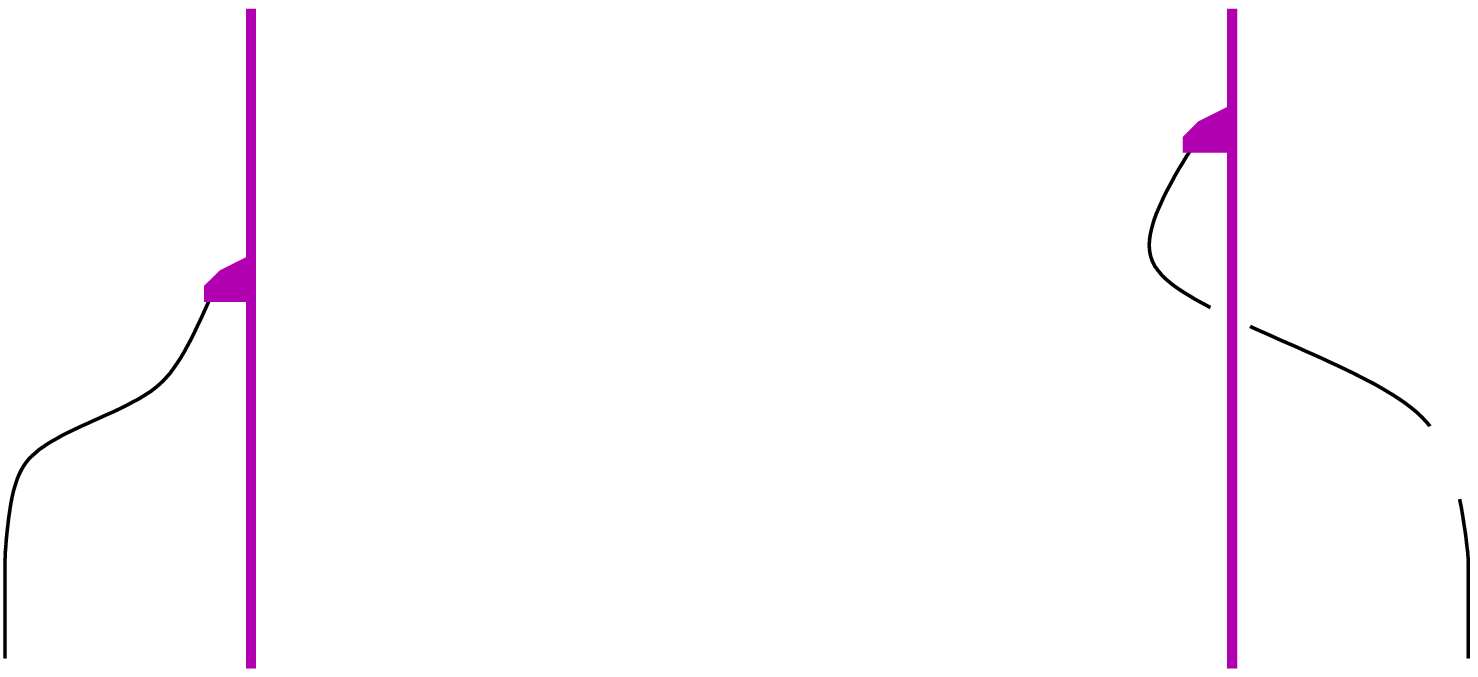}}
  \put(70,34)     {$ \longmapsto $}
  \setlength{\unitlength}{24810sp}
  \setlength{\unitlength}{1pt}
  \put(-2.5,-9.5)  {\scriptsize$A$}
  \put(22.5,-9.5)  {\scriptsize$\M$}
  \put(23,76.5)    {\scriptsize$\M$}
  \put(30,42.2)    {\scriptsize$\rho$}
  \put(131,-9.5)   {\scriptsize$\M$}
  \put(131.5,76.5) {\scriptsize$\M$}
  \put(137.5,59.3) {\scriptsize$\rho$}
  \put(156.5,21.5) {\scriptsize$\sigma$}
  \put(156.5,-9.5) {\scriptsize$A$}
  \epicture24 \labl{eq:right-mod-*}
Indeed one quickly verifies that the representation property
for right modules is satisfied:
  \bea \begin{picture}(300,94)(0,30)
  \put(0,0)     {\includeourbeautifulpicture{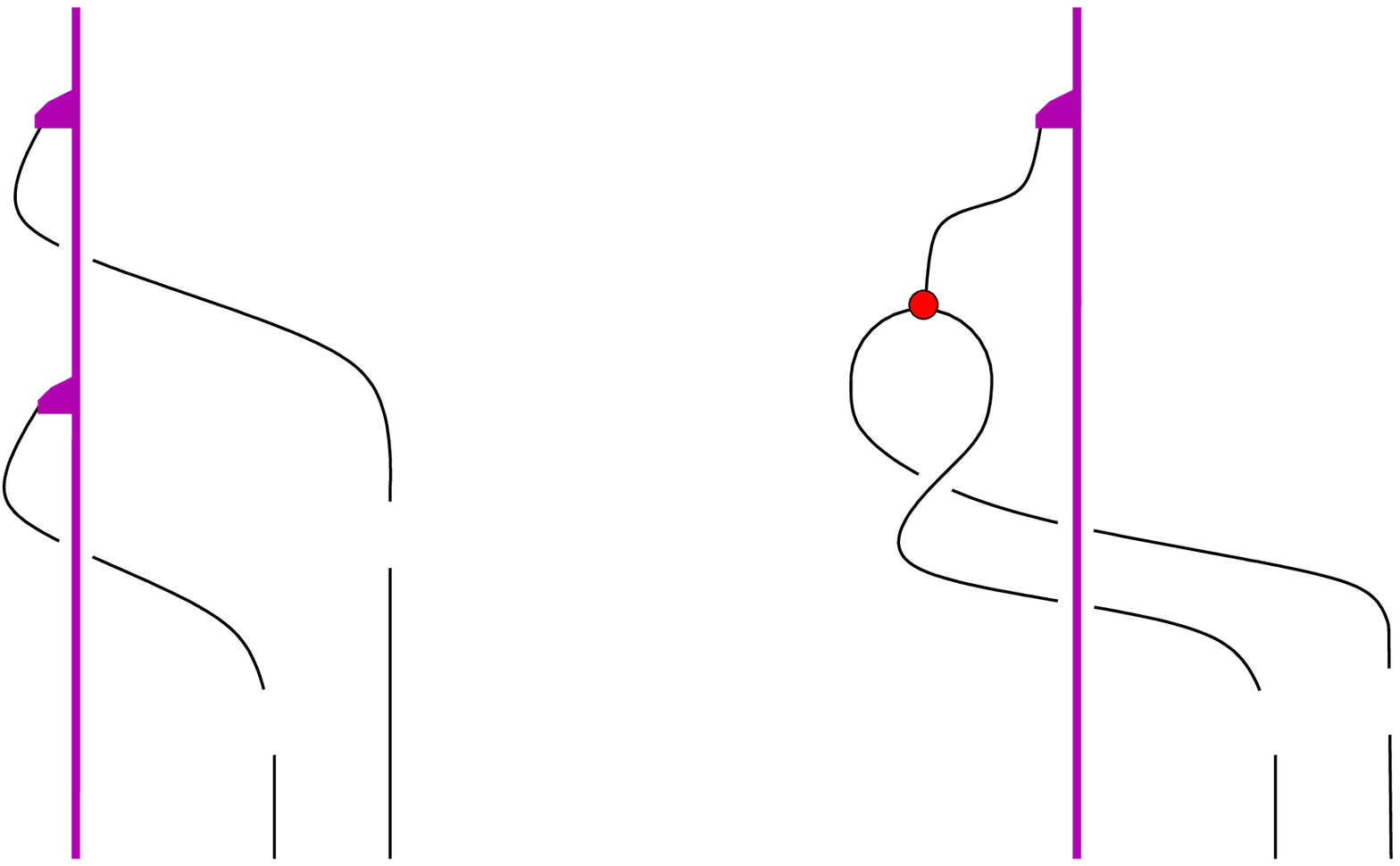}}
  \put(79,50)     {$=$}
  \put(203,50)    {$=$}
  \setlength{\unitlength}{24810sp}
  \setlength{\unitlength}{1pt}
  \put(5.5,-9.5)   {\scriptsize$\M$}
  \put(5.8,116)    {\scriptsize$\M$}
  \put(12,61.1)    {\scriptsize$\rho$}
  \put(12,99.5)    {\scriptsize$\rho$}
  \put(32.1,-8.8)  {\scriptsize$A$}
  \put(33.5,16.8)  {\scriptsize$\sigma$}
  \put(47.8,-8.8)  {\scriptsize$A$}
  \put(49.3,41.5)  {\scriptsize$\sigma$}
  \put(136,-9.5)   {\scriptsize$\M$}
  \put(136.5,116)  {\scriptsize$\M$}
  \put(143.1,98.5) {\scriptsize$\rho$}
  \put(162,-8.8)   {\scriptsize$A$}
  \put(163,17.1)   {\scriptsize$\sigma$}
  \put(178,-8.8)   {\scriptsize$A$}
  \put(179,19.1)   {\scriptsize$\sigma$}
  \put(237,-9.5)   {\scriptsize$\M$}
  \put(237.5,116)  {\scriptsize$\M$}
  \put(244.5,98.5) {\scriptsize$\rho$}
  \put(260.5,-8.8) {\scriptsize$A$}
  \put(272.5,54.8) {\scriptsize$\sigma$}
  \put(279,-8.8)   {\scriptsize$A$}
  \epicture18 \labl{eq:pic10}
There is also another way to obtain a right module, denoted by $M^\vee$, 
given a left module $M$, which does not involve a reversion, 
but rather the duality, see (I:4.60):
  \bea \begin{picture}(260,71)(0,23)
  \put(0,0)     {\includeourbeautifulpicture{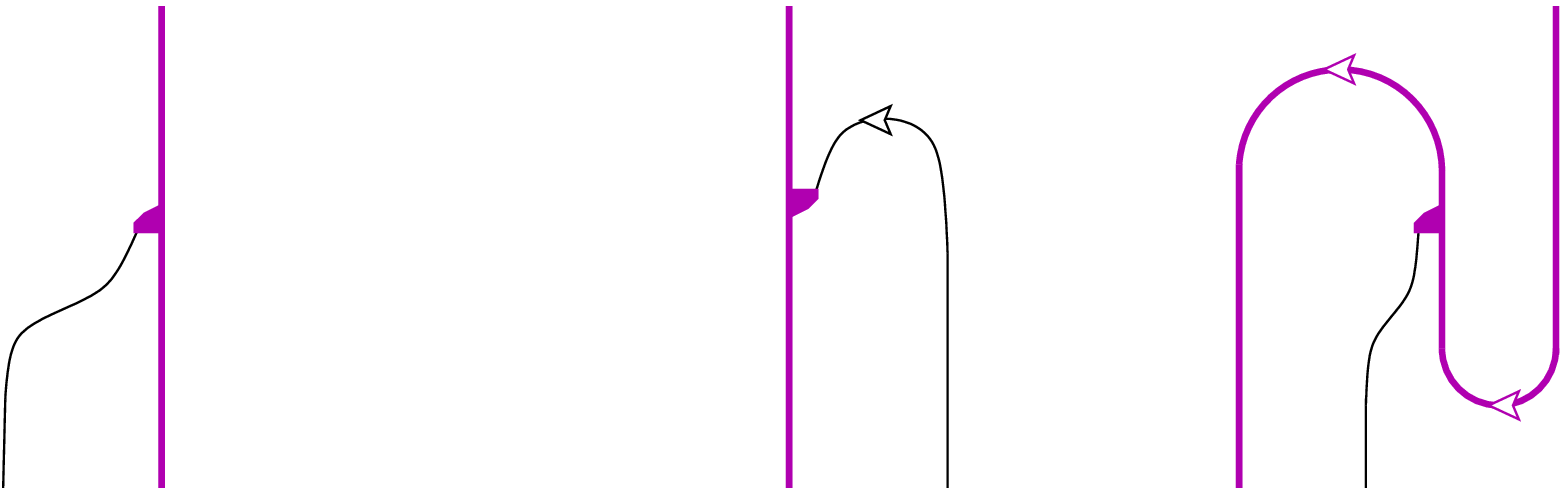}}
  \put(72,37)     {$ \longmapsto $}
  \put(182,37)    {$:=$}
  \setlength{\unitlength}{24810sp}
  \setlength{\unitlength}{1pt}
  \put(-2.9,-8.5)  {\scriptsize$A$}
  \put(22.5,-9.5)  {\scriptsize$\M$}
  \put(29.5,46)    {\scriptsize$\rho$}
  \put(23,87)      {\scriptsize$\M$}
  \put(128.5,-9.5) {\scriptsize$\M^\vee_{}$}
  \put(129.4,87)   {\scriptsize$\M^\vee_{}$}
  \put(205.5,-9.5) {\scriptsize$\M^\vee_{}$}
  \put(158,-9.5)   {\scriptsize$A$}
  \put(247.5,46)   {\scriptsize$\rho$}
  \put(229,-9.5)   {\scriptsize$A$}
  \put(260,87)     {\scriptsize$\M^\vee_{}$}
  \epicture10 \labl{eq:right-mod-no*} 
However, while \erf{eq:right-mod-*} defines a right module structure on
the same object $\M$, \erf{eq:right-mod-no*} defines it on the
dual object $\M^\vee$. 

Combining \erf{eq:right-mod-*} and \erf{eq:right-mod-no*} we
obtain a map from left modules to left modules:
  \bea \begin{picture}(350,97)(0,15)
  \put(0,0)     {\includeourbeautifulpicture{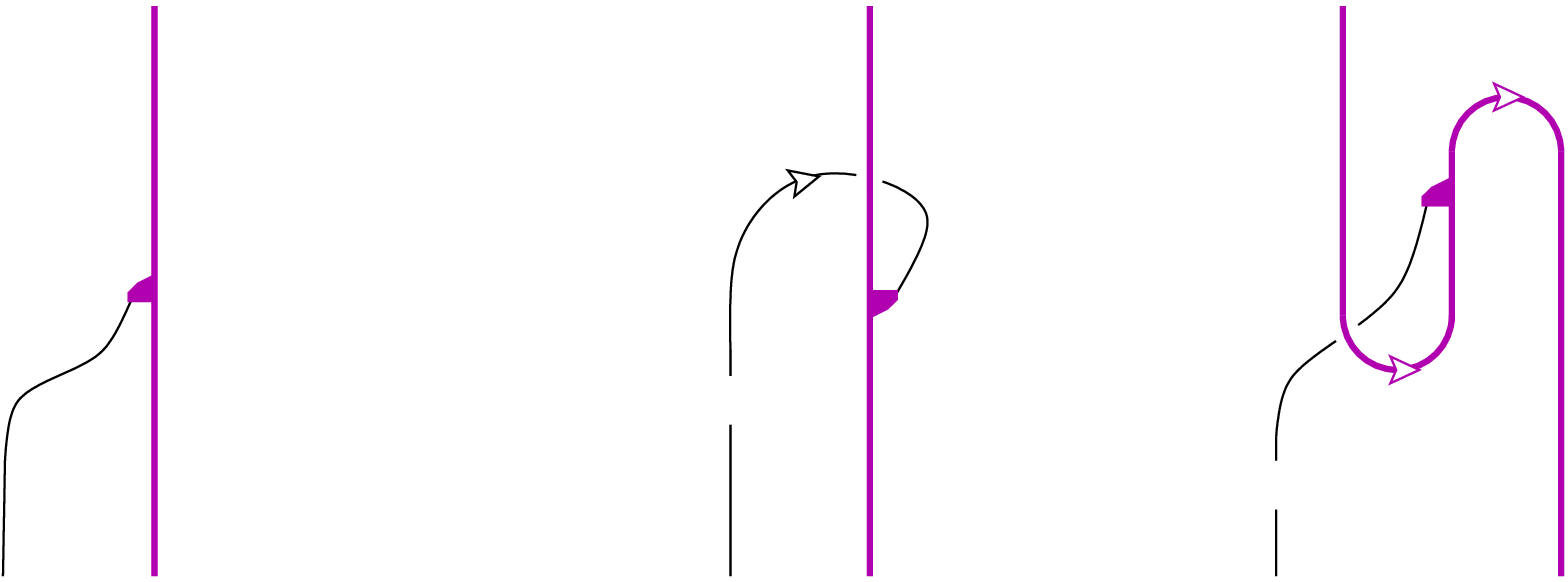}}
  \put(70,42)     {$ \longmapsto $}
  \put(188,42)    {$ \equiv $}
  \put(302,42)    {$=:\; \rho^\sigma \,. $}
  \setlength{\unitlength}{1pt}
  \put(-2.9,-8.5)  {\scriptsize$A$}
  \put(22.5,-9.5)  {\scriptsize$\M$}
  \put(29.7,51.5)  {\scriptsize$\rho$}
  \put(23.5,106)   {\scriptsize$\M$}
  \put(126,-8.5)   {\scriptsize$A$}
  \put(128,30.5)   {\scriptsize$\sigma$}
  \put(148.8,-9.5) {\scriptsize$\M^\vee{}$}
  \put(150.3,106)  {\scriptsize$\M^\vee{}$}
  \put(223,-8.5)   {\scriptsize$A$}
  \put(225,15.5)   {\scriptsize$\sigma$}
  \put(235,106)    {\scriptsize$\M^\vee{}$}
  \put(261,69.1)   {\scriptsize$\rho$}
  \put(272.3,-9.5) {\scriptsize$\M^\vee{}$}
  \epicture10 \labl{eq:rho-sigma}
This suggests the following definition.
\\[-2.3em]

\dtl{Definition}{df:conj-mod}
Let $A$ be an algebra. For a left $A$-module $M\eq(\M,\rho)$ the {\em 
conjugate\/} (left) {\em module\/} \wrt a reversion $\sigma$ on $A$ is
  \be
  M^\sigma := (\M^\vee , \rho^\sigma) \ee
with $\rho^\sigma$ given in \erf{eq:rho-sigma}.

\medskip

$M^\sigma$ is again a left $A$-module.
Note that $M^{\sigma\sigma}$ is in general not equal to $M$, just like
$U^{\vee\vee}$ is in general not equal to $U$ as an object of $\calc$. 
But $M^{\sigma\sigma}$ and $M$ are isomorphic as 
left $A$-modules. To see this recall that an isomorphism $\delta_U$
between $U$ and $U^{\vee\vee}$ can be given with the help of the left
and right dualities (cf.\ (I:2.13) and chapter 2.2 of \cite{BAki}):
  \be
  \delta_U = (\id_{U^{\vee\vee}} \oti d_U )
  \circ (\tilde b_{U^\vee} \oti \id_U) \,\in \Hom(U,U^{\vee\vee}) \,.  
  \labl{deltaU}
The morphisms $\delta_U$ satisfy
$\delta_{U\otimes V}\eq\delta_U \oti\delta_V$ and
$\delta_{U^\vee}\eq( (\delta_U)^\vee )^{-1}$. 
It can be checked by direct substitution (being careful
not to mix up the left and right dualities) that an
isomorphism between $M^{\sigma\sigma}$ and $M$ is
provided by $\delta_\M$, i.e.\ we have 
  \be
  \rho = \delta_\M^{-1} \circ \rho^{\sigma\sigma}\!
  \circ (\id_A \oti\delta_\M) \,.  \labl{eq:rho-sigsig}

As a word of warning, let us recall that \cite{fuSc16} with suitable
assumptions on the algebra $A$ it can happen that the category of left
$A$-modules has the structure of a tensor category with duality. In that case,
the conjugate module as introduced in definition \ref{df:conj-mod} is,
however, not necessarily equal to the dual module in the sense of this duality.

The following proposition lists some further properties of conjugate modules.
\\[-2.3em]

\dtl{Proposition}{pr:*mod}
Let $A$ be a \JA, with reversion $\sigma$.
\\[.3em]
(i)~\,\,The mapping $M\,{\mapsto}\,M^ \sigma$ and $f \,{\mapsto}\, f^\vee$ 
furnishes a contravariant endofunctor of \calca. Thus for any two $A$-modules
$M$, $N$, the mapping $f \,{\mapsto}\, f^\vee$ furnishes an isomorphism
$\HomA(M,N) \,{\cong}\,\HomA
   $\linebreak[0]$ 
(N^\sigma\!{,}\,M^\sigma)$.
\\[.3em]
(ii)~\,The square of this endofunctor is equivalent to the identity functor,
so that $\HomA(M,N^\sigma) \,{\cong}
   $\linebreak[0]$ 
\HomA(N,M^\sigma)$.
\\[.3em]
(iii)~For induced modules one has
$(\inda(U))^\sigma \,{\cong}\, \inda(U^\vee)$ as left $A$-modules.
In particular, $A^\sigma \,{\cong}\, A$.

\medskip\noindent
Proof:\\
(i)~\,\,Given a morphism $f \iN \HomA(M,N)$ we must verify that
$f^\vee \iN \HomA(N^\sigma\!{,}\,M^\sigma)$.
Writing out the definitions, this amounts to proving that
  \bea \begin{picture}(385,114)(0,22)
  \put(110,0)    {\includeourbeautifulpicture{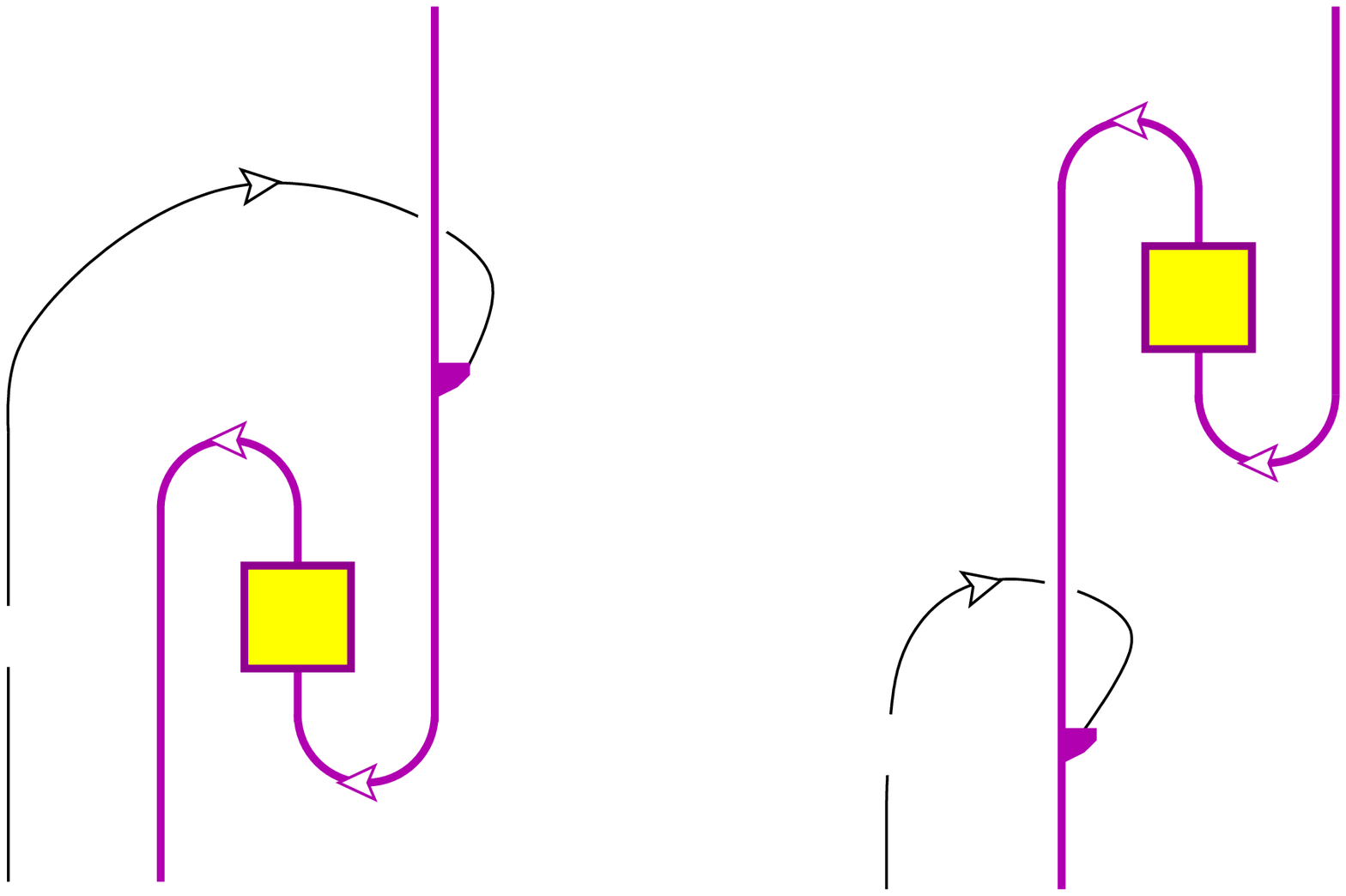}}
  \put(0,53)      {$ \rho^\sigma \circ (\id_A \otimes f^\vee) \;=$}
  \put(205,53)    {$=$}
  \put(320,53)    {$ =\; f^\vee \circ \rho^\sigma \,.$}
  \setlength{\unitlength}{1pt}
  \put(110.6,-8.5) {\scriptsize$A$}
  \put(112,35.3)   {\scriptsize$\sigma$}
  \put(130.5,-9.5) {\scriptsize$\dot N^\vee_{}$}
  \put(153.4,37.9) {\scriptsize$f$}
  \put(169.9,130)  {\scriptsize$\dot M^\vee_{}$}
  \put(235,-8.5)   {\scriptsize$A$}
  \put(236,20.3)   {\scriptsize$\sigma$}
  \put(258.5,-9.5) {\scriptsize$\dot N^\vee_{}$}
  \put(281,83.1)   {\scriptsize$f$}
  \put(297.9,130)  {\scriptsize$\dot M^\vee_{}$}
  \epicture13 \labl{eq:f^-check}
This equation can be seen to hold by deforming the $\M$- and
$\dot N$-ribbons to move $f$ close to the representation morphism
$\rho$ and using that $f$ is in $\HomA(M,N)$. That this mapping is
linear and invertible is obvious.
\\[.3em]
(ii)~\,We have already seen above that 
$M^{\sigma\sigma}\,{\cong}\,M$ via the morphism $\delta_U$ \erf{deltaU}.
To check the statement for morphisms, we note that
$f \,{\mapsto}\, f \cir \delta_{\dot N}$ provides an isomorphism from
$\HomA(N^{\sigma\sigma}\!,M)$ to $\HomA(N,M)$.
Combining this with the isomorphism from (i) gives
  \be  \HomA(M,N^\sigma) \cong \HomA(N^{\sigma\sigma}\!{,}\,M^{\sigma})
  \cong \HomA(N,M^\sigma) \,.  \ee

\vskip.3em\noindent
(iii)~Consider the morphism
$\varphi\iN\Hom( A{\otimes}U^\vee{,}\,U^\vee{\otimes}A^\vee)$ given by
  \be
  \varphi := (c^{}_{U^\vee,A^\vee})^{-1}
  \circ (\Phi_2 \oti\id_{U^\vee}) \circ (\sigma^{-1}\oti\id_{U^\vee}) \,.
  \ee
Its inverse is easily computed to be
  \be
  \varphi^{-1} = (\sigma \oti\id_{U^\vee})\circ
  (\Phi_2^{-1} \oti\id_{U^\vee}) \circ c^{}_{U^\vee,A^\vee} \ee
(as $A$ is symmetric, the morphism $\Phi_2$, defined in (I:3.33), is equal 
to $\Phi_1$, see above; $\Phi_2^{-1}$ is given explicitly in (I:3.35)).
Denote now by $\rho$ the representation morphism of $\inda(U^\vee)$.
To establish that $(\inda(U))^\sigma$ is isomorphic to
$\inda(U^\vee)$ it is enough to show that
$\rho\eq\varphi^{-1} \cir \rho^\sigma \cir (\id_A\oti\varphi)$.
This equality, in turn, can be seen as follows.
  \bea \begin{picture}(420,75)(15,32)
  \put(128,0)    {\includeourbeautifulpicture{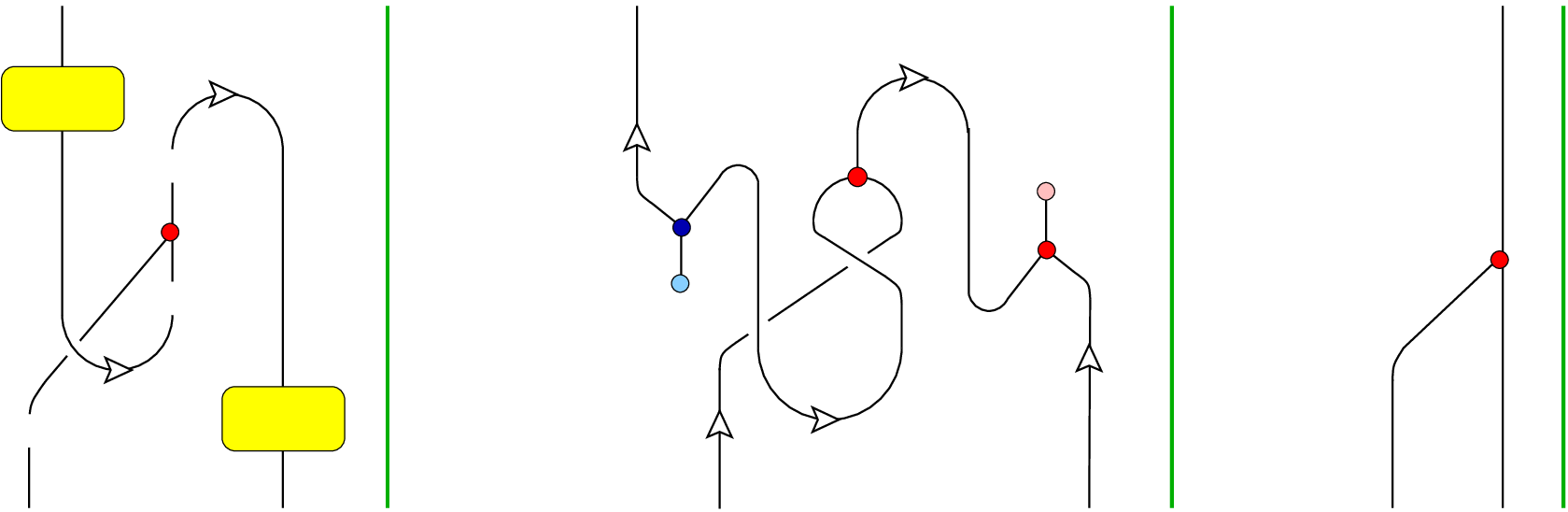}}
  \put(0,45)      {$ \varphi^{-1} \circ \rho^\sigma \circ
                   (\id_A \otimes \varphi) \;=$}
  \put(225,45)    {$ = $}
  \put(375,45)    {$ = $}
  \setlength{\unitlength}{1pt}
  \put(130,-8.5)   {\scriptsize$A$}
  \put(132,14.2)   {\scriptsize$\sigma$}
  \put(134,78.4)   {\scriptsize$\Phi_2^{\!-\!1}$}
  \put(137,103)    {\scriptsize$A$}
  \put(158.6,65.8) {\scriptsize$\sigma^{\!-\!1}$}
  \put(160,40.5)   {\scriptsize$\sigma$}
  \put(179,16)     {\scriptsize$\Phi_2$}
  \put(180,-8.5)   {\scriptsize$A$}
  \put(199,-9)     {\scriptsize$U^\vee$}
  \put(199,103)    {\scriptsize$U^\vee$}
  \put(249,103)    {\scriptsize$A$}
  \put(265,-8.5)   {\scriptsize$A$}
  \put(338,-8.5)   {\scriptsize$A$}
  \put(353,-9)     {\scriptsize$U^\vee$}
  \put(353,103)    {\scriptsize$U^\vee$}
  \put(397,-8.5)   {\scriptsize$A$}
  \put(418,-8.5)   {\scriptsize$A$}
  \put(418.6,103)  {\scriptsize$A$}
  \put(430,-9)     {\scriptsize$U^\vee$}
  \put(430,103)    {\scriptsize$U^\vee$}
  \epicture18 \ee
In the first step \erf{eq:sig-phi} together with the explicit expression for
$\varphi$ and $\varphi^{-1}$ is
inserted. The $U^\vee$-ribbon can then be disentangled, resulting in the first
picture. The second picture is obtained by using \erf{eq:*ssFA2} and
the explicit form of $\Phi_2$ and $\Phi_2^{-1}$. The last step amounts
to a deformation of the $A$-ribbons and then associativity, Frobenius,
unit and counit property. 
\\
The special case $A^\sigma \,{\cong}\, A$ is obtained when setting $U\eq\one$.
\qed

The mapping $M \,{\mapsto}\, M^\sigma$ gives rise to an involution on the set 
$\J\eq\{M_\kappa \,|\, \kappa\eq1,...\,,|\J|\}$ of
representatives of isomorphism classes of simple $A$-modules.
Some properties of this involution are collected in the following 
\\[-2.3em]

\dtl{Proposition}{pr:CB-prop}
(i)~~\,For $M_\rho$ a simple left $A$-module, the left $A$-module 
${(M_\rho)}^\sigma$ is simple, too.
\\[.3em]
Further, setting 
  \be
  C^\sigma_{\kappa\rho} := \dimc\,\HomA(M_\kappa,{(M_\rho)}^\sigma) \,,  
  \labl{eq:CB-def} 
we have\\[3pt]
(ii)~~\,$(M_\kappa)^\sigma\,{\cong}\,\bigoplus_{\rho\in\JJ}
    C^\sigma_{\kappa\rho} M_\rho\,$ as left $A$-modules.\\[3pt]
(iii)~\,The matrix $C^\sigma\eq(C^\sigma_{\kappa\rho})$ is symmetric and
    squares to the unit matrix.
\\[.3em]
(iv)~$\;C^\sigma$ is a permutation matrix of order two.
\\[.3em]
(v)~\,\,\,If $A$ is haploid, then it is simple as a module over itself,
$A\,{\cong}\,M_{\kappa_\circ}$ for some $\kappa_\circ\iN\J$, and
$C^\sigma_{\kappa_\circ\rho} \eq \delta_{\kappa_\circ,\rho}$.  

\medskip\noindent
Proof:\\
(i)~~\,The statement that $(M_\rho)^\sigma$ is simple is equivalent to
$\dimc\,\HomA((M_\rho)^\sigma,(M_\rho)^\sigma)\eq1$. The latter, in turn, 
is an immediate
consequence of $M_\rho$ being simple and proposition \ref{pr:*mod}(i). 
\\[.3em]
(ii)~~Since the category $\calca$ of left $A$-modules is semisimple, 
every $A$-module $M$ decomposes as an $A$-module as
$M\,{\cong}\,\bigoplus_{\kappa\in\JJ} \lra{M_\kappa}M M_\kappa$, abbreviating 
$\lra MN\,{\equiv}\, \dimc\,\HomA(M,N)$ as introduced in (I:4.5). When $M$ 
is simple, this yields (ii).
\\[.3em]
(iii)~Symmetry of $C^\sigma_{\kappa\rho}$ is a consequence of
proposition \ref{pr:*mod}(i).
\\
Specialising the general relation $\HomA(M,N)\,{\cong}\,\bigoplus_{\lambda\in\JJ}
\HomA(M,M_\lambda)\oti \HomA(M_\lambda, N)$ to the case $M\eq(M_\kappa)^\sigma$ and
$N\eq(M_\rho)^\sigma$ and comparing dimensions results in
$\delta_{\kappa,\rho}\eq\sum_{\lambda\in\JJ}
C^\sigma_{\kappa\lambda} C^\sigma_{\lambda\rho}$.
\\[.3em]
(iv)~\,Since $(M_\rho)^\sigma$ is again simple, it is isomorphic to
precisely one $M_\kappa$ with $\kappa\iN\J$. 
It follows that $C^\sigma$ is a permutation, which by (iii) has order two.
\\[.3em]
(v)~\,\,follows directly from proposition \ref{pr:*mod}(ii).
\qed

\dt{Remark}
(i)~\,Recall that simple $A$-modules correspond to elementary boundary conditions.
Thus the matrix $C^\sigma$ is precisely the {\em boundary conjugation matrix\/}
that was introduced, based on the idea of ``complex charges'' \cite{bisa4},
in \cite{prss3} (see also \cite{sosc}). Also recall that the $A$-modules give 
rise to a NIMrep of the fusion rules of the category \calc; the additional 
structure provided by the matrix $C^\sigma$ has been termed an `$S$-NIMrep' 
in \cite{sosc}. We will return to this point in the context of
annulus coefficients in section \ref{sec:ann}, where we also give an explicit
formula for $C^\sigma_{\kappa\rho}$ in terms of the reversion $\sigma$ on $A$.
\\[.3em]
(ii)~Proposition \ref{pr:*mod}(ii) implies that for {\em induced\/} modules the
effect of conjugation is independent of $\sigma$. As a consequence, a
characteristic feature of a reversion $\sigma$ is its action on submodules
of induced modules. Concretely, suppose that
  \be
  \inda(U_k) \cong M^{(k)}_1 \,{\oplus} \cdots {\oplus}\, M^{(k)}_{n_k} \ee
is the decomposition of the induced module $\inda(U_k)$ into simple
submodules. Because of $(U_k)^\vee \,{\cong}\, U_{\bar k}$, for any
$i\eq1,2,...\,,n_k$, we then have $(M^{(k)}_i)^\sigma\,{\cong}\,M^{(\bar k)}_j$
for some $j$. These conditions restrict the possible actions of $\sigma$ on
simple modules. For example, for $k\eq\bar k$ every $\sigma$
only permutes the simple submodules of $\inda(U_k)$ amongst themselves.
\\[.3em]
(iii)~When $A$ is not haploid, then it is not simple as a module over itself,
and boundary conjugation acts, in general, non-trivially on the simple
submodules of $A$.

\medskip

In order to discuss defect lines on non-orientable world sheets we
need a notion of conjugation also for bimodules. In fact we will
make use of three different types of conjugation. Our conventions for
$A$-$B$-bimodules are given in definition I:4.5.
An $A$-$B$-bimodule $X$ is a triple $X\eq(X,\rho,\tilde\rho)$, where
$\rho$ and $\tilde\rho$ are a left action of $A$ and a commuting right action
of $B$ on $X$, respectively. The following result is a straightforward
consequence of the definition of $\sigma$ and of a left/right action.
It can be established by moves analogous to \erf{eq:pic10}; we
omit the details of the proof.
\\[-2.2em]

\dtl{Proposition}{pr:bimod-conj}
Let $X\eq(\dot X, \rho , \tilde \rho)$ be an $A$-$B$-bimodule. Then
  \be
  X^s = (\dot X, \rho^s, \tilde \rho^s)
  \qquad {\rm and} \qquad
  X^v = (\dot X^\vee, \rho^v, \tilde \rho^v) \labl{eq:bimod-sv}
with
  \be\bearll
  \rho^s := \tilde\rho \circ c_{BX} \circ (\sigmaB \oti \id_X) \,, &
  \tilde\rho^s := \rho \circ c_{XA} \circ (\id_X\oti\sigmaA) \,,
  \\{}\\[-.6em]
  \rho^v := (\tilde d_B{\otimes}\id_{\dot X^\vee}) \circ
    (\id_B{\otimes} (\tilde\rho)^\vee) \,,\quad &
  \tilde\rho^v := (\id_{\dot X^\vee}{\otimes}d_A) \circ
    ((\tilde\rho)^\vee{\otimes}\id_A) \,,
  \eear\ee
are $B$-$A$-bimodules, and
  \be
  X^\sigma = (\dot X^\vee, \rho^\sigma, \tilde \rho^\sigma)
  \labl{eq:bimod-sig}
with
  \be\bearll
  \rho^\sigma := (\tilde d_A{\otimes}\id_{\dot X^\vee})
    \circ(\sigmaA{\otimes}(\rho\cir c_{XA})^\vee) \,, \quad &
  \tilde\rho^\sigma := (\id_{\dot X^\vee}{\otimes}d_B) \circ
    ((\tilde\rho\cir c_{BX})^\vee{\otimes}\sigmaB)
  \eear\ee
is an $A$-$B$-bimodule.

\medskip

Pictorially, the three types of conjugation look as follows:
  \bea \begin{picture}(420,99)(0,24)
  \put(45,0)     {\includeourbeautifulpicture{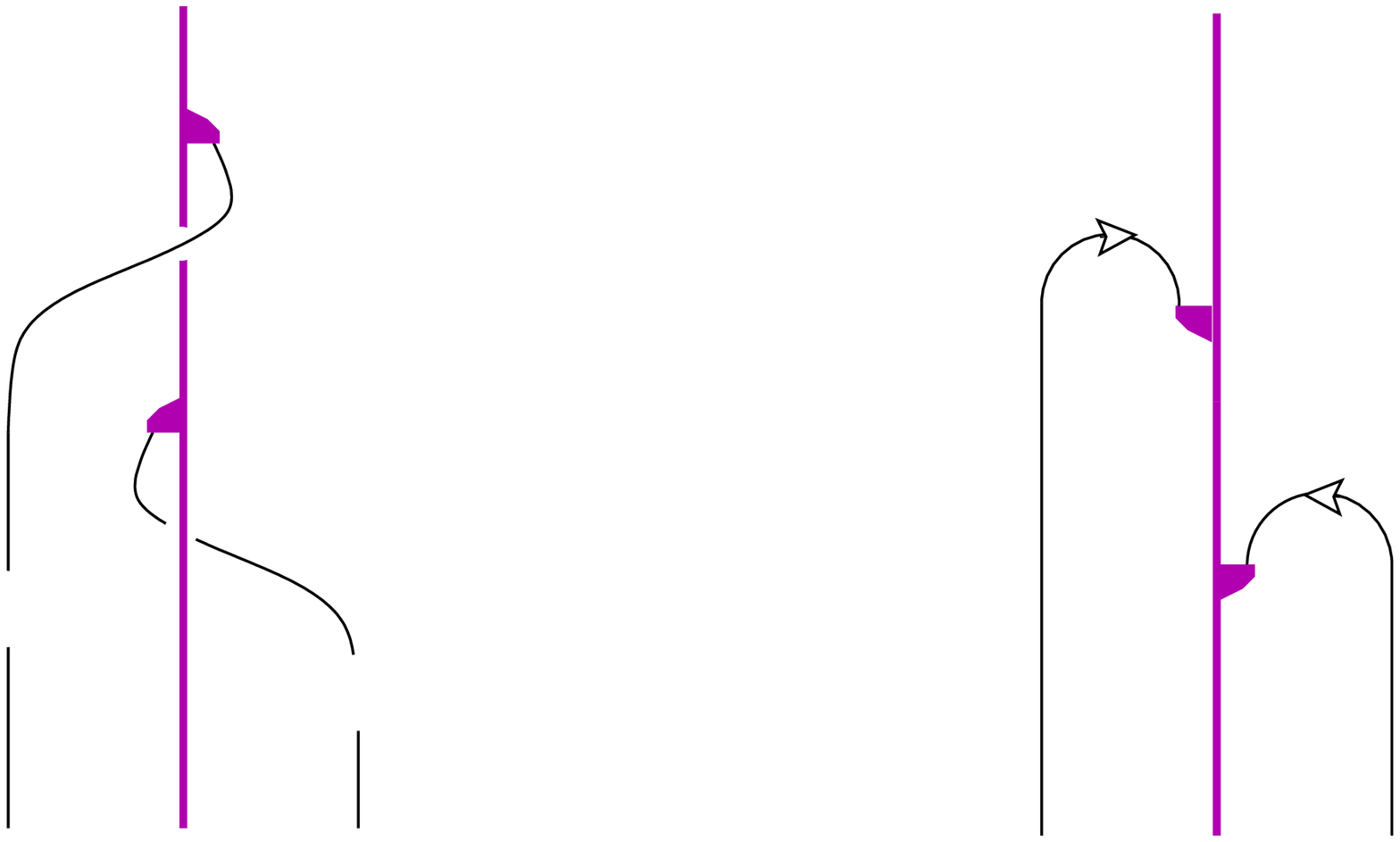}}
  \put(0,53)       {$ X^s \; :=$}
  \put(147,53)     {$ X^v \; :=$}
  \put(300,53)     {$ X^\sigma \; :=$}
  \setlength{\unitlength}{24810sp}
  \setlength{\unitlength}{1pt}
  \put(47,31.7)    {\scriptsize$\sigma_{\!B}$}
  \put(48,-9.2)    {\scriptsize$B$}
  \put(73,-9.2)    {\scriptsize$X$}
  \put(73,122)     {\scriptsize$X$}
  \put(96.7,20.4)  {\scriptsize$\sigma_{\!A}$}
  \put(97,-9.2)    {\scriptsize$A$}
  \put(194,-9.2)   {\scriptsize$B$}
  \put(218,-9.2)   {\scriptsize$X^\vee$}
  \put(218,122)    {\scriptsize$X^\vee$}
  \put(243,-9.2)   {\scriptsize$A$}
  \put(348,64.7)   {\scriptsize$\sigma_{\!A}$}
  \put(350.4,-9.2) {\scriptsize$A$}
  \put(373.5,-9.2) {\scriptsize$X^\vee$}
  \put(373.5,122)  {\scriptsize$X^\vee$}
  \put(397.8,21.7) {\scriptsize$\sigma_{\!B}$}
  \put(399,-9.2)   {\scriptsize$B$}
  \epicture16 \labl{eq:Xsvs-pics}

In relation with defect lines on non-orientable surfaces one
is lead to investigate the conjugate module $X^s$ in more
detail. The following proposition tells us how this conjugation
acts on $\alpha$-induced bimodules (recall the definition of
the latter from section I:5.4).
\\[-2.3em]

\dtl{Proposition}{pr:bimod-alphaind}
For $U$ an object of $\calc$, there are the isomorphisms 
  \be
  (\alpha^+(U))^s \cong \alpha^-(U) \qquad {\rm and} \qquad
  (\alpha^-(U))^s \cong \alpha^+(U) \,.  \ee
of $A$-$A$-bimodules

\medskip\noindent
Proof:\\
Let us demonstrate $(\alpha^+(U))^s \,{\cong}\, \alpha^-(U)$; the 
second identity can be seen in a similar fashion. Consider
the morphism $\psi\,{:=}\, c_{U\!A} \cir c_{AU} \cir (\sigma{\otimes}\id)
\in\Hom(A{\otimes}U,A{\otimes}U)$. Clearly, $\psi$ is invertible. 
Further, denoting by $\rho$ and $\tilde\rho$ the left/right 
representation of $A$ for $\alpha^-(U)$, the following moves
establish that $\psi$ commutes with the representation morphisms
of $\alpha^{\pm}(U)$:
  \begin{eqnarray}  \begin{picture}(420,115)(16,32)
  \put(105,0)   {\includeourbeautifulpicture{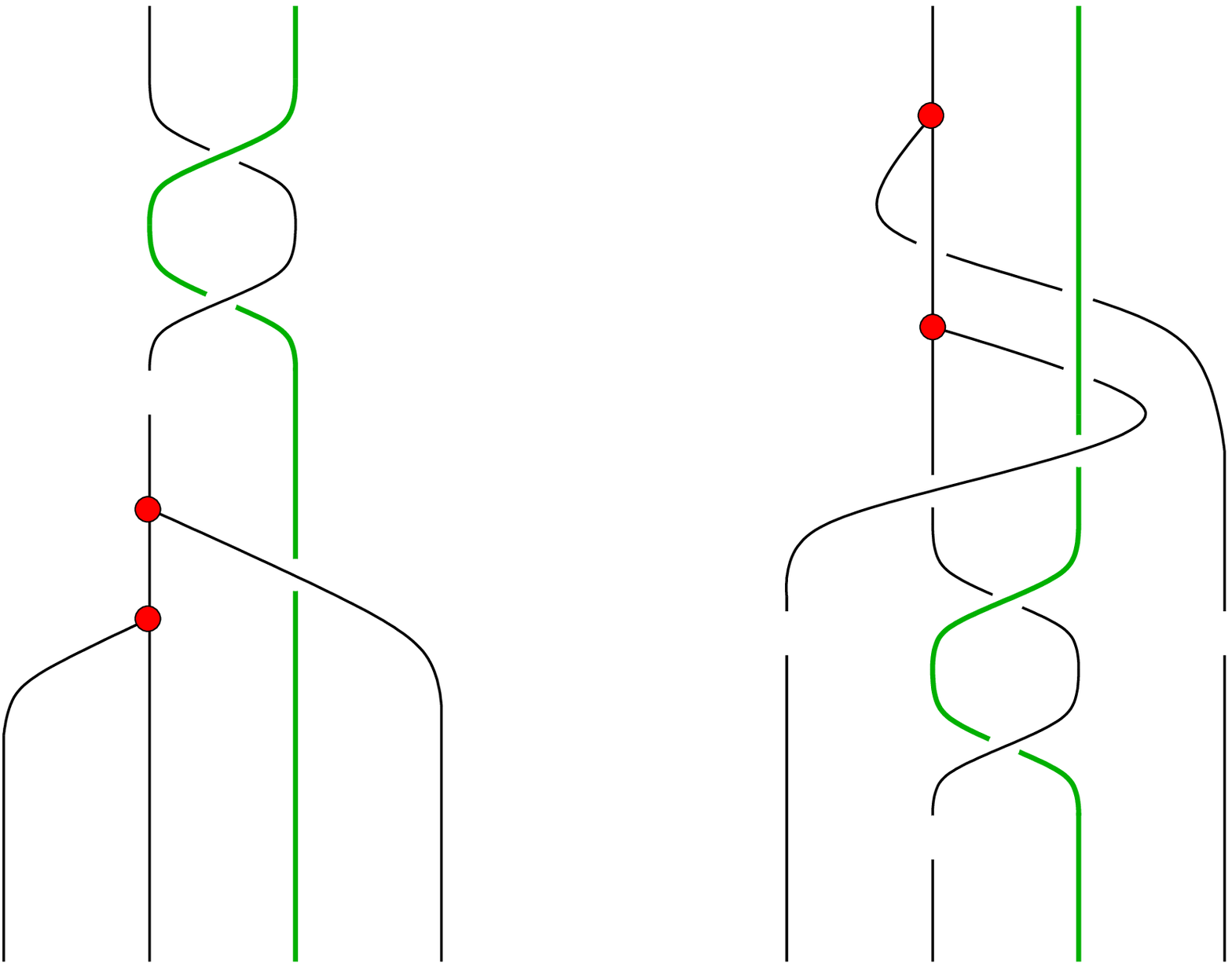}}
  \put(0,66)      {$ \psi \circ \tilde\rho \circ (\rho{\otimes}\id_A)\; =$}
  \put(190,66)    {$ = $}
  \put(300,66)    {$ =\; \tilde\rho^s \circ (\rho^s{\otimes}\id_A)
                     \circ (\id_A{\otimes}\psi{\otimes}\id_A) \,.$}
  \setlength{\unitlength}{24810sp}
  \setlength{\unitlength}{1pt}
  \put(102,-9.2)   {\scriptsize$A$}
  \put(124,-9.2)   {\scriptsize$A$}
  \put(124,146)    {\scriptsize$A$}
  \put(125,83.5)   {\scriptsize$\sigma$}
  \put(146,-9.2)   {\scriptsize$U$}
  \put(146,146)    {\scriptsize$U$}
  \put(167,-9.2)   {\scriptsize$A$}
  \put(218,-9.2)   {\scriptsize$A$}
  \put(219,48.1)   {\scriptsize$\sigma$}
  \put(239,-9.2)   {\scriptsize$A$}
  \put(239,146)    {\scriptsize$A$}
  \put(240.7,17.1) {\scriptsize$\sigma$}
  \put(261,-9.2)   {\scriptsize$U$}
  \put(261,146)    {\scriptsize$U$}
  \put(282,-9.2)   {\scriptsize$A$}
  \put(283.7,48.1) {\scriptsize$\sigma$}

  \end{picture} \nonumber\\[1.5em]{} \label{pic127}
  \\[-1.5em]{}\nonumber\end{eqnarray}
Thus we have $\psi\iN\Hom_{\!A|A}(\alpha^-(U),(\alpha^+(U))^s)$,
showing that indeed the two bimodules are isomorphic.
\qed


\subsection{Reversions on Morita equivalent \alg s}\label{sec:morita}

When studying the relationship between algebras and conformal field
theory, it is natural to introduce a concept of CFT-equivalence for
algebras. Loosely speaking:
\begin{quote}
  {\sl Two algebras are CFT-equivalent iff the associated full conformal 
  field theories have the same correlators, up to a suitable relabelling
  of fields, boundary conditions and defects.}
\end{quote}
It was already noted in \cite{fuRs4} that when considering the class 
of CFTs defined on oriented world sheets, two (symmetric special Frobenius) 
algebras are CFT-equivalent if they are Morita equivalent.
However, as was already mentioned above, the number of possible reversions 
for an algebra is not invariant under Morita equivalence. 
Thus when considering CFTs defined on unoriented world sheets
and \JA s, CFT-equivalence of algebras will amount to a certain
refinement of the notion of Morita equivalence.

The purpose of this section is to provide a sufficient criterion, given
in proposition \ref{pr:B-star}, for when a reversion on an algebra $A$ 
induces a reversion on a Morita equivalent algebra $B$. We expect
that two simple \JA s are CFT-equivalent if they are Morita equivalent and 
if the construction of proposition \ref{pr:B-star} is applicable. We will 
return to this issue in future work.

\medskip

Suppose we are given a symmetric special Frobenius algebra $A$ and a 
left $A$-module $M$. In this section we will in addition assume that $A$ is a 
{\em simple\/} algebra, which means that $Z(A)_{00}\eq1$. If this is not the 
case, $A$ can be written as a direct sum of several simple algebras, which 
corresponds to the superposition of several CFTs, see the discussion 
at the end of section I:3.2.

Below we will several times make use of subobjects that are defined
as images of idempotents. The conventions we use are summarised in
the following definition.
\\[-2.3em]

\dtl{Definition}{df:ImP}
Let $P\iN\Hom(U,U)$ be an idempotent, $P\cir P\eq P$. A triple $(S,e,r)$, 
consisting of an object $S\iN\objc$ and morphisms $e\iN\Hom(S,U)$ and 
$r\iN\Hom(U,S)$, is called a {\em retract\/} of $P$ iff
  \be
  e \circ r = P  \qquad {\rm and} \qquad r \circ e = \id_S \,.
  \labl{eq:er-prop}
The subobject $S$ of $U$ is then called the {\em image\/} of $P$.

\medskip
Since by assumption the modular tensor category $\calc$ is in particular 
semisimple and abelian, 
morphisms $e$ and $r$ satisfying \erf{eq:er-prop} exist for every idempotent 
$P$. Note further that $P \cir e\eq e \cir r \circ e\eq e$. Together with a 
similar argument for $r$, this leads to the relations
  \be
  P \circ e = e \qquad {\rm and} \qquad r \circ P = r \,.  \labl{eq:er-prop2}
Given an $A$-module $M$, consider the idempotent 
$P_{\otimes A} \iN \Hom(M^\vee{\otimes}M,M^\vee{\otimes}M)$ given by
  \bea \begin{picture}(120,35)(0,16)
  \put(58,0)    {\includeourbeautifulpicture{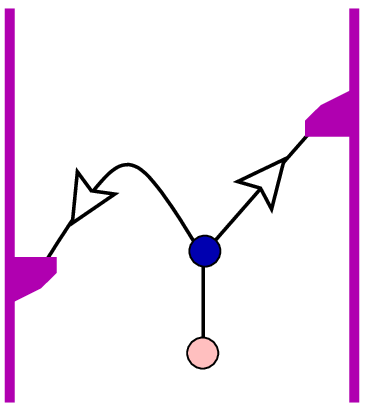}}
  \put(0,22)       {$ P_{\otimes A} \; :=$}
  \setlength{\unitlength}{24810sp}
  \setlength{\unitlength}{1pt}
  \put(53,-9.9)    {\scriptsize$\M^\vee$}
  \put(53,48)      {\scriptsize$\M^\vee$}
  \put(77,27)      {\scriptsize$A$}
  \put(92,-9.9)    {\scriptsize$\M$}
  \put(92,48)      {\scriptsize$\M$}
  \epicture02 \labl{eq:Ptensor}
That this in indeed an idempotent follows from the same moves as in (I:5.127).
The image of this idempotent is a subobject $B$ of $M^\vee{\otimes}M$, 
isomorphic to $M^\vee\OtA M$. The graphical notation we will use for $e$ and 
$r$ is
  \bea \begin{picture}(200,36)(0,17)
  \put(40,0)    {\includeourbeautifulpicture{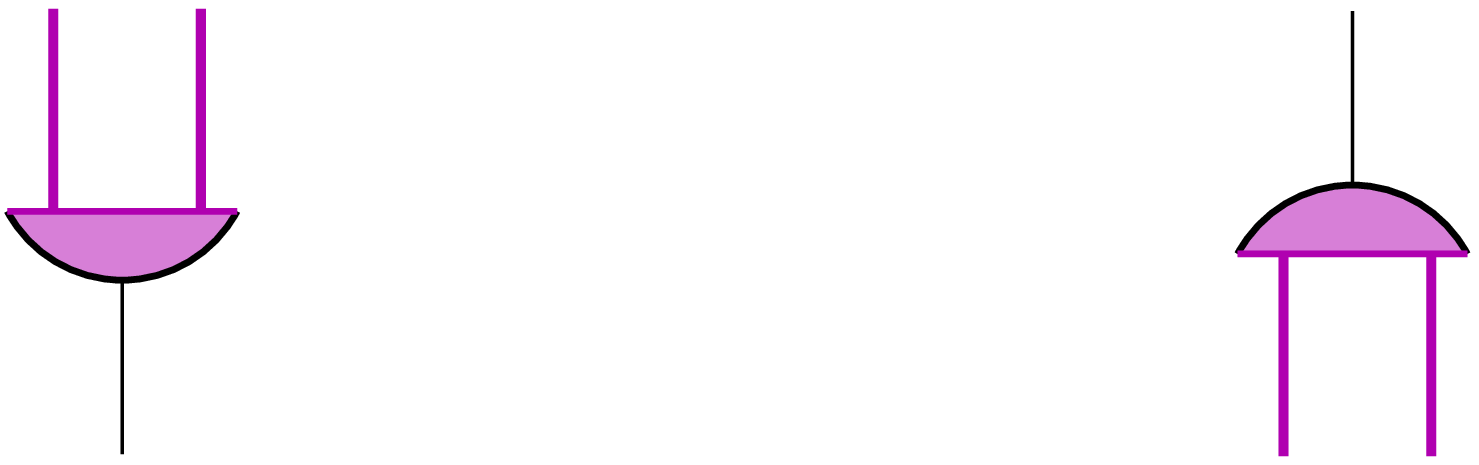}}
  \put(0,22)      {$ e \; :=$}
  \put(135,22)    {$ r \; :=$}
  \setlength{\unitlength}{24810sp}
  \setlength{\unitlength}{1pt}
  \put(41,54.5)    {\scriptsize$\M^\vee$}
  \put(49.3,-8.8)  {\scriptsize$B$}
  \put(59,54.5)    {\scriptsize$\M$}
  \put(176,-9.9)   {\scriptsize$\M^\vee$}
  \put(186,54.5)   {\scriptsize$B$}
  \put(194,-9.9)   {\scriptsize$\M$}
  \epicture02 \labl{eq:pic50}

\medskip
\dtl{Proposition}{pr:B-ssFA}
Let $A$ be a simple symmetric special Frobenius algebra, $M$ an $A$-module,
and let $B$ the image of the idempotent $P_{\otimes A}$ in \erf{eq:Ptensor}.
The object $B$ together with the morphisms $m, \eta, \Delta, \eps$ given by
  \bea \begin{picture}(360,181)(0,0)
  \put(40,2)    {\includeourbeautifulpicture{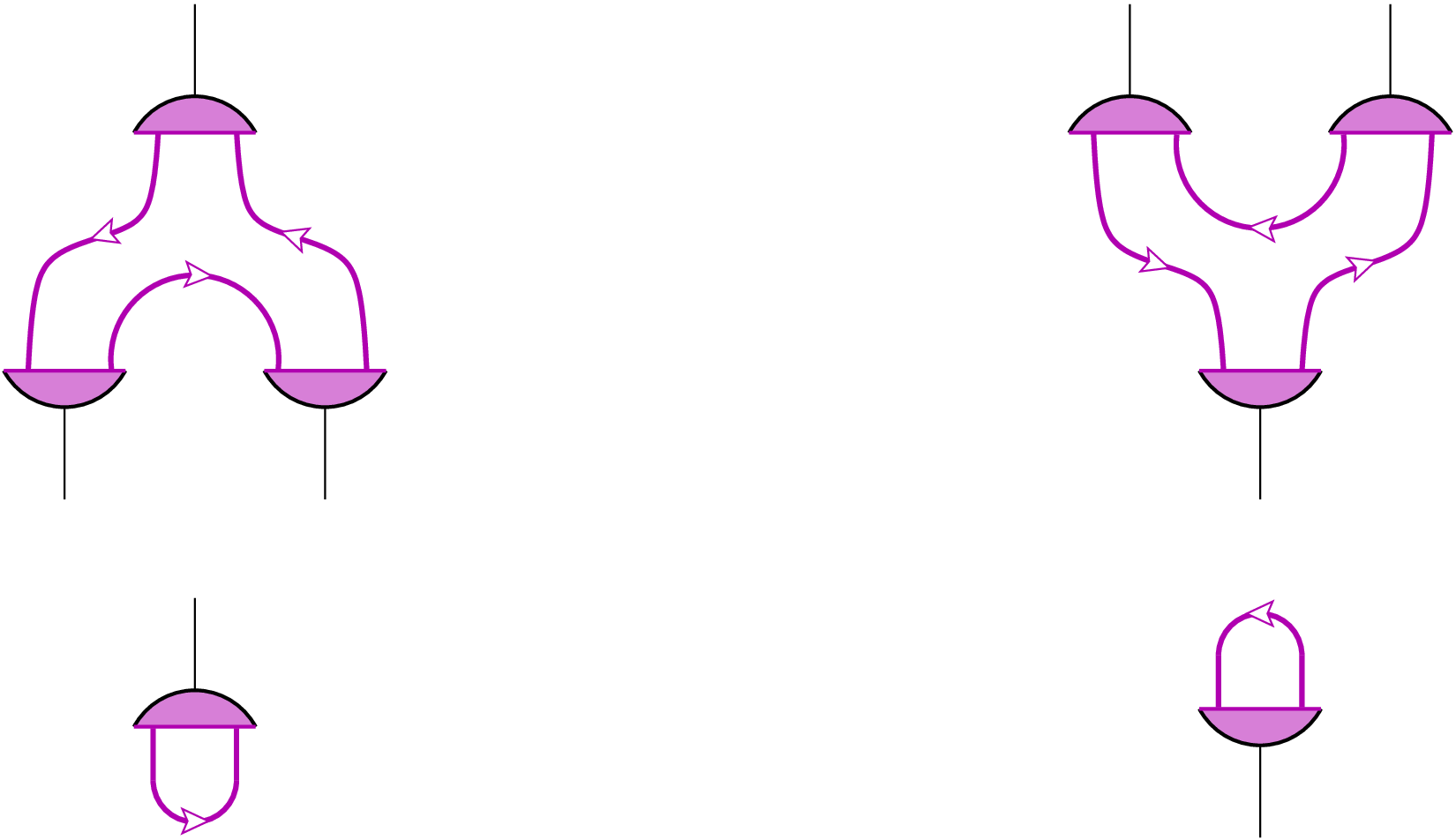}}
  \put(0,120)      {$ m \; :=$}
  \put(170,120)    {$ \Delta\; :=\ \dsty\frac{\dim(A)}{\dim(\M)} $}
  \put(30,25)      {$ \eta \; :=$}
  \put(195,25)     {$ \eps\; :=\ \dsty\frac{\dim(\M)}{\dim(A)} $}
  \setlength{\unitlength}{24810sp}
  \setlength{\unitlength}{1pt}
  \put(49,63.8)    {\scriptsize$B$}
  \put(65.5,104.2) {\scriptsize$\M$}
  \put(77,178.8)   {\scriptsize$B$}
  \put(104,63.8)   {\scriptsize$B$}
  \put(117.5,108.2){\scriptsize$\M$}
  \put(271,178.8)  {\scriptsize$B$}
  \put(286.4,140.2){\scriptsize$\M$}
  \put(297,63.8)   {\scriptsize$B$}
  \put(325,178.8)  {\scriptsize$B$}
  \put(338.4,138.2){\scriptsize$\M$}
  \put(77,54.8)    {\scriptsize$B$}
  \put(91.1,15.5)  {\scriptsize$\M$}
  \put(297.7,-6.8) {\scriptsize$B$}
  \put(312.5,35.5) {\scriptsize$\M$}
  \epicture03 \labl{eq:B-ssFA}
is a simple symmetric special Frobenius algebra.

\medskip\noindent
Proof:\\
As an example, let us demonstrate that $B$ is 
associative, special and simple; we omit the details of the analogous
arguments for the unit, counit, co-associativity, Frobenius and symmetry
properties. Consider the moves
  $$   \begin{picture}(420,157)(2,0)
  \put(0,0)    {\includeourbeautifulpicture{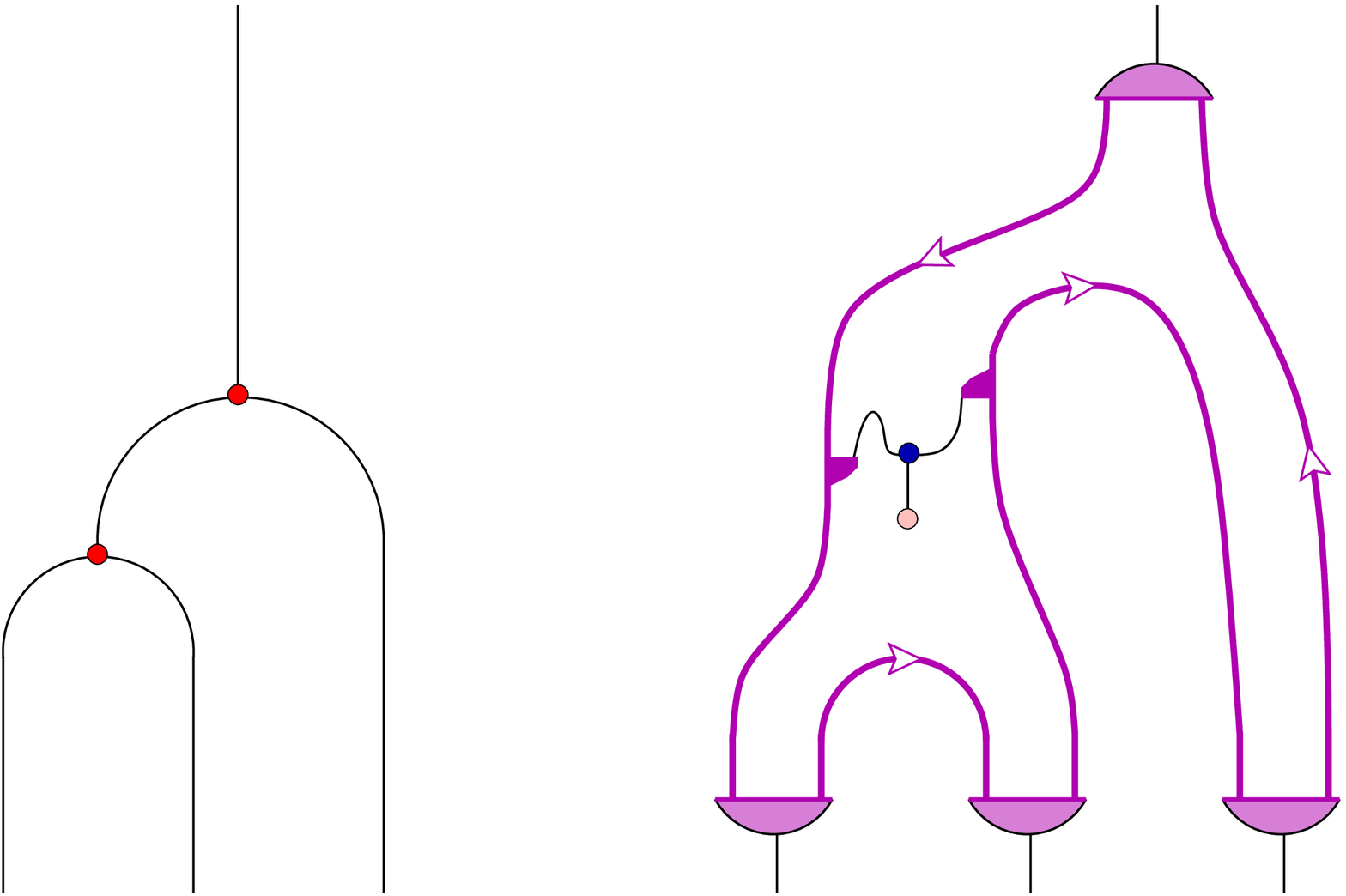}}
  \put(93,70)      {$ = $}
  \put(253,70)     {$ = $}
  \setlength{\unitlength}{24810sp}
  \setlength{\unitlength}{1pt}
  \put(-3.5,-8.6)  {\scriptsize$B$}
  \put(29.2,-8.6)  {\scriptsize$B$}
  \put(38.2,156)   {\scriptsize$B$}
  \put(60.7,-8.6)  {\scriptsize$B$}
  \put(127.7,-8.6) {\scriptsize$B$}
  \put(141.5,20.5) {\scriptsize$\M$}
  \put(153.5,78.6) {\scriptsize$A$}
  \put(171.5,-8.6) {\scriptsize$B$}
  \put(185.5,20.5) {\scriptsize$\M$}
  \put(194.2,156)  {\scriptsize$B$}
  \put(214.5,-8.6) {\scriptsize$B$}
  \put(228.5,20.5) {\scriptsize$\M$}
  \put(293.5,-8.6) {\scriptsize$B$}
  \put(306.5,20.5) {\scriptsize$\M$}
  \put(334.5,50.6) {\scriptsize$A$}
  \put(336.7,-8.6) {\scriptsize$B$}
  \put(348.5,20.5) {\scriptsize$\M$}
  \put(352.5,156)  {\scriptsize$B$}
  \put(379.4,-8.6) {\scriptsize$B$}
  \put(393.3,20.5) {\scriptsize$\M$}
  \end{picture} $$
  \bea \begin{picture}(0,103)(0,35)
  \put(58,0)    {\includeourbeautifulpicture{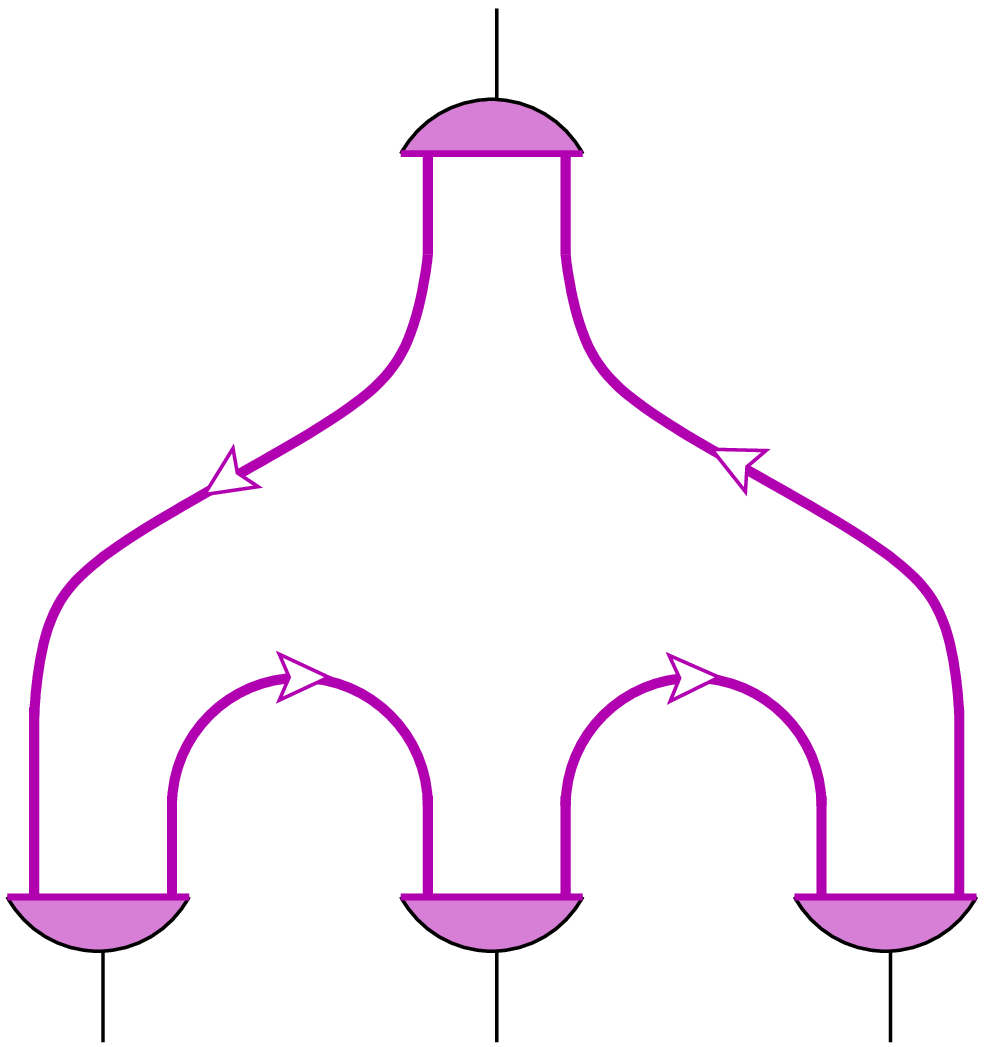}}
  \put(30,58)      {$ = $}
  \setlength{\unitlength}{24810sp}
  \setlength{\unitlength}{1pt}
  \put(65.3,-8.6)  {\scriptsize$B$}
  \put(78.4,20.5)  {\scriptsize$\M$}
  \put(108.7,-8.6) {\scriptsize$B$}
  \put(108.5,117)  {\scriptsize$B$}
  \put(121.7,20.5) {\scriptsize$\M$}
  \put(151.3,-8.6) {\scriptsize$B$}
  \put(164.7,20.5) {\scriptsize$\M$}
  \epicture22 \labl{eq:B-ssFA-x2}
In the first step the definition \erf{eq:B-ssFA} for the multiplication on 
$B$ is inserted as well as the relation \erf{eq:er-prop} to 
get the idempotent $P_{\otimes A}$. In the second step the two leftmost 
$e$'s have been replaced by $P_{\otimes A} \cir e$; using a series of 
moves involving the representation property of the module $M$ as well as 
the properties of $A$, the upper $A$ ribbon can be attached with both of 
its ends to the left $M$-ribbon. Using the representation property 
once more, together with the fact that $A$ is symmetric and special, 
the $A$-loop can be removed altogether; this has been done in step three. 
Comparison with the `mirrored' analogue of the above manipulations then 
shows that $B$ is associative.
\\
Specialness is a bit harder to see. Define the morphism 
$\phi\iN\Hom(\one,A)$ by
  \bea \begin{picture}(150,40)(0,25)
  \put(43,0)    {\includeourbeautifulpicture{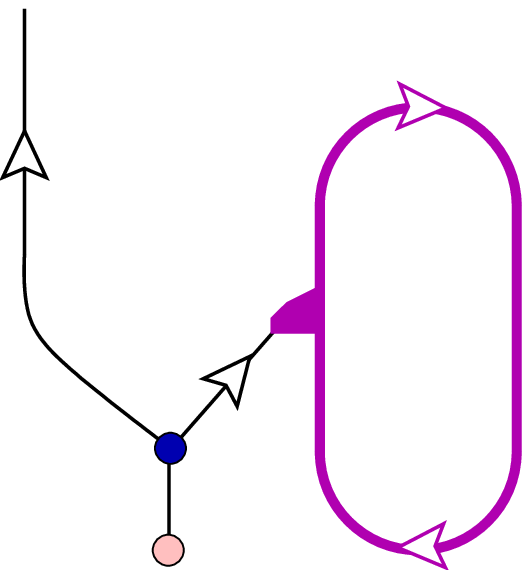}}
  \setlength{\unitlength}{24810sp}
  \setlength{\unitlength}{1pt}
  \put(0,27)       {$ \phi \; :=$}
  \put(42.5,65)    {\scriptsize$A$}
  \put(79.5,16)    {\scriptsize$\M$}
  \epicture09 \labl{eq:phi-def}
The morphism $\phi$ has the property $m\cir(\phi\oti\id)\eq m\cir(\id\oti\phi)$.
To see this, note the following moves on $m \cir (\phi\oti\id)$:
  \bea \begin{picture}(370,98)(0,28)
  \put(0,0)      {\includeourbeautifulpicture{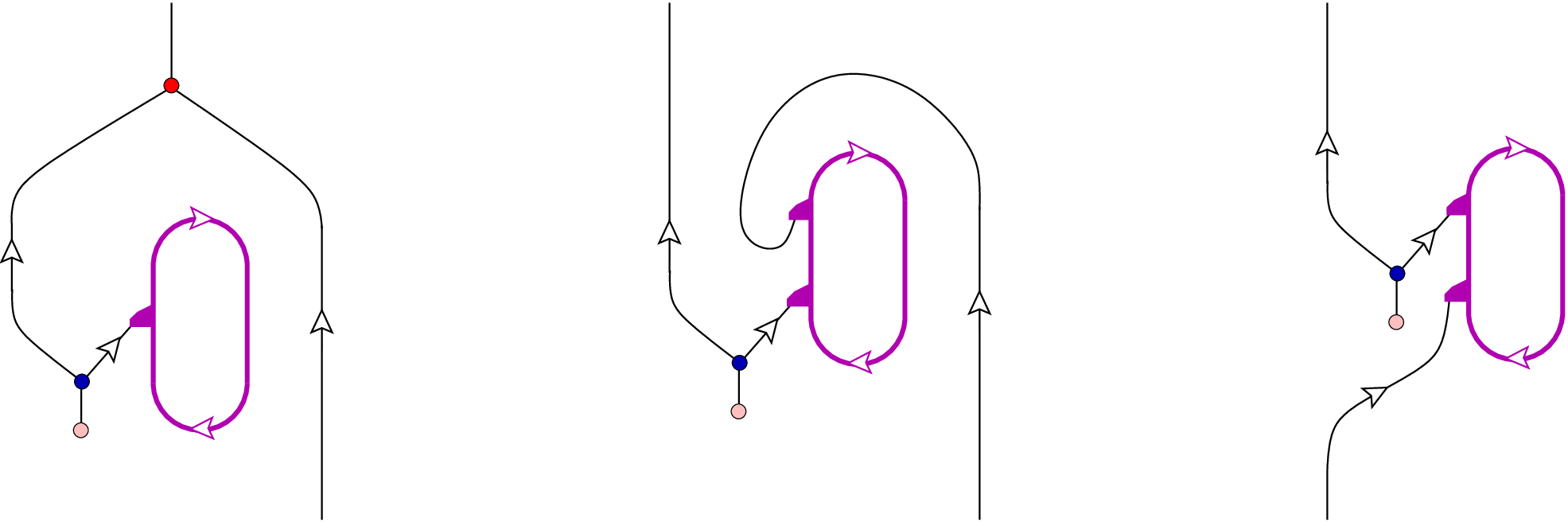}}
  \setlength{\unitlength}{24810sp}
  \setlength{\unitlength}{1pt}
  \put(36.1,124)   {\scriptsize$A$}
  \put(36.7,35)    {\scriptsize$\M$}
  \put(70.5,-9.2)  {\scriptsize$A$}
  \put(110,54)     {$ = $}
  \put(150.5,124)  {\scriptsize$A$}
  \put(188.5,61)   {\scriptsize$\M$}
  \put(221.5,-9.2) {\scriptsize$A$}
  \put(260,54)     {$ = $}
  \put(301.8,124)  {\scriptsize$A$}
  \put(301.5,-9.2) {\scriptsize$A$}
  \put(339.7,61)   {\scriptsize$\M$}
  \epicture15 \labl{eq:B-ssFA-x1}
The first step uses the various associativity properties and 
symmetry of $A$, as well as the representation property of the module $M$. In 
the second step the ribbon graph is deformed to take the upper representation 
morphism around the annular $M$-ribbon. By a similar set of moves one can 
arrive at the \rhs\ of \erf{eq:B-ssFA-x1} from $m \cir (\id \oti\phi)$, thus
establishing the stipulated equality.
\\
As in section I:3.4 we denote the vector space $\Hom(\one,A)$ by $\Atop$.
This space has a distinguished subspace denoted by ${\rm cent}_A(\Atop)$
(see definition I:3.15 ); the observation above just tells us that 
$\phi\iN{\rm cent}_A(\Atop)$. On the other hand, the unit $\eta$ of $A$ is 
always in ${\rm cent}_A(\Atop)$. Now by theorem I:5.1(iii) the 
dimension of this space is equal to $Z(A)_{00}$. In our case $A$ is simple,
so that $Z(A)_{00}\eq1$ and hence $\phi\eq\lambda \eta$ for some complex number
$\lambda$. Composing both sides with the counit $\eps$ we find that
$\lambda\eq\dimM / \dimA$, i.e.
  \be
  \phi = \Frac{\dimM}{\dimA} \, \eta \,.  \labl{eq:phi-eta}
After these prelimiaries we can show that
  \begin{eqnarray}  \begin{picture}(420,148)(13,0)
  \put(96,0)      {\includeourbeautifulpicture{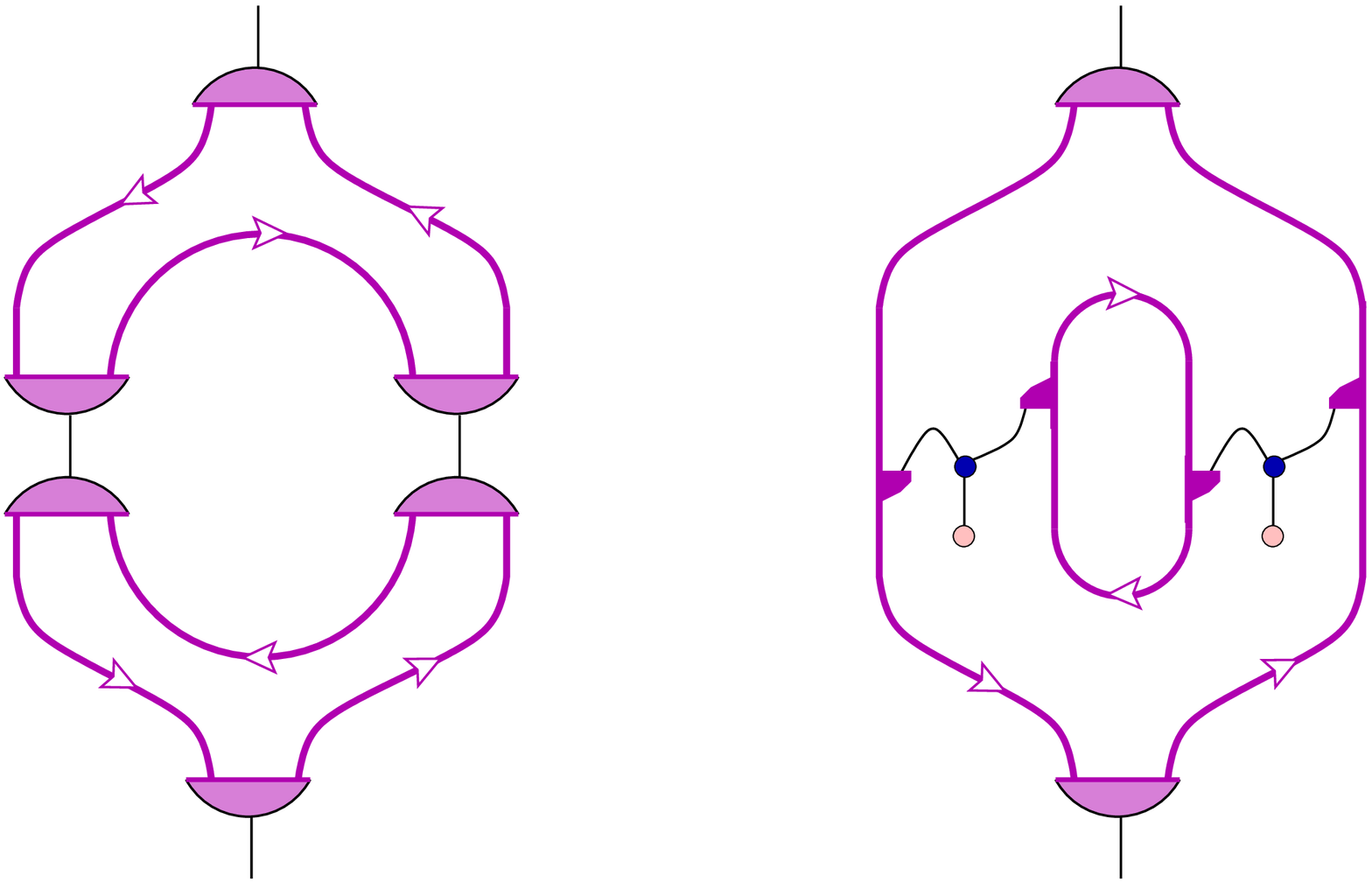}}
  \setlength{\unitlength}{24810sp}
  \setlength{\unitlength}{1pt}
  \put(0,68)       {$ \frac{\dim(\M)}{\dim(A)}\;m \cir \Delta \;= $}
  \put(108.3,68)   {\scriptsize$B$}
  \put(116.2,52)   {\scriptsize$\M$}
  \put(116.2,84)   {\scriptsize$\M$}
  \put(132.5,-9.2) {\scriptsize$B$}
  \put(134.1,145)  {\scriptsize$B$}
  \put(170.3,68)   {\scriptsize$B$}
  \put(178.6,52)   {\scriptsize$\M$}
  \put(178.6,84)   {\scriptsize$\M$}
  \put(200,68)     {$ = $}
  \put(249.5,71)   {\scriptsize$A$}
  \put(272.5,-9.2) {\scriptsize$B$}
  \put(273.1,145)  {\scriptsize$B$}
  \put(299.5,71)   {\scriptsize$A$}
  \put(316.1,52)   {\scriptsize$\M$}
  \put(338,68)     {$ = $}
  \put(400.1,85)   {\scriptsize$A$}
  \put(407.5,-9.2) {\scriptsize$B$}
  \put(408.2,145)  {\scriptsize$B$}
  \put(451.9,52)   {\scriptsize$\M$}
  \end{picture} \nonumber\\[-1.7em]{} \label{pic55}
  \\[-.6em]{}\nonumber\end{eqnarray}
In the first step the definitions \erf{eq:B-ssFA} are substituted and
in the second step the combinations $e \circ r$ have been replaced by
$P_{\otimes A}$ from \erf{eq:Ptensor}. The combination of the two $A$-ribbons
attached to the closed $M$-loop in the middle figure above can be
deformed to correspond to the \rhs\ of \erf{eq:B-ssFA-x1}. The third step
amounts to replacing it by the lhs of \erf{eq:B-ssFA-x1}. Finally using
\erf{eq:phi-eta} we can get rid of the annular $M$-ribbon altogether and 
are left with $\dimM/\dimA$ times the combination 
$r\cir P_{\otimes A} \cir e$ which is equal to $\id_B$.
The second part of the specialness property, i.e.\ $\eps\cir\eta\eq\dim(B)$,
follows from the identity (I:3.49), applied to the algebra $B$. Indeed,
composing (I:3.49) with $\eta$ and using
$m\cir\Delta\eq\id_B$, which was established above, gives $\eps\cir\eta$
on the right and $\dim(B)$ on the left hand side, respectively.
\\
Thus $B$ is special. Next we treat simplicity of $B$, i.e.\
show that $Z(B)_{00}\eq1$. It is a general principle (to
be explained in detail in a separate publication) that Morita equivalent
algebras lead to identical CFTs on oriented surfaces.
In particular we have $Z(A)\eq Z(B)$ when $A$ and $B$ are Morita equivalent;
it is this more general equality that will be checked in the sequel.
\\
Consider the ribbon graph for $Z(B)$, as given in figure
(I:5.24) with all $A$-ribbons replaced by $B$'s:
  \begin{eqnarray}  \begin{picture}(420,120)(16,0)
  \put(58,0)    {\includeourbeautifulpicture{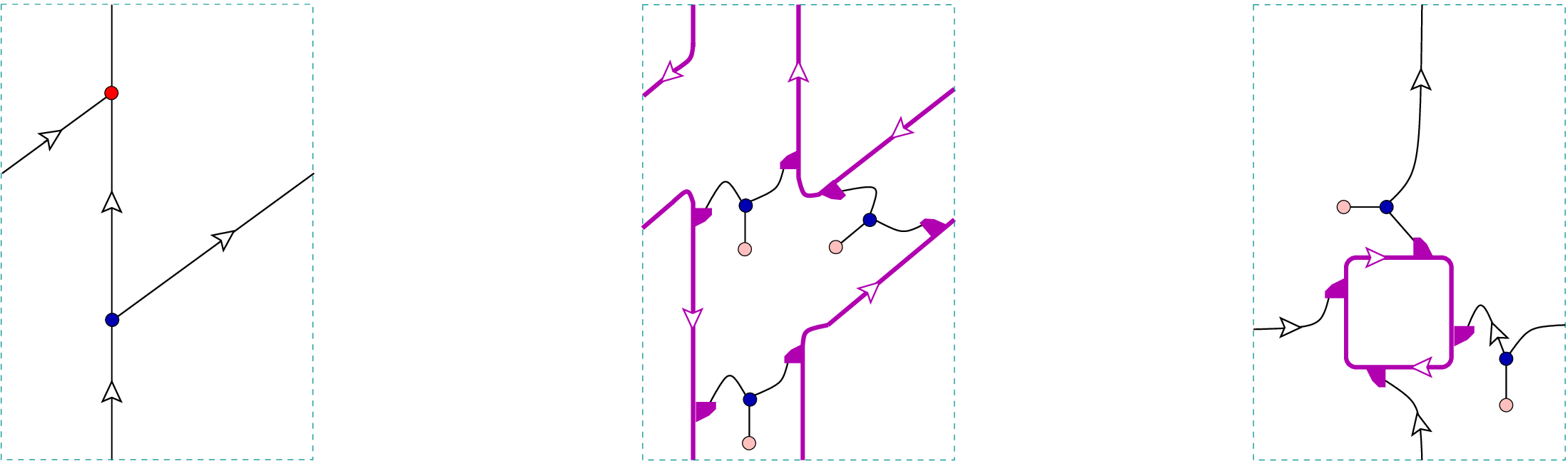}}
  \setlength{\unitlength}{24810sp}
  \setlength{\unitlength}{1pt}
  \put(0,57)       {$ Z(B) \;= $}
  \put(88,7)       {\scriptsize$B$}
  \put(126,60)     {\scriptsize$B$}
  \put(154,57)     {$ =\ \Frac{\dim(A)}{\dim(\M)} $}
  \put(248,70)     {\scriptsize$A$}
  \put(249,20)     {\scriptsize$A$}
  \put(264.7,10)   {\scriptsize$\M$}
  \put(264.7,106)  {\scriptsize$\M$}
  \put(284,63)     {\scriptsize$A$}
  \put(316,57)     {$ =\ \Frac{\dim(A)}{\dim(\M)} $}
  \put(404.5,32)   {\scriptsize$\M$}
  \put(423,85)     {\scriptsize$A$}
  \put(423.5,3)    {\scriptsize$A$}
  \put(447,26)     {\scriptsize$A$}
  \end{picture} \nonumber\\[.3em]{} \label{eq:B-ssFA-x3}
  \\[-1.3em]{}\nonumber\end{eqnarray}
Here in the 
second step the definition of $m$ and $\Delta$ as in \erf{eq:B-ssFA} is 
substituted and the morphisms $e$ and $r$ are pulled around a cycle of
the torus so as to combine them to $P_{\otimes A}$'s. In the second figure 
note that the top horizontal $A$-ribbon can be removed with a move similar 
to the one indicated in the second-to-last figure of \erf{eq:B-ssFA-x2}. 
The third equality amounts to shrinking the $M$-ribbon and shifting 
the fundamental domain of the torus such that the $M$-loop 
is in the center of the figure. Finally, making use of the representation 
property and the relation \erf{eq:phi-eta} above, the ribbon graph 
on the \rhs\ of \erf{eq:B-ssFA-x3} can be transformed into that of $Z(A)$.
\mbox{$\quad $}\qed

\medskip

So far we have established that $B\eq M^\vee \OtA M$ is again
a simple symmetric special Frobenius algebra. To establish Morita equivalence
of $A$ and $B$ we need $A$-$B$-bimodules ${}_AM_B$ and ${}_BM_A$ such that
  \be
  B \cong {}_BM_A \otA {}_AM_B \qquad {\rm and} \qquad 
  A \cong {}_AM_B \otB {}_BM_A   \labl{eq:morita}
as $B$-$B$- and $A$-$A$-bimodules, respectively. Below we will show that the 
left $A$-module $M$ is also a right $B$-module and that $M^\vee$ is a left 
$B$-module. Then it is checked that indeed $A\,{\cong}\,M \otB M^\vee$.
The left and right module structures on $M^\vee$ and $M$ are simply given by 
  \bea \begin{picture}(355,58)(0,31)
  \put(0,0)    {\includeourbeautifulpicture{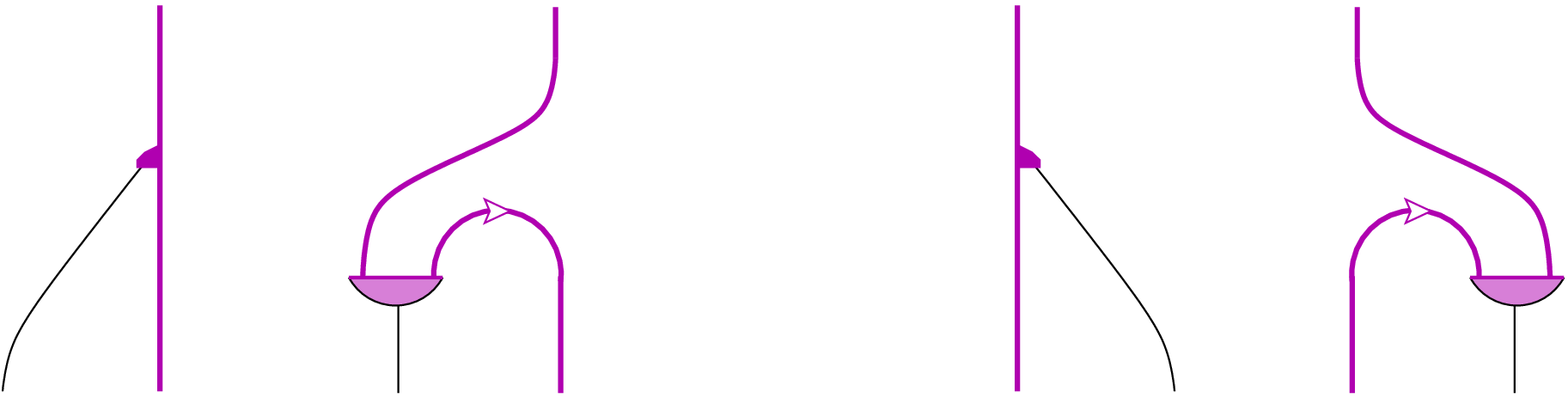}}
  \setlength{\unitlength}{24810sp}
  \setlength{\unitlength}{1pt}
  \put(-3.5,-8.8)  {\scriptsize$B$}
  \put(28,-9.9)    {\scriptsize$\M^\vee$}
  \put(29,86.5)    {\scriptsize$\M^\vee$}
  \put(51,38)      {$ := $}
  \put(81.1,-8.8)  {\scriptsize$B$}
  \put(114,-9.9)   {\scriptsize$\M^\vee$}
  \put(115,86.5)   {\scriptsize$\M^\vee$}
  \put(158,38)     {and}
  \put(211.5,-9.9) {\scriptsize$\M$}
  \put(213.1,86.5) {\scriptsize$\M$}
  \put(246.1,-8.8) {\scriptsize$B$}
  \put(261,38)     {$ := $}
  \put(283,-9.9)   {\scriptsize$\M$}
  \put(284.5,86.5) {\scriptsize$\M$}
  \put(319.5,-8.8) {\scriptsize$B$}
  \epicture16 \labl{eq:M-B-module}
It is easy to check that these indeed fulfill the representation properties
(I:4.2). The idempotent $P_{\otimes B}$ on $M \oti M^\vee$ whose image is
$M \otB M^\vee$ takes the form
  \bea \begin{picture}(240,46)(0,31)
  \put(50,0)   {\includeourbeautifulpicture{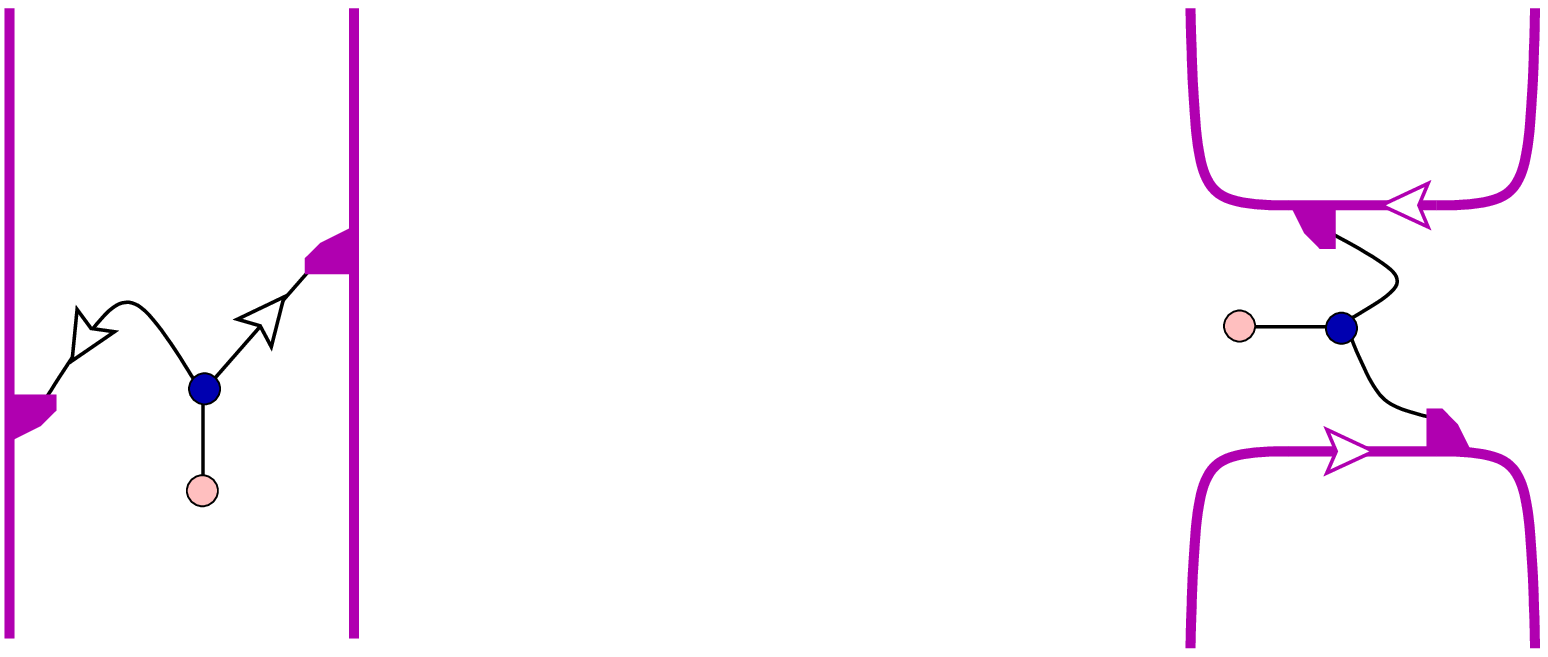}}
  \setlength{\unitlength}{24810sp}
  \setlength{\unitlength}{1pt}
  \put(0,35)       {$ P_{\otimes B} \;= $}
  \put(46,-9.9)    {\scriptsize$\M$}
  \put(47,74.2)    {\scriptsize$\M$}
  \put(69,35.8)    {\scriptsize$B$}
  \put(83,-9.9)    {\scriptsize$\M^\vee$}
  \put(84,74.2)    {\scriptsize$\M^\vee$}
  \put(109,35)     {$ = \ \dsty\frac{\dim(A)}{\dim(\M)} $}
  \put(176,-9.9)   {\scriptsize$\M$}
  \put(177,74.2)   {\scriptsize$\M$}
  \put(200.8,32.8) {\scriptsize$A$}
  \put(212,-9.9)   {\scriptsize$\M^\vee$}
  \put(213,74.2)   {\scriptsize$\M^\vee$}
  \epicture14 \labl{pic58}
To establish that the image of $P_{\otimes B}$ is indeed $A$ we give
explicitly the morphisms $\,e_B\iN\Hom(A,M
   $\linebreak[0]$
{\otimes} M^\vee)$ and $r_{\!B}\iN\Hom(M{\otimes} M^\vee\!,A)$:
  \bea \begin{picture}(320,39)(0,30) 
  \put(40,0)   {\includeourbeautifulpicture{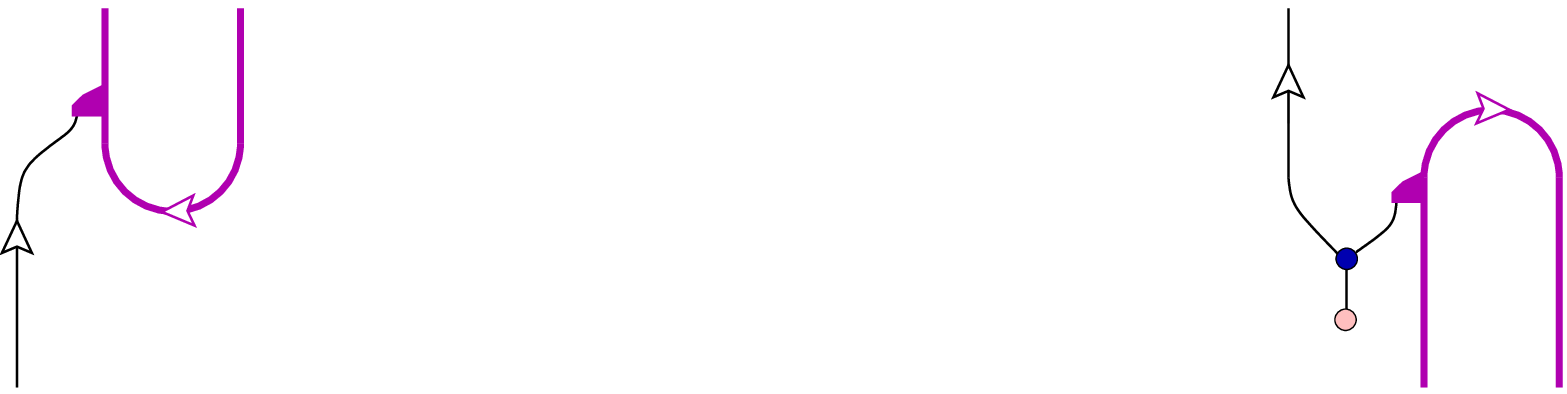}}
  \setlength{\unitlength}{24810sp}
  \setlength{\unitlength}{1pt}
  \put(0,30)       {$ e_B \;= $}
  \put(38.5,-8.8)  {\scriptsize$A$}
  \put(53,64.5)    {\scriptsize$\M$}
  \put(74,64.5)    {\scriptsize$\M^\vee$}
  \put(145,30)     {$ r_{\!B} \;=\; \dsty\frac{\dim(A)}{\dim(\M)} $}
  \put(242,64.5)   {\scriptsize$A$}
  \put(263,-9.9)   {\scriptsize$\M$}
  \put(283,-9.9)   {\scriptsize$\M^\vee$}
  \epicture17 \labl{eq:er-A-def}
One can check that these obey the relations \erf{eq:er-prop} and \erf{eq:er-prop2},
with $A$ replaced by $B$. Note that all this relies on $A$ being simple, 
for we need the relation \erf{eq:phi-eta} to remove the $M$-loops 
that occur in the calculation. One can further verify that the algebra
structure on $M\OtB M^\vee$ agrees with that of $A$.
Altogether we have established the following 
\\[-2.3em]

\dtl{Theorem}{th:AB-morita}
Let $A$ be a simple symmetric special Frobenius algebra in \calc.
Given a left $A$-module $M$ that is not a zero object,
set $B\,{:=}\,M^\vee \otA M$. The morphisms \erf{eq:B-ssFA} turn $B$ into
a simple symmetric special Frobenius algebra such that
$A \,{\cong}\, M \,{\otimes_B}\, M^\vee$ and $B \,{\cong}\, M^\vee \otA M$ as
$A$-$A$- and $B$-$B$-bimodules, respectively. In particular, $A$ and $B$
are Morita equivalent.

\medskip

As an aside, note that composing the identities (I:3.49) with the unit $\eta$,
replacing all $A$'s by $B$'s and recalling that $B$ is special, we find
the relation $\eps \cir\eta\eq\dim(B)$ for the unit and counit of $B$. 
On the other hand, starting from the definitions 
\erf{eq:B-ssFA} we find $\eps \cir\eta\eq(\dimM)^2/\dimA$. This leads to
\\[-2.3em]

\dt{Corollary}
If two simple symmetric special Frobenius algebras $A$ and $B$ are related 
by the isomorphisms \erf{eq:morita} with ${}_BM_A\eq({}_AM_B)^\vee$ then,
with $M\eq{}_AM_B$,
  \be
  \dim(A)\, \dim(B) = \big(\dim(\M)\big)^2 \,.  \labl{eq:morita-cond}

\medskip

Let us now turn to the question whether a reversion on $A$ gives 
rise to a reversion on $B$. 
\\[-2.2em]

\dtl{Proposition}{pr:B-star}
Let $A$ be a simple \JA, with reversion $\sigma$. Let $M$ be a left 
$A$-module and $B\eq M^\vee \OtA M$. Suppose there is an isomorphism
$g \iN \HomA(M,M^\sigma)$ such that
  \bea \begin{picture}(110,47)(0,30) 
  \put(0,0)   {\includeourbeautifulpicture{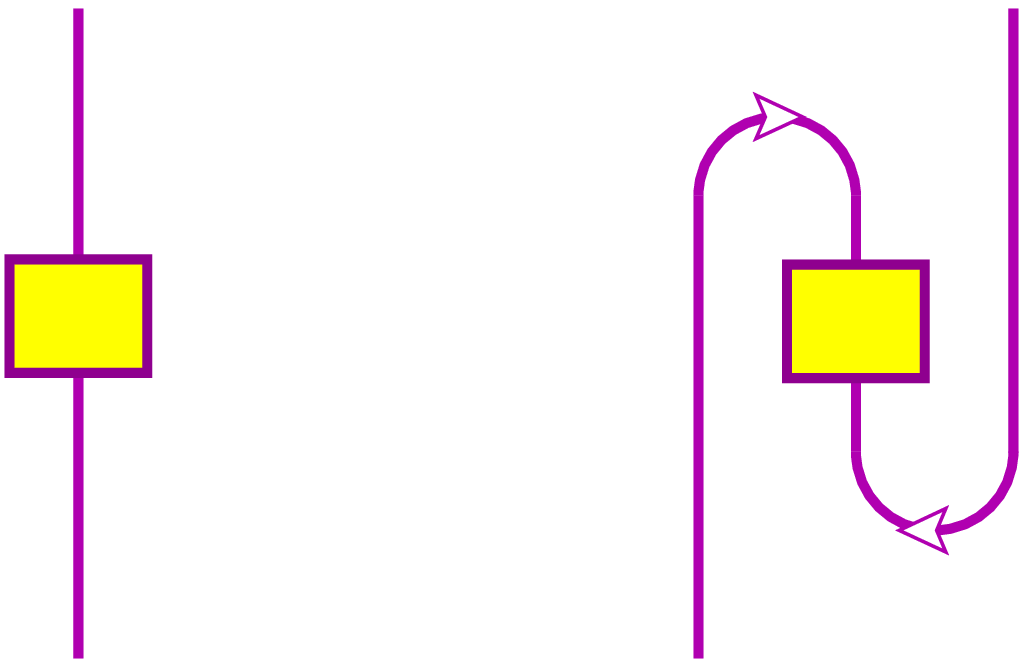}}
  \setlength{\unitlength}{24810sp}
  \setlength{\unitlength}{1pt}
  \put(3,75.2)     {\scriptsize$\M^\vee$}
  \put(4,-9.9)     {\scriptsize$\M$}
  \put(6,36.8)     {\scriptsize$g$}
  \put(33,35)      {$ =\; \eps_g $}
  \put(71.5,-9.9)  {\scriptsize$\M$}
  \put(91.5,36.8)  {\scriptsize$g$}
  \put(106.5,75.2) {\scriptsize$\M^\vee$}
  \epicture16 \labl{eq:g-prop}
Then $\eps_g\iN\{\pm 1\}$, and $\tilde\sigma_g\iN\Hom(B,B)$ given by%
   \foodnode{To avoid confusion about the labelling, let us recall some of the
   conventions of \cite{fuRs4} for the pictorial representation of morphisms
   and ribbon graphs. Labels for objects of $\calc$ that are placed above and
   below the ends of a ribbon denote the target and the source of the relevant
   morphism, respectively. In contrast, when a label for an object is placed
   {\em next\/} to a ribbon, it indicates the label, or `color', of that
   ribbon (see section I:2.3). Thus e.g.\ a ribbon
   with an upwards-pointing arrow labelled by $X$ represents the identity
   morphism in $\Hom(X,X)$, while a ribbon with a downwards-pointing arrow
   labelled by $X$ represents the identity morphism in $\Hom(X^\vee,X^\vee)$.}
  \bea \begin{picture}(90,75)(0,40)
  \put(59,0)   {\includeourbeautifulpicture{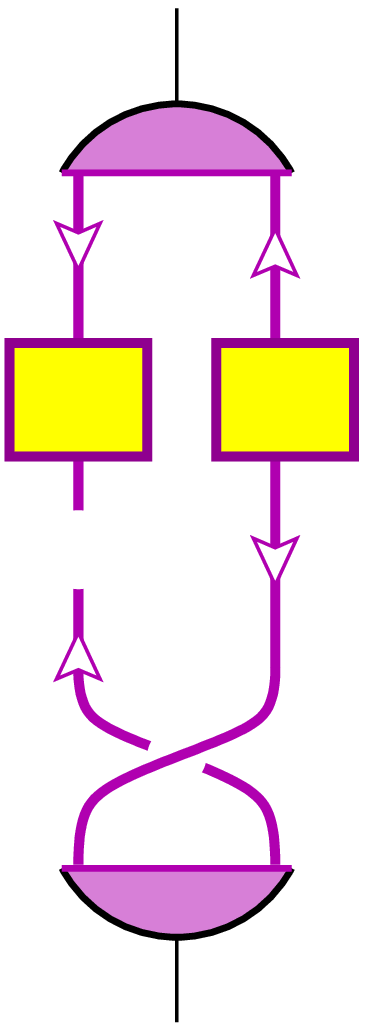}}
  \setlength{\unitlength}{24810sp}
  \setlength{\unitlength}{1pt}
  \put(0,53)       {$ \tilde\sigma_g \;:=\; \eps_g $}
  \put(56.1,86.5)  {\scriptsize$\M$}
  \put(58,21.5)    {\scriptsize$\M$}
  \put(65.6,49.9)  {\scriptsize$\theta_{\!\M}$}
  \put(65,67.5)    {\scriptsize$g$}
  \put(74.2,-8.5)  {\scriptsize$B$}
  \put(75.5,115)   {\scriptsize$B$}
  \put(85,66.5)    {\scriptsize$g^{\!-\!1}$}
  \put(90,21.5)    {\scriptsize$\M$}
  \put(91.7,86.5)  {\scriptsize$\M$}
  \epicture11 \labl{eq:B-star}
is a reversion on $B$.

\medskip\noindent
Proof:\\
For the sake of this proof we abbreviate $\tilde\sigma\,{\equiv}\,\tilde
\sigma_g$. That $\eps_g$ takes values in $\{\pm1\}$ follows immediately from
applying \erf{eq:g-prop} twice and deforming the resulting ribbon graph.
\\
Let $s \iN \Hom(M^\vee \oti M, M^\vee \oti M)$ be given
by figure \erf{eq:B-star} with $r$ and $e$ at the top and
bottom removed, such that $\tilde \sigma\eq r \cir s \cir e$. The moves
below establish the identity $s \cir P_{\otimes A}\eq P_{\otimes A} \cir s$:
  \bea \begin{picture}(420,118)(9,46)
  \put(78,0)   {\includeourbeautifulpicture{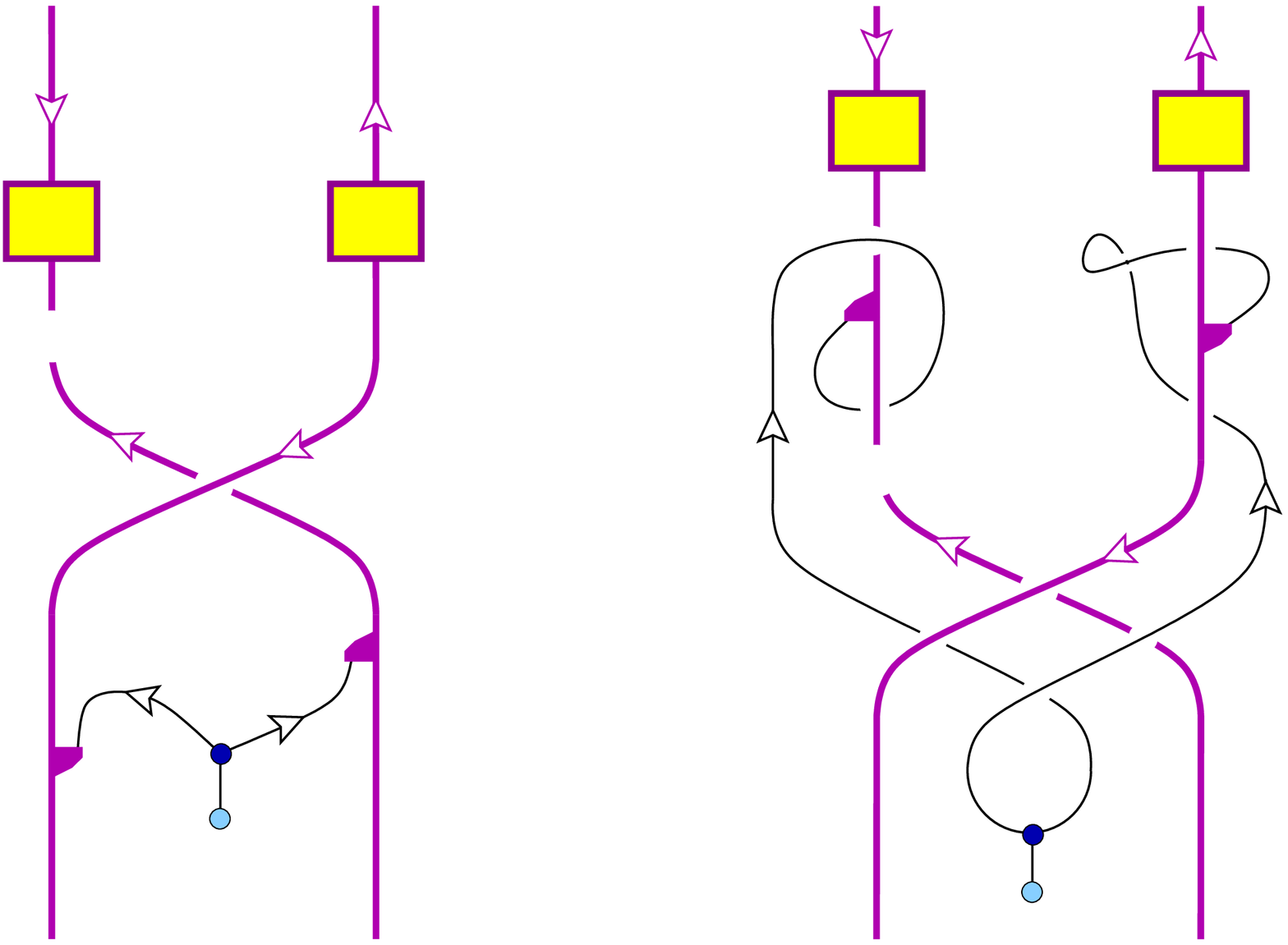}}
  \setlength{\unitlength}{24810sp}
  \setlength{\unitlength}{1pt}
  \put(0,77)       {$ \eps_g \, s \circ P_{\otimes A} \;= $}
  \put(81.7,-9.9)  {\scriptsize$\M^\vee$}
  \put(82.2,160)   {\scriptsize$\M^\vee$}
  \put(84.5,119.5) {\scriptsize$g$}
  \put(84.8,99.2)  {\scriptsize$\theta_{\!\M}$}
  \put(112,35.5)   {\scriptsize$A$}
  \put(135.5,118.3){\scriptsize$g^{\!-\!1}$}
  \put(136,-9.9)   {\scriptsize$\M$}
  \put(136.5,160)  {\scriptsize$\M$}
  \put(170,77)     {$ = $}
  \put(218.7,-9.9) {\scriptsize$\M^\vee$}
  \put(219.2,160)  {\scriptsize$\M^\vee$}
  \put(221.5,134.5){\scriptsize$g$}
  \put(223.3,77.2) {\scriptsize$\theta_{\!\M}$}
  \put(247,21.8)   {\scriptsize$A$}
  \put(273,-9.9)   {\scriptsize$\M$}
  \put(273.5,160)  {\scriptsize$\M$}
  \put(272.5,134.3){\scriptsize$g^{\!-\!1}$}
  \put(318,77)     {$ = $}
  \put(351.7,-9.9) {\scriptsize$\M^\vee$}
  \put(352.2,160)  {\scriptsize$\M^\vee$}
  \put(355.2,90.5) {\scriptsize$g$}
  \put(355.9,67.3) {\scriptsize$\theta_{\!\M}$}
  \put(373.2,114)  {\scriptsize$\sigma$}
  \put(382.4,78.2) {\scriptsize$A$}
  \put(394.2,105.3){\scriptsize$\thetaA$}
  \put(396.7,123)  {\scriptsize$\sigma^{\!-\!1}$}
  \put(406,-9.9)   {\scriptsize$\M$}
  \put(406.5,160)  {\scriptsize$\M$}
  \put(406.2,74.9) {\scriptsize$g^{\!-\!1}$}
  \epicture31 \labl{eq:sP=Ps}
The second equality is just a deformation of the ribbon
graph. In the third equality we use that $g\iN\HomA(M,M^\sigma)$,
which allows us to take the representation morphisms past 
$g$ and $g^{-1}$. Recall that the representation of
$A$ on $M^\sigma$ is of the form \erf{eq:rho-sigma}. 
The right-most graph in \erf{eq:sP=Ps} is equal to 
$\eps_g \, P_{\otimes A} \cir s$: the twist $\thetaA$
on the $A$-ribbon combines with $\sigma^{-1}$ to $\sigma$
so that the braiding in front of the coproduct can be removed
with the help of \erf{eq:sig-coprod}.
\\
It then follows that
  \be
  \tilde\sigma \circ \tilde\sigma
  = r \circ s \circ e \circ r \circ s \circ e
  = r \circ s \circ P_{\otimes A} \circ s \circ e
  = r \circ P_{\otimes A} \circ s \circ s \circ e \,.  \ee
Further we have $r \cir P_{\otimes A}\eq r$, and
one quickly checks that $s \cir s\eq\theta_{M^\vee\otimes M}$.
Functoriality of the twist thus implies
$\tilde\sigma\cir\tilde\sigma\eq r \cir\theta_{M^\vee\otimes M}\cir e
\eq r \cir e \cir\theta_B$. The identity $r\cir e\eq\id_B$ then
gives the desired relation $\tilde\sigma\cir\tilde\sigma\eq\theta_B$.
\\
To obtain $\tilde\sigma \cir m\eq m \cir c_{B,B} \cir
(\tilde\sigma\oti\tilde\sigma)$ one can use the following equalities:
  \bea \begin{picture}(410,120)(0,54)
  \put(0,0)   {\includeourbeautifulpicture{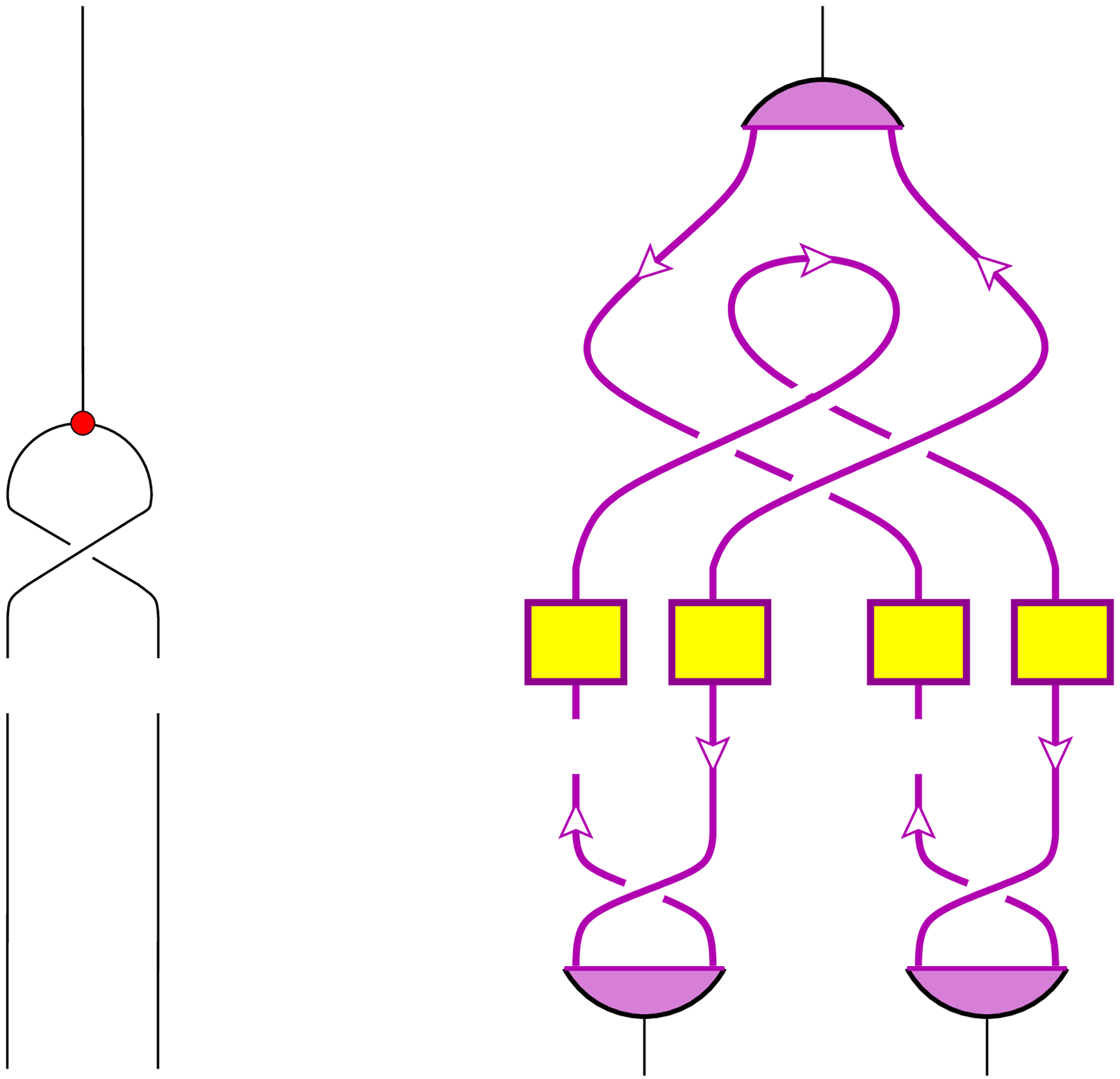}}
  \setlength{\unitlength}{24810sp}
  \setlength{\unitlength}{1pt}
  \put(.7,-8.5)    {\scriptsize$B$}
  \put(2.5,59.5)   {\scriptsize$\tilde\sigma$}
  \put(13,172.8)   {\scriptsize$B$}
  \put(24,-8.5)    {\scriptsize$B$}
  \put(25.8,59.5)  {\scriptsize$\tilde\sigma$}
  \put(53,85)      {$ = $}
  \put(84,20.5)    {\scriptsize$\M$}
  \put(92.4,50.2)  {\scriptsize$\theta_{\!\M}$}
  \put(92,67.5)    {\scriptsize$g$}
  \put(101.1,133.8){\scriptsize$\M$}
  \put(101.5,-8.5) {\scriptsize$B$}
  \put(111.5,66.5) {\scriptsize$g^{\!-\!1}$}
  \put(117,20.5)   {\scriptsize$\M$}
  \put(130,172.8)  {\scriptsize$B$}
  \put(138,20.5)   {\scriptsize$\M$}
  \put(146.4,50.2) {\scriptsize$\theta_{\!\M}$}
  \put(146,67.5)   {\scriptsize$g$}
  \put(155.5,-8.5) {\scriptsize$B$}
  \put(156.3,133.8){\scriptsize$\M$}
  \put(165.5,66.5) {\scriptsize$g^{\!-\!1}$}
  \put(171,20.5)   {\scriptsize$\M$}
  \put(203,85)     {$ = $}
  \put(228.6,20.5) {\scriptsize$\M$}
  \put(246.5,-8.5) {\scriptsize$B$}
  \put(256.2,145.8){\scriptsize$\M$}
  \put(259,62.6)   {\scriptsize$g$}
  \put(259.8,45.8) {\scriptsize$\theta_{\!\M}^{-\!1}$}
  \put(259.8,26.9) {\scriptsize$\theta_{\!\M}$}
  \put(266,134.6)  {\scriptsize$g$}
  \put(266.8,113.7){\scriptsize$\theta_{\!\M}$}
  \put(277.7,172.8){\scriptsize$B$}
  \put(283.4,20.5) {\scriptsize$\M$}
  \put(284.5,133.6){\scriptsize$g^{\!-\!1}$}
  \put(289.9,61.9) {\scriptsize$g^{\!-\!1}$}
  \put(291.2,145.8){\scriptsize$\M$}
  \put(300.5,-8.5) {\scriptsize$B$}
  \put(317.3,20.5) {\scriptsize$\M$}
  \put(339,85)     {$ = $}
  \put(369.7,-8.5) {\scriptsize$B$}
  \put(384.3,130.2){\scriptsize$\tilde\sigma$}
  \put(382.6,172.8){\scriptsize$B$}
  \put(393,-8.5)   {\scriptsize$B$}
  \epicture38 \labl{eq:tilde-sig-mult}
In the first step the definitions \erf{eq:B-ssFA} and
\erf{eq:B-star} for the multiplication and $\tilde \sigma$ are substituted.
The second step is just a deformation of the ribbon graph. The see the 
last equality, replace the morphism $g$ in the bottom part of the graph by 
the \rhs\ of \erf{eq:g-prop}; then it cancels against $g^{-1}$, and the factor 
$\eps_g$ combines with the resulting ribbon graph to $\tilde \sigma \cir m$.
\\
In view of proposition \ref{pr:*ssFA}, we have now established that
$\tilde \sigma$ is a reversion on the symmetric
special Frobenius algebra $B$.
\qed

Proposition \ref{pr:B-star} allows us to construct a reversion 
$\tilde\sigma_g$ on $B\eq M^\vee \OtA M$ if we are given an isomorphism 
$g\iN\HomA(M,M^\sigma)$ with property \erf{eq:g-prop}. In the remainder of 
this section we illustrate that this construction can be inverted, i.e.\ 
from $(B,\tilde\sigma_g)$ one can recover $(A,\sigma)$:
\\[-2.2em]

\dtl{Proposition}{pr:sigmatildetilde}
Applying the prescription \erf{eq:B-star} twice yields the identity
map, i.e.\ $\tilde{\tilde\sigma}\eq\sigma$.

\medskip\noindent
Proof:\\
Let us again abbreviate $\tilde\sigma\,{\equiv}\,\tilde\sigma_g$.
Recall from \erf{eq:M-B-module} that $M^\vee$ is a left $B$-module and 
that $A\,{\cong}\, M\OtB M^\vee$, with $A$ realised as a retract of
$M\Oti M^\vee$ according to \erf{eq:er-A-def}. Consider the morphism 
$\tilde g\,{:=}\,g^{-1}{\circ}\,\theta_{\M^\vee}^{-1}$. The equalities
  \bea \begin{picture}(420,114)(9,49)
  \put(0,0)    {\includeourbeautifulpicture{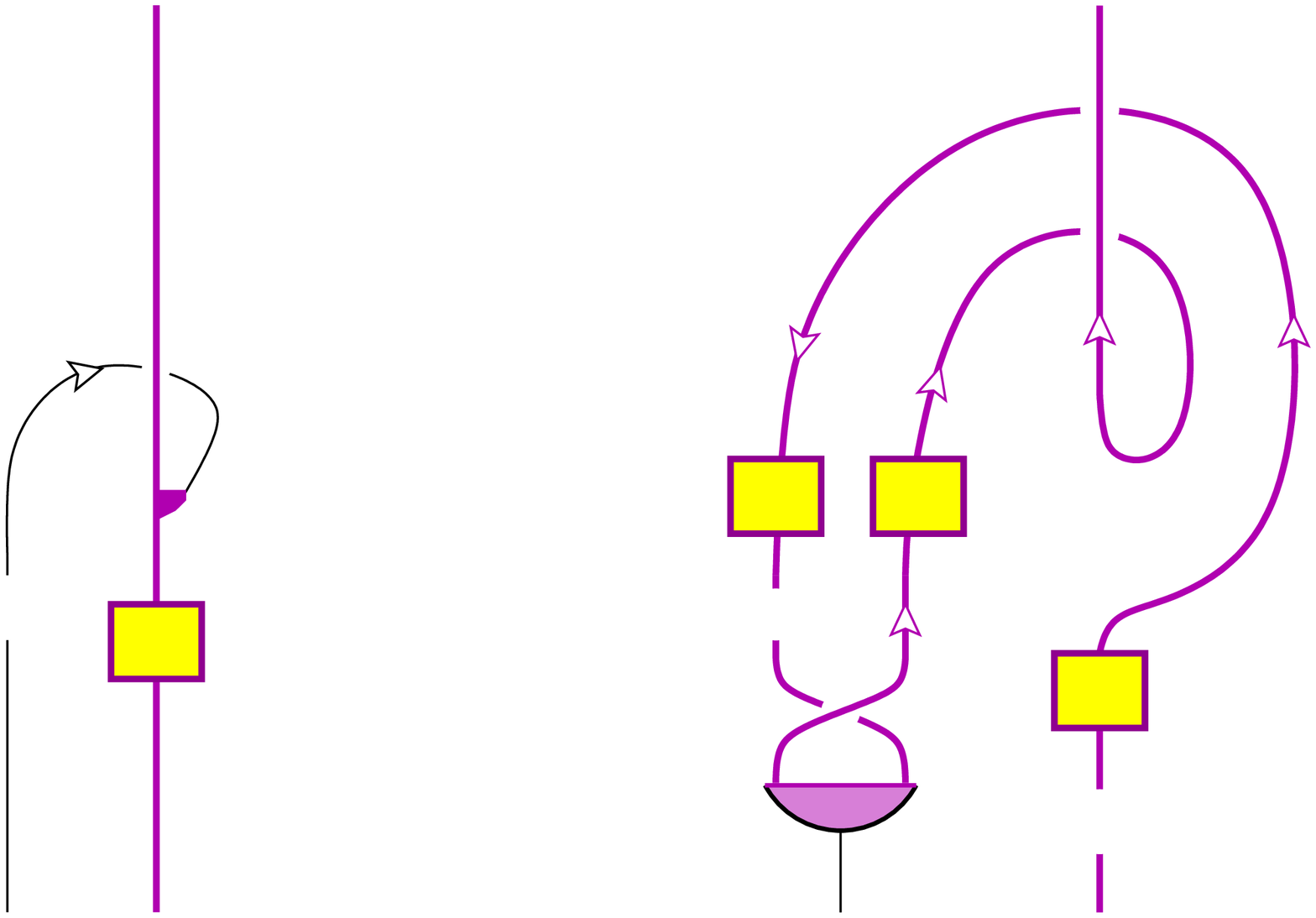}}
  \setlength{\unitlength}{24810sp}
  \setlength{\unitlength}{1pt}
  \put(1.1,-8.5)   {\scriptsize$B$}
  \put(2.9,49.2)   {\scriptsize$\tilde\sigma$}
  \put(24.7,-9.9)  {\scriptsize$\M^\vee_{}$}
  \put(26,156)     {\scriptsize$\M$}
  \put(28.5,44.2)  {\scriptsize$\tilde g$}
  \put(73,74)      {$ =\;\eps_g $}
  \put(139.1,-8.5) {\scriptsize$B$}
  \put(131.9,69.2) {\scriptsize$g$}
  \put(132.3,48.4) {\scriptsize$\theta_{\!\M}$}
  \put(152.1,68.1) {\scriptsize$g^{\!-\!1}$} 
  \put(182,-9.9)   {\scriptsize$\M^\vee_{}$}
  \put(183.7,156)  {\scriptsize$\M$}
  \put(182.7,35.9) {\scriptsize$g^{\!-\!1}$} 
  \put(185.9,13.7) {\scriptsize$\theta_{\!\M}^{\!-\!1}$} 
  \put(245,74)     {$ = $}
  \put(285.1,-8.5) {\scriptsize$B$}
  \put(316.7,-9.9) {\scriptsize$\M^\vee_{}$}
  \put(316.9,114.8){\scriptsize$g^{\!-\!1}$} 
  \put(318.3,156)  {\scriptsize$\M$}
  \put(320.5,96.3) {\scriptsize$\theta_{\!\M^\vee}^{\!-\!1}$} 
  \put(355,74)     {$ = $}
  \put(384.1,-8.5) {\scriptsize$B$}
  \put(411.7,-9.9) {\scriptsize$\M^\vee_{}$}
  \put(414.3,156)  {\scriptsize$\M$}
  \put(414.8,93.8) {\scriptsize$\tilde g$}
  \epicture37 \labl{pic64}
imply that $\tilde g \iN\Hom_B(M^\vee,(M^\vee)^{\tilde\sigma})$.
Here in the first step we substitute the definitions
\erf{eq:M-B-module}, \erf{eq:B-star} and use
\erf{eq:sP=Ps} to cancel the idempotent $P_{\otimes A}$
against $e$. In the second step \erf{eq:g-prop} is
inserted to combine a $g$ and a $g^{-1}$ and to remove
the prefactor $\eps_g$, and functoriality of the twist is used.
\\[.3em]
The morphism $\tilde g$ satisfies the analogue of \erf{eq:g-prop},
with all $g$'s replaced by $\tilde g$ and $\eps_g\eq\eps_{\tilde g}$.
This can be seen by composing both sides of \erf{eq:g-prop}
with $\theta$ and comparing their inverses.
We can now apply proposition \ref{pr:B-star} to $(B,\tilde\sigma)$ and 
$\tilde g$. This gives rise
to a reversion $\tilde{\tilde\sigma}$ on $A$ given by
  \begin{eqnarray} \begin{picture}(420,160)(14,0)
  \put(83,0)   {\includeourbeautifulpicture{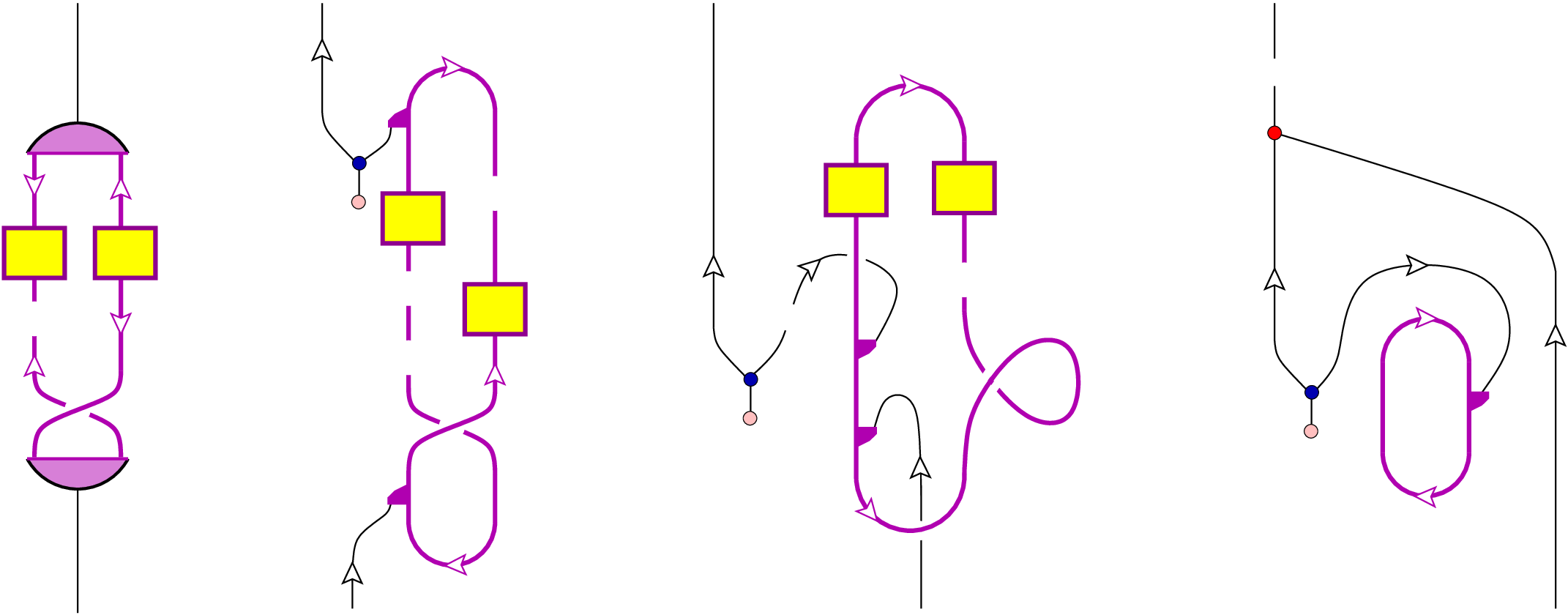}}
  \setlength{\unitlength}{24810sp}
  \setlength{\unitlength}{1pt}
  \put(-6,120)     {$ \eps_g\, \frac{\dim(\M)}{\dim(A)}\;
                      \tilde{\tilde\sigma} \,= $}
  \put(40,71)      {$ \frac{\dim(\M)}{\dim(A)} $}
  \put(82.4,42.5)  {\scriptsize$\M$}
  \put(89.4,89)    {\scriptsize$\tilde g$}
  \put(90.3,71.1)  {\scriptsize$\theta_{\!\M}$}
  \put(98.4,-8.5)  {\scriptsize$A$}
  \put(99.5,157)   {\scriptsize$A$}
  \put(109.4,88)   {\scriptsize$\tilde g^{\!-\!1}$}
  \put(115.5,42.5) {\scriptsize$\M$}
  \put(141,71)     {$ = $}
  \put(160.5,157)  {\scriptsize$A$}
  \put(167.4,-8.5) {\scriptsize$A$}
  \put(181.4,97)   {\scriptsize$g^{\!-\!1}$}
  \put(183.5,79.3) {\scriptsize$\theta_{\!\M^\vee}^{\!-\!1}$}
  \put(183.7,61.7) {\scriptsize$\theta_{\!\M^\vee}^{}$}
  \put(205.2,74.6) {\scriptsize$g$}
  \put(205.9,103)  {\scriptsize$\theta_{\!\M^\vee}$}
  \put(208.1,24.5) {\scriptsize$\M$}
  \put(208.1,121)  {\scriptsize$\M$}
  \put(233,71)     {$ = $}
  \put(258.5,157)  {\scriptsize$A$}
  \put(278.6,73.1) {\scriptsize$\sigma$}
  \put(291.9,104)  {\scriptsize$g^{\!-\!1}$}
  \put(309.2,-8.5) {\scriptsize$A$}
  \put(321.8,105)  {\scriptsize$g$}
  \put(322.6,81.3) {\scriptsize$\theta_{\!\M}^{}$}
  \put(325.5,33.5) {\scriptsize$\M$}
  \put(365,71)     {$ = \; \eps_g$}
  \put(398.5,157)  {\scriptsize$A$}
  \put(399.6,134)  {\scriptsize$\sigma$}
  \put(418.2,43.5) {\scriptsize$\M$}
  \put(467.8,-8.5) {\scriptsize$A$}
  \end{picture} \nonumber\\[.6em]{} \label{eq:tt-sigma}
  \\[-1.98em]{}\nonumber\end{eqnarray}
In the second step the morphisms $e_B$ and $r_{\!B}$ from \erf{eq:er-A-def} are
written out. The fourth equality follows when moving $\sigma$ past the
coproduct (with moves similar to \erf{eq:sig-phi}), canceling $g$ against
$g^{-1}$ with the help of \erf{eq:g-prop}, cancelling the two twists
on $\M$ and using the representation property on $\rho_M$. The $\M$-loop can 
be flipped from the right to the left (as is easily reproduced with actual 
ribbons, see also \erf{eq:trace-flip} below) so that
the $A$-ribbon now is attached to the outside of the loop.
In the right-most figure of \erf{eq:tt-sigma} one can recognise the morphism
$\phi$ given in \erf{eq:phi-def}. Since $A$ is taken to be simple we can
apply \erf{eq:B-ssFA-x1}. This produces a factor $\dim(\M)/\dim(A)$
and leaves us with $\tilde{\tilde\sigma}\eq\sigma$, i.e.\ we have
indeed recovered the original reversion on $A$.
\qed

\subsection{On the classification of reversions}\label{sec:*-class}

As already mentioned, the number of possible
reversions is not constant on a Morita class
$\mocla$ of simple symmetric special Frobenius algebras. Given one
such algebra $A$ one can pass to `larger and larger' representatives
$M^\vee \otA M$ of its class $\mocla$ by taking reducible
$A$-modules $M$ with an arbitrarily large number of simple submodules.
A priori it is very hard to control the number of solutions to the polynomial 
equations \erf{eq:*ssFA-bas} that a reversion has to fulfill. However, 
it turns out that it is actually possible to describe all reversions 
on a given Morita class $\mocla$ by solving the polynomial equations 
\erf{eq:*ssFA-bas} only for a finite number of
`small' representatives of $\mocla$. 

The algorithm can be conveniently summarised in the language of {\em module 
categories\/} \cite{pare11,beRn}. The category of left modules over
an algebra in a tensor category \calc\ has the structure of a module category 
\calm\ over \calc. Two algebras  give rise to equivalent module categories
precisely if they are Morita equivalent. Module
categories therefore allow one to formulate statements that are invariant under
Morita equivalence in an invariant way. More concretely, every algebra in a
given Morita class is isomorphic to the {\em internal End\/} 
\underline{{\rm End}}$(M)$ of some object $M$ of \calm. 
In our algorithm for finding representatives for \JA s, it is sufficient to 
restrict $M$ to be either a simple object of \calm\ or else the direct sum of 
two non-isomorphic simple objects, $M\,{\cong}\, M_1\,{\oplus}\, M_2$, that 
have the property that their internal Ends are isomorphic as opposite algebras, 
\underline{{\rm End}}$(M_1)\,{\cong}\,$\underline{{\rm End}}$(M_2)_{\rm op}$. 
Reversions need to be classified only on all algebras
\underline{{\rm End}}$(M)$ with $M$ of the form just described.

An explicit realisation of this algorithm is formulated in the following
     
\medskip\noindent
        {\bf Prescription}:\\[-1.6em]{}

  \begin{itemize}
  \item[1)] Choose a haploid representative $A$ of the Morita class $\mocla$. 
  \item[2)] Determine a complete set of (representatives of isomorphism
  classes of) simple modules $\{ M_\kappai \,{|}\, \kappai\iN\JJ \}$ of $A$.
  \end{itemize}
The next step consists of two parts:
  \begin{itemize}
\item[3a)] For each of the algebras
  $B^{\rm a}_\kappa\,{:=}\,M_\kappai^\vee\otA M_\kappai$ with $\kappai\iN\JJ$,
  find all solutions to the requirements \erf{eq:*ssFA-bas} that a 
  reversion must satisfy.
\item[3b)] For each of the algebras $B^{\rm b}_{\kappai\kappaj}\,{:=}\,
  (M_\kappai{\oplus}M_\kappaj)^\vee \otA (M_\kappai{\oplus}M_\kappaj)$ with
  $\kappai,\kappaj\iN\JJ$, $\kappai\,{\ne}\,\kappaj$
  and $M_\kappai^\vee\otA M_\kappai \,{\cong}\, M_\kappaj^\vee\otA M_\kappaj$
  as objects in $\calc$,
  find all solutions to \erf{eq:*ssFA-bas} that act as a permutation
  $(\alpha,\beta)\,{\mapsto}\,(\beta,\alpha)$ on $\complex\,{\oplus}\,
  \complex \,{\cong}\, B^{\rm b}_{\kappai\kappaj,{\rm top}}$.
  \end{itemize}
We denote by $\mathcal{B}\eq\{ (B_x, \sigma_x) \,{|}\, x\iN S \}$, for some 
index set $S$, the set of all \staralgebras\ (up to isomorphism) constructed in
steps 3a) and 3b). Then all other reversions on $\mocla$ can be
linked to one already appearing in $\mathcal{B}$ in the following manner:
  \begin{itemize}
\item[4)] Assume that $\sigma_C$ is a reversion on some $C\iN\mocla$.
  Then there exist $(B_x,\sigma_x)\iN\mathcal{B}$, together with a left
  $B_x$-module $N$ and a morphism $g\iN\Hom_{B_x}(N,N^{\sigma_x})$ fulfilling
  \erf{eq:g-prop} with $\eps_g\iN\{\pm 1\}$, such that $(C,\sigma_C)$ is 
  isomorphic as a \JA\ to $(N^\vee{\otimes_{B_x}}N,\tilde\sigma_g)$ with
  $\tilde\sigma_g$ defined as in \erf{eq:B-star}.
  \end{itemize}

\dt{Remark}
(i)~\,\,In short, the non-linear equations \erf{eq:*ssFA-bas} need to be 
solved only for a finite number of cases, namely for the
algebras $B^{\rm a}_{\kappai}$ and $B^{\rm b}_{\kappai\kappaj}$.
For all other algebras in $\mocla$ only a linear problem must be solved.
Furthermore, these particular representatives of $\mocla$ are `small',
so that this task is much easier to fulfill than for a generic representative.
Of particular convenience is that the \alg\ $A$ can be taken to be haploid
(by proposition \ref{th:sssFA}, this is always possible).
The algorithm could be easily adapted to non-haploid representatives, 
but the concrete calculations would then become more involved.
\\[.3em]
(ii)~\,In CFT terms the above procedure can be interpreted as follows. 
A full CFT is constructed from the algebra
of boundary fields on a given D-brane. In order to
define the full CFT on non-orientable surfaces, the algebra must
be equipped with a reversion. Now all distinct ways to define the full CFT 
on non-orientable surfaces can already be found by considering reversions 
on D-branes which are either elementary or else a superposition of 
two distinct elementary D-branes that have the same boundary field content.
\\[.3em]
(iii)~Note that in steps 1)\,--\,4) one is only concerned with reversions 
on {\em simple\/} symmetric special
Frobenius algebras. In general the algebra could be a direct sum
  \be
  A \;\cong\; A_1 \oplus A_2 \oplus \cdots \oplus A_n  \labl{eq:nonsimp-A}
of simple algebras $A_k$ in the sense of proposition I:3.21. In the category
of vector spaces treated in section \ref{sec:vect} below, this corresponds
to having a direct sum of several full matrix algebras. It does not add much
complication to extend the treatment to non-simple algebras. The reversion
$\sigma$ either maps a component algebra $A_k$ to itself, or else it
maps $A_k$ to the component $A_{\pi(k)}$ with $\pi(k)\,{\ne}\,k$ for some
order two permutation $\pi$. In the first case $\sigma$ must be of the form
described in the algorithm. In the second case, we may suppose that
$k\,{<}\,\pi(k)$. Then as will be shown in appendix \ref{approof}, with an
appropriate choice of basis $\sigma$ acts like the identity from $A_k$ to
$A_{\pi(k)}$ and like the twist from $A_{\pi(k)}$ to $A_{k}$.
More specifically, let $U\iN\objc$ be isomorphic to
$A_k$ and $A_{\pi(k)}$ as an object of \calc; then there are isomorphisms
$\varphi_1\iN\Hom(U,A_k)$ and $\varphi_2\iN \Hom(U,A_{\pi(k)})$ such that
$\varphi_2^{-1}\cir\sigma\cir\varphi_1^{}\eq\id_U$ and
$\varphi_1^{-1}\cir\sigma\cir\varphi_2^{}\eq\theta_U$. (That $\theta_U$
appears is required in order that $\sigma$ squares to the twist.)
\\
As discussed in remark I:5.4(i) and at the end of section I:3.2,
a direct sum $A\,{\cong}\, A_1 \,{\oplus}\, A_2$ of two algebras
corresponds to the direct sum of the two associated CFTs; this may be
written symbolically as ${\rm CFT}\eq{\rm CFT}_1\,{+}\,{\rm CFT}_2$. If
$A_2\,{\cong}\,A_{1,\rm op}$, then there exists a reversion on $A$ that
exchanges the two subalgebras. On the CFT side this implies, for instance,
that transporting a field in ${\rm CFT}_1$ around the circumference of a
M\"obius strip, one obtains a field in ${\rm CFT}_2$ and vice versa.

\medskip

The statements implicit in the steps 1)\,--\,4) above are proven in
appendix \ref{approof}. Here we just summarise the strategy of the proof.
First, the structure of the $\complex$-algebra $A_{{\rm top}}$ is analyzed.
It turns out to be a semisimple algebra over $\complex$ with an involution
(an involutive anti-algebra homomorphism) $\sigma_{{\rm top}}$.
Further, one sees that idempotents in $A_{{\rm top}}$ give rise to
idempotents on the whole algebra $A$. If $A$ is simple, then the image of
$A$ under such an idempotent turns out to be Morita equivalent to $A$.
This provides a direct way to show that every simple symmetric special
Frobenius algebra is Morita equivalent to a haploid algebra.

For a similar reasoning in the case of Jandl algebras, we use idempotents
in $A_{{\rm top}}$ that are left invariant by the involution
$\sigma_{{\rm top}}$. The image of $A$ under such an idempotent is then
equivalent, as a \JA, to $A$. Using the classification of
semisimple complex algebras with involution, one arrives at the
alternatives 3a) and 3b) in the description above. An additional argument,
which uses Morita equivalence of full matrix algebras over $\complex$,
finally excludes direct sums of two isomorphic simple modules.

\sect{Partition functions on unoriented world sheets}\label{sec:pno}

In this section we present the construction of the connecting
three-manifold and the embedded ribbon graph for unoriented world sheets 
$X$ without field insertions. $X$ may be non-orientable and may have a
non-empty boundary. The M\"obius strip and the Klein bottle will be treated 
explicitly.

For oriented world sheets the construction of the connecting manifold
and the ribbon graph has already been achieved in section I:5.1. The
procedure given below restricts to the one of section I:5.1 in the 
oriented case. There is, however, an important difference. The construction
in section I:5.1 works for any symmetric special Frobenius algebra $A$; it
requires both an orientable world sheet {\em and\/} the specification of
an orientation. The construction below, on the other hand, requires again
a symmetric special Frobenius algebra $A$, but in addition the
specification of a reversion on $A$. 
On the other hand, if $A$ admits a reversion at all, then
the correlation functions of the associated \cft\ on any oriented
world sheet do not depend on the actual choice of orientation,
provided that the change of orientation is accompanied by an appropriate 
conjugation on boundary conditions and fields.

The labelling of boundary segments is slightly different from the
oriented case, too. Instead of associating an individual left $A$-module 
to a boundary segment, we specify an equivalence class as
follows. We consider the pairs\foodnode{%
  The subscript `1' in $\orr_1$ reminds us that $\orr_1$ refers to
  the orientation of a one-dimensional manifold. Below we
  also encounter orientations $\orr_2$ of surfaces.
  }
$(M,\orr_1)$ consisting of a left $A$-module $M$ and an orientation 
$\orr_1$ of the boundary segment, which in the cases studied here -- 
no field insertions -- is topologically an $S^1$. Boundary
conditions are such pairs modulo the equivalence relation
  \be\begin{array}{rl}
  (M,\orr_1) \sim (M',\orr_1') \quad{\rm ~if~either~ }
  & \orr_1'\eq\orr_1 \ {\rm ~and~}\ M'\,{\cong}\, M \\[.2em]
  {\rm or~} & \orr_1'\eq{-}\orr_1 \ {\rm ~and~}\ M'\,{\cong}\, M^\sigma
  \,. \eear\labl{eq:bnd-class}
Thus the set of all boundary conditions is the quotient
  \be  \mathcal{B} = \{\,(M,\orr_1)\,\} \,{/}\, {\sim}\;
   =:  \{\,[M,\orr_1]\,\}\,.  \labl{eq:bc-B-def}

When displaying pictures we still like to draw lines instead of
ribbons, so we introduce some more notation to keep track of the
orientation and half-twists of the ribbons. The `white' side of a ribbon 
will still be drawn as a solid line, while for the opposite, `black',
side we use a dashed line. In other words, ribbons are 
parallel to the paper plane, and if a ribbon is drawn as a solid line,
then its orientation agrees with that of the paper plane (i.e.\ the 
standard orientation of $\reals^2$), while if it is drawn as a dashed 
line, then it has the opposite orientation. Half-twists will be drawn 
as follows:
  \bea \begin{picture}(260,20)(0,22)
  \put(0,0)     {\includeourbeautifulpicture{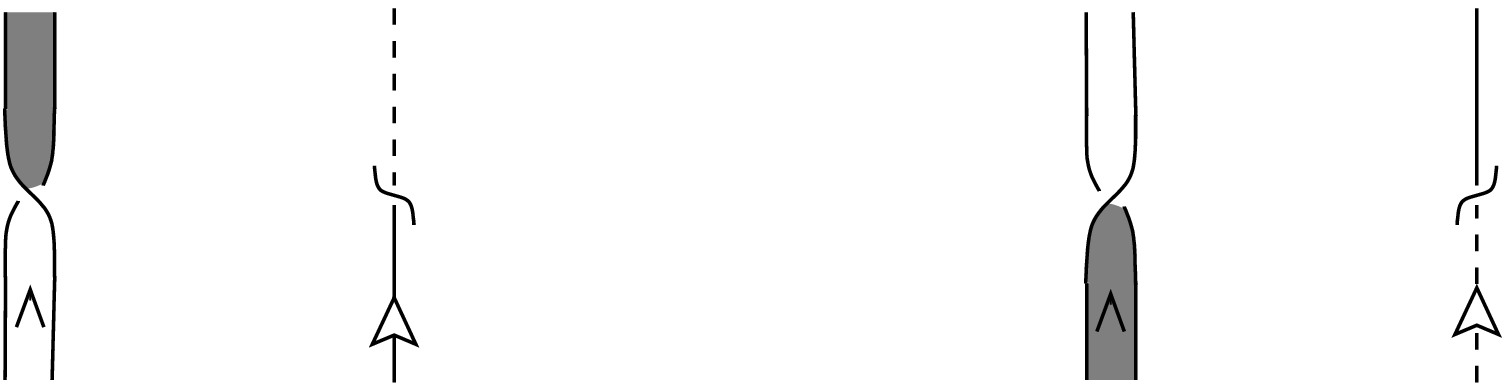}}
  \put(20,19)     {$ = $}
  \put(139,19)    {$ = $}
  \put(228,19)    {etc.}
  \epicture07 \labl{pic17}

Half-twists occur usually in combination with the morphism $\sigma$.
It is therefore convenient to abbreviate this combination, as we already
have done in \erf{eq:box-prop}; to save space, in the sequel we also use
the notation 
  \bea  \begin{picture}(290,51)(0,20) 
  \put(0,0)     {\includeourbeautifulpicture{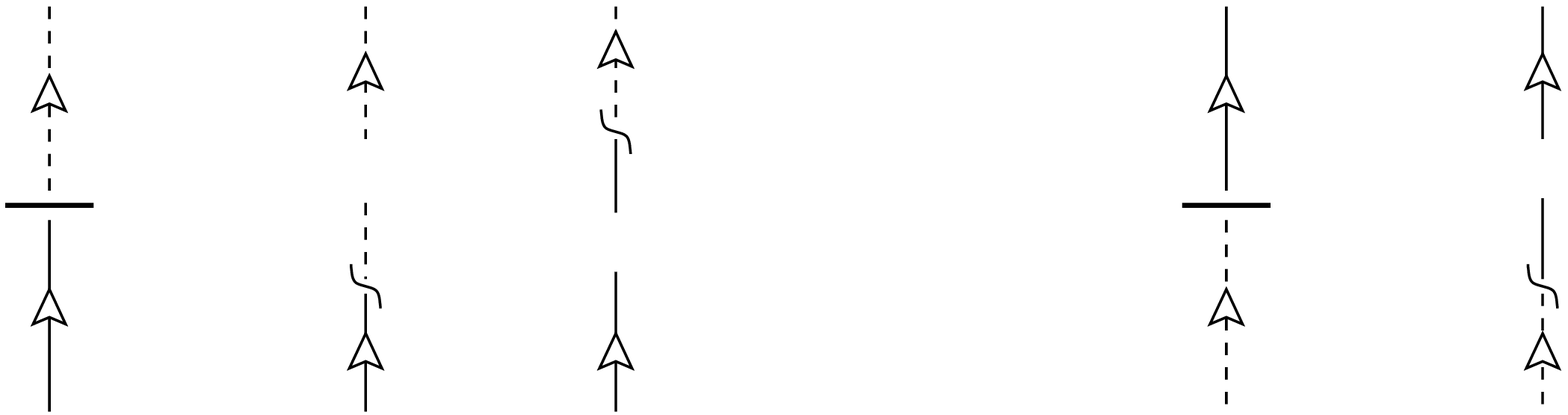}}
  \setlength{\unitlength}{24810sp}
  \setlength{\unitlength}{1pt}
  \put(3.7,-8.8)   {\scriptsize$A$}
  \put(4.2,63.9)   {\scriptsize$A$} 
  \put(23,27)      {$ := $}
  \put(50.2,-8.8)  {\scriptsize$A$}
  \put(50.7,63.9)  {\scriptsize$A$} 
  \put(51.7,34.7)  {\scriptsize$\sigma$}
  \put(67,27)      {$ = $}
  \put(86.5,-8.8)  {\scriptsize$A$}
  \put(87,63.9)    {\scriptsize$A$} 
  \put(88,23.9)    {\scriptsize$\sigma$}
  \put(176.7,-8.8) {\scriptsize$A$}
  \put(177.2,63.9) {\scriptsize$A$} 
  \put(198,27)     {$ := $}
  \put(222.7,-8.8) {\scriptsize$A$}
  \put(223.2,63.9) {\scriptsize$A$} 
  \put(224.2,34.6) {\scriptsize$\sigma$}
  \put(240,27)     {$ = $}
  \put(259.3,-8.8) {\scriptsize$A$}
  \put(259.8,63.9) {\scriptsize$A$} 
  \put(261.8,24.2) {\scriptsize$\sigma$}
  \epicture06 \labl{eq:bw-bar}
for the \boxelement\ that we introduced in \erf{eq:box-prop}. Also recall 
that the \boxelement\ is not a morphism; instead, it is to be understood as 
an abbreviation for a piece of ribbon graph embedded in a three-manifold.
Further, when the coupon representing a morphism has only dashed lines 
attached to it, it is implied that we are also looking at 
the black side of the coupon.

The \boxelement\ \erf{eq:bw-bar} satisfies
  \bea  \begin{picture}(420,49)(10,31)
  \put(0,0)     {\includeourbeautifulpicture{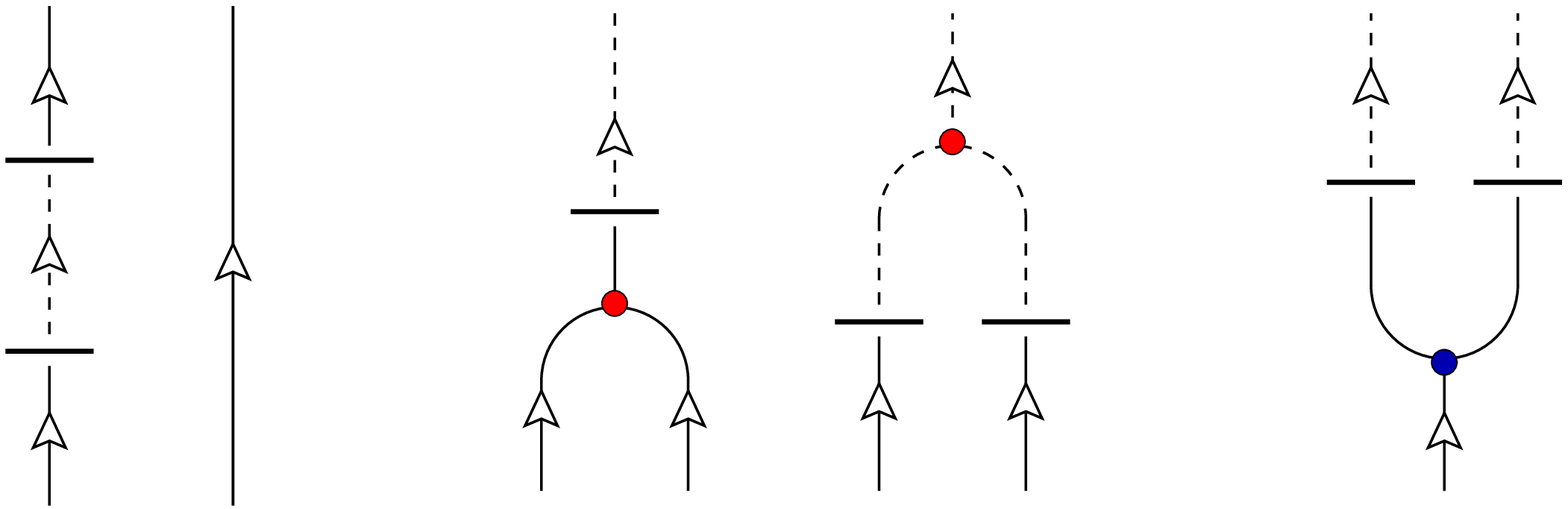}}
  \setlength{\unitlength}{24810sp}
  \setlength{\unitlength}{1pt}
  \put(3.3,-8.8)   {\scriptsize$A$}
  \put(3.8,77.2)   {\scriptsize$A$} 
  \put(16.3,34)    {$ = $}
  \put(30.3,-8.8)  {\scriptsize$A$}
  \put(30.8,77.2)  {\scriptsize$A$} 
  \put(75.5,-8.1)  {\scriptsize$A$}
  \put(86.7,77.2)  {\scriptsize$A$} 
  \put(97.4,-8.1)  {\scriptsize$A$}
  \put(109,34)     {$ = $}
  \put(125.5,-8.1) {\scriptsize$A$}
  \put(136.7,77.2) {\scriptsize$A$} 
  \put(147.4,-8.1) {\scriptsize$A$}
  \put(198.7,77.2) {\scriptsize$A$} 
  \put(208.5,-8.1) {\scriptsize$A$}
  \put(219.7,77.2) {\scriptsize$A$} 
  \put(238,34)     {$ = $}
  \put(248.7,77.2) {\scriptsize$A$} 
  \put(259.5,-8.1) {\scriptsize$A$}
  \put(271.7,77.2) {\scriptsize$A$} 
  \put(314.1,64.5) {\scriptsize$A$} 
  \put(327,34)     {$ = $}
  \put(341.3,64.5) {\scriptsize$A$} 
  \put(389.1,2.5)  {\scriptsize$A$} 
  \put(401,34)     {$ = $}
  \put(416.2,2.5)  {\scriptsize$A$} 
  \epicture16 \labl{eq:bar-prop}
as well as the corresponding relations with all 2-orientations reversed.
The properties \erf{eq:bar-prop} are direct consequences of \erf{eq:sig-def}
and \erf{eq:sig-coprod}. For example,
  \bea  \begin{picture}(298,55)(0,34) 
  \put(0,0)     {\includeourbeautifulpicture{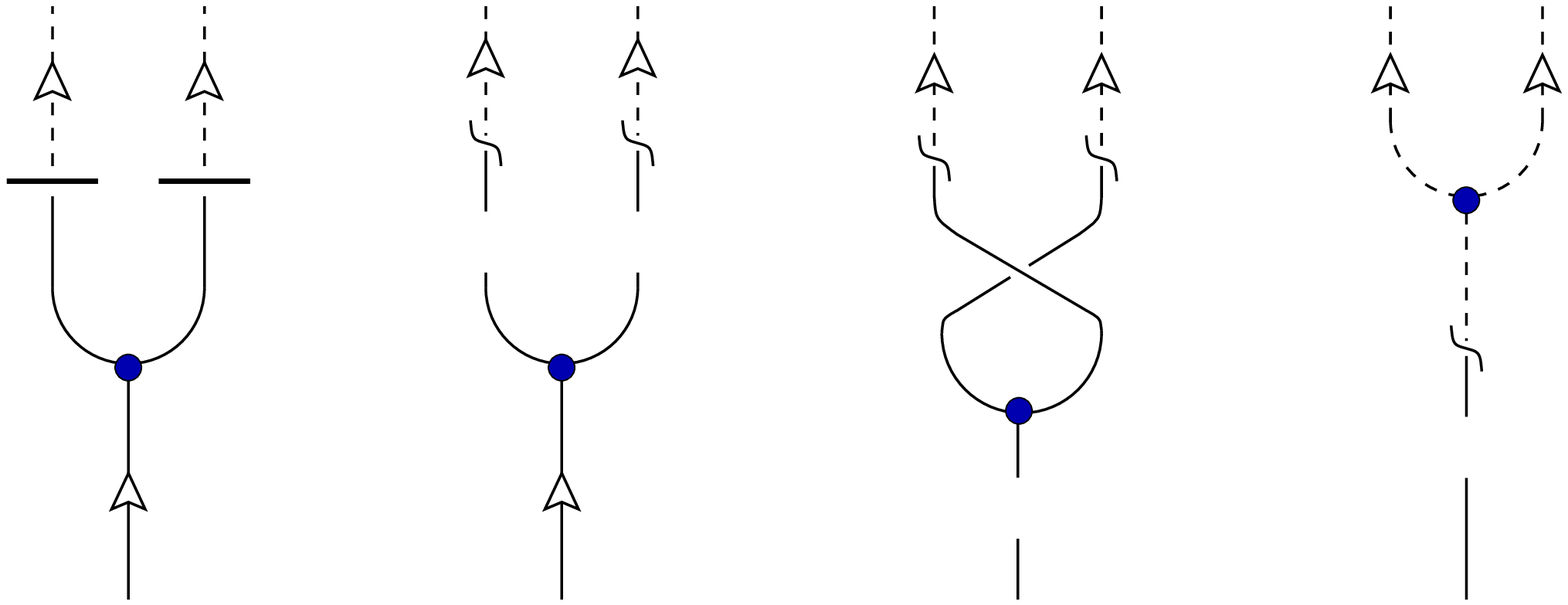}}
  \setlength{\unitlength}{24810sp}
  \setlength{\unitlength}{1pt} 
  \put(3.7,89.2)   {\scriptsize$A$}
  \put(14.2,-8.8)  {\scriptsize$A$}
  \put(25.7,89.2)  {\scriptsize$A$}
  \put(45,40)      {$ = $}
  \put(65.7,89.2)  {\scriptsize$A$}
  \put(66.9,49.2)  {\scriptsize$\sigma$}
  \put(76.5,-8.8)  {\scriptsize$A$}
  \put(87.7,89.2)  {\scriptsize$A$}
  \put(88.9,49.2)  {\scriptsize$\sigma$}
  \put(107,40)     {$ = $}
  \put(129,89.2)   {\scriptsize$A$}
  \put(139.9,-8.8) {\scriptsize$A$}
  \put(141.7,11.6) {\scriptsize$\sigma$}
  \put(153,89.2)   {\scriptsize$A$}
  \put(174,40)     {$ = $}
  \put(194,89.2)   {\scriptsize$A$}
  \put(204.5,-8.8) {\scriptsize$A$}
  \put(206.1,20.4) {\scriptsize$\sigma$}
  \put(216,89.2)   {\scriptsize$A$}
  \put(236,40)     {$ = $}
  \put(259,89.2)   {\scriptsize$A$}
  \put(269.5,-8.8) {\scriptsize$A$}
  \put(281,89.2)   {\scriptsize$A$}
  \epicture22 \ee

\subsection{The connecting manifold and the ribbon graph}\label{sec:crg}

As input for the construction we first must specify the rational CFT
we are considering. This is done by giving the modular tensor category 
$\calc$ which encodes the chiral data, plus a symmetric special
Frobenius algebra $A$ in $\calc$ and a reversion $\sigma$ on $A$.  
Second, we must specify the world sheet $X$ for which the correlator
is to be computed. Each connected component of the boundary $\partial
X$ must be labelled by a class $[M,\orr_1]\iN\mathcal{B}$.

The correlator for the world sheet $X$ is an element of the space
of conformal blocks on the double $\hat X$ of $X$.
The double is the orientation bundle modulo an equivalence relation:
  \be
  \hat X = {\rm Or}(X) \,{/}\,{\sim}
  \qquad {\rm with} \quad (x,\orr_2)\sim(x,-\orr_2)
  \;{\rm ~iff~}\; x\iN\partial X \,.  \ee
We denote the points of $\hat X$ as equivalence classes $[x,\orr_2]$.

The three-dimensional topological field theory associated to the
modular tensor category $\calc$ (see sections I:2.3 and I:2.4 for a short
overview and references) provides us with a state space $\calh(\hat X)$
and, for any three-manifold $\MX$ with boundary $\hat X$ and embedded ribbon
graph, a map $Z(\MX,\emptyset,\hat X){:}\; \complex\,{\to}\,\calh(\hat X)$.
The crucial point of the construction is to find a (universal)
prescription for $\MX$
such that the image of $1\iN\complex$ under the map $Z$ is precisely
the correlation function we are searching.

As a three-manifold, $\MX$ is the {\em connecting manifold\/} for $X$,
defined as \cite{hora7,fffs2,fffs3}
  \be
  \MX := \hat X \times [-1,1] \,{/}\,{\sim}
  \qquad {\rm with} \quad\ ([x,\orr_2],t)\sim([x,-\orr_2],-t) \,.
  \labl{eq:MX-def}
Again we denote equivalence classes by square brackets, i.e.\ write
$[x,\orr_2,t]$ for points of $\MX$. By construction,
the boundary of $\MX$ is indeed just the double $\hat X$, and
there is a natural embedding $\iota_X{:}\; X\,{\to}\,\MX$ as well as 
a projection $\pi_X{:}\; \MX\,{\to}\,X$, given by
  \be
  \iota_X( x ) = [ x,\pm \orr_2,0 ] \,, \qquad
  \pi_X( [x,\orr_2,t] ) = x \,.  \labl{eq:embed-X}
Both choices of sign in the definition of $\iota_X$ refer to the
same equivalence class.  
In fact, $\MX$ is nothing else than a `thickening' of the
world sheet $X$. Conversely, $X$ is a retract of
$\MX$, so that going from $X$ to $\MX$ does not 
introduce any new homotopical information.

To turn the double $\hat X$ of the world sheet into an extended surface
we still need to fix a Lagrangian subspace $\lambda_X\,{\subset}\, 
H_1(\hat X,\reals)$, see e.g.\ section IV.4.1 of \cite{TUra} for details. 
We take it to be the kernel of the inclusion homomorphism $H_1(\partial 
\MX,\reals) \,{\to}\, H_1(\MX,\reals)$. In short, $\lambda_X$ is spanned 
by all cycles in $H_1(\hat X,\reals)$ which are contractible within $\MX$.
\\[-2.3em]

\dtl{Remark}{remaslov}
An alternative definition of the Lagrangian subspace is given in lemma 
3.5 of \cite{fffs3}. Recall that $X$ can be regarded as the quotient of 
$\hat X$ by the orientation reversing involution $\sigma{:}\; 
\hat X\,{\to}\,\hat X$ (not to be confused with some reversion on
an algebra) that acts as $[x,\orr_2]\,{\mapsto}\,[x,-\orr_2]$, and that 
there is an induced action $\sigma_*$ of this
involution on $H_1(\hat X,\reals)$. In \cite{fffs3} the Lagrangian subspace 
is taken to be the eigenspace of $\sigma_*$ to the eigenvalue $-1$;
we denote this space by $\lambda_-(X)$.
\\
That the two definitions are equivalent, i.e.\ that 
$\lambda_-(X)\eq\lambda_X$, can be seen as follows.
Choose a basis of cycles in $\hat X$ that generates $H_1(\hat X,\zet)$.
Since the action $\sigma_*$ is induced by the action of $\sigma$ on cycles, 
in this basis $\sigma_*$ is given by a matrix with integral entries. It 
follows that we can also find vectors $\{ v^-_i \}$ with integral entries 
that form a basis of the eigenspace $\lambda_-(X)$ in $H_1(\hat X,\reals)$. 
Let $a$ be one of the basis vectors. By construction we have 
$\sigma_*(a)\eq{-}a$. Let the closed curve\,%
  \footnote{~In case that $a$ is represented by a union of
  several curves the proof works analogously.}
$\gamma{:}\; [0,1]\,{\to}\,\hat X$ be a representative of $a$, and denote
$\tilde\gamma(s)\,{:=}\,\sigma(\gamma(1{-}s))$. Owing to $\sigma_*(a)\eq{-}a$ 
the union $\gamma\,{\sqcup}\,\tilde\gamma$ is a representative of the cycle $2a$.
\\
Let us show that $\gamma\,{\sqcup}\,\tilde\gamma$ is contractible when embedded 
in $\MX$. The connecting manifold $\MX$ may be rewritten as
  \be
  \MX = \hat X \times [0,1] / \sim \qquad {\rm with} \qquad
  (x,0) \sim (\sigma(x),0) \,.  \labl{eq:MX-ident}
Via the embedding $\iota{:}\; \hat X\,{\to}\,\MX$ given by
$x \,{\mapsto}\, (x,1)$ we get two curves $\gamma_1(s)\eq(\gamma(s),1)$ and
$\tilde\gamma_1(s)\eq(\tilde\gamma(s),1)$. Denoting by $\iota_*$ the embedding
$H_1(\hat X,\zet)\,{\to}\, H_1(\MX,\zet)$ induced by $\iota$, it follows that 
$\gamma_1\,{\sqcup}\,\tilde\gamma_1$ is a representative of $\iota_*(2a)$. In 
fact, $\{\gamma_t\,{\sqcup}\,\tilde\gamma_t\}$ with $\gamma_t(s)\eq(\gamma(s),t)$ 
and $\tilde\gamma_t(s)\eq(\tilde\gamma(s),t)$ provides us with a whole family
of representatives of $\iota_*(2a)$. Now due to the identification in 
\erf{eq:MX-ident} we have $\gamma_0(s)\eq\tilde\gamma_0(1{-}s)$. Thus
$\gamma_0\,{\sqcup}\,\tilde\gamma_0$ is contractible in $\MX$, and hence
$\iota_*(2a)\eq 2 \iota_*(a) \eq 0$. Finally, we obtain $H_1(\,\cdot,\reals)$
by tensoring $H_1(\,\cdot,\zet)$ with ${\otimes_{\mathbb Z}}\reals$, and thus 
in $H_1(M_X,\reals)$ we indeed have $\iota_*(a) \eq 0$. It follows that the 
basis elements $v^-_i$ of $\lambda_-(X)$ are all contained in $\lambda_X$.
Since all Lagrangian subspaces have equal dimension, this implies
that $\lambda_-(X)\eq\lambda_X$.

\medskip

The ribbon graph embedded in $\MX$ is constructed in several
steps, which involve a number of arbitrary choices. We first
list all the steps and show afterwards that the outcome is independent
of all the choices made, i.e.\ they all lead to equivalent ribbon graphs.

\noindent
\begin{itemize}
\item[\Nxt] 
Choose  a representative $(M,\orr_1)$ in each of the equivalence classes 
$[M,\orr_1]$ that label the boundary components of $\partial X$ 
-- \underline{Choice \#1}.

\item[\Nxt] Choose a dual triangulation of $X$ -- \underline{Choice \#2}.

\item[\Nxt] Choose a local orientation around each vertex 
  of the triangulation that lies in the interior of $X$
  -- \underline{Choice \#3}.

\item[\Nxt] Each component of the boundary $\partial X$ is already 
oriented by the choice of $(M,\orr_1)$ made above. Around each vertex 
of the triangulation that lies on $\partial X$, choose the
orientation $\orr_2$ induced by $\orr_1$ and the inward pointing
normal (the local orientation then agrees with the one around a
boundary point on the upper half plane ${\rm Im}(z) \,{\ge}\, 0$).

\item[\Nxt] Think of the triangulated world sheet with local
orientations as embedded in $\MX$ according to \erf{eq:embed-X}. 
At each vertex in the interior place the (piece of) ribbon graph
  \bea  \begin{picture}(150,60)(0,37)
  \put(0,0)     {\includeourbeautifulpicture{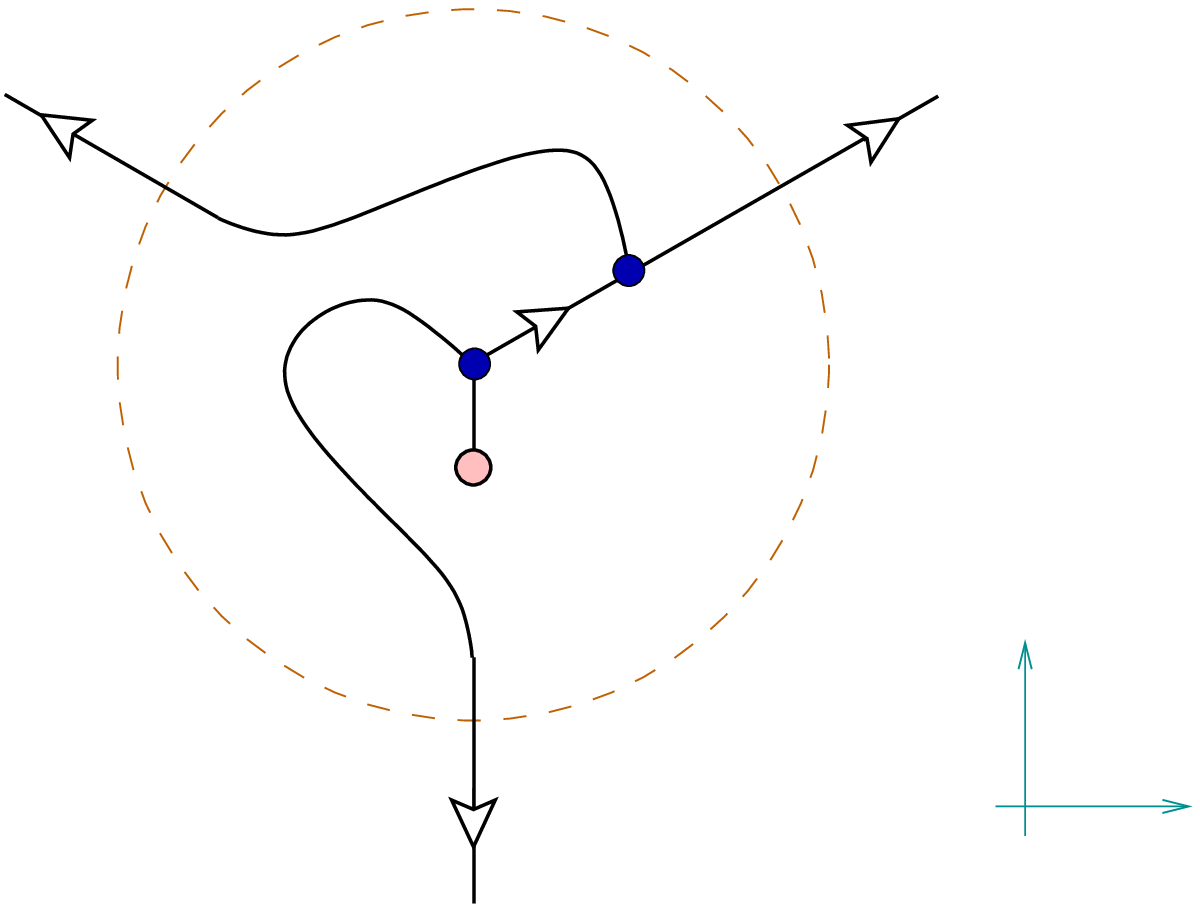}}
  \setlength{\unitlength}{24810sp}
  \setlength{\unitlength}{1pt}
  \put(-8.2,87.5)  {\scriptsize$A$}
  \put(47.9,-8.8)  {\scriptsize$A$}
  \put(104,87.5)   {\scriptsize$A$}
  \put(111,32.2)   {\tiny$y$}
  \put(132.2,10.1) {\tiny$x$}
  \epicture22 \labl{eq:bulk-vertex}
such that its orientation agrees with the local orientation 
$\orr_2$ around that vertex. The $x$-$y$-coordinates indicated 
in \erf{eq:bulk-vertex} encode the orientation by comparing
it to the standard orientation of $\reals^2$. The graph 
\erf{eq:bulk-vertex} can be inserted in three distinct ways, obtained 
by rotating the picture by an angle $\pm2\pi/3$ -- \underline{Choice \#4}.

\item[\Nxt]
On a boundary edge with chosen representative $(M,\orr_1)$
place an $M$-ribbon such that the orientation of its core
agrees with $\orr_1$ and the orientation of its surface agrees
with the local orientation on $X$ given by $\orr_1$ and the inward
pointing normal.

\item[\Nxt] At each vertex on a boundary 
component with chosen representative $(M,\orr_1)$ place the ribbon graph
  \bea  \begin{picture}(210,66)(0,32)
  \put(0,0)     {\includeourbeautifulpicture{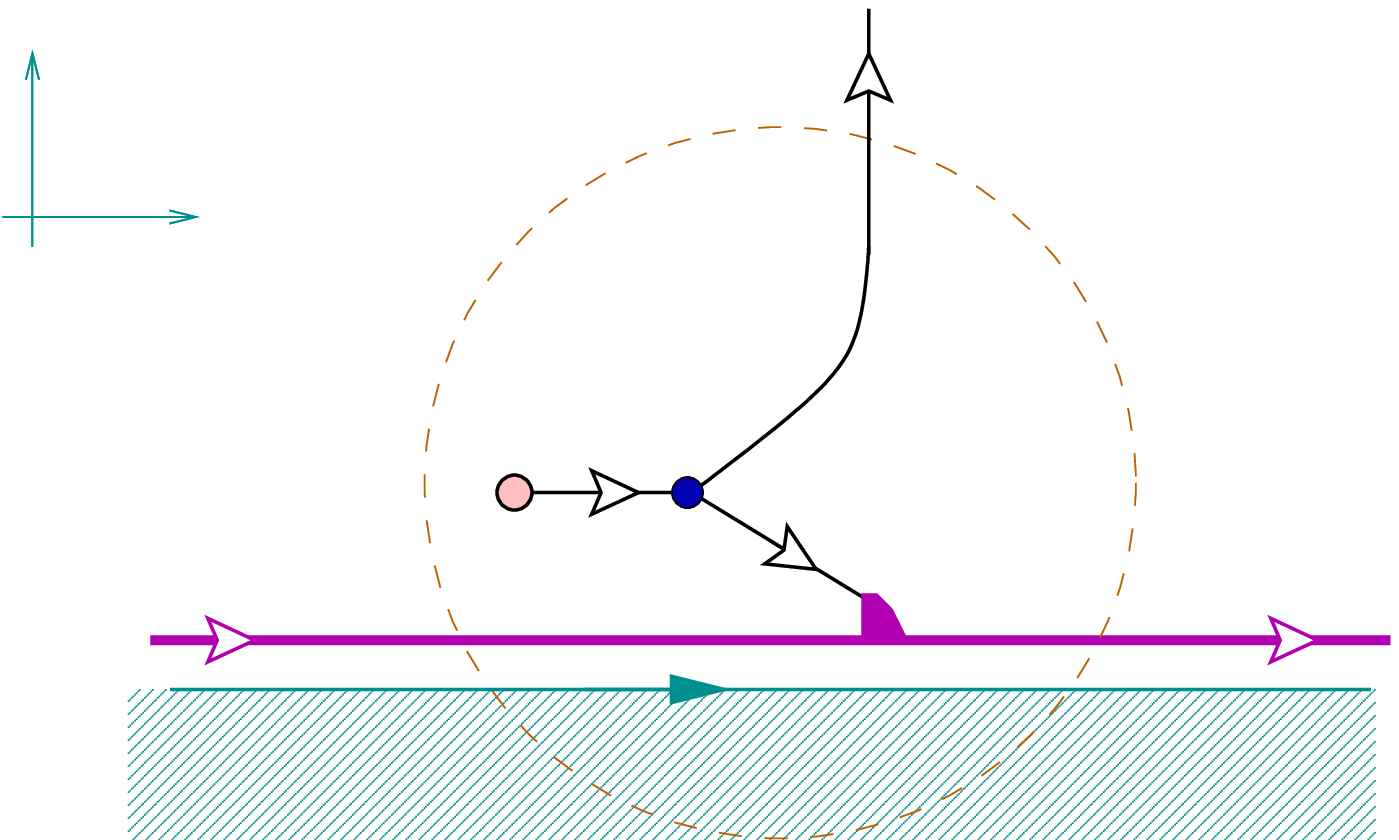}}
  \setlength{\unitlength}{24810sp}
  \setlength{\unitlength}{1pt}
  \put(1.7,89.5)    {\tiny$y$}
  \put(23.2,67.3)   {\tiny$x$}
  \put(12.3,24.8)   {\scriptsize$\M$}
  \put(92,94.3)     {\scriptsize$A$}
  \put(148.3,24.8)  {\scriptsize$\M$}
  \epicture16 \labl{eq:bnd-vertex}
in such a way that the orientation of bulk and boundary agree with that
around the vertex. The arrow on the boundary indicates its orientation.
(There is only a single way to place the graph \erf{eq:bnd-vertex}.)

\item[\Nxt] Two vertices $v,v'$ connected by an edge of the 
triangulation have the same orientation if $\orr_2(v)$, transported 
along the edge, agrees with $\orr_2(v')$. 
Note that whether two vertices possess the same orientation or not 
in general depends not only on the vertices, but also on the particular
edge that connects them.
\\[3pt]
At an edge in the interior of $X$ which connects two
vertices of the same orientation (\wrt that edge) place the
ribbon graph (\ref{eq:bulk-edge}~a) such that the orientation at either
end agrees with that around the vertices. 
At an edge in the interior of $X$ which connects two
vertices of opposite orientation place the ribbon graph
(\ref{eq:bulk-edge}\,b) again such that the orientations agree.\foodnode{%
   The second of the graphs (\ref{eq:bulk-edge}) differs slightly from
   \erf{eq:box-el} because the construction is formulated such
   that an edge in the interior of $X$ has two incoming $A$-ribbons,
   rather than one incoming and one outgoing.}
  \bea  \begin{picture}(390,48)(30,19)
  \put(0,0)     {\includeourbeautifulpicture{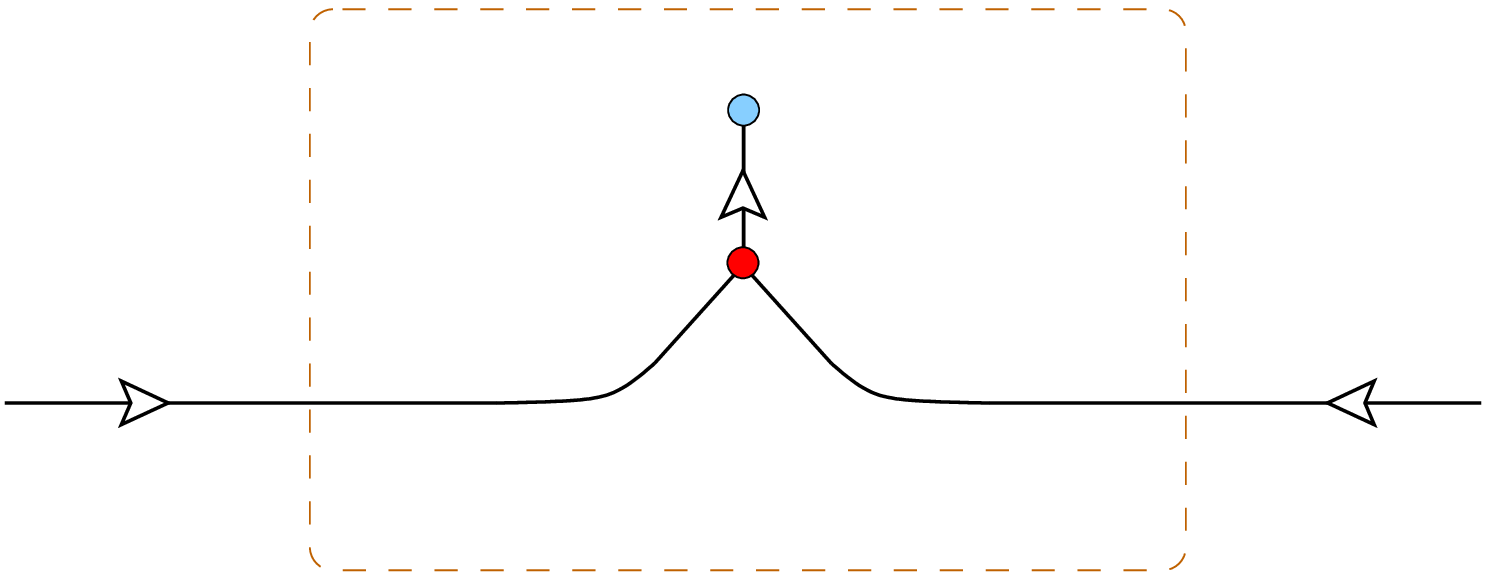}}
  \setlength{\unitlength}{24810sp}
  \setlength{\unitlength}{1pt}
  \put(-5.8,53.5)   { a) }
  \put(-2.3,20.9)   {\scriptsize$A$}
  \put(159.8,20.9)  {\scriptsize$A$}
  \put(237,53.5)    { b) }
  \put(240.5,20.9)  {\scriptsize$A$}
  \put(403,20.9)    {\scriptsize$A$}
  \epicture06 \labl{eq:bulk-edge}
In both cases there are two possibilities to place the corresponding
graph -- \underline{Choices \#5a} and \underline{\#5b}.
\end{itemize}

\noindent
This completes the prescription to construct the ribbon graph
embedded in $\MX$. We now proceed to show independence of the various
choices involved.
\\[5pt]
\nxt\underline{Choices \#4, \#5a}:\\[2pt]
Independence of these two choices was already shown in equations
(I:5.8) and (I:5.9).
\\[5pt]
\nxt\underline{Choice \#5b}:\\[2pt]
The ribbon graph is independent of this choice if
  \bea  \begin{picture}(390,47)(0,20)
  \put(0,0)     {\includeourbeautifulpicture{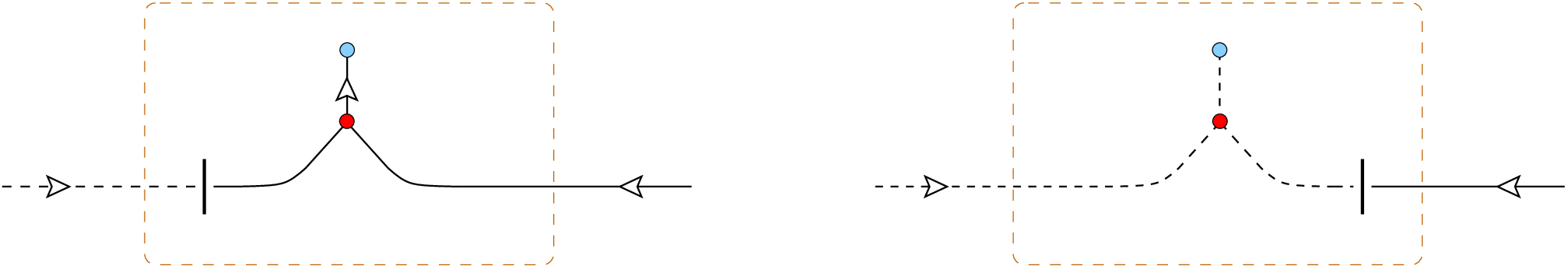}}
  \setlength{\unitlength}{24810sp}
  \setlength{\unitlength}{1pt}
  \put(-2.3,21.2)   {\scriptsize$A$}
  \put(159.3,21.2)  {\scriptsize$A$}
  \put(180,28.5)    {$ = $}
  \put(202.3,21.2)  {\scriptsize$A$}
  \put(364.2,21.2)  {\scriptsize$A$}
  \epicture08 \labl{eq:ch5a}
But this equality is an immediate consequence of the
properties \erf{eq:bar-prop} of the \boxelement\ \erf{eq:bw-bar}.
\\[5pt]
\nxt\underline{Choice \#3}:\\[2pt]
Let us show that reversing the orientation of a
vertex in the bulk yields equivalent ribbon graphs. 
\\
Pick a vertex $V$ in the bulk and denote its orientation by $+$. 
The vertex $V$ is joined by edges of the triangulation to three 
other vertices (some of which may lie on the boundary). 
Depending on the orientation of the surrounding vertices, there are 
a priori four equalities to be shown, each of which amounts
to reversing the orientation of $V$:
  \bea  \begin{picture}(290,80)(0,28)
  \put(0,0)     {\includeourbeautifulpicture{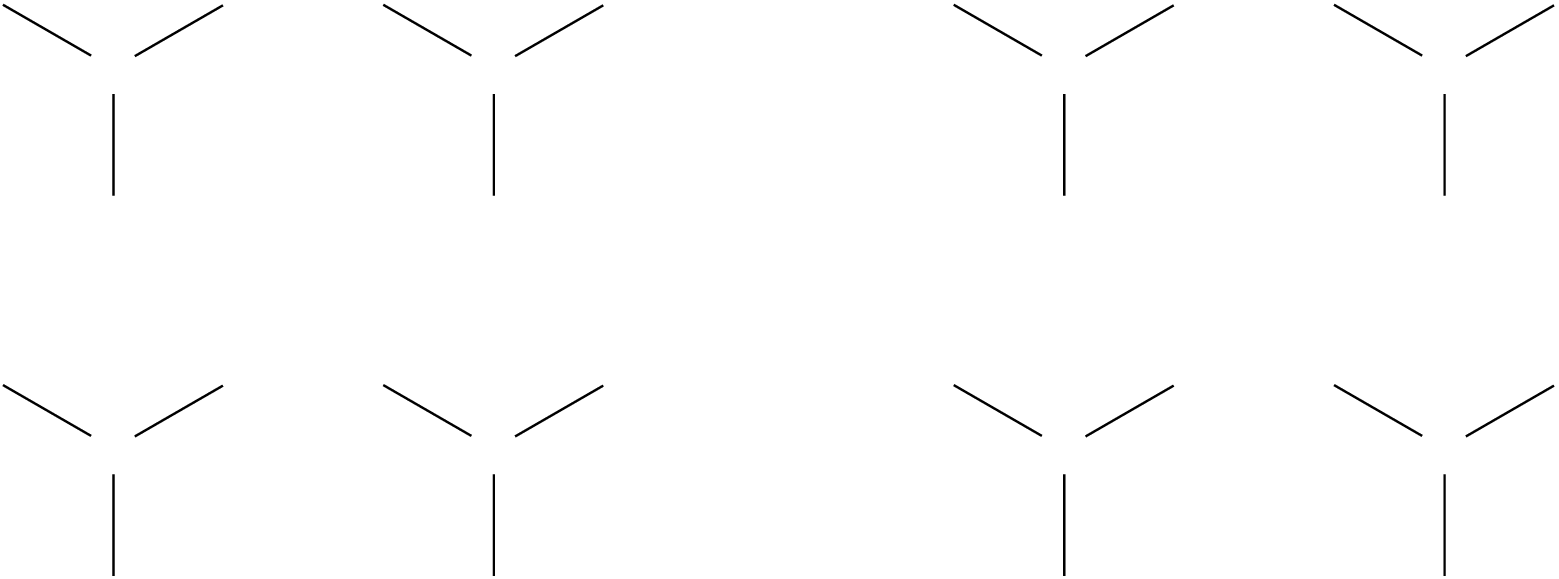}}
  \setlength{\unitlength}{24810sp}
  \setlength{\unitlength}{1pt}
  \put(-7.3,97)   {\scriptsize$+$}
  \put(16,57)     {\scriptsize$+$}
  \put(16,86)     {\scriptsize$+$}
  \put(40,97)     {\scriptsize$+$}
  \put(49,78)     {$=$}
  \put(57.3,97)   {\scriptsize$+$}
  \put(80.8,57)   {\scriptsize$+$}
  \put(80.8,84.6) {\scriptsize$-$}
  \put(104.6,97)  {\scriptsize$+$}
  \put(-7.3,32.6) {\scriptsize$-$}
  \put(16,-5.7)   {\scriptsize$-$}
  \put(16,19.6)   {\scriptsize$-$}
  \put(40,32)     {\scriptsize$-$}
  \put(49,13)     {$=$}
  \put(57.3,32.6) {\scriptsize$-$}
  \put(80.8,-5.7) {\scriptsize$-$}
  \put(80.8,21)   {\scriptsize$+$}
  \put(104.6,32.6){\scriptsize$-$}
  \put(154.8,97)  {\scriptsize$+$}
  \put(178.1,57)  {\scriptsize$+$}
  \put(178.1,86)  {\scriptsize$+$}
  \put(201.6,97.6){\scriptsize$-$}
  \put(210,78)    {$=$}
  \put(219.6,97)  {\scriptsize$+$}
  \put(242.9,57)  {\scriptsize$+$}
  \put(242.9,84.6){\scriptsize$-$}
  \put(266.9,97.6){\scriptsize$-$}
  \put(154.8,32)  {\scriptsize$+$}
  \put(178.1,-5.7){\scriptsize$-$}
  \put(178.1,19.6){\scriptsize$-$}
  \put(201.6,32)  {\scriptsize$-$}
  \put(210,13)    {$=$}
  \put(219.6,32)  {\scriptsize$+$}
  \put(242.9,-5.7){\scriptsize$-$}
  \put(242.9,21)  {\scriptsize$+$}
  \put(266.9,32)  {\scriptsize$-$}
  \epicture11 \labl{eq:ch3a}
In fact of these four relations only two are independent. 
The equations arranged in a column in \erf{eq:ch3a} are
the same; they are just looked at from the two different sides of the 
paper plane. 
\\
Let us demonstrate the top left equality in \erf{eq:ch3a}
as an example. Choosing one of the three ways to insert the
bulk vertex (due to independence of choice \#4 it does not matter which
one) and one of the two ways to put the ribbon graphs for the edges
(choices \#5a, \#5b above) the equality amounts to
  \bea  \begin{picture}(400,91)(0,56) 
  \put(0,0)     {\includeourbeautifulpicture{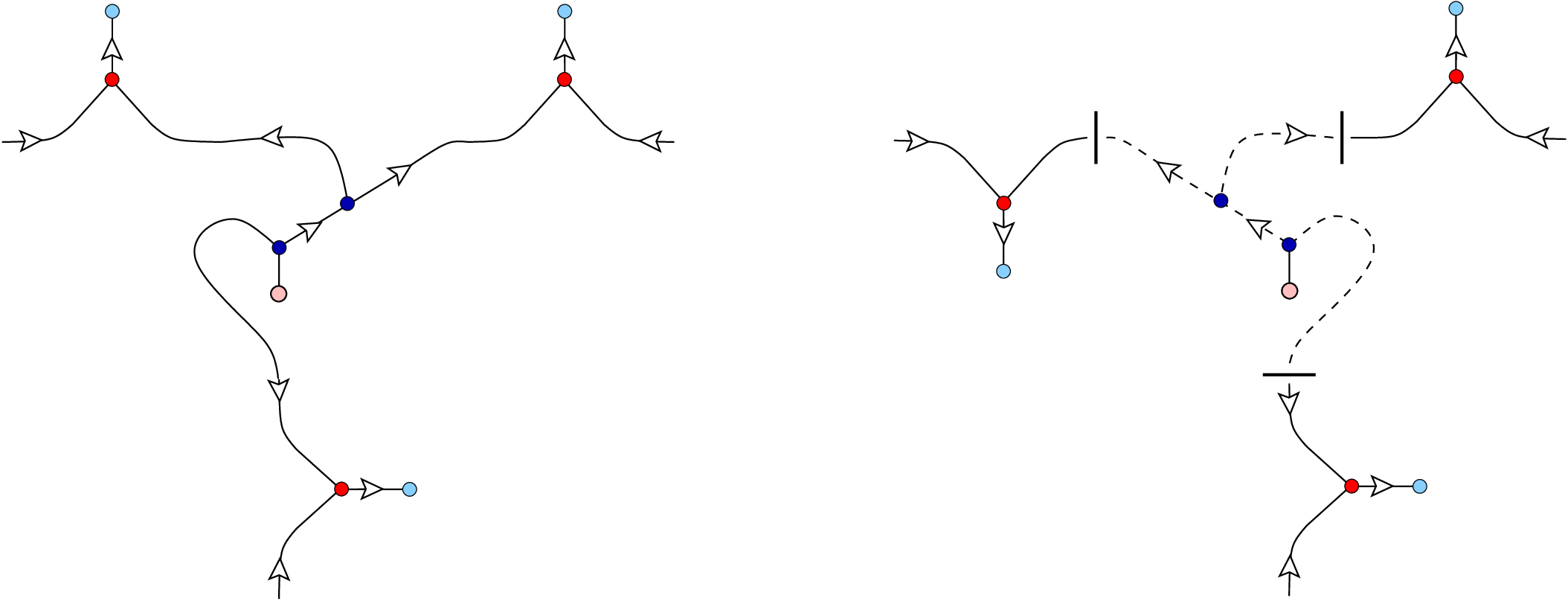}}
  \setlength{\unitlength}{24810sp}
  \setlength{\unitlength}{1pt}
  \put(-3.2,115.8) {\scriptsize$A$}
  \put(64.7,-8.8)  {\scriptsize$A$}
  \put(163.9,115.8){\scriptsize$A$}
  \put(189,71)     {$ = $}
  \put(215.6,115.8) {\scriptsize$A$}
  \put(313.1,-8.8) {\scriptsize$A$}
  \put(382.2,115.8){\scriptsize$A$}
  \epicture35 \labl{eq:ch3b}
This equality, in turn, is again an immediate consequence of the
properties \erf{eq:bar-prop}.
\\[5pt]
\nxt\underline{Choice \#1}:\\[2pt]
We must show that the choice of different representatives
from the classes labelling boundary components leads to
equivalent ribbon graphs. Pick a boundary component $D$. 
Let $(M,\orr_1)$ and $(M',\orr_1')$ be two representatives of the class
$[M,\orr_1]$ labelling $D$. By the definition \erf{eq:bnd-class} of
the equivalence relation there are then two possibilites:
\\[3pt]
(i) $\orr_1\eq\orr_1'$. Then $M' \,{\cong}\, M$. Let 
$\varphi\iN\HomA(M',M)$ be an invertible intertwiner, 
i.e.\ $\rho_{M'}\eq\varphi^{-1}\cir
   $\linebreak[0]$
\rho_M\cir(\id_A{\otimes}\varphi)$. 
The corresponding equality between ribbon graphs looks as
  \bea  \begin{picture}(340,41)(0,33) 
  \put(0,0)     {\includeourbeautifulpicture{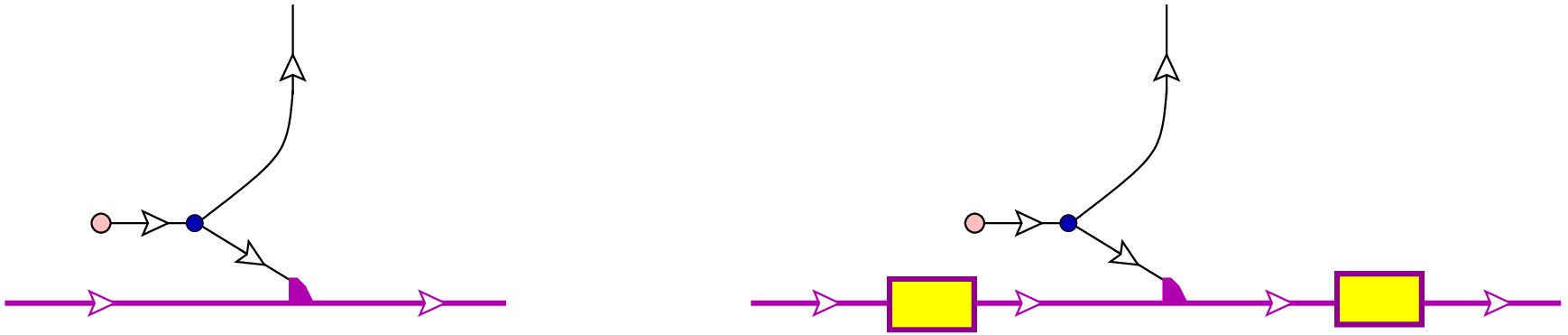}}
  \setlength{\unitlength}{24810sp}
  \setlength{\unitlength}{1pt} 
  \put(-3.1,9.3)    {\scriptsize$\M'$}
  \put(56.5,.6)     {\scriptsize$\r_{\!M'}^{}$} 
  \put(56.3,69.7)   {\scriptsize$A$} 
  \put(98.6,9.3)    {\scriptsize$\M'$}
  \put(124,23)      {$ = $}
  \put(148.2,9.3)   {\scriptsize$\M'$}
  \put(186.5,5.1)   {\scriptsize$\varphi$} 
  \put(216,-2.9)    {\scriptsize$\M$} 
  \put(233.4,69.4)  {\scriptsize$A$} 
  \put(233.7,.6)    {\scriptsize$\r_{\!M}^{}$} 
  \put(274,5.1)     {\scriptsize$\varphi^{\!-\!1}_{}$} 
  \put(312.5,9.3)   {\scriptsize$\M'$}
  \epicture18 \labl{eq:ch1a}
Starting from the ribbon graph obtained with the representative 
$(M',\orr_1')$, this identity can be substituted at each each vertex on 
$D$. Since $D$ is topologically a circle, each $\varphi$ will then cancel
against the $\varphi^{-1}$ from a neighbouring vertex. In the end
what is left is the ribbon graph for the representative $(M,\orr_1)$. 
\\[3pt]
(ii) $\orr_1\eq{-}\orr_1'$. Then $M'\,{\cong}\,M^\sigma$. Let 
$\varphi\iN\HomA(M',M^\sigma)$ be an invertible intertwiner, 
i.e.\ $\rho_{M'}\eq\varphi^{-1} \cir \rho_M^\sigma 
\cir (\id_A{\otimes}\varphi)$. We would like to show the identity
  \bea  \begin{picture}(360,36)(0,44) 
  \put(0,0)     {\includeourbeautifulpicture{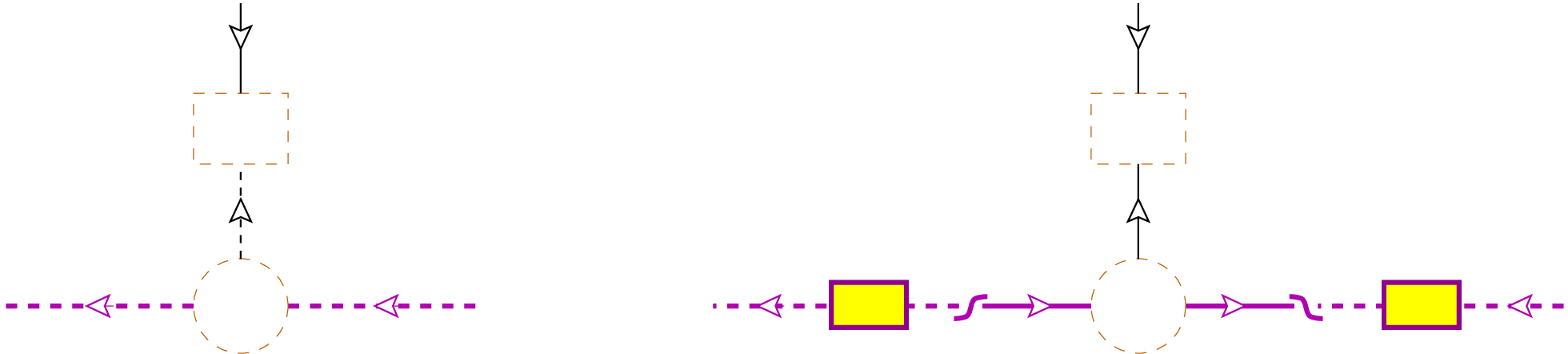}}
  \setlength{\unitlength}{24810sp}
  \setlength{\unitlength}{1pt}
  \put(-2.9,14.3)   {\scriptsize$\M'$}
  \put(56.3,25.5)   {\scriptsize$A$} 
  \put(56.3,67.4)   {\scriptsize$A$} 
  \put(104.9,14.3)  {\scriptsize$\M'$}
  \put(133,23)      {$ = $}
  \put(158.6,14.3)  {\scriptsize$\M'$}
  \put(193,9.6)     {\scriptsize$\varphi^{\!-\!1}_{}$} 
  \put(239,1.8)     {\scriptsize$\M$}
  \put(261.7,25.5)  {\scriptsize$A$} 
  \put(261.7,67.4)  {\scriptsize$A$} 
  \put(322.5,10.1)  {\scriptsize$\varphi$}
  \put(353.5,14.3)  {\scriptsize$\M'$}
  \epicture22 \labl{eq:ch1b}
respectively -- depending on the orientation of the vertex the 
outgoing $A$-ribbon connects to -- the corresponding identity with
the outgoing $A$-ribbon drawn dashed. Below we demonstrate 
\erf{eq:ch1b}; the other identity can be seen similarly.
If we substitute the equality \erf{eq:ch1b} at
every vertex on $D$, the half-twists as well as the morphisms 
$\varphi$, $\varphi^{-1}$ from neighbouring vertices cancel,
thus transforming the ribbon graph resulting from choosing the
representative $(M',\orr_1')$ into the one for $(M,\orr_1)$.
\\
To establish the relation \erf{eq:ch1b} consider the moves
  \bea  \begin{picture}(420,92)(20,70)
  \put(0,0)     {\includeourbeautifulpicture{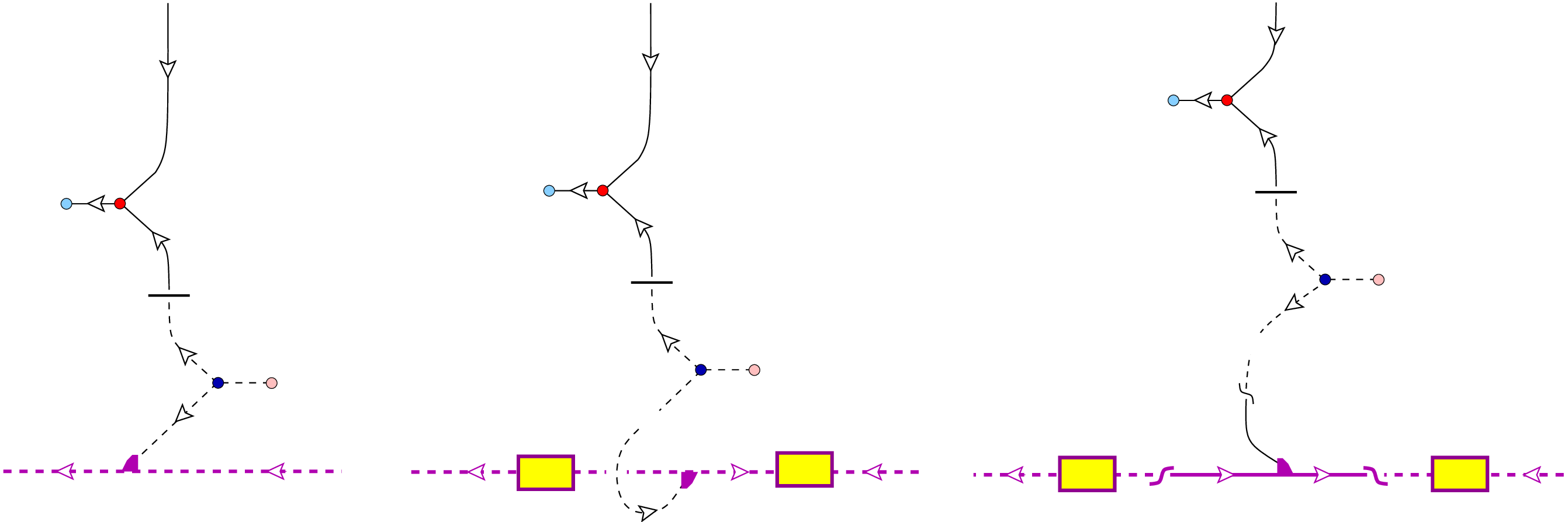}
  \setlength{\unitlength}{24810sp}
     \put(531,85)   {\scriptsize$\sigma$}
     \put(1031,143)   {\scriptsize$\sigma$}
  \setlength{\unitlength}{1pt}
  }
  \put(-3.5,18.8)   {\scriptsize$\M'$}
  \put(54.3,156.8)  {\scriptsize$A$}
  \put(102,18.8)    {\scriptsize$\M'$}
  \put(115,77)      {$ = $}
  \put(125,18.8)    {\scriptsize$\M'$}
  \put(164.4,13.1)  {\scriptsize$\varphi^{\!-\!1}_{}$} 
  \put(205.1,156.8) {\scriptsize$A$} 
  \put(217.4,9.5)   {\scriptsize$\r_{\!M}^{}$}
  \put(249.3,14.9)  {\scriptsize$\varphi$}
  \put(284.3,18.8)  {\scriptsize$\M'$}
  \put(293,77)      {$ = $}
  \put(300.5,18.8)  {\scriptsize$\M'$}
  \put(334.5,13.1)  {\scriptsize$\varphi^{\!-\!1}_{}$} 
  \put(401.1,156.8) {\scriptsize$A$} 
  \put(397,9.2)     {\scriptsize$\r_{\!M}^{}$}
  \put(454.5,14.3)  {\scriptsize$\varphi$}
  \put(485,18.8)    {\scriptsize$\M'$}
  \epicture42 \labl{eq:ch1c} 
The \lhs\ of \erf{eq:ch1c} is obtained by substituting the definitions 
\erf{eq:bnd-vertex} and \erf{eq:bulk-edge} into the \lhs\ of \erf{eq:ch1b}. 
Then in the first step the intertwiner between $M'$ and $M^\sigma$ is 
inserted, and the representation morphism $\rho^\sigma_M$ is expressed 
through $\rho_M$ as in \erf{eq:rho-sigma}. The second step amounts to a 
$180^\circ$-rotation of the segment of the ribbon graph around $\rho_M$.
On the \rhs\ of \erf{eq:ch1c} the combination of $\sigma$ and
half-twist can be replaced by the \boxelement\ \erf{eq:bw-bar}. 
Using \erf{eq:bar-prop} the two \boxelement s cancel; the 
resulting ribbon graph is equal to the \rhs\ of \erf{eq:ch1b}.
\\[5pt]
\nxt\underline{Choice \#2}:\\[2pt]
Independence of the triangulation follows by showing that the so-called 
{\em fusion\/} and {\em bubble\/} moves lead to equivalent ribbon graphs,
see (I:5.10). In the oriented case this amounts to showing
the identities (I:5.11) and (I:5.12). Since independence
of the local orientation around the bulk vertices and the
boundary components has already been established above, we
are free to choose the orientations in such a way that they
coincide with those used in the proof for the oriented case,
so that one can copy that proof literally.
(The proof is straightforward, but a bit lengthy, and accordingly has
not been written down in detail in \cite{fuRs4}.)

\medskip

This establishes the independence of the construction of all
the choices involved. In the sequel, after reconsidering
the annulus we will apply the
construction to the M\"obius strip and Klein bottle.

\subsection{The annulus revisited}\label{sec:ann}

The annulus coefficients ${\rm A}_{kMN}$ with two lower indices are those 
appearing in the open string partition function in type I string theory.
Here we will obtain these numbers from considering an annulus
whose boundary components possess the same orientation, as indicated in
figure (\ref{eq:annulus1}\,a) below. This should be contrasted with the 
construction in the orientable case that was given in section I:5.8. In the 
context studied there, the two boundaries of the annulus automatically have 
opposite orientations. But apart from this aspect 
the construction is very similar, so we can go through it rather quickly. 

Recall from section I:5.8 that the double $\hat X$ of an annulus $X$
is a torus, and the connecting manifold $\MX$ is a solid torus. Let the 
two boundary components of $X$ be labelled by the equivalence classes 
$[M,\orr_1]$ and $[N,\orr_1']$.  The orientations $\orr_1$ and $\orr_1'$ of 
the boundary components are as in figure (\ref{eq:annulus1}\,a), which also 
indicates the triangulation we use:
  \bea  \begin{picture}(260,45)(0,46)
  \put(17,0)     {\includeourbeautifulpicture{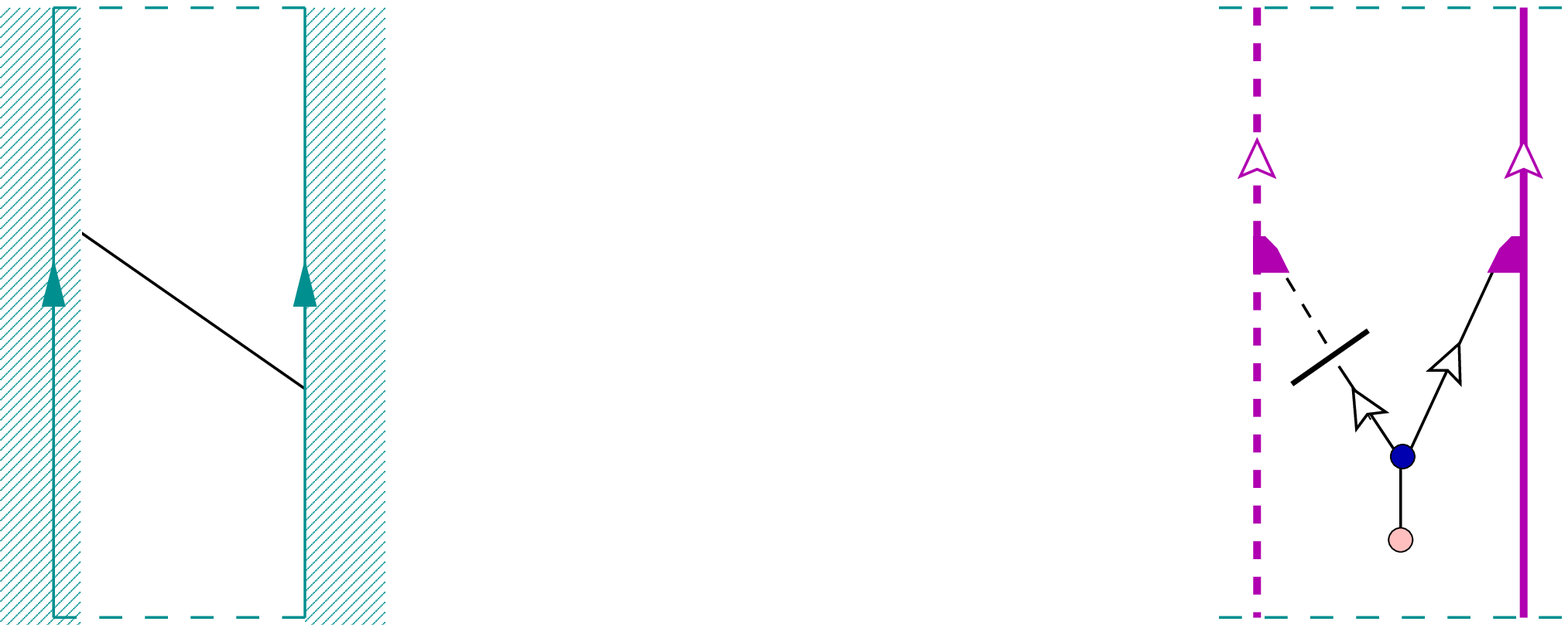}}
  \setlength{\unitlength}{24810sp}
  \setlength{\unitlength}{1pt}
  \put(-15,79)      { a) }
  \put(8.5,17)      {\scriptsize$N$}
  \put(68.3,17)     {\scriptsize$M$}
  \put(163,79)      { b) }
  \put(186.5,14)    {\scriptsize$\dot N$}
  \put(217.7,21)    {\scriptsize$A$}
  \put(234.5,14)    {\scriptsize$\dot M$}
  \epicture25 \labl{eq:annulus1} 
Figure (\ref{eq:annulus1}\,b) shows the resulting ribbon graph. It is
understood that the world sheet is embedded in $\MX$; the ribbon graph 
is then obtained inserting the elements \erf{eq:bnd-vertex} and 
\erf{eq:bulk-edge}. Turning
the $N$-ribbon in (\ref{eq:annulus1}\,b) by $180^\circ$ leads to
the following three-manifold and embedded ribbon graph:
  \bea  \begin{picture}(260,90)(0,58)
  \put(65,0)     {\includeourbeautifulpicture{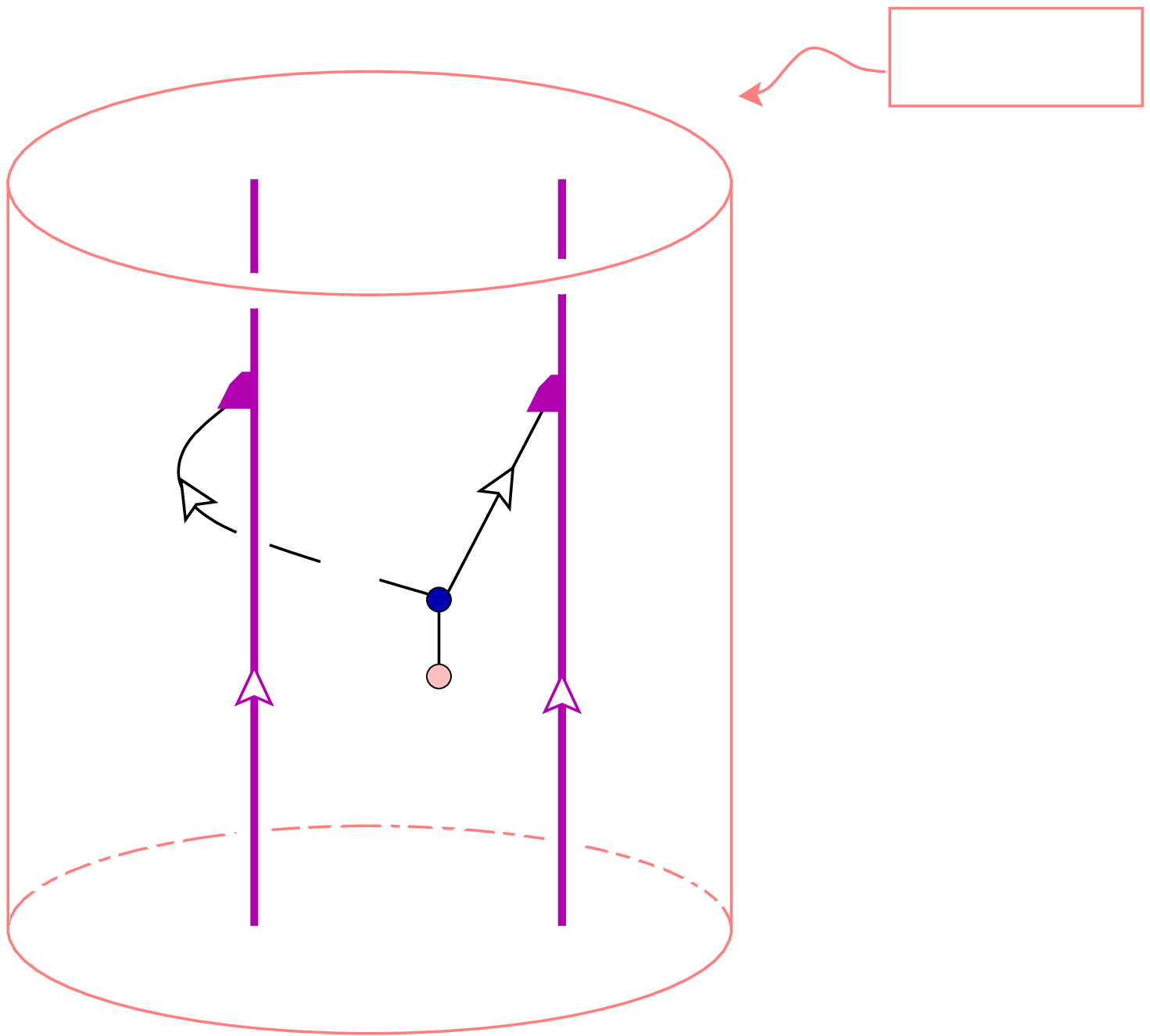}}
  \setlength{\unitlength}{24810sp}
  \setlength{\unitlength}{1pt}
  \put(0,66)        {$ {\rm A}_{MN} \;= $}
  \put(92.3,15.5)   {\scriptsize$\dot N$}
  \put(112.5,63.4)  {\scriptsize$\sigma$}
  \put(128.5,58.3)  {\scriptsize$A$}
  \put(145.6,15.5)  {\scriptsize$\dot M$}
  \put(191.9,133)   {$D{\times}S^1$}
  \put(233,66)      {$ \in \calh(\emptyset;\rmT) \;. $}
  \epicture42 \labl{AMNinH}
Here $\calh(\emptyset;\rmT)$ is the space of zero-point conformal blocks 
on the torus, see section I:5.2. Expanding \erf{AMNinH} in terms of 
characters $|\chii_k;\rmT\rangle$
as in (I:5.118) we obtain the annulus coefficients:
  \bea  \begin{picture}(260,88)(0,56)
  \put(65,0)     {\includeourbeautifulpicture{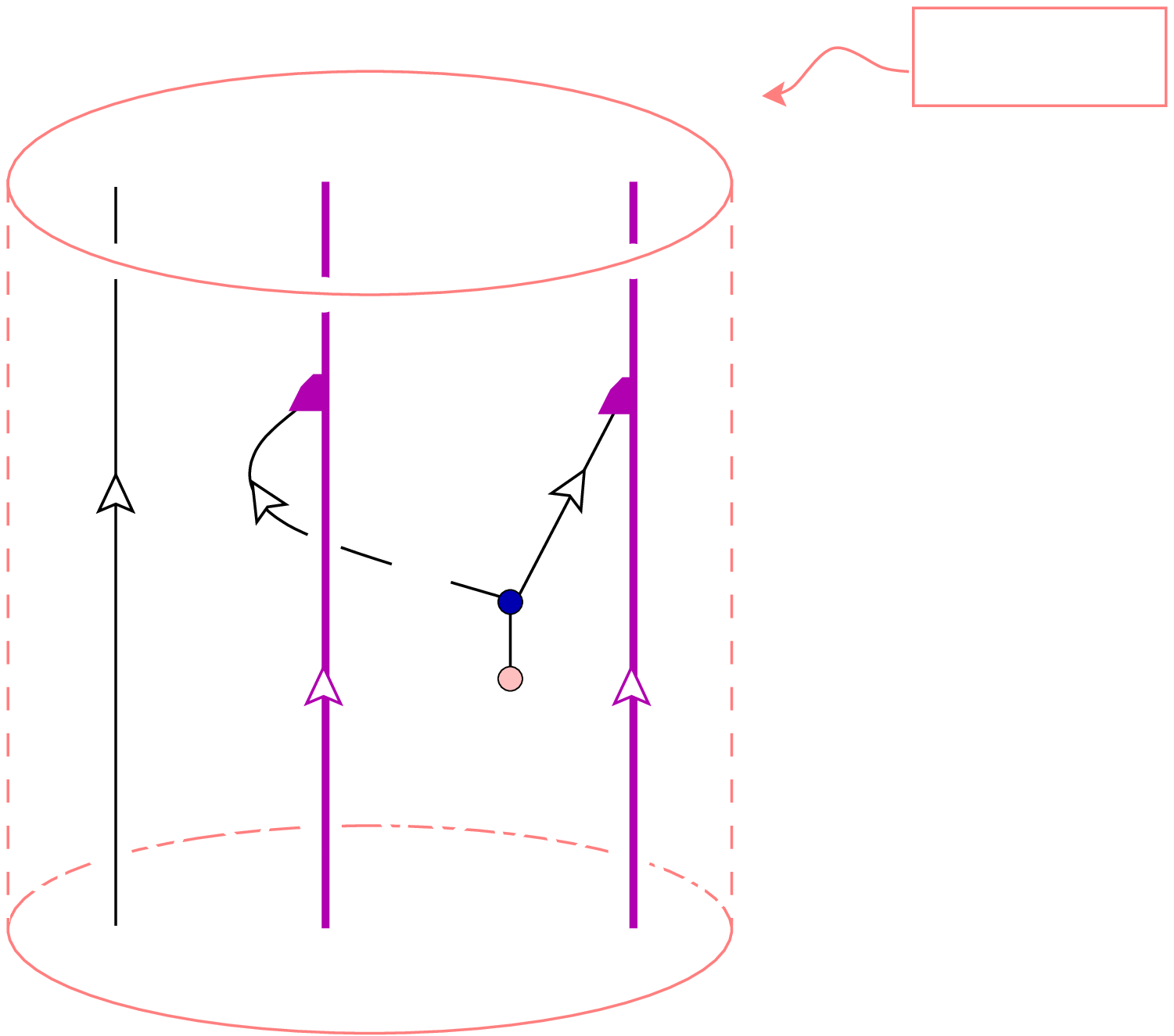}}
  \setlength{\unitlength}{24810sp}
  \setlength{\unitlength}{1pt}
  \put(0,66)        {$ {\rm A}_{kMN} \;= $}
  \put(75.3,15.9)   {\scriptsize$k$}
  \put(102.4,15.6)  {\scriptsize$\dot N$}
  \put(122.2,63.2)  {\scriptsize$\sigma$}
  \put(138.9,58.2)  {\scriptsize$A$}
  \put(144.2,15.6)  {\scriptsize$\dot M$}
  \put(193.8,133.1) {$S^2{\times}S^1$}
  \epicture39 \labl{eq:Akmn}
Note that upon moving the $M$- and $N$-ribbons past each other, rotating the 
$M$-ribbon by $360^\circ$ and using \erf{eq:sig-phi}, it follows that
  \be
  {\rm A}_{kMN} = {\rm A}_{kNM} \,, \labl{eq:Aknm-sym}
i.e.\ the annulus coefficients with two lower boundary indices are symmetric.
This is in contrast to the situation for the annulus coefficients with 
one lower and one upper boundary index, which according to (I:5.122) satisfy
${\rm A}_{kM}^{\ \ N}\eq {\rm A}_{\bar kN}^{\ \;M}$. Also, comparing the
graphs \erf{eq:Akmn} and \erf{eq:rho-sigma} to the graph (I:5.119) for
${\rm A}_{kM}^{\ \ N}$ yields the identity (compare \cite{sosc})
  \be  {\rm A}_{kMN}^{}\eq {\rm A}_{kM}^{\ \;N^\sigma} .  \labl{AkMNs}

As in the oriented case, we can cut the three-manifold \erf{eq:Akmn} along 
an $S^2$ so as to obtain a ribbon graph in $S^2 \,{\times}\,[0,1]$. The 
graphical representation of the resulting cobordism is the same as
in \erf{eq:Akmn}, except that we no longer identify top and bottom.
Let us denote the linear map associated to this cobordism by
  \be
  Q_{kMN} :\quad  \calh(k,\dot N,\M; S^2) \to \calh(k,\dot N,\M; S^2) \,.
  \labl{eq:QkMN}
$Q_{kMN}$ is a projector, so that upon gluing top and bottom of the 
cobordism $S^2\,{\times}\,[0,1]$ back together results in the relation
  \be
  {\rm A}_{kMN} = \Tr{\calH(k,\dot N,\M; S^2)} Q_{kMN} \in \zet_{\ge 0}
   \,.  \labl{eq:Akmn-trace}
The projector property can be shown in much the same way as in (I:5.127),
the only difference being that we also must use the equality \erf{eq:*ssFA2}.
The trace computes the dimension of the image of the projector and hence is 
a non-negative integer. We will see that the projector $Q_{kMN}$ also enters
in the computation of the M\"obius strip amplitude.

\dtl{Remark}{maslovo}
Note that in writing \erf{eq:Akmn} and \erf{eq:Akmn-trace} we have implicitly 
made use of the functoriality axiom (I:2.56). For example, in \erf{eq:Akmn}
functoriality reads
  \be
  \langle\chi_k;\rmT| {\rm A}_{MN} = \kappa^m\,
  Z({\rm A}_{kMN},\emptyset,\emptyset)\,1  \ee
with $\kappa\eq S_{0,0}\sum_{j\in\II}\theta_j^{-1}\dim(U_j)^2$.
This leads to the result \erf{eq:Akmn} provided that the factor
$\kappa^m$ arising form the framing anomaly is equal to 1.
Now recall from remark \ref{remaslov} that the Lagrangian subspace in
$\rmT\eq\partial{\rm A}_{MN}$ is spanned by the cycle contractible in
${\rm A}_{MN}$. The same holds for $\partial M^-_{\chiI_k}$ (see section
\ref{CcMK} below and appendix \ref{sec:app-tor}). Then the 
argument leading to \erf{eq:chi-chi-maslov} is applicable, and using it
one can show that the two Maslov indices determining $m$ are equal to zero.

For the same reason there will not appear any anomaly factor $\kappa^m$
in the equations \erf{eq:MkR} for the M\"obius strip, \erf{eq:Kk} for the
Klein bottle, \erf{eq:MkR-CkR}, \erf{eq:MR-bas-xfer} and \erf{eq:pic102} 
for the M\"obius strip in the crossed channel, and \erf{eq:pic109} for
the crossed-channel Klein bottle.
With a similar reasoning one can show that the anomaly $m$ vanishes
in the relations expressed by \erf{eq:Akmn-trace}, as well as in 
\erf{eq:MkR-trace} and \erf{eq:Kk-trace} below, i.e.\ the factor 
$\kappa^m$ that a priori appears in these formulas is equal to 1.
In general one has the trace formula (I:2.58).

\medskip

The boundary conjugation matrix $C^\sigma$ defined in proposition 
\ref{pr:CB-prop} coincides with the one described in \cite{prss3,sosc}, 
i.e.\ it can be expressed in terms of
annulus coefficients with two lower indices. This is stated in
\\[-2.3em]

\dtl{Lemma}{lem:C-A}
The boundary conjugation matrix \erf{eq:CB-def} satisfies
  \be  C^\sigma_{MN} = {\rm A}_{0MN}  \,.  \labl{Csigma=A0}
  
\medskip\noindent
Proof:\\
In the present approach this relation emerges as follows. According to 
formula \erf{AkMNs}, interpreting the annulus coefficients ${\rm A}_{kM}
^{\ \ N}$ as dimensions of spaces of $A$-morphisms as in (I:5.135) yields
  \be
  {\rm A}_{kMN} = \dimc \, \HomA(M \oti U_k , N^\sigma) \,.  \labl{A=dimhom}
Taking now $M$ and $N$ to be simple $A$-modules and setting $k\eq 0$, 
comparison with the definition \erf{eq:CB-def} of $C^\sigma$ gives 
\erf{Csigma=A0}.
\qed

The equality \erf{Csigma=A0} shows that it is indeed natural to use 
$C^\sigma$ and its inverse to lower and raise the indices of annulus
partition functions.

\medskip

Just like in \cite{fuRs4} we will present more explicit formulas only for 
the special case that $\N ijk \iN \{0,1\}$ and $\lra{U_k}A \iN \{0,1\}$ 
for all $i,j,k\in\I$.  The evaluation of the invariant ${\rm A}_{0MN}$ is 
analogous to the computation of ${\rm A}_{kM}^{\ \;N}$ in (I:5.144); we find
  \bea  \begin{picture}(310,48)(0,52)
  \put(188,0)     {\includeourbeautifulpicture{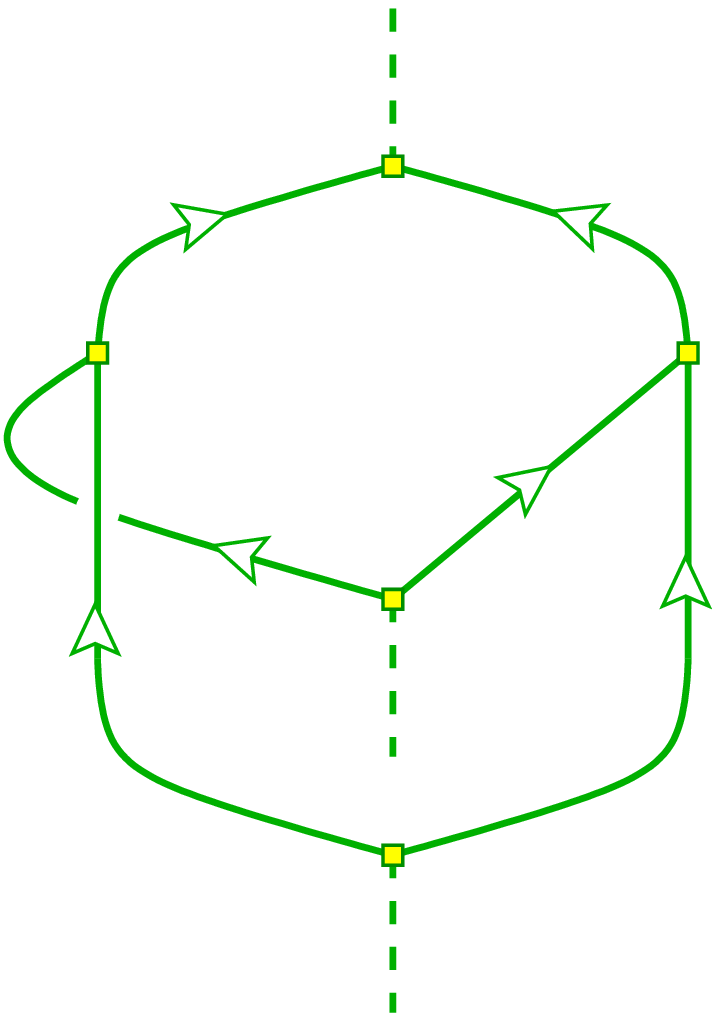}}
  \setlength{\unitlength}{24810sp}
  \setlength{\unitlength}{1pt}
  \put(0,53)        {$ {\rm A}_{0MN} \;=\; \dsty\sum_{a,n,\alpha,\beta}
                       \r_{a,n\alpha}^{N\,n\alpha}
                       \r_{\bar a,\bar n\beta}^{M\,\bar n\beta} \,
                       \sigma(a)\, \Delta^{a \bar a}_0 $}
  \put(203.3,29.8)  {\scriptsize$n$}
  \put(203.3,78.2)  {\scriptsize$n$}
  \put(205.2,48)    {\scriptsize$a$}
  \put(251.5,69)    {\scriptsize$\bar a$}
  \put(253.6,78.2)  {\scriptsize$\bar n$}
  \put(254.1,29.8)  {\scriptsize$\bar n$}
  \epicture33  \ee
Here the $a$-summation runs over simple subobjects of $A$, the $n$-summation 
over simple subobjects $U_n$ of the simple module $N$ such that $U_n^\vee$
is a subobject of $M$, and the labels $\alpha$, $\beta$ are multiplicity labels
for the occurrence of $U_n$ in $N$ and $U_n^\vee$ in $M$, respectively.
It follows that
  \be
  C^\sigma_{MN} = \sum_{a,n,\alpha,\beta}
  \rho_{a,n\alpha}^{N\,n\alpha} \rho_{\bar a,\bar n\beta}^{M\,\bar n\beta}\,
  \sigma(a)\, \Delta^{a \bar a}_0\,
  \Rs{}nan \Gs na{\bar a}nn0 \Fs n{\bar a}{\bar n}0{\bar n}n
  \labl{eq:conj-inv}

\newpage

\subsection{The M\"obius strip}\label{sec:moebius}

Like any world sheet $X$, the M\"obius strip is obtained from its complex
double $\hat X$ by dividing out an anticonformal involution $\sigma$.
The double $\hat X$ is a torus with complex structure parametrised by
$\tau\eq\frac12+\ii s$ with $s\iN\reals_{\ge0}$. The involution
acts as $\sigma(z)\eq1{-}\bar z$. Using that $z$ is identified
with $z{+}1$ and $z{+}\tau$ it follows that the fixed
points of $\sigma$ are $\frac12\,{+}\,\ii\reals$ and $1\,{+}\,\ii\reals$,
and that a fundamental domain $X$ for the action of $\sigma$ on $\hat X$
is given by the shaded region in figure (\ref{eq:Moebius}\,a):
  \bea  \begin{picture}(330,47)(0,53)
  \put(48,0)      {\includeourbeautifulpicture{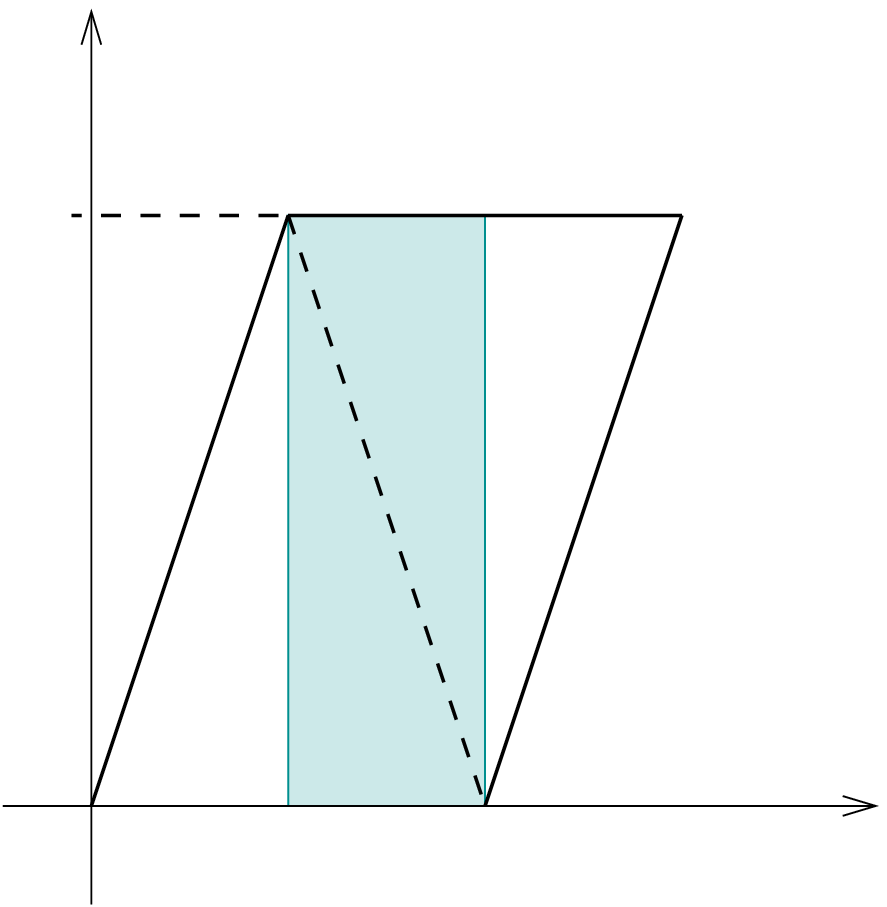}
  \setlength{\unitlength}{24810sp}
  \put(0,198)        {\tiny$\ii s$}
  \setlength{\unitlength}{1pt}
  }
  \put(254,15)    {\includeourbeautifulpicture{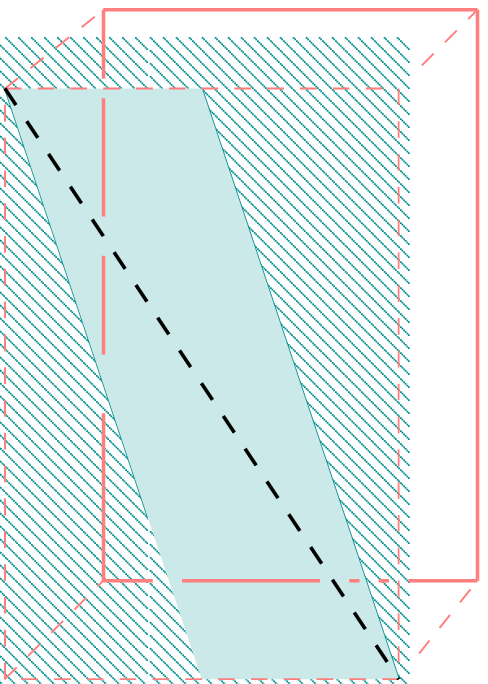}}
  \setlength{\unitlength}{24810sp}
  \setlength{\unitlength}{1pt}
  \put(0,47)        {$ \hat X \;= $ }
  \put(30,90)       { a) }
  \put(64,13.5)     {\scriptsize$E'$}
  \put(76.1,5.2)    {\tiny$\frac12$}
  \put(86,13.5)     {\scriptsize$E$}
  \put(89,67.5)     {\scriptsize$F$}
  \put(99.2,5.2)    {\tiny$1$}
  \put(107,67.5)    {\scriptsize$F'$}
  \put(193,47)      {$ \MX \;= $ }
  \put(224,80)      { b) }
  \put(255.7,35.5)  {\scriptsize$E'$}
  \put(274,28.5)    {\scriptsize$E$}
  \put(271,64.5)    {\scriptsize$F$}
  \put(283,70.5)    {\scriptsize$F'$}
  \put(299.3,13.8)  {\tiny$t=0$}
  \put(308,24.4)    {\tiny$t=1$}
  \epicture31  \labl{eq:Moebius}
It is easily checked that $X$ indeed has the topology of a M\"obius strip.

The connecting manifold $\MX$ is constructed according to
\erf{eq:MX-def}: $\MX\eq\hat X \,{\times}\, [-1,1] \,{/}{\sim}\,$. This is 
drawn in (\ref{eq:Moebius}\,b), where for convenience the quadrangle 
defining the torus as in (\ref{eq:Moebius}\,a) is drawn with right angles 
-- which we may do because when applying TFT arguments we are interested in 
$\MX$ only as a topological manifold. The top and bottom rectangles of 
(\ref{eq:Moebius}\,b), as well as the left and right side, are identified.  
Figure (\ref{eq:Moebius}\,b) also shows a fundamental domain for the action 
of the relation $\sim\,$. It consists of $\hat X\,{\times}\,[0,1]$, 
together with an identification of points on $X{\times}\{0\}\,$: points 
in the triangle $E$ are identified with their reflections 
in $E'$, and points in $F$ with their reflections in $F'$.

For a horizontal section of figure (\ref{eq:Moebius}\,b) the
identification yields the topology of a disk. 
For the bottom side of (\ref{eq:Moebius}\,b), this is
illustrated in figure (\ref{eq:Moebius2}\,a):
  \bea  \begin{picture}(420,73)(11,68)
  \put(24,24)     {\includeourbeautifulpicture{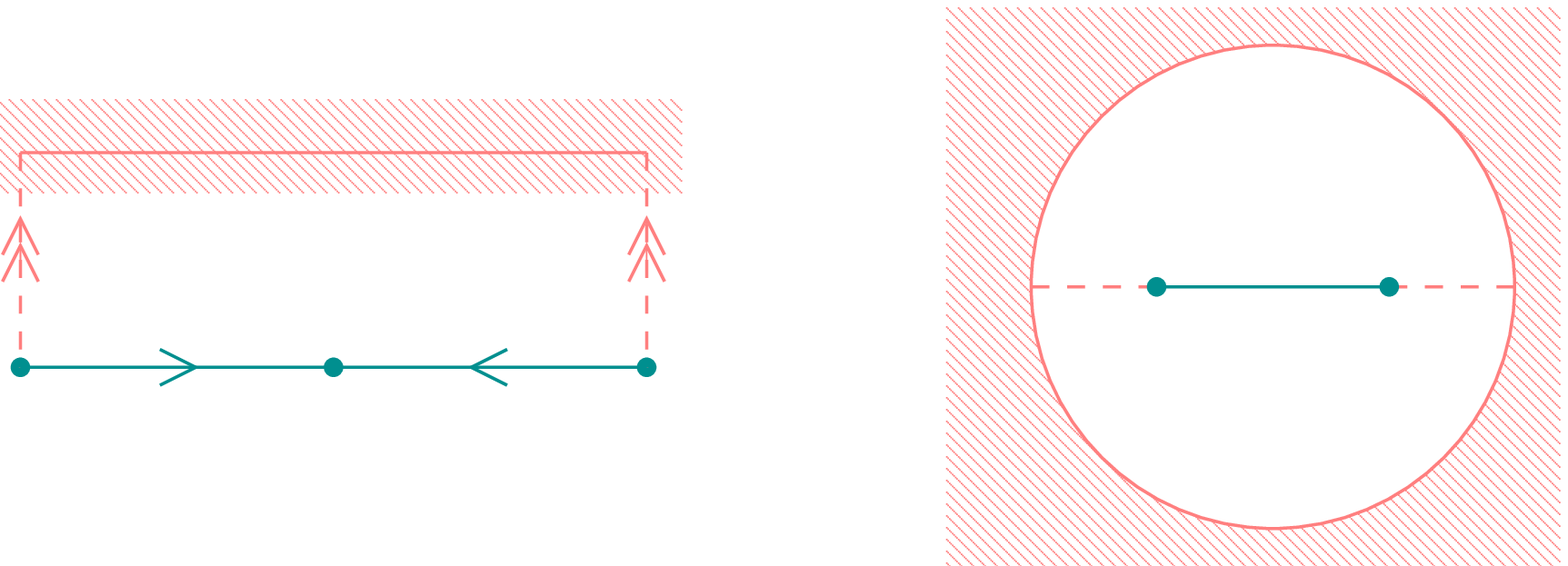}}
  \put(285,0)     {\includeourbeautifulpicture{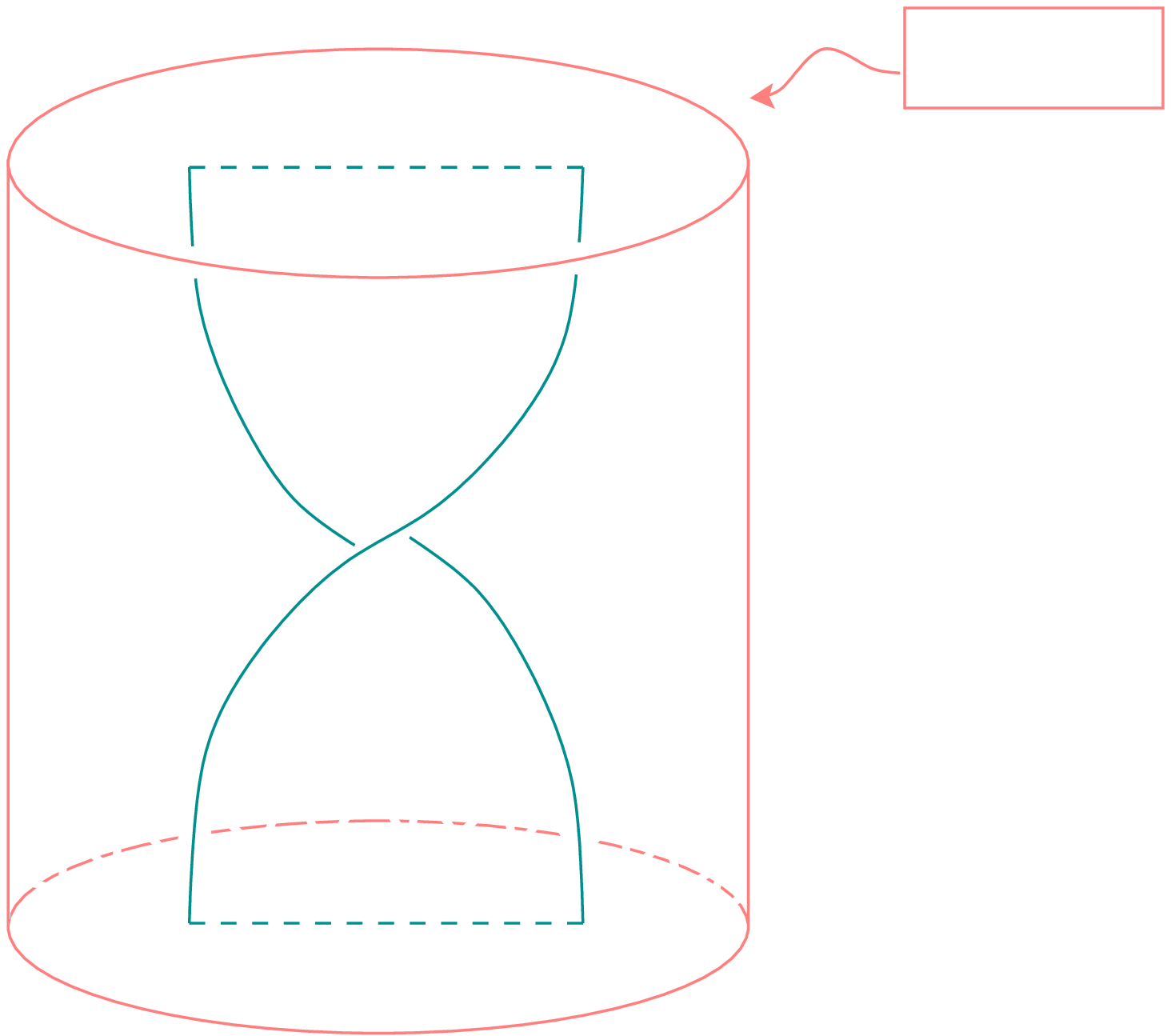}}
  \setlength{\unitlength}{24810sp}
  \setlength{\unitlength}{1pt}
  \put(0,79)        { a) }
  \put(31.5,40.5)   {\scriptsize$E$}
  \put(88.5,40.5)   {\scriptsize$E'$}
  \put(122,55)      {$ = $ }
  \put(173.5,48.8)  {\scriptsize$E$}
  \put(173.5,60.5)  {\scriptsize$E'$}
  \put(260,126)     { b) }
  \put(411.3,129.8) {$D{\times}S^1$}
  \epicture38  \labl{eq:Moebius2}
As we move upwards in (\ref{eq:Moebius}\,b), the identification
fixed points move half a period along the horizontal direction;
this results in figure (\ref{eq:Moebius2}\,b). The two-manifold
drawn in $\MX$ is thus the image of the embedding
$\iota_X{:}\; X\,{\to}\,\MX$ of the M\"obius strip
as in \erf{eq:embed-X}.

Next we construct the ribbon graph embedded in $\MX$. Let
the boundary condition be $[R,\orr_1]$, where the boundary
orientation $\orr_1$ is `up' in (\ref{eq:Moebius}\,a). 
As a triangulation to start with we choose one that is slightly
more complicated than necessary, in order to illustrate how the 
moves \erf{eq:bar-prop} simplify the resulting ribbon graph:
  \bea  \begin{picture}(385,36)(0,54)
  \put(0,0)       {\includeourbeautifulpicture{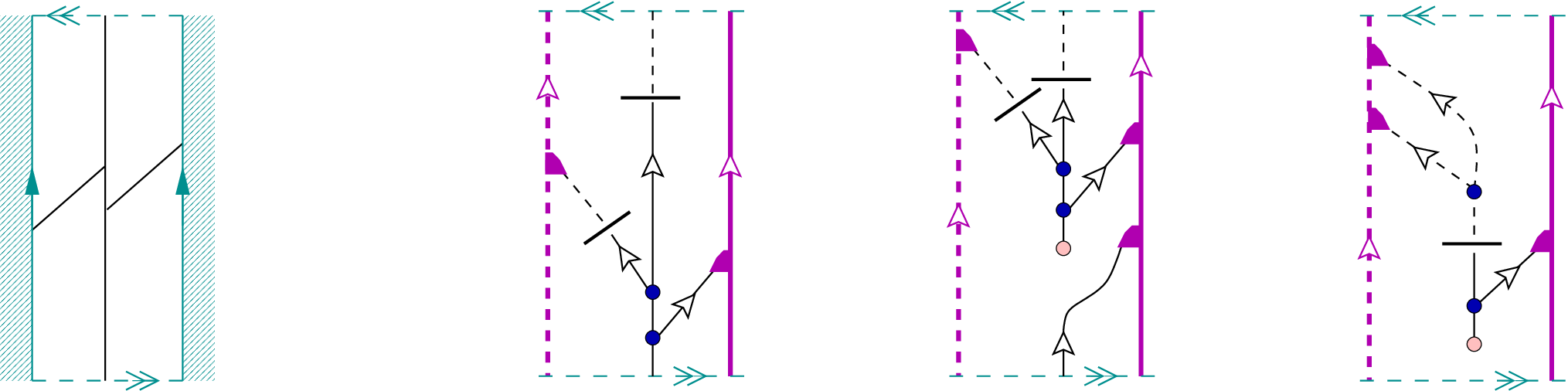}}
  \setlength{\unitlength}{24810sp}
  \setlength{\unitlength}{1pt}
  \put(-8.5,17)     {\scriptsize$R$}
  \put(51.5,17)     {\scriptsize$R$}
  \put(77,43)       {$ \longmapsto $}
  \put(120.6,13)    {\scriptsize$\dot R$}
  \put(145.4,19.4)  {\scriptsize$A$}
  \put(174.5,13)    {\scriptsize$\dot R$}
  \put(196,43)      {$ = $}
  \put(218.1,13)    {\scriptsize$\dot R$}
  \put(243.4,49)    {\scriptsize$A$}
  \put(257.5,17)    {\scriptsize$A$}
  \put(272.2,13)    {\scriptsize$\dot R$}
  \put(292,43)      {$ = $}
  \put(315.2,13)    {\scriptsize$\dot R$}
  \put(351.5,17)    {\scriptsize$A$}
  \put(368.8,13)    {\scriptsize$\dot R$}
  \epicture31  \labl{eq:pic31}
The left hand side shows the starting point -- the chosen triangulation 
of the M\"obius strip, together with a representative $(R,\orr_1)$ to 
label the boundary. In the other pictures the world sheet is regarded as 
being embedded in $\MX$. The ribbon graph in the second picture is 
obtained by choosing the orientation of the paper plane at the two 
interior vertices and inserting the graphs \erf{eq:bulk-vertex}, 
\erf{eq:bnd-vertex} and \erf{eq:bulk-edge}, as well as removing some 
unit and counit morphisms via the Frobenius property. The first equality 
is a consequence of the representation property of $\rho_R$ and the  
Frobenius property. To obtain the second equality, one takes the bottom 
$A$-ribbon of the third picture through the anti-periodic identification 
of the top and bottom.  In the last picture one can use the representation 
property on the dashed $A$-ribbons, together with specialness of $A$; 
the resulting ribbon graph, with the M\"obius strip embedded in 
$D\,{\times}\, S^1$ as in (\ref{eq:Moebius2}\,b), is
  \bea  \begin{picture}(270,84)(0,53)
  \put(58,0)     {\includeourbeautifulpicture{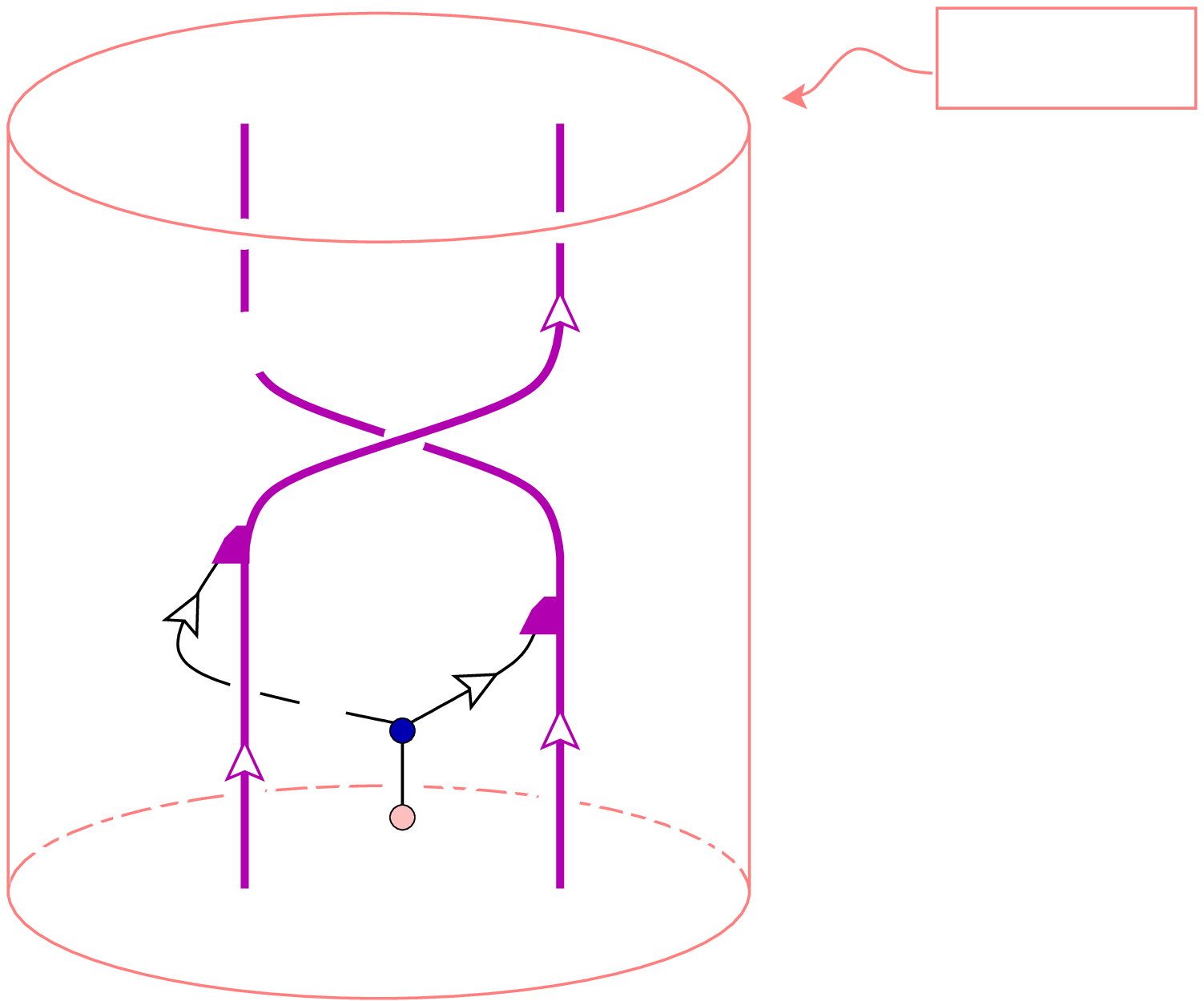}}
  \setlength{\unitlength}{24810sp}
  \setlength{\unitlength}{1pt}
  \put(0,66)        {$ {\rmM}_R \;= $}
  \put(83.4,16.1)   {\scriptsize$\dot R$}
  \put(90.9,88.3)   {\scriptsize$\theta_{\!\dot R}$}
  \put(100.6,38.3)  {\scriptsize$\sigma$}
  \put(115.2,33.6)  {\scriptsize$A$}
  \put(136.5,16.1)  {\scriptsize$\dot R$}
  \put(188.7,124.9) {$D{\times}S^1$}
  \put(210,66)      {$ \in\calh(\emptyset;\rmT) $}
  \epicture30  \ee
Here the twist $\theta_{\!\dot R}$ comes from the two half-twists of the 
boundary ribbon that are implicit in the identification of the boundary
segments in figure \erf{eq:pic31}.

The ribbon invariant $\rmM_R$ can be expanded in terms of the standard
basis of $\calh(\emptyset;\rmT)$ that was given in section I:5.2:
  \be
  {\rmM}_R = \sum_{l\in \II} {\rmM}_{l R}\, |\chii_l;\rmT\rangle
  \labl{eq:MR-basis}
To extract the coefficients ${\rmM}_{kR}$ we compose both sides of 
\erf{eq:MR-basis} with the dual basis $\langle \chii_k;\rmT|$. On the \rhs\ 
this yields $\delta_{k,l}$, while on the \lhs\ it amounts to gluing the 
manifold (I:5.18) to ${\rmM}_R$. The result is a ribbon graph in the 
closed three-manifold $S^2\,{\times}\, S^1$ whose invariant gives 
the coefficients ${\rmM}_{kR}$ of the M\"obius amplitude (there is no
framing anomaly, see remark \ref{maslovo}):
  \bea  \begin{picture}(230,85)(0,50)
  \put(58,0)     {\includeourbeautifulpicture{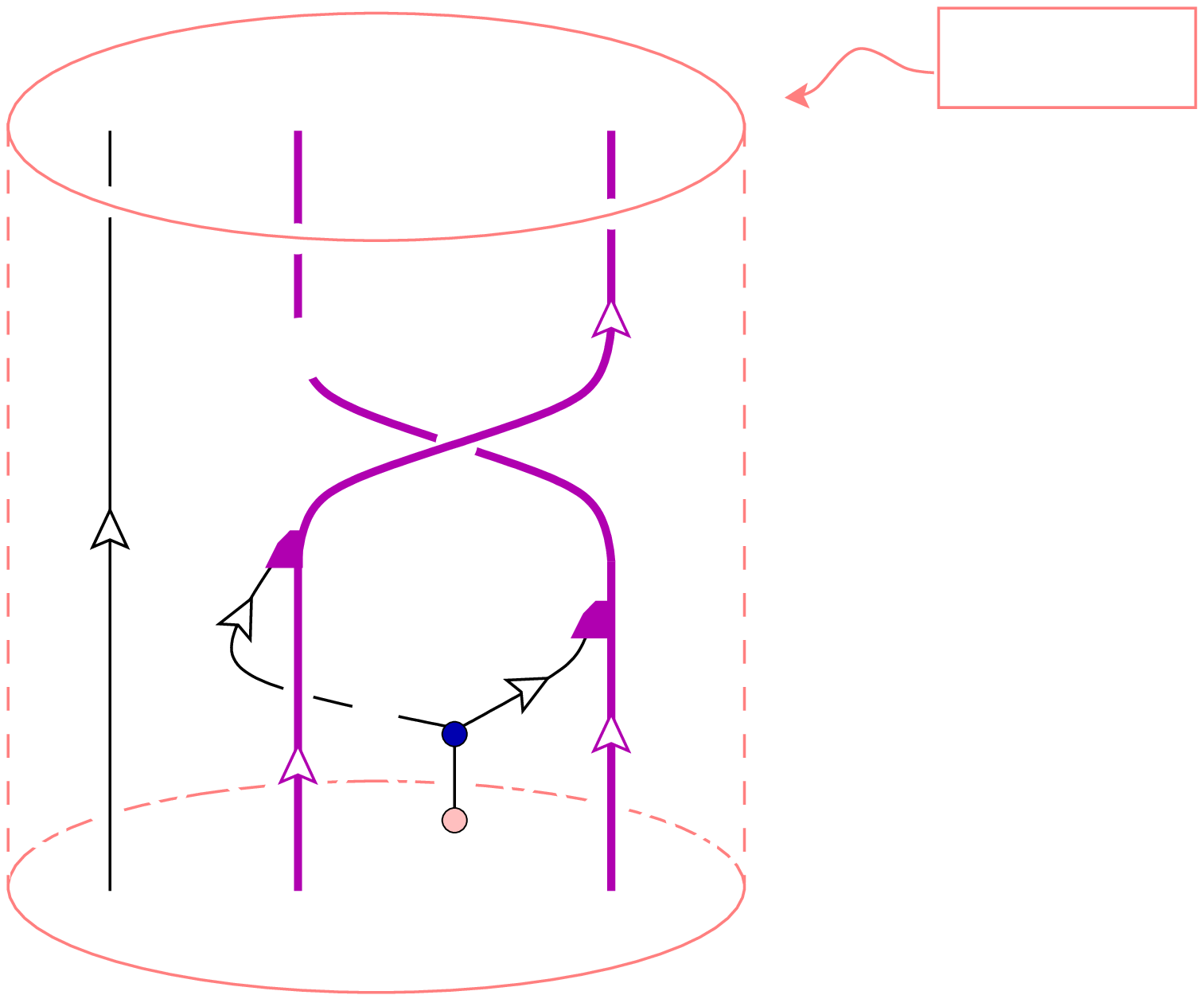}}
  \setlength{\unitlength}{24810sp}
  \setlength{\unitlength}{1pt}
  \put(0,66)        {$ {\rmM}_{kR} \;= $}
  \put(67.5,16.1)   {\scriptsize$k$}
  \put(90.8,15.8)   {\scriptsize$\dot R$}
  \put(98.8,88.1)   {\scriptsize$\theta_{\!\dot R}$}
  \put(108.4,37.7)  {\scriptsize$\sigma$}
  \put(123.1,33)    {\scriptsize$A$}
  \put(133.7,15.8)  {\scriptsize$\dot R$}
  \put(188.5,124.7) {$S^2{\times}S^1$}
  \epicture31  \labl{eq:MkR}
To establish some properties of the numbers ${\rmM}_{kR}$ 
it is helpful rewrite them as a trace:
  \be
  {\rmM}_{kR} = {\rm tr}_{\calH(k,R,R;S^2)} ( T_{kR}\, Q_{kRR} ) \,,
  \labl{eq:MkR-trace}
where $Q_{kRR}$ is the linear map defined in \erf{eq:QkMN} and
$T_{kR}$ is given by
  \bea  \begin{picture}(400,67)(0,43)
  \put(51,0)     {\includeourbeautifulpicture{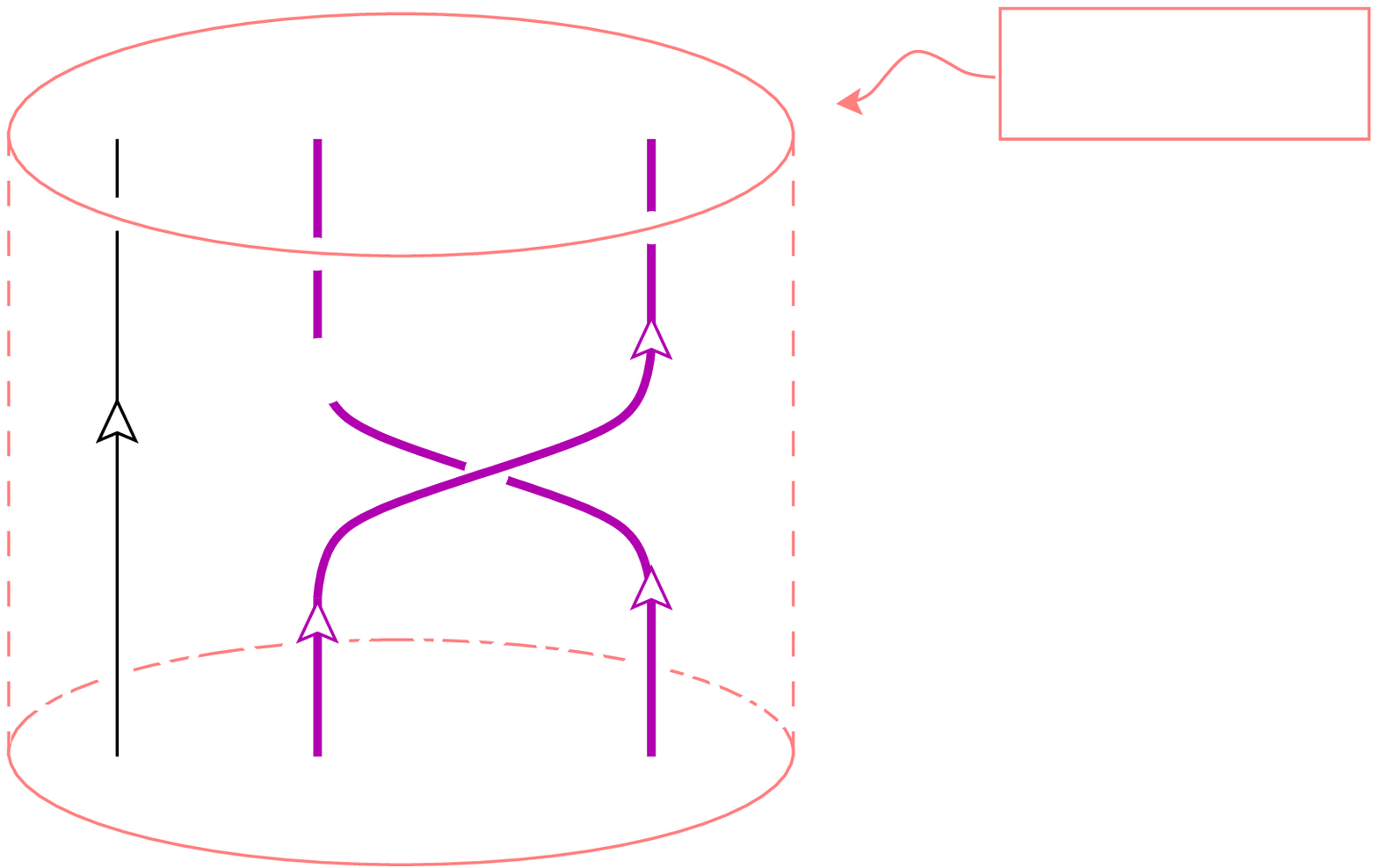}}
  \setlength{\unitlength}{24810sp}
  \setlength{\unitlength}{1pt}
  \put(0,53)        {$ T_{kR} \;= $}
  \put(67.8,15.5)   {\scriptsize$k$}
  \put(93.5,15.2)   {\scriptsize$\dot R$}
  \put(91.3,62.4)   {\scriptsize$\theta_{\!\dot R}$}
  \put(136.9,15.2)  {\scriptsize$\dot R$}
  \put(183.1,100)   {$S^2{\times}[0{,}1]$}
  \put(210,53)      {$ :\quad\calh(k,R,R;S^2)\,\to\,\calh(k,R,R;S^2) $}
  \epicture26  \labl{eq:TkR}
The linear maps $T$ and $Q$ fulfill the relations given in
\\[-2.3em]

\dtl{Lemma}{lem:TQ-prop}
The linear maps $T_{kR}$ and $Q_{kRR}$ defined in \erf{eq:QkMN} and
\erf{eq:TkR} satisfy $Q_{kRR}\,Q_{kRR}\eq Q_{kRR}$ as well as
  \be
  T_{kR}\,T_{kR} = \theta_k\,\id_{\calH(k,R,R;S^2)} \qquad{\rm and}\qquad
  T_{kR}\,Q_{kRR} = Q_{kRR}\,T_{kR} \,.   \labl{TQ-props}
It follows in particular that $T_{kR}$ and $Q_{kRR}$
can be diagonalised simultaneously.

\medskip\noindent
Proof:\\
The first property was already established in section \ref{sec:ann}. 
The first of the relations \erf{TQ-props} can be
seen as follows. It is one of the basic properties of a topological field
theory that when two cobordisms
are related by a homeomorphism that restricts to the identity
on the boundary, then the linear maps assiged to the cobordisms are identical.
The following figure shows two cobordisms from $S^2$ to $S^2$ that are
related in this manner:
  \bea  \begin{picture}(420,102)(18,51)
  \put(0,0)     {\includeourbeautifulpicture{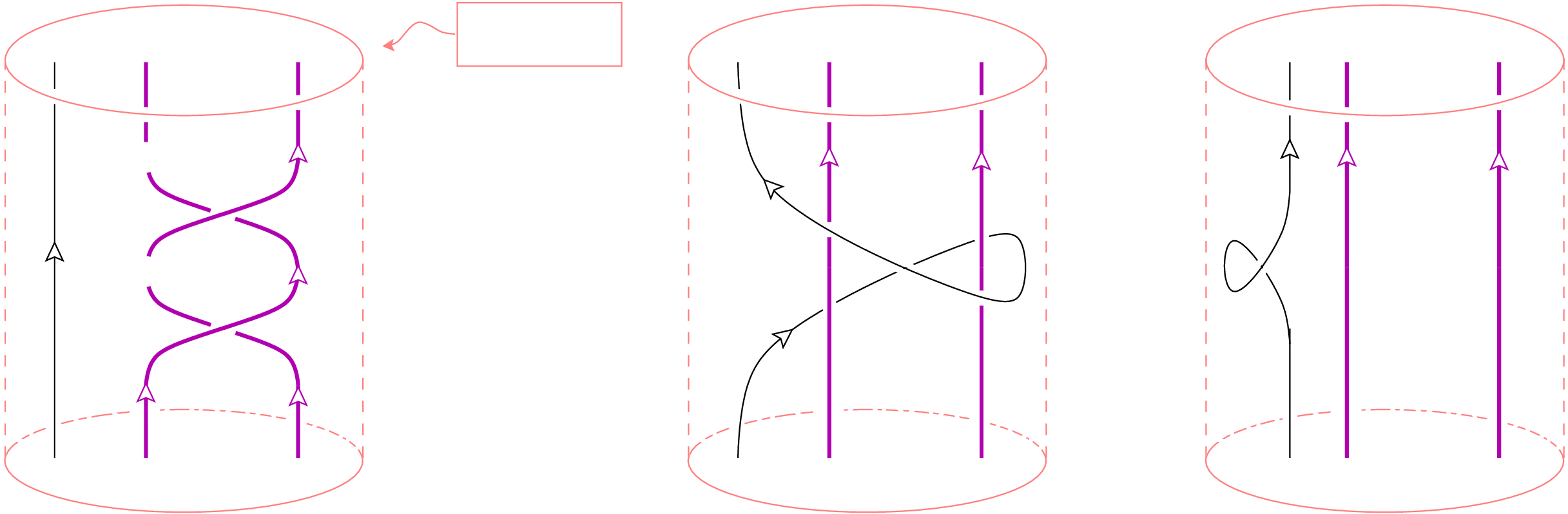}}
  \setlength{\unitlength}{24810sp}
  \setlength{\unitlength}{1pt}
  \put(16.7,16)     {\scriptsize$k$}
  \put(42.4,15.7)   {\scriptsize$\dot R$}
  \put(39.5,66.4)   {\scriptsize$\theta_{\!\dot R}$}
  \put(39.5,98.8)   {\scriptsize$\theta_{\!\dot R}$}
  \put(85.5,15.7)   {\scriptsize$\dot R$}
  \put(132,133) {$S^2{\times}[0{,}1]$}
  \put(140,67)      {$ \stackrel f\longmapsto $}
  \put(211.3,15.8)  {\scriptsize$k$}
  \put(237.2,15.8)  {\scriptsize$\dot R$}
  \put(280.3,15.8)  {\scriptsize$\dot R$}
  \put(315,67)      {$ = $}
  \put(368.2,15.8)  {\scriptsize$k$}
  \put(383.3,15.8)  {\scriptsize$\dot R$}
  \put(426.8,15.8)  {\scriptsize$\dot R$}
  \epicture31  \labl{eq:TT1}
If we take the vertical interval to be $[0,2\pi]$ and denote the vertical 
position by $\varphi$, the map $f$ corresponds to a clockwise rotation of the
horizontal $S^2$'s by the angle $\varphi$ around the central point in the 
drawing \erf{eq:TT1}. (This is best visualised with actual ribbons; the 
pictures above have already been converted back to blackboard framing).
The equality on the \rhs\ of \erf{eq:TT1} follows by wrapping
the $k$-ribbon around `infinity' on the horizontal $S^2$. Now the
cobordism on the \rhs\ of \erf{eq:TT1} is indeed equal to $\theta_k \id\,$.
\\[5pt]
That the second  relation in \erf{TQ-props} holds is already clear from the
fact that the two sides of the equality just correspond to different choices 
for drawing the horizontal triangulation line on the embedded M\"obius strip 
in figure (\ref{eq:Moebius2}\,b). Let us nonetheless check it explicitly;
consider the moves
  \begin{eqnarray}  \begin{picture}(420,158)(22,0)
  \put(71,0)     {\includeourbeautifulpicture{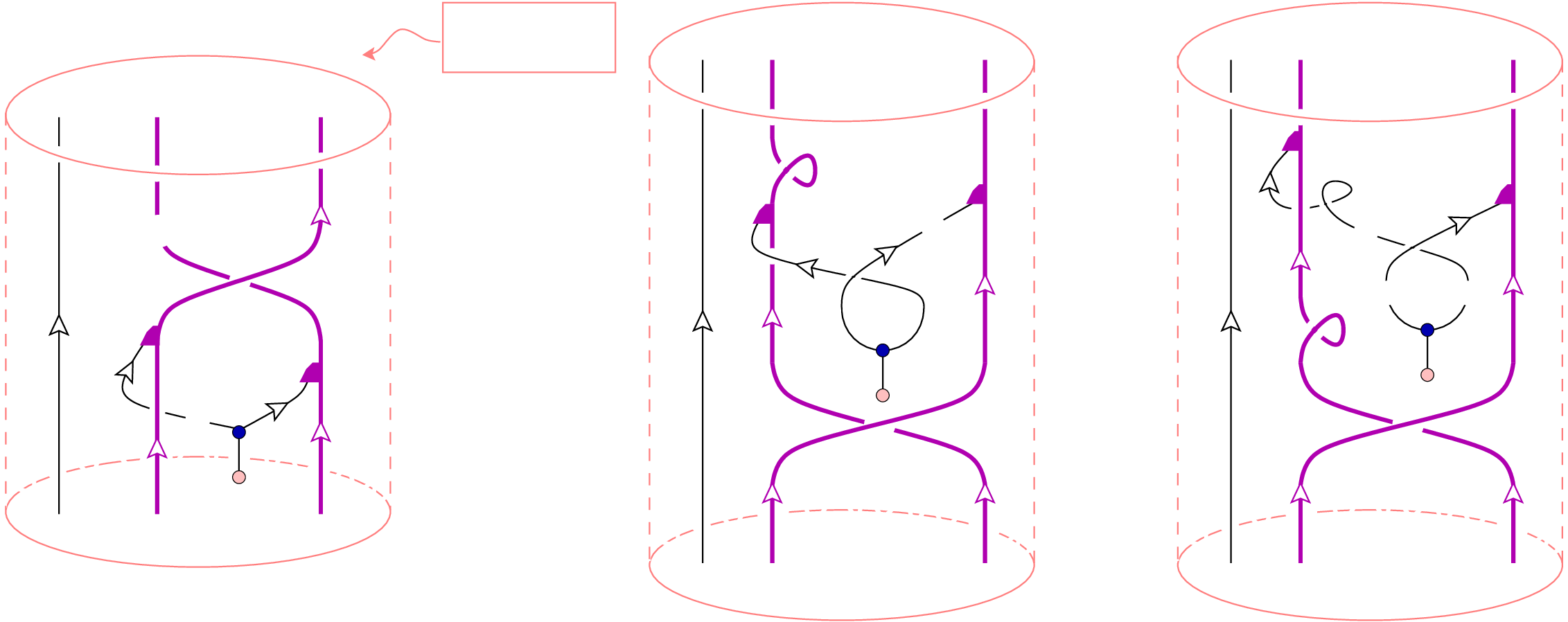}}
  \setlength{\unitlength}{24810sp}
  \setlength{\unitlength}{1pt}
  \put(0,78)        {$ T_{kR}\, Q_{kRR} \;= $}
  \put(87.8,29.1)   {\scriptsize$k$}
  \put(111.7,101.5) {\scriptsize$\theta_{\!\dot R}$}
  \put(113.7,28.8)  {\scriptsize$\dot R$}
  \put(120.7,51.1)  {\scriptsize$\sigma$}
  \put(134.9,45.4)  {\scriptsize$A$}
  \put(156.9,28.8)  {\scriptsize$\dot R$}
  \put(189.3,150.7) {$S^2{\times}[0{,}1]$}
  \put(210,78)      {$ = $}
  \put(258.2,15.9)  {\scriptsize$k$}
  \put(276.4,15.6)  {\scriptsize$\dot R$}
  \put(296.5,83.8)  {\scriptsize$A$}
  \put(314.8,102.8) {\scriptsize$\sigma$}
  \put(322.8,15.6)  {\scriptsize$\dot R$}
  \put(359,78)      {$ = $}
  \put(397.9,15.9)  {\scriptsize$k$}
  \put(415.7,15.6)  {\scriptsize$\dot R$}
  \put(428.8,101.2) {\scriptsize$\sigma^{\!-\!1}$}
  \put(434.6,85.5)  {\scriptsize$\sigma$}
  \put(440.8,90.8)  {\scriptsize$A$} 
  \put(456.4,86.5)  {\scriptsize$\sigma$}
  \put(462.5,15.6)  {\scriptsize$\dot R$}
  \end{picture} \nonumber\\[-.1em]{} 
  \\[-1.6em]{}\nonumber\end{eqnarray} 
In the first step the horizontal $A$-ribbon is moved past the braiding of the
two $R$-ribbons.
In the second step the representation morphism on the left
$R$-ribbon is taken through the twist, which leads to a twist $\thetaA$ on
the $A$-ribbon, and an identity $\id_A\eq\sigma^{-1} \cir\sigma$
is inserted. On the \rhs, the two $\sigma$ morphisms above 
the coproduct cancel, by \erf{eq:sig-coprod}, against the braiding of the 
$A$-ribbons, and the combination $\thetaA\cir\sigma^{-1}$ equals $\sigma$.
Implementing these identities leaves us with
the ribbon graph for $Q_{kRR}\,T_{kR}$. 
\\[5pt]
Finally, $Q_{kRR}$ can be diagonalised because it is a projector, and 
$T_{kR}$ can be diagonalised because it squares to a multiple of the
identity. Due to $T_{kR}\,Q_{kRR}\eq Q_{kRR}\,T_{kR}$, the
diagonalisation of both maps can be achieved simultaneously. 
\qed

It turns out that the numbers ${\rmM}_{kR}$ are not integral. 
But they are integral up to known phases. To see this,
let us choose once and for all a square root of the phases $\theta_k$ which
specify the action of the twist on simple objects $U_k$, i.e.\ choose numbers
  \be
  t_k \qquad {\rm such~that} \qquad t_0 = 1 \,, \quad {t_k}^2 = \theta_k 
  \quad\ {\rm and} \quad\ t_k = t_{\bar k} \,.  \labl{eq:tk-def}
If the modular tensor category is the representation category of a chiral 
algebra (which is the case we are interested in), then each object $U_k$
comes with a conformal weight $\Delta_k$ and one can take
  \be  t_k = \exp(-\pi \ii \Delta_k) \,.  \ee
The numbers
  \be \rmm_{kR} := {\rmM}_{kR} / t_k \labl{eq:m-def}
are indeed integers:
\\[-2.3em]

\dtl{Theorem}{thm:m-prop}
The numbers $\rmm_{kR}$ defined in \erf{eq:m-def} satisfy
  \be  \bearll
  \rm(i)  & \rmm_{kR} \in \zet \,, \\[.4em]
  \rm(ii) & \rmm_{kR} = \rmm_{\bar k R^\sigma}\,, \\[.4em]
  \rm(iii)& \frac12\, (\rmm_{kR} + {\rm A}_{kRR})
  \in \{ 0,1,...\,,{\rm A}_{kRR} \} \,.
  \eear\labl{eq:m-prop}

\medskip\noindent
Proof:\\
(i) According to \erf{eq:MkR-trace} the coefficient ${\rmM}_{kR}$ is the
trace of the linear map $T_{kR}\,Q_{kRR}$. Let $\{v_i\}$ be a common
eigenbasis of $Q_{kRR}$ and $T_{kR}$ (which exists by lemma \ref{lem:TQ-prop}).
The eigenvalues $\lambda_i^T$ of $T_{kR}$ in this basis square to
$\theta_k$, and hence $\lambda_i^T\eq{\pm} t_k$, while the eigenvalues
$\lambda_i^Q$ of the projector $Q_{kRR}$ are either zero or one.
As a consequence the combinations $\lambda_i^T\lambda_i^Q/t_k$ take values
in $\{0,\pm 1\}$, and hence $\rmm_{kR}\eq{\rmM}_{kR} / t_k\eq
\sum_i \lambda_i^T\lambda_i^Q / t_k$ is indeed an integer.
\\[5pt]
(ii) We show that ${\rmM}_{kR}\eq{\rmM}_{\bar k R^\sigma}$, which owing to
$t_k\eq t_{\bar k}$ in \erf{eq:tk-def} implies (ii). Let us turn the ribbon
graph \erf{eq:MkR} in $S^2 \,{\times}\, S^1$ whose invariant is ${\rmM}_{kR}$
upside down. We then have
  \bea  \begin{picture}(290,105)(18,57)
  \put(0,0)     {\includeourbeautifulpicture{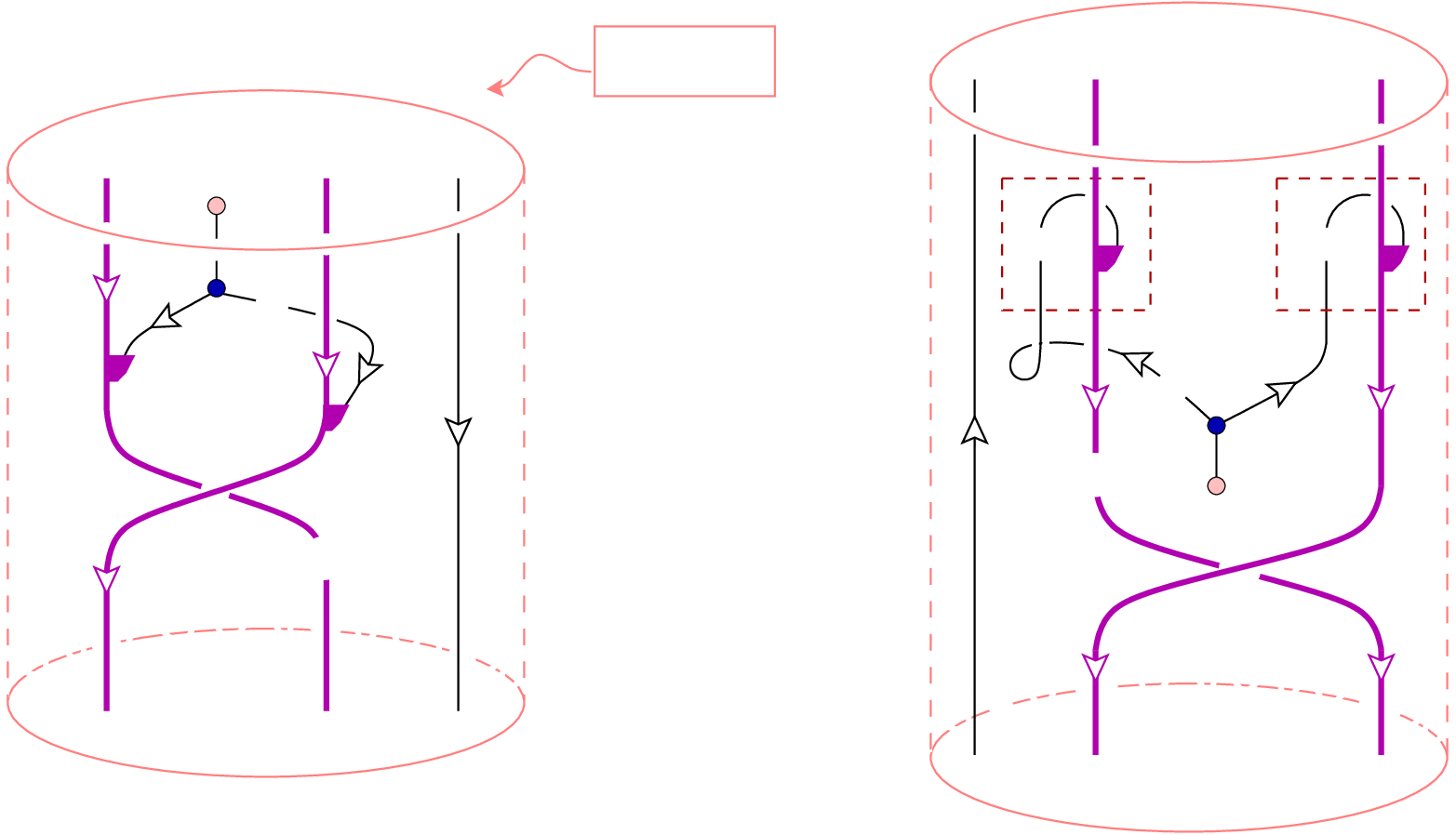}}
  \setlength{\unitlength}{24810sp}
  \setlength{\unitlength}{1pt}
  \put(12.2,24.7)   {\scriptsize$\dot R$}
  \put(38.4,98.7)   {\scriptsize$A$}
  \put(51.2,102.5)  {\scriptsize$\sigma$}
  \put(65.5,24.7)   {\scriptsize$\dot R$}
  \put(62.2,52.2)   {\scriptsize$\theta_{\!\dot R}$}
  \put(91.2,25.1)   {\scriptsize$k$}
  \put(117.6,148)   {$S^2{\times}S^1 $}
  \put(138,77)      {$ = $}
  \put(193.2,30)  {\scriptsize$\overline k$}
  \put(202.2,114.2) {\scriptsize$\sigma$}
  \put(214.2,67.5)  {\scriptsize$\theta_{\!\dot R}$}
  \put(217.2,15.5)  {\scriptsize$\dot R$}
  \put(229.1,87.2)  {\scriptsize$\sigma^{\!-\!1}$}
  \put(242.7,76.3)  {\scriptsize$A$}
  \put(258.2,114.2) {\scriptsize$\sigma$}
  \put(263.4,15.5)  {\scriptsize$\dot R$}
  \epicture34 \ee
which is derived by the following manipulations. First
the downwards pointing $k$-ribbon is replaced by an upwards directed
$\bar k$-ribbon, which is then moved to the left side of the $R$-ribbons.
Then symmetry in the form (I:3.35) is used on the coproduct so as to move the
counit below the coproduct.
Next a pair $\sigma \circ \sigma^{-1}$ is inserted on the $A$-ribbon that joins
the left $R$-ribbon. Now the part of the ribbon graph lying within the dashed 
boxes can be
recognised from \erf{eq:rho-sigma} as the representation morphism for $R^\sigma$.
The twist $\thetaA$ combines with $\sigma^{-1}$ to $\sigma$. Using periodicity in
the vertical direction we can shift the figure
such that the braiding is placed above
the $A$-ribbon. The resulting graph is precisely that of ${\rmM}_{\bar k R^\sigma}$.
\\[5pt]
(iii) Equations \erf{eq:Akmn-trace} and \erf{eq:MkR-trace} imply that
  \be
  \Frac 12\,(\rmm_{kR}+{\rm A}_{kRR}) = \tr \llb \Frac12 \, (T_{kR}/t_k + \id)\,
  Q_{kRR} \lrb \,.  \labl{eq:m+A}
Owing to lemma \ref{lem:TQ-prop} the combination
$\frac 12\,(T_{kR}/t_k \,{+}\, \id_{\calH(k,R,R;S^2)})$ is a projector and
commutes with $Q_{kRR}$. Thus the product appearing in the trace \erf{eq:m+A}
is a projector as well, and hence its trace equals the
dimension of its image and is thus a non-negative integer.
Furthermore the image of the product of the projectors
is contained in the image of $Q_{kRR}$, so that
$\frac 12 (\rmm_{kR}+{\rm A}_{kRR}) \le {\rm A}_{kRR}$.
\mbox{$\quad $} \qed

\subsection{The Klein bottle}\label{sec:klein}

The double $\hat X$ of the Klein bottle is a torus with modular paramenter 
$\tau\eq2\ii s$, as indicated in figure (\ref{eq:Klein}\,a):
  \bea  \begin{picture}(340,52)(0,59)
  \put(48,0)      {\includeourbeautifulpicture{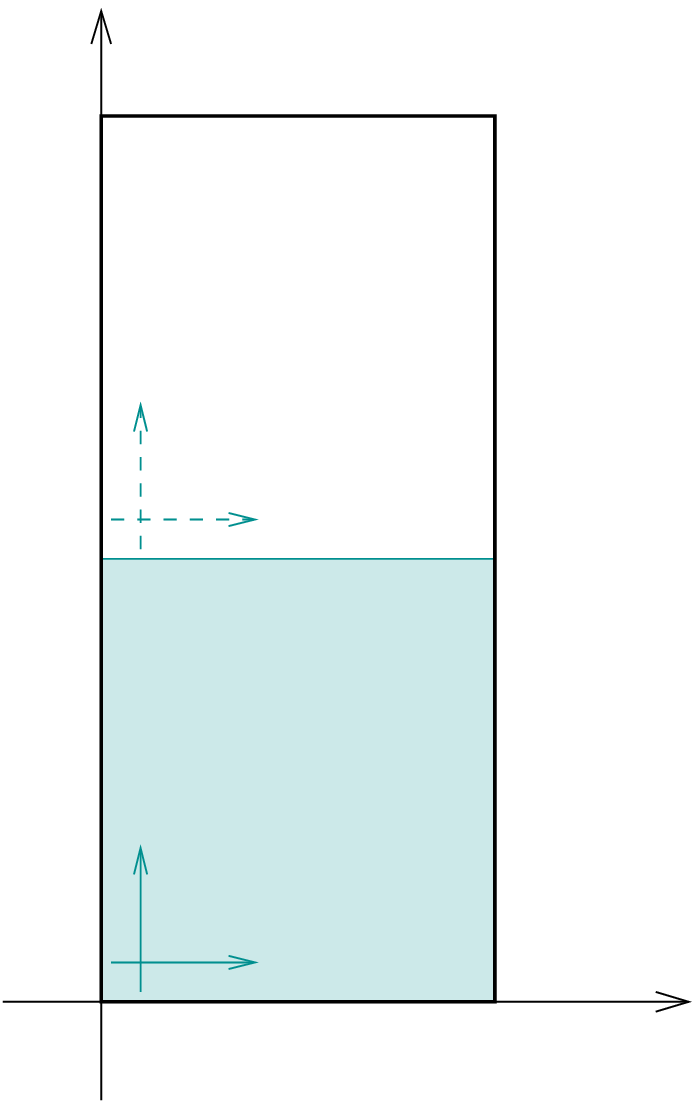}}
  \put(254,15)    {\includeourbeautifulpicture{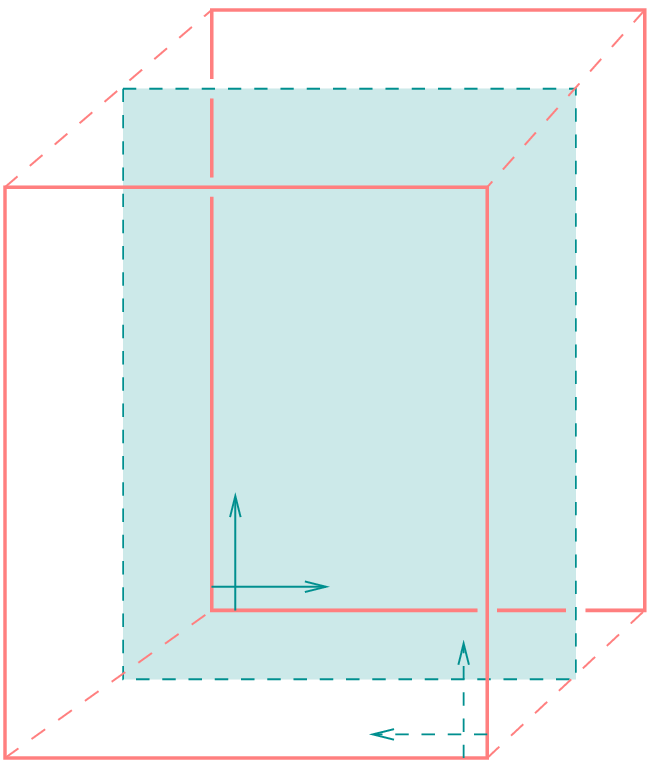}}
  \setlength{\unitlength}{24810sp}
  \setlength{\unitlength}{1pt}
  \put(0,57)        {$ \hat X \;= $ }
  \put(25,110)      { a) }
  \put(51.8,58.2)   {\tiny$\ii s$} 
  \put(48.5,106.5)  {\tiny$2\ii s$} 
  \put(100.7,5.2)   {\tiny$1$} 
  \put(193,47)      {$ \MX \;= $ }
  \put(224,84)      { b) }
  \put(310.4,14.1)  {\tiny$t{=}{-}{1}$}
  \put(319.3,22.5)  {\tiny$t{=}0$} 
  \put(326.4,30.7)  {\tiny$t{=}1$}
  \epicture38  \labl{eq:Klein}
The anticonformal involution on $\hat X$ is given by
$\sigma(z)\eq1{+}\ii s{-}\bar z$. The quotient $X\eq\hat X /\sigma$ has 
indeed the topology of a Klein bottle; a fundamental domain for the action 
of $\sigma$ on $\hat X$ is indicated by the shaded
region in figure (\ref{eq:Klein}\,a).

The connecting manifold $\MX\eq\hat X \,{\times}\,[-1,1]\,{/}\,{\sim}\,$,
with $(z,t)\,{\sim}\,(\sigma(z),-t)$, is presented in figure 
(\ref{eq:Klein}\,b). What is shown is a fundamental domain
  \be
  \MX = [0,1] \,{\times}\, [0,s] \,{\times}\, [-1,1] \,{/}\,{\sim}\,,  \ee
where now the equivalence relation $\sim$ identifies the left and right 
side of (\ref{eq:Klein}\,b) periodically, while top and bottom rectangles
are identified with a $180^\circ$ rotation. Explicitly,
  \be
  (0,y,t) \sim (1,y,t) \qquad {\rm and} \qquad
  (x,0,t) \sim (1{-}x,s,-t) \,.  \ee
The small perpendicular arrows `\,\raisebox{.15em}{$\uparrow
$}\hspace{-5pt}\raisebox{-.15em}{$\rightarrow$}\,' in figures 
(\ref{eq:Klein}\,a) and (\ref{eq:Klein}\,b) indicate how the shaded
and unshaded regions in figure (\ref{eq:Klein}\,a) for the double 
$\hat X$ are to be identified with the boundary of $\MX$. The shaded
plane in (\ref{eq:Klein}\,b) shows where the embedding
$\iota_X{:}\; X \,{\to}\,\MX$ places the world sheet inside $\MX$.

The next step is to choose a triangulation of the Klein bottle $X$. We
take the one displayed in (\ref{eq:Klein2}\,a):
  \bea  \begin{picture}(260,39)(0,57)
  \put(25,0)      {\includeourbeautifulpicture{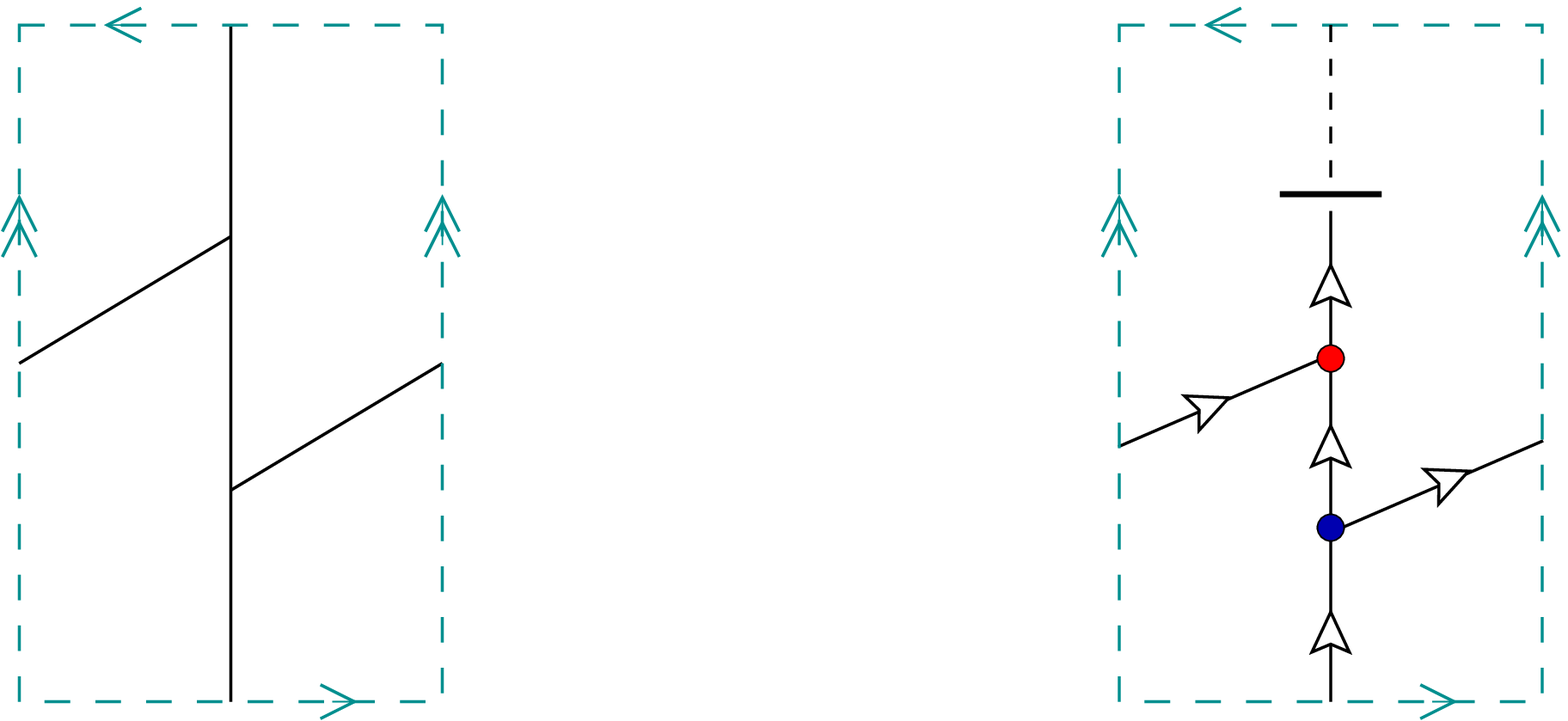}}
  \setlength{\unitlength}{24810sp}
  \setlength{\unitlength}{1pt}
  \put(0,80)        { a) }
  \put(115,43)      {$ \longmapsto $ }
  \put(140,80)      { b) }
  \put(187.4,38.4)  {\scriptsize$A$}
  \put(196.4,18.5)  {\scriptsize$A$}
  \epicture37  \labl{eq:Klein2}
Like for the M\"obius strip we choose the orientation of the paper plane at
the two vertices and insert the graphs \erf{eq:bulk-vertex},
\erf{eq:bulk-edge}. The resulting
ribbon graph can be simplified, leading to (\ref{eq:Klein2}\,b).
Picture (\ref{eq:Klein2}\,b) is to be understood as embedded in $\MX$
like in (\ref{eq:Klein}\,b), with the third direction suppressed.
Drawing also the latter, the three-manifold and ribbon graph
representing the Klein bottle amplitude are given by
  \bea  \begin{picture}(270,84)(0,50)
  \put(58,0)     {\includeourbeautifulpicture{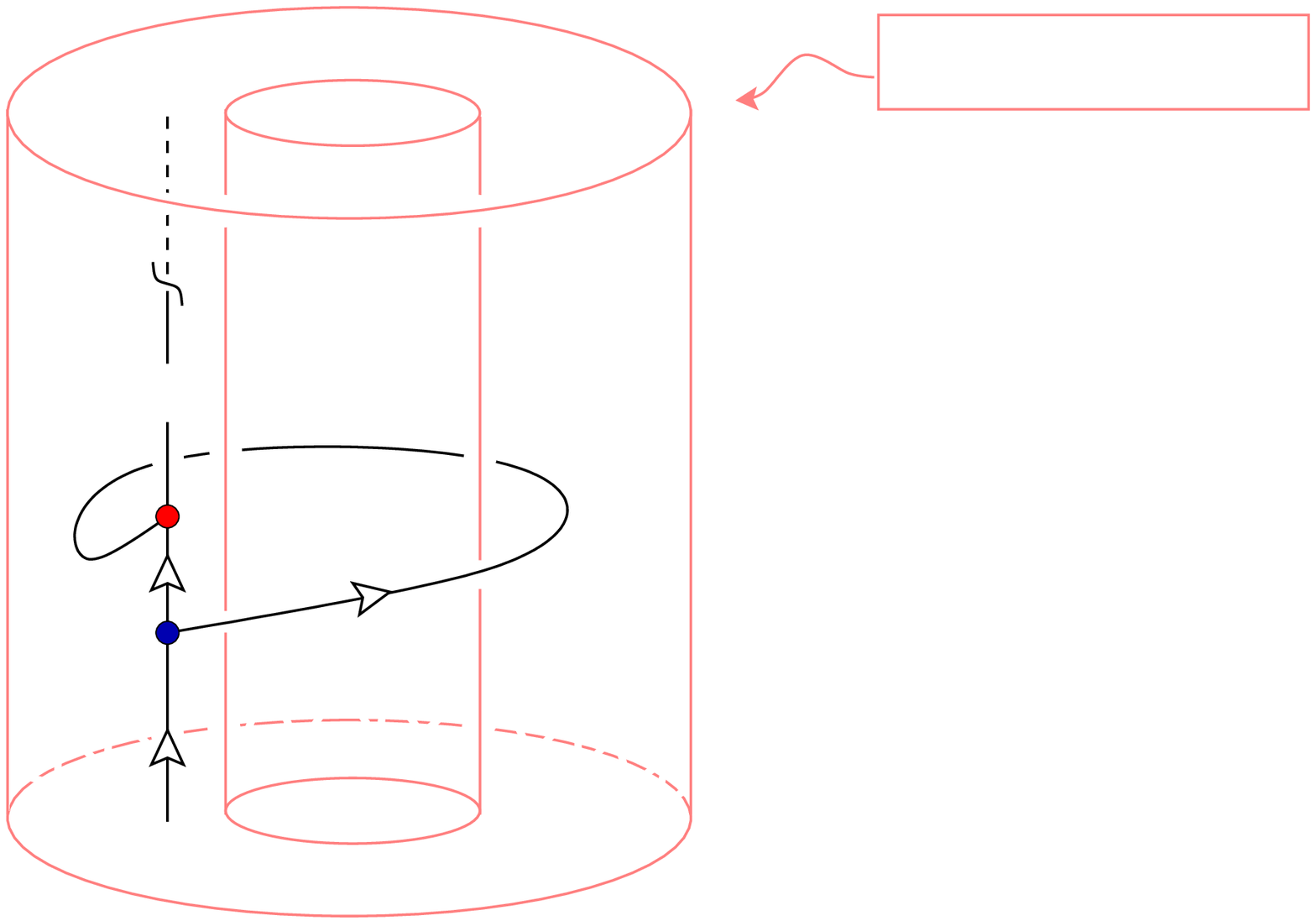}}
  \setlength{\unitlength}{24810sp}
  \setlength{\unitlength}{1pt}
  \put(0,66)        {$ \K \;= $}
  \put(73.8,49.1)   {\scriptsize$A$}
  \put(80.9,76)     {\scriptsize$\sigma$}
  \put(191.6,123.4) {$ I{\times}S^1{\times}I{/}{\sim} $}
  \put(210,66)      {$ \in\calh(\emptyset;\rmT) $}
  \epicture25  \labl{eq:K-ribbon}
In this picture top and bottom are identified via $(r,\phi)_{\rm top}
\,{\sim}\, (1/r,-\phi)_{\rm bottom}$, where $r$ and $\phi$ are radial and
angular coordinates in the horizontal plane and $I$ is an interval. Note 
that with this identification the boundary of \erf{eq:K-ribbon} is indeed a 
single torus. As usual, the coefficients ${\rm K}_k$ occuring in the expansion
  \be
  \K = \sum_{k\in\II} {\rm K}_k \, |\chii_k;\rmT\rangle  \labl{eq:K-exp}
of ${\rm K}$ in a basis are obtained by taking the inner product with
$\langle\chii_k;\rmT|$. On the \rhs\ this corresponds to gluing the
solid torus (I:5.18) to the boundary of ${\rm K}$:
  \bea  \begin{picture}(235,84)(0,50)
  \put(58,0)     {\includeourbeautifulpicture{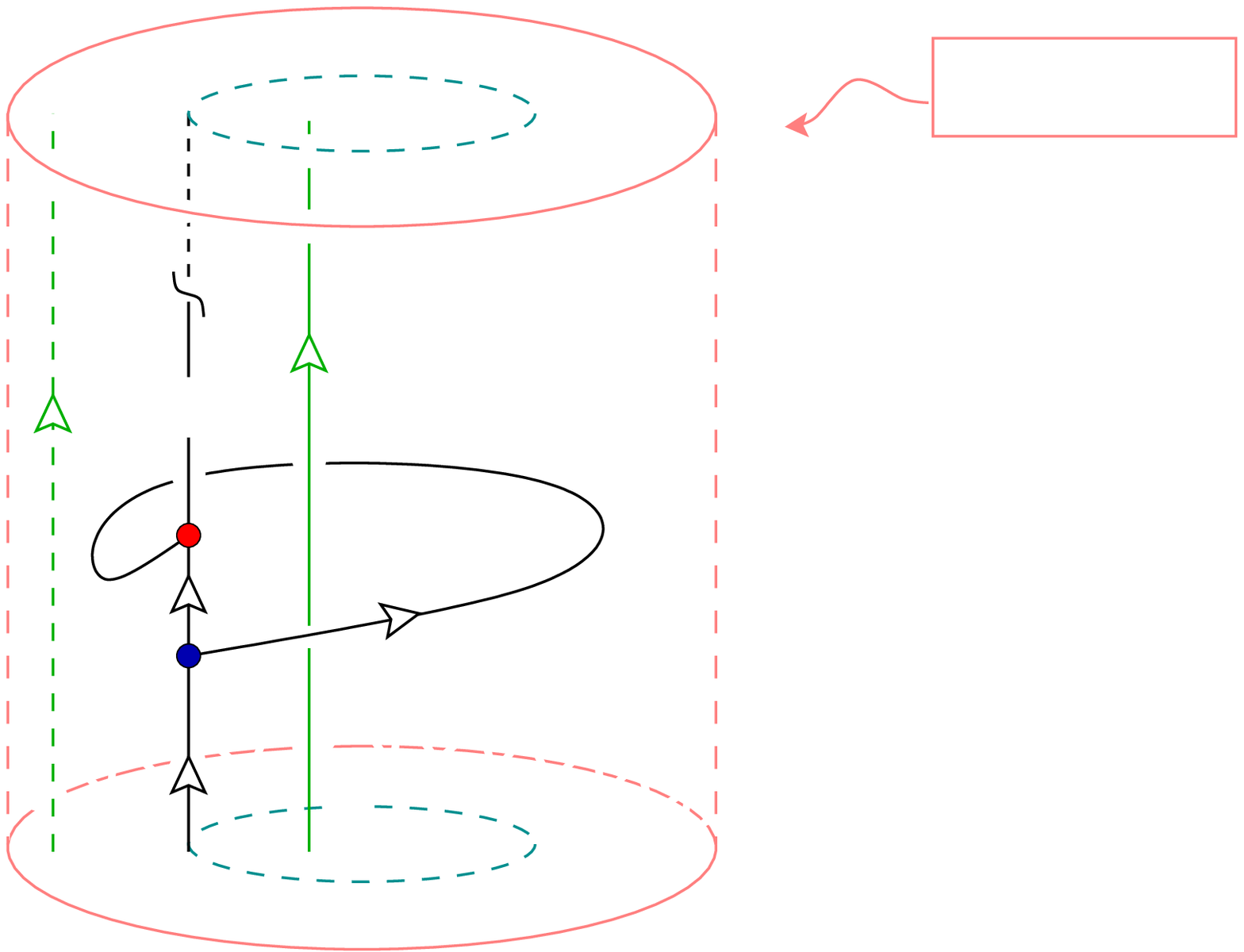}}
  \setlength{\unitlength}{24810sp}
  \setlength{\unitlength}{1pt}
  \put(0,66)        {$ \K_k \;= $}
  \put(67.3,14.4)   {\scriptsize$k$}
  \put(75.5,48.9)   {\scriptsize$A$}
  \put(83.2,76.4)   {\scriptsize$\sigma$}
  \put(103.7,14.4)  {\scriptsize$k$}
  \put(192.8,120.8) {$ S^2{\times}I\!{/}{\sim}$}
  \epicture26  \labl{eq:Kk}
The picture shows the closed three-manifold that is obtained from 
$I\times S^2$ via the identification $(r,\phi)_{\rm top}\,{\sim}\,(1/r,
-\phi)_{\rm bottom}$. The dashed circles indicate the intersection of the 
embedded world sheet with the boundary $S^2$'s of $I\,{\times}\,S^2$;
they are placed at $r\eq1$. With this identification the $U_k$-ribbon 
that hits the top-$S^2$ in the interior of the circle has its orientation
as a two-manifold reversed and is identified with the $U_k$-ribbon 
hitting the bottom-$S^2$ outside the circle (which accordingly is to be 
drawn as a dashed line).

To proceed we would like to rewrite \erf{eq:Kk} as a trace 
over $\calh(k,A,k;S^2)$. The corresponding linear map can be written as
the composition of two pieces. The first is the specialisation to $i\eq j$
of the map $P_{ij}$ from (I:5.37), which was used in (I:5.38) to express 
the coefficients $Z_{ij}$ of the torus partition function as a trace.
The second map, to be denoted by $G_k$, implements the gluing map 
identifying top and bottom in \erf{eq:Kk}: 
  \bea  \begin{picture}(420,85)(0,67)
  \put(50,0)     {\includeourbeautifulpicture{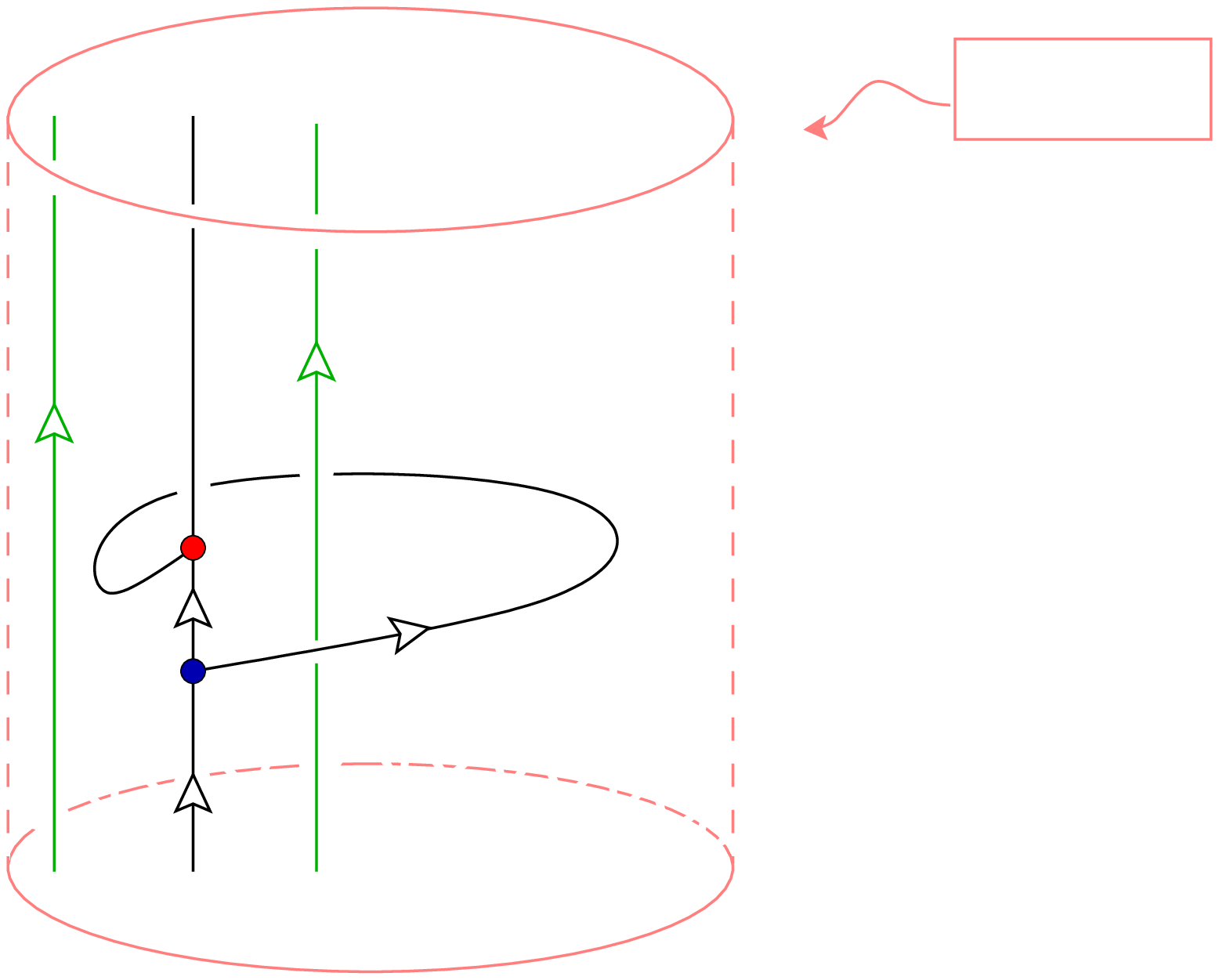}}
  \setlength{\unitlength}{24810sp}
  \setlength{\unitlength}{1pt}
  \put(0,76)        {$ P_{kk} \;= $}
  \put(59.4,27.4)   {\scriptsize$k$}
  \put(68.1,62.1)   {\scriptsize$A$}
  \put(96.1,27.4)   {\scriptsize$k$}
  \put(187.3,132.7) {$ S^2{\times}I $}
  \put(245,76)      {$ G_k \;= $}
  \put(305.8,95.7)  {\scriptsize$k$}
  \put(305.8,54.7)  {\scriptsize$k$}
  \put(318.8,26.8)  {\scriptsize$\sigma$}
  \put(321.7,95.5)  {\scriptsize$A$}
  \put(321.7,54.8)  {\scriptsize$A$}
  \put(337.7,95.7)  {\scriptsize$k$}
  \put(337.7,54.7)  {\scriptsize$k$}
  \put(396.1,98)    {\scriptsize$E$}
  \put(396.1,53.7)  {\scriptsize$F$}
  \epicture42  \labl{eq:GP-def}
In the ribbon graph for $G_k$ the two intermediate horizontal $S^2$'s 
(labelled $E$, $F$) are identified via $(r,\phi)_E \,{\sim}\, (1/r,-\phi)_F$. 
(Here again according to remark \ref{maslovo} the framing anomaly vanishes.)

Both $P_{kk}$ and $G_k$ are endomorphisms of $\calh(k,A,k;S^2)$. Gluing
the top-$S^2$ of the ribbon graph for $P_{kk}$ to the bottom of $G_k$ using
the identity map and vice versa, one arrives precisely at the closed 
three-manifold \erf{eq:Kk}. Thus
  \be  
  {\rm K}_k = \Tr{\calH(k,A,k;S^2)}G_k\,P_{kk} \,.  \labl{eq:Kk-trace}
Properties of the number ${\rm K}_k$ can be established along the
same lines as for the M\"obius strip. We start with 
\\[-2.3em]

\dtl{Lemma}{lem:GP-prop} 
The linear maps $G_k$ and $P_{kk}$ defined in \erf{eq:GP-def} 
satisfy $P_{kk}\,P_{kk}\eq P_{kk}$ as well as
  \be
  G_k\,G_k = \id_{\calH(k,A,k;S^2)}  \qquad{\rm and}\qquad
  G_k\, P_{kk} = P_{kk}\, G_k \,.   \labl{GP-props} 
 
\medskip\noindent 
Proof:\\
That $P_{kk}$ is a projector was already shown in (I:5.39). In the 
manipulations establishing \erf{GP-props} we will display only the 
cylinder at $r\eq1$ of the ribbon graphs \erf{eq:GP-def}, or in other
words, only the $A$-ribbons; the $k$-ribbons just play the role of
spectators. The abbreviations for $P_{kk}$ and $G_k$ obtained this way
look as
  \bea  \begin{picture}(360,38)(0,52)
  \put(50,0)     {\includeourbeautifulpicture{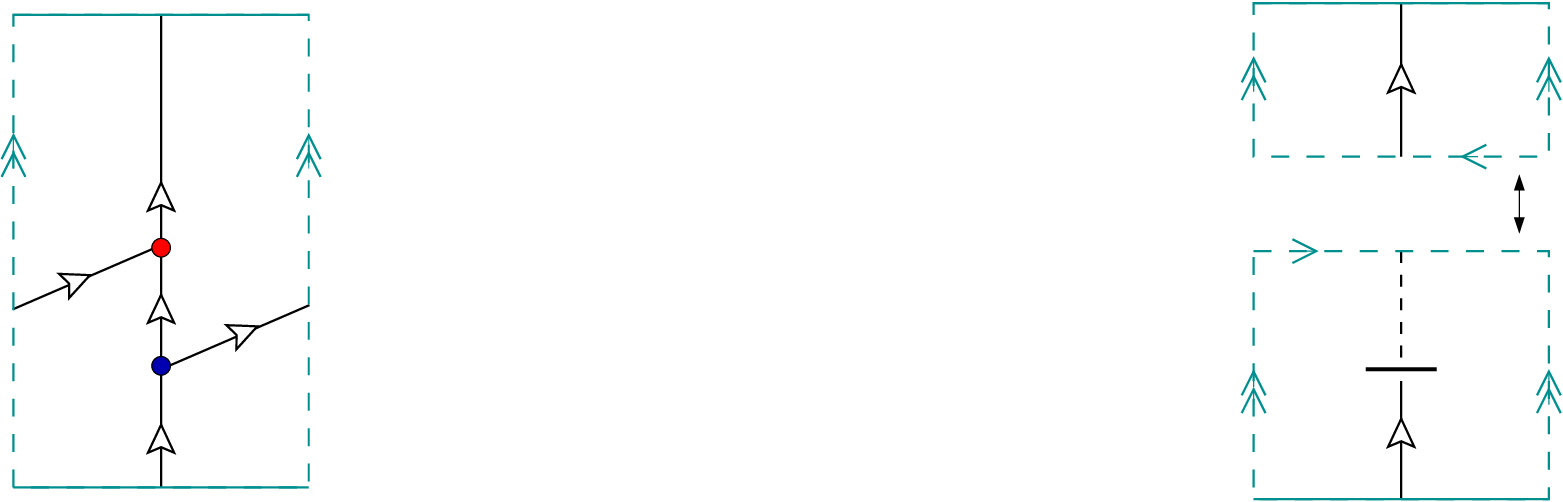}}
  \setlength{\unitlength}{24810sp}
  \setlength{\unitlength}{1pt}
  \put(0,44)        {$ P_{kk} \;= $}
  \put(71.7,38.4)   {\scriptsize$A$}
  \put(80.4,18.6)   {\scriptsize$A$}
  \put(228,44)      {$ G_k \;= $}
  \put(307.8,3.2)   {\scriptsize$A$}
  \put(307.8,83.4)  {\scriptsize$A$}
  \epicture28  \ee
That $G_k$ squares to the identity map follows from
  \bea  \begin{picture}(280,107)(0,74)
  \put(65,0)     {\includeourbeautifulpicture{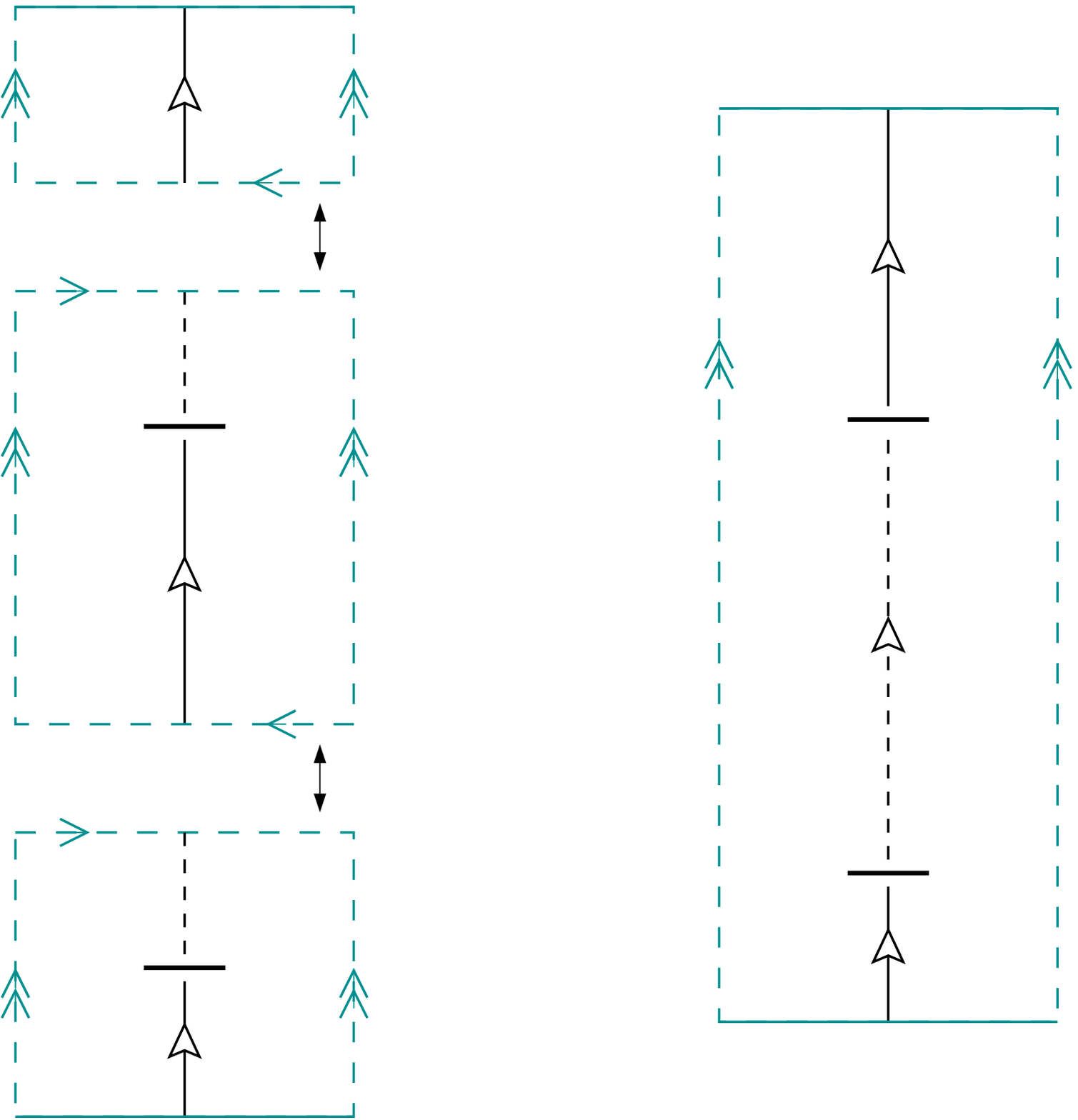}}
  \setlength{\unitlength}{24810sp}
  \setlength{\unitlength}{1pt}
  \put(0,86)        {$ G_k\,G_k \;= $}
  \put(96.1,3.2)    {\scriptsize$A$}
  \put(96.1,66.5)   {\scriptsize$A$}
  \put(96.1,152.5)  {\scriptsize$A$}
  \put(147,86)      {$ = $}
  \put(208.4,18.2)  {\scriptsize$A$}
  \epicture50 \labl{p47}
The second equality can be understood as a homeomorphism of three-manifolds
which acts as the identity on the top and bottom parts of the first 
picture, and as $(r,\varphi)\,{\to}\,(1/r,-\varphi)$ in the middle (recall 
that the $r$-direction is suppressed). This changes the two gluing maps in 
the first picture to identity maps, so that the second picture can be
drawn as a single piece. The two \boxelement s on the \rhs\ of \erf{p47}
cancel by \erf{eq:bar-prop}, 
leaving the identity map on $\calh(k,A,k;S^2)$.
\\
To see the second equality in \erf{GP-props}, consider the moves
  \bea  \begin{picture}(300,64)(0,69)
  \put(0,0)     {\includeourbeautifulpicture{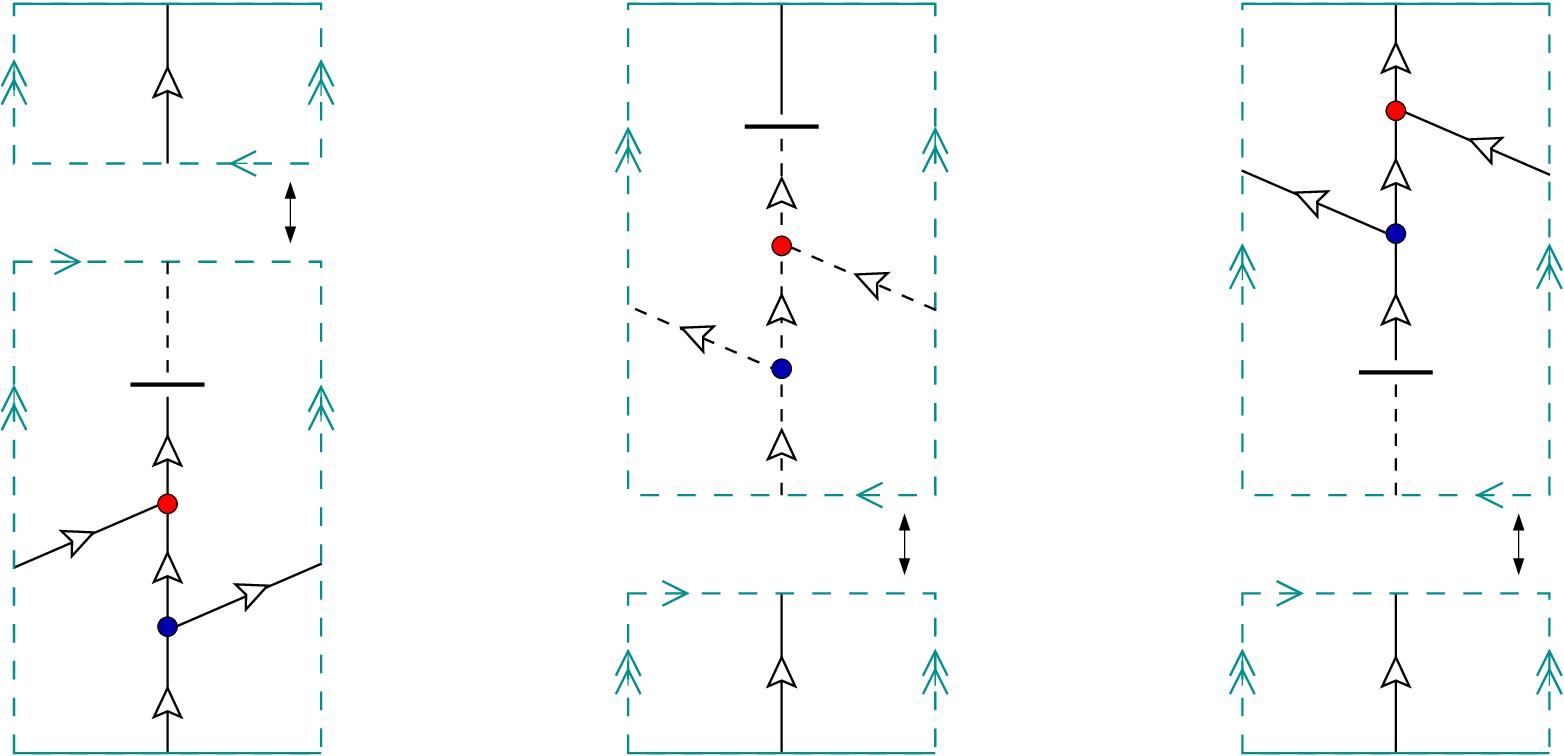}}
  \setlength{\unitlength}{24810sp}
  \setlength{\unitlength}{1pt}
  \put(22.2,36.2)   {\scriptsize$A$}
  \put(31.4,2.5)    {\scriptsize$A$}
  \put(31.4,125)    {\scriptsize$A$}

  \put(80,65)       {$ = $}
  \put(139.1,3.5)   {\scriptsize$A$}
  \put(139.1,125)   {\scriptsize$A$}
  \put(139.4,80.7)  {\scriptsize$A$}
  \put(187,65)      {$ = $}
  \put(238.7,85.7)  {\scriptsize$A$}
  \put(247.2,3.5)   {\scriptsize$A$}
  \put(247.2,125)   {\scriptsize$A$}
  \epicture44 \ee
The \lhs\ equals $G_k P_{kk}$. In the first step the horizontal section
along which the two pieces of the three-manifold are glued together is 
moved downwards. The identification
$(r,\varphi) \,{\to}\, (1/r,-\varphi)$ causes a reflection in the picture and a
change of the 2-orientation of the $A$-ribbon. The second step uses 
\erf{eq:bar-prop} to move the \boxelement\ through the $A$-ribbon that
winds around the cylinder. 
Upon moving the \boxelement\  through the identification plane as well
and applying a move like (I:5.25) so as to reverse the direction of the 
horizontal $A$-ribbon, one arrives at the ribbon graph for $P_{kk} G_k$.
\qed

It is now straightforward to establish the crucial properties of the 
numbers ${\rm K}_k$. They are summarised in 
\\[-2.3em]

\dtl{Theorem}{thm:K-prop}
The numbers ${\rm K}_k$ defined in \erf{eq:Kk} fulfill
  \be  \bearll
  \rm(i)  & {\rm K}_k \in \zet \,, \\[.4em]
  \rm(ii) & {\rm K}_k = {\rm K}_{\bar k} \,, \\[.4em]
  \rm(iii)& \frac12\,(Z_{kk}+{\rm K}_k)
  \in \{ 0,1,...\,,Z_{kk} \} \,.
  \eear\labl{eq:K-prop}

\medskip\noindent
Proof:\\
(i) is an immediate consequence of lemma \ref{lem:GP-prop}. The linear maps
$G_k$ and $P_{kk}$ commute, are diagonalisable and have integer eigenvalues. 
Thus the trace of their product is an integer.
\\[5pt]
(ii) can be seen analogously to theorem \ref{thm:m-prop}(ii). 
Start with drawing the three-manifold and embedded ribbon graph of \erf{eq:Kk}
upside down, then use the move (I:5.25) to reverse the direction of all 
$A$-ribbons, and finally move the combination of $\sigma$ and half-twist 
(the \boxelement) through the identification. The resulting graph is that 
for ${\rm K}_{\bar k}$.
\\[5pt]
To see (iii) recall formula (I:5.38) and write
  \be
  \Frac12\,(Z_{kk}+{\rm K}_k) = \Tr{\calH(k,A,k;S^2)} \llb \Frac12 \,
  (G_k+\id_{\calH(k,A,k;S^2)}) \, P_{kk} \lrb \,.  \ee
Lemma \ref{lem:GP-prop} shows that $(G_k+\id)/2$ is a projector that commutes
with $P_{kk}$. Thus their product is a projector, too, and its trace is a
non-negative integer. The upper bound $Z_{kk}$ follows since the image of
the product is contained in the image of $P_{kk}$.
\qed

\dt{Remark}
(i)~\,For the charge conjugation modular invariant, i.e.\ for the \JA\ 
$A\eq\one$ (having reversion $\sigma\eq\id_\one$), a TFT 
demonstration of the assertions of theorems \ref{thm:m-prop} 
and \ref{thm:K-prop} was already given in \cite{fffs3}.
\\[.3em]
(ii)~In unoriented closed and open string theory,
the properties (\ref{eq:m-prop}(iii)) and (\ref{eq:K-prop}(iii))
involving the four partition functions $Z$, ${\rm K}$, ${\rm A}$ and $\rmM$ 
arise as consistency conditions. The particular combinations appearing in 
these relations determine the spectra of (orientifold-projected)
open and closed string states, respectively, and therefore must be
non-negative integers.

\subsection{About ribbons in $\RP$}\label{sec:S3/Z2}

We now present the basic invariant of the three-manifold
$\RP\eq S^3/\zet_2$. Using this invariant, arbitrary ribbon
graphs in $\RP$ can easily be expressed as ribbon graphs in $S^3$.
This will be instrumental in the evaluation of the crossed channel 
amplitudes for the Klein bottle and the M\"obius strip. 
Other quantities arising in the computation of these amplitudes
will be introduced in section \ref{SASAGs}.  

To represent $\RP$ we take the upper half of $\reals^3$ modulo 
an identification of points on its boundary:
  \be
  \RP = \{ (x,y,z) \iN \reals^3{\cup}\{\infty\}\,|\,z\,{\ge}\,0 \} \,/{\sim}
  \,, \qquad  (x,y,0) \,{\sim}\, \Frac{-1}{x^2{+}y^2}\, (x,y,0) \,.
  \labl{eq:RP3-ident}
Similar to the pictorial conventions for ribbon graphs in $S^3$, 
$S^2{\times}S^1$ and $D^2{\times}S^1$ that we introduced in (I:5.13), 
ribbon graphs in $\RP$ will be drawn as indicated in the example of the
Hopf link as follows:
  \bea  \begin{picture}(170,56)(0,49)
  \put(0,0)     {\includeourbeautifulpicture{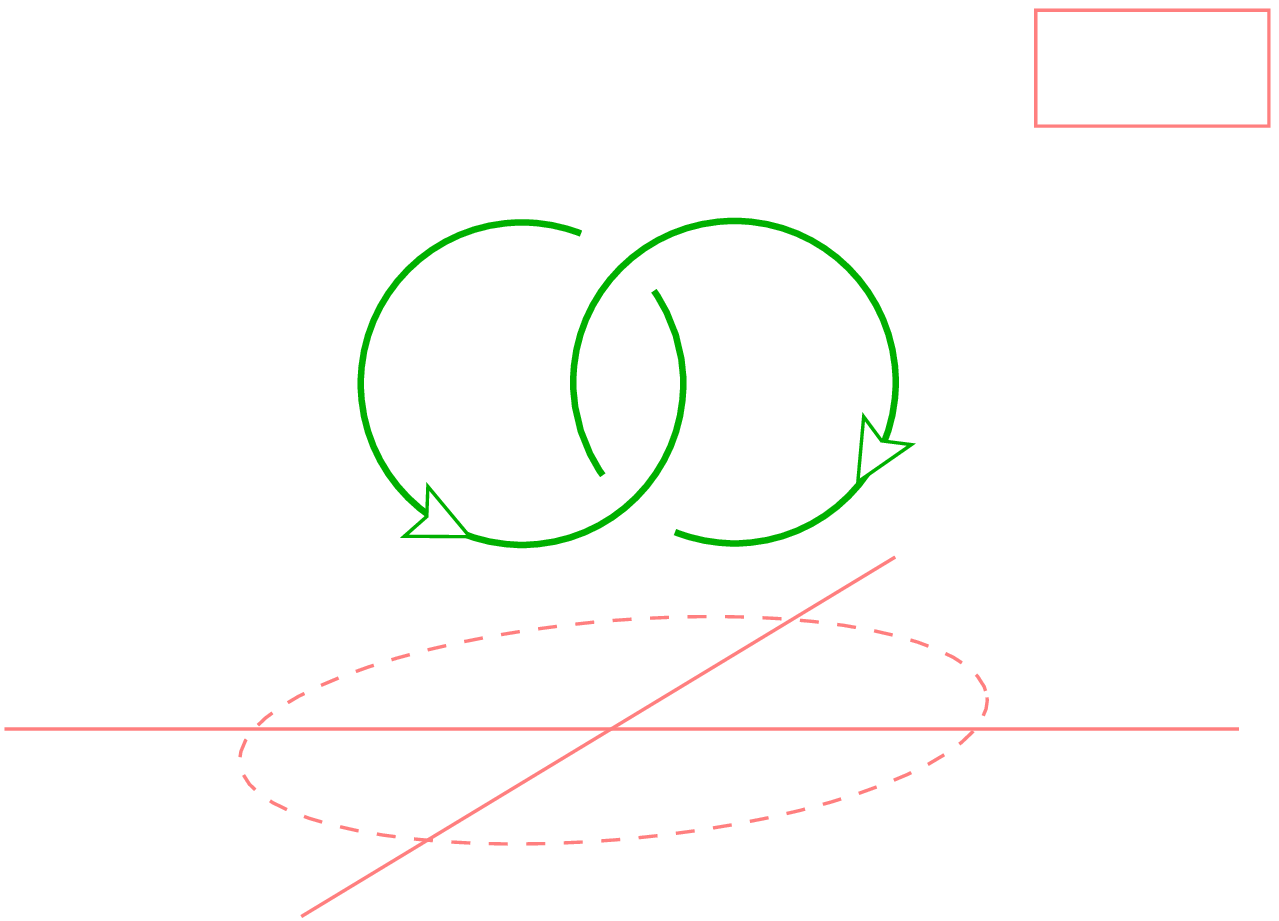}}
  \setlength{\unitlength}{24810sp}
  \setlength{\unitlength}{1pt}
  \put(116.1,88.6)  {$\RP$}
  \epicture27 \labl{eq:RP3-pic}
Here the dashed circle indicates the unit circle $x^2\,{+}\,y^2\,{=}\,1$ in 
the $x$-$y$ plane. The three-manifold $\RP$ can be obtained from surgery on 
the unknot in $S^3$ with framing $-2$, as is described in detail in section 
4.6 and appendix B of \cite{fffs3}. It follows that
  \bea  \begin{picture}(375,52)(0,50)
  \put(0,0)     {\includeourbeautifulpicture{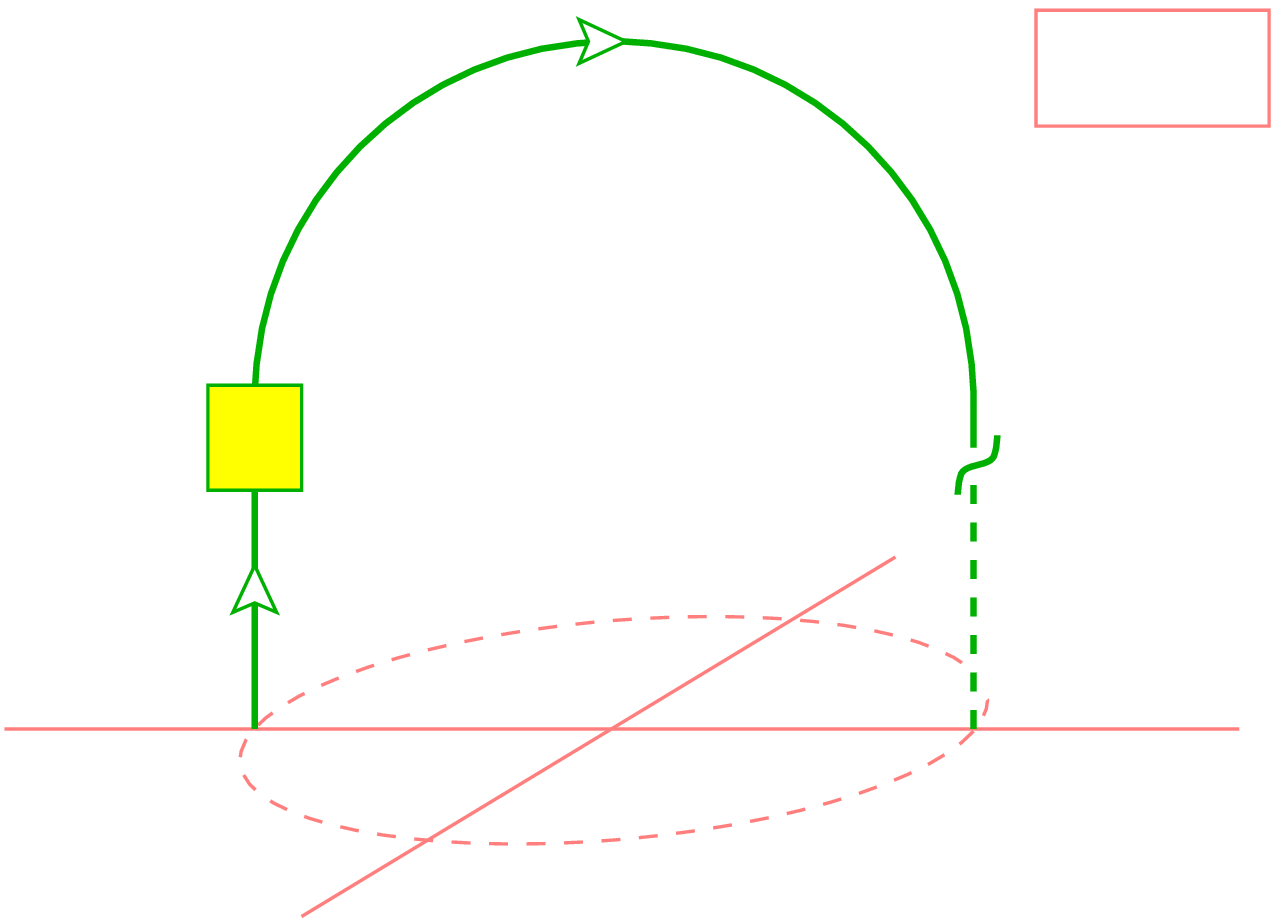}}
  \setlength{\unitlength}{24810sp}
  \setlength{\unitlength}{1pt}
  \put(19.8,25.5)   {\scriptsize$X$}
  \put(25.8,50.7)   {\scriptsize$\phi$}
  \put(45.5,83.5)   {\scriptsize$X$}
  \put(116.5,88.5)  {$\RP$}
  \put(169,47)      {$ =\; S_{00}\, \kappa^{-1}
                       \dsty\sum_{j\in\II}S_{0j}\,\theta_j^{-2} $}
  \put(294.5,43.8)  {\scriptsize$\phi$}
  \put(299.1,27.5)  {\scriptsize$X$}
  \put(299.3,63.5)  {\scriptsize$X$}
  \put(315.7,45.7)  {\scriptsize$j$}
  \put(357,89.5)  {$S^3$}
  \epicture27 \labl{eq:RP-1rib}
for every $\phi\iN\Hom(X,X)$. Here the number $\kappa$ is the {\em charge\/}
of the modular tensor category \calc, which (see section I:2.4) is given by
$\kappa\eq S_{00} \sum_{k\in\II} \theta_k^{-1} (\dim(U_k))^2$. In terms of
the central charge $c$ of the CFT, $\kappa$ is
expressed as $\kappa\eq\eE^{2 \pi \ii c/8}$.

\medskip

Note that the half-twist that is present in the graph in $\RP$ on the \lhs\ 
of \erf{eq:RP-1rib} must be chosen precisely as indicated; it cannot be 
replaced by the opposite half-twist.  As a check of this statement, let 
us study the particular case that $\phi\eq\id_X$ with $X\eq U_k$, and 
that the three-dimensional TFT is unitary in the sense of \cite{TUra}. 
Such TFTs have the property that orientation reversal of
a three-manifold results in complex conjugation of its invariant, see
sections II.5.4 and IV.11 of \cite{TUra}. We consider the two ribbon graphs
  \bea  \begin{picture}(420,56)(15,46)
  \put(35,0)     {\includeourbeautifulpicture{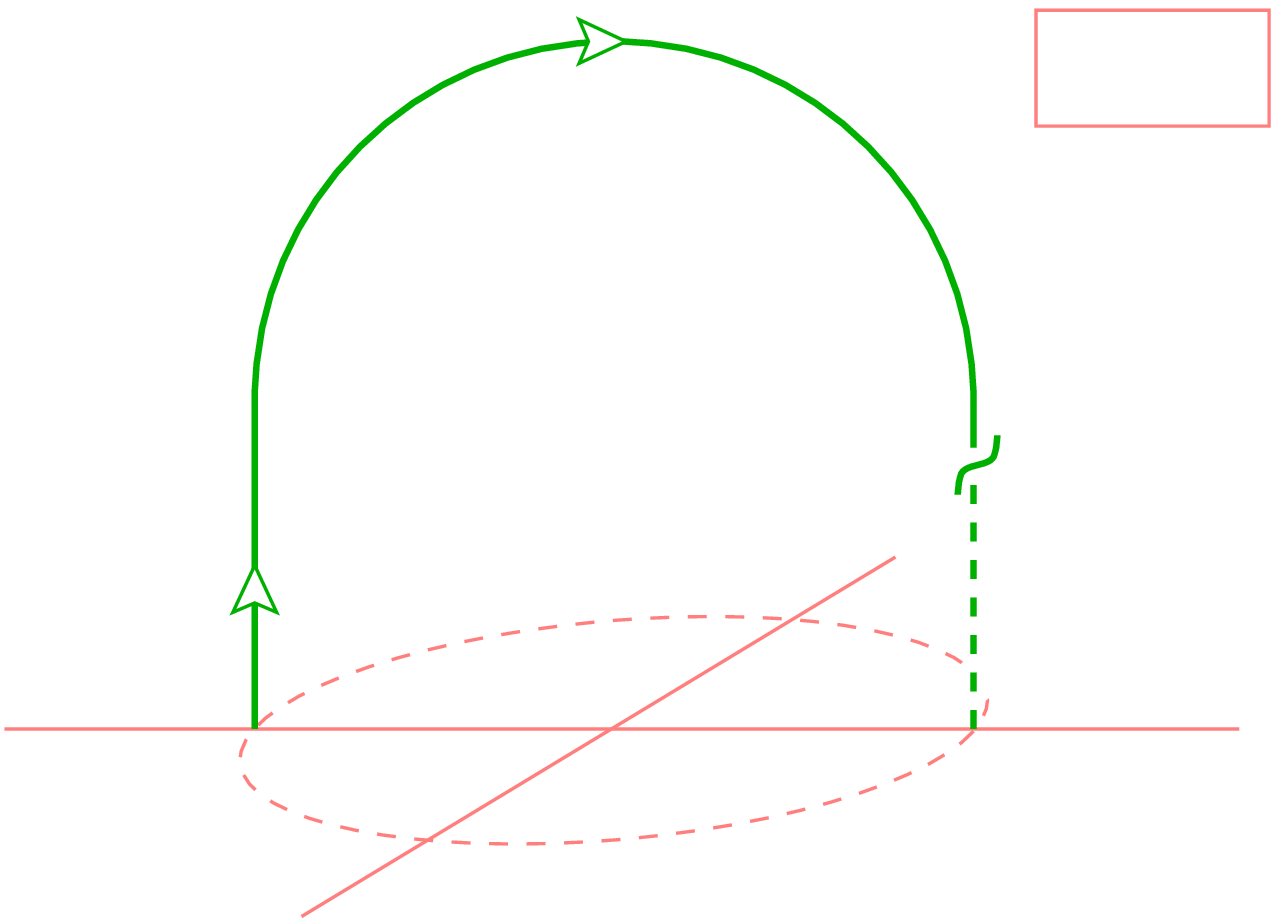}}
  \setlength{\unitlength}{24810sp}
  \setlength{\unitlength}{1pt}
  \put(0,47)        {$ \Gamma_{\!+} \;:= $}
  \put(64.5,44.5)   {\scriptsize$k$}
  \put(151.3,88.8)  {$\RP$}
  \put(210,47)      {and $\quad\ \Gamma_{\!-} \;:= $}
  \put(318.8,44.5)  {\scriptsize$k$}
  \put(405.7,88.8)  {$\RP$}
  \epicture28 \labl{eq:G+G-}
These graphs can be mapped to one another by an orientation
reversing homeomorphism, so that by assumption the associated invariants
satisfy $(\Gamma_+)^*\eq\Gamma_{\!-}$. Furthermore,
  \bea  \begin{picture}(420,58)(9,48)
  \put(35,0)     {\includeourbeautifulpicture{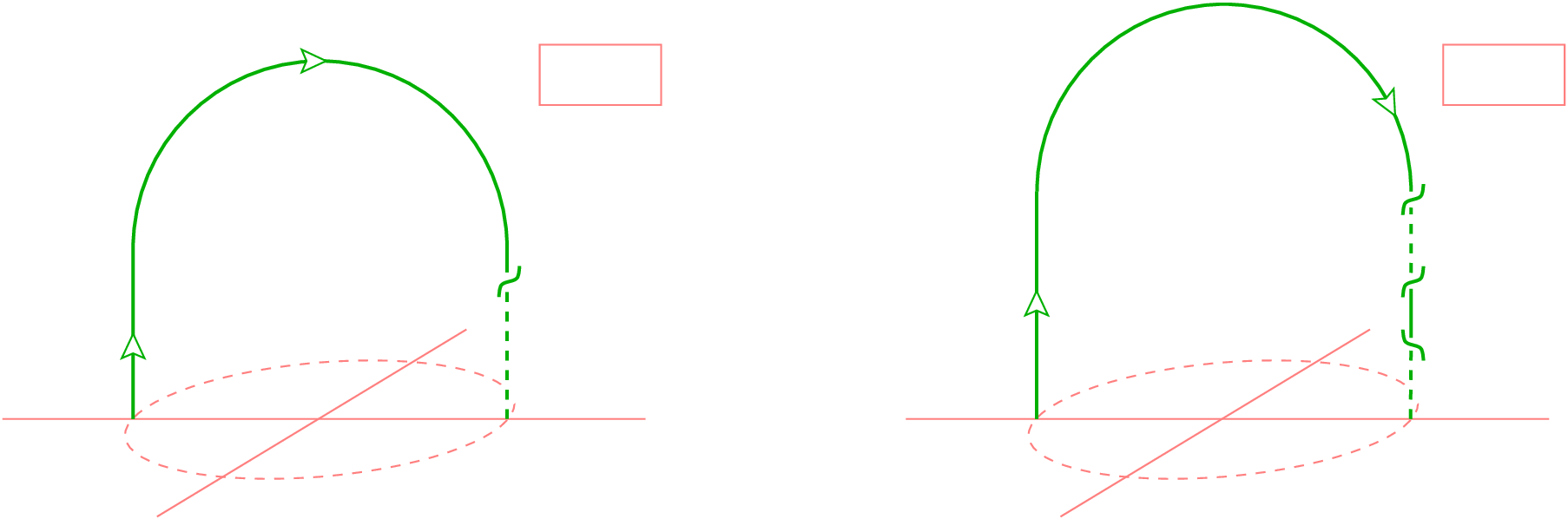}}
  \setlength{\unitlength}{24810sp}
  \setlength{\unitlength}{1pt}
  \put(0,47)        {$ \Gamma_{\!+} \;= $}
  \put(64.8,40.5)   {\scriptsize$k$}
  \put(150.7,88.5)  {$\RP$}
  \put(195,47)      {$ = $}
  \put(255.5,53.5)  {\scriptsize$k$}
  \put(341.4,88.5)  {$\RP$}
  \put(371,47)      {$ =\; \theta_k\,\Gamma_{\!-}\,. $}
  \epicture26 \ee
These relations may be rewritten as
  \be
  (\Gamma_{\!\epsilon)}^* = \Gamma_{\!-\epsilon} \qquad {\rm and} \qquad
  \Gamma_{\!\epsilon} = (\theta_k)^\epsilon\, \Gamma_{\!-\epsilon}
  \labl{eq:gamma-conj}
for $\epsilon\eq{\pm}$.

Now recalling the definition of $S_{ij}$ as the invariant of the
Hopf link in $S^3$ (see (I:2.22)), in the situation at hand the ribbon graph 
on the \rhs\ of \erf{eq:RP-1rib} is given by $s_{kj}\eq S_{kj}/S_{00}$, 
and hence according to \erf{eq:RP-1rib} we have
  \be  \Gamma_{\!\epsilon} = \kappa^{-1}
  \sum_{j\in\II} S_{0j} \theta_j^{-2} S_{kj} \labl{eq:gammaeps}
with $\epsilon\eq{+}$. As a consistency check, we now show by an independent 
argument that the index $\epsilon$ in formula \erf{eq:gammaeps} cannot be an 
arbitrary sign, but must be $\epsilon\eq{+}$.
The matrices $S$ and $T$ representing the modular group
satisfy $(ST)^3\eq C\eq S^2$ with $C$ the charge conjugation matrix.
Since $C$ commutes with $S$ and $T$ this implies $S^{-1} T^2 S^{-1}\eq 
C T^{-1} S T^{-2} S T^{-1}$.  Substituting $T_k\eq\kappa^{-1/3} 
\theta_k^{-1}$ into the $0\bar k$-component of this relation yields
  \be
  \kappa^{-1} \sum_{j\in\II} S_{0j} \theta^{-2}_j S_{jk} =
  \kappa \sum_{j\in\II} S_{0j} \theta^2_j S_{j\bar k} \theta_k \,,  \ee
which when combined with \erf{eq:gammaeps} leads to
  \be
  \big(\Gamma_\epsilon\big)^* = \big(\kappa^{-1}
  \!\sum_{j\in\II} S_{0j} \theta_j^{-2} S_{jk}\big)^*
  = \kappa
  \sum_{j\in\II} S_{0j} \theta_j^{2} S_{j\bar k}
  = \theta_k^{-1} \kappa^{-1}
  \!\sum_{j\in\II} S_{0j} \theta_j^{-2} S_{jk}
  = \theta_k^{-1} \Gamma_\epsilon \,.  \ee
Comparison with \erf{eq:gamma-conj} thus shows that indeed
$\epsilon\eq{+}$, as implied by \erf{eq:RP-1rib}.

\medskip

For the computations in the sequel it will be convenient to display 
separately the form that \erf{eq:RP-1rib} takes when $X\eq U\oti V$:
  \bea  \begin{picture}(410,59)(0,56)
  \put(0,0)     {\includeourbeautifulpicture{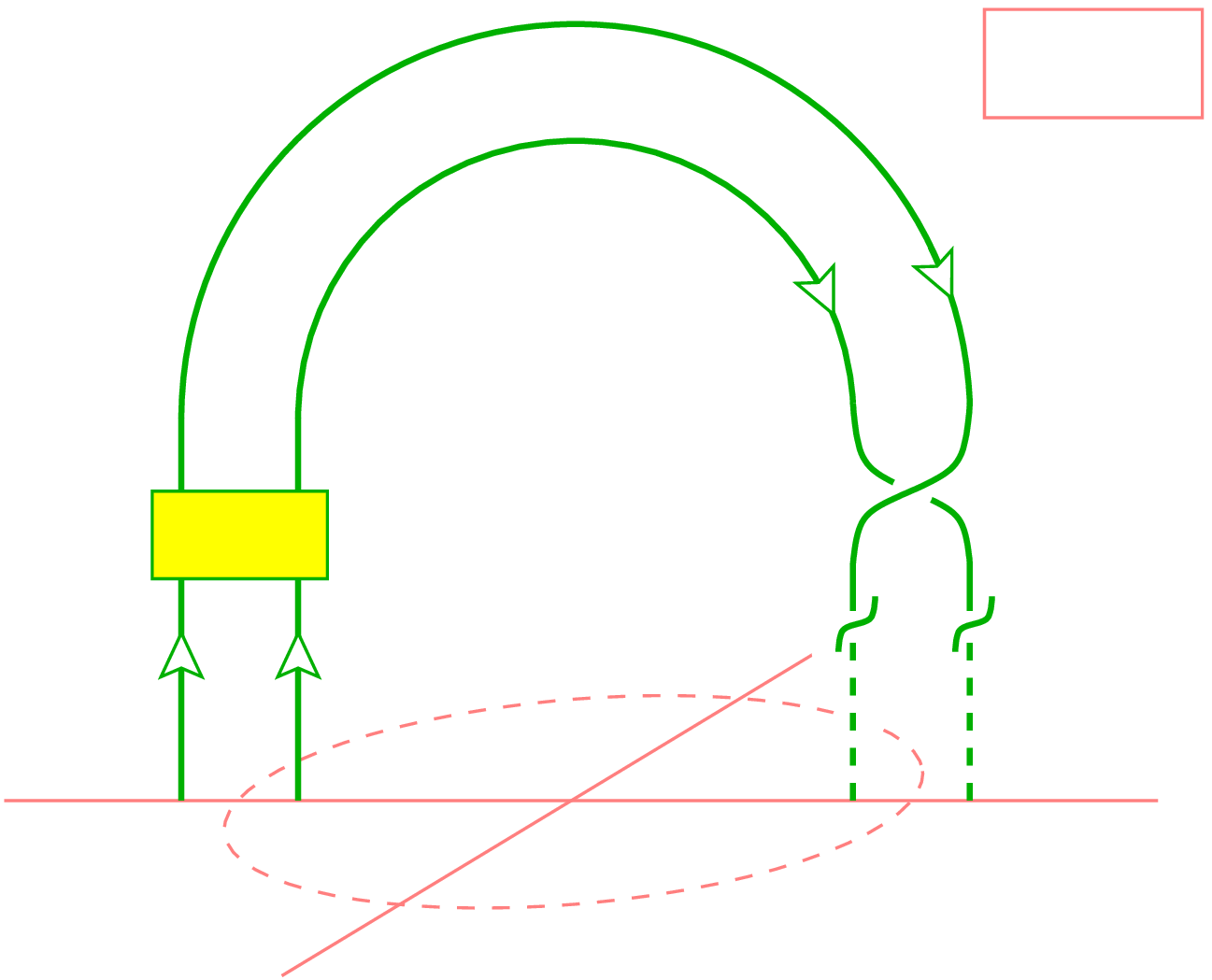}}
  \setlength{\unitlength}{24810sp}
  \setlength{\unitlength}{1pt}
  \put(13.8,23.5)   {\scriptsize$U$}
  \put(15.1,72.5)   {\scriptsize$U$}
  \put(26.5,50.5)   {\scriptsize$\phi$}
  \put(36.5,28.5)   {\scriptsize$V$}
  \put(38.5,72.5)   {\scriptsize$V$}
  \put(119,102.5) {$\RP$}
  \put(171,55)      {$ =\; S_{00} \kappa^{-1} \dsty\sum_{j\in\II}
                       S_{0j}\,\theta_j^{-2} $}
  \put(298.4,78.5)  {\scriptsize$U$}
  \put(310.5,60.5)  {\scriptsize$\phi$}
  \put(321.4,78.5)  {\scriptsize$V$}
  \put(339.8,62.8)  {\scriptsize$j$}
  \put(379.7,108.5) {$S^3$}
  \epicture30 \labl{eq:RP-2rib}
We would also like to point out an effect that arises when manipulating
ribbon graphs in $\RP$ in the presentation \erf{eq:RP3-pic} and that will play 
a role in some of the calculations below.  Recall that the ribbons touch 
the $x$-$y$-plane indeed not in a single point, as suggested by the 
pictures in blackboard framing convention, but in a little arc.  
When moving the pair of arcs at which a ribbon touches the $x$-$y$-plane 
one must take into account the identification \erf{eq:RP3-ident}. 
In general this introduces additional twists on the ribbons which are 
hard to see in the blackboard framing convention. For example, for the 
\lhs\ of \erf{eq:RP-1rib} one has
  \bea  \begin{picture}(340,56)(0,50)
  \put(0,0)     {\includeourbeautifulpicture{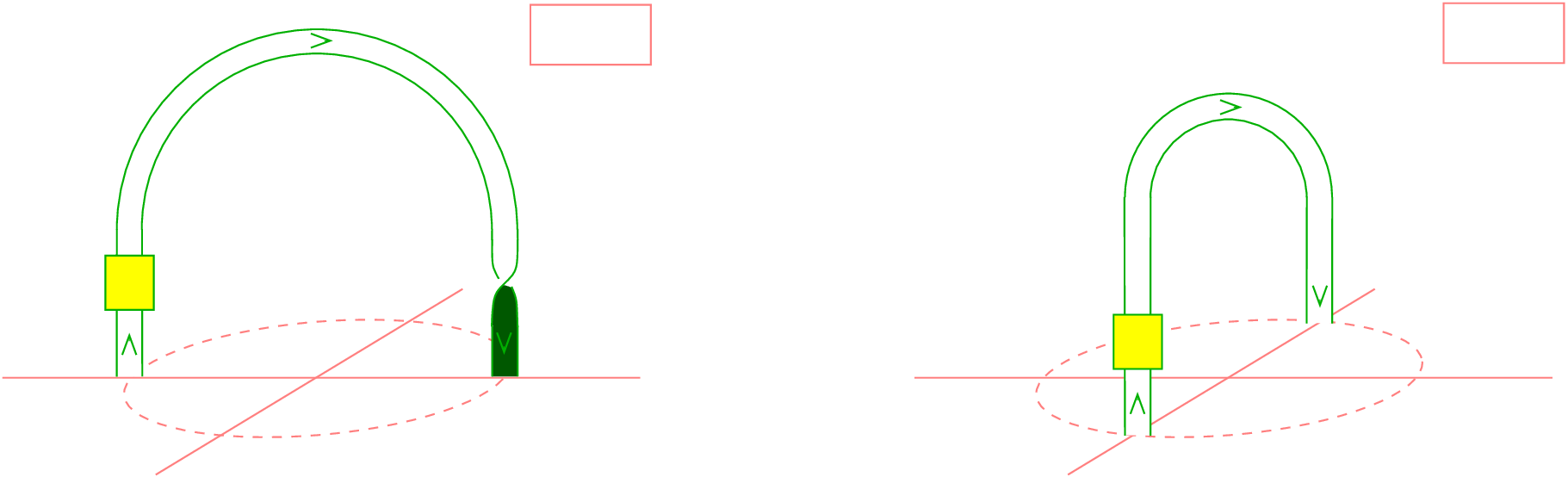}}
  \setlength{\unitlength}{24810sp}
  \setlength{\unitlength}{1pt}
  \put(21.3,73.5)   {\scriptsize$X$}
  \put(24.5,38.8)   {\scriptsize$\phi$}
  \put(115.3,89.4)  {$\RP$}
  \put(161,47)      {$ = $}
  \put(239.5,77.5)  {\scriptsize$X$}
  \put(239.5,26.5)  {\scriptsize$\phi$}
  \put(309.5,89.4)  {$\RP$}
  \epicture28 \labl{eq:pic106}

\subsection{The ingredients $S^A$, $\tilde S^A$ and $\Gamma^{\sigma}$}
\label{SASAGs}

In the closed string channel of 
annulus, M\"obius strip and Klein bottle partition functions, 
specific ribbon graphs $S^A$, $\tilde S^A$ and $\Gamma^{\sigma}$ appear 
as basic constituents. Before introducing them, we recall the definition 
  \bea  \begin{picture}(25,95)(0,37)
  \put(0,0)   {\includeourbeautifulpicture{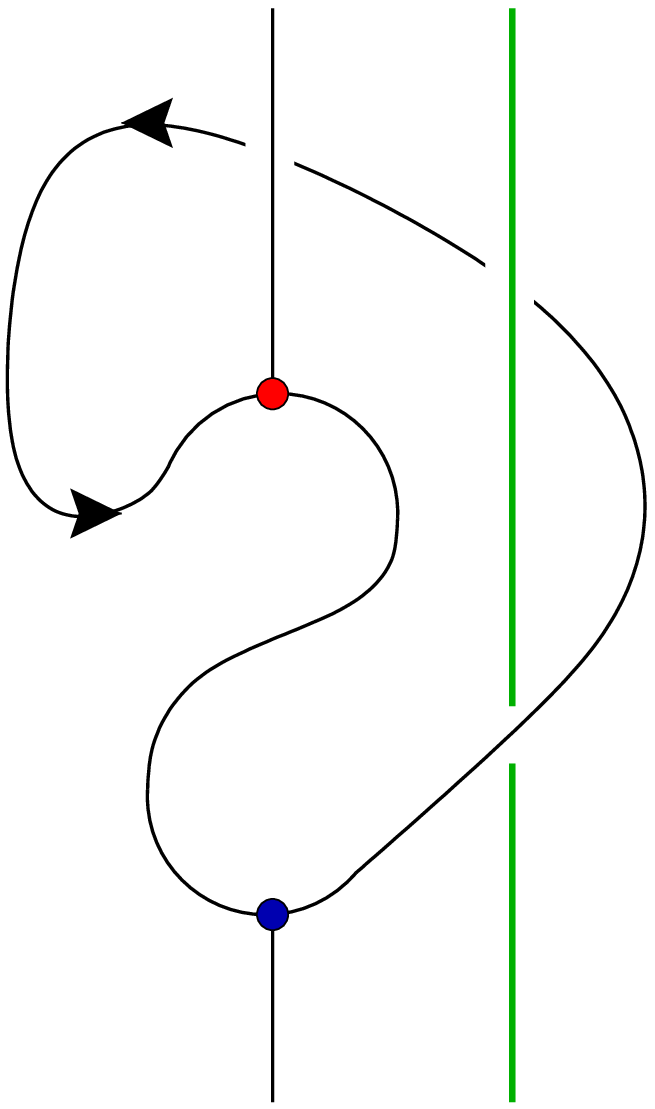}}
  \put(-48.6,57.1) {$P_X\;:=$}
  \put(8.9,32)     {\scriptsize$A$}
  \put(25.5,-7.9)  {\scriptsize$A$}
  \put(52.5,-7.9)  {\scriptsize$X$}
  \put(25.5,125)   {\scriptsize$A$}
  \put(53.3,125)   {\scriptsize$X$}
  \put(72,67)      {\scriptsize$A$}
  \epicture28\label{eq:PX-def} \labl{PX-def}
(see (I:5.34)) of the projector $P_X\iN\Hom(A{\otimes}X,A{\otimes}X)$.
This projector allows to select interesting subspaces
$\Loc(A{\otimes}X,Y)$ (already defined in section I:5.5) 
and $\LocUp(X,A{\otimes}Y)$ of $\Hom(A{\otimes}X,Y)$ and
$\Hom(X,A{\otimes}Y)$, respectively:
\\[-2.3em]

\dtl{Definition}{def:loc}
The subspaces of {\em local morphisms\/}
in $\Hom(A{\otimes}X,Y)$ and $\Hom(X,A{\otimes}Y)$ are
  \be\bearl
  \Loc(A{\otimes}X,Y) := \big\{\,\varphi\iN\Hom(A{\otimes}X,Y)
  \,|\, \varphi\,{\circ}\,P_X\eq\varphi \,\big\} \,,
  \\{}\\[-.7em]
  \LocUp(X,A{\otimes}Y) := \big\{\,\psi\iN\Hom(X,A{\otimes}Y)
  \,|\, P_Y\,{\circ}\,\psi\eq\psi \,\big\} \,.
  \eear\labl{eq:loc-def}

\medskip\noindent
We will also need two linear maps that can be restricted to 
bijections between the subspaces \erf{eq:loc-def}. 

\dtl{Definition}{df:lambda-omega}
We define two linear maps
  \be\bearll
  \Lambda_X^Y: & \Hom(A{\otimes}X,Y) \to \Hom(X,A{\otimes}Y) 
  \qquad{\rm and} \\{}\\[-.6em]
  \Omega_X^Y : & \Hom(A{\otimes}X,Y) \to \Hom(A{\otimes}Y^\vee,X^\vee)
  \eear\ee
by 
  \bea  \begin{picture}(350,41)(0,43)
  \put(62,0)     {\includeourbeautifulpicture{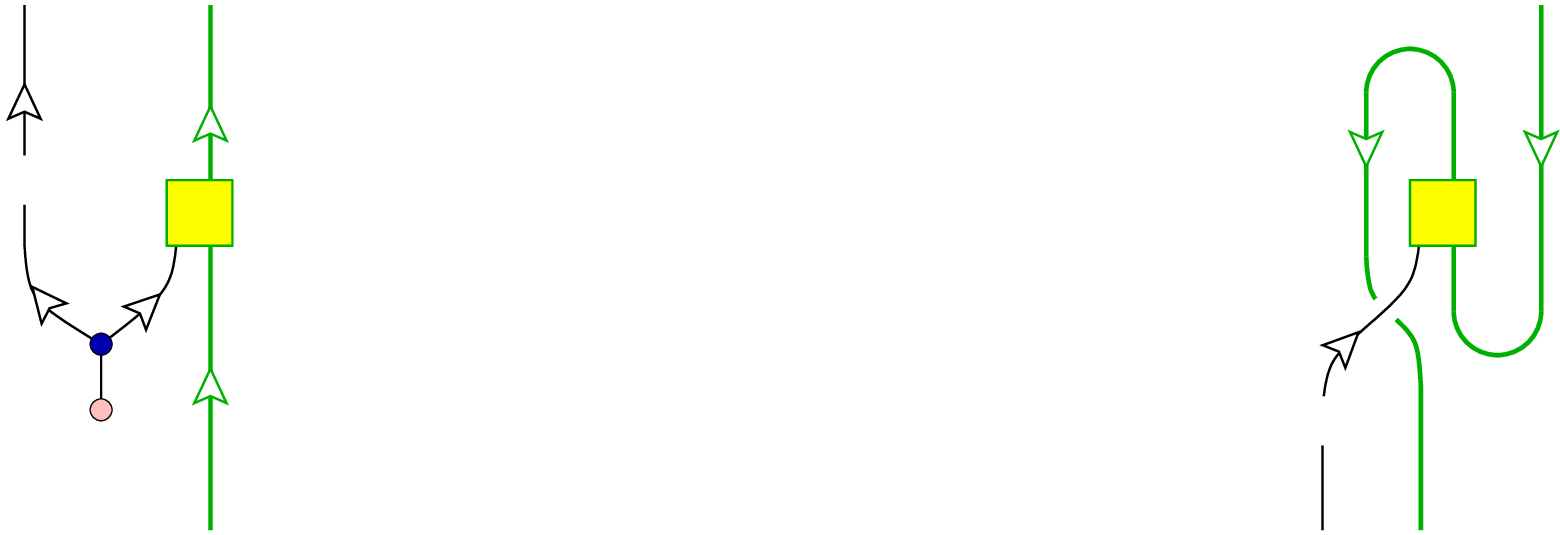}
  \setlength{\unitlength}{24810sp}
      \put(5,143) {\scriptsize$\sigma$}
      \put(75,130) {\scriptsize$\varphi$}
      \put(94.5,186) {\scriptsize$Y$}
      \put(94.5,79) {\scriptsize$X$}
      \put(16.5,194) {\scriptsize$A$}
      \put(592,130) {\scriptsize$\varphi$}
      \put(541,42) {\scriptsize$\sigma$}
      \put(537,84) {\scriptsize$A$}
      \put(605,89) {\scriptsize$X$}
      \put(543,169) {\scriptsize$Y$}
  \setlength{\unitlength}{1pt}
  }
  \put(0,39)    {$ \Lambda_X^Y(\varphi) \;:= $}
  \put(165,39)  {and $\quad\ \Omega_X^Y(\varphi) \;:= $}
  \epicture18 \labl{eq:lambda-omega-def}
for $\varphi\iN\Hom(A{\otimes}X,Y)$.

\dtl{Proposition}{pr:lambda-omega-def}
The maps $\Lambda_X^Y$ and $\Omega_X^Y$ \erf{eq:lambda-omega-def}
restrict to bijections
  \be\bearll
  \Lambda_X^Y :& \Loc(A{\otimes}X,Y) \stackrel\cong\to
  \LocUp(X,A{\otimes}Y) \,,
  \\{}\\[-.8em]
  \Omega_X^Y  :& \Loc(A{\otimes}X,Y) \stackrel\cong\to
  \Loc(A{\otimes}Y^\vee,X^\vee) \,.
  \eear\labl{eq:lam-om-domain}

\medskip\noindent
Proof:\\
For any $\varphi\iN\Hom(A{\otimes}X,Y)$ the equalities
  \bea  \begin{picture}(420,66)(5,65)
  \put(40,0)     {\includeourbeautifulpicture{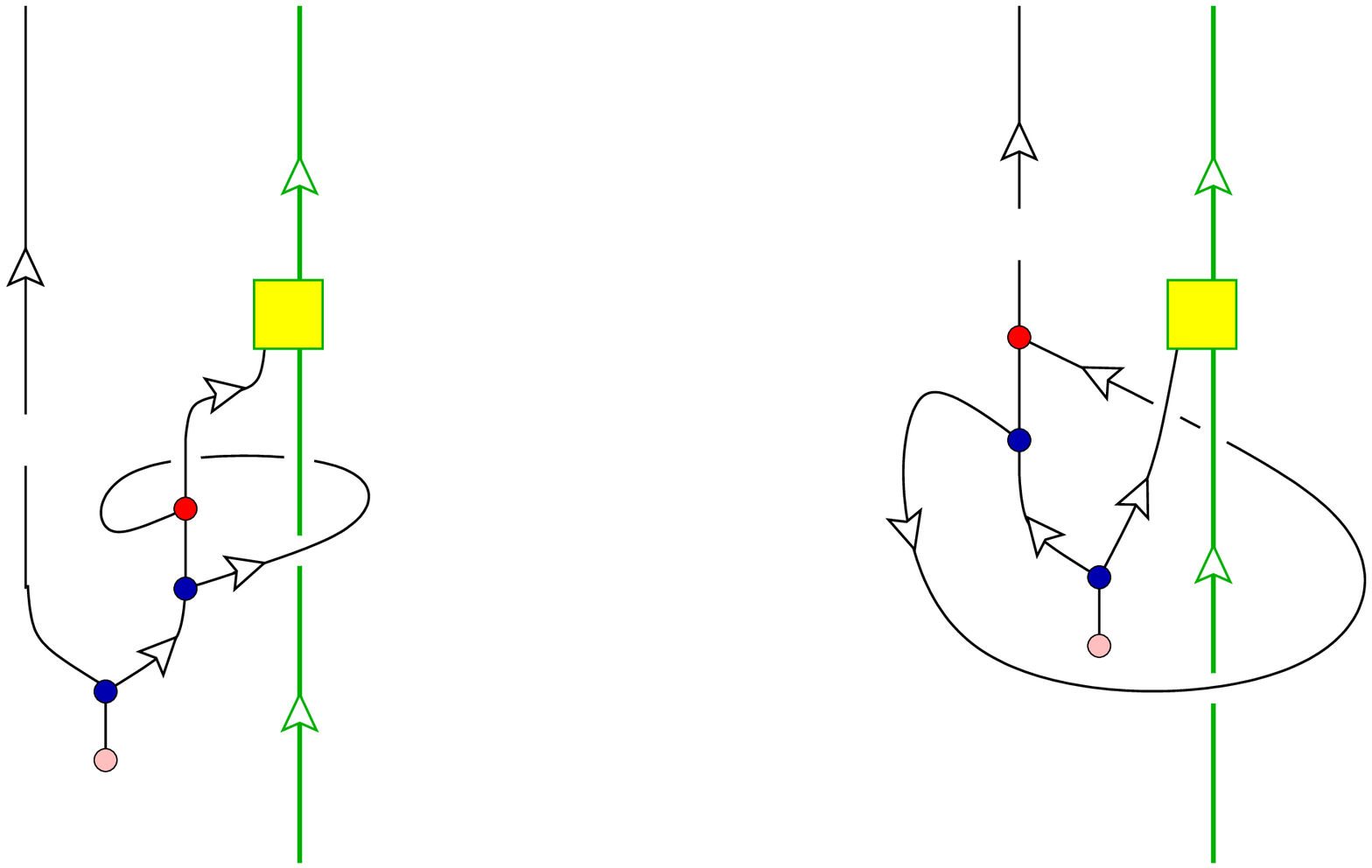}
  \setlength{\unitlength}{24810sp}
      \put(107,215) {\scriptsize$\varphi$}
      \put(5,164) {\scriptsize$\sigma$}
      \put(14,304) {\scriptsize$A$}
      \put(125,304) {\scriptsize$Y$}
      \put(125,21) {\scriptsize$X$}
      \put(55,171) {\scriptsize$A$}
      \put(398,246) {\scriptsize$\sigma$}
      \put(470,215) {\scriptsize$\varphi$}
      \put(408,304) {\scriptsize$A$}
      \put(486,304) {\scriptsize$Y$}
      \put(486,21) {\scriptsize$X$}
      \put(545,105) {\scriptsize$A$}
      \put(912,304) {\scriptsize$Y$}
      \put(912,21) {\scriptsize$X$}
      \put(838,228) {\scriptsize$A$}
      \put(769,152) {\scriptsize$A$}
      \put(895,129) {\scriptsize$\varphi$}
      \put(823,128) {\scriptsize$\sigma$}
      \put(785,221.5) {\scriptsize$\sigma^{\!-\!1}$}
      \put(865,254) {\scriptsize$\sigma$}
  \setlength{\unitlength}{1pt}
  }
  \put(130,62)  {$ = $}
  \put(282,62)  {$ = $}
  \epicture42 \ee
hold. The \lhs\ is equal to $\Lambda_X^Y(\varphi\circ P_X)$. In the second
step $\sigma$ is taken through the multiplication and comultiplication.
The resulting $\sigma$ and $\sigma^{-1}$ on the $A$-loop cancel;
afterwards the $A$-loop can be deformed into the projector $P_Y$
(the move involves two twists $\thetaA$ and $\thetaA^{-1}$, which
cancel). Altogether we find
  \be
  \Lambda_X^Y(\varphi\cir P_X) = P_Y \circ \Lambda_X^Y(\varphi) \,.
  \labl{eq:lam-P}
In particular, if $\varphi$ is already in $\Loc(A{\otimes}X,Y)$, then 
$\Lambda_X^Y(\varphi)$ lies in $\LocUp(A{\otimes}X,Y)$.
\\[.3em]
Next consider, again for any $\varphi\iN\Hom(A{\otimes}X,Y)$, the moves
  \bea  \begin{picture}(420,72)(2,64)
  \put(40,0)     {\includeourbeautifulpicture{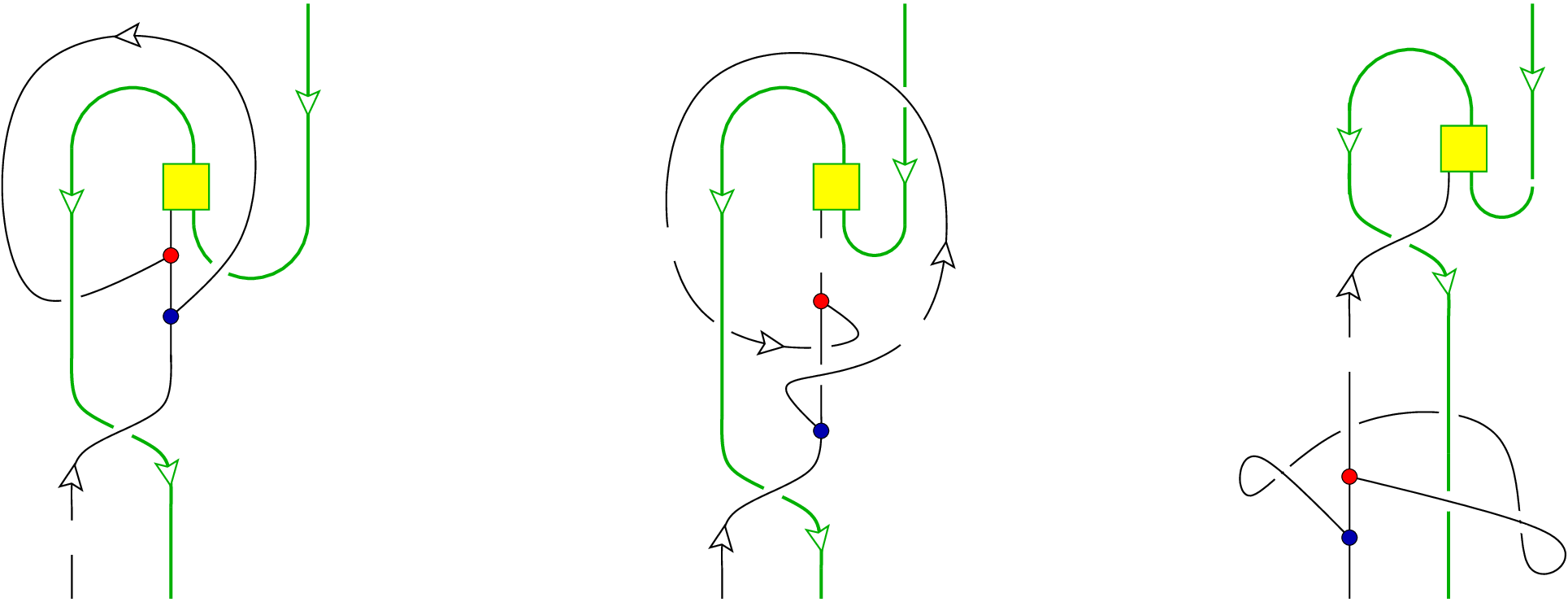}
  \setlength{\unitlength}{24810sp}
      \put(105,242) {\scriptsize$\varphi$}
      \put(38,34) {\scriptsize$\sigma$}
      \put(25,80) {\scriptsize$A$}
      \put(109,44) {\scriptsize$Y$}
      \put(3,313) {\scriptsize$A$}
      \put(189,243) {\scriptsize$X$}
      \put(489,241) {\scriptsize$\varphi$}
      \put(482,201) {\scriptsize$\sigma$}
      \put(536,156) {\scriptsize$\sigma$}
      \put(385,206) {\scriptsize$\sigma^{\!-\!1}$}
      \put(542,320) {\scriptsize$X$}
      \put(567,226) {\scriptsize$A$}
      \put(409,98) {\scriptsize$Y$}
      \put(404,37) {\scriptsize$A$}
      \put(796,143) {\scriptsize$\sigma$}
      \put(862,266) {\scriptsize$\varphi$}
      \put(913,265) {\scriptsize$X$}
      \put(862,148) {\scriptsize$Y$}
      \put(777,187) {\scriptsize$A$}
      \put(901,75) {\scriptsize$A$}
  \setlength{\unitlength}{1pt}
  }
  \put(146,66)  {$ = $}
  \put(287,66)  {$ = $}
  \epicture41 \ee
Moving the $A$-loop down behind the morphism $\varphi$ on the
\lhs\ one checks that the \lhs\ equals $\Omega_X^Y(\varphi\cir P_X)$.
On the \rhs\ the two twists on the $A$-loop cancel and the resulting
morphism is $\Omega_X^Y(\varphi)\cir P_{Y^\vee}$, so that
  \be
  \Omega_X^Y(\varphi\circ P_X) = \Omega_X^Y(\varphi) \circ P_{Y^\vee}\,.
  \labl{eq:om-P}
The equalities \erf{eq:lam-P} and \erf{eq:om-P} establish that
the $\Lambda_X^Y$ and $\Omega_X^Y$ indeed map local morphisms to
local morphisms as claimed in \erf{eq:lam-om-domain}.
\\[.3em] 
It remains to be shown that 
$\Lambda_X^Y$ and $\Omega_X^Y$ are invertible. To this end define
the maps $\tilde\Lambda_X^Y$ and $\tilde\Omega_X^Y(\beta)$ by
  \bea  \begin{picture}(350,42)(0,42)
  \put(67,0)     {\includeourbeautifulpicture{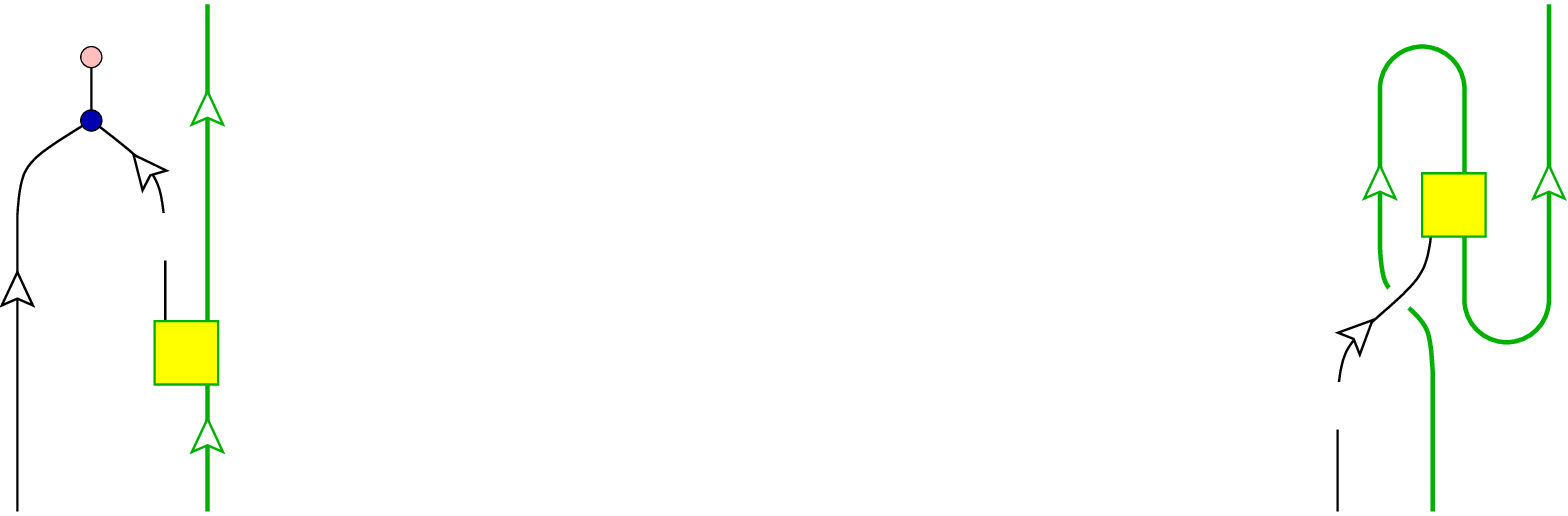}
  \setlength{\unitlength}{24810sp}
      \put(15,8) {\scriptsize$A$}
      \put(96,8) {\scriptsize$X$}
      \put(96,192) {\scriptsize$Y$}
      \put(52,113) {\scriptsize$\sigma^{\!-\!1}$}
      \put(74,63) {\scriptsize$\alpha$}
      \put(568,42) {\scriptsize$\sigma$}
      \put(618,8) {\scriptsize$X$}
      \put(668,192) {\scriptsize$Y$}
      \put(618,127) {\scriptsize$\beta$}
      \put(577,8) {\scriptsize$A$}
  \setlength{\unitlength}{1pt}
  }
  \put(0,39)    {$ \tilde\Lambda_X^Y(\alpha) \;:= $}
  \put(177,39)  {and $\quad\ \tilde\Omega_X^Y(\beta) \;:= $}
  \epicture23 \labl{eq:lam-om-tilde}
for $\alpha\iN\LocUp(X,A{\otimes}Y)$ and $\beta\iN\Loc(A{\otimes}Y^\vee\!,
X^\vee)$. An argument similar to the one given above for $\Lambda_X^Y$ and 
$\Omega_X^Y$ shows that they restrict to maps
  \be\bearll
  \tilde\Lambda_X^Y :& \LocUp(X,A{\otimes}Y) \to\Loc(A{\otimes}X,Y) \,,
  \\{}\\[-.8em]
  \tilde\Omega_X^Y  :& \Loc(A{\otimes}Y^\vee,X^\vee)\to\Loc(A{\otimes}X,Y)
  \,.  \eear\ee
Finally one easily verifies that $\Lambda_X^Y$ and $\tilde\Lambda_X^Y$,
respectively $\Omega_X^Y$ and $\tilde\Omega_X^Y$,
are left and right inverse to one another.
\qed

Note that while the definition of the morphism
$\tilde\Omega_X^Y(\beta)$ in \erf{eq:lam-om-tilde} looks similar to that
of $\Omega_{Y^\vee}^{X^\vee}(\beta)$, the former results in a morphism
in $\Loc(A{\otimes}X,Y)$ while the latter gives an element of
$\Loc(A{\otimes}X^{\vee\vee},Y^{\vee\vee})$. The two maps can be
related via an isomorphism
  \be
  \tilde\Omega_X^Y(\beta) = \delta_Y^{-1} \circ
  \Omega_{Y^\vee}^{X^\vee}(\beta) \circ (\id_A{\otimes}\delta_X) \,,
  \ee
with $\delta_U \iN \Hom(U,U^{\vee\vee})$ as defined above (I:2.13).
When evaluating $\tilde \Omega_X^Y$ in a basis, this can give rise to
additional sign factors.

\medskip

In the manipulations below we will use two specific moves on
several occasions. It is helpful to list them explicitly:
\\[-2.3em]

\dt{Lemma}
(i)\,~For every object $U$, any endomorphism $\phi\iN\Hom(U,U)$ has
equal left and right trace,
  \bea  \begin{picture}(310,49)(0,30)
  \put(40,0)     {\includeourbeautifulpicture{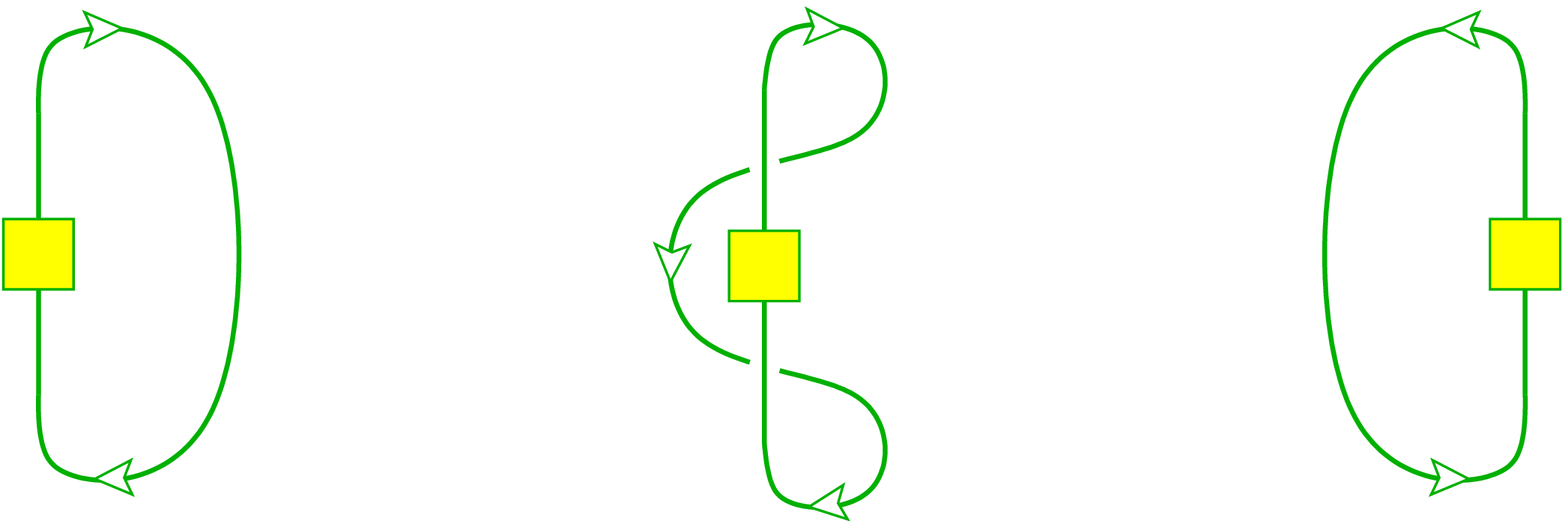}
  \setlength{\unitlength}{24810sp}
      \put(9,98) {\scriptsize$\phi$}
      \put(99,117) {\scriptsize$U$}
      \put(289,94) {\scriptsize$\phi$}
      \put(234,90) {\scriptsize$U$}
      \put(518,95) {\scriptsize$U$}
      \put(584,98) {\scriptsize$\phi$}
  \setlength{\unitlength}{1pt}
  }
  \put(102,35)  {$ = $}
  \put(198,35)  {$ = $}
  \epicture22 \labl{eq:trace-flip}
(ii) For any three objects $U,\,V,\,W$ and morphisms $\alpha,\,\beta$
(with source and range as indicated in the figure) we have
  \bea  \begin{picture}(420,36)(24,36)
  \put(40,0)     {\includeourbeautifulpicture{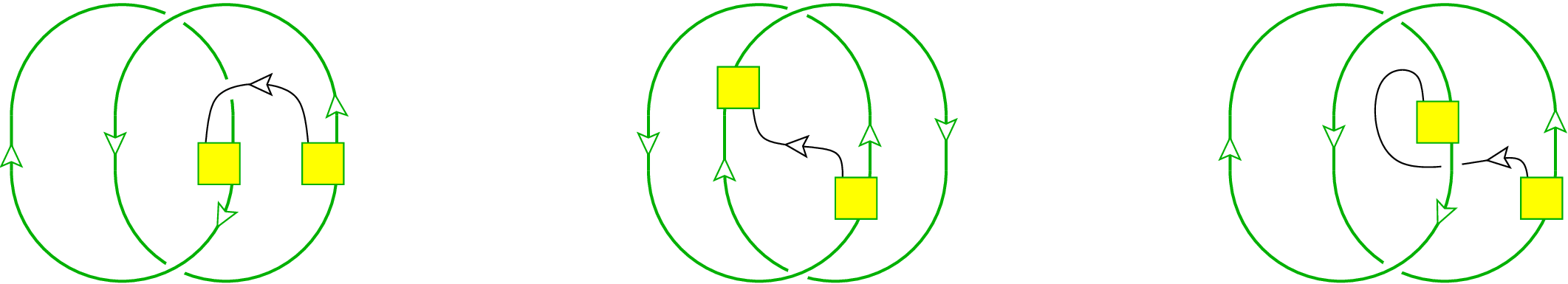}
  \setlength{\unitlength}{24810sp}
      \put(137,75) {\scriptsize$\alpha$}
      \put(205,74) {\scriptsize$\beta$}
      \put(17,101) {\scriptsize$U$}
      \put(212,153) {\scriptsize$V$}
      \put(170,140) {\scriptsize$W$}
      \put(474,124) {\scriptsize$\alpha$}
      \put(553,49) {\scriptsize$\beta$}
      \put(408,132) {\scriptsize$V$}
      \put(613,147) {\scriptsize$U$}
      \put(516,101) {\scriptsize$W$}
      \put(933,101) {\scriptsize$\alpha$}
      \put(1000,49) {\scriptsize$\beta$}
      \put(808,102) {\scriptsize$U$}
      \put(971,93) {\scriptsize$W$}
      \put(1003,155) {\scriptsize$V$}
  \setlength{\unitlength}{1pt}
  }
  \put(158,27)  {$ = $}
  \put(306,27)  {$ = $}
  \epicture22 \labl{eq:Hopf-rot}

\medskip\noindent
Proof:\\
(i)\,~The two twists $\theta_U$ and $\theta_U^{-1}$ introduced 
on the $U$-ribbon in the middle figure cancel. 
\\[3pt]
(ii) The left and right figures are obtained from the middle one by keeping
$\beta$ and the $V$-ribbon fixed and rotating the coupon that represents
the morphism $\alpha$ (dragging along also the attached ribbons) by 
$\pm 180^\circ$.
\qed

We call the move \erf{eq:trace-flip} the {\em trace flip\/} on $U$,
and the moves \erf{eq:Hopf-rot} the clockwise (right equality) and
counter-clockwise (left equality) {\em Hopf turn\/} of $U$.
That left and right traces coincide means that a \mtc\ $\calc$ is
in particular spherical.

\medskip

We now present the definition of $S^A$, $\tilde S^A$ and $\Gamma^\sigma$.
\\[-2.3em]

\dtl{Definition}{df:SSG-def}
For $X$ an object, $M$ a left $A$-module, $\varphi\iN
\Hom(A{\otimes}X,X)$ and $\psi\iN\Hom(X,A{\otimes}X)$ we set
  \bea  \begin{picture}(420,80)(14,0)
  \put(114,0)     {\includeourbeautifulpicture{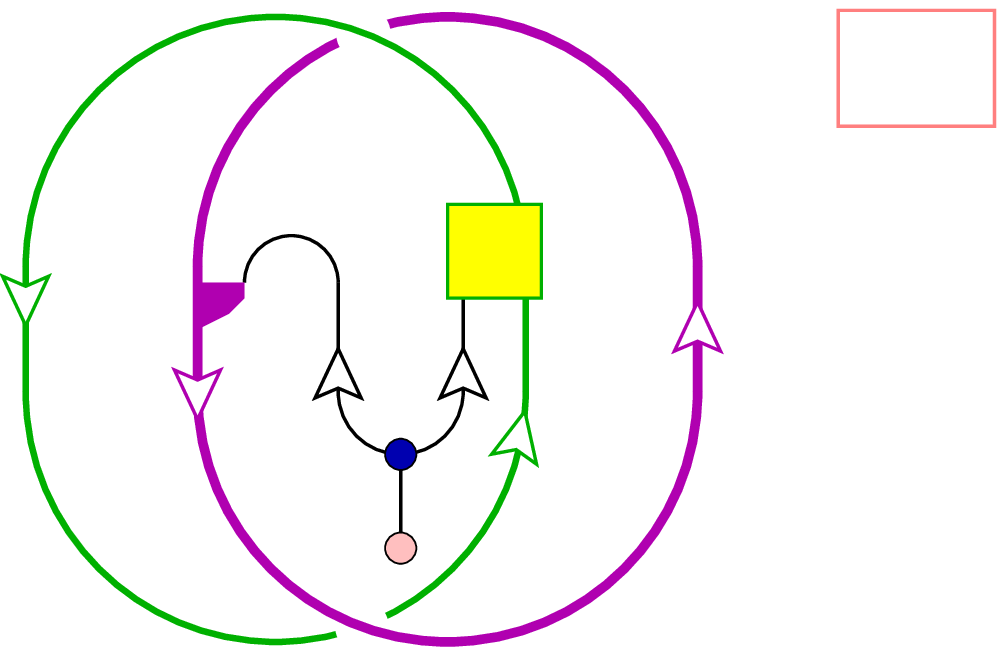}
  \setlength{\unitlength}{24810sp}
      \put(134,113) {\scriptsize$\varphi$}
      \put(85,124) {\scriptsize$A$}
      \put(201,132) {\scriptsize$M$}
      \put(158,71) {\scriptsize$X$}
  \setlength{\unitlength}{1pt}
  }
  \put(354,0)     {\includeourbeautifulpicture{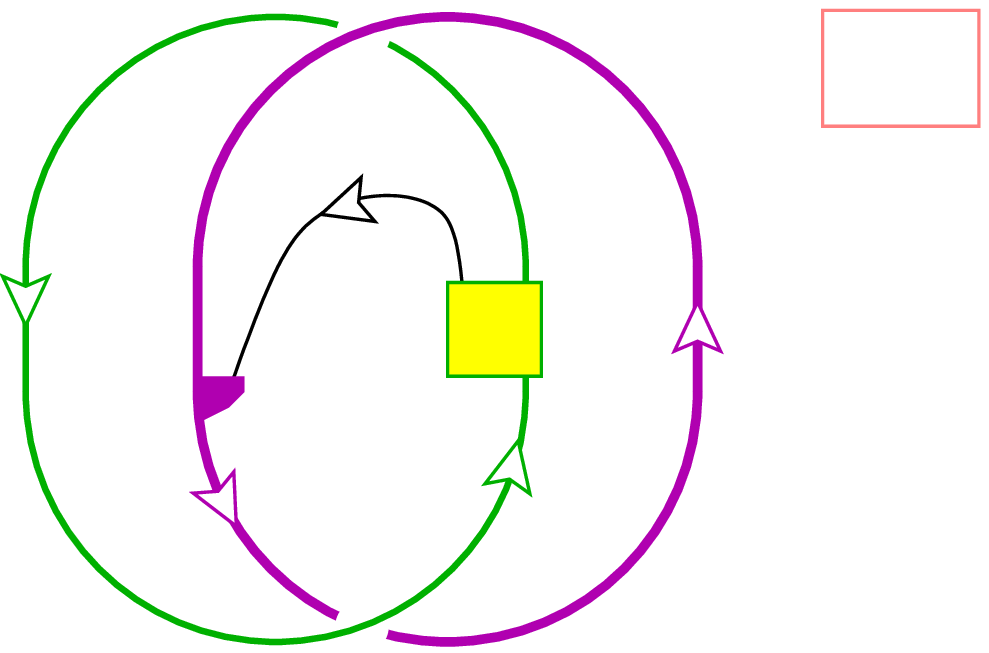}
  \setlength{\unitlength}{24810sp}
      \put(135,89) {\scriptsize$\psi$}
      \put(90,104) {\scriptsize$A$}
      \put(199,137) {\scriptsize$M$}
      \put(155,132) {\scriptsize$X$}
  \setlength{\unitlength}{1pt}
  }
  \put(0,34)    {$ S^A(M;X,\varphi) \;:=\; S_{00} $}
  \put(208.3,59.5){$S^3$}
  \put(240,34)  {$ \tilde S^A(X,\psi;M)\;:=\; S_{00} $}
  \put(447,59.5){$S^3$}
  \epicture-6 \labl{eq:SS-def}
and
  \bea  \begin{picture}(250,50)(0,48)
  \put(140,0)     {\includeourbeautifulpicture{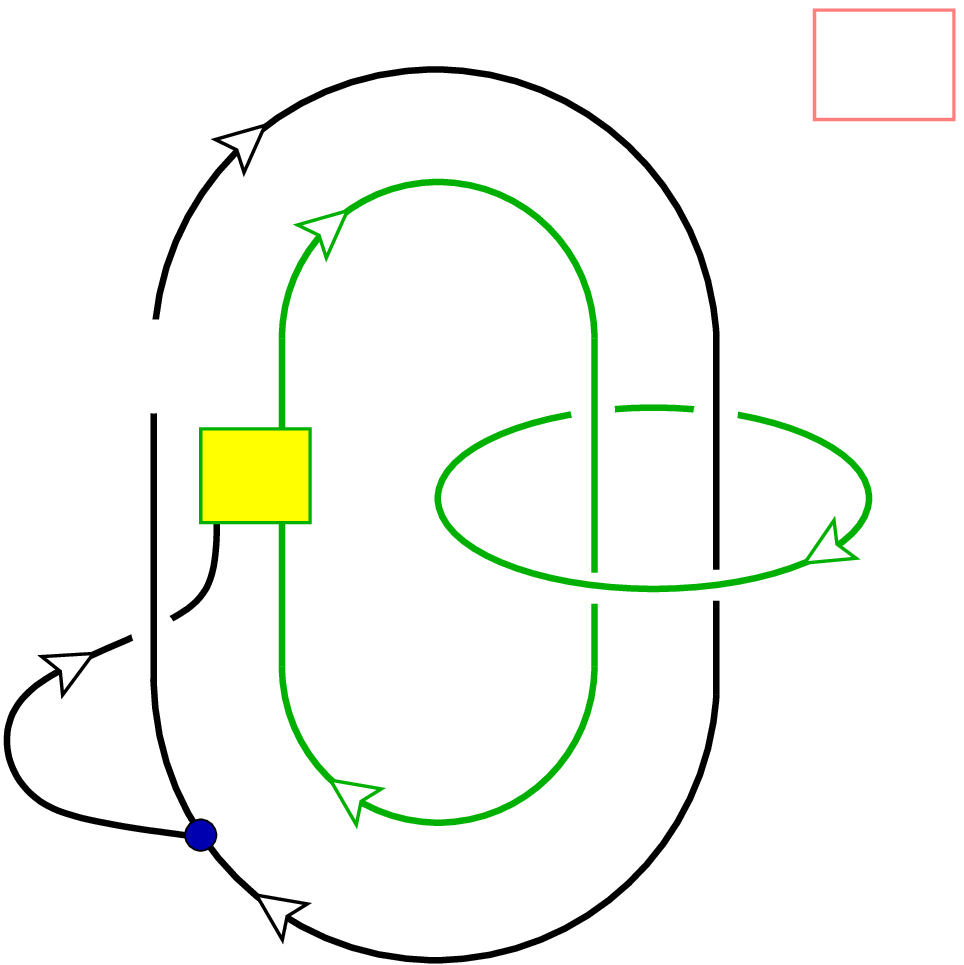}
  \setlength{\unitlength}{24810sp}
      \put(35,168) {\scriptsize$\sigma^{\!-\!1}$}
      \put(65,137) {\scriptsize$\varphi$}
      \put(89,83) {\scriptsize$X$}
      \put(71,255) {\scriptsize$A$}
      \put(-18,59) {\scriptsize$A$}
      \put(257,132) {\scriptsize$U_j$}
      \put(242,249) {$S^3$}
      \put(,) {\scriptsize$$}
  \setlength{\unitlength}{1pt}
  }
  \put(-30,46)    {$ \Gamma^\sigma(X,\varphi) \;:=\; S_{00} \kappa^{-1}
                   \dsty\sum_{j\in\II} S_{0j}\,\theta_j^{-2} $}
  \epicture25 \labl{eq:Ga-def}

\medskip

Alternatively one can represent the invariant $\Gamma^\sigma(X,\varphi)$
directly in $\RP$; for example one has
  \bea  \begin{picture}(420,64)(14,52)
  \put(75,0)     {\includeourbeautifulpicture{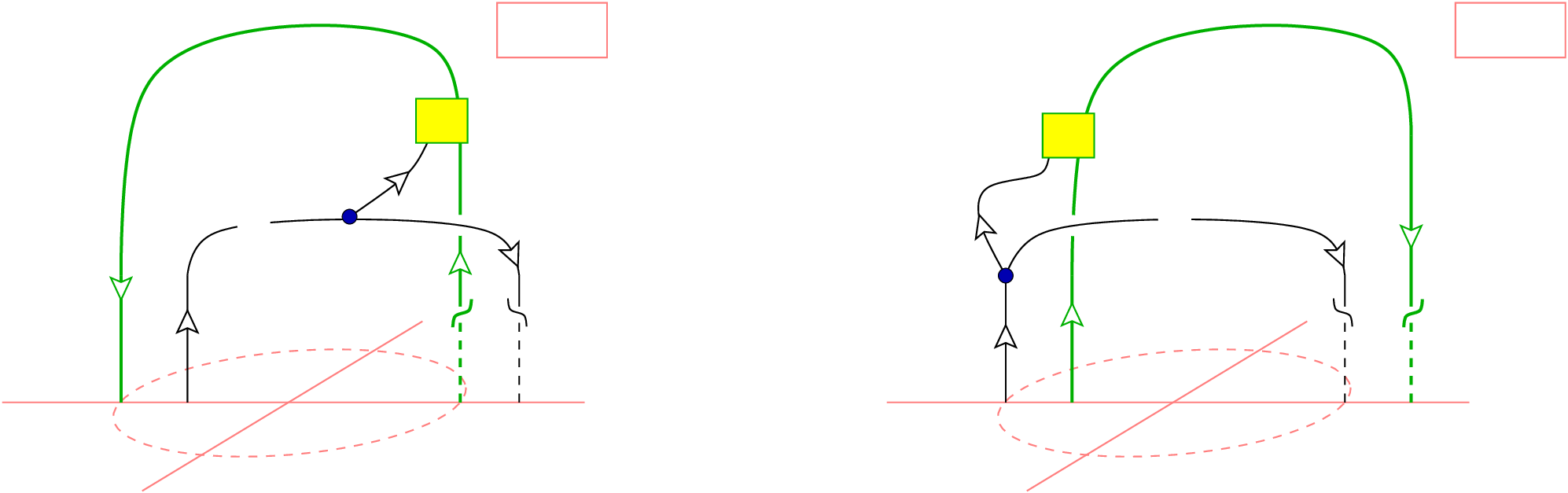}
  \setlength{\unitlength}{24810sp}
      \put(313,273.5){$\RP$}
      \put(263,223) {\scriptsize$\varphi$}
      \put(149,161) {\scriptsize$\sigma$}
      \put(289,188) {\scriptsize$X$}
      \put(123,111.5) {\scriptsize$A$}
      \put(649,216) {\scriptsize$\varphi$}
      \put(715,163) {\scriptsize$\sigma$}
      \put(662,116) {\scriptsize$X$}
      \put(772,173) {\scriptsize$A$}
      \put(904,273.5){$\RP$}
  \setlength{\unitlength}{1pt}
  }
  \put(1,53)    {$ \Gamma^\sigma(X,\varphi) \;= $}
  \put(247,53)  {$ = $}
  \epicture27 \labl{eq:Ga-RP}
The equality of the first and second graph in \erf{eq:Ga-RP}
amounts to moving the point at which the core of the left-most
$X$-ribbon touches the $x$-$y$-plane counter-clockwise along the dashed
circle for half a circumference. Due to the identification this moves
the right-most intersection point of the $X$-ribbon with the horizontal
plane counter-clockwise as well. In this manipulation, we must keep in mind
the effect mentioned at the end of section \ref{sec:S3/Z2}. Afterwards the 
$X$-ribbon is displaced slightly to the right, while the $A$-ribbon is 
shifted to the left. 
\\
The equality of \erf{eq:Ga-RP} and \erf{eq:Ga-def} is obtained by applying 
\erf{eq:RP-2rib}: To bring the second graph in \erf{eq:Ga-RP} to the form 
\erf{eq:RP-2rib}, lift the $A$-ribbon above the $X$-ribbon. One also must 
reverse the half-twist on the $A$-ribbon, giving rise to a twist 
$\thetaA^{-1}$, which combines with $\sigma$ to $\sigma^{-1}$.

The relation of \erf{eq:SS-def} to the similar quantities
(I:5.97) and (I:5.132) is via a choice of basis, as
will be discussed in more detail at the end of this section.

\dtl{Proposition}{pr:lamda-omega-action}
For $X$ an object, $\varphi\iN\Hom(A{\otimes}X,X)$, 
$\psi\iN\Hom(X,A{\otimes}X)$ and $M$ a left $A$-module, the quantities
$S^A$, $\tilde S^A$ and $\Gamma^\sigma$ defined in \erf{eq:SS-def}
and \erf{eq:Ga-def} fulfill
  \begin{eqnarray}
  {\rm (i)} && S^A(M^\sigma;X,\varphi)
  = \tilde S^A(X,\Lambda_X^X(\varphi);M) \,, \label{eq:SA-StA}
  \\\nonumber{}\\[-.8em]
  {\rm (ii)} && S^A(M^\sigma;X,\varphi)
  = S^A(M;X^\vee,\Omega_X^X(\varphi)) \,,    \label{eq:SA-StA2}
  \\\nonumber{}\\[-.8em]
  {\rm (iii)} && \Gamma^\sigma(X,\varphi)
  = \Gamma^\sigma(X^\vee,\Omega_X^X(\varphi)) \,,
  \label{eq:Ga-GaO}\\\nonumber{}\\[-.8em]
  {\rm (iv)} &&
  S^A(M;X,\varphi)=S^A(M;X,\varphi\,{\circ}P_X)\,, \qquad
  \Gamma^\sigma(X,\varphi) = \Gamma^\sigma(X,\varphi\,{\circ}P_X)\,,
  \quad  \nonumber \\\nonumber{}\\[-1em]&&
  \tilde S^A(X,\psi;M) = \tilde S^A(X,P_X{\circ}\,\psi;M)\,.
  \end{eqnarray}

\medskip\noindent
Proof:\\
For the proof of (i)\,--\,(iii) we use the trace flip \erf{eq:trace-flip}
and the Hopf turn \erf{eq:Hopf-rot}.
\\[5pt]
(i) Consider the moves
  \bea  \begin{picture}(420,24)(11,45)
  \put(40,0)     {\includeourbeautifulpicture{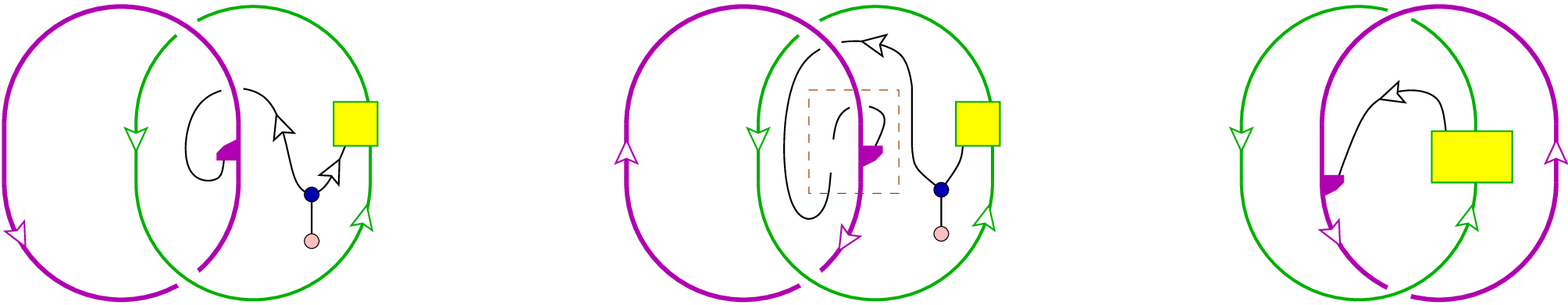}
  \setlength{\unitlength}{24810sp}
      \put(221,156) {\scriptsize$X$}
      \put(85,191) {\scriptsize$M^\sigma$}
      \put(173,125) {\scriptsize$A$}
      \put(212,107) {\scriptsize$\varphi$}
      \put(394,109) {\scriptsize$M$}
      \put(603,157) {\scriptsize$X$}
      \put(566,101) {\scriptsize$A$}
      \put(510,87) {\scriptsize$\sigma$}
      \put(598,109) {\scriptsize$\varphi$}
      \put(886,86) {\tiny$\Lambda_X^X\!(\!\varphi\!)$}
      \put(838,92) {\scriptsize$A$}
      \put(762,154) {\scriptsize$X$}
      \put(967,111) {\scriptsize$M$}
  \setlength{\unitlength}{1pt}
  }
  \put(153,32)  {$ = $}
  \put(298,32)  {$ = $}
  \epicture24 \ee
The \lhs\ is obtained from $S^A(M^\sigma;X,\varphi)$ by
an counter-clockwise Hopf turn on $M^\sigma$. In the first step the
definition \erf{eq:rho-sigma} of $\rho^\sigma$ is substituted
and the second step amounts to a trace flip on $M$ as well as
insertion of the definition \erf{eq:lambda-omega-def}
of $\Lambda_X^X(\varphi)$. The \rhs\ equals
$\tilde S^A(X,\Lambda_X^X(\varphi);M)$.
\\[5pt]
(ii) Consider the moves
  \bea  \begin{picture}(420,50)(9,57)
  \put(0,0)     {\includeourbeautifulpicture{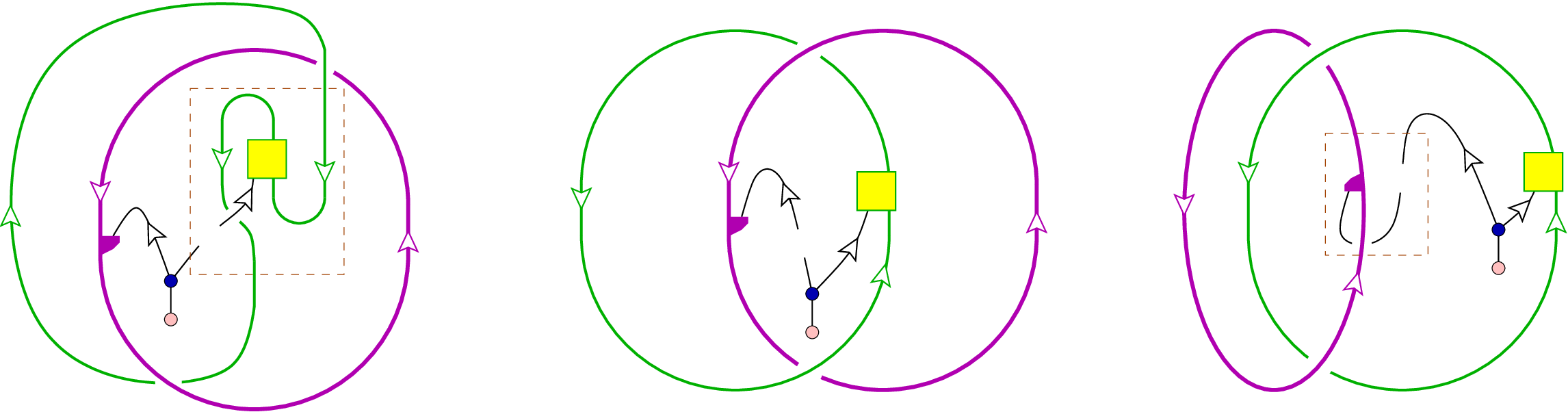}
  \setlength{\unitlength}{24810sp}
      \put(181,175) {\scriptsize$\varphi$}
      \put(140,120) {\scriptsize$\sigma$}
      \put(105,142) {\scriptsize$A$}
      \put(280,202) {\scriptsize$M$}
      \put(21,190) {\scriptsize$X$}
      \put(560,114) {\scriptsize$\sigma$}
      \put(611,150) {\scriptsize$\varphi$}
      \put(727,203) {\scriptsize$M$}
      \put(624,202) {\scriptsize$X$}
      \put(551,169) {\scriptsize$A$}
      \put(982,159) {\scriptsize$\sigma$}
      \put(1021,206) {\scriptsize$A$}
      \put(1081,166) {\scriptsize$\varphi$}
      \put(1073,236) {\scriptsize$X$}
      \put(951,51) {\scriptsize$M$}
  \setlength{\unitlength}{1pt}
  }
  \put(128,53)  {$ = $}
  \put(292,53)  {$ = $}
  \epicture32 \ee
The \lhs\ shows the definition of
$S^A(\,M;X^\vee,\Omega_X^X(\varphi))$. The first step amounts
to a clockwise Hopf turn of $X$ followed by a trace flip on $X$.
Also, in the first step the identity \erf{eq:sig-phi} is used
to move $\sigma$ past the coproduct.
The second step is a clockwise Hopf turn of $M$. By definition
\erf{eq:rho-sigma} of $\rho^\sigma$ the \rhs\ is equal to
$S^A(M^\sigma;X,\varphi)$.
\\[5pt]
(iii) Consider the moves
  \bea  \begin{picture}(390,188)(0,68)
  \put(40,0)     {\includeourbeautifulpicture{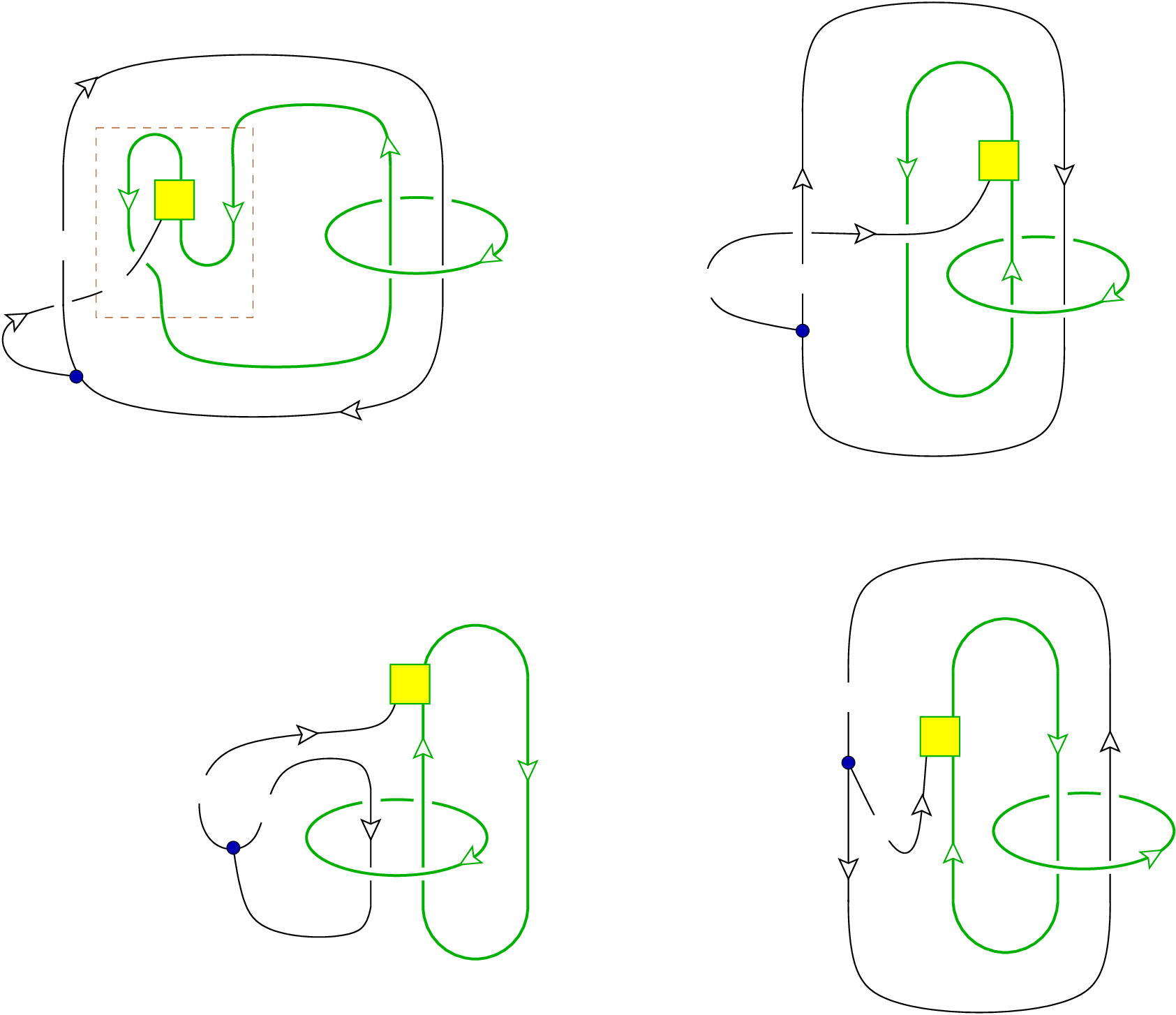}
  \setlength{\unitlength}{24810sp}
      \put(113,559) {\scriptsize$\varphi$}
      \put(681,587) {\scriptsize$\varphi$}
      \put(275,225) {\scriptsize$\varphi$}
      \put(642,188) {\scriptsize$\varphi$}
      \put(31,527) {\scriptsize$\sigma^{\!-\!1}$}
      \put(74,504) {\scriptsize$\sigma$}
      \put(487,502) {\scriptsize$\sigma$}
      \put(546,503) {\scriptsize$\sigma^{\!-\!1}$}
      \put(131,151) {\scriptsize$\sigma$}
      \put(170,135) {\scriptsize$\sigma^{\!-\!1}$}
      \put(574,213) {\scriptsize$\sigma^{\!-\!1}$}
      \put(603,123) {\scriptsize$\sigma$}
      \put(22,583) {\scriptsize$A$}
      \put(-20,468) {\scriptsize$A$}
      \put(559,608) {\scriptsize$A$}
      \put(157,192) {\scriptsize$A$}
      \put(771,238) {\scriptsize$A$}
      \put(335,559) {\scriptsize$U_j$}
      \put(782,502) {\scriptsize$U_j$}
      \put(320,147) {\scriptsize$U_j$}
      \put(819,120) {\scriptsize$U_j$}
      \put(240,462) {\scriptsize$X$}
      \put(653,435) {\scriptsize$X$}
      \put(303,273) {\scriptsize$X$}
      \put(639,244) {\scriptsize$X$}
  \setlength{\unitlength}{1pt}
  }
  \put(61,57)   {$ = $}
  \put(195,202) {$ = $}
  \put(216,57)  {$ = $}
  \epicture43 \ee
The first step is a deformation of the $X$-ribbon. The second step consists
of a trace flip on $X$ combined with a deformation of the $A$-ribbon. In 
the third step the $A$-loop is rotated by $180^\circ$ clockwise around its 
center and then a trace flip is executed on it. Afterwards one can use
the identity (I:5.25) to turn around the direction of the $A$-ribbon.
This also changes the product into a coproduct. Finally the morphism 
$\sigma^{-1}$ is moved clockwise around the $A$-loop; upon use of 
\erf{eq:sig-coprod} this cancels the $\sigma$ in front of $\varphi$ and 
introduces the correct braiding to get \erf{eq:Ga-def}.
When multiplied with $\kappa^{-1} S_{00}S_{0j}\theta_j^{-2}$
and summed over $j\iN\II$, the \lhs\ gives
$\Gamma^\sigma(X^\vee,\Omega_X^X(\varphi))$, while the \rhs\ yields
(after reversing the direction of the $U_j$-ribbon, which is allowed
owing to the $j$-summation) $\Gamma^\sigma(X,\varphi)$.
\\[5pt]
(iv) The two equalities involving $S^A$ and $\tilde S^A$ can be
demonstrated by moves similar to the ones in (I:5.99). To see the 
equality for $\Gamma^\sigma$ one can start from the first representation 
of $\Gamma^\sigma(X,\varphi)$ in \erf{eq:Ga-RP}. To simplify the pictures 
we will only show a two-dimensional slice of $\RP$ that does not 
intersect the $X$-ribbon. In the representation \erf{eq:RP3-ident}, 
\erf{eq:RP3-pic} of $\RP$ the slice is the set $D\,{:=}\,\{\, (x,y,z) 
\,|\, x^2{+}y^2{=}1, z\iN[0,1] \,\}$.  The set $D$ can be mapped to a 
crosscap with a hole, which we will identify with the set 
$C_\circ\eq\{ x\iN\reals^2 \,|\, |x|\iN[1/2,1] \} /{\sim}$ with 
$x \,{\sim}\, {-}x$ iff $|x|\eq 1$. 

In \erf{eq:Ga-RP} we can deform all the $A$-ribbons to lie within 
the set $D$, except for the final part of the $A$-ribbon that is
attached to the coupon for the morphism $\varphi$;
after mapping $D$ to $C_\circ$ we then obtain
  \bea  \begin{picture}(140,66)(0,65)
  \put(0,0)     {\includeourbeautifulpicture{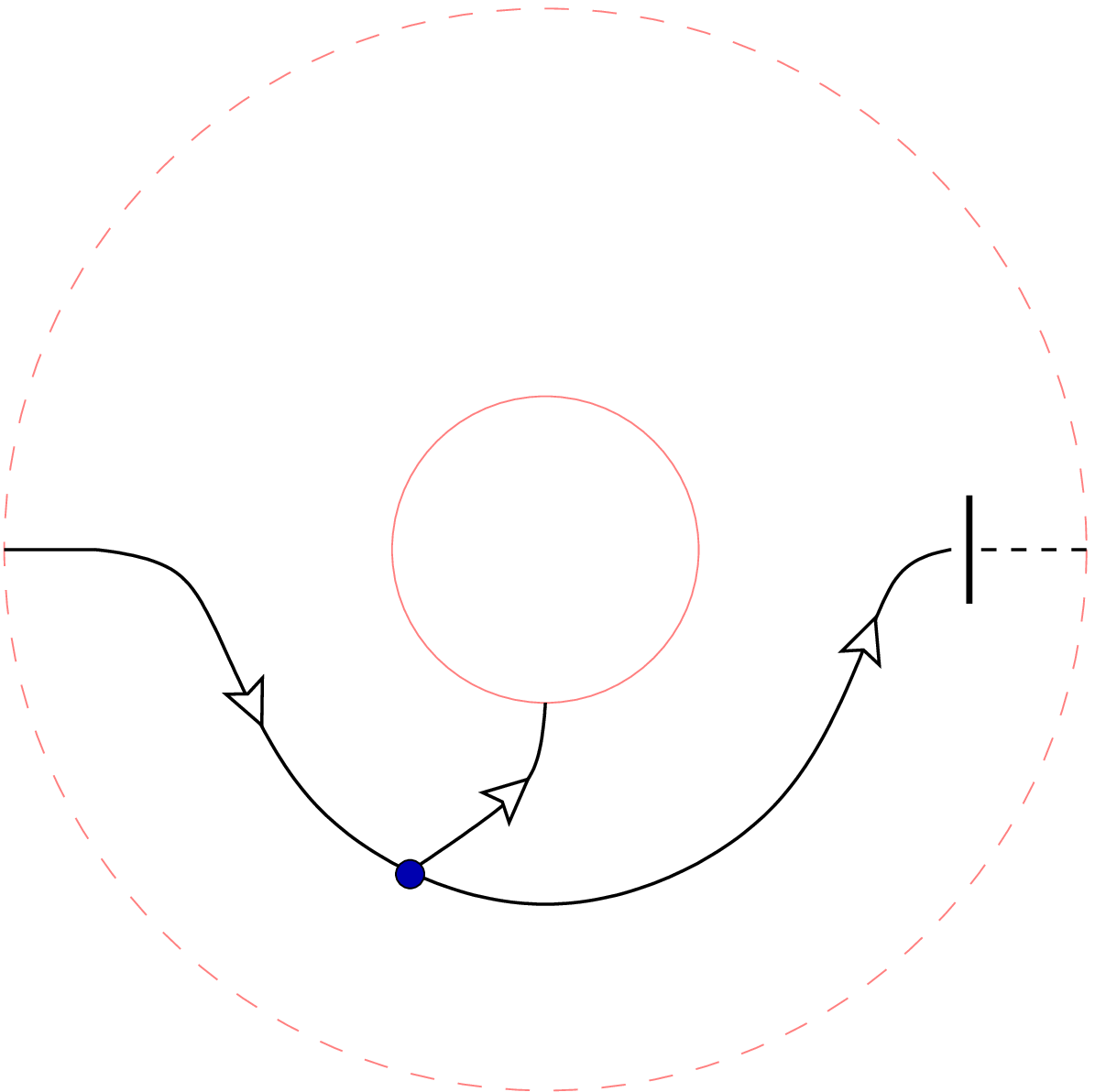}
  \setlength{\unitlength}{24810sp}
      \put(229,61) {\scriptsize$A$}
  \setlength{\unitlength}{1pt}
  }
  \epicture44 \labl{eq:pic132}
Here the dashed circle corresponds to the dashed circle in
\erf{eq:Ga-RP}, and opposite points on this circle are identified.
The $A$-ribbon ending on the solid circle in the center continues
to the morphism $\varphi$ in the full picture \erf{eq:Ga-RP}. In addition,
the morphism $\sigma$ and the half-twist in \erf{eq:Ga-RP} have been
combined to the \boxelement\ \erf{eq:bw-bar}.

In the representation \erf{eq:pic132}, consider now the following moves on
the $A$-ribbons
  \bea  \begin{picture}(400,208)(0,87)
  \put(40,0)     {\includeourbeautifulpicture{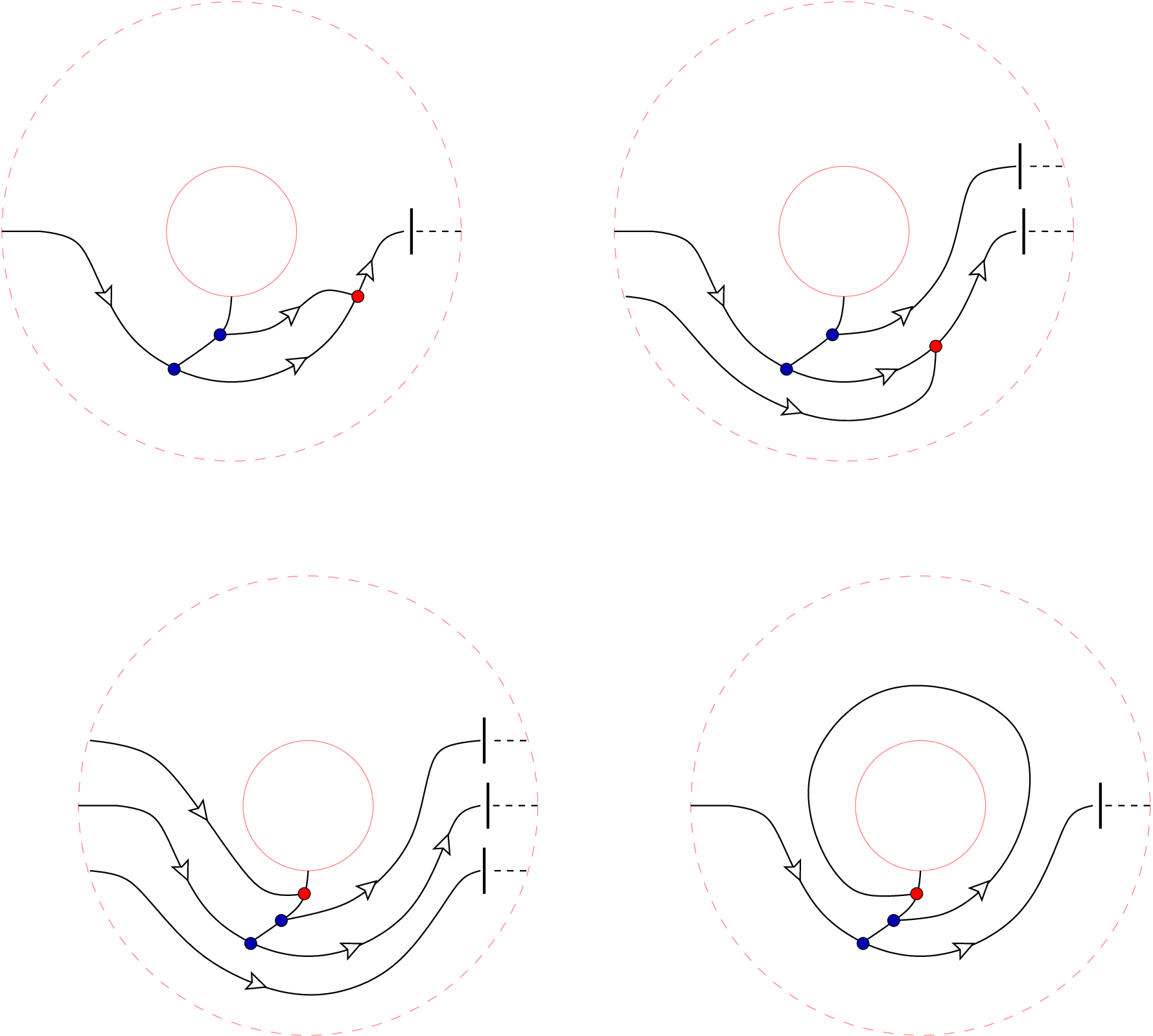}
  \setlength{\unitlength}{24810sp}
      \put(62,593) {\scriptsize$A$}
      \put(526,580) {\scriptsize$A$}
      \put(130,200) {\scriptsize$A$}
      \put(549,175) {\scriptsize$A$}
  \setlength{\unitlength}{1pt}
  }
  \put(36,61)   {$ = $}
  \put(187,224) {$ = $}
  \put(209,61)  {$ = $}
  \epicture64 \labl{eq:pic133}
The \lhs\ is equal to \erf{eq:pic132} by coassociativity and specialness.
In the first step the multiplication morphism is taken past the \boxelement\ 
using \erf{eq:bar-prop} and is then moved further along the $A$-ribbon,
using also the Frobenius property to move it past the comultiplication.
The second step is similar to the previous one. In the third step the lower
segment of the $A$-ribbon is taken through the dotted circle, reemerging
at the top in the way indicated. Now drawing the \rhs\ of \erf{eq:pic133}
in the full picture \erf{eq:Ga-RP}, we see that what we achieved is
precisely to insert a projector $P_X$ in front of $\varphi$.
\qed

As already pointed out in section I:5.4 and
will be described in detail elsewhere, bulk fields are labelled by
local morphisms in the spaces $\Hom(A{\otimes}U_j, U_i^\vee)$. For example,
by lemma I:5.6 the dimension of the local subspace is exactly $Z_{ij}$. 
The quantities $S^A$, $\tilde S^A$ and $\Gamma^\sigma$ are related to
one-point functions on the disk and on the crosscap. 
Proposition \ref{pr:lamda-omega-action}(iv) implies that when we choose an 
eigenbasis of the projector $P_{U_j}$, only `physical' labels give nonzero 
results. Such a basis and its dual were introduced in (I:5.55) and (I:5.56),
denoting the bases of $\Hom(A{\otimes}U_l,U_{\bar k})$ and
$\Hom(U_{\bar k},A{\otimes}U_l)$ by $\{\mu_\alpha^{kl}\}$ and 
$\{\bar\mu_\alpha^{kl}\}$, respectively.
Let us order the basis vectors such that the subsets
  \be
  \{\mu_\alpha^{kl} \,|\, \alpha = 1,\dots,Z_{kl} \}
  \qquad {\rm and} \qquad
  \{\bar\mu_\alpha^{kl} \,|\, \alpha = 1,\dots,Z_{kl} \} \labl{eq:mu-basis}
are dual bases of the subspaces $\Loc(A{\otimes}U_l,U_{\bar k})$ and 
$\LocUp(U_{\bar k},A{\otimes}U_l)$, respectively. With these choices
the relation between $S^A$ and $\tilde S^A$ from 
\erf{eq:SS-def} and the quantities given in (I:5.97) and (I:5.132) is
  \be
  S^A_{\kappa,p\alpha} = S^A(M_\kappa;U_p,\mu_\alpha^{\bar p p} )
  \qquad {\rm and} \qquad
  \tilde S^A_{p\alpha,\kappa}
  = \tilde S^A(U_p,\bar\mu_\alpha^{\bar p p};M_\kappa ) \,.  \ee
It was shown in propositions I:5.16 and I:5.17 that the matrices
$S^A_{\kappa,p\alpha}$ and $\tilde S^A_{p\alpha,\kappa}$ are each
others' inverses. Further we set
  \be
  \Gamma^\sigma_{p\alpha} 
  := \Gamma^\sigma(U_p, \mu_\alpha^{\bar p p}) \qquad {\rm and} \qquad 
  \gamma_{p\alpha}^\sigma := t_p^{-1} \Gamma^\sigma_{p\alpha}
  \labl{eq:Ga-bas}
with $t_p$ the numbers introduced in \erf{eq:tk-def}. The combinations 
$\gamma^\sigma_{p\alpha}$ are introduced because the crossed channel 
amplitudes look a bit simpler when expressed in terms of these combinations. 
They are also used frequently in the literature, since
they arise when working with phase rotated characters 
$\hat{\chii}_k(\tau)\eq \eE^{\pi \ii (\Delta_k - c/24)} \chii_k(\tau)$,
as is convenient for the M\"obius strip amplitude (see e.g.\ \cite{prad}).

The quantitites $S^A_{\kappa,p\alpha}$, $\tilde S^A_{p\alpha,\kappa}$
and $\Gamma^\sigma_{p\alpha}$ can be expressed explicitly in terms of
fusing and braiding matrices. For brevity, as in paper~I we present
these expressions only for the case that both ${N_{ij}}^k\iN\{0,1\}$
and $\dimc\,\Hom(U_k,A) \iN\{0,1\}$ for all $i,j,k\iN\II$. Also recall that
when one of the labels $i,j,k$ is $0$, the choice of basis elements in 
$\Hom(U_i{\otimes}U_j,U_k)$ is prescribed by formula (I:2.33).

The multiplication $m$ and comultiplication $\Delta$ of the algebra $A$ 
are expressed in this basis according to (I:3.7) and (I:3.82).
The reversion on $A$ is given in terms of numbers $\sigma(a)$ 
introduced in formula \erf{eq:*bas}. For the representation
morphism $\rho$ of an $A$-module $M$, the expansion in a basis looks like
in (I:4.61). As a final ingredient, the local morphisms
in $\Hom(A\oti U_l,U_{\bar k})$ and its dual are given in a basis as
  \bea  \begin{picture}(340,51)(0,38)
  \put(0,0)     {\includeourbeautifulpicture{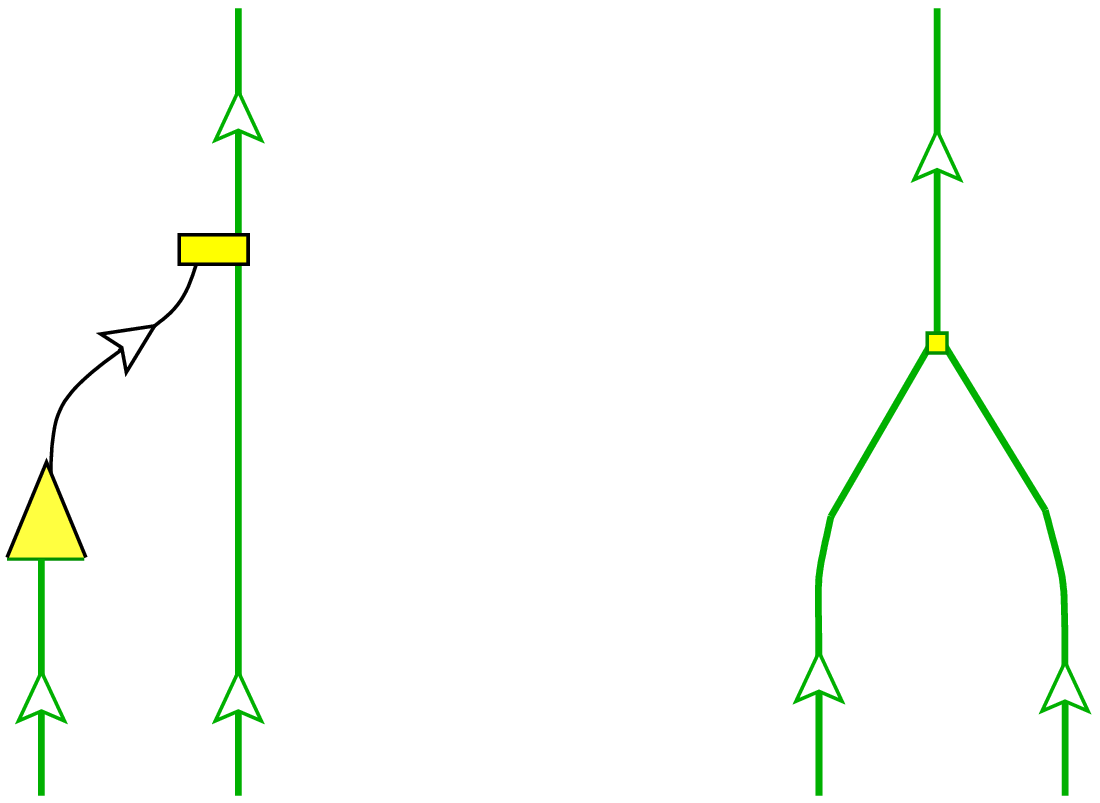}
  \setlength{\unitlength}{24810sp}
      \put(19,4) {\scriptsize$a$}
      \put(77.3,4) {\scriptsize$l$}
      \put(9.5,126) {\scriptsize$A$}
      \put(78,156) {\scriptsize$\alpha$}
      \put(76,213) {\scriptsize$\overline k$}
      \put(278,213) {\scriptsize$\overline k$}
      \put(242,4) {\scriptsize$a$}
      \put(313,4) {\scriptsize$l$}
  \setlength{\unitlength}{1pt}
  }
  \put(230,0)   {\includeourbeautifulpicture{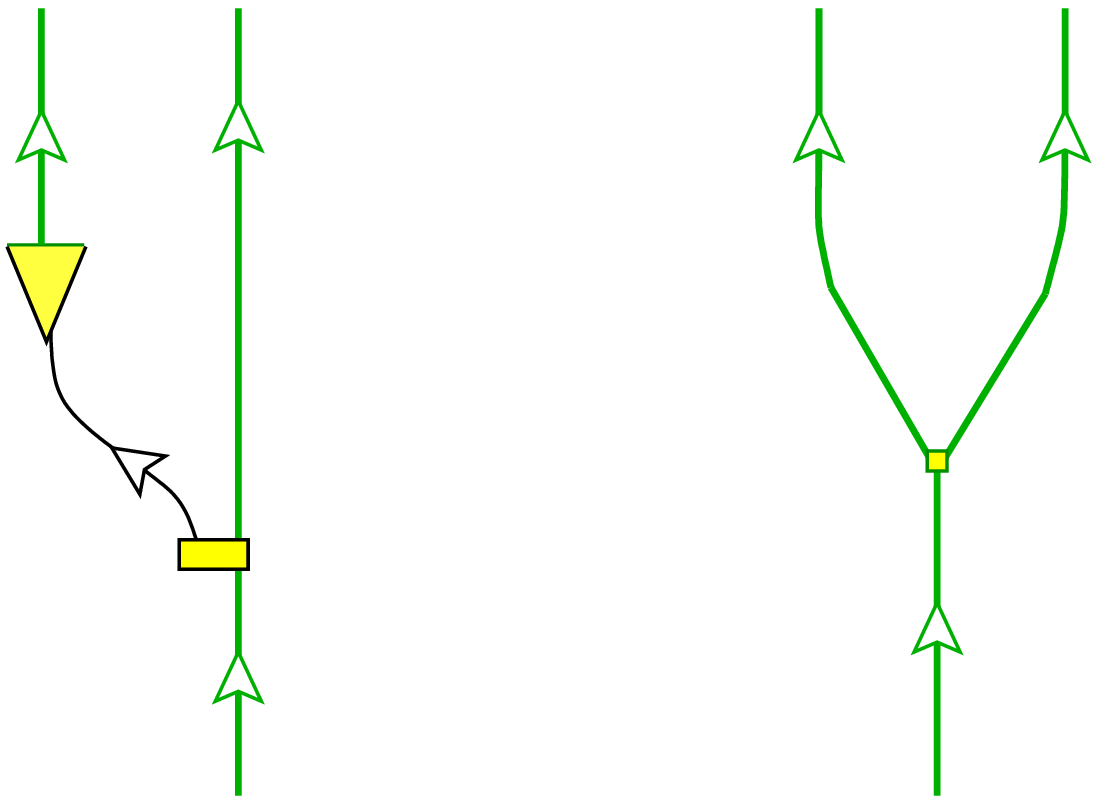}
  \setlength{\unitlength}{24810sp}
      \put(18,213) {\scriptsize$a$}
      \put(76,213) {\scriptsize$l$}
      \put(29.3,109) {\scriptsize$A$}
      \put(77.2,4) {\scriptsize$\overline k$}
      \put(80,66) {\scriptsize$\overline \alpha$}
      \put(243.6,213) {\scriptsize$a$}
      \put(314,213) {\scriptsize$l$}
      \put(278,4) {\scriptsize$\overline k$}
  \setlength{\unitlength}{1pt}
  }
  \put(43,41)   {$ =\; \mu_{\alpha;a}^{kl} $}
  \put(167,41)  {and}
  \put(273,41)  {$ =\; \bar\mu_{\alpha;a}^{kl} $}
  \epicture18 \labl{eq:mu-expand}
The link invariant for $S^A_{\kappa,p\alpha}$ can now be evaluated as
  \bea  \begin{picture}(420,93)(9,70)
  \put(58,0)     {\includeourbeautifulpicture{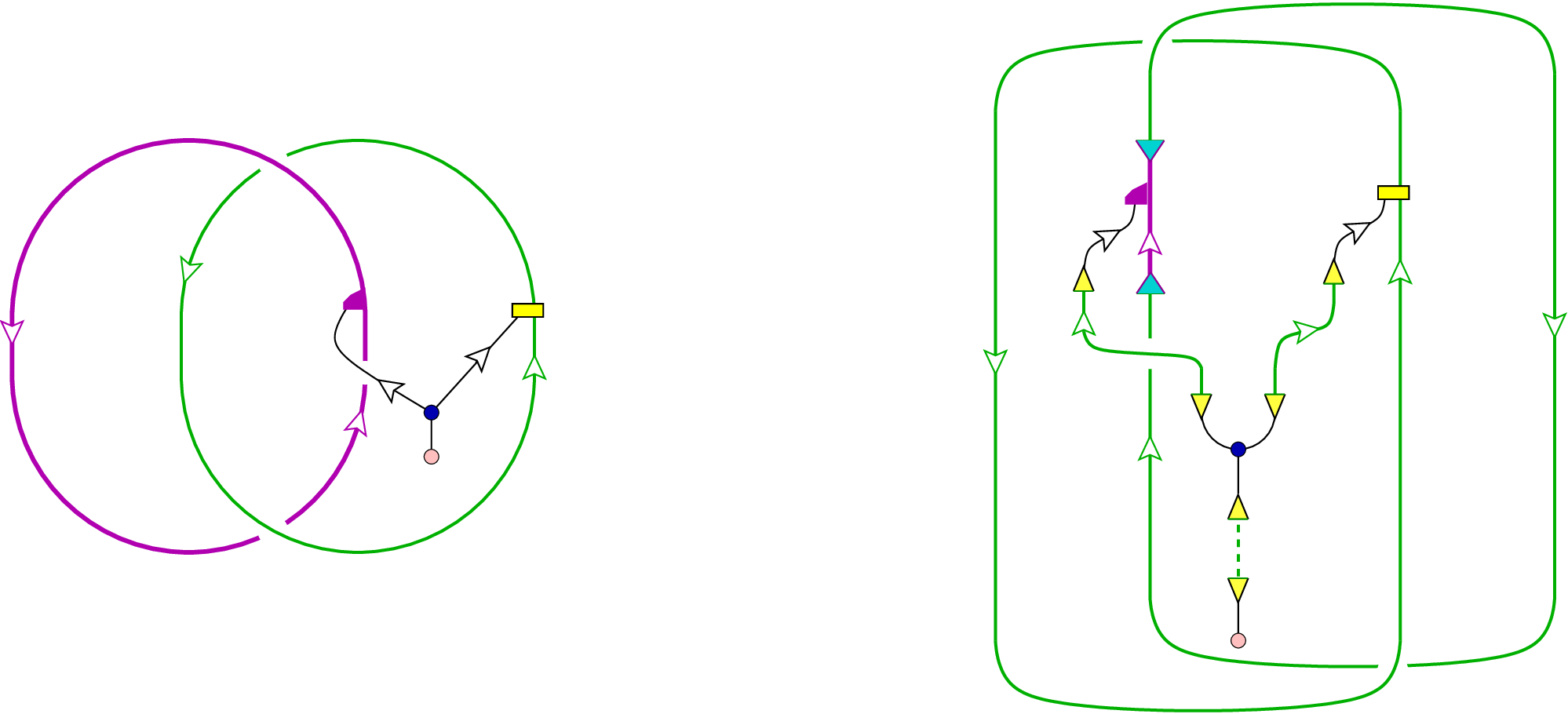}
  \setlength{\unitlength}{24810sp}
      \put(340,240) {\scriptsize$\alpha$}
      \put(316,312) {\scriptsize$p$}
      \put(90,361) {\scriptsize$M_\kappa$}
      \put(274,223) {\scriptsize$A$}
      \put(649,287) {\scriptsize$A$}
      \put(713,303) {\scriptsize$\M_\kappa$}
      \put(828,443) {\scriptsize$m$}
      \put(869,314) {\scriptsize$\alpha$}
      \put(728,225) {\scriptsize$a$}
      \put(770,232) {\scriptsize$\overline a$}
      \put(814,303) {\scriptsize$A$}
      \put(867,117) {\scriptsize$p$}
      \put(721,343) {\scriptsize$\overline\nu$}
      \put(721,263) {\scriptsize$\nu$}
  \setlength{\unitlength}{1pt}
  }
  \put(0,82)    {$ \dsty\frac{S^A_{\kappa,p\alpha}}{S_{00}} \;= $}
  \put(199,82)  {$ =\; \dsty\sum_{(m\nu)\prec M_\kappa}\sum_{a\prec A} $}
  \epicture42 \ee
where the first figure is related to the definition of $S^A$ in 
\erf{eq:SS-def} by a Hopf turn \erf{eq:Hopf-rot} of $M_\kappa$. In the 
second step we have executed a trace flip on $M_\kappa$ and inserted the 
bases (I:3.4) and (I:4.21) so as to decompose $A$ and $M_\kappa$ into 
simple objects.  Recalling also the normalisation convention (I:3.79) for 
the basis of $\Hom(\one,A)$ and $\Hom(A,\one)$, as well as the notation 
$a\IN A$ indicating that $U_a$ is a simple subobject of $A$ (see (I:3.78)) 
and the analogous notation $(m\nu)\IN M$ in which $\nu$ labels
possible multiplicities of $U_m$ in $\M$, we find
  \be
  S^A_{\kappa,p\alpha} = \sum_{a \In A} \sum_{(m\nu)\In M_\kappa}
  \rho_{a,m\nu}^{\kappa\,m\nu}\, \Delta_{0}^{a\bar a}
  \mu_{\alpha;\bar a}^{\bar pp} \, I_S(p,a,m) \labl{eq:SA-inv-1}
with $I_S(p,a,m)$ given by
  \bea  \begin{picture}(420,243)(9,91)
  \put(113,203)  {\includeourbeautifulpicture{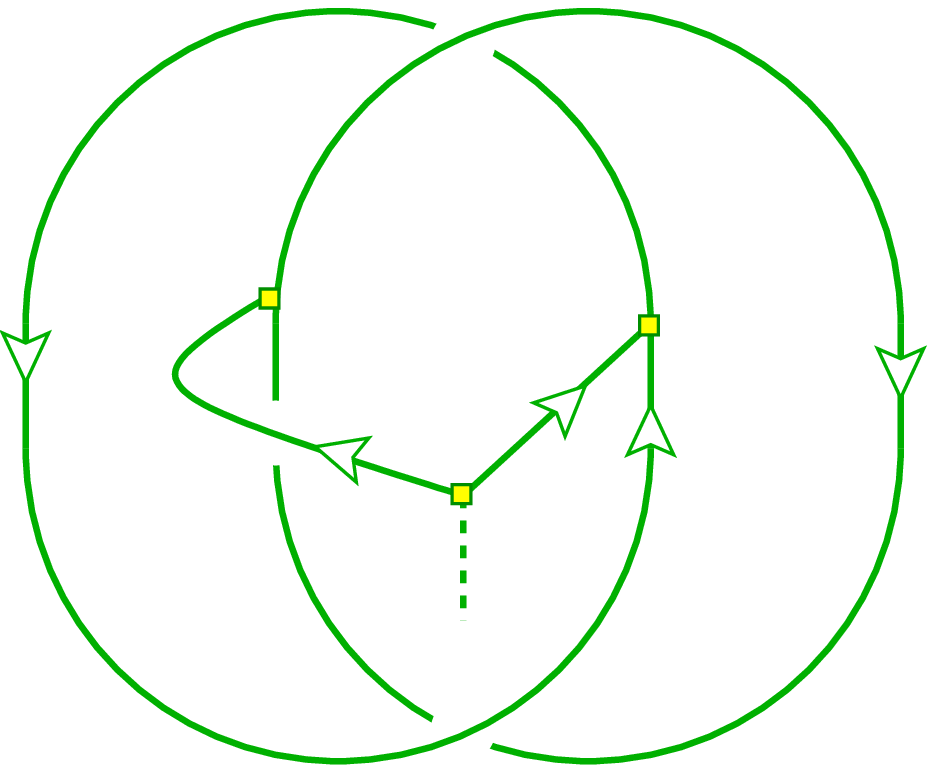}
  \setlength{\unitlength}{24810sp}
      \put(107,101) {\scriptsize$a$}
      \put(150,110) {\scriptsize$\overline a$}
      \put(99,169) {\scriptsize$m$}
      \put(189,165) {\scriptsize$p$}
  \setlength{\unitlength}{1pt}
  }
  \put(290,155)  {\includeourbeautifulpicture{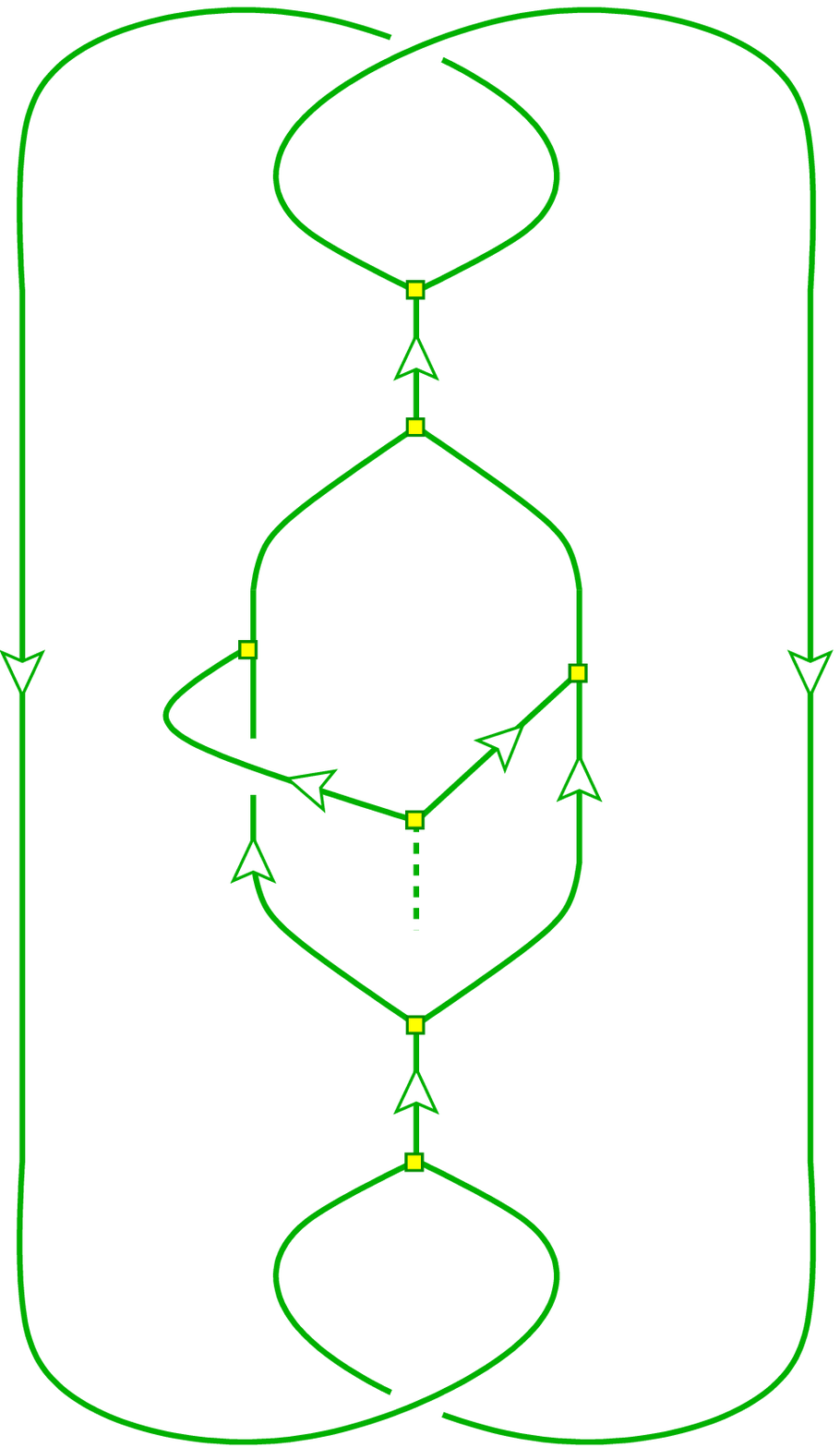}
  \setlength{\unitlength}{24810sp}
      \put(67,414) {\scriptsize$m$}
      \put(191,421) {\scriptsize$p$}
      \put(152,355) {\scriptsize$r$}
      \put(77,314) {\scriptsize$m$}
      \put(185,316) {\scriptsize$p$}
      \put(109,230) {\scriptsize$a$}
      \put(156,243) {\scriptsize$\overline a$}
      \put(79,156) {\scriptsize$m$}
      \put(198,186) {\scriptsize$p$}
      \put(150,113) {\scriptsize$r'$}
      \put(70,66) {\scriptsize$m$}
      \put(190,72) {\scriptsize$p$}
  \setlength{\unitlength}{1pt}
  }
  \put(338,0)    {\includeourbeautifulpicture{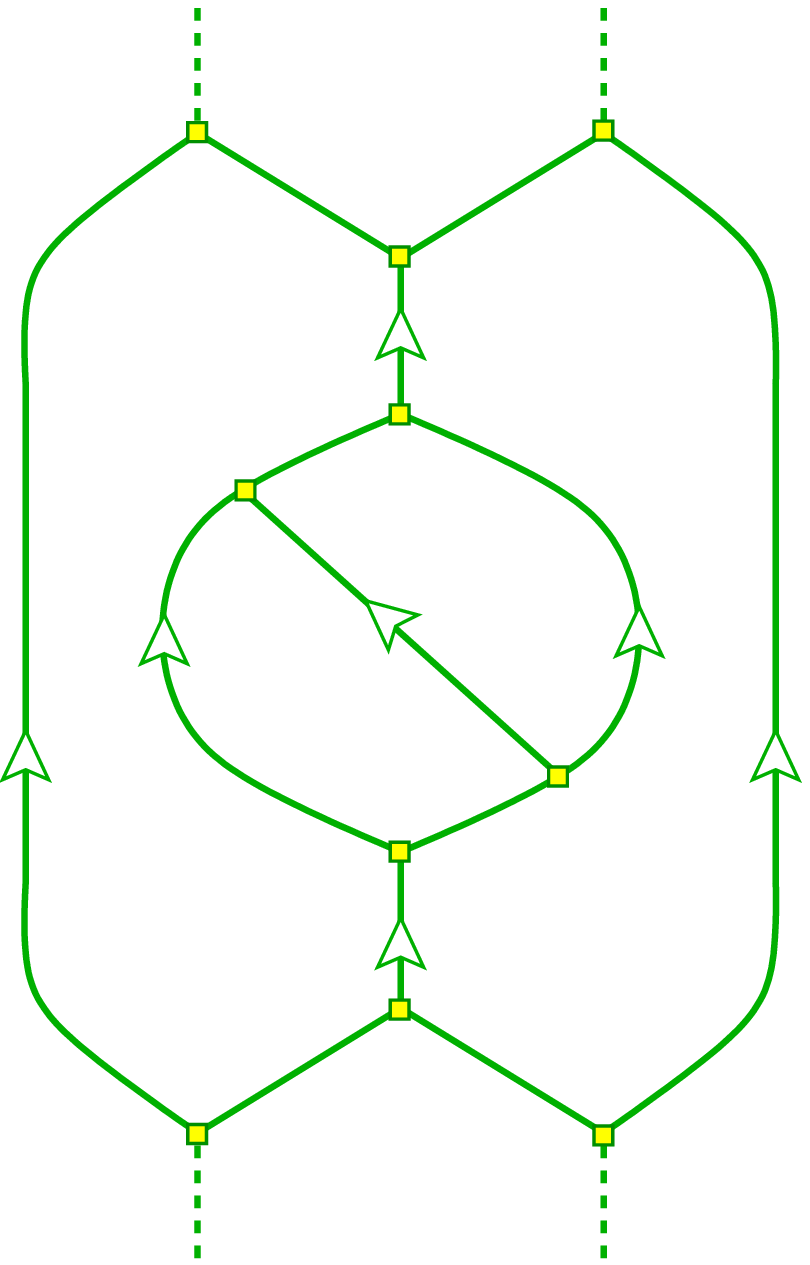}
  \setlength{\unitlength}{24810sp}
      \put(87,319) {\scriptsize$p$}
      \put(131,319) {\scriptsize$m$}
      \put(16,240) {\scriptsize$\overline p$}
      \put(198,259) {\scriptsize$\overline m$}
      \put(75,240) {\scriptsize$m$}
      \put(161,233) {\scriptsize$p$}
      \put(130,182) {\scriptsize$a$}
      \put(73,147) {\scriptsize$m$}
      \put(135,115) {\scriptsize$p$}
      \put(125,262) {\scriptsize$r$}
      \put(125,89) {\scriptsize$r$}
      \put(76,61) {\scriptsize$p$}
      \put(144,63) {\scriptsize$m$}
  \setlength{\unitlength}{1pt}
  }
  \put(0,243)   {$ I_S(p,a,m) / S_{00} \;:= $}
  \put(235,243) {$ =\,\ \dsty\sum_{r,r'}$}
  \put(46,67)   {$ =\; \dsty\sum_r \Rs{}mpr \Rs{}pmr \Rs-mam\, \dim(U_p)\,
                    \dim(U_m)\,\Fs a{\bar a}ppp0 $}
  \epicture69 \labl{eq:SA-calc-1}
In the first step two complete bases of intermediate simple objects
$U_r,U_{r'}$ are inserted. Since a morphism from $U_r$ to $U_{r'}$ is zero
unless $r\eq r'$, this reduces to a single sum over $r$. In the second
step the definitions (I:2.41) of \RR\ and the formula (I:2.60) for \FF\
as well as the relation (I:2.35) between duality and basis morphisms
are inserted. In the last graph in \erf{eq:SA-calc-1} the definition
(I:2.40) of $\GG\,{\equiv}\,\FF^{-1}$ can be substituted. The resulting
graph already appeared in a similar calculation in (I:2.63). Altogether
we find
  \be
  I_S(p,a,m) = \frac{S_{0p}\, S_{0m}}{S_{00}}\, \Rs-mam\, \Fs a{\bar a}ppp0
  \sum_r \frac{\theta_r}{\theta_m\theta_p}\, \Gs maprmp
  \Gs{\bar p}pmm0r\, \Fs{\bar p}pmmr0 \,.  \labl{eq:SA-inv-2}

The analogous calculation for $\tilde S^A_{p\alpha,\kappa}$
and $\Gamma^\sigma_{p\alpha}$ yields
  \be
  \tilde S^A_{p\alpha,\kappa} =
  \sum_{a \In A} \sum_{(m\nu)\In M_\kappa} \rho_{a,m\nu}^{\kappa\,m\nu}
  \,\bar\mu_{\alpha;a}^{\bar pp} \, I_{\tilde S}(p,a,m)  \ee
and
  \be
  \Gamma^\sigma_{p\alpha} =
  \sum_{a,b \In A} [\sigma(a)]^{-1}\, \Delta_a^{ba} \mu^{\bar pp}_{\alpha;b}
  \, I_\Gamma(p,a,b)  \labl{eq:Gamma-inv-1}
with constants $I_{\tilde S}(p,a,m)$ and $I_\Gamma(p,a,b)$ that can be
evaluated in a way similar to \erf{eq:SA-calc-1}, yielding
  \bea  \begin{picture}(340,150)(0,0)
  \put(108,70)    {\includeourbeautifulpicture{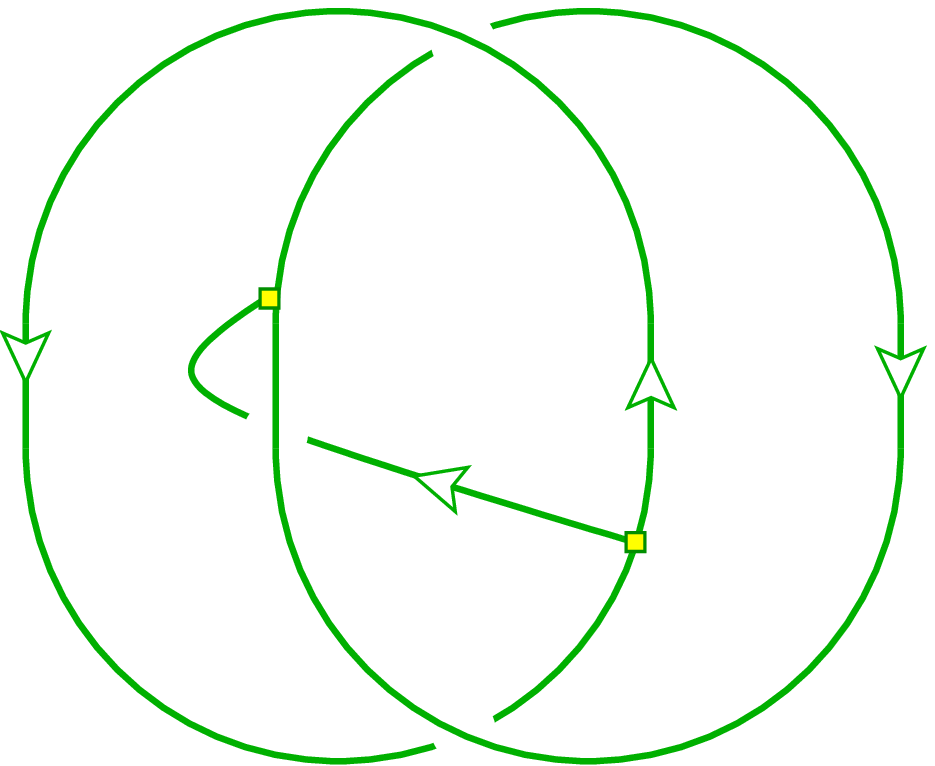}
  \setlength{\unitlength}{24810sp}
      \put(130,90) {\scriptsize$a$}
      \put(194,146) {\scriptsize$p$}
      \put(65,168) {\scriptsize$m$}
  \setlength{\unitlength}{1pt}
  }
  \put(0,109)   {$ I_{\tilde S}(p,a,m) \;=\; S_{00} $}
  \put(61,39)   {$ =\; \dsty\frac{S_{0p}\, S_{0m}}{S_{00}}\,\Rs{}mam
                   \dsty \sum_r \frac{\theta_r}{\theta_m\theta_p}\,
                   \Gs maprmp \Gs{\bar p}pmm0r \,\Fs{\bar p}pmmr0 $}
  \epicture{-2}9 \ee
and
  \begin{eqnarray}  \begin{picture}(420,52)(14,40)
  \put(170,0)    {\includeourbeautifulpicture{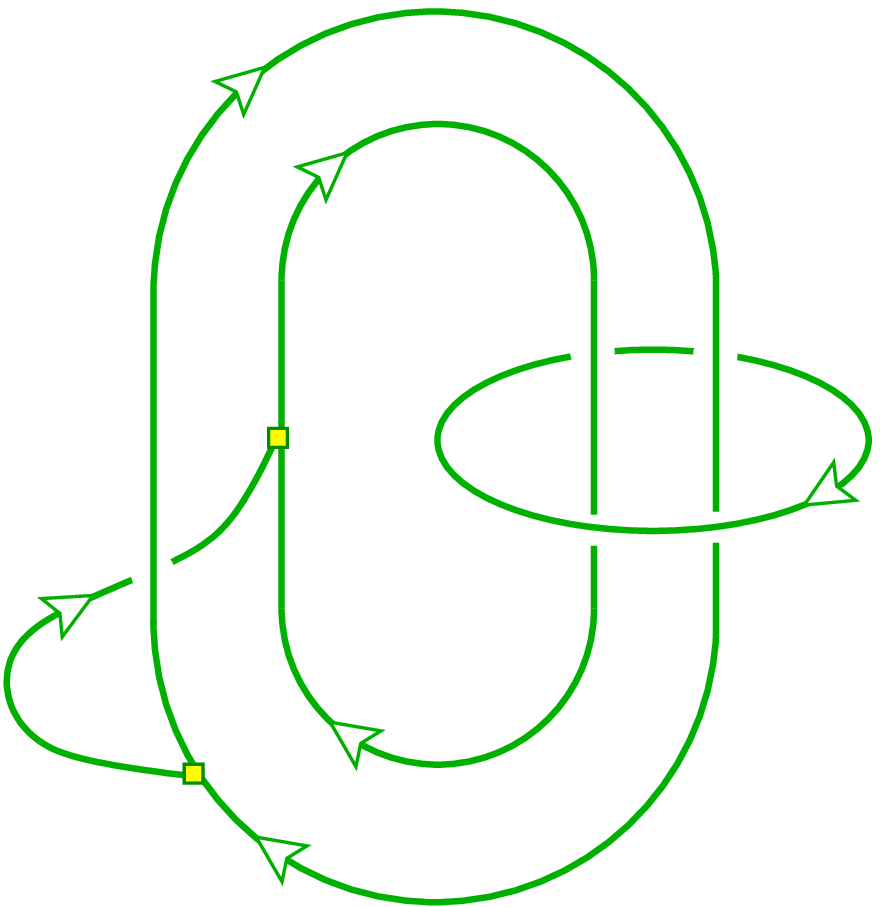}
  \setlength{\unitlength}{24810sp}
      \put(32,194) {\scriptsize$a$}
      \put(88,172) {\scriptsize$p$}
      \put(246,155) {\scriptsize$j$}
      \put(6,94) {\scriptsize$b$}
  \setlength{\unitlength}{1pt}
  }
  \put(0,47)    {$ I_\Gamma(p,a,b) \;=\; \kappa^{-1} S_{00}
                   \dsty\sum_j S_{0j}\, \theta_j^{-2} $}
  \put(280,47)  {$ =\; \dsty \kappa^{-1}\,\Rs-baa \sum_{j,r} S_{0j}
                   \,\theta_j^{-2} S_{jr}\,\Fs abprpa. $}
  \end{picture} \nonumber\\[1.4em]{} \label{eq:Gamma-inv-2}
  \\[-1.3em]{}\nonumber\end{eqnarray}
Here a summation over a complete basis of morphisms with intermediate
object $U_r$ is inserted in the two $U_a\Oti U_p\,$-ribbons. Substituting
(I:2.37) gives rise to the $\FF$-matrix element. What remains
is a Hopf link for a $U_r$- and a $U_j$-ribbon, whose invariant is $s_{j,r}$.

For the computations in the next section we also need to express the
inverse of the map $\Lambda_X^Y$ defined in \erf{eq:lambda-omega-def}
in a basis. We define the matrix $g^{kl}_{\alpha\beta}$ via
  \be
  (\Lambda_{\bar k}^l)^{-1}(\bar \mu^{kl}_\alpha) = \sum_\beta
  g^{kl}_{\alpha\beta} \, \dim(U_k) \, \mu^{\bar l\,\bar k}_\beta \,.
  \labl{eq:mug-def}

As a consequence of the bijections \erf{eq:lam-om-domain},
if $\bar \mu^{kl}_\alpha$ is local, then so is the \lhs\ of \erf{eq:mug-def},
and thus in this case the sum on the \rhs\ can be restricted to local
$\mu^{\bar l\bar k}_\beta$. Also by proposition \ref{pr:lambda-omega-def},
the map $(\Lambda_{\bar k}^l)^{-1}$ is invertible on the local
morphisms, so that $g^{kl}_{\alpha\beta}$ is invertible
as a matrix in $\alpha$,$\beta$.
Composing both sides of \erf{eq:mug-def} with $\bar \mu^{kl}_\beta$
and using the description \erf{eq:lam-om-tilde}
of the inverse of $\Lambda_X^Y$ we find
  \bea  \begin{picture}(230,44)(0,27)
  \put(132,0)     {\includeourbeautifulpicture{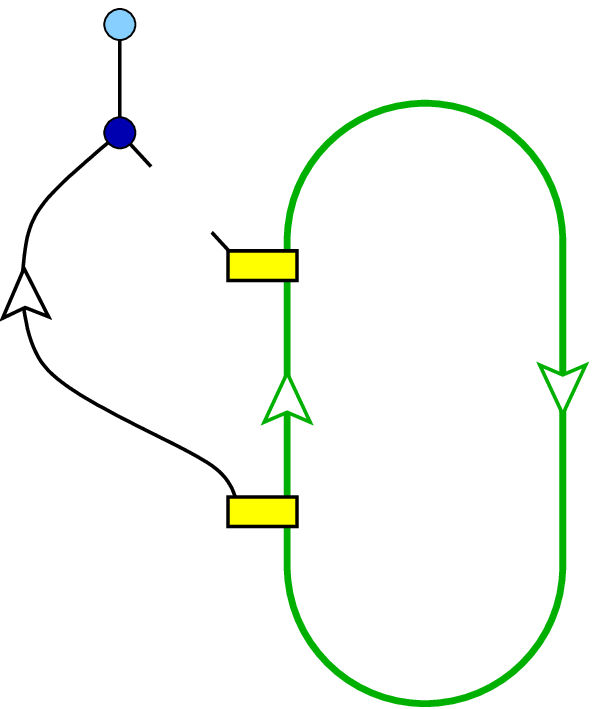}
  \setlength{\unitlength}{24810sp}
      \put(93,91) {\scriptsize$\overline k$}
      \put(169,116) {\scriptsize$l$}
      \put(92,124) {\scriptsize$\alpha$}
      \put(92,55) {\scriptsize$\beta$}
      \put(42,142) {\scriptsize$\sigma^{\!-\!1}$}
      \put(32,90) {\scriptsize$A$}
  \setlength{\unitlength}{1pt}
  }
  \put(0,33)    {$ \dim(U_k)\, \dim(U_l) \, g^{kl}_{\alpha\beta} \;= $}
  \epicture04 \labl{eq:gab-graph}
Taking in this graph the morphism $\bar \mu^{kl}_\alpha$ around the
$U_l$-loop results in a graph that
(using symmetry of $A$, as well as \erf{eq:sig-phi}) can be deformed into
the one for $\dim(U_l)\,\dim(U_k)\,g^{\bar l \bar k}_{\beta\alpha}$.
Thus we get the relation
  \be
  g^{kl}_{\alpha\beta} = g^{\bar l\,\bar k}_{\beta\alpha} \,,  \labl{eq:g-symm}
and in particular
  \be
  g^{\bar k k}_{\alpha\beta} = g^{\bar k k}_{\beta\alpha} \,.  \ee
Evaluating the invariant \erf{eq:gab-graph} in the case
$\N ijk \iN \{0,1\}$ and $\lr{U_k}{A} \iN \{0,1\}$ we find
  \bea  \begin{picture}(380,53)(0,27)
  \put(271,0)     {\includeourbeautifulpicture{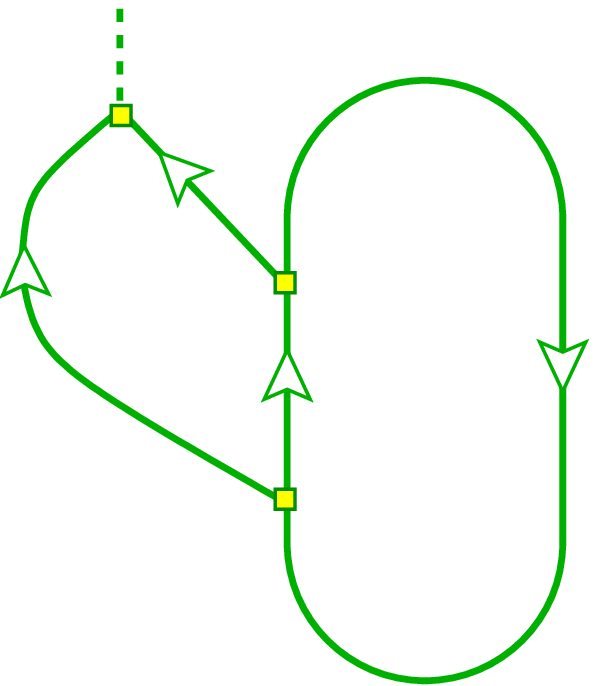}
  \setlength{\unitlength}{24810sp}
      \put(47.7,121.5) {\scriptsize$a$}
      \put(17,65) {\scriptsize$\overline a$}
      \put(93,80) {\scriptsize$\overline k$}
      \put(169.5,40) {\scriptsize$l$}
  \setlength{\unitlength}{1pt}
  }
  \put(16,33)    {$ g^{kl}_{\alpha\beta} \;=\; \dsty\sum_{a\In A}
                   \frac{\dim(A)}{\dim(U_k)\,\dim(U_l)}
                   \,m_{\bar a\,a}^{\;\; 0} \,\frac{1}{\sigma(a)}\,
                   \bar\mu_{\alpha,a}^{kl}
                    \,\bar\mu_{\beta,\bar a}^{\bar l\,\bar k} $}
  \epicture12 \labl{pic113}
(The factor $\dim(A)$ appears when inserting a basis of morphisms (I:3.4) in
front of the counit and using (I:3.79) with $\beta_\one\eq\dim(A)$.)
With the help of the identities (I:2.60) this can be rewritten as
  \be
  g^{kl}_{\alpha\beta} = \sum_{a\In A}
  \dim(A)\, \frac{m_{\bar a\,a}^{\;\; 0}}{\sigma(a)} \,
  \bar\mu_{\alpha,a}^{kl} \, \bar\mu_{\beta,\bar a}^{\bar l\,\bar k} \,
  \frac{\Gs{\bar a}all0{\bar k}}{\dim(U_k)} \,.  \labl{eq:g-inv}

\subsection{Crossed channel for M\"obius strip and Klein bottle} \label{CcMK}

In the two previous sections we have gathered the necessary ingredients
to compute the coefficients of the M\"obius strip and Klein bottle
amplitudes in the crossed channel. The purpose of this section is to 
establish compatibility of the M\"obius strip and Klein bottle amplitudes
with the crossed channel, which amounts to the two identities
  \be
  \rmm_{kR} = \frac{1}{S_{00}}\sum_{l\in\II}\sum_{\alpha,\beta=1}^{Z_{\bar l,l}}
  P_{kl}\, S^A_{R,l\alpha}\,
  g^{\bar l\,l}_{\alpha\beta} \; \gamma^\sigma_{l\beta} \labl{eq:M-crossed}
and
  \be
  \K_k =  \frac{1}{S_{00}} \sum_{l\in\II}\sum_{\alpha,\beta=1}^{Z_{\bar l,l}}
  S_{kl}\, \gamma^\sigma_{l\alpha} \, g^{\bar l\,l}_{\alpha\beta} \,
  \gamma^\sigma_{l\beta} \,.  \labl{eq:K-crossed}
Here $P$ is the $P$-matrix of \cite{bisa3}: Defining the matrix $\hat T$ by
  \be
  \hat T_{ij} := \delta_{i,j}\, \theta_i^{-1} \labl{eq:Th-def}
and choosing a square root of $\hat T$ according to 
$(\hat T^{1/2})_{kl}\,{:=}\, \delta_{kl} t_k^{-1}$ with $t_k$ as in 
\erf{eq:tk-def}, the $P$-matrix is given by $P \eq \kappa^{-1} \hat P$, where
  \be
  \hat P_{ij} = (\hat T^{1/2} S \hat T^2 S \hat T^{1/2})_{ij} 
  \labl{eq:Ph-def}
and $\kappa$ is the charge of \calc, i.e.\ the number described 
after \erf{eq:RP-1rib}.

The identities \erf{eq:M-crossed} and \erf{eq:K-crossed} reflect the fact 
that in the crossed channel the M\"obius amplitude $\rmM_R$ can be expressed 
as the inner product of a boundary state and a crosscap state, while the Klein 
bottle amplitude ${\rm K}$ is given by the inner product of two crosscap 
states.
This amounts to a consistency condition between the closed string spectrum
and the coefficients $\rmm_{kR}$ and ${\rm K}_k$ which is satisfied when
the identities \erf{eq:M-crossed} and \erf{eq:K-crossed} hold
\cite{scst,sosc}.
In our approach the relations \erf{eq:M-crossed} and \erf{eq:K-crossed} are
deduced, rather than postulated, i.e.\ the consistency between closed string
spectrum and M\"obius strip and Klein bottle amplitudes is satisfied
automatically.
Also recall that the annulus coefficients provide us with a NIM-rep
of the fusion rules; NIM-reps which obey the consistency condition just
mentioned, as well as the integrality properties (\ref{eq:m-prop}(iii))
and (\ref{eq:K-prop}(iii)), have been termed `U-NIMreps' in \cite{sosc}.

\medskip

In section \ref{sec:moebius} we have obtained the three-manifold and
ribbon graph ${\rmM}_R$ for the M\"obius strip amplitude. We denote the
boundary $\partial_+{\rmM}_R$ by $Y$; by construction, $Y$ is a torus.

A convenient basis in the space $\calh(\emptyset ; Y)$ of zero-point conformal 
blocks on the M\"obius strip with boundary condition $Y$ can be selected as 
follows. We start from the basis $|\chii_k;\rmT\rangle$ in (I:5.15).
Denoting the three-manifold in (I:5.15) by $M_{\chiI_k}^+$ 
(in section I:5.2, the notation $M_1$ was used instead), by definition
$|\chii_k;\rmT\rangle\eq Z(M_{\chiI_k}^+,\emptyset,\rmT) 1$. Similarly, for 
the dual basis one chooses the three-manifold $M_{\chiI_k}^-$ as in (I:5.18)
(where it was called $M_2$) and puts 
$\langle\chii_k;\rmT|\eq Z(M_{\chiI_k}^-,\rmT,\emptyset)\iN 
\calh(\emptyset;\rmT)^*_{}$.  
The basis in $\calh(\emptyset;\rmT)$ can be transported to 
$\calh(\emptyset;Y)$ by specifying a homeomorphism
$f{:}\;\rmT\,{\to}\,Y$. Let $C_Y\,{:=}\,Y{\times}\,[0,1]$ be the cylinder 
over $Y$. For a homeomorphism
$g{:}\; \partial_+M_2\,{\to}\,\partial_-M_1$, denote by $M_1{\glue g}M_2$ 
the three-manifold obtained by glueing $\partial_+M_2$ to $\partial_-M_1$ 
using $g$. Then $Z(C_Y{\glue f}M_{\chiI_k}^+,\emptyset,Y)1$, with $k$
ranging over $\I$, provides a basis of $\calh(\emptyset;Y)$. 

An essential point below will be to compare two bases obtained by 
different choices of $f$. On the torus $T$ fix a homology basis of 
cycles $a,b$ such that the $a$-cycle is contractible in $M_{\chiI_k}^+$ 
and the $b$-cycle is the cycle that is obtained when moving the $U_k$-ribbon
in $M_{\chiI_k}^+$ to the boundary and reversing the direction. In terms of 
the picture (I:5.15), the $a$-cycle winds counter-clockwise horizontally
around the torus, while the $b$-cycle runs vertically from bottom to top,
as is indicated in figure (\ref{eq:pic99}\,a) below.

We now introduce two homeomorphisms $f_{a,b}{:}\;\rmT\,{\to}\,Y$, whose
homotopy class is determined via the images of the $a$- and $b$-cycle in 
$Y$; figure (\ref{eq:pic99}\,b) indicates the action of $f_{a}$, while 
(\ref{eq:pic99}\,c) shows the action of $f_{b}$:
  \bea  \begin{picture}(420,46)(12,59)
  \put(23,0)      {\includeourbeautifulpicture{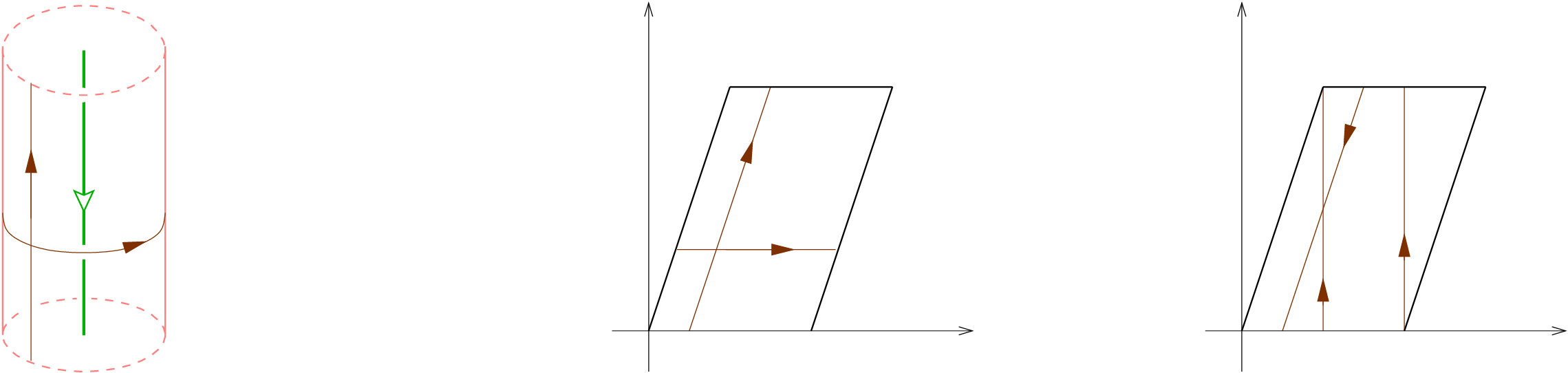}
  \setlength{\unitlength}{24810sp}
      \put(10.5,163) {\scriptsize$b$}
      \put(95,99) {\scriptsize$a$}
      \put(64.5,147) {\scriptsize$k$}
      \put(533.3,156) {\scriptsize$b$}
      \put(559.8,93.4) {\scriptsize$a$}
      \put(932.2,70) {\scriptsize$a$}
      \put(891.7,49) {\scriptsize$b$}
      \put(989.3,141) {\scriptsize$a$}
  \setlength{\unitlength}{1pt}
  }
  \put(-5,90)   { a) }
  \put(160,90)  { b) }
  \put(320,90)  { c) }
  \epicture41  \labl{eq:pic99}
We denote the two corresponding bases of $\calh(\emptyset;Y)$ by
  \be
  |a_k;Y\rangle := Z(C_Y{\glue{f_a}}M_{\chiI_k}^+,\emptyset,Y) 1
  \qquad {\rm and} \qquad
  |b_k;Y\rangle := Z(C_Y{\glue{f_b}}M_{\chiI_k}^+,\emptyset,Y) 1
  \,, \labl{eq:X-basis}
respectively. This way we get two expansions of the vector
$Z({\rmM}_R,\emptyset,Y) 1 \iN \calh(\emptyset;Y)$ (as before, by abuse of 
notation we will use the symbol ${\rmM}_R$ also for this vector),
  \be
  {\rmM}_R = \sum_{l\in\II} {\rmM}_{lR}\, |a_l;Y\rangle
  = \sum_{l\in\II}\widetilde{\rmM}_{lR}\, |b_l;Y\rangle \,.
  \labl{eq:MR-2bases}
The basis $\{|a_l;Y\rangle\}$ is chosen such that the coefficients 
${\rmM}_{lR}$ appearing here are precisely those given
in formula \erf{eq:MkR}. The coefficients $\widetilde{\rmM}_{lR}$ in the 
basis $\{|b_l;Y\rangle\}$ will be computed below. 

First, however, let us derive the relation between the two expansions.
One quickly checks that the bases dual to \erf{eq:X-basis} are given by
  \be
  \langle a_k;Y| = Z(M_{\chiI_k}^-{\glue{f_a^{-1}}}C_Y,Y,\emptyset) 
  \qquad {\rm and} \qquad
  \langle b_k;Y| = Z(M_{\chiI_k}^-{\glue{f_b^{-1}}}C_Y,Y,\emptyset)  
  \,, \ee
respectively. It then follows from \erf{eq:MR-2bases} that
  \be
  {\rmM}_{kR} = \sum_{l\in\II}\langle a_k;Y|b_l;Y\rangle\, \widetilde{\rmM}_{lR}
  \,.  \labl{eq:MkR-CkR}
By construction, the inner products $\langle a_k;Y|b_l;Y\rangle$ are ribbon
invariants:
  \be
  \langle a_k;Y|b_l;Y\rangle = 
  Z( M_{\chiI_k}^-\!{\glue{f_a^{-1}\circ f_b}}\! M_{\chiI_l}^+ , 
  \emptyset , \emptyset )1 \,.  \labl{eq:MR-bas-xfer}
More generally, let us represent the $a$-cycle by $\ppvector 10$ and
the $b$-cycle by $\ppvector 01$, and let $g{:}\; T\,{\to}\,T$ be an invertible 
homeomorphism acting on the homology basis as $\ppmatrix p q r s $, with
$ps{-}qr\eq1$. Then we are interested in the matrix $M(g)$ with entries
  \be
  M(g)^{}_{kl} = M\!\left[\ppmatrix p q r s \right]_{kl} := 
  Z( M_{\chiI_k}^-\!{\glue{g}}M_{\chiI_l}^+ ,\emptyset ,\emptyset )1 \,.  \ee
The matrices $M$ form an $|\I|$-dimensional projective representation of the 
modular group SL$(2,\zet)$. One way to compute them is to combine
  \be
  M\!\left[\ppmatrix 0 {-1} 1 0 \right]_{kl} = S_{kl}  \qquad{\rm and}\qquad
  M\!\left[\ppmatrix 1   1  0 1 \right]_{kl}
                             = \hat T_{kl} = \delta_{kl}^{}\,\theta_k^{-1} 
  \ee
with the recursion relations
  \bea
  M\!\left[\ppmatrix 1 1 0 1  \ppmatrix a b c d \right]_{kl}
    = \sum_r \hat T_{kr}\,  M\!\left[\ppmatrix a b c d \right]_{rl} ,
  \\{}\\[-.5em]
  M\!\left[\ppmatrix 0 {-1} 1 0  \ppmatrix a b c d \right]_{kl}
    = \kappa^{-{\rm sign}(ac)}
  \sum_r S_{kr}\, M\!\left[\ppmatrix a b c d \right]_{rl}
  .  \eear\labl{eq:rec-rel}
A detailed derivation of these relations is given in appendix
\ref{sec:app-tor}. 

Note that the composition $f_a^{-1}{\cir}\,f_b$ in \erf{eq:MR-bas-xfer} 
takes the cycle $b$ to $-b$ and $a$ to $-a{+}2b$.
Applying \erf{eq:rec-rel} to \erf{eq:MR-bas-xfer} we thus compute
  \be
  \langle a_k;Y|b_l;Y\rangle = M\left[\ppmatrix {-1} 0 2 {-1} \right]_{kl} 
  = \kappa^{-1} (S \hat T^2 S)_{kl} \,. \ee

After working out the matrix describing the change of basis in \erf{eq:MkR-CkR} 
we proceed to evaluate the link invariant $\widetilde {\rmM}_{kR}$. To 
construct the three-manifold we start from a different fundamental domain for 
the torus, namely the shaded region in the first of the following pictures:
  \bea  \begin{picture}(350,74)(0,55)
  \put(15,0)      {\includeourbeautifulpicture{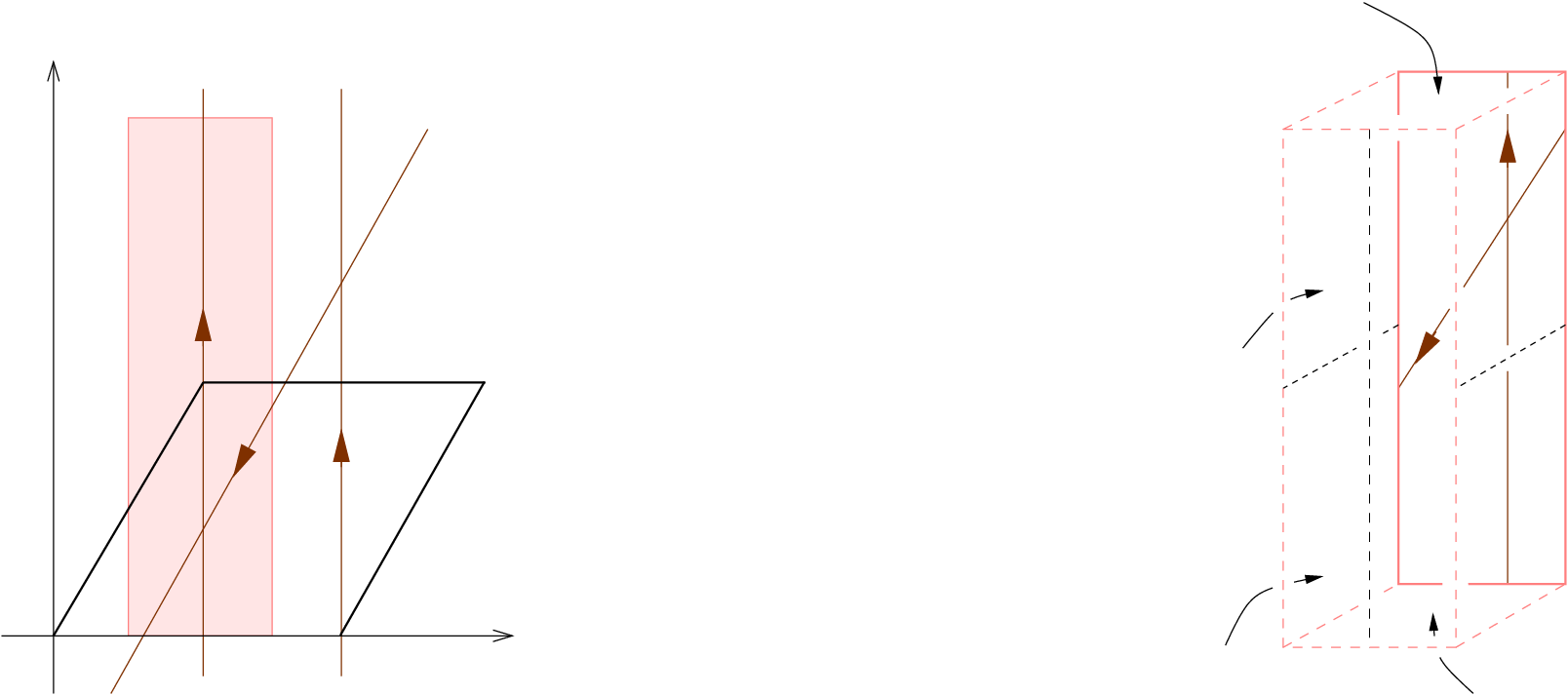}
  \setlength{\unitlength}{24810sp}
      \put(106,232) {\scriptsize$a$}
      \put(176,101) {\scriptsize$a$}
      \put(199,236) {\scriptsize$b$}
      \put(636,256) {\scriptsize$A$}
      \put(699,256) {\scriptsize$A'$}
      \put(654,333) {\scriptsize$B$}
      \put(597,153) {\scriptsize$D'$}
      \put(590,2) {\scriptsize$C'$}
      \put(735,-9) {\scriptsize$B'$}
      \put(723,34) {\scriptsize$D$}
      \put(730,17) {\tiny$t{=}0$}
      \put(778,51) {\tiny$t{=}1$}
      \put(724,264) {\scriptsize$C$}
      \put(700,182) {\scriptsize$b$}
      \put(749,194) {\scriptsize$a$}
  \setlength{\unitlength}{1pt}
  }
  \put(317,84)  {  }
  \epicture34  \labl{eq:pic100}
The connecting manifold ${\rmM}_R$ is constructed as in \erf{eq:MX-def}. 
A fundamental domain of ${\rmM}_R$ is shown in the second of the pictures 
\erf{eq:pic100}. This picture is to be understood as follows. The top and 
bottom faces are to be identified periodically, $B \,{\sim}\, B'$; among 
the lateral faces one must identify $C\,{\sim}\, C'$ and $D\,{\sim}\, D'$; 
on the face at $t\eq0$ the left half $A$ is to be identified with the right 
half $A'$ via reflection about the dashed line.
The face at $t\eq1$ is the actual boundary of ${\rmM}_R$ (a torus), and
we have indicated on it the images of the $a$- and $b$-cycles.
After implementing the identification $A\,{\sim}\,A'$ explicitly, ${\rmM}_R$
is described by the first of the following pictures:
  \bea  \begin{picture}(380,67)(0,61)
  \put(24,0)      {\includeourbeautifulpicture{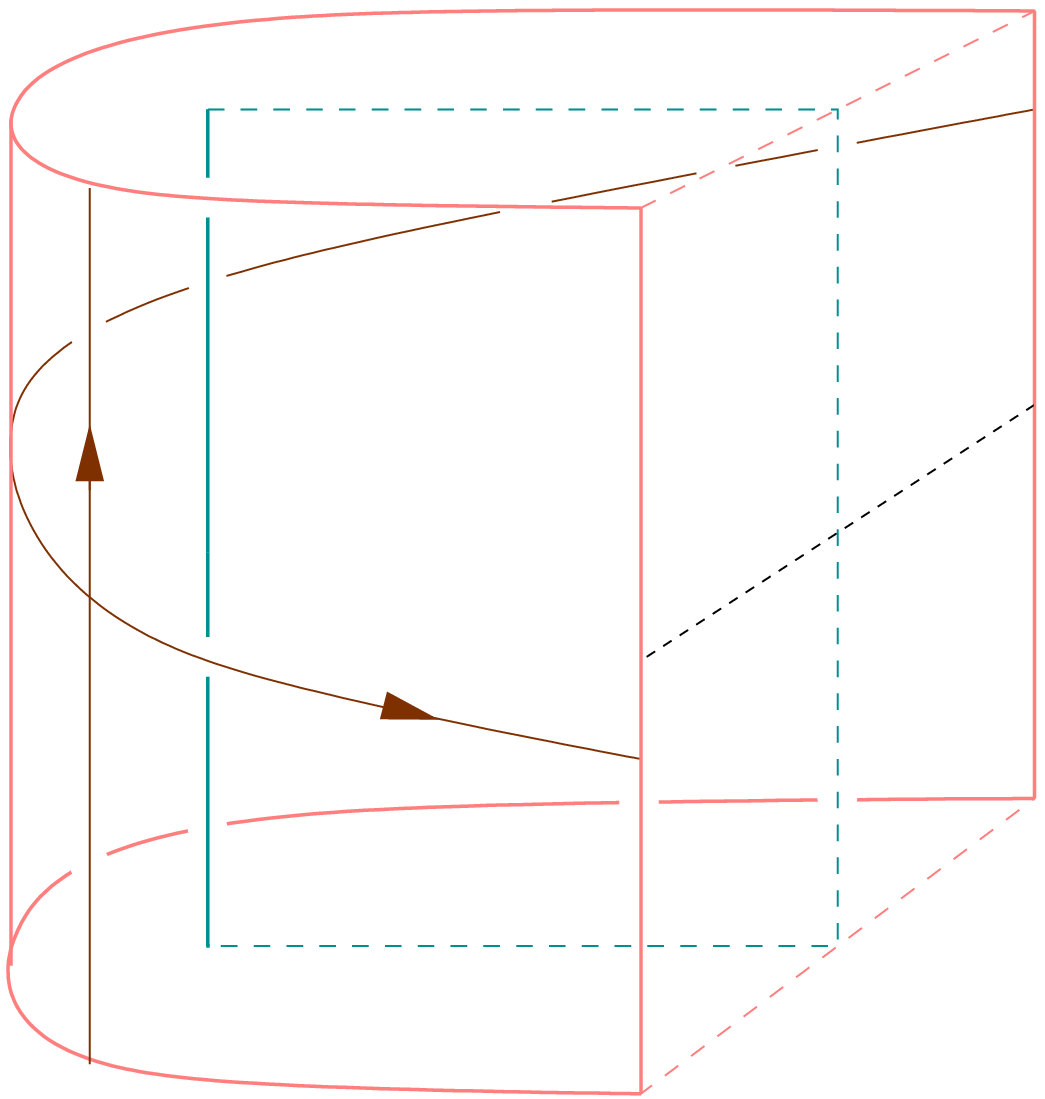}
  \setlength{\unitlength}{24810sp}
      \put(34,187) {\scriptsize$a$}
      \put(123,119) {\scriptsize$b$}
      \put(191,239) {\scriptsize$D'$}
      \put(279,287) {\scriptsize$C$}
      \put(189,17) {\scriptsize$C'$}
      \put(273,82) {\scriptsize$D$}
      \put(710,41) {\scriptsize$\dot R$}
      \put(772,158) {\scriptsize$A$}
      \put(820,174) {\scriptsize$A$}
  \setlength{\unitlength}{1pt}
  }
  \put(-5,110)  { a) }
  \put(250,91)  { b) }
  \epicture38  \labl{eq:pic101}
Here again the $a$- and $b$-cycles are drawn on the boundary
$\partial_+{\rmM}_R$; the dashed lines show where the world sheet is to be 
embedded in ${\rmM}_R$, with the left vertical line indicating its boundary. 
Let us stress that this picture describes the
same three-manifold as \erf{eq:Moebius2}, though this is not immediately 
obvious from the pictures.  The second figure in \erf{eq:pic101} represents 
the world sheet, i.e.\ the M\"obius strip. Here top and bottom are
identified while for the right boundary -- thinking of it as an $S^1$
of circumference $2\pi$ -- we have $\phi \sim \phi + \pi$, i.e.\ a crosscap.
This picture also contains the ribbon graph to be embedded in ${\rmM}_R$;
one can convince oneself that this ribbon graph is equivalent to the one
in \erf{eq:pic31}.

To obtain the invariant for $\widetilde{\rmM}_{kR}$ via
\erf{eq:MR-2bases}, we must evaluate $\langle b_k ; Y | {\rmM}_R$. In terms
of the left figure in \erf{eq:pic101}, this amounts to glueing the solid
torus (I:5.18) along the boundary in accordance with the location of the
cycles. The resulting three-manifold is an $\RP$. In order to reproduce
the ribbon graph including all twists it is helpful to draw ribbons instead
of the reduced blackboard framing notation. One finds
  \bea  \begin{picture}(240,116)(0,91)
  \put(60,0)      {\includeourbeautifulpicture{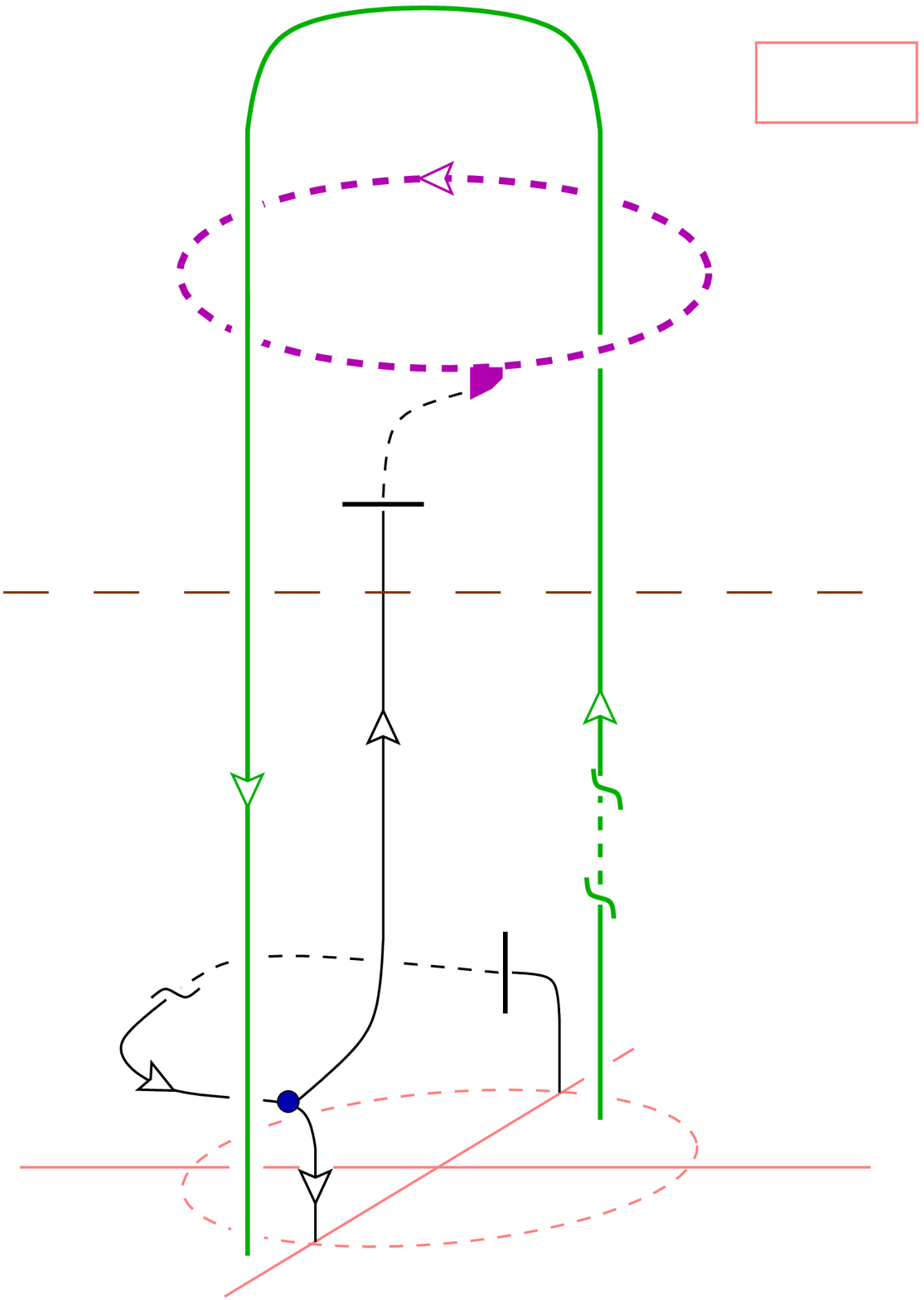}
  \setlength{\unitlength}{24810sp}
      \put(327,500.7) {$\RP$}
      \put(167,198) {\scriptsize$A$}
      \put(148,477) {\scriptsize$\dot R$}
      \put(110,250) {\scriptsize$k$}
      \put(258,321) {\scriptsize$k$}
      \put(31,97) {\scriptsize$A$}
  \setlength{\unitlength}{1pt}
  }
  \put(-10,100)   {$ \widetilde {\rmM}_{kR} \;= $}
  \epicture60  \labl{eq:pic102}
In the representation of $\RP$ that we use here, horizontal sections
(like the one indicated by the long dashes) correspond to $S^2$'s. To 
proceed we cut \erf{eq:pic102} along the dashed line and insert the identity
  \bea  \begin{picture}(320,36)(0,61)
  \put(195,0)      {\includeourbeautifulpicture{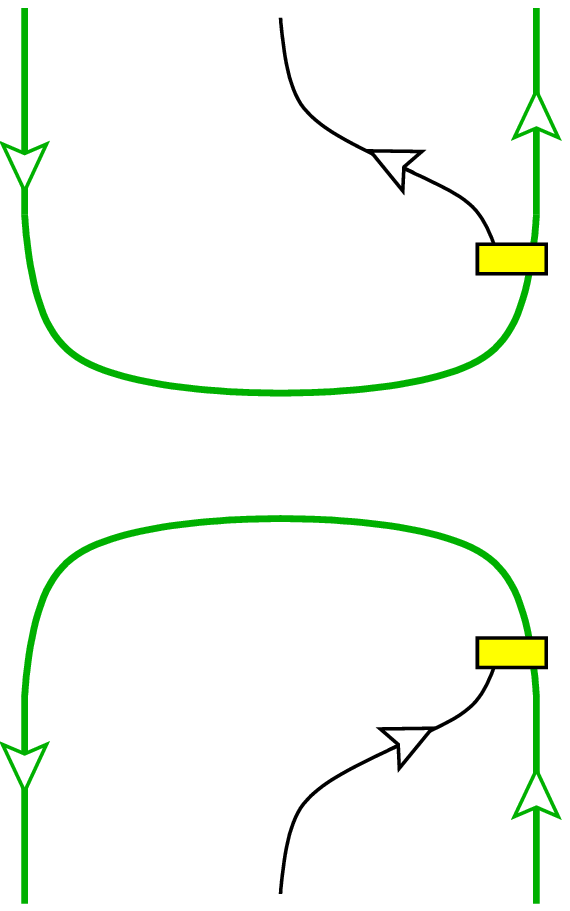}
  \setlength{\unitlength}{24810sp}
      \put(165,185) {\scriptsize$\overline\alpha$}
      \put(164,234) {\scriptsize$k$}
      \put(15,181) {\scriptsize$k$}
      \put(13,58) {\scriptsize$k$}
      \put(165,72) {\scriptsize$\alpha$}
      \put(164,34) {\scriptsize$k$}
      \put(75,211) {\scriptsize$A$}
      \put(75,36) {\scriptsize$A$}
  \setlength{\unitlength}{1pt}
  }
  \put(0,50)    {$ \id_{\calH((k,-),A,k;S^2)} \;=\; \dsty
                   \frac{1}{S_{00}\,\dim(U_k)}\; \sum_\alpha $}
  \epicture36  \labl{eq:kAk-id}
This is an identity of endomorphisms of $\calH((k,-),A,k;S^2)$. The
\rhs\ consists of two three-manifolds, which are solid
three-balls: one of them contains the top component of the ribbon graph, 
and the other one the bottom component.  The boundary of this 
(disconnected) three-manifold is given by $S^2\,{\sqcup}\,S^2$. As usual, 
we implicitly apply the functor $(Z,\calh)$, so that the \rhs\ is an 
endomorphism of $\calH((k,-),A,k;S^2)$, too.
The factor $S_{00}\,\dim(U_k)$ is the invariant of an $S^3$ with an
embedded $U_k$-loop, which appears when one verifies that \erf{eq:kAk-id} 
obeys $\id \cir \id\eq\id$. This leads to the following decomposition
of the invariant \erf{eq:pic102}:
  \bea  \begin{picture}(420,59)(9,57)
  \put(150,0)      {\includeourbeautifulpicture{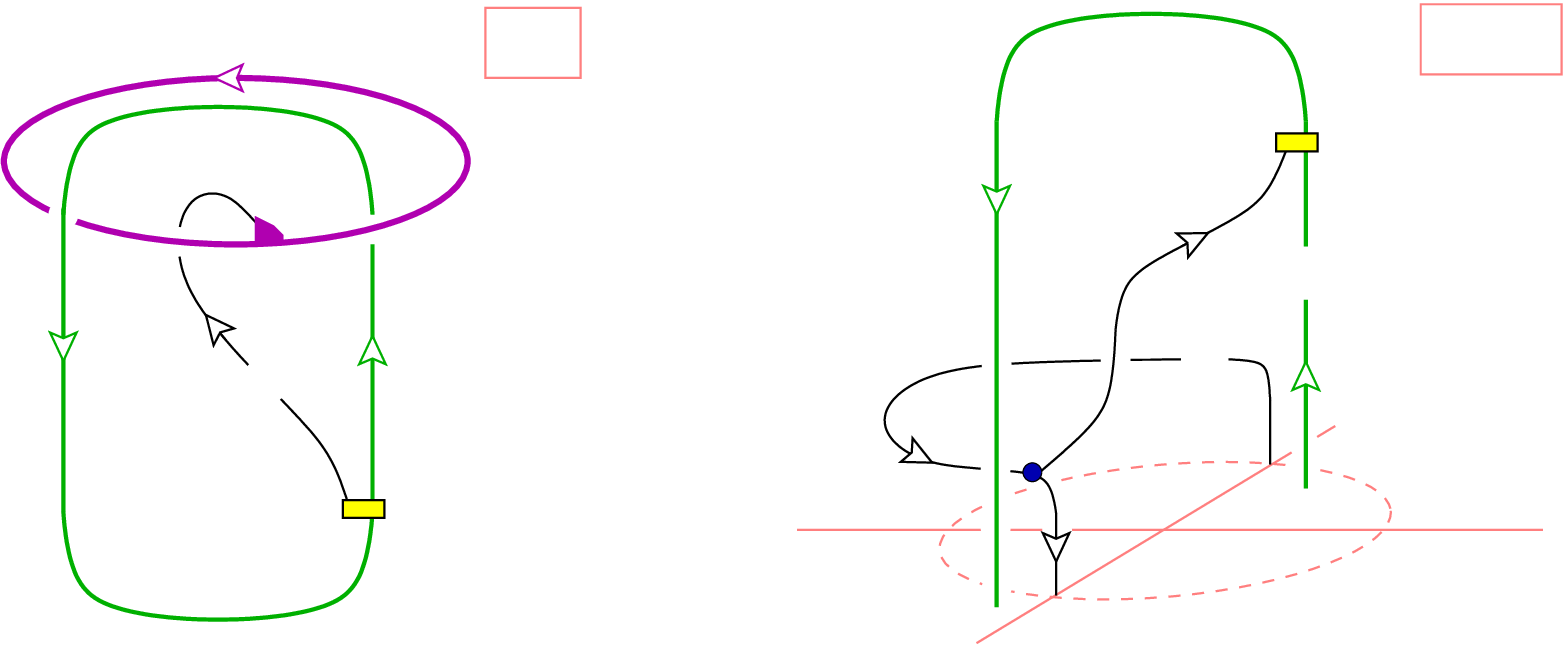}
  \setlength{\unitlength}{24810sp}
      \put(240,279) {$S^3$}
      \put(137,279) {\scriptsize$\dot R$}
      \put(121,120) {\scriptsize$\sigma$}
      \put(37,81) {\scriptsize$k$}
      \put(192,60) {\scriptsize$\overline\alpha$}
      \put(77,148) {\scriptsize$A$}
      \put(689,279.5) {$\RP$}
      \put(571,131) {\scriptsize$\sigma$}
      \put(618,169) {\scriptsize$\theta^{-1}$}
      \put(533,184) {\scriptsize$A$}
      \put(636,235) {\scriptsize$\alpha$}
      \put(483,247) {\scriptsize$k$}
      \put(404,99) {\scriptsize$A$}
  \setlength{\unitlength}{1pt}
  }
  \put(0,53)    {$ \widetilde {\rmM}_{kR} \;=\; \dsty
                   \frac{1}{S_{00}\,\dim(U_k)} \sum_\alpha $}
  \epicture37  \ee
By comparison with the results \erf{eq:SA-StA}, \erf{eq:SS-def} and
\erf{eq:rho-sigma}, the ribbon graph in $S^3$ can be recognised to be
$\tilde S^A(U_k,\bar\mu^{\bar kk}_\alpha ; R^\sigma)$.
The ribbon graph in $\RP$ can be deformed so as to become equal to
$\theta_k^{-1} \Gamma^\sigma( U_k, \mu^{\bar kk}_\alpha)$. To see this,
rotate the points at which the $A$-ribbon touches the identification plane
by $90^\circ$ counter-clockwise, and the $U_k$-ribbons
by the same amount clockwise, keeping in mind the subtlety \erf{eq:pic106}.
Up to the twist $\theta_k^{-1}$, one then indeed obtains the first
picture in \erf{eq:Ga-RP}.

With the help of \erf{eq:SA-StA} and \erf{eq:mug-def},
the $\tilde S^A$-factor can be rewritten as
  \be
  \tilde S^A(U_k,\bar\mu^{\bar kk}_\alpha ; R^\sigma)
  = S^A((R^\sigma)^\sigma ; U_k , (\Lambda_k^k)^{-1}(\bar\mu^{\bar k k}_\alpha))
  = \sum_\beta \dim(U_k) \, g^{\bar k k}_{\alpha\beta} \,
  S^A((R^\sigma)^\sigma ; U_k , \mu^{\bar k k}_\beta)) \,.  \ee
Recalling from \erf{eq:rho-sigsig} that $(R^\sigma)^\sigma \,{\cong}\, R$
as $A$-modules, it follows that altogether we have
  \be
  \widetilde {\rmM}_{kR} = \frac{1}{S_{00}} \sum_{\alpha,\beta} \,
  g^{\bar k k}_{\alpha\beta} \, S^A_{R,k\beta} \, \theta_k^{-1} \,
  \Gamma^\sigma_{k\alpha} \,.  \ee
The relation between ${\rmM}_{kR}$ and $\widetilde {\rmM}_{kR}$ thus becomes
\be
  {\rmM}_{kR} = \sum_l \kappa^{-1} (S \hat T^2 S)_{kl} \, \widetilde{\rmM}_{lR}
  =   \frac{1}{S_{00}}
  \sum_{l,\alpha,\beta} \kappa^{-1} \, t_k \, \hat P_{kl} \, t_l 
  \, \theta_l^{-1} \, S^A_{R,l\alpha} \, g^{\bar l\, l}_{\alpha\beta} \,
  \Gamma^\sigma_{l\beta} \,.  \ee
Using the definitions \erf{eq:m-def} and \erf{eq:Ga-bas} and the relations
\erf{eq:Th-def} and \erf{eq:Ph-def}, the above equation gives the relation 
\erf{eq:M-crossed}. The summation range for $\alpha$ and $\beta$
in \erf{eq:M-crossed} (and likewise in \erf{eq:K-crossed}) follows from 
proposition \ref{pr:lamda-omega-action}(iv) together with the convention 
\erf{eq:mu-basis}.

\medskip

Let us now turn to the corresponding calculation for the Klein bottle, 
which we present in less detail. The cobordism $\K$ 
for the Klein bottle amplitude was constructed in \erf{eq:K-ribbon}. Denote 
the boundary $\partial_+\K$ by $Y$, which is again a 2-torus. 
The expansion \erf{eq:K-exp} uses the basis 
$|a_k;Y\rangle \iN \calh(\emptyset;Y)$ that is obtained by
transporting the basis $|\chii_k;T\rangle$ from $\calh(\emptyset;T)$ 
using a map $f_a{:}\; T \,{\to}\, Y$ acting on cycles as indicated
in the figure on the \lhs\ of \erf{eq:pic107}, while the basis 
$|b_k;Y\rangle$ which we want to expand in below is given by transporting
$|\chii_k;T\rangle$ with a map $f_b{:}\; T \,{\to}\,Y$ acting on cycles as
in the figure on the \rhs\ of \erf{eq:pic107}:
  \bea  \begin{picture}(420,47)(0,40)
  \put(26,0)      {\includeourbeautifulpicture{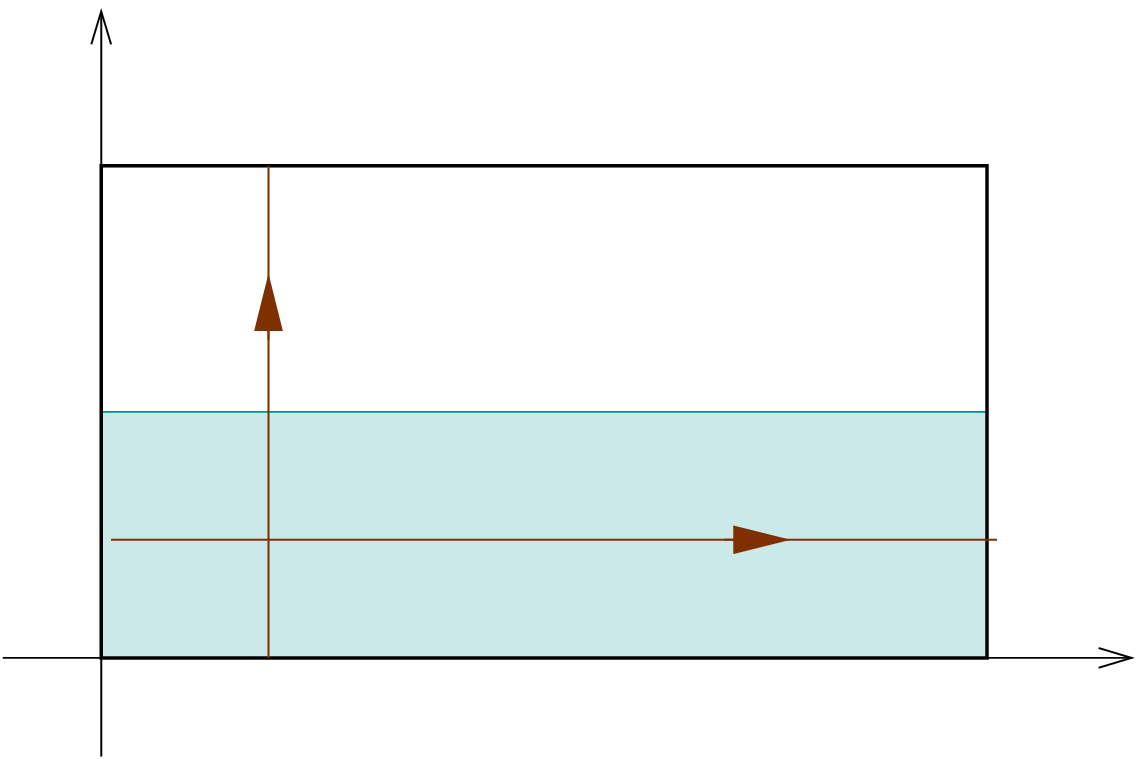}
  \setlength{\unitlength}{24810sp}
      \put(85,134) {\scriptsize$b$}
      \put(170,70) {\scriptsize$a$}
      \put(-4,166) {\scriptsize$2\ii s$}
      \put(6,96) {\scriptsize$\ii s$}
      \put(279,8) {\scriptsize$1$}
      \put(636,166) {\scriptsize$2\ii s$}
      \put(794,5) {\scriptsize$\frac12$}
      \put(725,124) {\scriptsize$a$}
      \put(832,72) {\scriptsize$b$}
      \put(921,8) {\scriptsize$1$}
  \setlength{\unitlength}{1pt}
  }
  \put(0,74)    { a) }
  \put(243,74)  { b) }
  \epicture25  \labl{eq:pic107}
The two expansions of $\K \iN \calh(\emptyset;Y)$ read
  \be
  \K = \sum_{l\in\II}\K_l\, |a_l;Y\rangle
  = \sum_{l\in\II} \tilde\K_l\, |b_l;Y\rangle \,.  \ee
This defines implicitly the coefficients $\tilde\K_l$ which we will evaluate as a link
invariant. Before doing so, note that by the same argument as above the coefficients
$\K_k$ and $\tilde \K_l$ are related by the matrix $M[\,\cdot\,]_{kl}$. 
But this time the two bases differ just by an $S$-transformation, so that
  \be
  \K_k = \sum_{l\in\II} M\!\left[ \ppmatrix 0 {-1} 1 0 \right]_{kl} \tilde \K_l
  =  \sum_{l\in\II} S_{kl}\, \tilde \K_l \,.  \labl{eq:K-Kt}
In the left figure of \erf{eq:pic107} the shaded area indicates the base 
that was used in \erf{eq:Klein} to give a fundamental domain of the 
connecting manifold $\K$. Here we use a different realisation
of the same manifold $\K$, with the shaded
region in the right figure of \erf{eq:pic107} as a base, above
which we take the intervals $[-1,1]$. The resulting manifold is shown in the
first of the following figures:
  \bea  \begin{picture}(380,47)(0,40)
  \put(33,0)      {\includeourbeautifulpicture{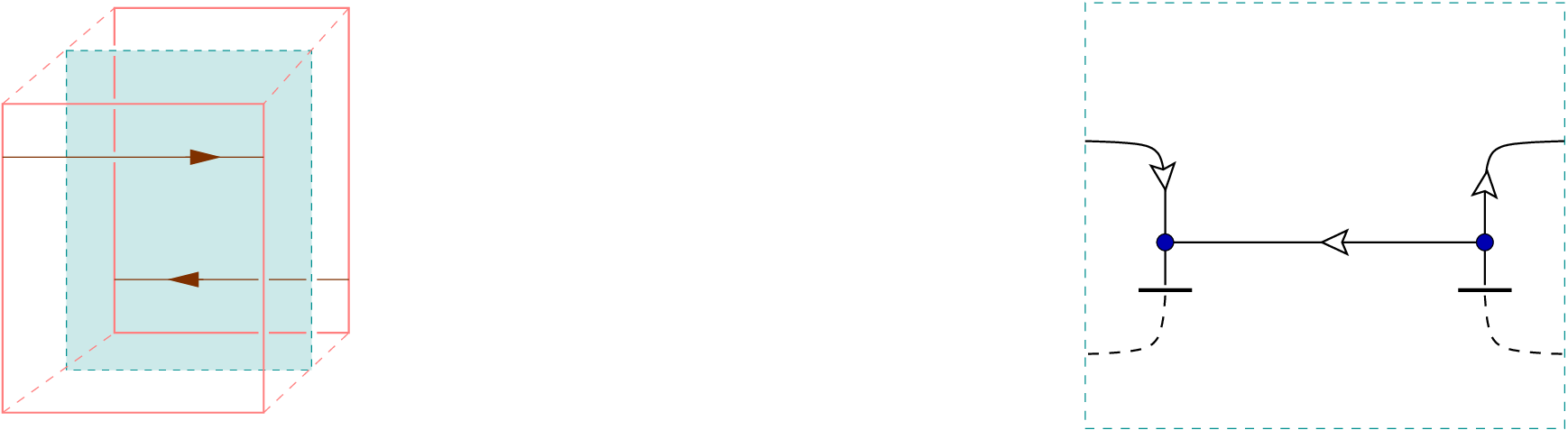}
  \setlength{\unitlength}{24810sp}
      \put(94,195) {\scriptsize$A$}
      \put(26,112) {\scriptsize$D$}
      \put(161,112) {\scriptsize$C$}
      \put(86,27) {\scriptsize$A'$}
      \put(145,0) {\tiny$t{=}{-}1$}
      \put(172,22) {\tiny$t{=}0$}
      \put(193,45) {\tiny$t{=}1$}
      \put(99,124) {\scriptsize$b$}
      \put(89,88) {\scriptsize$b$}
      \put(665,107) {\scriptsize$A$}
      \put(602,160) {\scriptsize$A$}
      \put(799,160) {\scriptsize$A$}
  \setlength{\unitlength}{1pt}
  }
  \put(0,74)    { a) }
  \put(218,74)  { b) }
  \epicture25  \labl{eq:pic108}
In this figure, the top and bottom faces are identified periodically, 
$A\,{\sim}\,A'$. On the face $C$ points are identified via 
$(x{=}1/2,y,t)\,{\sim}\, (x{=}1/2,y{+}s,-t)$,
and similarly on the face $D$, $(x{=}0,y,t)\,{\sim}
    $\linebreak[0]$%
(x{=}0,y{+}s,-t)$.
The front and back faces of the figure constitute the boundary $Y$ of $\K$. The 
position of the $b$-cycle from the right figure of \erf{eq:pic107} is indicated.
The second of the figures \erf{eq:pic108} shows the world sheet; the place at
which it is embedded in the manifold in the first figure is indicated by the
shaded square, which is the plane $t\eq0$. Also shown in the second figure is
the ribbon graph that one has on the world sheet; it can be
seen to be equivalent to the one in \erf{eq:Klein2}.

The ribbon graph for the coefficient $\tilde\K_k\eq\langle b_k ; Y |\K$
is obtained by glueing the graph for $\langle \chii_k ; \rmT|$ in (I:5.18)
to the boundary $Y$ of $\K$ as prescribed by the cycles in the left figure
of \erf{eq:pic108}. One finds
  \bea  \begin{picture}(290,93)(0,66)
  \put(50,0)      {\includeourbeautifulpicture{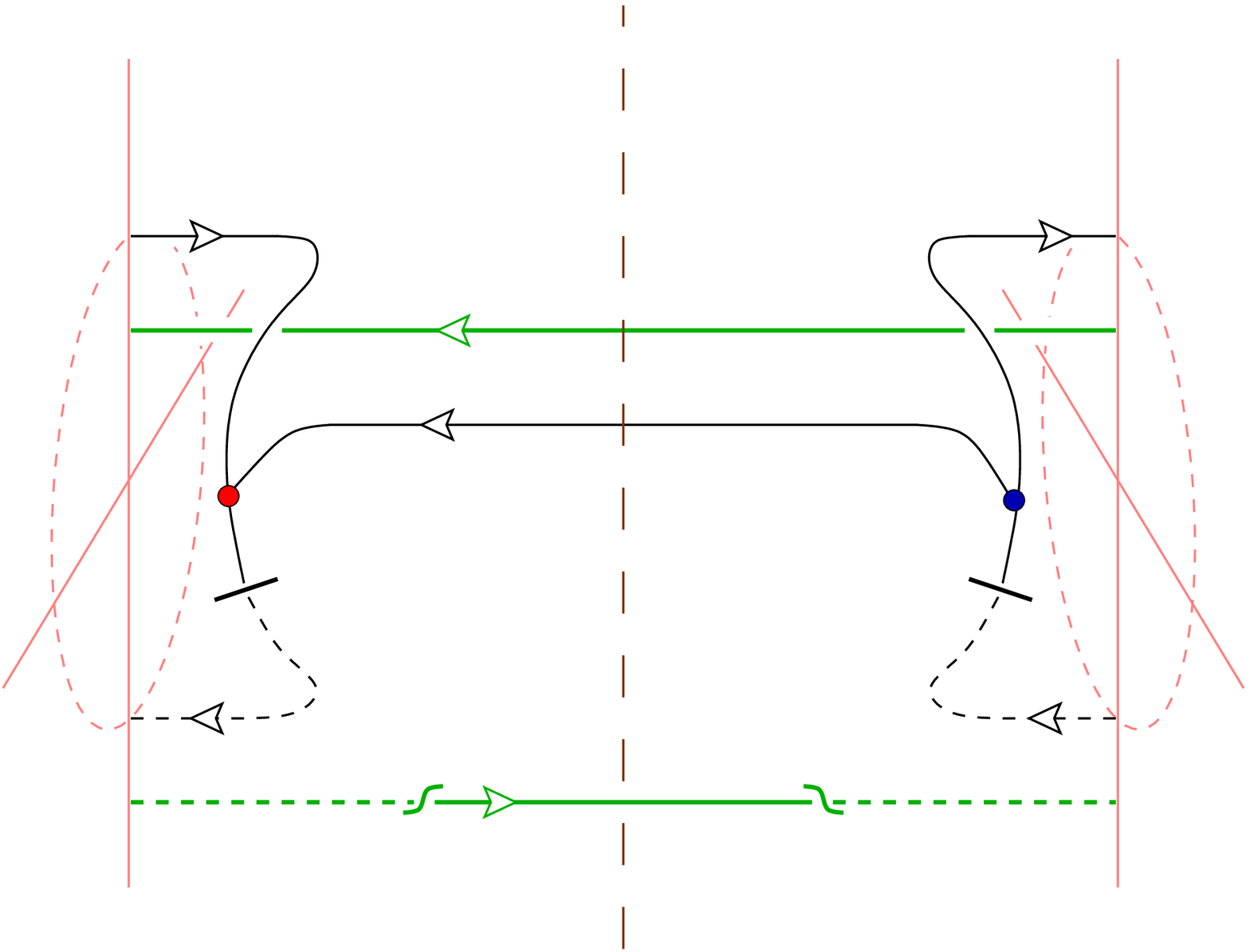}
  \setlength{\unitlength}{24810sp}
      \put(305,276) {\scriptsize$k$}
      \put(289,72) {\scriptsize$k$}
      \put(315,206) {\scriptsize$A$}
      \put(106,315) {\scriptsize$A$}
      \put(386,120) {\scriptsize$A$}
  \setlength{\unitlength}{1pt}
  }
  \put(0,72)    {$ \tilde\K_k \;= $}
  \epicture40  \labl{eq:pic109}
This three-manifold drawn here has $S^2$'s as vertical sections and is bounded
by two crosscaps. At the dashed vertical line we now insert the identity 
\erf{eq:kAk-id}; this leads to a decomposition into products of
invariants of graphs in $\RP$:
  \bea  \begin{picture}(420,60)(13,56)
  \put(126,0)      {\includeourbeautifulpicture{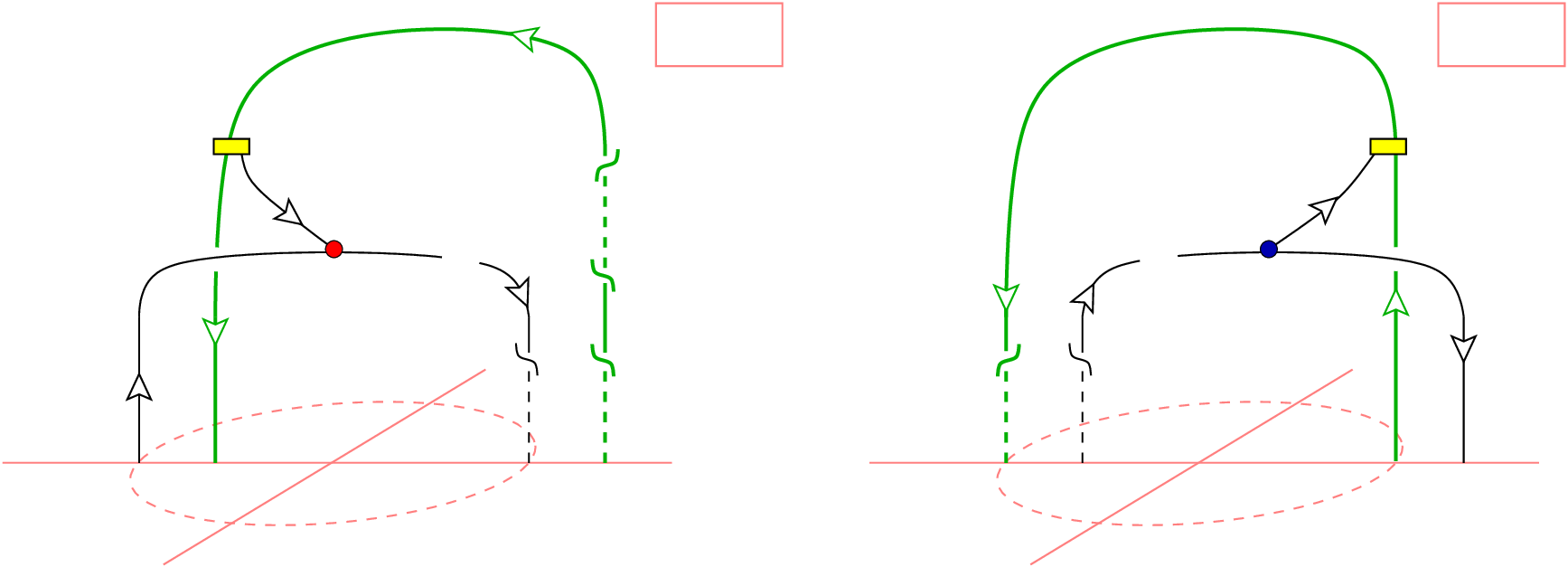}
  \setlength{\unitlength}{24810sp}
      \put(359,273.1) {$\RP$}
      \put(329,172) {\scriptsize$k$}
      \put(140,221) {\scriptsize$\overline\alpha$}
      \put(123,98) {\scriptsize$k$}
      \put(238,159) {\scriptsize$\sigma$}
      \put(55,109) {\scriptsize$A$}
      \put(775.7,273.1) {$\RP$}
      \put(611,160) {\scriptsize$\sigma$}
      \put(785,138) {\scriptsize$A$}
      \put(544,186) {\scriptsize$k$}
      \put(755,218) {\scriptsize$\alpha$}
  \setlength{\unitlength}{1pt}
  }
  \put(0,57)    {$ \tilde\K_k \;=\; \dsty
                     \frac{1}{S_{00}\,\dim(U_k)}\; \sum_\alpha $}
  \epicture33  \labl{eq:pic110}
The second of these $\RP$-invariants has the same form as the second graph
in \erf{eq:Ga-RP}, hence it is equal
to $\Gamma^\sigma(U_k,\mu_\alpha^{\bar kk})$. Also, after substituting the
definitions \erf{eq:lambda-omega-def} and \erf{eq:lam-om-tilde},
by comparison with the second graph in \erf{eq:Ga-RP}
the first $\RP$-invariant is found to be equal to
  \be
  \theta_k^{-1} \Gamma^\sigma(U_k^\vee, \Omega_k^k(
  {\Lambda_k^k}^{-1}(\bar\mu_\alpha^{\bar kk}) ) )
  = \theta_k^{-1} \Gamma^\sigma(U_k, {\Lambda_k^k}^{-1}(\bar\mu_\alpha^{\bar kk}))
  = \theta_k^{-1} \dim(U_k)\! \sum_\beta g^{\bar k k}_{\alpha\beta}
  \Gamma^\sigma(U_k, \mu_\beta^{\bar kk}) \,.  \ee
The first equality follows by \erf{eq:Ga-GaO}, while in the second
the expansion \erf{eq:mug-def} is inserted. Putting these results together 
we arrive at
  \be
  \tilde\K_k = \frac{ \theta_k^{-1}}{S_{00}}
  \sum_{\alpha,\beta} \Gamma^\sigma_{k\alpha}\,
  g^{\bar k k}_{\alpha\beta}\, \Gamma^\sigma_{k\beta}  \,.  \ee
Together with \erf{eq:Ga-bas} and \erf{eq:K-Kt} this establishes
\erf{eq:K-crossed}.

\dt{Remark} 
The notation in \erf{eq:M-crossed} and \erf{eq:K-crossed}
has been chosen so as to resemble formulas (2.14) and (2.15) of \cite{sosc}.
More explicitly, the quantities appearing in \cite{sosc} -- to be 
decorated with a hat -- are related to the ones appearing here as follows:
  \be
  \hat\Gamma_{m\alpha} = x_m  \gamma^{\sigma}_{m\alpha} \,, \qquad
  \hat g^m_{\alpha\beta} = (S_{00}^{} x_m^2)^{-1} g^{\bar m m}_{\alpha\beta}
  \qquad{\rm and}\qquad  \hat B_{(m\alpha)a} = x_m  S^A_{a,m\alpha}  \ee
for some normalisation constants $x_m$; we can choose the
bases \erf{eq:mu-basis} such that $x_m\eq 1$. The identities
proven in the setting of this paper can be employed
to reproduce some of the expressions in \cite{sosc}.
For example (2.19) of \cite{sosc} follows from writing out
\erf{eq:Ga-GaO} in a basis, with $\hat\Omega\eq\Omega^{-1}$.
Furthermore, since $S^A$ and $\tilde S^A$ are each other's inverse, 
(2.3) in \cite{sosc} forces $\tilde S^A_{j\alpha,b}\eq S_{0j}\,
\hat B_{(j\alpha)b}^*$. Finally, comparing (2.8) in \cite{sosc} to a 
combination of \erf{eq:SA-StA} and \erf{eq:SA-StA2} yields 
$\hat C\eq 1/S_{00}\, \Omega g$, in agreement
with (2.9) and (2.10) of \cite{sosc}. In particular 
\erf{eq:g-symm} shows that $\hat g$ is indeed symmetric.


\subsection{Defect lines on non-orientable surfaces}\label{sec:defect}

One may think of a defect as a prescription for how to join
two world sheets along their boundaries (or two boundary components of a
single world sheet) by some kind of boundary condition. The joint boundary
component constitutes the one-dimensional defect. The theories on the two 
sides of the defect need in general not be the same. Indeed, one could even 
interpret a physical world sheet boundary as a defect linking the world 
sheet theory to the trivial theory with only a single state. In the CFT 
context defects have been studied in several situations, see e.g.\
\cite{osaf,osaf2,koLe2,lelu,jelu,watt8,pezu5,pezu7,coSc,bddo,qusc,cohu},
and for a lattice based analysis also \cite{cmop,grim2,chmp}.

The defects we study here are not of the most general kind. 
First of all, we require the defect to be conformally invariant, i.e.\ 
correlators involving the stress tensors $T$ or $\bar T$ vary smoothly as we
move the insertion point across a defect. This implies that the generators
$L_m$, $\bar L_m$ of infinitesimal conformal transformations commute with the
defect. As a consequence, the defect can be deformed continuously without 
changing the value of the correlation function. In this sense conformal 
defects are {\em tensionless\/}.

The defects we consider form a subclass of conformal defects. We are working
with a given chiral algebra and describe all conformal defects that
join two CFTs having this chiral algebra in common, requiring also that the
currents in the chiral algebra commute with the defect. Thus these defects 
are {\em rational conformal defects\/}. Rational conformal defects with the 
same CFT on either side have been studied in \cite{pezu5,pezu7,coSc,cohu}.

Defects on orientable world sheets have been discussed in \cite{fuRs4}, 
see specifically sections I:4.4, I:5.10 and remark I:5.19(ii). 
Below we discuss defects with the same CFT on either side on 
surfaces that are not necessarily orientable. We will see that wrapping 
defects around a non-orientable cycle\,%
  \footnote{~By a non-orientable cycle we mean a cycle on the world sheet 
  that does not possess an orientable neighbourhood.}
singles out a special subclass of defects. These defects can be
wrapped around such cycles without the need to insert a defect
changing field, i.e.\ they can still be moved freely on the
world sheet.

\medskip

The treatment of defect lines parallels that of boundary conditions
in section \ref{sec:crg}. For the moment, we take the defect line to be the
embedding of a circle $S^1$ into the world sheet such that it wraps
an orientable cycle; non-orientable cycles will be treated afterwards.\,%
  \footnote{~Using TFT tools we
  can also describe two defect lines joining into a single one or, more
  generally, networks of defect lines embedded in the world sheet.
  In the present paper such more general situations will not be considered.}
Such defects are labelled by
equivalence classes of triples $(X, \orr_1, \orr_2)$ where
$X$ is an $A$-$A$-bimodule, $\orr_1$ is an orientation of the defect
line, and $\orr_2$ is an orientation of a neighbourhood of the defect
in the world sheet $X$. The equivalence relation is given by
  \be  \!\!\!\!\!\!\!\!\!
  (X, \orr_1, \orr_2) \sim (X', \orr_1', \orr_2') {\rm ~~iff~}
  \left\{\begin{array}{rlllll} \!\!\!{\rm either~}
  &\orr_1' = \orr_1  &\!{\rm and}\!& \orr_2' = \orr_2
                     &\!{\rm and}\!& X' \cong X
  \,, \\[3pt] {\rm or~}
  &\orr_1' = \orr_1  &\!{\rm and}\!& \orr_2' = -\orr_2
                     &\!{\rm and}\!& X' \cong X^s
  \,, \\[3pt] {\rm or~}
  &\orr_1' = -\orr_1 &\!{\rm and}\!& \orr_2' = \orr_2
                             &\!{\rm and}\!& X' \cong X^v
  \,, \\[3pt] {\rm or~}
  &\orr_1' = -\orr_1 &\!{\rm and}\!& \orr_2' = -\orr_2 
                     &\!{\rm and}\!& X' \cong X^\sigma
  \,. \eear\right. \labl{eq:defect-equiv}
As we will see below, graphs describing correlators with
defect lines that are identified by this equivalence relation 
are equivalent as ribbon graphs. We denote
the set of all defect line labels by
  \be
  \mathcal{D} := \{\,(X,\orr_1,\orr_2)\,\} \,{/}{\sim}\,
   =:  \{\,[X,\orr_1,\orr_2]\,\}\,.  \ee

The starting point of the construction of the ribbon graph and three-manifold
for a correlator with defect lines
is a world sheet $X$ with a set of marked non-intersecting circles,
such that the cycles wrapped by the circles are orientable. 
The boundary components of $X$ are labelled by elements of
$\mathcal{B}$ (see \erf{eq:bc-B-def}), while the circles
are labelled by elements of $\mathcal{D}$. 

The three-manifold $\MX$ is constructed precisely as in section \ref{sec:crg}. 
The construction of the ribbon graph is identical to the construction given
there, apart from the following modification:

\noindent
\begin{itemize}
\item[\Nxt] 
Choose a representative $(X,\orr_1,\orr_2)$ for each equivalence class 
$[X,\orr_1,\orr_2]$ labelling a circular defect line -- \underline{Choice \#1$'$}.

\item[\Nxt]
The dual triangulation of $X$ is chosen in such a way that (just like the 
boundary components) the defect circles are part of the 
triangulation -- \underline{Choice \#2$'$}.

\item[\Nxt]
At each vertex of the triangulation that lies on on a defect circle labelled
by $[X,\orr_1,\orr_2]$, place one of the (pieces of) ribbon graphs
  \bea  \begin{picture}(340,69)(0,65)
  \put(0,0)      {\includeourbeautifulpicture{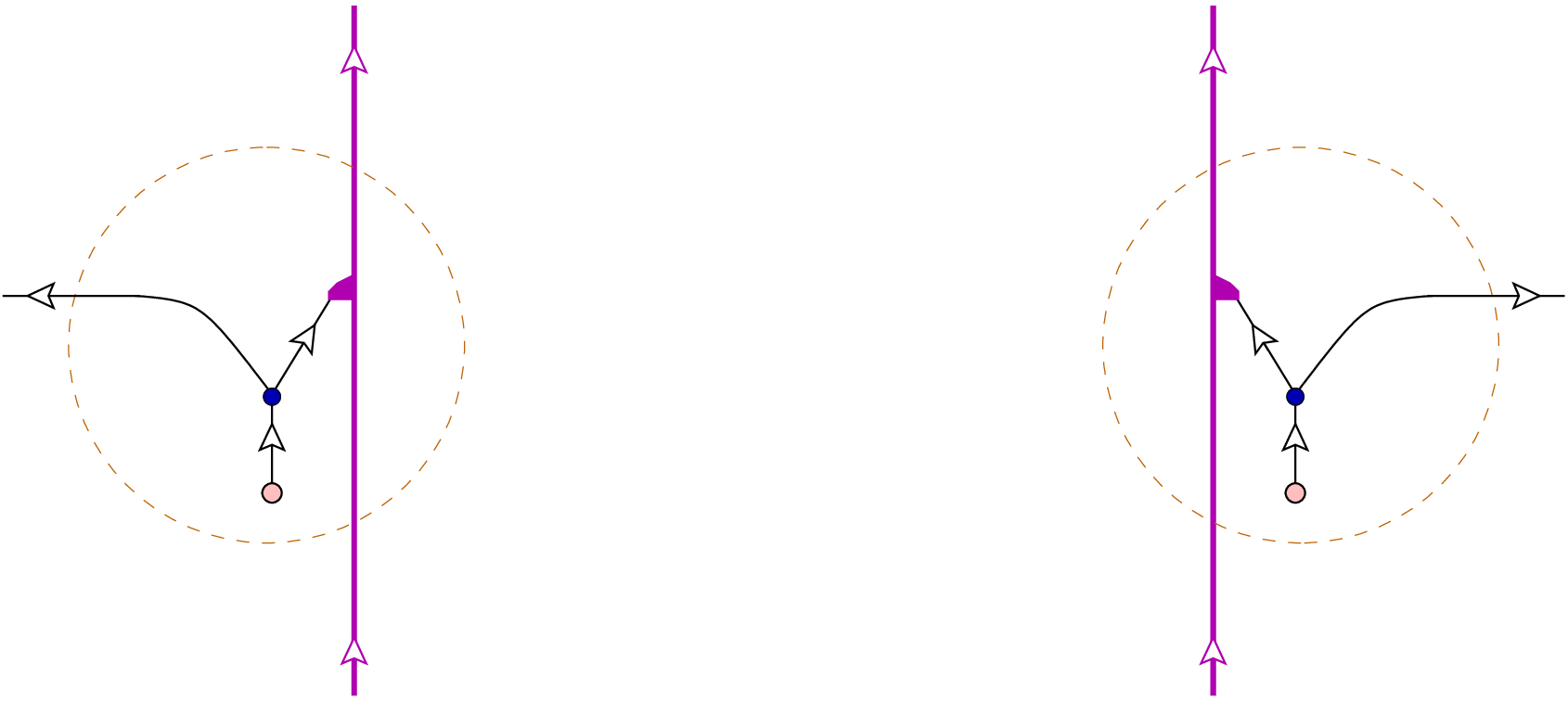}
  \setlength{\unitlength}{24810sp}
      \put(70,215) {\scriptsize$A$}
      \put(192,136) {\scriptsize$X$}
      \put(730,215) {\scriptsize$A$}
      \put(608,136) {\scriptsize$X$}
  \setlength{\unitlength}{1pt}
  }
  \put(0,57)    {$ $}
  \epicture38  \ee
depending on to which side of the defect the triangulation edge lies. The 
representative $(X,\orr_1,\orr_2)$ chosen before gives an
orientation $\orr_1$ to the defect circle and $\orr_2$ to a
neighbourhood of the circle. The ribbons are placed such that the
orientation of the ribbon matches $\orr_2$ and such that the orientation of
the core of the $X$-ribbon matches $\orr_1$.
\end{itemize}

\noindent
No further modifications to the prescription in 
section \ref{sec:crg} are needed. Let us show that also after these
modifications the correlator is independent of the choices made.
\\[5pt]
\nxt\underline{Choice \#2$'$}:\\[2pt]
Independence of the triangulation in the presence of defect lines
can be seen in much that same way as was the case for
choice \#2 in section \ref{sec:crg}. The crucial additional
ingredient is that the left and right action of the algebra $A$
on a bimodule $X$ commute with each other. This allows to move
two vertices of the triangulation that lie on the defect past each
other, as indicated in the following picture:
  \bea  \begin{picture}(260,39)(0,52)
  \put(0,0)      {\includeourbeautifulpicture{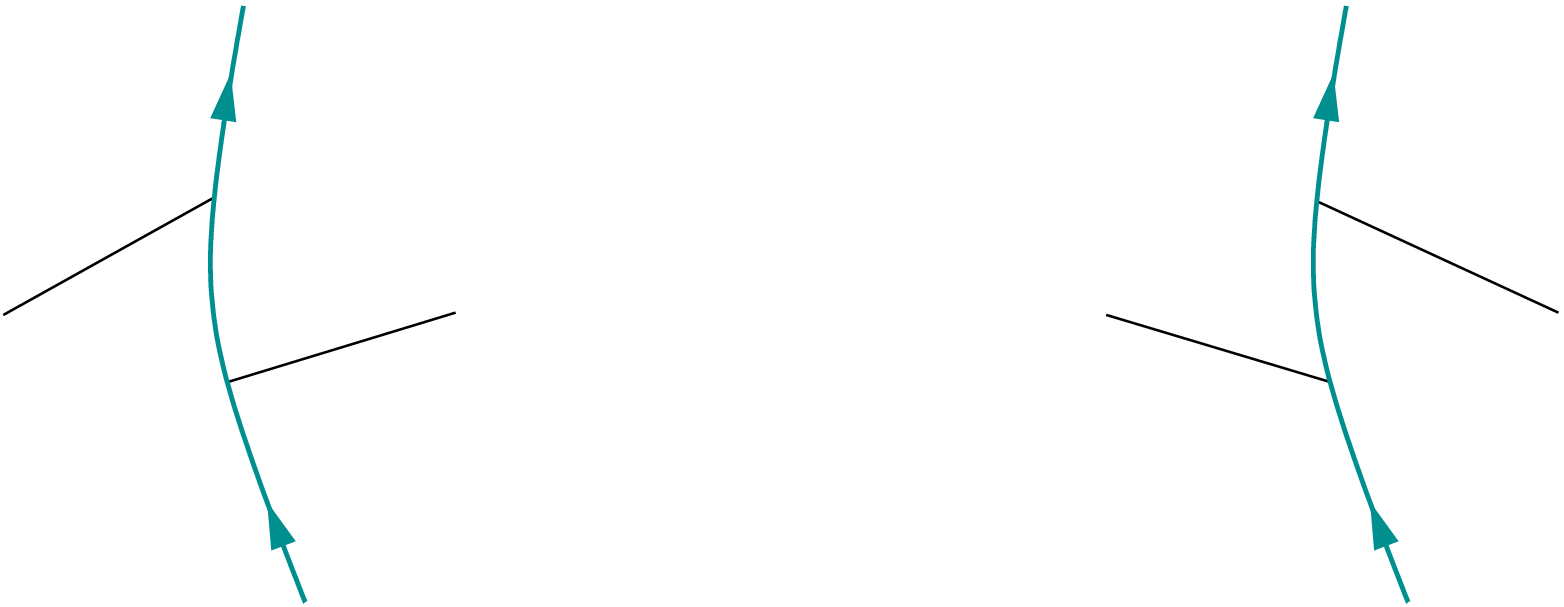}
  \setlength{\unitlength}{24810sp}
      \put(126,34) {\scriptsize$X$}
      \put(582,34) {\scriptsize$X$}
  \setlength{\unitlength}{1pt}
  }
  \put(123,47)  {$ = $}
  \epicture31  \ee
\nxt\underline{Choice \#1$'$}:\\[2pt]
Consider a defect circle labelled by the equivalence class
$[X,\orr_1,\orr_2]$. We must show that any two representatives
$(X,\orr_1,\orr_2)$ and $(X',\orr_1',\orr_2')$ of this class
lead to equivalent ribbon graphs. Analogously to choice \#1 in section 
\ref{sec:crg} this can be established by treating one by one  
the cases arising in the equivalence relation 
\erf{eq:defect-equiv}. Supposing that the ribbon graph has been 
constructed for the representative $(X',\orr_1',\orr_2')$, we
thus proceed as follows.
\\[4pt]
(i)\, If $\orr_1'\eq\orr_1$ and $\orr_2'\eq\orr_2$, then $X'\,{\cong}\,X$.
\\
Choose an isomorphism $\varphi \iN \Hom_{\!A|A}(X,X')$ and
insert the identity $\id_{X'}\eq\varphi \cir \varphi^{-1}$
on the $X'$-ribbon. By definition, $\varphi$ commutes with 
the left and right action of $A$ on $X'$; therefore we can move
$\varphi$ around the defect circle until it reaches its inverse.
Using $\varphi^{-1}\cir\varphi\eq\id_X$, it follows that
the net effect of these moves is that we have replaced the
annular $X'$-ribbon by an $X$-ribbon with the same orientation
of core and surface.
\\[4pt]
(ii)\, If $\orr_1'\eq\orr_1$ and $\orr_2'\eq{-}\orr_2$, then
$X'\,{\cong}\,X^s$.
\\
Choose an isomorphism $\varphi \iN \Hom_{\!A|A}(X^s,X')$ and insert 
$\id_{X'}\eq\varphi \cir\varphi^{-1}$ on the $X'$-ribbon. Then use the moves
  \bea  \begin{picture}(390,90)(0,70)
  \put(0,0)      {\includeourbeautifulpicture{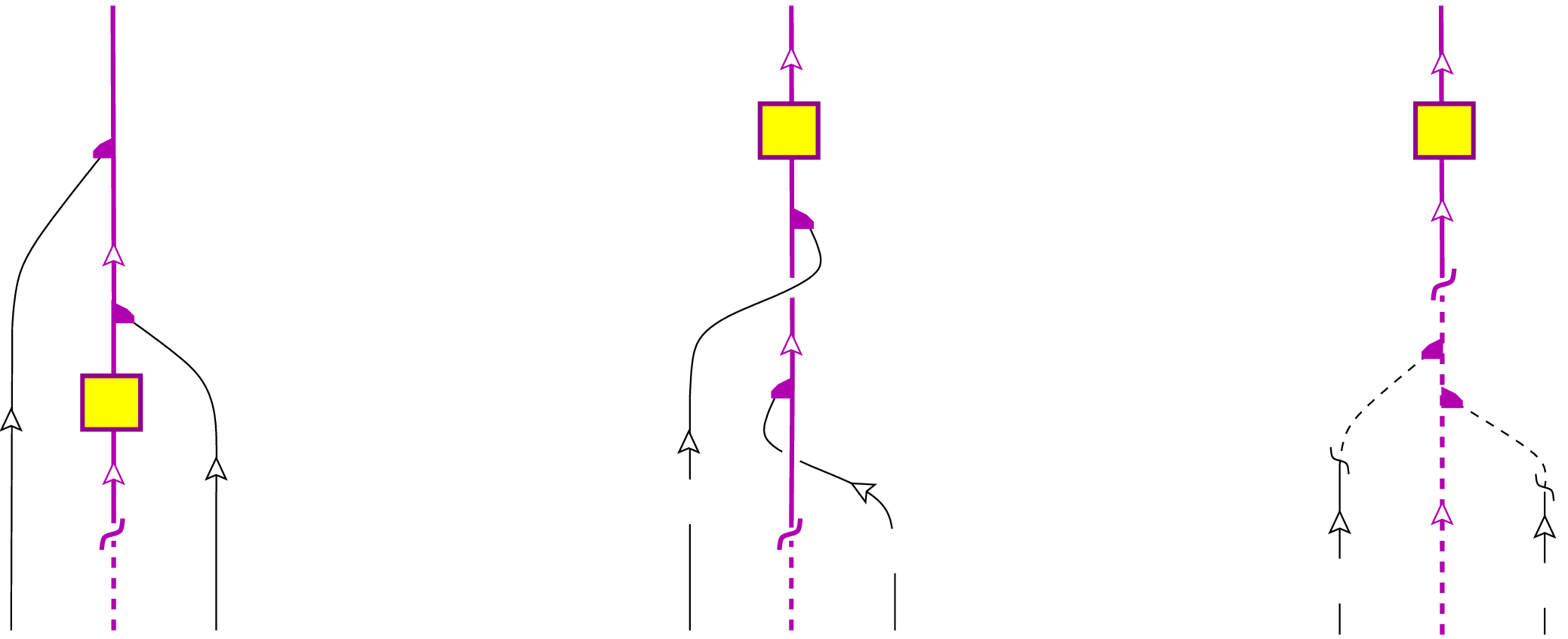}
  \setlength{\unitlength}{24810sp}
      \put(12,78) {\scriptsize$A$}
      \put(63,147) {\scriptsize$\varphi$}
      \put(81,278) {\scriptsize$X'$}
      \put(47,85) {\scriptsize$X$}
      \put(134,166) {\scriptsize$A$}
      \put(497,321) {\scriptsize$\varphi$}
      \put(433,82) {\scriptsize$\sigma$}
      \put(564,50) {\scriptsize$\sigma$}
      \put(514,174) {\scriptsize$X$}
      \put(515,362) {\scriptsize$X'$}
      \put(424,185) {\scriptsize$A$}
      \put(564,94) {\scriptsize$A$}
      \put(919,322) {\scriptsize$\varphi$}
      \put(930,364) {\scriptsize$X'$}
      \put(932,222) {\scriptsize$X$}
      \put(850,31) {\scriptsize$\sigma$}
      \put(982,28) {\scriptsize$\sigma$}
      \put(849,144) {\scriptsize$A$}
      \put(984,121) {\scriptsize$A$}
  \setlength{\unitlength}{1pt}
  }
  \put(105,75)  {$ = $}
  \put(266,75)  {$ = $}
  \epicture46  \labl{eq:X-Xs}
to take $\varphi$ around the defect circle. Here we used the fact $\varphi$
is an intertwiner and substituted the definition \erf{eq:Xsvs-pics}
of the action of $A$ on $X^s$ and then rotated the $X$-ribbon
clockwise by $180^\circ$. The combination of half-twist and
$\sigma$ in the graph on the \rhs\ can be recognised as the
\boxelement\ \erf{eq:bw-bar}. Combining finally $\varphi$ with
$\varphi^{-1}$ to $\id_X$, we have replaced the
$X'$-ribbon by an $X$-ribbon with opposite orientation of
the surface and the same orientation of the core, as required.
\\[4pt]
(iii)\, If $\orr_1'\eq{-}\orr_1$ and $\orr_2'\eq\orr_2$, then
$X'\,{\cong}\,X^v$.
\\
Choose an isomorphism $\varphi \iN \Hom_{\!A|A}(X^v,X')$ and
follow the same procedure as in point (ii), but this time using the moves
  \bea  \begin{picture}(270,55)(0,56)
  \put(0,0)      {\includeourbeautifulpicture{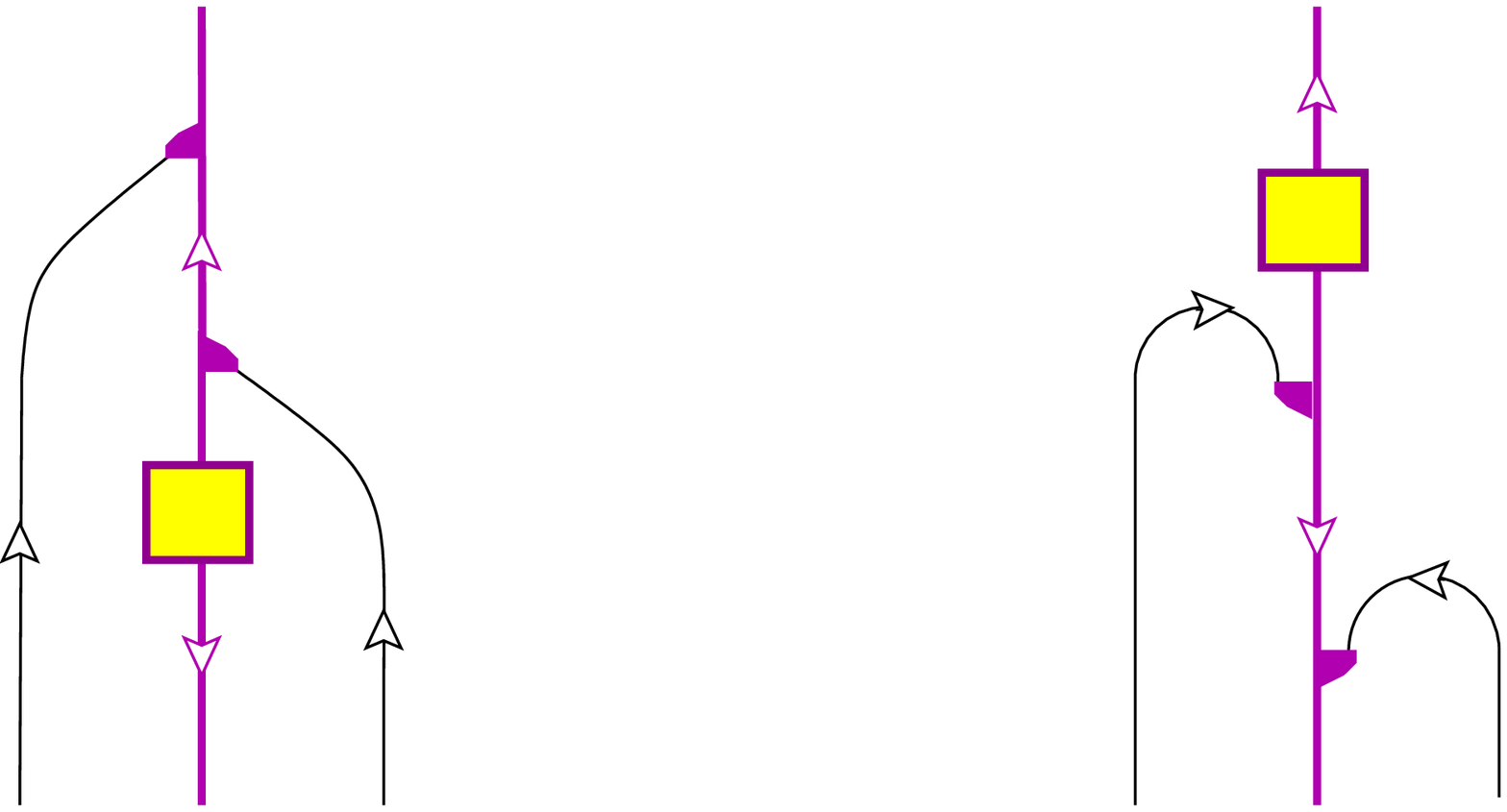}
  \setlength{\unitlength}{24810sp}
      \put(14,39) {\scriptsize$A$}
      \put(80,39) {\scriptsize$X$}
      \put(148,39) {\scriptsize$A$}
      \put(63,106) {\scriptsize$\varphi$}
      \put(80,219) {\scriptsize$X'$}
      \put(464,209) {\scriptsize$\varphi$}
      \put(413,80) {\scriptsize$A$}
      \put(480,113) {\scriptsize$X$}
      \put(484,251) {\scriptsize$X'$}
      \put(547,42) {\scriptsize$A$}
  \setlength{\unitlength}{1pt}
  }
  \put(98,52)   {$ = $}
  \epicture35  \ee
in place of \erf{eq:X-Xs}.
Then taking $\varphi$ around the defect circle replaces the $X'$-ribbon
by an $X$-ribbon with opposite orientation of the core.
\\[4pt]
(iv)\, If $\orr_1'\eq{-}\orr_1$ and $\orr_2'\eq{-}\orr_2$, then
$X'\,{\cong}\,X^\sigma$.
\\
Choose an isomorphism $\varphi$ in $\Hom_{\!A|A}(X^\sigma,X')$ and take 
it around the defect circle, using the equalities
  \bea  \begin{picture}(390,78)(0,70)
  \put(0,0)      {\includeourbeautifulpicture{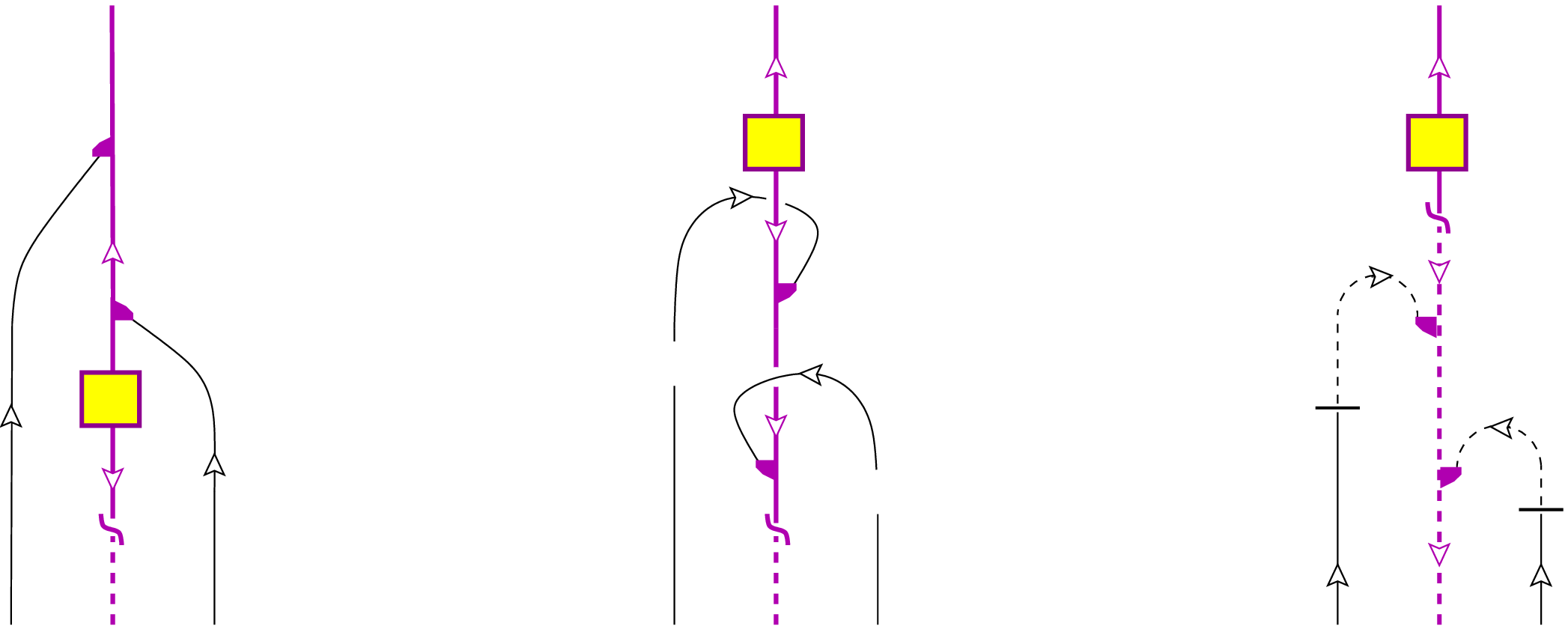}
  \setlength{\unitlength}{24810sp}
      \put(81,337) {\scriptsize$X'$}
      \put(14,25) {\scriptsize$A$}
      \put(81,25) {\scriptsize$X$}
      \put(145,25) {\scriptsize$A$}
      \put(65,144) {\scriptsize$\varphi$}
      \put(440,25) {\scriptsize$A$}
      \put(506,25) {\scriptsize$X$}
      \put(569,25) {\scriptsize$A$}
      \put(510,352) {\scriptsize$X'$}
      \put(492,308) {\scriptsize$\varphi$}
      \put(428,165) {\scriptsize$\sigma$}
      \put(560,84) {\scriptsize$\sigma$}
      \put(870,25) {\scriptsize$A$}
      \put(934,25) {\scriptsize$X$}
      \put(1003,25) {\scriptsize$A$}
      \put(918,309) {\scriptsize$\varphi$}
      \put(935,351) {\scriptsize$X'$}
  \setlength{\unitlength}{1pt}
  }
  \put(103,72)  {$ = $}
  \put(265,72)  {$ = $}
  \epicture48  \ee
Again this results in the ribbon graph obtained for
the representative $(X,\orr_1,\orr_2)$.

\medskip

Having described the construction of the ribbon graph for defects
that wrap orientable cycles, let us now turn to the case of non-orientable
cycles. Consider the M\"obius strip as an example, and take the circle
running parallel to the boundary in the middle of the strip. When trying
to place a bimodule ribbon $X$ along such a cycle, at some point one
must join the `white' side of the $X$-ribbon to its `black' side.
The most general way to do this is to pick some morphism 
$\varphi\iN\Hom(X,X)$ and take
  \bea  \begin{picture}(60,42)(0,54)
  \put(0,0)      {\includeourbeautifulpicture{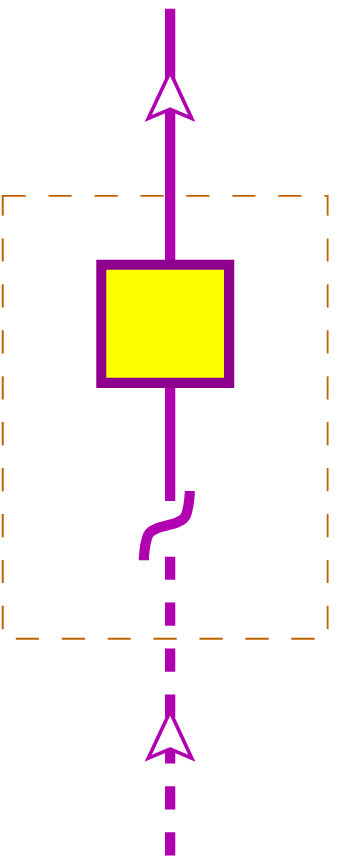}
  \setlength{\unitlength}{24810sp}
      \put(40,152) {\scriptsize$\varphi$}
      \put(61,209) {\scriptsize$X$}
      \put(58,45) {\scriptsize$X$}
  \setlength{\unitlength}{1pt}
  }
  \epicture32  \ee
Choosing the other possible half-twist differs from the above by a twist, 
which can be absorbed into the definition of $\varphi$. However we still need
to impose the requirement that different triangulations of the world sheet 
lead to equivalent ribbon graphs. This amounts to the property
  \bea  \begin{picture}(230,92)(0,61)
  \put(0,0)      {\includeourbeautifulpicture{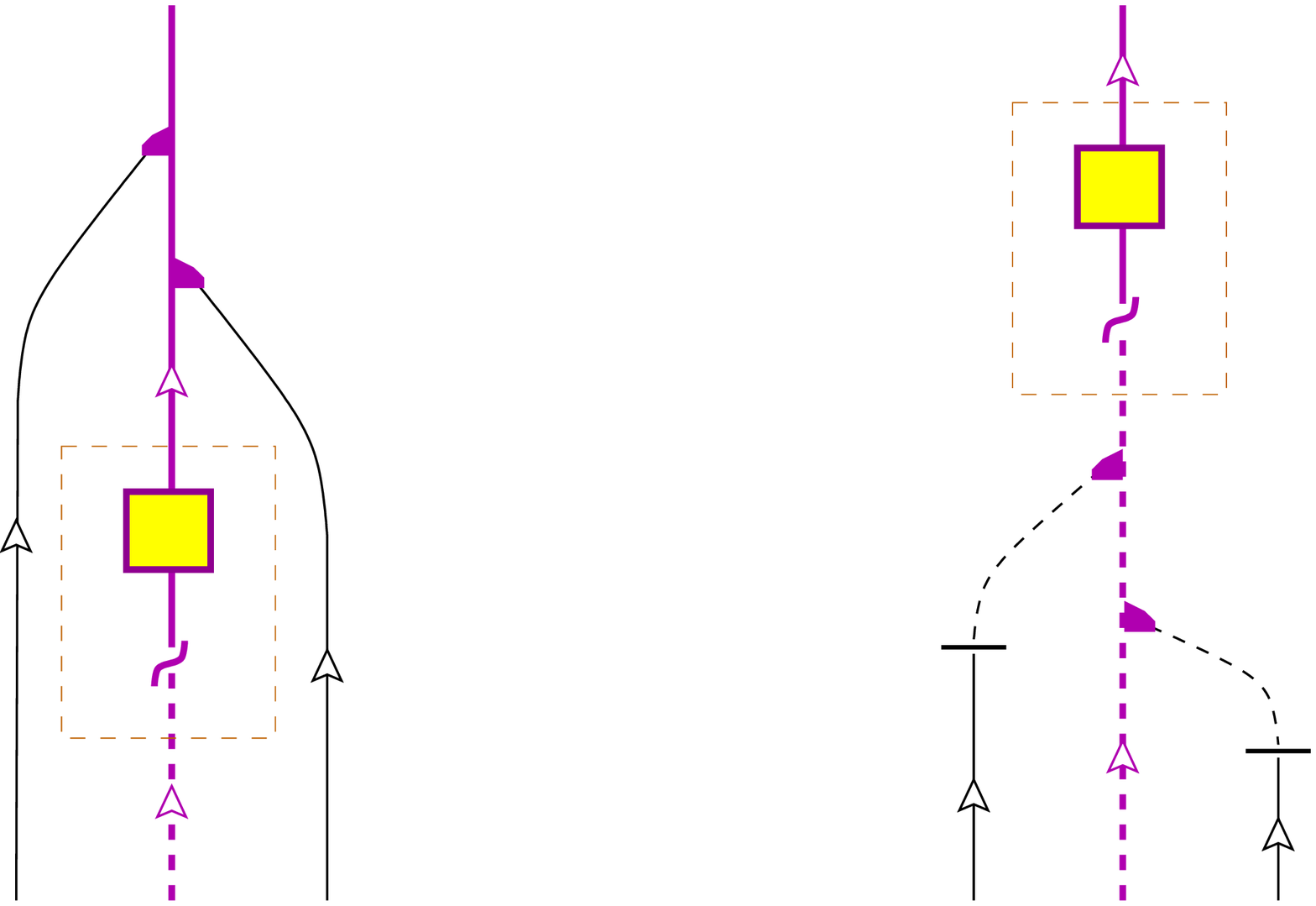}
  \setlength{\unitlength}{24810sp}
      \put(12,47) {\scriptsize$A$}
      \put(84,47) {\scriptsize$X$}
      \put(149,47) {\scriptsize$A$}
      \put(64,162) {\scriptsize$\varphi$}
      \put(84,341) {\scriptsize$X$}
      \put(504,359) {\scriptsize$X$}
      \put(485,313) {\scriptsize$\varphi$}
      \put(499,162) {\scriptsize$X$}
      \put(432,69) {\scriptsize$A$}
      \put(542,109) {\scriptsize$A$}
  \setlength{\unitlength}{1pt}
  }
  \put(103,70)  {$ = $}
  \epicture39  \ee
Comparison with \erf{eq:X-Xs} shows that independence of triangulation
requires $\varphi \iN \Hom_{\!A|A}(X^s,X)$. In general there need
not exist a bimodule homomorphism between $X^s$ and $X$. In this case
it would be necessary to insert a suitable defect field (if it exists at all)
at some point of the defect so as to join the two ends of the $X$-ribbon, 
thereby pinning down the location of the defect at this point.\,%
 \footnote{~Since it commutes with the Virasoro generators $L_m$ and $\bar L_m$,
 a defect line can be deformed continously, on the world sheet minus the
 positions of field insertions, without changing the value of the correlator.
 But field insertions do not commute with $L_m,\, \bar L_m$, and hence moving
 a defect field insertion does in general change the value of a correlator.
 In this sense
 a defect line gets pinned down at the insertion point of a defect field.}
In contrast, if we can find a nonzero  $\varphi \iN \Hom_{\!A|A}(X^s,X)$, then
a defect described by the bimodule $X$ can be wrapped around a non-orientable
cycle without fixing it at any point via insertion of a defect field,
so that it can still be deformed freely on the world sheet.

Computing the dimension of the space $\Hom_{\!A|A}(X^s,X)$ amounts
to computing the invariant
  \be
  \dimc\,\Hom_{\!A|A}(X^s,Y) = Z^{X^s|Y}_{00} \,,  \labl{eq:defect-Hom}
see (I:5.151) for its definition. The equality can be demonstrated
by combining part (iv) of theorem I:5.23 with propositions I:4.6 and I:5.22.

In the case $\N ijk \iN \{0,1\}$ and
$\lra{U_k}A \iN \{0,1\}$ for all $i,j,k\iN\I$, the invariant
$Z^{X^s|Y}_{00}$ can be evaluated as follows. First of all, inserting the 
definition of $X^s$ in \erf{eq:Xsvs-pics} into the ribbon graph (I:5.151) 
(with $k\eq l\eq0$) and slightly deforming the resulting graph gives
  \bea  \begin{picture}(220,79)(0,64)
  \put(55,0)      {\includeourbeautifulpicture{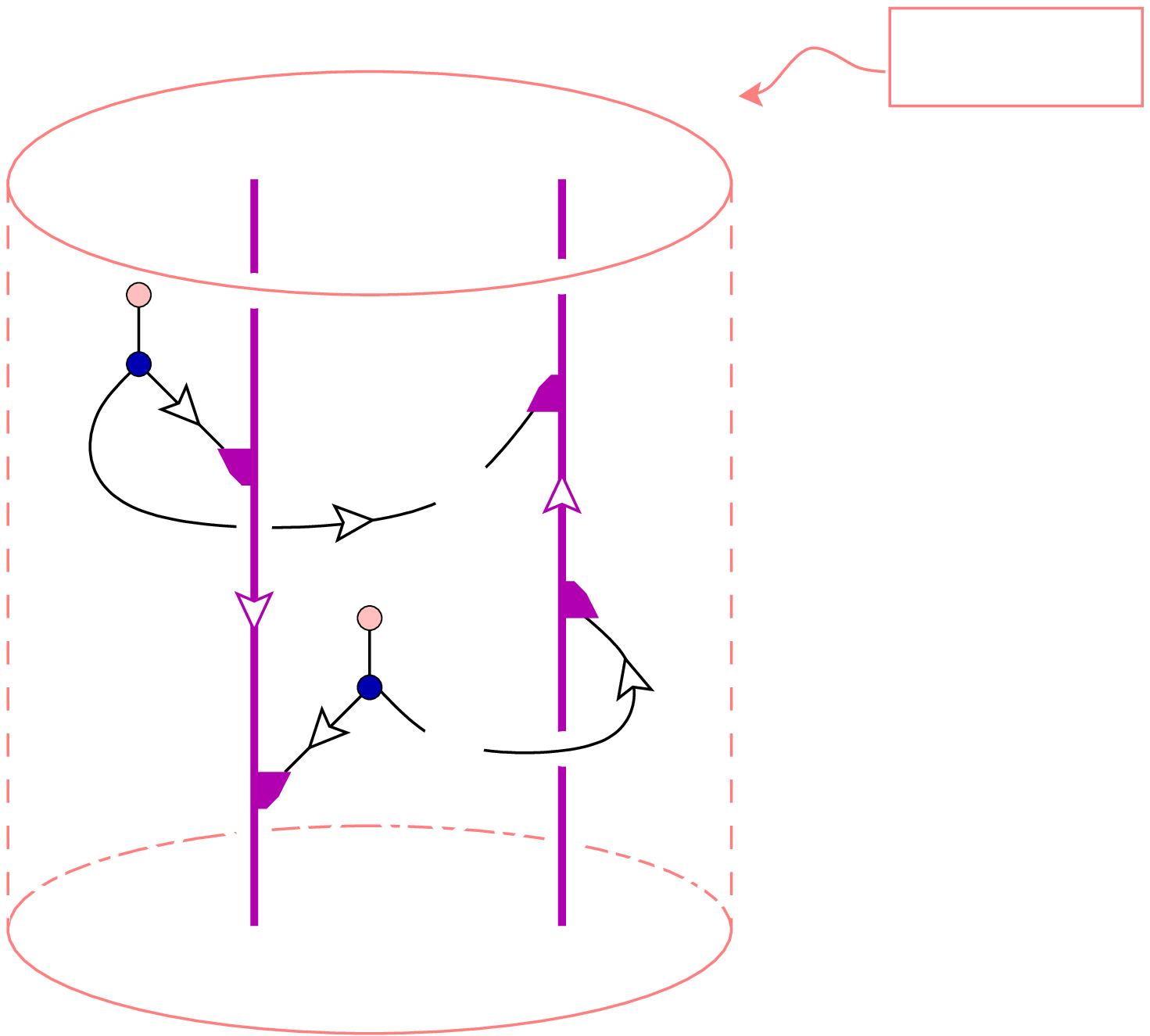}
  \setlength{\unitlength}{24810sp}
      \put(334,353) {$S^2{\times}S^1$}
      \put(165,199) {\scriptsize$\sigma$}
      \put(162,104) {\scriptsize$\sigma$}
      \put(65,145) {\scriptsize$Y$}
      \put(218,197) {\scriptsize$X$}
      \put(123,201) {\scriptsize$A$}
      \put(244,133) {\scriptsize$A$}
  \setlength{\unitlength}{1pt}
  }
  \put(0,77)    {$ Z^{X^s|Y}_{00} \;= $}
  \epicture41  \ee
When substituting the expressions for comultiplication and the
representation morphisms in a basis this leads to
  \bea \dsty
  Z^{X^s|Y}_{00} = 
  \sum_{a,b\In A} \sigma(a)\, \sigma(b)\,
  \Delta^{a\bar a}_0\, \Delta^{\bar bb}_0
  \\[-6pt] \dsty \mbox{\hspace{6.5em}}
  \sum_{x,r\In X} \sum_{y,s\In Y} \sum_{\alpha,\beta,\mu,\nu}
  \rho_{b,x\alpha}^{X\;r\mu} \,
  \tilde\rho_{a,r\mu}^{X\;x\alpha} \,
  \rho_{\bar a,y\beta}^{X\;s\nu} \,
  \tilde\rho_{\bar b,s\nu}^{X\;y\beta}
  \, \Gamma_{X^s Y}(a,b,r,s,x,y)
  \,, \eear\ee
where $\alpha$, $\beta$, $\mu$ and $\nu$ run from 1 to
$\lr{U_x}{X}$, $\lr{U_y}{Y}$, $\lr{U_r}{X}$ and $\lr{U_s}{Y}$,
respectively. The invariant $\Gamma_{X^s Y}(a,b,r,s,x,y)$ is defined as
  \begin{eqnarray}  \begin{picture}(420,79)(19,79)
  \put(132,0)      {\includeourbeautifulpicture{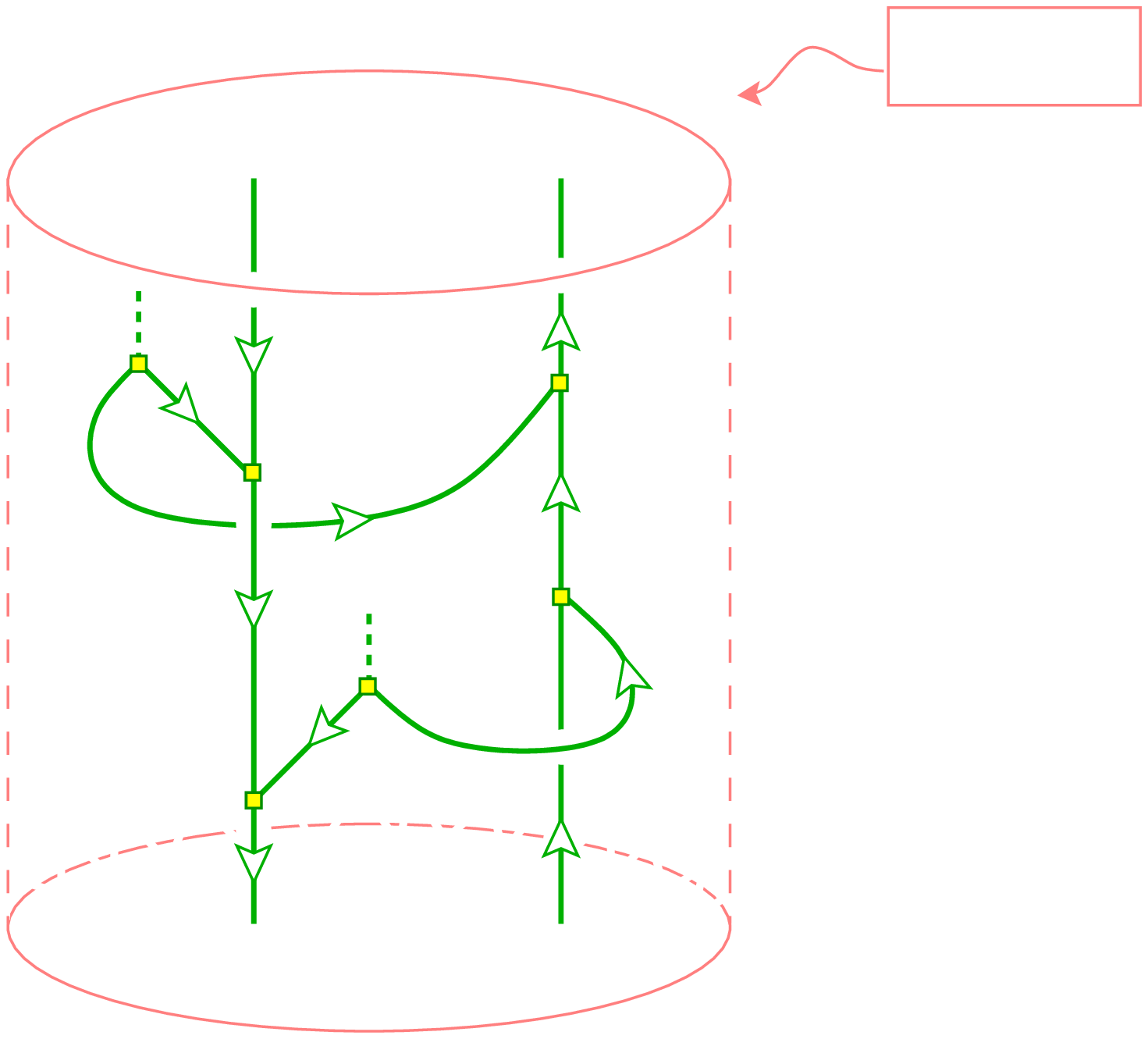}
    \setlength{\unitlength}{24810sp}
      \put(99,318) {\scriptsize$s$}
      \put(213,325) {\scriptsize$r$}
      \put(152,231) {\scriptsize$b$}
      \put(59,239) {\scriptsize$\overline b$}
      \put(81,149) {\scriptsize$y$}
      \put(119,123) {\scriptsize$\overline a$}
      \put(173,140) {\scriptsize$a$}
      \put(219,222) {\scriptsize$x$}
      \put(677,295) {\scriptsize$r$}
      \put(864,191) {\scriptsize$b$}
      \put(794.5,325) {\scriptsize$x$}
      \put(766,233) {\scriptsize$r$}
      \put(730,183) {\scriptsize$a$}
      \put(747,139) {\scriptsize$\overline a$}
      \put(794,118) {\scriptsize$y$}
      \put(811,83) {\scriptsize$\overline b$}
      \put(334.8,383) {$S^2{\times}S^1$}
      \put(865.5,408.5) {$S^3$}
    \setlength{\unitlength}{1pt}
    }
  \put(0,64)    {$ \Gamma_{X^s Y}(a,b,r,s,x,y) \;:= $}
  \put(308,64)  {$ =\; \dsty\frac{\delta_{rs}}{\dim(U_r)} $}
  \end{picture} \nonumber\\[6.5em]{} 
  \\[-1.9em]{}\nonumber\end{eqnarray}
In the second equality we first apply dominance on the vertical $U_s$- and
$U_r$-ribbons; only the tensor unit survives in the intermediate channel,
compare the manipulations in (I:5.101). Next the three-point vertices on
the left-most vertical ribbon are dragged around the trace. The resulting
invariant in $S^3$ is easily evaluated to give
  \be
  \Gamma_{X^s Y}(a,b,r,s,x,y) = \delta_{x,y} \, \delta_{r,s} \,
  \Fs a{\bar a}xxr0 \, \Gs r{\bar b}brx0 \, \Rs{}arx \, \Rs{}xbr .  \ee

\newpage
\sect{Examples for reversions}\label{sec:ex}

\subsection{The category of complex vector spaces}\label{sec:vect}

The simplest example of a \mtc\ is the category $\Vect_\complex$
of \findim\ complex vector spaces. It has only a single isomorphism class of
simple objects, the one containing the tensor unit $\one\,{\cong}\,\complex$. 
The braiding is symmetric and simply given by
$c_{U,V}(x{\otimes}y)\eq y{\otimes}x$.

Algebras in $\Vect_\complex$ are just ordinary algebras over the
complex numbers. As we will see, special symmetric Frobenius algebras are 
semisimple, and hence isomorphic to direct sums of matrix algebras.
\JA s, in turn, are semisimple algebras with involution. 
This simple form allows us to find all reversions explicitly \cite{SChA}.  
\medskip

For $\calc\eq\Vect_\complex$ one has by definition (see section I:3.4)
$A \,{\cong}\, \Atop$. Lemma \ref{le:Atop-mat} then tells us that every
symmetric special Frobenius algebra in $\Vect_\complex$ is isomorphic to 
a direct sum of matrix algebras, and hence semisimple. Without loss of 
generality we may thus take
  \be
  A = \bigoplus_{i=1}^r {\rm Mat}_{n_i}(\complex) \,.  \labl{eq:vec-Adef}
The defining properties \erf{eq:*ssFA1} and \erf{eq:*ssFA2} of a 
reversion $\sigma$ reduce to the statement that $\sigma$
is an involutive anti-automorphism,
  \be
  \sigma^2 = \id_A  \qquad {\rm and} \qquad
  \sigma(\vec X \vec Y) = \sigma(\vec Y)\, \sigma(\vec X)
  \quad{\rm for\ all~~}\vec X,\vec Y\iN A \,,  \ee 
i.e.\ $A$ is an \alg\ with involution.

It is also easy to write down a reversion on $A$. Let us write an
element $\vec X$ of $A$ as $\vec X\eq(X_1,...\,,X_r)$ and abbreviate by
$\vec X^{\rm t}$ the element $(X_1^{\rm t},...\,,X_r^{\rm t})$. Then
$\sigma_\circ$ defined by
  \be
  \sigma_\circ(\vec X) := \vec X^{\rm t} \ee
is a reversion on $A$. Furthermore, as discussed in section 
\ref{sec:str}, any two reversions can be related by an automorphism 
of $A$,
  \be
  \sigma = \omega \circ \sigma_\circ \,.  \ee
Thus the classification of all reversions on $A$ 
reduces to the standard problem of finding all automorphisms of $A$.

To describe this classification,
denote the $i$th simple matrix block of $A$ by $A_i$ and let ${\rm e}_i$
be the corresponding primitive idempotent, i.e.\ the unit matrix in 
$A_i$. The ${\rm e}_i$ span the center of $A$ and obey
  \be
  \id_A = \sum_{i=1}^r {\rm e}_i \qquad {\rm and}
  \qquad {\rm e}_i\, {\rm e}_j = \delta_{i,j}\, {\rm e}_i \,.
  \labl{eq:vec-idem}
Furthermore, for every idempotent $p$ in the center of $A$,
${\rm e}_i\, p\eq p$ implies that either $p\eq{\rm e}_i$ or $p\eq0$.

Let $\omega$ be an automorphism of $A$. It is easily verified that the 
set $\{ \omega({\rm e}_i)\,|\, i\eq1,...\,,r \}$ satisfies the properties 
\erf{eq:vec-idem}. Since the set of primitive idempotents is unique, it 
follows that there is a permutation $\pi\iN S_r$ such that 
$\omega({\rm e}_i)\eq {\rm e}_{\pi^{-1}(i)}$. Because of $x {\rm e}_i\eq x$ 
for $x \iN A_i$, this implies that $\omega(x)\, {\rm e}_{\pi^{-1}(i)}\eq
\omega(x)$, so that $\omega(x)\iN A_{\pi^{-1}(i)}$. Thus the automorphism 
$\omega$ induces an isomorphism $\omega_i{:}\; A_i\,{\to}\,A_{\pi^{-1}(i)}$.
This is only possible if the two matrix blocks have the same size, 
$n_i\eq n_{\pi^{-1}(i)}$, and then $\omega_i$ is an automorphism of
${\rm Mat}_{n_i}(\complex)$.

By the Skolem\hy Noether theorem (see e.g.\ section I.3 of \cite{FAde})
all automorphisms of the simple algebra $A_i$ are inner,
$\omega^{}_i(X)\eq U_i^{} X U_i^{-1}$ for some invertible $U_i\iN A_i$. The
automorphism $\omega$ thus acts on a general element $\vec X \iN A$ as
  \be
  \omega(\vec X)
  = (U^{}_{\pi(1)} X^{}_{\pi(1)} U_{\pi(1)}^{-1} ,
  ...\, ,
  U^{}_{\pi(r)} X^{}_{\pi(r)} U_{\pi(r)}^{-1} ) \,.  \labl{eq:vec-sig}

However, while $\erf{eq:*ssFA2}$ is true for all choices of $U_i$ in 
\erf{eq:vec-sig}, the requirement $\sigma^2\eq\id_A$ gives the extra
condition that
  \be
  \sigma(\sigma(\vec X)) = (
  U^{}_{\pi(1)} (U_{\pi(\pi(1))}^{-1})^{\rm t} X^{}_{\pi(\pi(1))}
  U_{\pi(\pi(1))}^{\rm t} U_{\pi(1)}^{-1} ,...\,, ) \ee
must be equal to $\vec X$ for all $\vec X\iN A$.
This implies, first of all, that $\pi$ is a
permutation of order two. Second, it follows that
the matrix $U_i^{\rm t} U_{\pi(i)}^{-1}$
commutes with all matrices $X_i$, for $i\eq1,...\,,r$,
and hence is a multiple of the identity matrix, so that
$U_i^{\rm t} U_{\pi(i)}^{-1}\eq\lambda_i\onemat_{n_i\times n_i}$, i.e.
  \be
  U_i^{\rm t} = \lambda_i\, U_{\pi(i)} \,.  \ee
Relabelling $U_{\pi(i)}\,{\to}\, U_i$,
we conclude that all reversions on $A$ are of the form
  \be
  \sigma^\pi_{U_1,\dots,U_m}(\vec X) =
  (U^{}_1 X_{\pi(1)}^{\rm t}U_1^{-1},...\,,U^{}_m X_{\pi(m)}^{\rm t}U_m^{-1})
  \,, \labl{eq:vec-ex-sigma}
where $\pi$ is a permutation of order two
and $U_{\pi(i)}\eq\lambda_i^{} U_i^{\rm t}$ for some $\lambda_i\iN\complex$.

\medskip

We would now like to bring \erf{eq:vec-ex-sigma} to a standard form by 
choosing an appropriate isomorphism of algebras with involution.
Consider the automorphism $\omega$ of the algebra $A$ given by
  \be
  \omega(\vec X) =
  (V^{}_1 X^{}_1 V_1^{-1}, ...\,, V^{}_m X^{}_m V_m^{-1}) \,.  \ee
Note that this is in general only an automorphism of algebras, but
not of algebras with involution. Indeed, under $\omega$ the reversion
$\sigma$ changes to $\tilde\sigma\eq\omega \cir \sigma \cir \omega^{-1}$.
The action of $\tilde\sigma$ is given by
  \be
  \tilde\sigma(\vec X) =
  ( V_1 U_1 V_{\pi(1)}^{\rm t} X_{\pi(1)}^{\rm t}
  (V_1 U_1 V_{\pi(1)}^{\rm t})^{-1}, ...\, ) \,.  \ee
Now given $i\iN\{1,2,...\,,r\}$ with $\pi(i)\,{\ne}\,i$, take the smaller of 
the two values $i$, $\pi(i)$. For concreteness, assume this is $i$.
Choose $V_i\eq U_i^{-1}$ and $V_{\pi(i)}\eq\onemat$. Then on the
$i$th component we have $\tilde\sigma(\vec X)|_i\eq X_{\pi(i)}^{\rm t}$
while on the $\pi(i)$th component we find
  \be
  \tilde\sigma(\vec X)|_{\pi(i)} =
  V_{\pi(i)} U_{\pi(i)} V_i^{\rm t} X_i^{\rm t}
  (V_{\pi(i)} U_{\pi(i)} V_i^{\rm t})^{-1} = X_i^{\rm t} \,, \ee
where we also used $U_{\pi(i)}\eq\lambda_i U_i^{\rm t}$.

On the other hand, for $\pi(i)\eq i$ the condition
$U_{\pi(i)}\eq\lambda_i U_i^{\rm t}$ implies that
$U_i\eq{\pm} U_i^{\rm t}$, and the $i$th component of $\tilde\sigma$ reads
  \be
  \tilde\sigma(\vec X)|_{i} =
  V^{}_{i} U^{}_{i} V_i^{\rm t} X_i^{\rm t}
  (V^{}_{i} U^{}_{i} V_i^{\rm t})^{-1} \,.  \ee
By an appropriate choice of $V_i$ we can ensure that the combination 
$V_{i} U_{i} V_i^{\rm t}$ is either equal to the identity in 
${\rm Mat}_{n_i}(\complex)$ or else is built out of antisymmetric 
$2{\times} 2$ blocks along the diagonal.

\medskip

We conclude that given a $\complex$-algebra $(A,\sigma)$ with involution, we
can find an isomorphism $\omega \iN \Hom(A,A)$ which is an isomorphism, as 
algebras with involution, of $(A,\sigma)$ and $(A,\tilde\sigma)$ with
$\tilde\sigma\eq\omega \cir \sigma \cir \omega^{-1}$ of the special form
  \be
  \tilde\sigma = \sigma^\pi_{D_1,\dots,D_r} \,, \ee
where $\pi$ is again a permutation of order two and the
$D_i \iN {\rm Mat}_{n_i}(\complex)$ are the identity
matrix whenever $\pi(i)\,{\ne}\, i$. For $\pi(i)\eq i$ the matrix $D_i$
is either the identity matrix or the antisymmetric matrix with blocks
\,\ppmatrix01{-1}0\, along the diagonal.

\subsection{The Ising model}\label{sec:ising-ex}

In this section we treat the example of the critical Ising model in some 
detail. The results obtained in the section are well known; see 
e.g.\ chapter 12 of \cite{DIms} and \cite{bips,prss,osaf,osaf2,yAma}
for an (incomplete) list of references treating the Ising model from a 
\cft\ point of view. The two different M\"obius strip and Klein bottle 
amplitudes for the Ising model were first written down in \cite{prss}.
In \cite{kaOk,luwu2,otpe} the Ising model on a lattice
has been analysed on the M\"obius strip and Klein bottle.
What is new in our presentation is the entirely systematic
manner in which the various data are obtained.

Below we start by determining all haploid algebras in the
category of $c\eq\frac12$ Virasoro representations. Next we run
through the steps 1)\,--\,3) of section \ref{sec:*-class} so as to find all
possible reversions. Then the algebra $A\eq\one\,{\oplus}\,\Eps$, 
which allows for two reversions, is considered. For this algebra
the modules and bimodules are worked out and the 
M\"obius and Klein bottle coefficients are computed.

\subsubsection*{Algebras and induced modules}

Let $\Ising$ be the modular category given by the chiral data of the
two-dimensional critical Ising model. We will define it in more detail 
below. For now just note that $\Ising$ has precisely three non-isomorphic
simple objects $\one$, $\sigma$ and $\Eps$, with fusion rules
  \be
  \sigma\oti\sigma \cong \one \,{\oplus}\, \Eps \,, \qquad
  \sigma\oti\Eps \cong \sigma \,, \qquad 
  \Eps\oti\Eps \cong \one  \,.  \ee
{}From this information alone we can already drastically reduce the 
number of objects in $\Ising$ that can carry the structure of a simple
symmetric special Frobenius algebra. {}From proposition \ref{th:sssFA} we 
know that each Morita class of such algebras has a haploid representative, 
so that for the moment we restrict our attention to the case $\lr\one A\eq1$
(recall the notations (I:2.3) and (I:4.5)). Thus we would like to turn
the object
  \be
  A = \one \oplus \sigma^{\oplus n_\sigma} \oplus \Eps^{\oplus n_\Eps}
  \labl{eq:ising-alg-aux}
into a symmetric special Frobenius algebra. 

Let us suppose that $A$ is such an algebra, and compute the embedding 
structure of the induced $A$-modules, using the reciprocity relation 
$\lra{\inda(U)}{\inda(V)}\eq\lr{U}{A{\otimes}V}$. We
find the following table for the numbers $\lra{\inda(U)}{\inda(V)}$:
  \be  \mbox{
  \begin{tabular}{c|ccc}
           & $\one$     & $\sigma$   & $\Eps$ \\[1.3pt] \hline \\[-10pt]
  $\one$   &  $\,1$       & $n_\sigma$ & $n_\Eps$ \\
  $\sigma$ & $\,n_\sigma$ & $1+n_\Eps$ & $n_\sigma$ \\
  $\Eps$   & $\,n_\Eps$   & $n_\sigma$ & $1$
  \end{tabular}
  } \labl{n-table}
The diagonal entries equal to $1$ tell us that $\inda(\one)$ and $\inda(\Eps)$ 
are simple $A$-modules. It follows that 
$n_\Eps\eq\lra{\inda(\one)}{\inda(\Eps)}$ must be either one or zero, 
depending on whether the two induced modules are isomorphic or not.
\\[3pt]
(a)~\,Suppose that $n_\Eps\eq0$. Then also $\inda(\sigma)$ is simple, so that
$\lra{\inda(\one)}{\inda(\sigma)} \le 1$. Suppose $n_\sigma=1$. Then from table
\erf{n-table} we read off that both $\,\inda(\one)\,{\cong}\,\inda(\sigma)$ and 
$\,\inda(\sigma)\,{\cong}\,\inda(\Eps)$, whereas $n_\Eps\eq0$ implies that 
$\inda(\one)$ is not isomorphic to $\inda(\Eps)$. This contradiction tells us
that for $n_\Eps\eq0$ we also have $n_\sigma\eq0$.
\\[3pt]
(b)~\,Suppose that $n_\Eps\eq1$. Then $\inda(\one)\,{\cong}\,\inda(\Eps)$ and
$\inda(\sigma)\,{\cong}\,M_1 \,{\oplus}\, M_2$ with $M_{1,2}$ non-iso\-mor\-phic 
simple $A$-modules. Thus at most one of the $M_{1,2}$ can be isomorphic to $\one$,
and inspecting again table \erf{n-table} we deduce that $n_\sigma\,{\le}\, 1$.

\smallskip

The analysis of induced $A$-modules thus proves to be surprisingly restrictive.
It reduces the list \erf{eq:ising-alg-aux} to the possibilities
  \be
  A = \one \qquad{\rm or}\qquad
  A = \one \oplus \Eps  \qquad{\rm or}\qquad
  A = \one \oplus \sigma \oplus \Eps \,.  \ee
The object $\one$ is trivially a simple symmetric special 
Frobenius algebra. For $A\eq\one\,{\oplus}\,\Eps$ we note that the structure
constant $m_{\Eps\Eps}^\one$ (we use the notation introduced in section 
I:3.6) is merely a normalisation and can be set to one. Since
$\Eps\oti\Eps\,{\cong}\,\one$ it follows that the multiplication on
$\one\,{\oplus}\,\Eps$ is unique. Furthermore, since 
$\one\,{\oplus}\,\Eps\,{\cong}\,\sigma^\vee{\otimes}\,\sigma$,
and since any object of the form $U^\vee{\otimes}\,U$ can be endowed
with the structure of a simple symmetric special Frobenius
algebra, the multiplication on $\one\,{\oplus}\,\Eps$ indeed
exists. The third case $A\eq\one\,{\oplus}\,\sigma\,{\oplus}\,\Eps$, on the
other hand, does not allow for a special Frobenius algebra
structure. But in order to see this we still need a better
knowledge of the category $\Ising$.

\subsubsection*{Chiral data}

Let us summarise the chiral data $\Ising$. The Virasoro central charge 
of the Ising model is $c\eq\frac12$, so that $\kappa\eq \eE^{\pi\ii /8}$.
There are three isomorphism classes of simple objects, from which we
choose representatives $\one$, $\sigma$ and $\Eps$. Their conformal 
weights and quantum dimensions are
  \be
  \Delta_\one = 0 \,,\quad
  \Delta_\sigma = 1/16 \,,\quad
  \Delta_\Eps = 1 \qquad {\rm and} \qquad
  \dim(\one) = 1 = \dim(\Eps)\,,\quad
  \dim(\sigma) = \sqrt{2} \,.  \ee
In terms of the weights $\Delta_k$ the
twist and the braiding of $\Ising$ can be expressed as
  \be
  \theta_k = \eE^{-2 \pi\ii \Delta_k} \qquad {\rm and} \qquad
  \Rs{}abc = \eE^{-\pi\ii(\Delta_c-\Delta_a-\Delta_b)} \,.
  \labl{eq:Is-th-R}
The $S$-matrix and the $P$-matrix (see before \erf{eq:Ph-def}) 
are given by
  \be
  S \equiv \left(\begin{array}{ccc}
      S_{\one\one}   &\! S_{\one\sigma}   \!& S_{\one\Eps} 
    \\{}\\[-.98em]
      S_{\sigma\one} &\! S_{\sigma\sigma} \!& S_{\sigma\Eps} 
    \\{}\\[-.98em]
      S_{\Eps\one}   &\! S_{\Eps\sigma}   \!& S_{\Eps\Eps}
    \end{array}\right)
  = \left(\begin{array}{ccc}
    \tfrac 12 &\! \tfrac 1{\sqrt{2}} \!& \tfrac 12 
    \\{}\\[-.98em]
    \tfrac 1{\sqrt{2}}    &\! 0 \!&     -\tfrac1{\sqrt{2}} 
    \\{}\\[-.98em]
    \tfrac 12 &\! -\tfrac1{\sqrt{2}} \!& \tfrac12 \end{array}\right)
  \quad\; {\rm and} \quad\;
  P = \left(\begin{array}{ccc}
    \cos \tfrac \pi 8 & 0 & \sin \tfrac \pi 8 \\{}\\[-.98em]
    0 & 1 & 0 \\{}\\[-.98em]
    \sin \tfrac \pi 8 & 0 &\! -\cos \tfrac \pi 8 \end{array}\right) .\
  \ee
Finally, the fusion matrices can be described as follows.
We will only consider $\FF$-matrix elements allowed by fusion
(all others are zero). All those $\Fs abcdpq$ for which one or
more of $a,b,c,d$ is $\one$ are equal to one. The other $\FF$'s read,
with a suitable choice of gauge,
  \be \bearll
  \Fs\Eps\Eps\Eps\Eps\one\one = 1 \,, &
  \Fs\sigma\Eps\sigma\Eps\sigma\sigma 
  = \Fs\Eps\sigma\Eps\sigma\sigma\sigma = -1 \,, \\{}\\[-.5em]
  \Fs\sigma\sigma\Eps\Eps\sigma\one 
  = \Fs\Eps\Eps\sigma\sigma\sigma\one = 2 \,,\qquad &
  \Fs\Eps\sigma\sigma\Eps\one\sigma 
  = \Fs\sigma\Eps\Eps\sigma\one\sigma = \tfrac 12 \,, \\{}\\[-.5em]
  \multicolumn2l {
  \Fs\sigma\sigma\sigma\sigma xy = 
  \pmatrix{ 
  \FF_{\one\one} & \FF_{\one\Eps} \cr
  \FF_{\Eps\one} & \FF_{\Eps\Eps} }
  = \pmatrix{ \tfrac1{\sqrt{2}} & \tfrac1{2 \sqrt{2}} \cr 
              \sqrt{2} & -\tfrac1{\sqrt{2}} } . }
  \eear\ee
The inverse matrices $\GG$ are related to $\FF$ via (I:2.61), which
in the present case simplifies to
  \be
  \Gs abcdpq = \Fs cbadpq \,.  \ee

\subsubsection*{Classification of algebras}

We now show that there is no symmetric special Frobenius algebra 
structure on the object $A\eq\one\,{\oplus}\,\sigma\,{\oplus}\,\Eps$. 
To this end we consider the associativity constraint (I:3.78) 
for three choices of the quadruple $(abc)d$:
  \be\begin{array}{rll}
  ({\rm i}):~& (\Eps\sigma\sigma)\Eps \ \   \Rightarrow \ \,  &
    m_{\Eps\sigma}^\sigma \, m_{\sigma\sigma}^\Eps 
    = m_{\sigma\sigma}^\one \, m_{\Eps\one}^\Eps \,
      \Fs\Eps\sigma\sigma\Eps\one\sigma \,,
  \\[8pt]
  ({\rm ii}):~& (\sigma\Eps\sigma)\Eps \ \   \Rightarrow \ \,  &
    m_{\sigma\Eps}^\sigma \, m_{\sigma\sigma}^\Eps
    = m_{\Eps\sigma}^\sigma \,m_{\sigma\sigma}^\Eps \, 
      \Fs\sigma\Eps\sigma\Eps\sigma\sigma \,,
  \\[8pt]
  ({\rm iii}):~& (\sigma\sigma\Eps)\Eps \ \   \Rightarrow \ \,  &
    m_{\sigma\sigma}^\one \, m_{\one\Eps}^\Eps
    = m_{\sigma\Eps}^\sigma \, m_{\sigma\sigma}^\Eps \, 
      \Fs\sigma\sigma\Eps\Eps\sigma\one\,.
  \eear\labl{(abc)d}
Since $A$ is required to be special, the morphisms (I:3.80) must
be invertible. This forces $m_{aa}^\one$ to be nonzero; we normalise it 
to $m_{aa}^\one\eq1$. With the explicit values for the $\FF$'s we then 
find from (\ref{(abc)d}(i)) and (\ref{(abc)d}(iii)) that 
$m_{\sigma\sigma}^\Eps\,{\neq}\,0$
and that $m_{\sigma\Eps}^\sigma\eq m_{\Eps\sigma}^\sigma\,{\neq}\,0$.
On the other hand, upon inserting the value of 
$\Fs\sigma\Eps\sigma\Eps\sigma\sigma$ and using that 
$m_{\sigma\sigma}^\Eps\,{\neq}\,0$, (\ref{(abc)d}(ii)) gives
$m_{\sigma\Eps}^\sigma\eq{-}m_{\Eps\sigma}^\sigma$, so that we have
produced a contradiction. Thus there is no associative multiplication
on $\one{\oplus}\sigma{\oplus}\Eps$ that is also special.

The complete list of (isomorphism classes of) haploid
symmetric special Frobenius algebras in $\Ising$ thus just consists of
$\one$ and the algebra
$A\eq\one\,{\oplus}\,\Eps$ with multiplication $m_{\Eps\Eps}^\one\eq1$.
Moreover, since the algebra structure on $A$ is unique, we have
$A\,{\cong}\,\sigma^\vee{\otimes}\,\sigma$ not only as objects, but also as 
algebras; thus $A$ is in the Morita class of $\one$. 

To summarise, there is only a single Morita class of simple
symmetric special Frobenius algebras in $\Ising$, and each
such algebra is isomorphic to an algebra of the form 
$U^\vee{\otimes}\,U$ for a (not necessarily simple) object $U$ of $\Ising$.

\subsubsection*{Classification of reversions}

To classify the possible reversions, we work through
the steps 1)\,--\,3) of section \ref{sec:*-class}.
\\[3pt]
Step 1): As seen above, there is only a single Morita class of
simple symmetric special Frobenius algebras in 
$\Ising$. We choose the representative $A\eq\one$.
\\[3pt]
Step 2): For $A\eq\one$ the simple modules are just the simple objects
of $\Ising$, i.e.\ we have three simple $A$-modules
  \be
  M_\one = \one \,,\qquad M_\sigma = \sigma \,,\qquad M_\Eps = \Eps \,.
  \ee
\\[3pt]
Step 3a): We must solve \erf{eq:*ssFA-bas-simple} for the three algebras
\be
  B^a_\one = \one \,,\qquad 
  B^a_\sigma = \sigma^\vee{\otimes}\,\sigma \cong \one \oplus \Eps \,,\qquad 
  B^a_\Eps = \Eps^\vee{\otimes}\,\Eps \cong \one \,.  \ee
On $B^a_\one$ and $B^a_\Eps$ there is the unique choice $\sigma(\one)\eq1$
for the reversion. For $B^a_\sigma$ we set 
$\sigma(k)\,{=:}\,s_k\eE^{-\pi\ii\Delta_k}$ for $k\iN\{\one,\Eps\}$.
Then the first condition in \erf{eq:*ssFA-bas-simple} forces $s_k\iN\{\pm1\}$,
while with the special form of $\RR$ in \erf{eq:Is-th-R}
the second condition reads
  \be
  m_{ij}^{~k} \, s_k = s_i \, s_j \, m_{ji}^{~k} \,.  \ee
These relations are solved by $s_\one\eq1$ and $s_\Eps\eq{\pm} 1$. We have
thus found two distinct reversions on $B^a_\sigma$.
\\[3pt]
Step 3b): The only case we need to investigate is 
$B^b_{\one\Eps}\eq(\one{\oplus}\Eps)^{\!\vee}{\otimes}\,(\one{\oplus}\Eps)$. 
It turns out that this leads again to two possible reversions, 
both of which are related to the $s_\Eps\eq{-}1$ reversion on
$B^a_\sigma$ via proposition \ref{pr:B-star}.
The slightly lengthy details are presented in appendix \ref{sec:ising-star}.

\subsubsection*{The algebra $A\eq\one{\oplus}\Eps$}

We will now investigate the algebra $A\eq\one{\oplus}\Eps$ in more detail.
The multiplication and comultiplication (recall equation (I:3.83)) are
given by
  \be
  m_{\one\one}^\one = m_{\one\Eps}^\Eps = m_{\Eps\one}^\Eps = m_{\Eps\Eps}^\one = 1
  \qquad {\rm and} \qquad
  \Delta^{\one\one}_\one = \Delta^{\one\Eps}_\Eps = 
  \Delta^{\Eps\one}_\Eps = \Delta^{\Eps\Eps}_\one = \tfrac12 \,, 
  \labl{eq:Is-Amult}
respectively. The two possible reversions on $A$ are
  \be
  \sigma(\one) = 1 \,,\qquad \sigma(\Eps) = s_\Eps / \ii 
  \quad~{\rm with~~} s_\Eps = \pm 1 \,.  \labl{eq:A-star-se}

Before working out the left $A$-modules and $A$-$A$-bimodules, let us compute 
the partition function to see how many of them there are. Applying 
formula (I:5.85) gives $Z_{ij}\eq\delta_{i,j}$, as expected. {}From theorem 
I:5.18 and remark I:5.19(ii) we then see that
  \bea
  \#\{\mbox{isom.\,classes\,of\,simple\,$A$-modules}\} = \tr Z = 3 
  \qquad{\rm and}\\[5pt]
  \#\{\mbox{isom.\,classes\,of\,simple\,$A$-$A$-bimodules}\}
   = \tr(ZZ^{\rm t}) = 3 \,.  \eear\labl{eq:ising-modcount}
Consider the induced modules
$\inda(U)$. The dimensions $\lra{\inda(U)}{\inda(V)}$ are found to be
  \be
  \begin{tabular}{c|ccc}
           \,& $\;\one$     & $\sigma$   & $\Eps$ \\[1.3pt] \hline \\[-10pt]
  $\one$   \,& $\;1$ & $0$ & $1$ \\
  $\sigma$ \,& $\;0$ & $2$ & $0$ \\
  $\Eps$   \,& $\;1$ & $0$ & $1$
  \end{tabular} \ee

\noindent
Thus $\inda(\one)\,{\cong}\,\inda(\Eps)$ and 
$\inda(\sigma)\,{\cong}\,F^+ \,{\oplus}\,F^-$ with two simple modules
$F^\pm$. Since as an object we have $\inda(\sigma)\,{\cong}\,\sigma\,
{\oplus}\,\sigma$, this means that the simple object $\sigma$ can be 
turned into an $A$-module in two distinct ways. Let us abbreviate
$\rho_a\,{:=}\,\rho_{a,\sigma}^{F^{\pm}\;\sigma}$, where the latter
notation is as introduced in (I:4.61). The representation
property (I:4.62) then reads
  \be
  \rho_\one = 1 \qquad {\rm and} \qquad
  \rho_a \, \rho_b \, \Fs ab\sigma\sigma\sigma c = 
  \delta_{c\In A} \, m_{ab}^c \, \rho_c \,,  \ee
where $\delta_{c\In A}\eq1$ if $U_c$ is a subobject of $A$ and 
$\delta_{c\In A}\eq0$ else.
The only non-trivial condition arises from $a\eq b\eq \Eps$. Using the
explicit values of $\FF$ it reads $(\rho_\Eps)^2\eq\tfrac12$. Altogether
we get the following result for the simple $A$-modules:
  \be
  \begin{tabular}{ccl}
  $A$-module & as object & representation morphisms \\[1.3pt] \hline \\[-8.5pt]
  $A$   & $\one\,{\oplus}\,\Eps$ & $\rho_{a,b}^{A~c}\eq m_{ab}^c$ \\{}\\[-.6em]
  $F^+$ & $\sigma$ & $\rho_{\one,\sigma}^{F^+\,\sigma} \eq 1 \,,~~
        \rho_{\Eps,\sigma}^{F^+\,\sigma} \eq \tfrac1{\sqrt{2}}$  \\{}\\[-.6em]
  $F^-$ & $\sigma$ & $\rho_{\one,\sigma}^{F^-\,\sigma} \eq 1 \,,~~
        \rho_{\Eps,\sigma}^{F^-\,\sigma} \eq {-}\tfrac1{\sqrt{2}}$ 
  \end{tabular} \ee
The effect of conjugation $M \,{\mapsto}\, M^\sigma$ on the simple
modules is encoded in the boundary conjugation matrix. Evaluating
\erf{eq:conj-inv} yields
  \be
  C^\sigma = \left(\begin{array}{ccc}
    C^\sigma_{F^+F^+} & C^\sigma_{F^+A} & C^\sigma_{F^+F^-}
  \\{}\\[-.98em]
    C^\sigma_{A F^+}  & C^\sigma_{A A}  & C^\sigma_{A F^-}
  \\{}\\[-.98em]
    C^\sigma_{F^-F^+} & C^\sigma_{F^-A} & C^\sigma_{F^-F^-} 
  \end{array}\right)
  = \left(\begin{array}{ccc}
    \delta_{s_\Eps,1} & 0 & \delta_{s_\Eps,-1}
  \\{}\\[-.98em]
    0 & 1 & 0 
  \\{}\\[-.98em]
    \delta_{s_\Eps,-1} & 0 & \delta_{s_\Eps,1} \end{array}\right).
  \labl{eq:ising-conj}
Thus the module $A$ is fixed under conjugation for both
reversions, as already shown in general in proposition
\ref{pr:*mod}(ii), while for $s_\Eps\eq{-}1$ the conjugation exchanges
the modules $F^+$ and $F^-$.

To obtain the $A$-$A$-bimodules, in general one would have to
decompose the induced $A{\otimes}A_{\rm op}$-left modules. In the present 
case it turns out to be sufficient to look at the $\alpha$-induced
bimodules. Let us recall the formulas (I:5.64), (I:5.65) as well as 
proposition 2.36 of \cite{ffrs}:
  \be\bearl
  \dimc\,\Hom_{\!A|A}(\alpha^-(U_k),\alpha^+(U_l)) = Z_{\bar l,k} \,,\\{}\\[-.6em]
  \dimc\,\Hom_{\!A|A}(\alpha^+(U_k),\alpha^+(U_l)) =  
    \dimc\,\Hom( C_l(A){\otimes}U_k,U_l ) \,,\\{}\\[-.6em]
  \dimc\,\Hom_{\!A|A}( \alpha^-(U_k), \alpha^-(U_l) ) = 
    \dimc\,\Hom( C_r(A){\otimes}U_k,U_l ) \,. 
  \eear\labl{dimUU}
The left and right centers $C_{l/r}(A)$ of $A$ are both equal to $\one$.
Evaluating the dimensions \erf{dimUU} for the six $\alpha$-induced bimodules
$\alpha^\pm(U_k)$ obtained from $U_k\eq\one,\sigma,\Eps$ gives three 
iso\-mor\-phism classes of simple bimodules,
  \be
  \alpha^+(\one) \cong \alpha^-(\one) \,,\qquad
  \alpha^+(\sigma) \cong \alpha^-(\sigma) \,,\qquad
  \alpha^+(\Eps) \cong \alpha^-(\Eps) \,.  \labl{eq:is-bimod}
The counting argument \erf{eq:ising-modcount} shows that in this way we
have found a representative for each isomorphism class of simple
$A$-$A$-bimodules.

Recall from section \ref{sec:defect} that $A$-$A$-bimodules label 
defect lines, and that there is a subclass of defects which can be 
wrapped around a non-orientable cycle without marking a point. Such 
defects are labelled by bimodules $X$ with $X^s \,{\cong}\,X$. Combining 
\erf{eq:is-bimod} with proposition \ref{pr:bimod-alphaind} we see
that all defects in the Ising model are of this type.

Finally we also need bases of local morphisms $\mu_\alpha^{kl} \iN 
\Loc(A{\otimes}U_l,U_{\bar k})$, see \erf{eq:mu-basis}. The dimensions
of these spaces are $Z_{kl}\eq\delta_{k,l}$. Now the morphism
spaces $\Hom(A{\otimes}\one,\one)$ and $\Hom(A{\otimes}\Eps,\Eps)$
are already one-dimensional, so that we can choose
$\mu^{\one\one}$ and $\mu^{\Eps\Eps}$ to be any
nonzero element in the respective spaces. For the two-dimensional space
$\Hom(A{\otimes}\sigma,\sigma)$ the situation is more complicated since 
only a one-dimensional subspace is local. To find this subspace
one must determine the image of the projector $P_\sigma$ as introduced
in \erf{eq:PX-def}. 

Expanding $\mu_\alpha^{kl}$ and its dual
as in \erf{eq:mu-expand}, altogether we find the local morphisms
to be given by
  \be \begin{array}{lll}
    \mu^{\one\one}_{\one} = \sqrt{2} \,,\qquad & 
    \mu^{\sigma\sigma}_{\one} = 0 \,,\qquad & 
    \mu^{\Eps\Eps}_{\one} = -\sqrt{2} \,, \\{}\\[-.6em]
    \mu^{\one\one}_{\Eps} = 0 \,,\qquad & 
    \mu^{\sigma\sigma}_{\Eps} = \eE^{\pi \ii /4} \,,\qquad & 
    \mu^{\Eps\Eps}_{\Eps} = 0 \,. 
  \eear \labl{eq:ising-mu}
The nonzero constants are normalisations and can be chosen
at will. We have picked them in such a way that the $S^A$-matrix
computed below coincides with the $S$-matrix. Among the vanishing
constants, the only non-trivial case is $\mu^{\sigma\sigma}_\one\eq0$. 
The corresponding dual local morphisms are
  \be \begin{array}{lll}
    \bar\mu^{\one\one}_{\one} = \tfrac1{\sqrt{2}} \,,\qquad & 
    \bar\mu^{\sigma\sigma}_{\one} = 0 \,,\qquad & 
    \bar\mu^{\Eps\Eps}_{\one} = -\tfrac1{\sqrt{2}} \,,\\{}\\[-.6em]
    \bar\mu^{\one\one}_{\Eps} = 0 \,,\qquad & 
    \bar\mu^{\sigma\sigma}_{\Eps} = \eE^{-\pi \ii /4} \,,\qquad & 
    \bar\mu^{\Eps\Eps}_{\Eps} = 0 \,. 
  \eear \ee

\subsubsection*{M\"obius and Klein bottle amplitudes for $A$}

The coefficients of the M\"obius and Klein bottle amplitude can be obtained 
with the crossed channel relations \erf{eq:M-crossed} and \erf{eq:K-crossed}.
To this end we must compute the quantities $S^A_{R,k}$, $\gamma^\sigma_{k}$ 
and $g^{\bar kk}$. 

Let us start with $S^A_{R,k}$ as given in \erf{eq:SA-inv-1}
and \erf{eq:SA-inv-2}. {}From the chiral data for $\Ising$ one computes
  \bea
  I_S(\one,\one,\one) = I_S(\one,\one,\Eps) = 
    I_S(\Eps,\one,\Eps) = I_S(\Eps,\one,\one) = \tfrac12 \,,\\[7pt]
  I_S(\one,\one,\sigma) = \tfrac1{\sqrt{2}} \,,\qquad
  I_S(\Eps,\one,\sigma) = -\tfrac1{\sqrt{2}} \,,\qquad
  I_S(\sigma,\Eps,\sigma) = 2 \, \eE^{-\pi \ii /4} \,, \eear \ee
which together with the data derived from the algebra
$A$ results in the matrix
  \be
  S^A \equiv \left(\begin{array}{ccc}
  S^A_{F^+,\one} & S^A_{F^+,\sigma} & S^A_{F^+,\Eps} 
  \\{}\\[-.98em]
  S^A_{A,\one} & S^A_{A,\sigma} & S^A_{A,\Eps} 
  \\{}\\[-.98em]
  S^A_{F^-,\one} & S^A_{F^-,\sigma} & S^A_{F^-,\Eps} \end{array}\right)
  = \left(\begin{array}{ccc}
    \tfrac 12 & \tfrac 1{\sqrt{2}} & \tfrac 12
  \\{}\\[-.98em]
    \tfrac 1{\sqrt{2}} & 0 & -\tfrac1{\sqrt{2}}
  \\{}\\[-.98em]
    \tfrac 12 & -\tfrac1{\sqrt{2}} & \tfrac12 \end{array}\right) .
  \ee
With normalisations as chosen in \erf{eq:ising-mu}, $S^A$ thus
coincides with the ordinary $S$-matrix of $\Ising$. While generically 
such a choice is not possible, in the present situation such a 
normalisation is guaranteed to exist because $A$ is 
Morita equivalent to $\one$.

Next we turn to the crosscap coefficients, for which we need the
invariants \erf{eq:Gamma-inv-2}. Inserting the chiral data gives
  \be
  I_\Gamma(\one,\one,\one) = I_\Gamma(\Eps,\Eps,\one) 
    = \cos \tfrac{\pi}8 \,, \qquad
  I_\Gamma(\one,\Eps,\one) = I_\Gamma(\Eps,\one,\one) 
    = -\ii \sin \tfrac{\pi}8  \,.  \ee
Combining this with \erf{eq:Gamma-inv-1} and \erf{eq:Ga-bas} and noting 
the trigonometric identities $\cos\tfrac\pi8 \,{+}\, \sin\tfrac\pi8 
    $\linebreak[0]$%
\eq \sqrt{2} \cos\tfrac\pi8$ and $\cos\tfrac\pi8 \,{-}\, \sin\tfrac\pi8
\eq \sqrt{2} \sin\tfrac\pi8$ results in
  \be
  \gamma_\one^\sigma = \cases{
    \cos\tfrac{\pi}{8} &{\rm for}\,\ $s_\Eps\eq1\,$, \cr
    \sin\tfrac\pi8 &{\rm for}\,\ $s_\Eps\eq{-}1\,$, }
  \qquad
  \gamma_\sigma^\sigma = 0 
  \,, \qquad
  \gamma_\Eps^\sigma = \cases{
    \sin\tfrac\pi8 &{\rm for}\,\ $s_\Eps\eq1\,$, \cr
    -\cos\tfrac\pi8 &{\rm for}\,\ $s_\Eps\eq{-}1\,$. }
  \ee
Finally the numbers $g^{\bar kk}$ are obtained from \erf{eq:g-inv}; we get
  \be
  g^{\one\one} = 1 \,, \qquad
  g^{\sigma\sigma} = \tfrac{s_\Eps}{\sqrt{2}} \,, \qquad
  g^{\Eps\Eps} = 1 \,.  \ee

We have now gathered all ingredients to evaluate the formulas 
\erf{eq:M-crossed} and \erf{eq:K-crossed}. A short calculation yields
  \be\begin{array}{lll}
  \rmm_{\one A} = 1  \,, \qquad & 
  \rmm_{\sigma A} = 0 \,, \qquad & 
  \rmm_{\Eps A} = s_\Eps \,, \\ [5pt]
  \rmm_{\one F^{\pm}} = \delta_{s_\Eps,1}  \,, \qquad & 
  \rmm_{\sigma  F^{\pm}} = 0 \,, \qquad & 
  \rmm_{\Eps  F^{\pm}} = \delta_{s_\Eps,-1} 
  \eear\labl{eq:Ising-Moebius}
for the coefficients of the M\"obius amplitude and
  \be
  \K_\one = 1 \,, \qquad  \K_\sigma = s_\Eps \,, \qquad 
  \K_\Eps = 1  \labl{eq:Ising-Klein}
for the Klein bottle \cite{prss}.

\subsubsection*{Lattice interpretation}

The appearance of two different reversions for the Ising model can be given
an intuitive lattice interpretation. Consider the Ising model realised on a
`square lattice', i.e.\ on a quadrangulation of the world sheet, with a
`spin' variable, which can take one of two possible values, placed at each
lattice site. We would like to think of the situation in terms of two
different geometric realisations of this spin variable.
In the first the two values are given by a small and a large circle,
respectively, while in the second they describe the direction of a
spin vector orthogonal to the lattice plane:
  \bea  \begin{picture}(410,42)(14,24)
  \put(49,0)     {\includeourbeautifulpicture{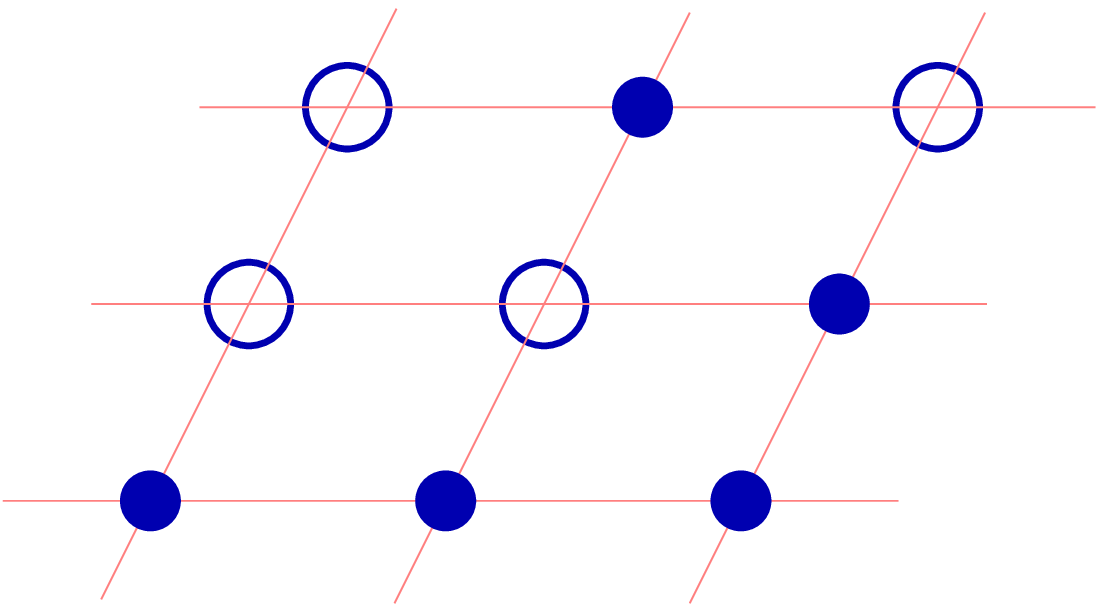}}
  \setlength{\unitlength}{1pt}
  \put(38,53)    { (1) }
  \put(271,53)   { (2) }
  \epicture09 \labl{pic128}
This visualisation suggests two different rules for transporting the spin 
variable around a non-orientable cycle (compare section \ref{sec:defect}):
if we think of the spin variable as in case (1) of \erf{pic128}, then we 
obtain back the same value, whereas in case (2) the two values get 
exchanged. As long as we are only interested in orientable surfaces, the
two geometric realisations are indistiguishable, but as we will see
below, if we include  
     lattices 
on non-orientable surfaces, alternative (1) is 
described by $s_\Eps\eq1$ while (2) corresponds to $s_\Eps\eq{-}1$. 

Let us start by considering the possible boundary conditions. 
In the case of description (1) we have `fixed small ($s$)',
`fixed large ($l$)' and `free ($f$)' boundary conditions,
while for description (2) one can take `fixed up ($u$)', 
`fixed down ($d$)' and `free ($f$)'.
On the M\"obius strip one must choose a single boundary condition
$R$. Equation \erf{eq:ch1c} tells us that a segment of the
M\"obius strip looks like a strip with
boundary conditions $R$ and $R^\sigma$. The effect of boundary
conjugation can thus be read off by `transporting' the boundary
condition half-way along the boundary of the M\"obius strip.
In the description (1) this has no effect, i.e.\
  \be  s\mapsto s\,, \qquad  l\mapsto l\,, \qquad  f\mapsto f \,, \ee
while in (2) the direction of the spin in the fixed boundary conditions
is inverted,
  \be  u\mapsto d\,, \qquad  d\mapsto u\,, \qquad  f\mapsto f \,. \ee
Identifying $u$ and $s$ with $F^+$, $d$ and $l$ with $F^-$, and $f$ 
with $A$, this is precisely the behaviour encoded in formula
\erf{eq:ising-conj}.

Next consider the Klein bottle amplitude, with $\tau\eq2\ii L/R$,
  \be
  \K(R,L) = \K_\one \, \chii_\one(\tau) + \K_\sigma \, \chii_\sigma(\tau)
  + \K_\Eps \, \chii_\Eps(\tau) \,.  \ee
This gives the partition function of the critical Ising model
on a rectangular lattice of width $R$ and length $L$ for which the
two vertical boundaries are periodically identified, while for the
two horizontal boundaries antiperiodic boundary conditions are chosen. 
Furthermore the antiperiodic identification exchanges 
$\uparrow$ with $\downarrow$ in realisation (2). Configurations of low 
energy in (1) tend to have all spins aligned, while (2) effectively 
has an additional line of frustration at which the interaction between
neighbouring spins is reversed. Thus configurations with long range
order will tend to have a higher energy for (2) than for (1). For 
example, the state on the rectangular lattice with all spin values equal
has zero energy for (1), but has nonzero energy for (2), originating 
from the interaction across the horizontal boundary. From the lattice
point of view we therefore expect that
  \be
  \K^{(1)} > \K^{(2)} \,.  \ee
According to \erf{eq:Ising-Klein}, this is indeed precisely what 
happens, since $\K^{(1)} \,{-}\, \K^{(2)}\eq 2\chii_\sigma(\tau)\,{>}\,0$.

Similar considerations can be repeated for the partition function 
$\rmM(R,L)$ on a M\"obius strip of width $R$ and circumference $L$ 
with free boundary condition. Again the realisation (2) has
effectively an additional line of frustration, so that we
expect $\rmM^{(1)} \,{>}\, \rmM^{(2)}$. The free boundary condition
is labelled by $A$, and the partition function reads
\be
  \rmM(R,L) = \rmm_{A\one} \, \hat{\chii}_\one(\tau)
  + \rmm_{A\sigma} \, \hat{\chii}_\sigma(\tau)
  + \rmm_{A\Eps} \, \hat{\chii}_\Eps(\tau) \,, \ee
where $\tau\eq\tfrac12(1{+}\ii L/R)$ and 
$\hat{\chii}(\tau)\eq \eE^{-\pi\ii(\Delta - c/24)}\chii(\tau)$.
Substituting the coefficients \erf{eq:Ising-Moebius} then indeed
results in $M^{(1)} \,{-}\, M^{(2)}\eq 2\hat{\chii}_\Eps(\tau)\,{>}\,0$.

\newpage
\appendix \sect{Appendix}

\subsection{Fusing and braiding moves}\label{sec:app-moves}

For convenience we list here a collection of useful fusing and braiding moves.
\vskip .7em

\begin{tabular}{|c|c|}
\cline{1-2}
\begin{picture}(200,57)(0,25)       \apppicture{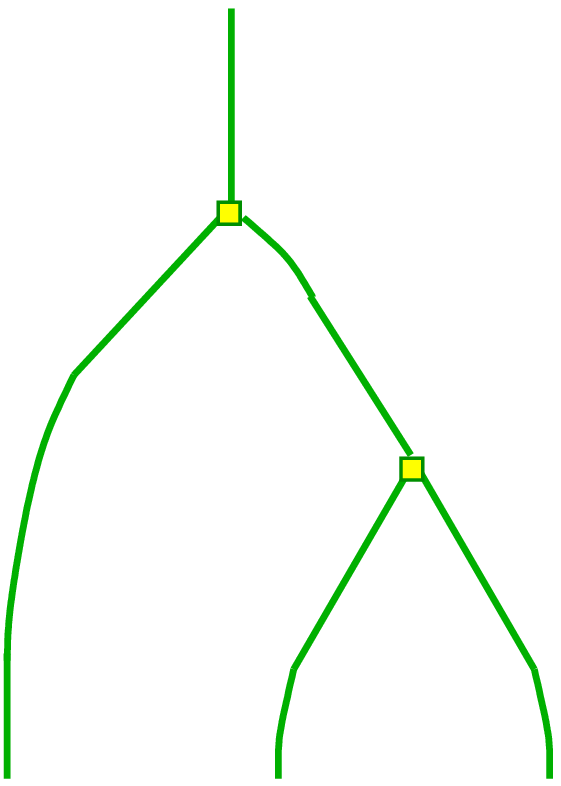}  {7}
                                    \apppicture{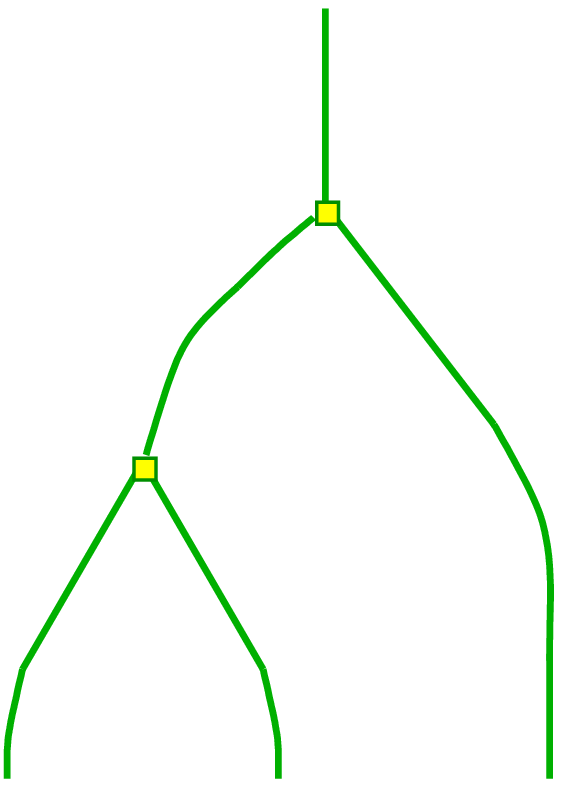}  {137}
  \put(5.8,-8)  {\scriptsize$i$}
  \put(25.5,68) {\scriptsize$l$}
  \put(28,-8)   {\scriptsize$j$}
  \put(37,38)   {\scriptsize$p$}
  \put(51,-8)   {\scriptsize$k$}
  \put(65,29)   {$=\; \dsty\sum_p \Fs ijklpq$}            
  \put(136,-8)  {\scriptsize$i$}
  \put(147,38)  {\scriptsize$q$}
  \put(158,-8)  {\scriptsize$j$}
  \put(163.5,68){\scriptsize$l$}
  \put(181,-8)  {\scriptsize$k$}
  \end{picture}
& 
\begin{picture}(200,57)(0,25)       \apppicture{F2}  {7}
                                    \apppicture{F1}  {137}
  \put(5.8,-8)  {\scriptsize$i$}
  \put(17,38)   {\scriptsize$p$}
  \put(28,-8)   {\scriptsize$j$}
  \put(33.5,68) {\scriptsize$l$}
  \put(51,-8)   {\scriptsize$k$}
  \put(65,29)   {$=\; \dsty\sum_p \Gs ijklpq$} 
  \put(136,-8)  {\scriptsize$i$}
  \put(155.5,68){\scriptsize$l$}
  \put(158,-8)  {\scriptsize$j$}
  \put(167,38)  {\scriptsize$q$}
  \put(181,-8)  {\scriptsize$k$}           
  \end{picture}
\\ \begin{picture}(0,33){}\end{picture}&\\ \cline{1-2}
\begin{picture}(200,57)(0,25)       \apppicture{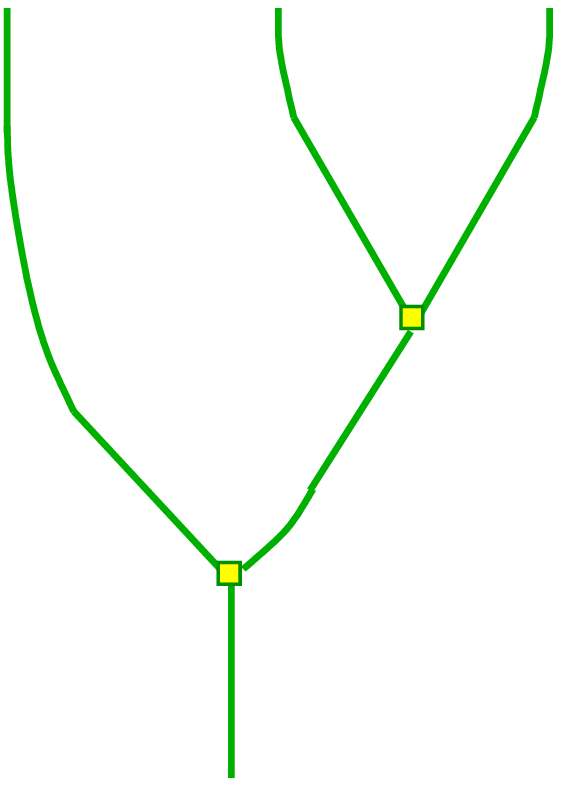}  {7}
                                    \apppicture{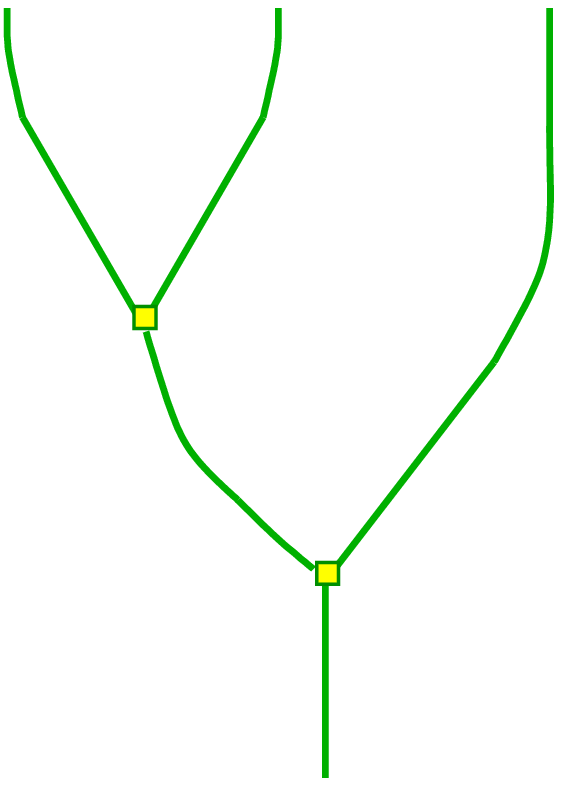}  {137}
  \put(6.5,68)    {\scriptsize$i$}
  \put(24.8,-8.8) {\scriptsize$l$}
  \put(28.4,69.5) {\scriptsize$j$}
  \put(37,24)     {\scriptsize$p$}
  \put(51,68)     {\scriptsize$k$}
  \put(65,29)     {$=\; \dsty\sum_p \Gs ijklqp$}  
  \put(136.4,68)  {\scriptsize$i$}
  \put(147,24)    {\scriptsize$q$}
  \put(158.8,69.5){\scriptsize$j$}
  \put(162.1,-8.8){\scriptsize$l$}
  \put(181,68)    {\scriptsize$k$}   
  \end{picture}
&
\begin{picture}(200,57)(0,25)       \apppicture{G2}  {7}
                                    \apppicture{G1}  {137}
  \put(6.5,68)    {\scriptsize$i$}
  \put(17,24)     {\scriptsize$p$}
  \put(28.4,69.5) {\scriptsize$j$}
  \put(32.8,-8.8) {\scriptsize$l$}
  \put(51,68)     {\scriptsize$k$}
  \put(65,29)     {$=\; \dsty\sum_p \Fs ijklqp$}
  \put(136.4,68)  {\scriptsize$i$}
  \put(155.1,-8.8){\scriptsize$l$}
  \put(158.8,69.5){\scriptsize$j$}
  \put(167,24)    {\scriptsize$q$}
  \put(181,68)    {\scriptsize$k$}
  \end{picture}
\\ \begin{picture}(0,33){}\end{picture}&\\ \cline{1-2}
\begin{picture}(200,53)(8,25)       \apppicture{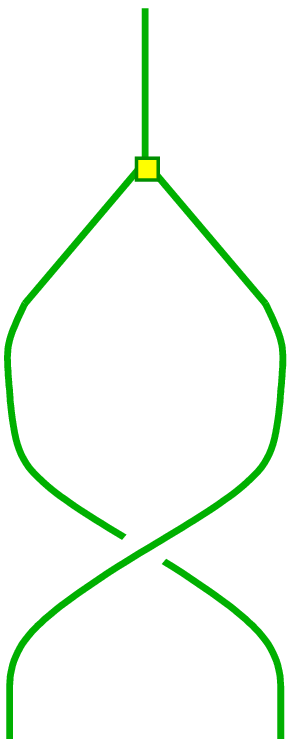}  {49}
                                    \apppicture{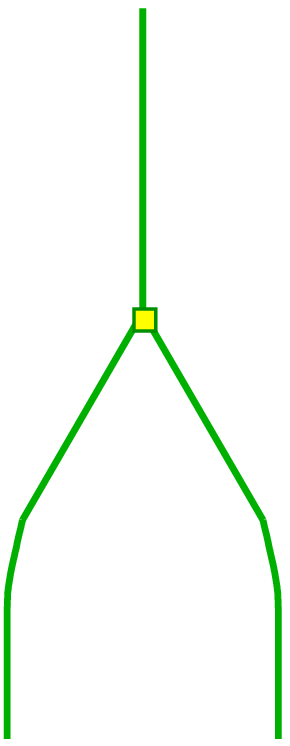}  {141}
  \put(48.5,-8)   {\scriptsize$i$}
  \put(59.5,65)   {\scriptsize$k$}
  \put(70.1,-8)   {\scriptsize$j$}
  \put(87,26)     {$=\; \Rss ijk$}
  \put(139.6,-8)  {\scriptsize$i$}
  \put(151.3,65)  {\scriptsize$k$}
  \put(162.2,-8)  {\scriptsize$j$}
  \end{picture}
&
\begin{picture}(200,53)(8,25)       \apppicture{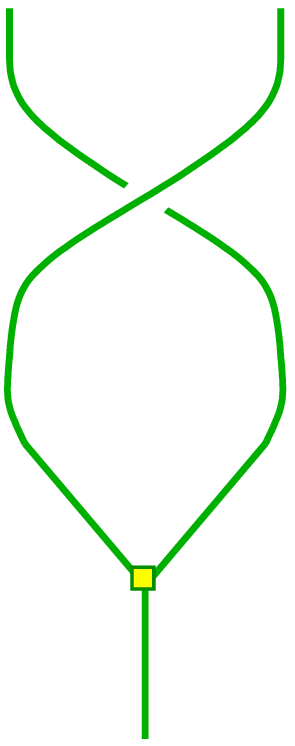}  {49}
                                    \apppicture{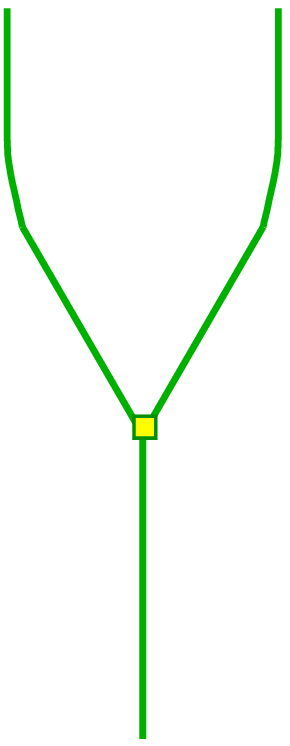}  {141}
  \put(48.5,65)   {\scriptsize$i$}
  \put(58.1,-8)   {\scriptsize$k$}
  \put(70.9,66.5) {\scriptsize$j$}
  \put(87,26)     {$=\; \Rss jik$}
  \put(140.5,65)  {\scriptsize$i$}
  \put(150.8,-8)  {\scriptsize$k$}
  \put(162.6,66.5){\scriptsize$j$}
  \end{picture}
\\ \begin{picture}(0,34){}\end{picture}&\\ \cline{1-2}
\begin{picture}(200,57)(0,25)       \apppicture{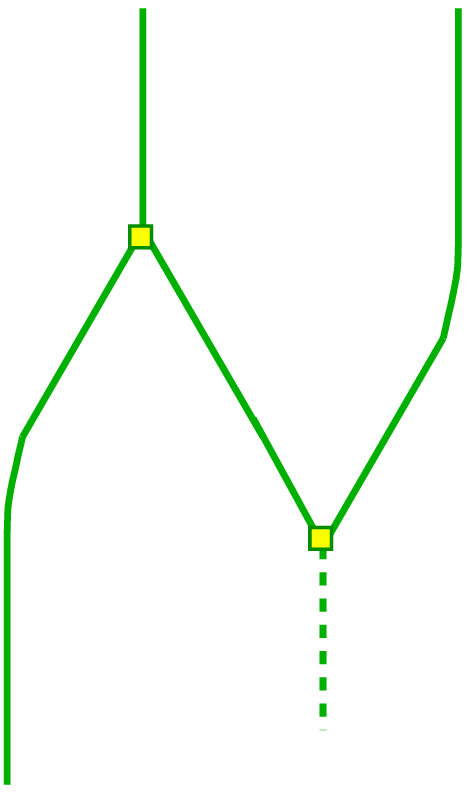}  {24}
                                    \apppicture{S2}  {141}
  \put(21.7,-8)   {\scriptsize$k$}
  \put(34.5,69)   {\scriptsize$i$}
  \put(44.5,36)   {\scriptsize$\overline\jmath$}
  \put(60.5,70.5) {\scriptsize$j$}
  \put(68,29)     {$=\; \Gs k{\overline\jmath}jki0$}
  \put(140.5,65)  {\scriptsize$i$}
  \put(150.8,-8)  {\scriptsize$k$}
  \put(162.2,66.5){\scriptsize$j$}
  \end{picture}
&
\begin{picture}(200,57)(0,25)       \apppicture{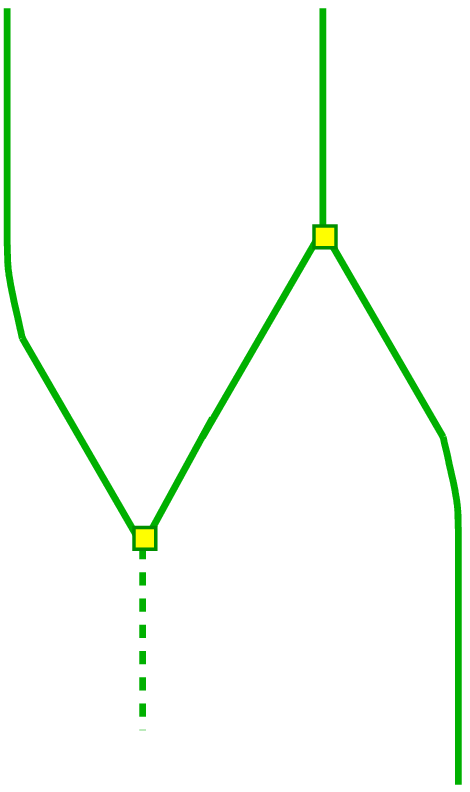}  {24}
                                    \apppicture{S2}  {141}
  \put(23.3,69)   {\scriptsize$i$}
  \put(40.5,36)   {\scriptsize$\overline\imath$}
  \put(48.5,70.5) {\scriptsize$j$}
  \put(60.5,-8)   {\scriptsize$k$}
  \put(76,29)     {$=\; \Fs i{\overline\imath}kkj0$}
  \put(140.5,65)  {\scriptsize$i$}
  \put(150.8,-8)  {\scriptsize$k$}
  \put(162.6,66.5){\scriptsize$j$}
  \end{picture}
\\ \begin{picture}(0,33){}\end{picture}&\\ \cline{1-2}
\begin{picture}(200,57)(0,25)       \apppicture{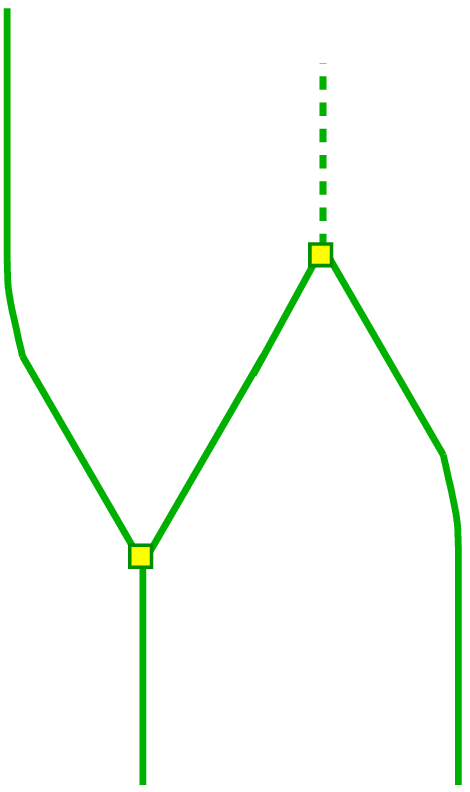}  {24}
                                    \apppicture{S1}  {141}
  \put(22.5,69)   {\scriptsize$k$}
  \put(34.5,-8)   {\scriptsize$i$}
  \put(38.5,33)   {\scriptsize$\overline\jmath$}
  \put(60.2,-8)   {\scriptsize$j$}
  \put(76,29)     {$=\; \Fs k{\overline\jmath}jk0i$}
  \put(139.9,-8)  {\scriptsize$i$}
  \put(151.1,65)  {\scriptsize$k$}
  \put(161.7,-8)  {\scriptsize$j$}
  \end{picture}
& 
\begin{picture}(200,57)(0,25)       \apppicture{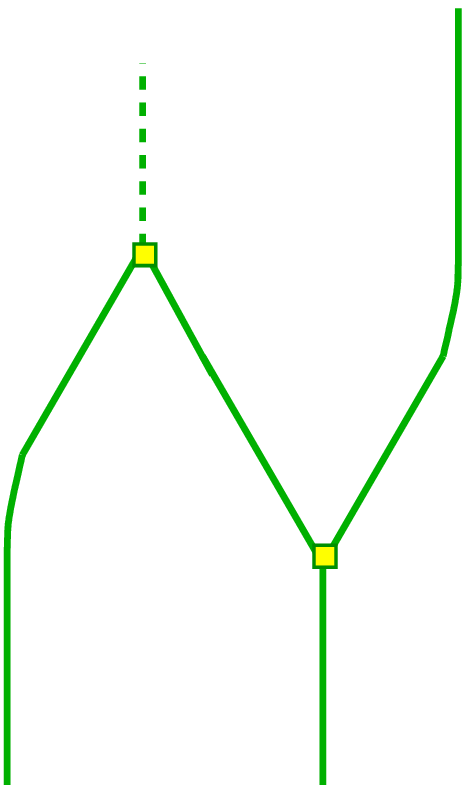}  {24}
                                    \apppicture{S1}  {141}
  \put(23.3,-8)   {\scriptsize$i$}
  \put(45.2,32)   {\scriptsize$\overline\imath$}
  \put(48.5,-8)   {\scriptsize$j$}
  \put(60.5,69)   {\scriptsize$k$}
  \put(76,29)     {$=\; \Gs i{\overline\imath}kk0j$}
  \put(139.9,-8)  {\scriptsize$i$}
  \put(150.8,65)  {\scriptsize$k$}
  \put(162.2,-8)  {\scriptsize$j$}
  \end{picture}
\\ \begin{picture}(0,33){}\end{picture}&\\ \cline{1-2}
\begin{picture}(200,57)(-10,25)     \apppicture{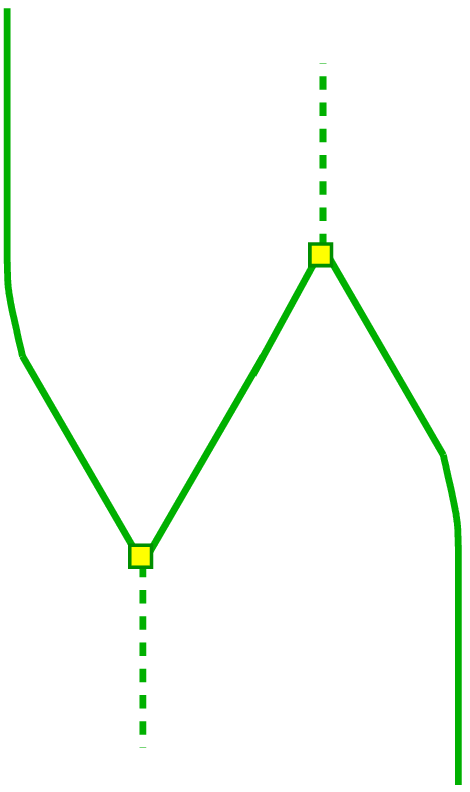}  {24}
                                    \apppicture{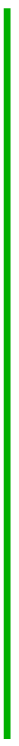}  {144}
  \put(22.5,69)   {\scriptsize$k$}
  \put(38.1,33)   {\scriptsize$\overline k$}
  \put(60.2,-8)   {\scriptsize$k$}
  \put(76,29)     {$=\; \Fs k{\overline k}kk00$}
  \put(142.1,-8)  {\scriptsize$k$}
  \put(142.1,65)  {\scriptsize$k$}
  \end{picture}
&
\begin{picture}(200,57)(-10,25)     \apppicture{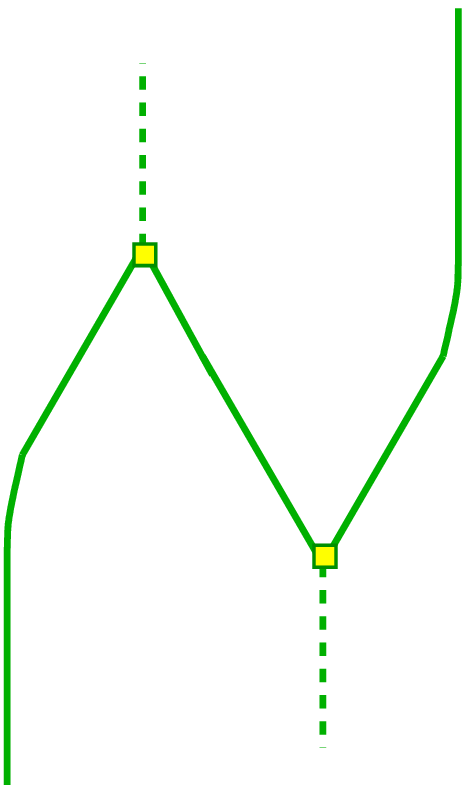}  {24}
                                    \apppicture{S3}  {144}
  \put(23.3,-8)   {\scriptsize$k$}
  \put(45.5,32)   {\scriptsize$\overline k$}
  \put(60.5,69)   {\scriptsize$k$}
  \put(76,29)     {$=\; \Gs k{\overline k}kk00$}
  \put(142.1,-8)  {\scriptsize$k$}
  \put(142.1,65)  {\scriptsize$k$}
  \end{picture}
\\[-.7em]& \\& \\& \\ \cline{1-2}
\end{tabular}

\noindent
These moves are displayed for the case that $\N ijk \iN \{0,1\}$.
Otherwise the three-point vertices are labelled by basis elements
of the respective morphism spaces, and analogous labels appear in
the coefficients.

\subsection{Proof of the algorithm for finding reversions}\label{approof}

In this appendix we present a detailed proof of the algorithm described 
in section \ref{sec:*-class}. First we recall from section I:3.4 that the 
vector space $\Atop\eq\Hom(\one,A)$ is a $\complex$-algebra, with product
  \be  \mtop(a,b) := m\cir(a\oti b)  \labl{mtop}
for $a,b\iN\Atop$, where $m$ is the multiplication morphism on $A$. 
The reversion $\sigma$ induces an involutive anti-automorphism on this
$\complex$-algebra: writing $\sigma a\,{:=}\,\sigma\cir a$ for $a\iN\Atop$,
we have
\\[-2.3em]

\dtl{Lemma}{sigmatop}
For $\sigma$ a reversion on a \ssFA\ $A$, we have
$\sigma^2 a\eq a$ for all $a\iN\Atop$, and
  \be  \mtop(\sigma a,\sigma b) = \sigma\,\mtop(b,a)  \labl{mtops}
for all $a,b\iN\Atop$.
\\
In particular, for $p$ an idempotent in $\Atop$, i.e.\ $\mtop(p,p)\eq p$, 
also $\sigma p$ is an idempotent, for ${\rm e}_i$ a primitive idempotent 
also $\sigma{\rm e}_i$ is a primitive idempotent,
and for $a\iN\Atop$ invertible, also $\sigma a$ is invertible.

\medskip\noindent
Proof:\\
We have $\sigma\cir\sigma\cir a\eq \thetaA\cir a$ by \erf{eq:*ssFA1},
and this equals $a$ owing to functoriality of the twist.
The equality \erf{mtops} follows by combining the property 
\erf{eq:*ssFA2} of $\sigma$ with the functoriality of the braiding.
\\
For $p\iN\Atop$ an idempotent, \erf{mtops} implies
$\mtop(\sigma p,\sigma p)\eq \sigma\cir\mtop(p,p)\eq\sigma\cir p$,
i.e.\ $\sigma p$ is again an idempotent. Similarly, using the same arguments 
as after \erf{eq:vec-idem}, one checks that $\{\sigma{\rm e}_i\}$ satisfies 
the defining properties of the set of primitive idempotents.
Finally, for $\mtop(a,b)\eq 1_{\ATop}\,{\equiv}\,\eta$, also
$\mtop(\sigma a,\sigma b)\eq\sigma\,\mtop(b,a)\eq\sigma\eta\eq\eta$.
\qed

Now for any idempotent $p\iN\Atop$ 
we can define two idempotents $P_B,\, P_M \iN \Hom(A,A)$ as
  \bea \begin{picture}(300,35)(0,29)
  \put(50,0)    {\includeourbeautifulpicture{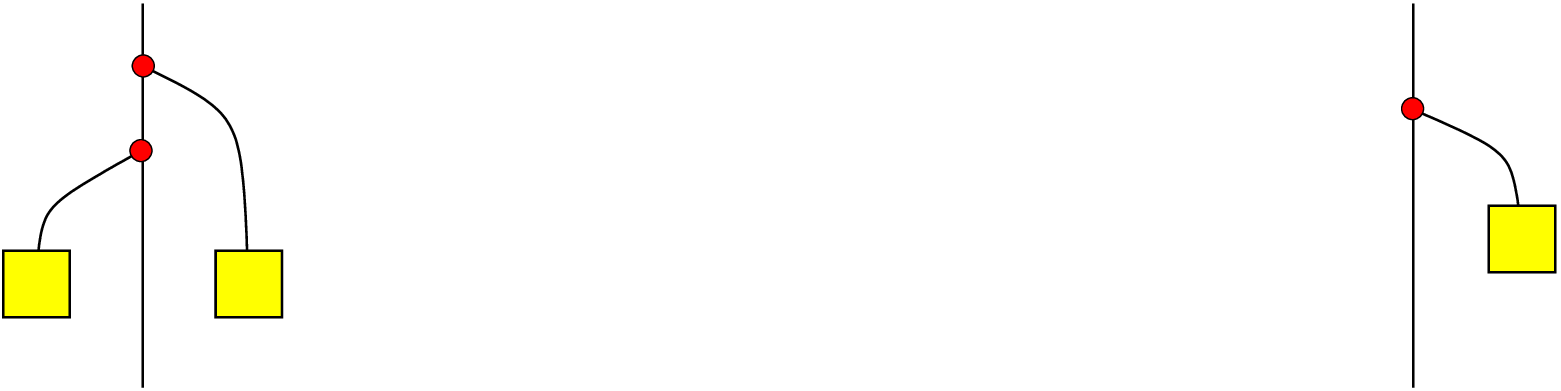}}
  \setlength{\unitlength}{24810sp}
  \setlength{\unitlength}{1pt}
  \put(0,26)       {$ P_B \;:= $}
  \put(53.4,15.5)  {\scriptsize$p$}
  \put(68.2,-8.5)  {\scriptsize$A$}
  \put(68.9,63.5)  {\scriptsize$A$}
  \put(86.8,15.5)  {\scriptsize$p$}
  \put(159,26)     {and $\qquad\quad P_M \;:= $}
  \put(265.1,-8.5) {\scriptsize$A$}
  \put(265.8,63.5) {\scriptsize$A$}
  \put(283.7,22.5) {\scriptsize$p$}
  \epicture15 \labl{eq:PMB-def}
The image of the idempotent $P_B$ can again be made into an algebra:
\\[-2.3em]

\dtl{Lemma}{le:ImP-sssFA}
For $A$ a simple symmetric special Frobenius algebra, let
$p\iN\Atop$ be a nonzero idempotent and $P_M,\, P_B$ as in
\erf{eq:PMB-def}. Denote by $(M,e_M,r_M)$ the image of $P_M$
and by $(B,e_B,r_{\!B})$ the image of $P_B$. Then
\\[3pt]
(i)~\,\,$(B,m_B, \eta_B, \Delta_B, \eps_B)$ with
  \be\bearll
  m_B := r_{\!B} \circ m_A \circ (e_B\oti e_B) \,,
  & \eta_B := r_{\!B} \circ \eta_A \,,
  \\{}\\[-.7em]
  \Delta_B :=
  \Frac{\dimA}{\dimM}\, (r_{\!B}\oti r_{\!B}) \circ \Delta \circ e_B\,,
  \ \ & \eps_B := \Frac{\dimM}{\dimA}\, \eps_A \circ e_B
  \eear\labl{eq:PB-alg}
is a simple symmetric special Frobenius algebra.
\\[3pt]
(ii)~\,The topological algebra $B\top$ associated to $B$ has dimension
  \be
  \dimc\,B\top \le \dimc\,\Atop \,,  \labl{eq:BtopAtop}
with equality if and only if $p$ is the unit $\eta_A$ of the algebra $A$.
\\[3pt]
(iii)~\,$M$ is a left $A$-module with the property that
$B\,{\cong}\, M^\vee\OtA M$ as symmetric special Frobenius algebras. 
Thus $B$ is Morita equivalent to $A$.

\medskip\noindent
Proof:\\
(i)~\,The associativity, coassociativity, unit and counit
properties, as well as symmetry and the Frobenius property of $B$,
all reduce to the corresponding properties of $A$ by a short calculation. 
In these manipulations it is helpful to use the identities
  \be
  m \circ (\id_A \oti p) \circ e_B = e_B =
  m \circ (p \oti \id_A) \circ e_B  \labl{eq:cancel-p}
and analogous ones for $r_{\!B}$, which follow directly from
\erf{eq:er-prop2} and the fact that $p$ is an idempotent. 
\\
The proof of specialness of $B$ works along the same lines as in
proposition \ref{pr:B-ssFA}. Define the element $\phi\iN\Atop$ by
  \bea \begin{picture}(90,44)(0,37)
  \put(38,0)    {\includeourbeautifulpicture{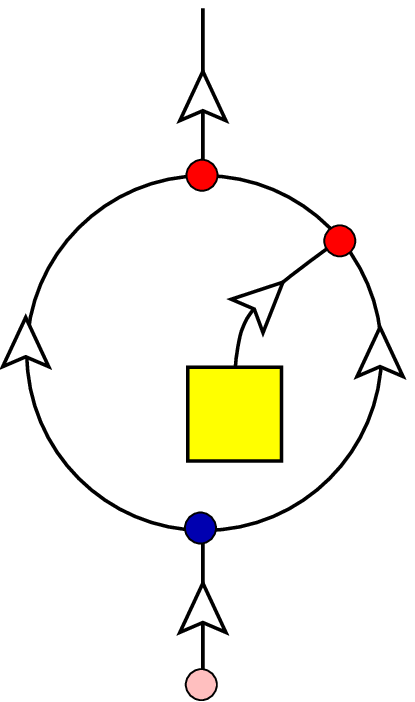}}
  \setlength{\unitlength}{24810sp}
  \setlength{\unitlength}{1pt}
  \put(0,33)       {$ \phi \;:= $}
  \put(57.1,80.2)  {\scriptsize$A$}
  \put(61.8,30.8)  {\scriptsize$p$}
  \epicture19 \labl{pic68} 
One checks that $\phi\iN{\rm cent}_A(\Atop)$. Since $A$ is by
assumption simple, ${\rm cent}_A(\Atop)$ is one-dimen\-si\-o\-nal, and hence 
$\phi\eq\xi\eta_A$ for some complex number $\xi$, which can be
determined by composing with $\eps_{\!A}^{}$; one finds 
  \be
  \phi = \Frac{\dimM}{\dimA}\, \eta_A \,.  \ee
Using this identity one can verify that indeed $m_B\cir\Delta_B\eq\id_B$. 
\\
Simplicity of $B$ will follow
from (iii) below, in combination with proposition \ref{pr:B-ssFA}.
\\[5pt]
(ii)~\,Define the linear maps $\varphi{:}\; B\top\,{\to}\,\Atop$ and
$\psi{:}\; \Atop\,{\to}\,B\top$ as $\varphi(x)\,{:=}\,e_B \cir x$ and
$\psi(y){:=}\,r_{\!B}\cir y$. By the identities \erf{eq:er-prop} we have
$\psi\cir\varphi\eq\id_{B\top}$; thus in particular $\varphi$ is injective.
Next note that $\psi(\eta_A{-}p)\eq\psi(P_B \cir\eta_A) - \psi(p)\eq0$,
which is seen with the help of \erf{eq:er-prop2} and $P_B\cir\eta_A\eq p$.
Now suppose there is an $x\iN B\top$ with $\varphi(x)\eq\eta_A\,{-}\,p$.
Applying $\psi$ to this equality yields $x\eq0$. But $\varphi(0)\eq0$,
so that an $x$ with the assumed property can exist only if $p\eq\eta_A$.
Thus for $p\neq\eta_A$ the dimension of the image of $\varphi$ is strictly
smaller than the dimension of $\Atop$. Since $\varphi$ is injective this
implies $\,\dimc\,B\top\,{<}\,\dimc\,\Atop$.
\\[5pt]
(iii) It is straightforward to check that
  \be
  \rho = r_M \circ m \circ (\id_A \oti e_M)  \labl{eq:A-M-action}
(i.e.\ the action of $A$ on $M$ by multiplication) defines a
representation of $A$. Let $P_{\otimes A}$ be the idempotent in \erf{eq:Ptensor}
and $(C,e_C,r_C)$ be its image. By proposition \ref{eq:B-ssFA}, $C$ with
morphisms \erf{eq:B-ssFA} is a simple symmetric special Frobenius algebra.
Consider now $\varphi\iN\Hom(B,C)$ and $\psi\iN\Hom(C,B)$ given by
  \bea \begin{picture}(320,80)(0,30)
  \put(42,0)    {\includeourbeautifulpicture{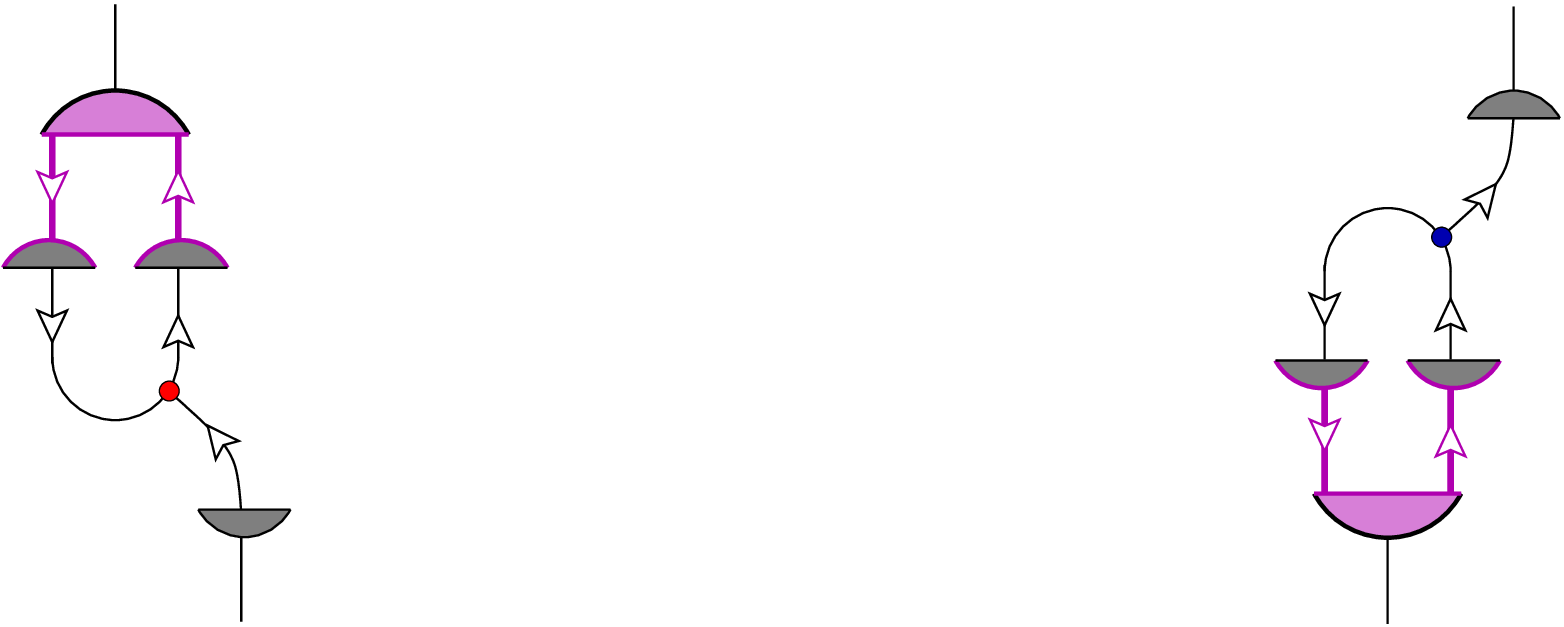}}
  \setlength{\unitlength}{24810sp}
  \setlength{\unitlength}{1pt}
  \put(0,49)       {$ \varphi \;:= $}
  \put(39.6,78.5)  {\scriptsize$\M$}
  \put(58.1,87.1)  {\scriptsize$r_{\!C}^{}$}
  \put(59.1,110.5) {\scriptsize$C$}
  \put(73.9,39.5)  {\scriptsize$A$}
  \put(74,78.5)    {\scriptsize$\M$}
  \put(79.7,-8.5)  {\scriptsize$B$}
  \put(83.2,62.5)  {\scriptsize$r_{\!\M}^{}$}
  \put(154,49)     {and}
  \put(222,49)     {$ \psi \;:= $}
  \put(259.5,24)   {\scriptsize$\M$}
  \put(275.5,18.3) {\scriptsize$e_C$}
  \put(276.5,-8.5) {\scriptsize$C$}
  \put(292.1,63.2) {\scriptsize$A$}
  \put(293.1,24)   {\scriptsize$\M$}
  \put(299,110.2)  {\scriptsize$B$}
  \put(301.2,43.2) {\scriptsize$e_{\M}^{}$}
  \epicture15 \labl{eq:B-MM-iso}
One checks that $\psi \cir \varphi\eq\id_B$ and
$\varphi \cir\psi\eq\id_C$. For example,
  \bea \begin{picture}(260,99)(0,38)
  \put(57,0)    {\includeourbeautifulpicture{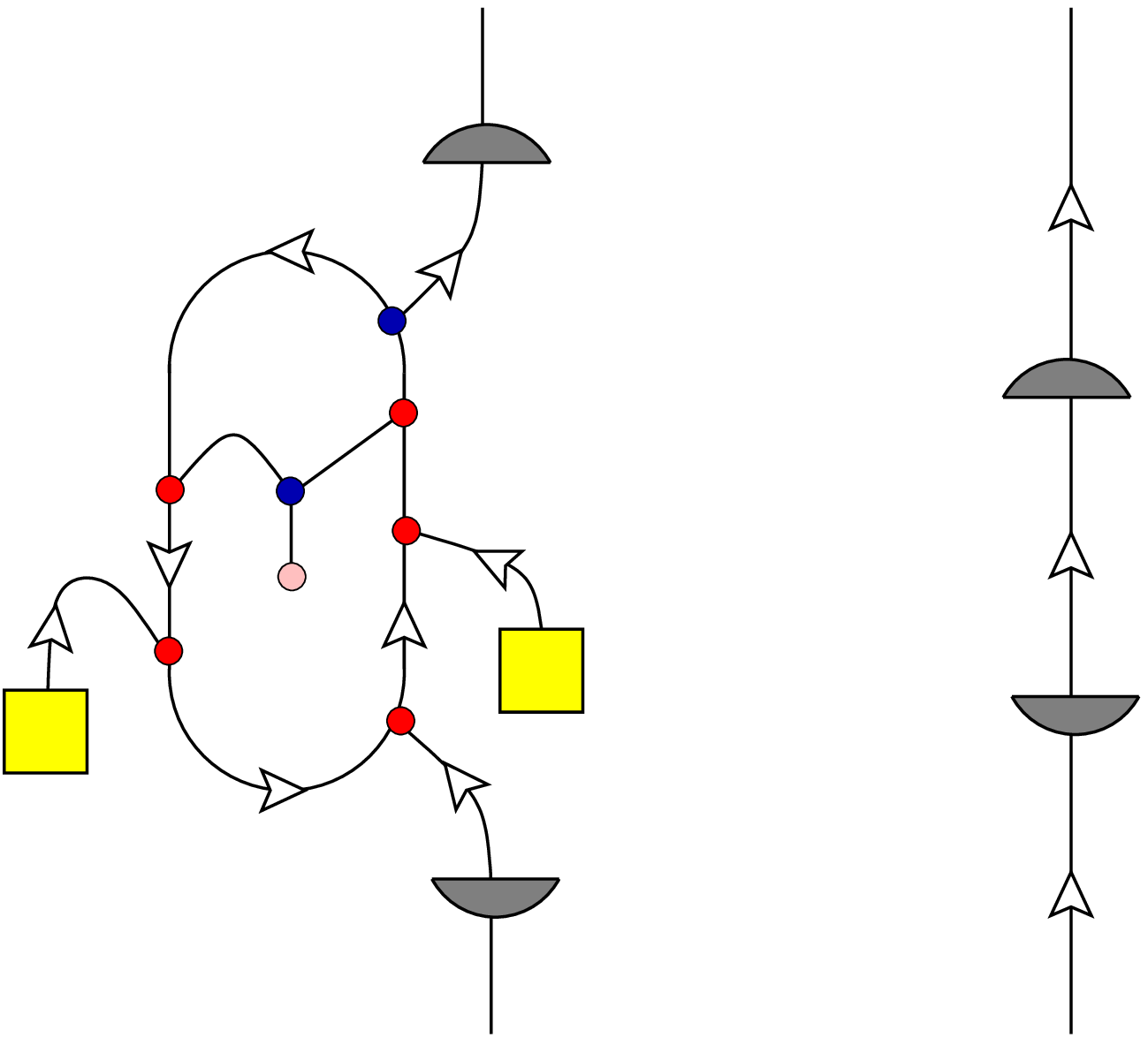}}
  \setlength{\unitlength}{24810sp}
  \setlength{\unitlength}{1pt} 
  \put(0,61)       {$ \psi \circ\varphi \;= $}
  \put(61,37)      {\scriptsize$p$}
  \put(102.7,30.3) {\scriptsize$A$}
  \put(107.8,85.1) {\scriptsize$A$}
  \put(114,-8.5)   {\scriptsize$B$}
  \put(113.8,131.5){\scriptsize$B$}
  \put(122.4,44.5) {\scriptsize$p$}
  \put(152,61)     {$ = $}
  \put(186,-8.5)   {\scriptsize$B$}
  \put(186.8,131.5){\scriptsize$B$}
  \put(191.3,48)   {\scriptsize$A$}
  \put(216,61)     {$ =\; \id_B \,.$}
  \epicture24 \labl{pic70}
Here we use that by definition of the action \erf{eq:A-M-action}
of $A$ on $M$ we have $e_M\iN\HomA(M,A)$ and $r_M \iN\HomA(A,M)$;
also, the $A$-ribbons from the idempotents $p$ are moved so as to  
merge directly with the $A$-ribbon from $e_M$, so that they can be
removed by use of \erf{eq:cancel-p}.
\\
The morphisms $\varphi$ and $\psi$ are actually algebra isomorphisms between
$B$ and $C$; for instance, for the multiplication we find
  \begin{eqnarray} \begin{picture}(420,224)(11,0) 
  \put(110,0)    {\includeourbeautifulpicture{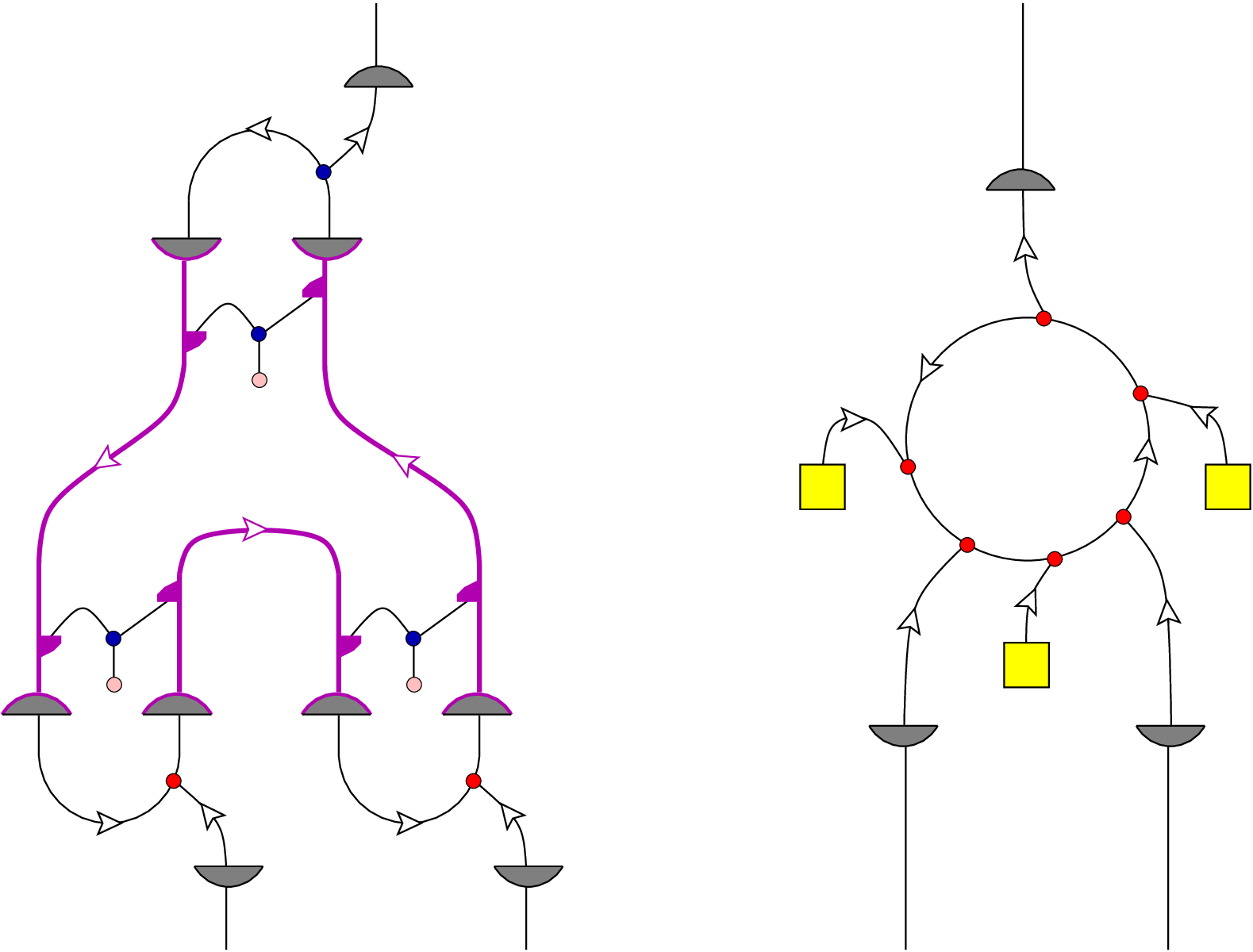}}
  \setlength{\unitlength}{24810sp}
  \setlength{\unitlength}{1pt}
  \put(0,107)      {$ \psi \circ m_C \circ (\varphi\oti\varphi) \;= $}
  \put(134.5,77)   {\scriptsize$A$}
  \put(142.5,147)  {\scriptsize$\M$}
  \put(146.7,30.7) {\scriptsize$A$}
  \put(153,62)     {\scriptsize$\M$}
  \put(158,-8.5)   {\scriptsize$B$}
  \put(167.9,147)  {\scriptsize$A$}
  \put(178.1,62)   {\scriptsize$\M$}
  \put(186.3,136)  {\scriptsize$\M$}
  \put(193.5,222)  {\scriptsize$B$}
  \put(203.5,77)   {\scriptsize$A$}
  \put(215.7,30.7) {\scriptsize$A$}
  \put(228.5,-8.5) {\scriptsize$B$}
  \put(252,107)    {$ = $}
  \put(297.4,106)  {\scriptsize$p$}
  \put(314.8,-8.5) {\scriptsize$B$}
  \put(322.3,91.1) {\scriptsize$A$}
  \put(341.3,147.8){\scriptsize$A$}
  \put(342.5,222)  {\scriptsize$B$}
  \put(344.4,65)   {\scriptsize$p$}
  \put(364.2,90.9) {\scriptsize$A$}
  \put(375.8,-8.5) {\scriptsize$B$}
  \put(391.2,106)  {\scriptsize$p$}
  \put(422,107)    {$ =\; m_B \,.$}
  \end{picture} \nonumber\\[-1.0em]{} \label{pic71}
  \\[-1.3em]{}\nonumber\end{eqnarray}
This establishes that indeed $B \,{\cong}\, M^\vee \OtA M$ as
Frobenius algebras.
\qed

It follows that, given a simple \ssFA\ $A$ as in lemma \ref{le:ImP-sssFA},
we can find a Morita equivalent algebra with a smaller
associated topological algebra, provided that $\Atop$ 
contains a non-trivial idempotent $p$. The following lemma
tells us that the structure of $\Atop$ cannot be complicated -- it
can be written as a direct sum of matrix algebras.

\dtl{Lemma}{le:Atop-mat}
Let $A$ be a special Frobenius algebra and let $\{M_\kappai \,{|}\,\kappai
\iN\JJ\}$ be a set of representatives for the simple left $A$-modules. 
Let $A$, regarded as a left module over itself, decompose as
$\bigoplus_{\kappai\in\JJ} (M_\kappai)^{\oplus n_\kappai}$ into simple
$A$-modules. Then
  \be
  \Atop \,\cong\, \bigoplus_{\kappai\in\JJ} {\rm Mat}_{n_\kappai}
  (\complex) \,. \ee

\noindent
Proof:\\
Define the map $\varphi{:}\; \Atop\,{\to}\,\HomA(A,A)$ by
$\varphi(x)\,{:=}\,m\cir(\id_A\oti x)$. The space $\HomA(A,A)$
is an algebra, with multiplication given by the composition. 
One checks that $\varphi$ is an algebra anti-homomorphism
  \be
  \varphi(x y) = \varphi(y)\, \varphi(x) \,.  \ee
Furthermore, $\varphi(x)\cir\eta_A\eq x$, implying that $\varphi$
is injective. Since by reciprocity (see e.g.\ proposition I:4.12) we 
have $\dimc\,\Hom(\one,A)\eq\dimc\,\HomA(A,A)$, it follows that 
$\varphi$ is an isomorphism.
The algebra structure of $\HomA(A,A)$ now follows from the identities
  \be  \bearll
  \HomA(A,A) \!& \cong \!\dsty\bigoplus_{\kappai,\kappaj\in\JJ}\!\!
  \HomA(M_\kappai^{\oplus n_\kappai}{,}\,M_\kappaj^{\oplus n_\kappaj})
  \\{}\\[-.7em]
  & \cong \dsty\bigoplus_{\kappai\in\JJ}
  \HomA( M_\kappai^{\oplus n_\kappai}{,}\,M_\kappai^{\oplus n_\kappai} )
  \cong \bigoplus_{\kappai\in\JJ} {\rm Mat}_{n_\kappai}(\complex) \,.
  \eear \ee
Here we substituted the assumption that $A$ decomposes as
$\bigoplus_{\kappai\in\JJ}\! M_\kappai^{\oplus n_\kappai}$ as an $A$-module.
Because of $\HomA(M_\kappai,M_\kappai)\,{\cong}\,\complex$, in the 
last step we obtain a direct sum of matrix algebras. In combination with
$\varphi$ this provides us with an algebra anti-isomorphism
$f$ from $\Atop$ to a direct sum of matrix algebras. 
Thus the transpose $f^{\rm t}$ is an algebra isomorphism, establishing
the lemma.
\qed

The only direct sum of matrix algebras that does not contain an idempotent 
$p\,{\ne}\,0$ different from the identity is the algebra
$\complex$ itself. Thus starting from a given algebra $B$ we can
always pass to a Morita equivalent algebra $A$ with smaller associated
topological algebra, $\dimc\,\Atop \,{<}\, \dimc\,B\top$, as long as
$\dimc\,B\top\,{>}\,1$. Together the lemmata \ref{le:ImP-sssFA} and
\ref{le:Atop-mat} amount to
(see also section\,3.3 of \cite{ostr} and remark I:3.5(iii))
\\[-2.3em]

\dtl{Proposition}{th:sssFA}
Every simple symmetric special Frobenius algebra $B$ is isomorphic to
$M^\vee\OtA M$, where $A$ is a haploid simple symmetric special Frobenius
algebra, $M$ is a left $A$-module, and the algebra structure on
$M^\vee\OtA M$ is as given in \erf{eq:B-ssFA}.
\\
$A$ is haploid iff it is a simple module over itself.

\bigskip

So far we have learned that the Morita class of each simple symmetric
special Frobenius algebra $A$ contains at least one haploid representative.
An analogous statement does, however, not hold for algebras $(A,\sigmaA)$
with reversion. But proposition \ref{pr:B-star} provides a sufficient
condition for whether $\sigmaA$ can be transported to another
representative in the Morita class. This can be translated into a
condition on the idempotents $p\iN\Atop$:
\\[-2.3em]

\dtl{Lemma}{le:PB-star}
Let $A$ be a simple \JA, with reversion $\sigmaA$, and let $p\iN\Atop$ 
be an idempotent such that  the idempotent $\sigmaA p$ is equal to $p$. 
Let $M$ and $B\,{\cong}\,M^\vee\OtA M$ be as in lemma \ref{le:ImP-sssFA}. Then
\\[3pt]
(i)~\,\,The morphism $\sigmaB\,{:=}\,r_{\!B} \cir\sigmaA \cir e_B$ defines 
a reversion on $B$.
\\[3pt]
(ii) There exists a $g\iN\HomA(M,M^\sigma)$ obeying \erf{eq:g-prop} with 
$\eps_g\eq1$, such that $\sigmaB$ is of the form \erf{eq:B-star}.

\medskip\noindent
Proof:\\
(i) By lemma \ref{le:ImP-sssFA} we can describe $B$ equivalently as 
$M^\vee\OtA M$ or as the image of $P_B$ with morphisms \erf{eq:PB-alg}.
Note that $P_B \cir\sigmaA\eq\sigmaA \circ P_B$, as can be seen
from properties \erf{eq:sig-def} and $\sigmaA p\eq p$. Using that
$P_B$ and $\sigmaA$ commute one checks that $\sigmaB\cir\sigmaB 
\eq\theta_B$ and $\sigmaB\cir m\eq m \cir c_{B,B}\cir(\sigmaB 
\oti\sigmaB)$. Thus
by proposition \ref{pr:*ssFA} $\sigmaB$ is a reversion on $B$.
\\[5pt]
(ii) Consider the morphism $g\iN\Hom(\M,\M^\vee)$ given by
  \be
  g := (e_M)^\vee \circ \Phi_2 \circ \sigmaA^{-1} \circ e_M \,,
  \labl{eq:g-def}
where $(M,e_M,r_M)$ is the image of $P_M$ as in lemma \erf{le:ImP-sssFA}
and $\Phi_2$ is taken from (I:3.33). The moves
  \bea \begin{picture}(420,107)(42,31)
  \put(57,0)    {\includeourbeautifulpicture{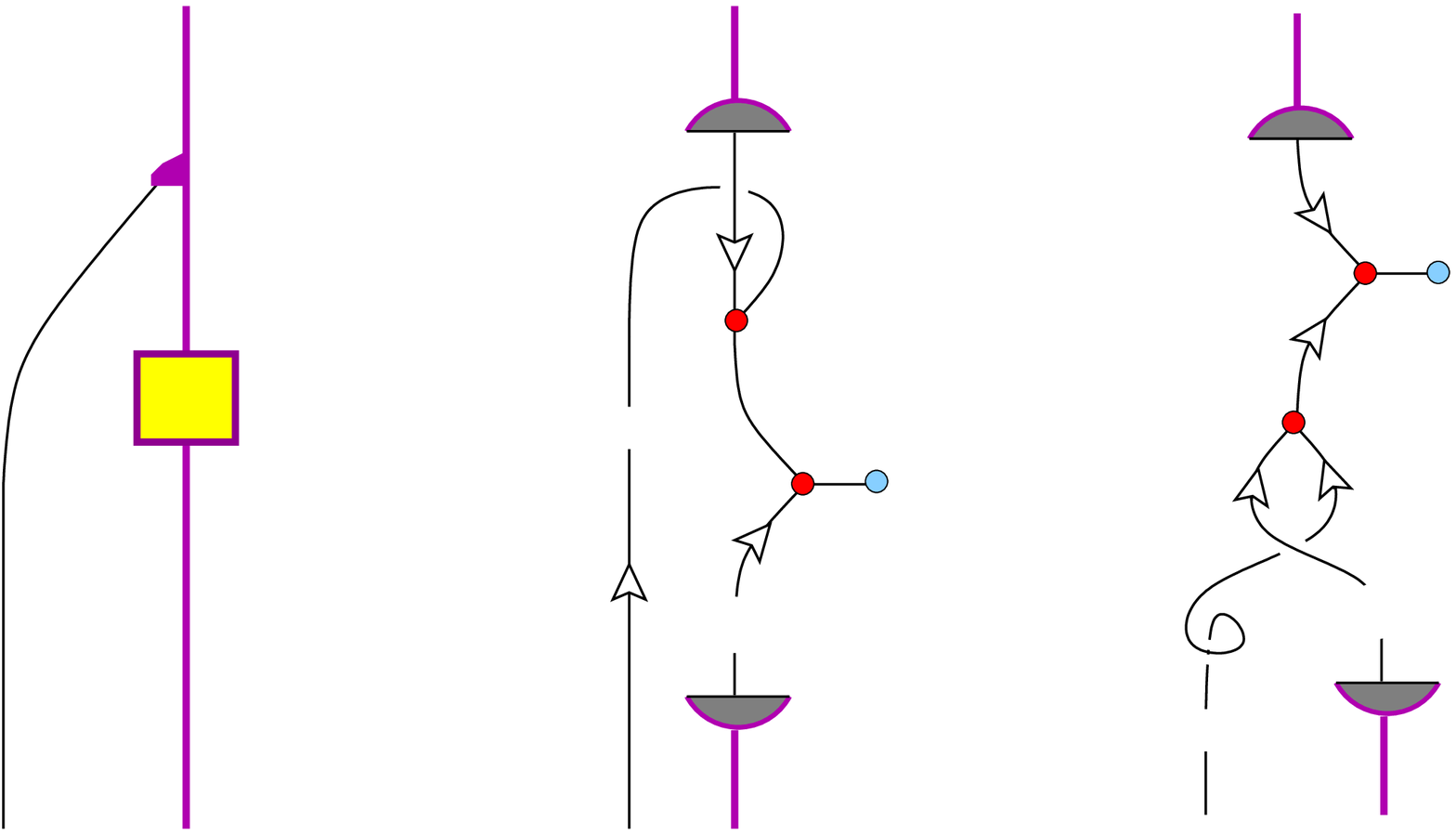}}
  \setlength{\unitlength}{1pt}
  \put(52.7,-8.7)  {\scriptsize$A$}
  \put(80.3,-9.9)  {\scriptsize$\M$}
  \put(81,131)     {\scriptsize$\M^\vee_{}$}
  \put(83.5,65.8)  {\scriptsize$g$}
  \put(87.3,102.7) {\scriptsize$\r_{M^\sigma}^{}$}
  \put(115,61)     {$ = $}
  \put(151.8,61.2) {\scriptsize$\sigma$}
  \put(149.7,-8.7) {\scriptsize$A$}
  \put(166.8,29.6) {\scriptsize$\sigma^{\!-\!1}$}
  \put(164.7,-9.9) {\scriptsize$\M$}
  \put(165,131)    {\scriptsize$\M^\vee_{}$}
  \put(170.6,51.6) {\scriptsize$A$}
  \put(212,61)     {$ = $}
  \put(237.7,-8.7) {\scriptsize$A$}
  \put(239.8,14.5) {\scriptsize$\sigma$}
  \put(251.7,131)  {\scriptsize$\M^\vee_{}$}
  \put(257,84.4)   {\scriptsize$A$}
  \put(265.2,-9.9) {\scriptsize$\M$}
  \put(265.8,32.4) {\scriptsize$\sigma^{\!-\!1}$}
  \put(298,61)     {$ = $}
  \put(320.5,-8.7) {\scriptsize$A$}
  \put(335.9,131)  {\scriptsize$\M^\vee_{}$}
  \put(337.5,62.1) {\scriptsize$\sigma^{\!-\!1}$}
  \put(335.5,-9.9) {\scriptsize$\M$}
  \put(341.8,81.3) {\scriptsize$A$}
  \put(383,61)     {$ = $}
  \put(398.7,-8.7) {\scriptsize$A$}
  \put(428,-9.9)   {\scriptsize$\M$}
  \put(427.3,131)  {\scriptsize$\M^\vee_{}$}
  \put(430.3,94.9) {\scriptsize$g$}
  \put(435,64.7)   {\scriptsize$\r_{M}^{}$}
  \epicture21 \labl{eq:pic72}
show that $g\iN\HomA(M,M^{\sigmaA})$. Moreover, $g$ has the properties
required in proposition \erf{pr:B-star}. Indeed, \erf{eq:g-prop}
holds with $\eps_g\eq1$, as can be seen by using
\erf{eq:sig-phi} and $\Phi_1\eq\Phi_2$ (i.e.\ $A$
symmetric, see definition I:3.4).
\\
Next we must check that $\sigmaB$ equals $\tilde\sigma_g$
as defined in \erf{eq:B-star}. To this end we use the explicit isomorphisms
\erf{eq:B-MM-iso} between $B$ and $M^\vee\OtA M$; we have
  \bea \begin{picture}(305,146)(0,43)
  \put(87,0)    {\includeourbeautifulpicture{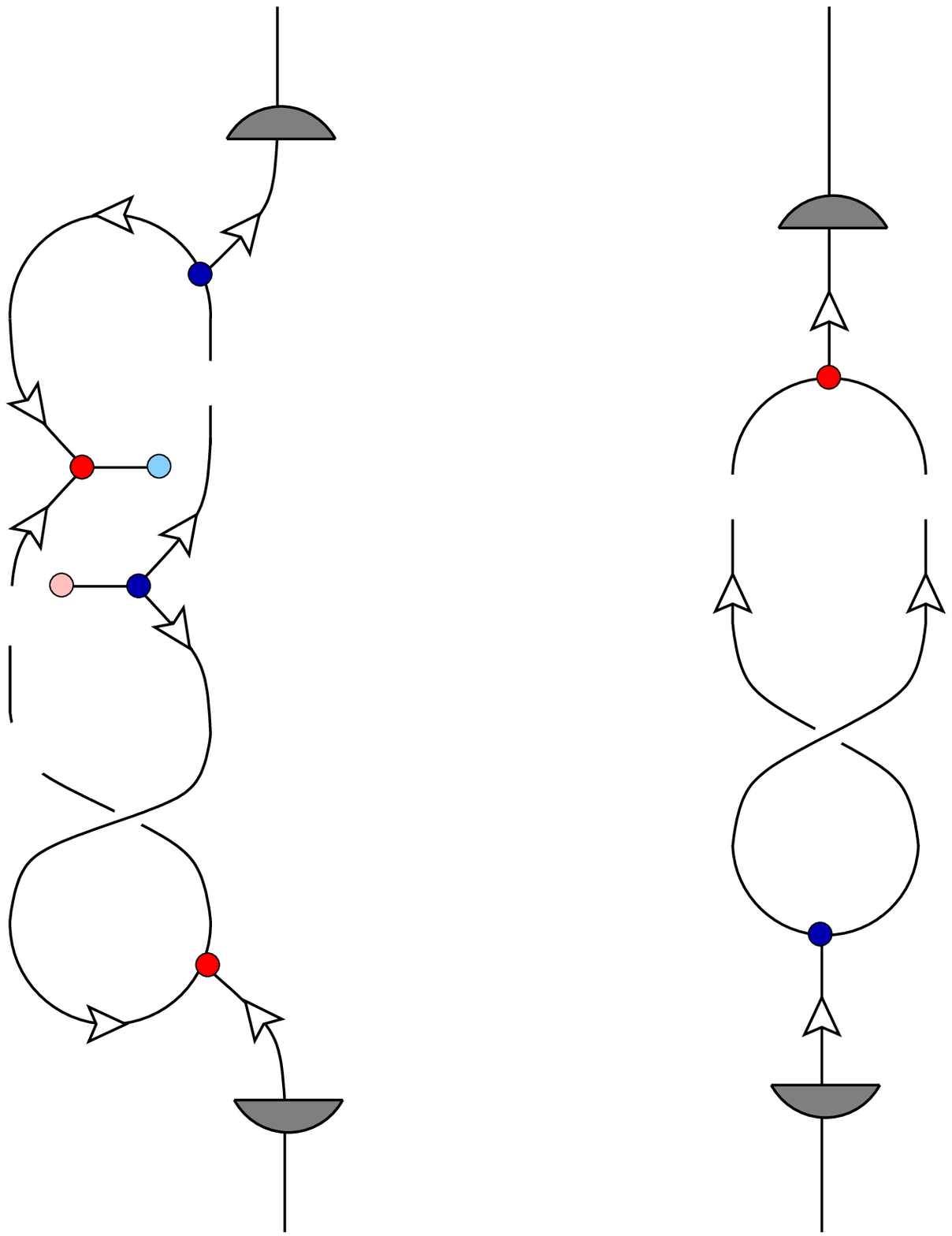}}
  \setlength{\unitlength}{24810sp}
  \setlength{\unitlength}{1pt}
  \put(0,82)       {$ \psi \circ \tilde\sigma_g \circ \varphi \;= $}
  \put(88.4,67.8)  {\scriptsize$\theta_{\!\!A}^{}$} 
  \put(85.9,87.4)  {\scriptsize$\sigma^{\!-\!1}$} 
  \put(118.6,122.5){\scriptsize$\sigma$} 
  \put(121.7,135.6){\scriptsize$A$} 
  \put(122.7,37.5) {\scriptsize$A$} 
  \put(127.2,-8.5) {\scriptsize$B$} 
  \put(127.3,182.6){\scriptsize$B$} 
  \put(158,82)     {$ = $}
  \put(194.3,105.9){\scriptsize$\sigma$} 
  \put(205.4,-8.5) {\scriptsize$B$} 
  \put(206.9,182.6){\scriptsize$B$} 
  \put(207,75.9)   {\scriptsize$A$} 
  \put(222.5,105.9){\scriptsize$\sigma$} 
  \put(247,82)     {$ =\; \sigmaB \,.$}
  \epicture30 \labl{pic73}
In the first step one inserts the definitions \erf{eq:B-star},
\erf{eq:B-MM-iso}, \erf{eq:g-def} and gets rid of all idempotents
using \erf{eq:sP=Ps} and \erf{eq:cancel-p}. The last step 
uses \erf{eq:sig-coprod} and specialness of $A$, as well
as the definition of $\sigmaB$ in part (i) of the present lemma.
\qed

For the next theorem we need more detailed information about the action
of $\sigma$ on the associated topological algebra $\Atop$. Let us summarise
the results of section \ref{sec:vect} below, 
where the case $\calc\eq\Vect_{\complex}$ is treated in more detail. 

Consider a direct sum $A\eq\bigoplus_{i=1}^m\!{\rm Mat}_{n_i}(\complex)$
of matrix algebras over $\complex$.
All anti-isomorphisms $\sigma$ squaring to the identity on $A$
are of the form
  \be
  \sigma^\pi_{U_1,\dots,U_m}(X_1,...\,,X_m) = 
  (U_1 X_{\pi(1)}^{\rm t}U_1^{-1},...\,,U_m X_{\pi(m)}^{\rm t}U_m^{-1})
  \labl{eq:vect-sigma}
for $X_i\iN{\rm Mat}_{n_i}(\complex)$, where $\pi$ is a permutation of order 
two with $n_{\pi(i)}\eq n_i$ and the $U_i \iN {\rm Mat}_{n_i}(\complex)$ are 
invertible and satisfy $U_{\pi(i)}\eq\lambda_iU_i^{\rm t}$ for some 
$\lambda_i\iN\complex$.
The linear maps $\omega{:}\; A\,{\to}\,A$ given by
  \be
  \omega_{V_1,\dots,V_m}(X_1,...\,,X_m) = 
  (V_1 X_1 V_1^{-1}, ...\,, V_m X_m V_m^{-1}) \labl{eq:omega}
are algebra automorphisms of $A$. Given any $\sigma$ on $A$ we can
find an $\omega$ of the form \erf{eq:omega} such that 
  \be
  \omega \circ \sigma \circ \omega^{-1} = \sigma^\pi_{\X_1,\dots,\X_m} \,,
  \labl{eq:general-vect-star}
where $\pi$ is again a permutation of order two and the 
$\X_i \iN {\rm Mat}_{n_i}(\complex)$ are the identity
matrix whenever $\pi(i)\,{\ne}\, i$. For $\pi(i)\eq i$ the matrix $\X_i$
is either the identity matrix or the anti-symmetric matrix with blocks
\,\ppmatrix01{-1}0\,
along the diagonal.

\medskip

Let us return to the case of a general modular tensor category $\calc$. 
Below it will be instrumental to deal with an algebra automorphism of
$A$ whose restriction to $\Atop$ is of the form \erf{eq:omega}. 
Such an automorphism can be constructed as follows.

For $A$ a symmetric special Frobenius algebra and $M$ a left $A$-module,
let $M\,{\cong}\,\bigoplus_{\kappai\in\JJ}\! M_\kappai^{\oplus n_\kappai}$
be its decomposition into the direct sum of simple $A$-modules, as
in lemma \ref{le:Atop-mat}. This means that there are morphisms
  \be
  x_{\kappai\alpha} \in \HomA(M_\kappai,M)
  \qquad {\rm and} \qquad
  \bar x_{\kappai\alpha} \in \HomA(M,M_\kappai) \ee
with $\kappai\iN\JJ$ and $\alpha\eq1,2,...\,,n_\kappai$ such that
  \be
  \bar x_{\kappai\alpha} \cir x_{\kappaj\alpha}
  = \delta_{\kappai,\kappaj}\, \delta_{\alpha,\beta}\, \id_{M_\kappai}
  \qquad {\rm and} \qquad
  \sum_{\kappai\in\JJ}\sum_{\alpha=1}^{n_\kappai} x_{\kappai\alpha} \cir
  \bar x_{\kappai\alpha} = \id_M \,.  \ee
A basis of $\HomA(M,M)$ is given by
  \be
  e_\kappai^{\alpha\beta} := x_{\kappai\alpha} \cir \bar x_{\kappai\beta}
  \;\in \HomA(M,M)  \labl{eq:app-e-basis}
with $\kappai\iN\JJ$ and $\alpha,\beta\eq1,2,...\,,n_\kappai$.
The basis elements obey the composition rule
  \be
  e_\kappai^{\alpha\beta} \circ e_\kappaj^{\gamma\delta}
  = \delta_{\kappai,\kappaj}\, \delta_{\beta\gamma}\,
  e_\kappai^{\alpha\delta}  \,.  \ee
One checks that in this basis the invertible elements of
$\HomA(M,M)$ are precisly those
  \be
  f = \sum_{\kappai\in\JJ}\sum_{\alpha,\beta=1}^{n_\kappai}
  f^\kappai_{\alpha\beta}\, e_\kappai^{\alpha\beta} \labl{eq:f-for-omega}
for which the matrices $f^\kappai\iN{\rm Mat}_{n_\kappai}(\complex)$
are invertible.

Invertible elements of $\HomA(M,M)$ can be used to construct
algebra automorphisms of $M^\vee \OtA M$. The following
result is a straightforward application of the definitions
\erf{eq:Ptensor} and \erf{eq:B-ssFA}.
\\[-2.3em]

\dtl{Lemma}{lem:HomA}
For $A$ a symmetric special Frobenius algebra and $M$ a left $A$-module,
denote by $(M^\vee \OtA M,e,r)$ the retract associated to the idempotent
$P_{\otimes A}$, as in \erf{eq:Ptensor}. If $f \iN \HomA(M,M)$ is invertible,
then
  \be
  \omega_f := (f^{-1})^\vee \otA f
  = r \circ ( (f^{-1})^\vee \oti f ) \circ e \labl{eq:omega-from-f}
is an algebra automorphism of the symmetric Frobenius algebra $M^\vee \OtA M$.

\medskip

Using the isomorphism $\HomA(M,M) \,{\cong}\, \Hom(\one,M^\vee \OtA M)$
one obtains a basis $\{b_\kappai^{\alpha\beta}\}$ of $B\top$
from the basis \erf{eq:app-e-basis} of $\Hom_A(M,M)$:
  \be
  b_\kappai^{\alpha\beta} := r \circ (\id_{M^\vee} \oti
  e_\kappai^{\beta\alpha} ) \circ \tilde b_M
  \ \in \Hom(\one,M^\vee \OtA M) \,.  \labl{eq:Btop-basis}
Here the order of the multiplicity indices in $b_\kappai^{\alpha\beta}$ is 
chosen in such a way that the product of two such morphisms obeys
  \bea \begin{picture}(220,79)(0,39)
  \put(105,0)    {\includeourbeautifulpicture{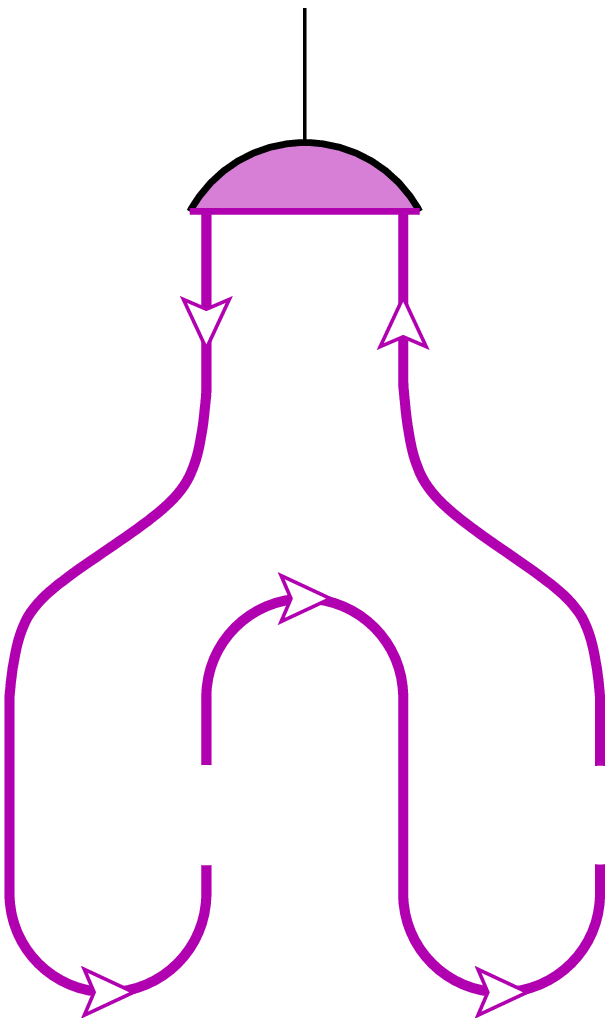}}
  \setlength{\unitlength}{24810sp}
  \setlength{\unitlength}{1pt}
  \put(-87,51)     {$\mtop(b_\kappai^{\alpha\beta},b_\kappaj^{\gamma\delta})
                    \,\equiv\, m \circ (b_\kappai^{\alpha\beta}
                    \oti b_\kappaj^{\gamma\delta}) \;= $}
  \put(117,81)     {\scriptsize$\M$}
  \put(124.5,19.7) {\scriptsize$e_\kappai^{\beta\alpha}$}
  \put(135.3,113.7){\scriptsize$B$}
  \put(150.6,81)   {\scriptsize$\M$}
  \put(167.8,20.2) {\scriptsize$e_\kappaj^{\delta\gamma}$}
  \put(194,51)     {$ =\; \delta_{\kappai,\kappaj}\, \delta_{\beta,\gamma}\,
                    b_\kappai^{\alpha\delta} \,.$}
  \epicture20 \labl{pic74}
In fact, this basis provides us with an isomorphism $\varphi$ of
$\complex$-algebras between $\bigoplus_{\kappai\in\JJ}\!
{\rm Mat}_{n_\kappai}(\complex)$ and $B\top$, given by
  \be
  \varphi :\quad (\X^1,...\,,\X^{|\JJ|}) \,\longmapsto\,
  \sum_{\kappai\in\JJ} \sum_{\alpha,\beta=1}^{n_\kappai}
  \X^\kappai_{\alpha\beta}\, b_\kappai^{\alpha\beta} \,.
  \labl{eq:mat-iso}
Indeed it is easy to check that $\varphi(\vec\X\vec\X')\eq
\varphi(\vec \X)\, \varphi(\vec\X')$, where we use
the shorthand notation $\vec \X\,{\equiv}\,(\X^1,...\,,\X^{|\JJ|})$.

Next let us compute the action of $\omega_f$ from \erf{eq:omega-from-f},
with $f$ as in \erf{eq:f-for-omega}, on $B\top$. One finds
  \be
  \omega_f \circ \varphi(\vec \X) = \sum_{\kappai,\alpha,\beta}
  \X^\kappai_{\alpha\beta} \, r \circ [ \id_{M^\vee} \oti
  (f \cir e_\kappai^{\beta\alpha} \cir f^{-1}) ] \circ \tilde b_M
  = \varphi((\vec f^{-1})^{\rm t} \vec \X \vec f^{\,\rm t}) \;.
\labl{eq:omega-action}
Thus the action of $\omega_f$ on $B\top$ is indeed
of the form \erf{eq:omega}.

\medskip

We have now gathered the necessary ingredients to establish the following
result, which can be regarded as a generalisation\foodnode%
   {What is generalised here is the case of a single matrix block, $m\eq1$.
   For a generalisation in the case $m{>}1$ see the discussion around
   equation \erf{eq:nonsimp-A}.}
of the classification \erf{eq:general-vect-star}
from $\Vect_{\complex}$ to a general modular tensor category $\calc$.

\dtl{Proposition}{th:*sssFA}
Let $(C,\sigma_C)$ be a simple \JA\ in a modular tensor category $\calc$. 
Then one can find
$A$, $M$, $B$, $\sigmaB$, $N$ and $g$ with the following properties:
\\[3pt]
(a)~$A\iN\objc$ is a haploid simple symmetric special 
Frobenius algebra; it is Morita equivalent to $C$.
\\[3pt]
(b)~$M$ is a left $A$-module such that $B$ is isomorphic to $M^\vee \OtA M$
as an algebra. $M$ is either a simple $A$-module, or else it is the direct 
sum $M\,{\cong}\,M_1\,{\oplus}\,M_2$ of two simple $A$-modules $M_{1,2}$
obeying $M_1^\vee\otA M_1\,{\cong}\,(M_2^\vee \otA M_2)_{\rm op}$ as 
algebras. (This implies in particular that
$M_1^\vee\otA M_1\,{\cong}\,M_2^\vee \otA M_2$ as objects in $\calc$.)
\\[3pt]
(c)~$\sigmaB\iN\End(B)$ is a reversion on the algebra $B$,
and $N$ is a left $B$-module.
\\[3pt]
(d)~$(C,\sigma_C)$ and $(N^\vee\OtB N,\tilde\sigma_g)$ are isomorphic
as \JA s, where $\tilde\sigma_g$ is defined as in \erf{eq:B-star} and 
$g\iN\Hom_B(N,N^{\sigmaB})$ fulfills \erf{eq:g-prop} with $\eps_g\eq1$.

\medskip\noindent
Proof:\\
(i)~\,{\em Construction of the idempotent $p$ in $C\top$}:\\
By proposition \ref{th:sssFA} we can find a haploid algebra $D$ and a 
$D$-module $R$ such that $C$ is isomorphic to $R^\vee{\otimes_D}R\,{=:}\,C'$.
Let then $\alpha\iN\Hom(C,C')$ be an algebra isomorphism between $C$ and 
$C'$; it induces a reversion $\sigma'_C\eq\alpha\cir\sigma_C \cir\alpha^{-1}$
on $C'$, such that $(C,\sigma_C) \,{\cong}\, (C',\sigma'_C)$ as \staralgebras.
\\
Decompose $R$ as $R \,{\cong}\, \bigoplus_{\kappai\in\JJ} \!
R_\kappai^{\otimes n_\kappai}$, with distinct simple $D$-modules $R_\kappai$.
When applied to $C'\top$, the construction of the isomorphism $\varphi$ 
in \erf{eq:mat-iso} shows that $C'\top$ is isomorphic to 
$\bigoplus_{\kappai=1}^{|\JJ|} {\rm Mat}_{n_\kappai}(\complex)$. 
Furthermore, since the twist is trivial on the tensor unit $\one$, 
$\sigma'_C$ acts on $C'\top$ as an anti-isomorphism that squares to 
the identity. Formulas \erf{eq:general-vect-star} and \erf{eq:omega-action}
imply that we can find an automorphism $\omega_f$ of $C'$ such that 
$\omega_f^{-1} \circ \sigma'_C \circ \omega_f$ acts as
$\sigma^\pi_{\X_1\dots \X_{|\JJ|}}$ on $C'\top$. Let ${b'}^{\alpha\beta}_
{\!\kappai}$ be a basis of $C'\top$ constructed as in \erf{eq:Btop-basis}. 
We choose an element $p'$ of $C'\top$ as 
  \be 
  p' := \left\{ \bearll
  \omega_f \circ {b'}^{11}_{\!1}
  & {\rm for}\;\ \X_1\eq\id \ \;{\rm and}\;\ \pi(1)\eq1 \,,  \\{}\\[-.8em]
  \omega_f \circ ({b'}^{11}_{\!1} \,{+}\, {b'}^{11}_{\!\pi(1)}) 
  & {\rm for}\;\ \X_1\eq\id \ \;{\rm and}\;\ \pi(1)\,{\ne}\,1 \,, \\{}\\[-.8em]
  \omega_f \circ ({b'}^{11}_{\!1} \,{+}\, {b'}^{22}_{\!1})  
  & \mbox{if $\X_1$ is antisymmetric}\,.
  \eear\right.\labl{p123}
We can transport $p'$ back to $C$ by defining 
$p\,{:=}\,\alpha^{-1} \cir p'\iN C\top$. One verifies that in all three 
cases $m \cir (p\oti p)\eq p$ and $\sigma_C \cir p\eq p$.
\\[5pt]
(ii)~\,{\em Construction of the data $B$, $\sigmaB$, $N$ and $g$}\,:\\
We can now apply lemma \ref{le:PB-star} to $C$ and the idempotent $p\iN C\top$. 
This yields an \staralgebra\ $(B,\sigmaB)$ and a left $C$-module $N'$, as well as a 
morphism $g'\iN\Hom_C(N',{N'}^{\sigma_C})$ with $\eps_{g'}\eq1$ such that
$B\eq{N'}^\vee{\otimes_C}N'$ and $\sigmaB$ is of the form \erf{eq:B-star}. 
As described at the end of section \ref{sec:morita} the construction
of $(B,\sigmaB)$ can be inverted. Thus there is a left $B$-module 
$N \,{\cong}\, {N'}^\vee$ and $g \iN \Hom_B(N,N^{\sigmaB})$ with
$\eps_g\eq1$ such that $(C,\sigma_C)$ is isomorphic as a \JA\
to $(N^\vee{\otimes_B}N, \tilde\sigma_g)$.
\\[5pt]
(iii)~\,{\em Construction of $A$ and $M$}:\\
The algebra $B$ is by definition the image of the idempotent $P_B \iN \Hom(C,C)$
introduced in \erf{eq:PMB-def}. It follows that $B\top$ is isomorphic as a
$\complex$-algebra to the image of the idempotent $p\iN C\top$ from part
(i) above. Moreover, $\sigmaB$ restricts to an anti-isomorphism 
$\sigmaB\raisebox{-2pt}{$|$}_{B\top}$ of $B\top$. We find that 
  \be
  (B\top, \sigmaB\raisebox{-2pt}{$|$}_{B\top}) \,\cong\, \left\{ \bearll
  (\complex, a{\mapsto} a)  
  & {\rm for}\;\ \X_1\eq\id \ \;{\rm and}\;\ \pi(1)\eq1 \,,  \\{}\\[-.8em]
  (\complex\,{\oplus}\,\complex , (a,b){\mapsto}(b,a) )
  & {\rm for}\;\ \X_1\eq\id \ \;{\rm and}\;\ \pi(1)\,{\ne}\,1 \,, \\{}\\[-.8em]
  ({\rm Mat}_2(\complex), \ppmatrix abcd {\mapsto} \ppmatrix d{\!-b}{\!-c}a\,)
  & \mbox{if $\X_1$ is antisymmetric}\,.
  \eear\right.\labl{b123} 
In the first case $B$ is already haploid and we can choose $A\eq M\eq B$.
Otherwise we can take the nontrivial idempotent 
$b^{11}_1\iN B\top$ according to the basis \erf{eq:Btop-basis}. Applying lemma 
\ref{le:ImP-sssFA} we obtain an algebra $A$ and a left $B$-module $M'$ 
such that $A \,{\cong}\, {M'}^\vee{\otimes_B}M'$. Conversely
we have $B \,{\cong}\, M'{\otimes_A}{M'}^\vee$, so that we can choose 
$M\eq{M'}^\vee$. The algebra $A$ is haploid by lemma \ref{le:ImP-sssFA}(ii)
and the fact that $b^{11}_1$ is nontrivial.
\\
We can construct an isomorphism between $B\top$ and 
$\bigoplus_\kappai\!{\rm Mat}_{n_\kappai}(\complex)$ as in \erf{eq:mat-iso}.
One finds that in order to recover the correct algebra structure on $B\top$,
the module $M$ must be simple in the first case of formulas \erf{p123} and
\erf{b123}, be of
the form $M \,{\cong}\, M_1 \,{\oplus}\, M_2$ with $M_{1,2}$ distinct 
simple modules in the second case, and
$M \,{\cong}\, M_1 \,{\oplus}\,M_1$ with $M_1$ simple in the third case.
\\[5pt]
(iv)~\,{\em Proof of $M_1^\vee \otA M_1 \,{\cong}\, 
(M_2^\vee \otA M_2)_{\rm op}$}:\\
When $B \,{\cong}\, (M_1{\oplus}M_2)^\vee \otA (M_1{\oplus}M_2)$ with
$M_1$ and $M_2$ two simple non-isomorphic left $A$-modules, then the
reversion $\sigmaB$ acts on $B\top$ as in the second case of \erf{b123}. 
Thus we can find a basis $\{p_1, p_2\}\,{\subset}\,\Hom(\one,B)$ of 
$B\top$ such that $\mtop(p_i,p_j)\eq\delta_{i,j}p_i$ and such that 
$\sigmaB$ acts on this basis as $\sigmaB p_1\eq p_2$, $\sigmaB p_2\eq p_1$.
\\
Denote by $P_i$ the
idempotent $P_B$ of \erf{eq:PMB-def} with $p_i$ in place of $p$ and
$A$ in place of $B$. It is easy to check that 
$\sigmaB \cir P_{1,2}\eq P_{2,1} \cir \sigmaB$. We can choose
the labelling of the idempotents $P_i$ such that the image
$(S_1,e_1,r_1)$ of $P_1$ satisfies $S_1 \,{\cong}\, M_1^\vee{\otimes_A}M_1$ 
(as an object in $\calc$), and analogously for $P_2$. 
\\
The objects $S_1$ and $S_2$ can be turned into algebras via
lemma \ref{le:ImP-sssFA}. We would like to show that $S_1 \,{\cong}\, 
S_{2,\rm op}$. To this end we first show that the morphisms
  \be
  f := r_2 \cir \sigmaB \cir e_1 \in \Hom(S_1,S_2)
  \qquad {\rm and} \qquad
  g := r_1 \cir \sigmaB^{-1} \cir e_2 \in \Hom(S_2,S_1) \,.  \ee
are inverse to one another; indeed we have
  \be
  f \circ g = r_2 \circ \sigmaB \circ P_1 \circ \sigmaB^{-1} \circ e_2
  = r_2 \circ P_2  \circ e_2 = \id_{S_2} \,,  \ee
and analogously for $g\cir f$. Next note the identities
  \be
  f \circ m_1 = r_2 \cir \sigma_B \cir m \cir (e_1{\otimes}e_1)
  = r_2 \cir m \circ c_{B,B} \cir (\sigma_B{\otimes}\sigma_B) \circ
  (e_1{\otimes}e_1) = m_2 \circ c_{S_2,S_2} \circ (f{\otimes}f) \,.  \ee
Here $m_{1,2}$ denote the multiplications on $S_{1,2}$, and 
in the last step we used that $P_2 \cir\sigma\eq\sigma \circ P_1$.
We thus see that $f$ is an algebra isomorphism from $S_1$ to
$S_2^{(1)} \cong S_{2,\rm op}^{}$.
\qed

\medskip

We are now almost done with the proof of the algorithm stated in steps 
1)\,--\,4) in section \ref{sec:*-class}.
It only remains to explain
why the case that $M_1\,{\cong}\, M_2$ (as $A$-modules) is not excluded in 
statement (b) of proposition \ref{th:*sssFA}, but nevertheless does not occur 
in step 3b) of the algorithm. This is achieved by proposition \ref{pr:M+M_to_M} 
below, where it is shown that this situation is Morita equivalent to an
\staralgebra\ of the form $M^\vee\OtA M$ with $M$ a simple $A$-module,
and hence is already treated in step 3a).

The following three lemmata prepare the proof of the proposition.

\dtl{Lemma}{le:Mat2iso}
For every left module $M$ over a symmetric special Frobenius algebra $A$ 
there is an isomorphism
  \be
  (M{\oplus}M)^\vee \otA (M{\oplus}M) \,\cong\,
  {\rm Mat}_2(\complex) \otimes (M^\vee\OtA M) \ee
of symmetric special Frobenius algebras. Here ${\rm Mat}_2(\complex)$ denotes 
the algebra structure on $\one^{\oplus 4}$ given by matrix multiplication.

\medskip\noindent
Proof:\\
Let $U\eq\one{\oplus}\one$, $A\eq M^\vee\OtA M$
and $B\eq(M{\oplus}M)^\vee \otA (M{\oplus}M)$. 
Let $\varphi$ be an invertible element of $\HomA(M{\oplus}M , M{\otimes}U)$. 
Define the morphism $f\iN\Hom(B,U^\vee{\otimes}U{\otimes}A)$ as
  \bea \begin{picture}(105,95)(0,50)
  \put(41,0)    {\includeourbeautifulpicture{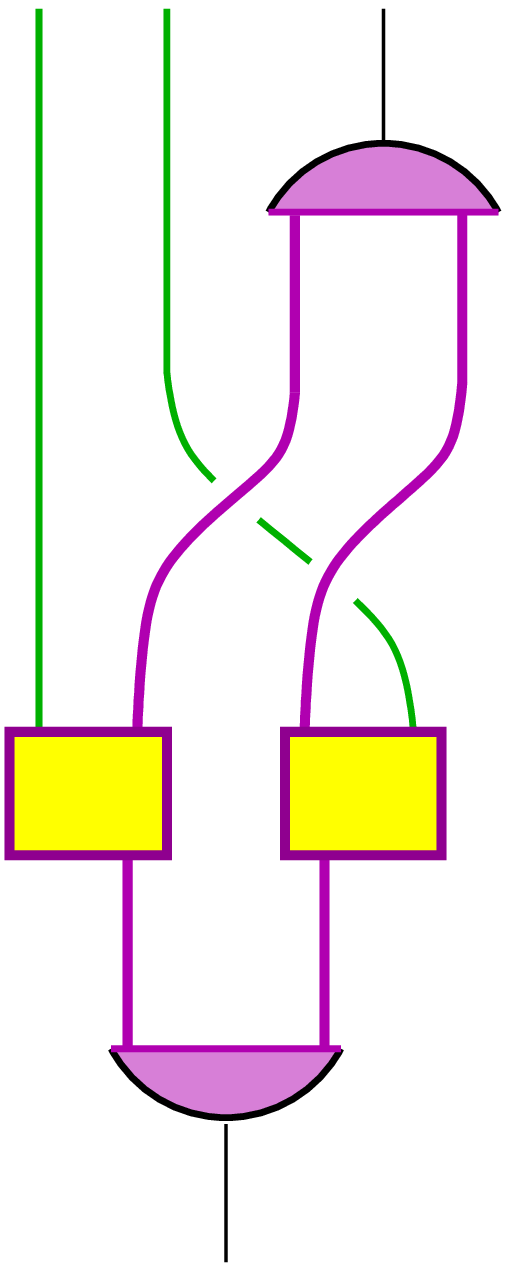}}
  \setlength{\unitlength}{24810sp}
  \setlength{\unitlength}{1pt}
  \put(-15,70)     {$ f \;:= $}
  \put(18.6,30)    {\scriptsize$(\M{\oplus}\M)^{\!\vee}_{}$} 
  \put(40.5,141.3) {\scriptsize$U^\vee_{}$} 
  \put(43.3,49.1)  {\scriptsize${\varphi^{\!-\!1}}
                    _{}^{\scriptscriptstyle\!\vee}$} 
  \put(56.7,141.3) {\scriptsize$U$} 
  \put(62.2,-8.5)  {\scriptsize$B$} 
  \put(78.6,30)    {\scriptsize$\M{\oplus}\M$} 
  \put(78.4,50.3)  {\scriptsize$\varphi$} 
  \put(79.7,141.3) {\scriptsize$A$} 
  \epicture31 \labl{pic114} 
We endow $C\,{:=}\,U^\vee{\otimes}U{\otimes}A$ with the algebra structure
of the tensor product of the algebras $U^\vee{\otimes}U$ 
and $A$, see proposition I:3.22. It is a fairly straightforward application 
of the definitions to check that $f$ is invertible and obeys
$f\cir\eta_B\eq\eta_C$ and $f\cir m_B\eq m_C\cir(f{\otimes}f)$.
Thus $B$ and $C$ are isomorphic as algebras. Since the algebra structure
determines the symmetric special Frobenius structure uniquely, 
$B$ and $C$ are isomorphic also as symmetric special Frobenius algebras.
\qed 

\dtl{Lemma}{le:Mat2_to_C}
Consider the algebra ${\rm Mat}_2(\complex)$ equipped with the reversion
$\sigma_2$ defined as
  \be
  \sigma_2 \ppmatrix abcd := \ppmatrix d{\!-b}{\!-c}a \,. \labl{eq:sig2-action}
The pair $({\rm Mat}_2(\complex), \sigma_2)$ is Morita equivalent to
$(\complex, \id)$ in the sense of proposition \ref{pr:B-star},
with $\eps_g\eq{-}1$.

\medskip\noindent
Proof:\\
In the category of vector spaces we choose the quantities $(A,\sigma)$,
$M$ and $g$ entering in proposition \ref{pr:B-star} as follows.
For the algebra $A$ we take $A\eq\complex$, with trivial reversion
$\sigma\eq\id_\complex$.
The module $M$ is a two-dimensional vector space $V$ on which $A$ acts
by multiplication. Let $\{e_1,e_2\}$ be a basis of $V$. The dual module
$M^\vee$ is the dual vector space $V^*$ with dual basis $\{e_1^*,e_2^*\}$.
Since the action of $A$ is multiplication by complex
numbers and $\sigma$ is the identity we have
$\Hom(M,M^\sigma)\eq{\rm Hom}(V,V^*)$. We set
  \be
  g := \sum_{p,q=1}^2 \epsilon_{pq} \, e_p^*{\otimes}e_q^*
  \equiv e_1^*{\otimes}e_2^* - e_2^*{\otimes}e_1^*
  \,. \labl{eq:Mat2C-1}
Note that the inverse of $g$ is
$g^{-1}\eq\sum_{pq} \epsilon_{pq}\, e_q{\otimes}e_p$.
\\[2pt]
To verify that $g$ fulfills the property \erf{eq:g-prop} one must check
that $g(e_k)\eq\eps_g \sum_m e_k( g(e_m) ) \, e_m^*$ for $k\eq 1,2$ with a 
suitable sign $\eps_g$. Substituting the definition \erf{eq:Mat2C-1} gives
$g(e_k)\eq\epsilon_{\bar k k} e_{\bar k}^*$ for the \lhs\ (where,
for the sake of this proof, we use the symbol $\bar p$ to denote the 
index such that $\epsilon_{p\bar p}\,{\ne}\,0$, i.e.\ $\bar p\eq3{-}p$),
while for the \rhs\ one gets the expression
  \be
  \eps_g \sum_{m,p,q} \epsilon_{pq} \big( (e_p^* e_k)(e_q^* e_m) \big) 
  \, e_m^* = \eps_g \epsilon_{k\bar k}\, e_{\bar k}^*  \,,  \ee
so that $\eps_g\eq{-}1$.
Thus proposition \ref{pr:B-star} can be applied, which provides us
with a reversion $\tilde\sigma_g$ on 
$B\eq V^*{\otimes}V \,{\cong}\, {\rm Mat}_2(\complex)$. Computing the
action of $\tilde\sigma_g$ on a basis yields
  \be
  \tilde\sigma_g(e_p^*{\otimes}e_q)
  = (-1) \cdot g(e_q){\otimes}g^{-1}(e_p^*)
  = \epsilon_{p\bar p} \epsilon_{q \bar q}\,
  e_{\bar q}^* {\otimes} e_{\bar p} \,.  \ee
This reproduces precisely the definition of $\sigma_2$
in formula \erf{eq:sig2-action}.
\qed 

\dtl{Lemma}{le:*-MatxA}
Let $A$ be an algebra, and let $\sigma$ be a reversion on 
$B\eq{\rm Mat}_2(\complex) \oti A$, with
${\rm Mat}_2(\complex)$ the algebra structure on
$\one^{\oplus 4}$ given by matrix multiplication and the
algebra structure on the tensor product defined according to
proposition I:3.22. Assume further that $\sigma$ satisfies
  \be
  \sigma \circ (\id_{\rm Mat}\oti\eta_A)
  = \sigma_2 \otimes \eta_A \, \ee
with $\sigma_2$ the reversion on
${\rm Mat}_2(\complex)$ introduced in \erf{eq:sig2-action}. Then 
$\sigma$ can be written as
  \be
  \sigma = \sigma_2 \otimes \sigma_{\!A}  \labl{eq:MatxA-sig-decomp}
for some reversion $\sigma_{\!A}$ on $A$.

\medskip\noindent
Proof:\\
According to the proof of lemma \ref{le:Mat2_to_C} (and with
the notations used there, in particular for 
$\epsilon_{pq}\eq{-}\epsilon_{qp}$ and $\bar p\eq3{-}p$),
in terms of the basis $e_{pq}\,{\equiv}\, e_p^*{\otimes}e_q$
the product and the action of $\sigma_2$ on ${\rm Mat}_2(\complex)$ read
  \be
  m \circ (e_{pq}{\otimes}e_{rs}) = \delta_{q,r} \, e_{ps}
  \qquad {\rm and} \qquad
  \sigma_2 \circ e_{pq} = 
  \eps_{p\bar p} \eps_{q\bar q}\, e_{\bar q\bar p} \,.  \ee
Denote by $\imath_{i\alpha}^A$ a basis of 
$\Hom(U_i,A)$, as in (I:3.4). Then a basis of $B$ is given by
the morphisms $e_{pq} \oti \imath_{i\alpha}^A$.
The action of $\sigma$ on this basis can be written as
  \be
  \sigma(e_{pq} \Oti \imath_{i\alpha}^A)
  = \sum_{r,s} e_{rs} \oti {(\imath_{i\alpha}^A)_{pq}}^{rs}
  \ee
for some morphisms ${(\imath_{i\alpha}^A)_{pq}}^{rs} \iN \Hom(U_i,A)$.
\\[3pt]
Next we compose the property $\sigma \cir m_B\eq m_B \cir c_{B,B} \cir 
(\sigma{\otimes}\sigma)$ of the reversion $\sigma$ with the morphism
 $(e_{pq}{\otimes}\eta_A) \oti (e_{rs}{\otimes}\imath_{i\alpha}^A)$.
Using that $e_{pq}{\otimes}\eta_A\iN \Hom(\one,B)$ amounts to having
trivial braiding, we find
  \bea
  \quad\ \ \delta_{q,r}\, \sigma (e_{ps}{\otimes}\imath_{i\alpha}^A)
  = \epsilon_{p\bar p} \epsilon_{q\bar q} \, m \cir  [ 
  \sigma(e_{rs}{\otimes}\imath_{i\alpha}^A) \oti   
  (e_{\bar q\bar p}{\otimes}\eta_A) ] 
  \\{}\\[-.6em]
  \Leftrightarrow\;
  \delta_{q,r} \sum_{m,n} e_{mn}{\otimes}
  {(\imath_{i\alpha}^A)_{ps}}^{\! mn} = 
  \epsilon_{p\bar p} \epsilon_{q\bar q}
  \sum_m e_{m\bar p}{\otimes} {(\imath_{i\alpha}^A)_{rs}}^{\! m\bar q} 
  \\{}\\[-.6em]
  \Leftrightarrow\;
  \delta_{q,r}  {(\imath_{i\alpha}^A)_{ps}}^{\! mn} 
  = \epsilon_{pn}\epsilon_{q\bar q}\, {(\imath_{i\alpha}^A)_{rs}}^{\! m\bar q}
  \eear\labl{eq:MatxA-rel1}
Repeating the calculation with
 $(e_{rs}{\otimes}\imath_{i\alpha}^A) \oti (e_{pq}{\otimes}\eta_A)$
in place of
 $(e_{pq}{\otimes}\eta_A) \oti (e_{rs}{\otimes}\imath_{i\alpha}^A)$
yields the analogous set of relations
  \be
  \delta_{p,s} {(\imath_{i\alpha}^A)_{rq}}^{\! mn}
  = \epsilon_{p\bar p} \epsilon_{qm}\, {(\imath_{i\alpha}^A)_{rs}}^{\! \bar pn}
  \,.  \labl{eq:MatxA-rel2}
It follows that 
  \be
  {(\imath_{i\alpha}^A)_{pq}}^{\! rs} = \delta_{p,\bar s} \delta_{q,\bar r}\,
  {(\imath_{i\alpha}^A)_{pq}}^{\! \bar q\bar p}  \,.  \ee
Setting in addition $m\eq n\eq \bar p$ in \erf{eq:MatxA-rel1} and 
$m\eq n\eq \bar q$ in \erf{eq:MatxA-rel2} leads to
  \be
  {(\imath_{i\alpha}^A)_{pp}}^{\! \bar p\bar p} 
  = \eps_{p\bar p} \eps_{q \bar q}\,
  {(\imath_{i\alpha}^A)_{qp}}^{\! \bar p\bar q} 
  = {(\imath_{i\alpha}^A)_{qq}}^{\! \bar p\bar p} \,.  \ee
Altogether we thus have
  \be
  {(\imath_{i\alpha}^A)_{pq}}^{\! rs}
  = \epsilon_{ps} \epsilon_{qr}\, (\imath_{i\alpha}^A)^\sigma \ee
for some morphism $(\imath_{i\alpha}^A)^\sigma \iN \Hom(U_i,A)$. Defining
the morphism $\sigma_{\!A} \iN \Hom(A,A)$ via its action on the
basis $\imath_{i\alpha}^A$ as 
$\sigma_{\!A}(\imath_{i\alpha}^A)\,{:=}\,(\imath_{i\alpha}^A)^\sigma$ 
we find that
  \be
  \sigma(e_{pq} \oti \imath_{i\alpha}^A)
  = \sum_{r,s} \epsilon_{ps} \epsilon_{qr}\,
     e_{rs} \oti \sigma_{\!A}(\imath_{i\alpha}^A)
  = \sigma_2(e_{pq}) \oti \sigma_{\!A}(\imath_{i\alpha}^A) \,.  \ee
This establishes equation \erf{eq:MatxA-sig-decomp}. 
\\[3pt]
It remains to show that $\sigma_{\!A}$ is a reversion on $A$.
By the compatibility of $\sigma$ with $m_B$ we have
  \be
  \sigma \cir m_B \cir [
  (e_{11}{\otimes}\id_A) \oti (e_{11}{\otimes}\id_A) ]
  = m_B \cir c_{B,B} \cir [
  \sigma(e_{11}{\otimes}\id_A) \oti \sigma(e_{11}{\otimes}\id_A) ] \,.
  \ee
The \lhs\ of this identity reduces to $\sigma\cir(e_{11}\oti m_A)\eq e_{22}
\oti (\sigma_{\!A} \cir m_A)$\,, while the \rhs\ can be rewritten as
  \be
  m_B \cir c_{B,B} \cir [ (e_{22}{\otimes}\sigma_{\!A})
  \oti (e_{22}{\otimes}\sigma_{\!A}) ]
  = e_{22} \otimes [m_A \cir c_{A,A} \cir (\sigma_{\!A}{\otimes}\sigma_{\!A})]
  \,. \ee
It follows that $\sigma_{\!A} \cir m_A\eq
m_A \cir c_{A,A} \cir (\sigma_{\!A}{\otimes}\sigma_{\!A})$, as required.
That the condition $\sigma_{\!A}\cir\sigma_{\!A}\eq\theta_{\!A}$ is satisfied
as well can be seen by composing both sides of the identity
$\sigma\cir\sigma\eq\theta_B$ with $e_{11}{\otimes}\id_A$ and noting that
$\theta_B\cir(e_{11}{\otimes}\id_A)\eq e_{11} {\otimes} \theta_{\!A}$.
\qed

\dtl{Proposition}{pr:M+M_to_M}
For $A$ a symmetric special Frobenius algebra, let $M$ be a simple left
$A$-module and $B$ the algebra $(M{\oplus}M)^\vee \otA (M{\oplus}M)$.
Suppose $\sigmaB$ is a reversion on $B$ acting on
$B\top \,{\cong}\, {\rm Mat}_2(\complex)$ according to formula
\erf{eq:sig2-action}.
Then there exist $g$ and $\tilde\sigma_g$ such that $(B,\sigmaB)$ is
Morita equivalent to $(M^\vee\OtA M,\tilde\sigma_g)$ in the sense of
proposition \ref{pr:B-star}.

\medskip\noindent
Proof:\\
Abbreviate $M^\vee\OtA M\,{=:}\,C$ and ${\rm Mat}_2(\complex)\oti C\,{=:}\,D$.
By lemma \ref{le:Mat2iso} we have $B \,{\cong}\, D$ as algebras.
Via this isomorphism the reversion $\sigmaB$ induces a reversion
$\sigma$ on $D$. By assumption, $\sigma$ acts on $D\top$ as $\sigma_2$ in
$\erf{eq:sig2-action}$. According to lemma \ref{le:*-MatxA} we can then
write $\sigma\eq\sigma_2{\otimes}\sigma_C$ for some reversion $\sigma_C$
on $C$.
\\
Next we apply the construction in proposition \ref{pr:B-star} separately
to both factors of $D$. That is, we take the module $V{\otimes}C$, and for
the morphism $g$ we choose $g\eq g_2{\otimes}g_C$, with $g_2$ obtained by 
inverting the constrution of lemma \ref{le:Mat2_to_C} (recall proposition 
\ref{pr:sigmatildetilde}) and  $g_C \iN \Hom_C(C,C^{\sigma_C}_{})$ equal to 
$\Phi_2\cir\sigma_C^{-1}$ (compare to the calculation in \erf{eq:pic72}). 
This reduces the ${\rm Mat}_2(\complex)$-part of $D$ to $\one$
as in lemma \ref{le:Mat2_to_C} and leaves $(C,\sigma_C)$ invariant.
Thus altogether $(D,\sigma)$, and hence also $(B,\sigmaB)$, is Morita
equivalent to $(C,\sigma_C)$ in the sense of proposition \ref{pr:B-star}.
\qed

This completes the proof of the classification algorithm for reversions that 
was described in steps 1)\,--\,4) in section \ref{sec:*-class}.

\medskip

In steps 1)\,--\,4) one is only concerned with reversions on {\em simple\/} 
symmetric special Frobenius algebras. In general, the algebra could be 
of the form \erf{eq:nonsimp-A}, though. In the sequel we extend the
construction to include also direct sums of simple algebras.

Let $e_i \iN \Hom(A_i,A)$ and $r_i \iN \Hom(A,A_i)$ denote the embedding 
and restriction for the subobject $A_i$ of $A$. The $e_i$ and $r_i$ are 
algebra homomorphisms.  A reversion $\sigma$ on $A$ can permute the 
components $A_i$. Thus suppose that
  \be
  r_l \circ \sigma \circ e_k = \delta_{l,\pi(k)} \,
  r_{\pi(k)} \circ \sigma \circ e_k \ee
for some permutation $\pi$ with $\pi \cir\pi\eq\id$. 
This implies in particular that $A_{\pi(k)} \,{\cong}\, A_{k,\rm op}$.
\\[5pt]
{\underline{Suppose $\pi(k)\eq k$}}. Then $\sigma$ restricts to
a reversion on $A_k$ via $r_k \circ \sigma \circ e_k$. It must
therefore be in the list obtained by steps 1)\,--\,4). 
\\[5pt]
{\underline{Suppose $\pi(k)\,{\ne}\,k$}}. Then we can choose
correlated bases for $A_k$ and $A_{\pi(k)}$ in which
$\sigma$ takes a particularly simple form. Choose dual bases 
  \be  
  f^{k}_{a,\alpha} \in \Hom(U_a,A_k) 
  \qquad {\rm and} \qquad \bar f^{k}_{a,\alpha} \in \Hom(A_k,U_a) \,.
  \ee
Using these, define dual bases for the subalgebra $A_{\pi(k)}$ as
  \be  \bearll
  f^{\pi(k)}_{a,\alpha} := r_{\pi(k)} \circ \sigma
  \circ e_k \circ f^{k}_{a,\alpha} \in \Hom(U_a,A_{\pi(k)}) 
  \qquad {\rm and}
  \\{}\\[-.6em]
  \bar f^{\pi(k)}_{a,\alpha} := \bar f^{k}_{a,\alpha} \circ
  r_{k} \circ \sigma^{-1} \circ e_{\pi(k)} \in \Hom(A_{\pi(k)},U_a) \,.
  \eear \labl{eq:fpi-def}
Using 
$e_{\pi(k)} \cir r_{\pi(k)} \cir\sigma\eq\sigma \cir e_k \cir r_k$ 
the components \erf{eq:*bas} of the
reversion $\sigma$ then take the form
  \be
  \sigma(a)_{k,\alpha}^{~~l,\beta}
  = \delta_{l,\pi(k)} \, \bar f^{\pi(k)}_{a,\beta}
  \circ r_{\pi(k)} \circ \sigma \circ e_k \circ f^{k}_{a,\alpha}
  = \delta_{l,\pi(k)} \, \delta_{\alpha,\beta} \,. \ee
Similarly we get
  \be
  \sigma(a)_{\pi(k),\alpha}^{~~~~l,\beta}
  = \delta_{l,k} \, \bar f^{k}_{a,\beta}
  \circ r_{k} \circ \sigma \circ e_{\pi(k)} \circ f^{\pi(k)}_{a,\alpha}
  = \theta_a \, \delta_{l,k} \, \delta_{\alpha,\beta} \,.  \ee
We can thus conclude that for all subalgebras $A_k$ of $A$ which
are not fixed under $\sigma$, i.e.\ $\pi(k)\,{\ne}\,k$, one can
always choose a basis such that the components of $\sigma$
take the particular form
  \be
  \sigma(a)_{k,\alpha}^{~~\pi(k),\beta} = \delta_{\alpha,\beta}
  \qquad{\rm and}\qquad
  \sigma(a)_{\pi(k),\alpha}^{~~~~k,\beta} = 
  \theta_a \, \delta_{\alpha,\beta} \,.  \ee
Applied to the category 
$\Vect_\complex$, the simple form of $\sigma$ corresponds
to $D_i\eq\id$ if $\pi(i)\,{\ne}\,i$ in the expression
\erf{eq:general-vect-star}.

\subsection{Invariants for glueing tori}\label{sec:app-tor} 

This appendix provides some details about the recursion relations 
\erf{eq:rec-rel}. 
To demonstrate these relations we use the methods of \cite{TUra},
especially sections II.3.9, IV.3 and IV.5 (the recursion
relations do not appear explicitly in \cite{TUra}, though). Some basic 
notions of three-dimensional topological field theory and references 
to the literature can be found in sections I:2.3 and I:2.4.

Define the extended surface $T$ to be a torus without field
insertions, with lagrangian subspace $\lambda_T$. We select a basis $a,b$ 
of cycles in $H_1(T,\reals)$ with intersection number $\omega(a,b)\eq1$
and take $\lambda_T$ to be the span of the cocycle $a$.

Define the cobordism $M^+_{\chiI_k}$ to be a solid torus with
$\partial_+ M^+_{\chiI_k}\eq T$ and $\partial_- M^+_{\chiI_k}\eq\emptyset$
such that the $a$-cycle of $T$ is contractible in $M^+_{\chiI_k}$. Inside 
$M^+_{\chiI_k}$ place a ribbon labelled by $U_k$ running parallel to the 
surface $T$ along the $b$-cycle and with core oriented opposite to the 
$b$-cycle, see figure (\ref{eq:pic99}\,a). Similarly define $M^-_{\chiI_k}$ 
to be a solid torus with
$\partial_+ M^-_{\chiI_k}\eq\emptyset$ and $\partial_- M^-_{\chiI_k}\eq T$ 
such that the $a$-cycle of $T$ is contractible in $M^-_{\chiI_k}$. There 
is again a ribbon labelled by $U_k$ following the $b$-cycle of $T$, but 
this time the orientation is the same way as the one of $b$.
Given two cobordisms $M_1$, $M_2$ and an invertible homeomorphism
$g{:}\ \partial_+ M_1 \,{\to}\, \partial_- M_2$, define
$M_2{\glue{g}}M_1$ to be the cobordism obtained by glueing
$M_1$ to $M_2$ using $g$. 

Identify $H_1(T,\reals)$ with $\reals^2$ by choosing $a\eq(1,0)$ and 
$b\eq(0,1)$. Let $g{:}\ T\,{\to}\,T$ be an invertible homeomorphism, with 
induced action on $H_1(T,\reals)$ given by the matrix $\ppmatrix pqrs$.  
The aim of this appendix is to establish recursions relations that allow 
us to compute the invariants
  \be
  M\left[\ppmatrix pqrs \right]_{kl} := 
  Z( M_{\chiI_k}^-{\glue{g}}M_{\chiI_l}^+,\emptyset,\emptyset)\, 1 \,.  
  \labl{eq:app-rec}
We start by defining Maslov indices and computing them for the torus. 
These are needed to formulate the framing anomaly in the functoriality 
property of the topological field theory. The functoriality property 
will then provide the recursion relations \erf{eq:rec-rel}.  
\medskip

In the description of the Maslov indices and the framing 
anomaly we follow closely the review in section 2.7 of \cite{fffs3}. 
Let $H$ be a real vector space with symplectic form $\omega$.
For any triple of lagrangian subspaces $\lambda_1, \lambda_2, 
\lambda_3\,{\subset}\,H$ define a 
quadratic form $q(x,y)$ 
on the subspace $( \lambda_1{+}\lambda_2){\cap}\lambda_3$ 
via $q(x,y)\,{:=}\,\omega(x_2,y)$, where $x\,{=:}\,x_1\,{+}\,x_2$ 
with $x_{1,2} \iN \lambda_{1,2}$. One verifies
that $q$ is independent of the choice of decomposition of $x$. 
The {\em Maslov index} $\mu(\lambda_1,\lambda_2,\lambda_3)$
is defined to be the signature of $q$; it
is antisymmetric in all three arguments.

We only need Maslov indices for $H\eq H_1(T,\reals)$, which has
dimension 2, so that the lagrangian subspaces $\lambda_{1,2,3}$ 
have dimension 1. We select vectors $v_1,v_2,v_3 \iN H$ such that
$\lambda_i$ is the linear span of $v_i$, for $i\eq1,2,3$. If any two 
of the $v_i$ are collinear, then $\mu(\lambda_1,\lambda_2,\lambda_3)\eq0$
by the antisymmetry of $\mu$; otherwise we have
  \be
  (\lambda_1 \,{+}\, \lambda_2) \,{\cap}\, \lambda_3 = \lambda_3 \ee
and there exist $\alpha,\beta \iN \reals$ such that 
$v_3\eq\alpha v_1\,{+}\,\beta v_2$. The quadratic form $q$ evaluated on 
$v_3$ is given by
$q(v_3,v_3)\eq\omega(\beta v_2 , v_3)\eq\alpha \beta \, \omega(v_2,v_1)$.
Since the subspace $\lambda_3$ on which $q$ is defined is one-dimensional,
the signature of $q$ is simply the sign of $q(v_3,v_3)$. Thus 
  \be
  \mu(\lambda_1,\lambda_2,\lambda_3)= {\rm sign}(\alpha\beta\,\omega(v_2,v_1))
  \,.  \labl{eq:mu-abo}
Now fix a basis $a,b$ of cycles with $\omega(a,b)\eq1$ and expand the vectors
$v_i$ as $v_i = x_i a + y_i b$. To determine $\alpha$, $\beta$ in
$v_3\eq\alpha v_1\,{+}\,\beta v_2$ we must solve the linear system
  \be
  \pmatrix{ x_3 \cr y_3 } = \pmatrix{ x_1 & x_2 \cr y_1 & y_2 } 
  \pmatrix{ \alpha \cr \beta } .  \ee
Noting that $\omega(v_i,v_j)\eq\omega(x_i a\,{+}\,y_i b, x_j a\,{+}\,y_j b)
\eq x_i y_j\,{-}\,x_j y_i$ one verifies that the linear system is solved by
$\alpha\eq{-}\omega(v_2,v_3) / \omega(v_1,v_2)$ and 
$\beta\eq\omega(v_1,v_3) / \omega(v_1,v_2)$. Inserting this result in
\erf{eq:mu-abo}, and using the convention that ${\rm sign}(0)\eq0$, finally 
gives the Maslov index for the torus as
  \be
  \mu(\lambda_1,\lambda_2,\lambda_3) = {\rm sign}[ \, \omega(v_1,v_2) \, 
  \omega(v_1,v_3) \, \omega(v_2,v_3) \, ] \,.  \labl{eq:maslov-torus}
While this formula was derived for the case that no two of the $v_i$
are collinear, it continues to hold in the collinear
case as well, since it then gives zero as required.

\medskip

For any cobordism $M$ a map $N_*$ from the set $\Lambda(\partial_- M)$ 
of lagrangian subspaces of $H_1(\partial_- M, \reals)$ to 
$\Lambda(\partial_+ M)$ can be constructed as follows. For 
$\lambda\iN\Lambda(\partial_- M)$ the vector
$x \iN H_1(\partial_+ M, \reals)$ is in $N_* \lambda$
iff there exists an $x' \iN \lambda$ such that 
$x\,{-}\,x'$ is homologous to zero as a cycle in $M$.
The map $N^*{:}\ \Lambda(\partial_+ M)
\,{\to}\,\Lambda(\partial_- M)$ is defined similarly.

The functoriality axiom of the topological field theory reads
  \be
  Z(M_2{\glue{f}}M_1, \partial_- M_1, \partial_+ M_2)
  =  \kappa^m \, Z(M_2, \partial_- M_2, \partial_+ M_2) 
  \cir f_\sharp \cir    Z(M_1, \partial_- M_1, \partial_+ M_1) \,,
  \ee
with the integer $m$ defined as the sum
  \be
  m := \mu(f_* N_* \lambda(\partial_- M_1),
  f_* \lambda(\partial_+ M_1), N^* \lambda(\partial_+ M_2))
  + \mu(f_* \lambda(\partial_+ M_1),
  \lambda(\partial_- M_2), N^* \lambda(\partial_+ M_2)) \labl{mmumu}
of Maslov indices.
The following schematic rewriting of formula \erf{mmumu} facilitates
keeping track of the arguments of the Maslov indices:
  \be\begin{array}{l} \mbox{}\\[-1.7em]
  \hspace*{150pt} {}^{M_2} \hspace*{160pt} {}^{M_1} \\[-8pt]
  \hspace*{110pt} \overbrace{\hspace*{90pt}}
  \hspace*{80pt} \overbrace{\hspace*{90pt}}\\
  \begin{array}{rcccccccl}
  &\Lambda(\partial_+ M_2) 
  &\stackrel{N^*}{\longrightarrow} &
  \Lambda(\partial_- M_2)
  &\stackrel{f_*}{\longleftarrow} &
  \Lambda(\partial_+ M_1) 
  &\stackrel{N_*}{\longleftarrow} &
  \Lambda(\partial_- M_1) & \\[5pt]
     m =  \qquad\  \mu\;( & 3 &&   && 2 && 1 &) \\[3pt]
  + \quad \mu\;( & 3 && 2 && 1 &&   &) 
  \end{array}\end{array}\ee
Both Maslov indices are evaluated in $\Lambda(\partial_- M_2)$; the numbers
indicate as which argument of $\mu(\cdot,\cdot,\cdot)$ the corresponding 
lagrangian subspace appears.

\medskip

Next we define the two sets of vectors $|\chii_k;T\rangle$ and 
$\langle\chii_k;T|$ as
  \be\begin{array}{ll}
  |\chii_k;T\rangle := Z(M_{\chiI_k}^+ , \emptyset , T )\,1 
    &\in \calh(T ; \emptyset)     \,,\\[4pt]
  \langle\chii_k;T| := Z(M_{\chiI_k}^+ , T , \emptyset ) 
    &\in \calh(T ; \emptyset)^*   \,.
  \eear\ee
As a warmup let us compute the composition of two of these vectors.
We have
  \be
  \delta_{k,l}
  = Z( M_{\chiI_k}^-{\glue{\idsmall}}M_{\chiI_l}^+,\emptyset,\emptyset )\,1
  = \kappa^m\, \langle\chii_k;T| \cir \id_\sharp \cir |\chii_l;T\rangle \,.
  \ee
The value $\delta_{k,l}$ for the invariant follows from the normalisation axiom
for cobordisms of the form $X \,{\times}\, [0,1]$, see (I:2.55).
The integer $m$ is given by the following combination of Maslov indices:
  \be\begin{array}{l} \mbox{}\\[-1.7em]
  \hspace*{125pt} {}^{M^-_{\chiI_k}} \hspace*{115pt} {}^{M^+_{\chiI_l}} \\[-7pt]
  \hspace*{102pt} \overbrace{\hspace*{60pt}}
  \hspace*{67pt} \overbrace{\hspace*{60pt}}\\[-4pt]
  \begin{array}{rcccccccl}
  &\Lambda(\emptyset) 
  &\stackrel{N^*}{\longrightarrow} &
  \Lambda(T)
  &\stackrel{\idsmall_*}{\longleftarrow} &
  \Lambda(T) 
  &\stackrel{N_*}{\longleftarrow} &
  \Lambda(\emptyset) & \\[5pt]
    m =  \qquad\    \mu\;( & 3 &&   && 2 && 1 &) \\[3pt]
  + \quad \mu\;( & 3 && 2 && 1 &&   &) 
  \end{array}\end{array}\labl{eq:chi-chi-maslov}
All six lagrangian subspaces entering in the 
computation of $m$ are in fact identical and equal to $\lambda_T$, 
so that $m\eq0$. To see this first recall that $\lambda_T$ is spanned by 
the $a$-cycle in $T$, and that $M_{\chiI_l}^+$ was chosen such that
the $a$-cycle is contractible in $M_{\chiI_l}^+$. It follows that $N_*$ maps 
$0 \iN \Lambda(\emptyset)$ to $\lambda_T$ in $\Lambda(T)$. The same holds for
$N^*$. Furthermore $\id_*$ acts as the identity on $\Lambda(T)$.
Thus as expected we have
  \be
  \langle\chii_k;T\,|\,\chii_l;T\rangle = \delta_{k,l} \,.  \labl{eq:chi-chi-dual}
The set $\{ |\chii_k;T\rangle \,|\, k \iN \I \}$ forms a basis of
$\calh(T ; \emptyset)$, while
$\{\langle\chii_k;T| \,|\, k \iN \I \}$ is a basis of
$\calh(T ; \emptyset)^*$. The result \erf{eq:chi-chi-dual} shows
that these two bases are dual to each other.

\medskip

Denote by $C_T$ the cylinder $T \,{\times}\, [0,1]$. We have $Z(C_T,T,T)\eq
\id_{\calH(T;\emptyset)}$ (by the normalisation axiom), so that we can write
  \be
  Z(C_T,T,T) = \sum_{r\in\II} |\chii_r;T\rangle \, \langle\chii_r;T| \,.  \ee
Let $f_1$, $f_2$ be two invertible homeomorphisms from $T$ to $T$. The
two three-manifolds $M_{\chiI_k}^-{\glue{\!f_1 \circ f_2\!}}M_{\chiI_l}^+$
and $M_{\chiI_k}^-\!{\glue{f_1}}C_T{\glue{f_2}}M_{\chiI_l}^+$
are homeomorphic. Employing functoriality we obtain the series of equalities
  \bea
  Z(M_{\chiI_k}^- {\glue{\!f_1 \circ f_2\!}} M_{\chiI_l}^+,\emptyset,\emptyset)
  \;=\; 
  Z(M_{\chiI_k}^-\!{\glue{f_1}}C_T{\glue{f_2}} M_{\chiI_l}^+,\emptyset,\emptyset) 
  \\[2pt] \qquad\qquad=\, 
  \kappa^{m_1} \, Z(M_{\chiI_k}^- , T , \emptyset) \circ {f_1}_\sharp \circ
  Z(C_T {\glue{f_2}} M_{\chiI_l}^+ , \emptyset , T)
  \\[6pt] \qquad\qquad=\, 
  \kappa^{m_1+m_2} \, \langle\chii_k;T|\, {f_1}_\sharp \circ 
  Z(C_T , T, T) \circ {f_2}_\sharp \circ Z(M_{\chiI_l}^+ , \emptyset, T)
  \\[6pt] \qquad\qquad=\, 
  \kappa^{m_1+m_2}  \sum_{r\in\II}
  \langle\chii_k;T| {f_1}_\sharp  |\chii_r;T\rangle \, \langle\chii_r;T|
  {f_2}_\sharp |\chii_l;T\rangle 
  \\[4pt] \qquad\qquad=\, 
  \kappa^{m_1+m_2-\tilde m_1 - \tilde m_2}  \sum_{r\in\II} 
  Z( M_{\chiI_k}^-\!{\glue{f_1}}M_{\chiI_r}^+ , \emptyset , \emptyset ) \,
  Z( M_{\chiI_r}^-\!{\glue{f_2}}M_{\chiI_l}^+ , \emptyset , \emptyset )   
  \,.  \eear\ee
The integers $m_1$, $m_2$, $\tilde m_1$, $\tilde m_2$ are given by
the following combinations of Maslov indices: First,
  \be\begin{array}{l} \mbox{}\\[-1.7em]
  \hspace*{130pt} {}^{M^-_{\chiI_k}} \hspace*{101pt} 
  {}^{C_T \glue{f_2} M^+_{\chiI_l}} \\[-7pt]
  \hspace*{107pt} \overbrace{\hspace*{60pt}}
  \hspace*{67pt} \overbrace{\hspace*{60pt}}\\[-3pt]
  \begin{array}{rcccccccl}
  &\Lambda(\emptyset) 
  &\stackrel{N^*}{\longrightarrow} &
  \Lambda(T)
  &\stackrel{(f_1)_*}{\longleftarrow} &
  \Lambda(T) 
  &\stackrel{N_*}{\longleftarrow} &
  \Lambda(\emptyset) & \\[5pt]
    m_1 =  \qquad\      \mu\;( & 3 &&   && 2 && 1 &) \\[3pt]
  + \quad \mu\;( & 3 && 2 && 1 &&   &) 
  \end{array}\end{array}\ee
The first Maslov index reads 
$\mu\big((f_1 f_2)_* \lambda_T, \,(f_1)_* \lambda_T,\,\lambda_T\big)$;
the second index is zero owing to $N^* 0\eq\lambda_T$, as
already discussed in \erf{eq:chi-chi-maslov}. Second,
  \be\begin{array}{l} \mbox{}\\[-1.9em]
  \hspace*{130pt} {}^{C_T} \hspace*{115pt} 
  {}^{M^+_{\chiI_l}} \\[-7pt]
  \hspace*{107pt} \overbrace{\hspace*{60pt}}
  \hspace*{67pt} \overbrace{\hspace*{60pt}}\\
  \begin{array}{rcccccccl}
  &\Lambda(T) 
  &\stackrel{N^*}{\longrightarrow} &
  \Lambda(T)
  &\stackrel{(f_2)_*}{\longleftarrow} &
  \Lambda(T) 
  &\stackrel{N_*}{\longleftarrow} &
  \Lambda(\emptyset) & \\[5pt]
    m_2 =  \qquad\      \mu\;( & 3 &&   && 2 && 1 &) \\[3pt]
  + \quad \mu\;( & 3 && 2 && 1 &&   &) 
  \end{array}\end{array}\ee
For the cylinder $C_T$ the map $N^*$ acts as identity on $\Lambda(T)$,
so that the first index vanishes because of
$N^* \lambda_T\eq\lambda_T$; the second index is zero as well,
since $N_* 0\eq\lambda_T$, again as in \erf{eq:chi-chi-maslov}. Finally,
  \be\begin{array}{l} \mbox{}\\[-1.7em]
  \begin{array}{l}
  \hspace*{130pt} {}^{M^-_{\chiI_k}} \hspace*{115pt} 
  {}^{M^+_{\chiI_r}} \\[-7pt]
  \hspace*{107pt} \overbrace{\hspace*{60pt}}
  \hspace*{67pt} \overbrace{\hspace*{60pt}}\\
  \begin{array}{rcccccccl}
  &\Lambda(\emptyset) 
  &\stackrel{N^*}{\longrightarrow} & \Lambda(T)
  &\stackrel{(f_1)_*}{\longleftarrow} & \Lambda(T) 
  &\stackrel{N_*}{\longleftarrow} & \Lambda(\emptyset) & \\[5pt]
    \tilde m_1 =  \qquad\      \mu\;( & 3 &&   && 2 && 1 &) \\[3pt]
  + \quad \mu\;( & 3 && 2 && 1 &&   &) 
  \end{array}\end{array}
\\{}\\[-.5em]
  \begin{array}{l}
  \hspace*{130pt} {}^{M^-_{\chiI_r}} \hspace*{115pt} 
  {}^{M^+_{\chiI_l}} \\[-7pt]
  \hspace*{107pt} \overbrace{\hspace*{60pt}}
  \hspace*{67pt} \overbrace{\hspace*{60pt}}\\
  \begin{array}{rcccccccl}
  &\Lambda(\emptyset) 
  &\stackrel{N^*}{\longrightarrow} & \Lambda(T)
  &\stackrel{(f_2)_*}{\longleftarrow} & \Lambda(T) 
  &\stackrel{N_*}{\longleftarrow} & \Lambda(\emptyset) & \\[5pt]
    \tilde m_2 =  \qquad\      \mu\;( & 3 &&   && 2 && 1 &) \\[3pt]
  + \quad \mu\;( & 3 && 2 && 1 &&   &) 
  \end{array}\end{array}\end{array}\ee
All four Maslov indices appearing in $\tilde m_1$ and $\tilde m_2$ are
zero, since all of them have two coinciding arguments,
due to either $N_* 0\eq \lambda_T$ or $N^* 0\eq \lambda_T$.
Altogether we obtain
  \be
  Z(M_{\chiI_k}^-{\glue{f_1 \circ f_2}}\!M_{\chiI_l}^+ , \emptyset, \emptyset )
  = \kappa^{\mu((f_1 f_2)_* \lambda_T, \,(f_1)_* \lambda_T,\,\lambda_T)}\!
  \sum_{r\in\II} ^{}
  Z( M_{\chiI_k}^-{\glue{f_1}}M_{\chiI_r}^+ , \emptyset , \emptyset ) \,
  Z( M_{\chiI_r}^-{\glue{f_2}}M_{\chiI_l}^+ , \emptyset , \emptyset )   
  \,.  \labl{eq:app-rsum}

The next step in obtaining the recursion relation is
the computation of the two invariants
  \be
  R(\mathcal{S})_{kl} := Z( M_{\chiI_k}^-{\glue{f_{\mathcal{S}}}}
    M_{\chiI_r}^+, \emptyset , \emptyset )
  \qquad{\rm and}\qquad
  R(\mathcal{T})_{kl} := Z( M_{\chiI_k}^-{\glue{f_{\mathcal{T}}}}
    M_{\chiI_r}^+ , \emptyset , \emptyset ) \,.
  \ee
Here $\mathcal{S}$ and $\mathcal{T}$ are the matrices
  \be
  \mathcal{S} = \pmatrix{ 0 & -1 \cr 1 & 0}
  \qquad{\rm and}\qquad
  \mathcal{T} = \pmatrix{ 1 & 1 \cr 0 & 1} \,,  \ee
while $f_{\mathcal{S}}{:}\ T\,{\to}\,T$ is an invertible homeomorphism 
such that the induced action on $H_1(T,\reals)$ is given by
the matrix $\mathcal{S}$ (in the same basis that was used in \erf{eq:app-rec}),
and similarly for $f_{\mathcal{T}}{:}\ T\,{\to}\,T$. We have
  \bea  \begin{picture}(420,96)(4,53)
  \put(64,0)     {\includeourbeautifulpicture{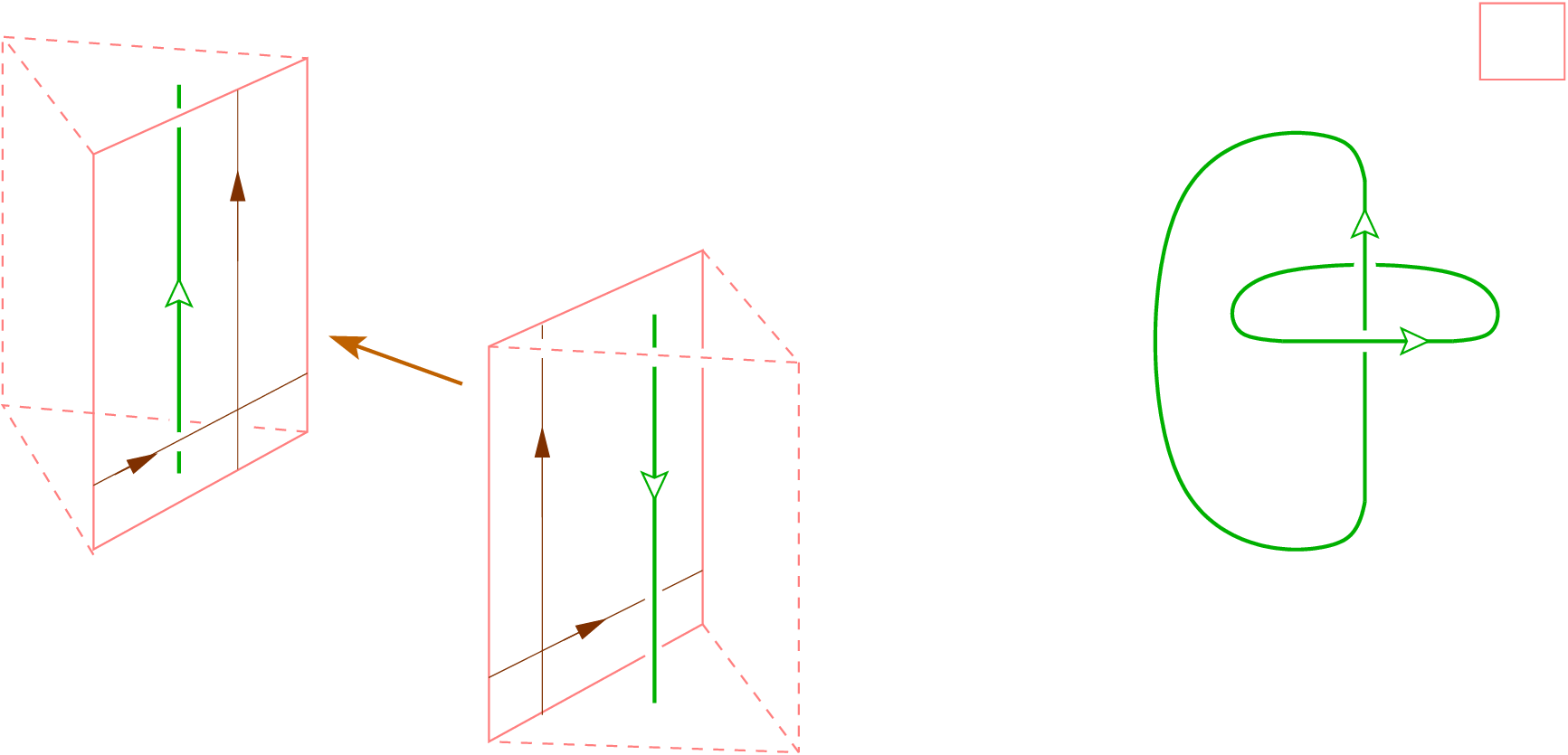}}
  \setlength{\unitlength}{24810sp}
  \setlength{\unitlength}{1pt}
  \put(0,78)       {$R(\mathcal{S})_{kl} \;=$}
  \put(94.2,109)   {\scriptsize$k$}
  \put(113.5,96)   {\scriptsize$b$}
  \put(118.5,77)   {\scriptsize$a$}
  \put(143.5,83.5) {\scriptsize$f_{\mathcal S}$}
  \put(174.9,48)   {\scriptsize$b$}
  \put(188.5,32.2) {\scriptsize$a$}
  \put(192.1,64)   {\scriptsize$l$}
  \put(249,78)     {$=\; S_{00}$}
  \put(341.6,60)   {\scriptsize$k$}
  \put(366.3,94.5) {\scriptsize$l$}
  \put(366.3,140)  {$S^3$}
  \put(380,78)     {$=\; S_{kl}\,.$}
  \epicture28 \labl{eq:app-RS}
In the first step the two tori $M^+_{\chiI_l}$ and $M^-_{\chiI_k}$ are
drawn as wedges of which the left and right dashed sides, as well
as top and bottom, are
identified. The map $f_{\mathcal{S}}$ glues the $a$-cycle to $b$ and the 
$b$-cycle to $-a$. This results in the three-manifold $S^3$ with ribbons 
as displayed on the \rhs. Similarly, for $R(\mathcal{T})$ we find
  \begin{eqnarray}  \begin{picture}(220,105)(62,45)
  \put(64,0)     {\includeourbeautifulpicture{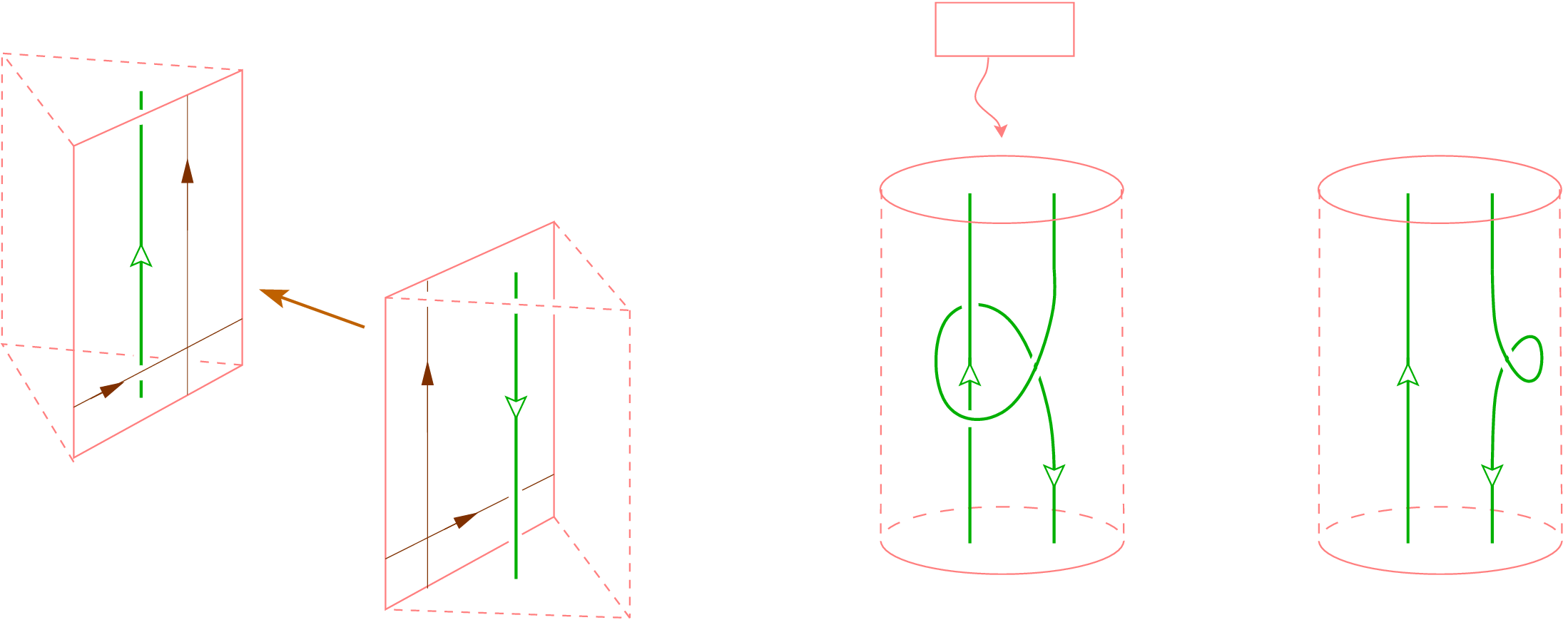}}
  \setlength{\unitlength}{24810sp}
  \setlength{\unitlength}{1pt}
  \put(0,78)       {$R(\mathcal{T})_{kl} \;=$}
  \put(94,109)     {\scriptsize$k$}
  \put(113.3,96)   {\scriptsize$b$}
  \put(118.5,77)   {\scriptsize$a$}
  \put(143.5,83.5) {\scriptsize$f_{\mathcal T}$}
  \put(174.8,48)   {\scriptsize$b$}
  \put(188.5,32.2) {\scriptsize$a$}
  \put(192.1,64)   {\scriptsize$l$}
  \put(252,78)     {$=$}
  \put(305.3,147)  {$S^2{\times}S^1$}
  \put(306.5,35)   {\scriptsize$k$}
  \put(335.7,48)   {\scriptsize$l$}
  \put(375,78)     {$=$}
  \put(418.9,35)   {\scriptsize$k$}
  \put(449.4,48)   {\scriptsize$l$}
  \end{picture} \nonumber\\[3.7em]{}
  && \qquad =\, \delta_{k,l}^{} \theta_l^{-1} \,=\, \hat T_{kl} \,.
  \label{eq:app-RT} \\[-1.3em]{}\nonumber\end{eqnarray}
The second step is most easily seen by drawing actual ribbons instead
of using the blackboard framing convention. The third step amounts
to taking the $U_l$-ribbon around the horizontal $S^2$ so that
(in this description of $S^2\,{\times}\, S^1$) it no longer wraps around the
$U_k$-ribbon.

\medskip

Finally we must compute the two Maslov indices 
  \be
  \mu_{\mathcal{S}} = 
  \mu\big((f_{\mathcal{S}}{\circ}f)_*^{} \lambda_T, 
    (f_{\mathcal{S}})_*^{} \lambda_T,\,\lambda_T\big)
  \qquad {\rm and} \qquad
  \mu_{\mathcal{T}} =
  \mu\big((f_{\mathcal{T}}{\circ}f)_*^{} \lambda_T, 
    (f_{\mathcal{T}})_* \lambda_T,\,\lambda_T\big) \,, \ee
where $f$ is taken to induce the action $\ppmatrix\alpha\beta\gamma\delta$ 
on $H_1(T,\reals)$. We obtain
  \be
  (f_{\mathcal{S}} \cir f)_*^{} \lambda_T = 
  \pmatrix{0&-1\cr1&0} \pmatrix{\alpha&\beta\cr\gamma&\delta} 
  \pmatrix{1\cr0} = \pmatrix{-\gamma\cr\alpha} .  \ee
Similarly,
  \be
  (f_{\mathcal{T}} \cir f)_*^{} \lambda_T = \pmatrix{\alpha{+}\gamma\cr\gamma}
  , \qquad
  (f_{\mathcal{S}})_*^{} \lambda_T = \pmatrix{0\cr1}
  , \qquad
  (f_{\mathcal{T}})_*^{} \lambda_T = \pmatrix{1\cr0} .  \ee
In these formulas, as well as in the sequel, we abuse notation by using a 
vector to describe both an element in $H_1(T,\reals)$ 
and, via its linear span, an element of $\Lambda(T)$.
Using the result \erf{eq:maslov-torus} we find
  \bea
  \mu_{\mathcal{S}} = \mu(-\gamma a{+}\alpha b,  b, a) =
  {\rm sign}\big( \omega( -\gamma a {+} \alpha b,b)
  \omega(-\gamma a {+} \alpha b,a) \omega(b,a) \big)
  = - {\rm sign}(\alpha\gamma) \,, \\{}\\[-.6em]
  \mu_{\mathcal{T}} = \mu( (\alpha{+}\gamma)a{+}\gamma c, a, a) = 0 \,.
  \eear\labl{eq:app-muST}
Evaluating \erf{eq:app-rsum} for $f_1\eq f_{\mathcal{S},\mathcal{T}}$ and
substituting \erf{eq:app-RS}, \erf{eq:app-RT} as well as
\erf{eq:app-muST} results in the relations \erf{eq:rec-rel}.

\subsection{More on reversions in the Ising model}\label{sec:ising-star} 

In this appendix we give some details on how to find the reversions on
the algebra $(\one{\oplus}\Eps)^\vee{\otimes}\, (\one{\oplus}\Eps)$ in step 3b) 
of our prescription for the classification of reversions. We will then see 
how these reversions are related to the ones on $A\eq\one{\oplus}\Eps$.

\medskip

Set $X\,{:=}\,\one{\oplus}\Eps$. For step 3b) we would like to find 
reversions on $B\,{:=}\,X^\vee{\otimes}X$ that act on a basis of 
$B\top$ via permutation of the two basis vectors. Define the morphisms
  \bea  \begin{picture}(280,60)(0,24)
  \put(49,0)     {\includeourbeautifulpicture{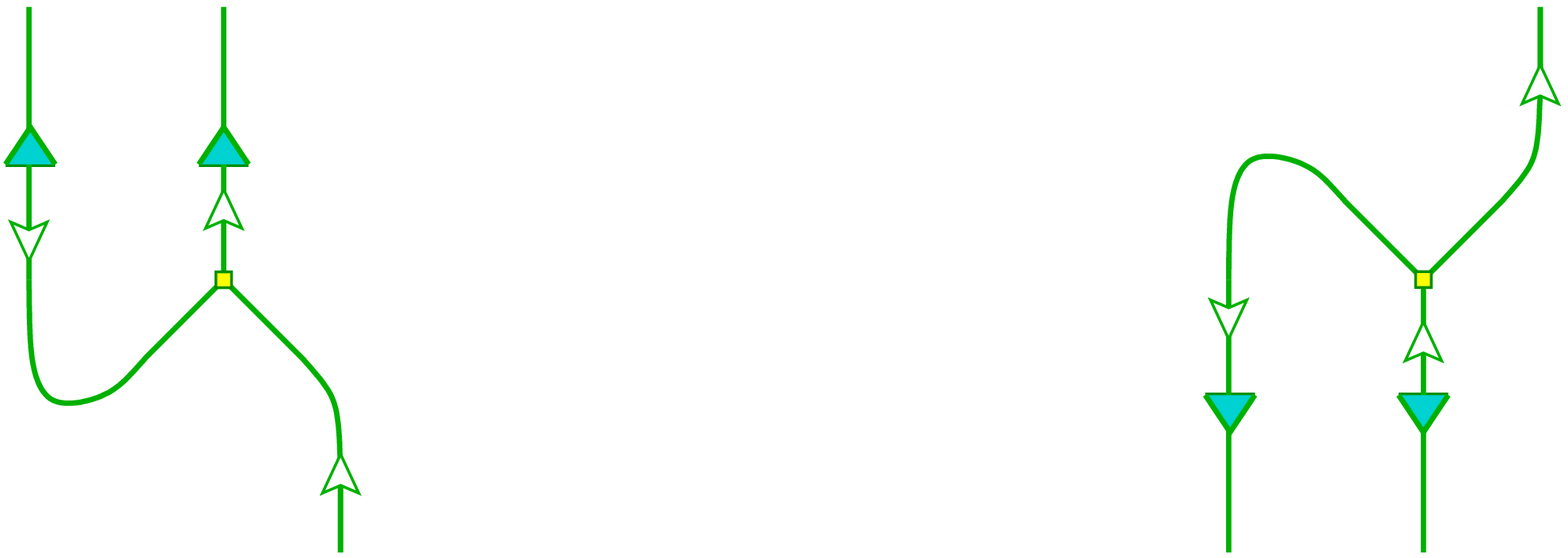}}
  \setlength{\unitlength}{24810sp}
  \setlength{\unitlength}{1pt}
  \put(0,34)       {$e_{ab} \;:=$}
  \put(48.5,79.8)  {\scriptsize$X^\vee$}
  \put(74.5,26.5)  {\scriptsize$a$}
  \put(77.2,79.8)  {\scriptsize$X$}
  \put(82.5,40.2)  {\scriptsize$b$}
  \put(90.2,-8.8)  {\scriptsize$[ab]$}
  \put(167,34)     {$\bar e_{ab} \;:=$}
  \put(214.2,-8.8) {\scriptsize$X^\vee$}
  \put(232.3,42.5) {\scriptsize$a$}
  \put(242.6,-8.8) {\scriptsize$X$}
  \put(257.2,42.3) {\scriptsize$b$}
  \put(256.3,80.2) {\scriptsize$[ab]$}
  \epicture12 \labl{eq:is3b-XXbas}
where $a,b,\iN\{\one,\Eps\}$ and $[ab]\,{\cong}\,a{\otimes}b$ is a
short-hand standing for $[\one\one]\eq[\Eps\Eps]\eq\one$ and
$[\one\Eps]\eq[\Eps\one]\eq\Eps$.  
We introduce the alternative labelling
  \be
  e^\one_1 := e_{\one\one} \,,\quad\
  e^\one_2 := e_{\Eps\Eps} \,,\quad\
  e^\Eps_1 := e_{\one\Eps} \,,\quad\
  e^\Eps_2 := e_{\Eps\one} \,,  \labl{newbas}
and similarly for $\bar e^a_\alpha$ and $\bar e_{ab}$. Note that as an 
object $B\cong \one{\oplus}\one{\oplus}\Eps{\oplus}\Eps$.
With the labelling above the morphisms $e^a_\alpha$ 
thus provide bases for the spaces $\Hom(a,B)$ and
$\bar e^a_\alpha$ for $\Hom(B,a)$. One can check that
the bases $e_\alpha^a$ and $\bar e_\alpha^a$ are dual to each other.

Recall that the multiplication on $X^\vee{\otimes}X$ is
given by \erf{eq:B-ssFA}  (with $A$ set to $\one$ and $M$ set to $X$). 
Thus to work out the multiplication on $B$ in the basis \erf{newbas}, 
following (I:3.7) we must compute the constants
$m^{~~c\gamma}_{a\alpha,b\beta}$ in
  \be
  \bar e^c_\gamma \circ m \circ (e^a_\alpha \oti e^b_\beta)  
  = m^{~~c\gamma}_{a\alpha,b\beta} ~ \lambda_{(ab)c} \,, \ee
where $\lambda_{(ab)c}$ refers to the basis element chosen in $\Hom(a{\otimes}
b,c)$ as in (I:2.29). With $\bar\beta\,{\equiv}\,3{-}\beta$ this yields
  \be
  m^{~~c\gamma}_{a\alpha,b\beta} = N_{ab}^{~c} \cdot \cases{
  \delta_{\alpha\gamma} \delta_{\alpha\beta} & for $\;a\eq\one \,,$ \cr
  \delta_{\alpha\gamma} \delta_{\alpha\bar\beta} & for $\;a\eq\Eps \,.$ }
  \labl{eq:is3b-mXX}
Next, the reversion evaluated in a basis takes the form \erf{eq:*bas},
  \be
  \bar e^a_\beta \circ \sigma \circ e^a_\alpha = 
  \sigma(a)_\alpha^{~\beta}\, \id_{U_a} \,.  \ee
The matrices $\sigma(a)_\alpha^{~\beta}$ are required to obey the 
constraints \erf{eq:*ssFA-bas}. Thus we must find solutions
to the $2{\times}2$ matrix equations
  \be
  \sigma(\one)\, \sigma(\one) = \mbox{\small $1\!\!$}1_{2{\times}2} \,,
  \qquad
  \sigma(\Eps)\, \sigma(\Eps) = - \mbox{\small $1\!\!$}1_{2{\times}2}
  \labl{eq:is3b-theta}
as well as the conditions
  \be
  \sum_{\rho=1,2} m^{~~c\rho}_{a\alpha,b\beta}
  \,\sigma(c)_\rho^{~\gamma}
  = \sum_{\mu,\nu=1,2} \sigma(a)_\alpha^{~\mu}\, \sigma(a)_\beta^{~\nu}
  \,m^{~~c\gamma}_{b\nu,a\mu} \,\Rs{}abc \,.  \labl{eq:is3b-*2}
Step 3b) of the classification algorithm demands $\sigma(\one)$ to act as
a permutation; we thus set
  \be
  \sigma(\one)_\alpha^{~\beta} = \delta_{\beta,\bar\alpha} \,.
  \labl{eq:is3b-perm}
In order to solve \erf{eq:is3b-*2} we evaluate the condition
for all choices $a,b,c$ allowed by fusion. A short calculation
shows that with \erf{eq:is3b-perm} and \erf{eq:is3b-mXX},
the condition \erf{eq:is3b-*2} is identically fulfilled for
$(ab)c\eq (\one\one)\one$, while the other cases give the
following restrictions:
  \be\bearll
  (A) \qquad (ab)c = (\one\Eps)\Eps \;\   &\Longrightarrow \;\   
  \delta_{\alpha\beta} \, \sigma(\Eps)_\beta^{~\gamma} 
  = \delta_{\alpha\gamma} \, \sigma(\Eps)_\beta^{~\alpha} \,, \\[5pt]
  (B) \qquad (ab)c = (\Eps\one)\Eps \;\   &\Longrightarrow \;\   
  \delta_{\alpha\bar\beta} \, \sigma(\Eps)_\alpha^{~\gamma}
  = \delta_{\gamma\bar\beta} \, \sigma(\Eps)_\alpha^{~\bar\beta} \,, \\[5pt] 
  (C) \qquad (ab)c = (\Eps\Eps)\one \;\   &\Longrightarrow \;\   
  \delta_{\alpha\bar\beta} \delta_{\alpha\bar\gamma} =
  - \sigma(\Eps)_\alpha^{~\bar\gamma} \, \sigma(\Eps)_\beta^{~\gamma} 
  \,. \eear\ee
{}From $(A)$ the matrix $\sigma(\Eps)$ can be deduced to be
diagonal, $\sigma(\Eps)_\alpha^{~\beta}\eq\delta_{\alpha\beta}\,
\sigma(\Eps)_\alpha^{~\alpha}$.
Then condition $(B)$ is identically fulfilled, while condition $(C)$ is
equivalent to $\sigma(\Eps)_\alpha^{~\alpha}\,
\sigma(\Eps)_{\bar\alpha}^{~\bar\alpha}\eq{-}1$.
Thus $(A)$\,--\,$(C)$ are equivalent to $\sigma(\Eps)$ being of the form
  \be
  \sigma(\Eps) = \ppmatrix a00{-a^{-1}} .  \ee
Relation \erf{eq:is3b-theta} restricts this further to $a\eq{\pm}\ii$. 
Thus altogether we find two possible reversions on $B$, given by
$\sigma^{(\pm)}(\one)_{\alpha}^{~\beta}\eq\delta_{\beta\bar\alpha}$ and
  \be
  \sigma^{(+)}(\Eps) = \ppmatrix \ii 00\ii \,, \qquad
  \sigma^{(-)}(\Eps) = \ppmatrix{-\ii}00{-\ii}  \,.  \labl{eq:is3b-2stars}

It turns out that both of these reversions are related to
the reversion on $A$ given in \erf{eq:A-star-se} 
with $s_\Eps\eq{-}1$ via propostion \ref{pr:B-star}. Let us see
how this comes about.

We are going to apply propostion \ref{pr:B-star} for the
choices $A\eq\sigma^\vee{\otimes}\sigma$, $M\eq\sigma^\vee{\otimes}X$ 
and $B\eq X^\vee{\otimes}X$. The action of $A$ on $M$ is defined as
  \be  
  \rho := \id_{\sigma^\vee} \oti \tilde d_\sigma \oti \id_X
  \in \Hom(\sigma^\vee\!{\otimes}\sigma{\otimes}\sigma^\vee\!{\otimes}X,
  \sigma^\vee\!{\otimes}X) \,.  \labl{eq:is3b-rhoA} 
Let us first check that indeed $B$ is isomorphic to $M^\vee\OtA M$. 
To this end we introduce morphisms $e\iN\Hom(B,M^\vee{\otimes}M)$
and $r\iN\Hom(M^\vee{\otimes}M,B)$ by
  \be
  e := \id_{X^\vee} \oti b_\sigma \oti\id_X \quad , \quad
  r := \tfrac1{\dim(\sigma)} \,  \id_{X^\vee} \oti
    \tilde d_\sigma \oti \id_X \,.  \ee
One verifies that $e$ and $r$ are indeed embedding and restriction 
morphisms as drawn in \erf{eq:pic50}, i.e.\ satisfy
$r\cir e\eq\id_{B}$ and $e \cir r\eq P_{\otimes A}$ as in \erf{eq:Ptensor}. 

For the following calculations we need to introduce two bases 
in addition to \erf{eq:is3b-XXbas}, one for the morphism spaces
$\Hom(U_a,A)$ and one for $\Hom(\sigma,\M)$, together with their duals. 
We choose 
  \bea  \begin{picture}(280,155)(0,24)
     \put(0,97) {\begin{picture}(0,0)
  \put(44,0)     {\includeourbeautifulpicture{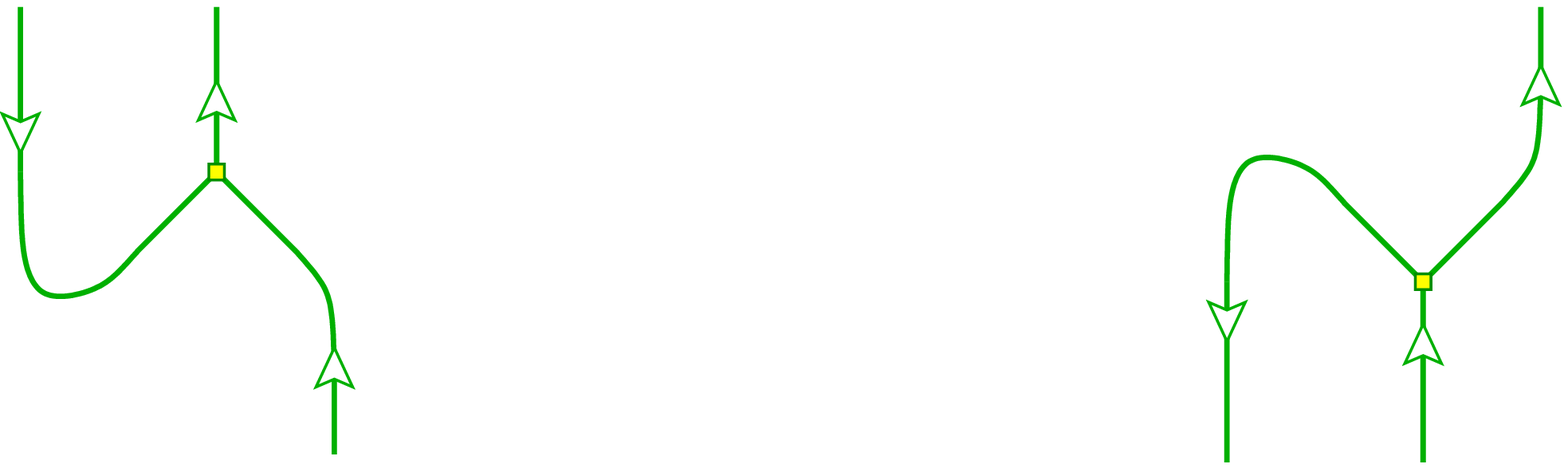}}
  \setlength{\unitlength}{24810sp}
  \setlength{\unitlength}{1pt}
  \put(0,31)       {$\alpha_a \;:=$}
  \put(43.3,66.5)  {\scriptsize$\sigma^{\!\vee}$}
  \put(71.9,66.5)  {\scriptsize$\sigma$}
  \put(86.5,-7.8)  {\scriptsize$U_a$}
  \put(144,31)     {$\bar\alpha_a \;:=\; \dsty\frac1{\sqrt2}$}
  \put(209.2,-8.5) {\scriptsize$\sigma^{\!\vee}$}
  \put(237.6,-7.6) {\scriptsize$\sigma$}
  \put(252.8,66.9) {\scriptsize$U_a$}
      \end{picture}} \put(0,0) {\begin{picture}(0,0)
  \put(49,0)     {\includeourbeautifulpicture{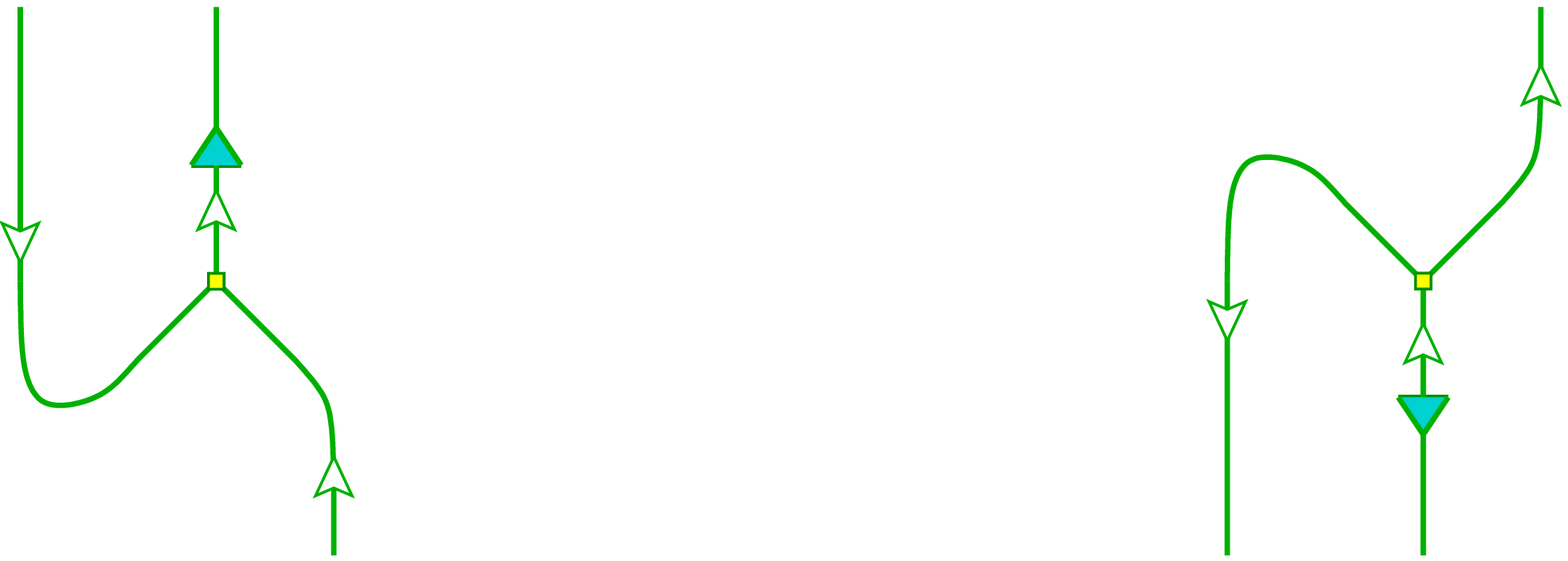}}
  \setlength{\unitlength}{24810sp}
  \setlength{\unitlength}{1pt}
  \put(0,35)       {$f_k \;:=$}
  \put(48.5,80.2)  {\scriptsize$\sigma^{\!\vee}$}
  \put(75.4,79.9)  {\scriptsize$X$}
  \put(81.3,39.7)  {\scriptsize$U_k$}
  \put(92.6,-7.8)  {\scriptsize$\sigma$}
  \put(147,35)     {$\bar f_k \;:=\; \sqrt2$}
  \put(214.7,-8.5) {\scriptsize$\sigma^{\!\vee}$}
  \put(241.5,-8.5) {\scriptsize$X$}
  \put(247.1,32.3) {\scriptsize$U_k$}
  \put(259.6,80)   {\scriptsize$\sigma$}
      \end{picture}} 
  \epicture15 \labl{pic130/131}
Note that we have $m_{\Eps\Eps}^\one\eq\alpha_\one \cir m_A \cir 
(\alpha_\Eps{\otimes}\alpha_\Eps)\eq2$,
i.e.\ the basis $\alpha_a$ is normalised slightly different from
\erf{eq:Is-Amult}. 

Next we would like to describe all module morphisms $g\iN\HomA(M,M^\sigma)$. 
Since these form a subspace of $\Hom(\M,\M^\vee)$ we can expand $g$ as
  \be
  g = \sum_{k,l\in\{\one,\Eps\}} G_{kl} \;
  (\lambda_{(\sigma\sigma)\one} \oti (\bar f_l)^\vee) 
  \cir (\bar f_k \oti b_\sigma) \,.  \labl{eq:is3b-G-expand}
For $g$ to be an intertwiner for the $A$-action we need
  \be
  (f_l)^\vee \circ
  g \circ \rho \circ (\alpha_k \otimes f_j)  = 
  (f_l)^\vee \circ \rho^\sigma \circ (\id_A \otimes g)
  \circ (\alpha_k \otimes f_j) 
  \labl{eq:is3b-gA-cond}
to hold for all values of $j,k,l$. Here $\rho$ is given by 
\erf{eq:is3b-rhoA} and $\rho^\sigma$ is defined in \erf{eq:rho-sigma}. 
A short calculation shows that \erf{eq:is3b-gA-cond} is equivalent to
  \be
  G_{jl} \, [ (-1)^{\delta_{k,\Eps} \delta_{j,\Eps}} -
  s_k\, (-1)^{\delta_{k,\Eps}\delta_{l,\Eps}} ] = 0 
  \qquad {\rm for~all}\quad j,k,l\iN\{\one,\Eps\} \,, \ee
where $s_k$ characterises the reversion $\sigma$ on $A$, i.e.\
$s_\one\eq1$ and $s_\Eps\eq{\pm}1$ as in \erf{eq:A-star-se}. Thus every
$g \iN \HomA(M,M^\sigma)$ can be written as in \erf{eq:is3b-G-expand} with
  \be
  G = \ppmatrix a00b {\rm ~~for~}\; s_\Eps = 1
  \qquad {\rm and} \qquad
  G = \ppmatrix 0ab0 {\rm ~~for~}\; s_\Eps = -1 \,, \ee
respectively, for some $a,b\iN\complex$.
In order to apply proposition \ref{pr:B-star} the morphism $g$ has
to fulfill \erf{eq:g-prop} for some $\eps_g$. In terms of the 
matrix $G$ one finds that this is equivalent to $G\eq\eps_g G^{\rm t}$.
This allows for three classes of solutions:
  \be\bearll
  s_\Eps=1\,:& ~~ G = \ppmatrix a00b \,,~~\eps_g\eq1 \,,\\[7pt] 
  s_\Eps=-1\,:& ~~
    G = \ppmatrix 0aa0 \,,~~\eps_g\eq1 
    \qquad {\rm or} \qquad
    G = \ppmatrix 0a{-a}0 \,,~~ \eps_g\eq{-}1 \,.
\eear\labl{eq:is3b-3G}
We have now fulfilled the conditions to apply proposition
\ref{pr:B-star} and compute the reversion $\tilde\sigma_g$ 
resulting from these three classes of $g$. In order to
do so we need the inverse of the morphism $g$.  We can expand 
$g^{-1}\iN\Hom(M^\vee,M)$ in a basis similar to \erf{eq:is3b-G-expand},
  \be
  g^{-1} = \sum_{k,l} \tilde G_{k,l} \, (d_\sigma \oti f_l) \circ
  ((f_k)^\vee \oti \Upsilon^{(\sigma\sigma)\one}) \,, \ee
with $\Upsilon^{(\sigma\sigma)\one} \iN \Hom(\one,\sigma{\otimes}\sigma)$ 
the basis element dual to $\lambda_{(\sigma\sigma)\one}$, as in (I:2.29).
The matrix $\tilde G$ is inverse to $G$ up
to a constant, $\tilde G\eq\sqrt{2}\,G^{-1}$.

The morphism $\tilde\sigma_g$ in \erf{eq:B-star} 
turns out to be independent of the actual
values of $a,b \iN \complex^\times$ in \erf{eq:is3b-3G},
so that we may choose $a,b\eq1$. With this choice we
obtain $\tilde G_{kl}\eq\sqrt{2}\eps_gG_{kl}$, for
all three cases in \erf{eq:is3b-3G}.
Evaluating \erf{eq:B-star}, with some effort one finds
  \be\bearll
  \bar e_{\one\one} \circ \tilde\sigma_g \circ e_{\one\one}
  = (g_{\one\one})^2 \, , \qquad &
  \bar e_{\one\one} \circ \tilde\sigma_g \circ e_{\Eps\Eps}
  = (g_{\Eps\one})^2 \, , \\[5pt]
  \bar e_{\Eps\Eps} \circ \tilde\sigma_g \circ e_{\one\one}
  = (g_{\one\Eps})^2 \, , \qquad &
  \bar e_{\Eps\Eps} \circ \tilde\sigma_g \circ e_{\Eps\Eps}
  = (g_{\Eps\Eps})^2 
  \eear\ee
for the elements $\tilde\sigma_g(\one)_{\alpha}^{~\beta}$ and 
  \be\bearll
  \bar e_{\one\Eps} \circ \tilde\sigma_g \circ e_{\one\Eps}
  =  -\ii\,g_{\one\Eps}\,g_{\Eps\one}    \, , \qquad &
  \bar e_{\one\Eps} \circ \tilde\sigma_g \circ e_{\Eps\one}
  =  -2\ii\,g_{\one\one}\,g_{\Eps\Eps}  \, ,  \\[5pt]
  \bar e_{\Eps\one} \circ \tilde\sigma_g \circ e_{\one\Eps}
  =  -\tfrac{\ii}{2}\,g_{\one\one}\,g_{\Eps\Eps}    \, , \qquad &
  \bar e_{\Eps\one} \circ \tilde\sigma_g \circ e_{\Eps\one}
  =  -\ii\,g_{\one\Eps}\,g_{\Eps\one}  
  \eear\ee
for the elements $\tilde\sigma_g(\eps)_{\alpha}^{~\beta}$.
Substituting the three possible forms of $g$ in \erf{eq:is3b-3G},
one finds that only the two cases with $s_\eps\eq{-}1$ act
as a permutation on $B\top$. Moreover, in these two
cases the matrix $\tilde\sigma_g(\eps)_{\alpha}^{~\beta}$ 
reproduces precisely the two choices in \erf{eq:is3b-2stars}. 

In the case of the Ising model,
the two reversions obtained in step 3b) are thus not
new, in the sense that via proposition \ref{pr:B-star}
they are both related to the reversion \erf{eq:A-star-se} 
on $A$ with $s_\Eps\eq{-}1$.

 \newpage
 
\newcommand\wb{\,\linebreak[0]} \def\wB {$\,$\wb}
 \newcommand\Bi       {\bibitem}
 \renewcommand\J[5]     {{\sl #5\/}, {#1} {#2} ({#3}) {#4} }
 \renewcommand\K[6]     {{\sl #6\/}, {#1} {#2} ({#3}) {#4}}
 \newcommand\PhD[2]   {{\sl #2\/}, Ph.D.\ thesis (#1)}
 \newcommand\Prep[2]  {{\sl #2\/}, pre\-print {#1}}
 \newcommand\BOOK[4]  {{\sl #1\/} ({#2}, {#3} {#4})}
 \newcommand\inBO[7]  {{\sl #7\/}, in:\ {\sl #1}, {#2}\ ({#3}, {#4} {#5}), p.\ {#6}}
 \newcommand\iNBO[7]  {{\sl #7\/}, in:\ {\sl #1} ({#3}, {#4} {#5}) }
 \newcommand\webb[2]  {{\sl #2\/}, available at http:/$\!$/#1}
 \newcommand\webp[2]  {{\sl #2\/}, available at\\ http:/$\!$/#1}
 \newcommand\Erra[3]  {\,[{\em ibid.}\ {#1} ({#2}) {#3}, {\em Erratum}]}
 \def\jf    {J.\ Fuchs}
 \def\dim   {dimension}  
 \def\comp  {Com\-mun.\wb Math.\wb Phys.}
 \def\cpma  {Com\-pos.\wb Math.}
 \def\fiic  {Fields\wB Institute\wB Commun.}
 \def\foph  {Fortschritte\wB d.\wb Phys.}
 \def\ijmp  {Int.\wb J.\wb Mod.\wb Phys.\ A}
 \def\jams  {J.\wb Amer.\wb Math.\wb Soc.}
 \def\jgap  {J.\wb Geom.\wB and\wB Phys.}
 \def\jhep  {J.\wb High\wB Energy\wB Phys.}
 \def\jomp  {J.\wb Math.\wb Phys.}
 \def\jopa  {J.\wb Phys.\ A}
 \def\josp  {J.\wb Stat.\wb Phys.}
 \def\mpla  {Mod.\wb Phys.\wb Lett.\ A}
 \newcommand\nqma[2] {\inBO{Non-perturbative QFT Methods and Their Applications}
            {Z.\ Horv\'ath and L.\ Palla, eds.} \WS\Si{2001} {{#1}}{{#2}} }
 \newcommand\nqft[2] {\inBO{
            Nonperturbative \QFT} {G.\ 't Hooft, A.\ Jaffe, G.\ Mack, P.K.\
            Mitter, and R.\ Stora, eds.} \PL\NY{1988} {{#1}}{{#2}} }
 \def\nuci  {Nuovo\wB Cim.}
 \def\nupb  {Nucl.\wb Phys.\ B}
 \def\phla  {Phys.\wb Lett.\ A}
 \def\phlb  {Phys.\wb Lett.\ B}
 \def\phrl  {Phys.\wb Rev.\wb Lett.}
 \def\phrp  {Phys.\wb Rep.}
 \def\phrd  {Phys.\wb Rev.\ D}
 \def\phre  {Phys.\wb Rev.\ E}
 \def\prtp  {Progr.\wb Theor.\wb Phys.}
 \def\trgr  {Trans\-form.\wB Groups}
 \def\AMS    {{American Mathematical Society}}
 \def\PL     {{Plenum Press}}
 \def\SV     {{Sprin\-ger Ver\-lag}}
 \def\WS     {{World Scientific}}
 \def\PR     {{Providence}}
 \def\Si     {{Singapore}}
 \def\NY     {{New York}}

\end{document}